%% file: rank2main.tex
\documentclass[11pt,a4paper]{article}
\pdfoutput=1
\usepackage{jheppub}
\usepackage{amsthm,amsbsy,amsfonts,mathrsfs,enumerate,float,wrapfig,amsmath}
\usepackage[utf8]{inputenc}
\usepackage{subfigure}

\newcommand{\nn}{\nonumber}
\newcommand{\be}{\begin{equation}}
\newcommand{\ee}{\end{equation}}
\newcommand{\bea}{\begin{eqnarray}}
\newcommand{\eea}{\end{eqnarray}}
\newcommand{\bF}{\mathbf{F}}
\newcommand{\bS}{\mathbf{S}}
\newcommand{\AS}{\mathbf{AS}}

\usepackage{tikz}
\usetikzlibrary{positioning}
\usetikzlibrary{arrows}

\title{
Dualities and 5-brane webs for 5d rank 2 SCFTs}
\author[a]{Hirotaka Hayashi,}
\author[b,c]{Sung-Soo Kim,}
\author[d]{Kimyeong Lee,}
\author[e,f]{and Futoshi Yagi}

\affiliation[a]{Department of Physics, School of Science, Tokai University,\\ 4-1-1 Kitakaname, Hiratsuka-shi, Kanagawa 259-1292, Japan}
\affiliation[b]{School of Physics, University of Electronic Science and Technology of China, \\No.4, Section 2, North Jianshe Road, Chengdu 610054, China}
\affiliation[c]{Institute of Fundamental and Frontier Sciences, \\University of Electronic Science and Technology of China, Chengdu 610054, China}
\affiliation[d]{School of Physics, Korea Institute for Advanced Study, \\
85 Hoegi-ro Dongdaemun-gu, Seoul 02455, Korea}
\affiliation[e]{School of Mathematics, Southwest Jiaotong University,\\ 
West zone, High-tech district, Chengdu, Sichuan 611756, China}
\affiliation[f]{Department of Physics, Technion - Israel Institute of Technology,\\ Haifa 32000, Israel}

\emailAdd{h.hayashi@tokai.ac.jp}
\emailAdd{sungsoo.kim@uestc.edu.cn}
\emailAdd{klee@kias.re.kr}
\emailAdd{futoshi\_yagi@swjtu.edu.cn}

\abstract{
We consider Type IIB 5-brane configurations for 5d rank 2 superconformal theories which are classified recently by geometry in \cite{Jefferson:2018irk}. We propose all the 5-brane web diagrams for these rank 2 theories and show dualities between some of different gauge theories with  explicit duality map of  mass parameters and Coulomb branch moduli.   In particular, we explicitly construct 5-brane configurations for $G_2$ gauge theory with six flavors and its dual $Sp(2)$ and $SU(3)$ gauge theories. We also present 5-brane webs for $SU(3)$ theories of Chern-Simons level greater than 5. 
}

\begin{document}
\preprint{
\begin{flushright}
\tt 
%preprint number
KIAS-P18047\\
\end{flushright}
}

\maketitle
%=================================================================

%=================================================================

\input{sec1.tex}

%====================================================================
\bigskip

\input{G2SU3Sp2.tex}

%====================================================================
\bigskip
\input{SU3Sp2.tex}

%====================================================================
\bigskip
\input{Sp23AS.tex}

%====================================================================
\bigskip
\input{SU3CS9.tex}
%
%====================================================================
%\bigskip
\input{concl.tex}

%====================================================================

\acknowledgments
We thank Hee-Cheol Kim and Gabi Zafrir for useful discussions. SSK would like to gratefully acknowledge the Tsinghua Sanya International Mathematics Forum (TSIMF) for hosting the workshop on SCFTs in dimension 6, also like to acknowledge ITP-CAS as well as YMSC at Tsinghua university for kind hospitality for his visit.
SSK is supported by the UESTC Research Grant A03017023801317. KL would like to gratefully acknowledge KITP Santa Barbara, SCGP Stony Brook and GGI Florence for his visit to their workshops where some of this work is done.  KL is supported in part  by the National Research Foundation of Korea Grant NRF-2017R1D1A1B06034369 and also by the National Science Foundation under Grant No. NSF PHY11-25915. FY is supported in part by Israel Science Foundation under Grant No. 352/13 and Grant No. 1390/17, by NSFC grant No. 11501470 and No. 11671328, and by Recruiting Foreign Experts Program No. T2018050 granted by SAFEA.

\bigskip
\appendix 

%====================================================================
%\bigskip
\input{readoffm0.tex}

%====================================================================
\input{app2-1.tex}

\input{app2-2.tex}

%====================================================================
\bigskip
\bigskip
\clearpage
%%%%%%%%%%%%%%%%%%%%%%%%%%%%%%%%%%%%%%%%%%%%%%%%%%
\bibliographystyle{JHEP}
\bibliography{ref}

\end{document}

%% file: sec1.tex
\section{Introduction}\label{sec:introduction}

Higher-dimensional gauge theories are in general not renormalizable as the gauge coupling becomes infinitely strong at high energies. However, those theories may make sense in the ultraviolet (UV) region when they have a non-trivial fixed point in UV. Such a phenomenon has arisen in the context of five-dimensional (5d) gauge theories with eight supercharges. From the field theoretic point of view, a necessary condition for the existence of UV complete 5d theories is that the metric of the Coulomb branch moduli space, which may be computed from the effective prepotential, must be non-negative \cite{Seiberg:1996bd, Intriligator:1997pq}. The condition revealed a possibility of the existence of some UV complete 5d gauge theories and their existence has been also confirmed by explicitly constructing the 5d gauge theories from M-theory compactifications on non-compact Calabi-Yau threefolds \cite{Seiberg:1996bd, Morrison:1996xf, Douglas:1996xp, Intriligator:1997pq} and also from 5-brane webs in type IIB string theory \cite{Aharony:1997ju, Aharony:1997bh}. 

However, it had turned out that 5-brane web diagrams can realize 5d gauge theories that lie  beyond the bound given in \cite{Intriligator:1997pq}. For example, one can add the hypermultiplets in the fundamental representation (flavors) up to $N_f = 2N+4$ for an $SU(N)$ gauge theory \cite{Bergman:2014kza, Hayashi:2015fsa}\footnote{The same conclusion was also obtained from the instanxton operator analysis in \cite{Yonekura:2015ksa, Gaiotto:2015una}.} although the original bound was $N_f = 2N$. Recently, the field theoretic condition for the existence of UV complete 5d gauge theories has been revisited and it was claimed in \cite{Jefferson:2017ahm} that the original condition discussed in \cite{Seiberg:1996bd, Intriligator:1997pq} should be relaxed. Namely, the metric of the Coulomb branch moduli space should  be non-negative only on a ``physical'' Coulomb branch moduli space where the tension of monopole strings is non-negative. The $SU(N)$ gauge theory with $N_f=2N+4$ flavors, which has the 5-brane web realization, indeed satisfies the new criteria. Not only that,  the new condition in fact led to a large class of new UV complete 5d theories in \cite{Jefferson:2017ahm}.  Although the criteria is a necessary condition for the existence of UV complete 5d gauge theories, most of the new rank 2 gauge theories found in \cite{Jefferson:2017ahm} have been also constructed geometrically using M-theory compactifications on non-compact Calabi-Yau threefolds in \cite{Jefferson:2018irk}, which confirms their existence. Furthermore, the geometric construction implies intriguing dualities including 5d $G_2$ gauge theories. For example, the identical physics is described by  the $G_2$ gauge theory with six flavors, the $SU(3)$ gauge with six flavors and the Chern-Simons (CS) level $4$ and the $Sp(2)$ gauge theory with 4 flavors and two hypermultiplets in the antisymmetric representation.

It is then natural to ask if the 5d rank 2 gauge theories constructed by geometries in \cite{Jefferson:2018irk} also admit a realization by 5-brane web diagrams in type IIB string theory. In this paper, we propose 5-brane web diagrams for {\it all} the 5d rank 2 gauge theories whose existence is geometrically confirmed in \cite{Jefferson:2018irk}. In particular, we start with a 5-brane web for the pure $G_2$ gauge theory \cite{Hayashi:2018bkd}, and add more flavors to explicitly construct new 5-brane web diagrams for the $G_2$ gauge theory with six flavors, the $SU(3)$ gauge theory with six flavors and the CS level $4$ and the $Sp(2)$ gauge theory with four flavors and two antisymmetric hypermultiplets. 
 Their 5-brane diagrams have a periodic direction implying a 6d UV fixed point. Furthermore, the dualities among the $G_2$, $SU(3)$ and $Sp(2)$ gauge theories may be understood from S-duality by rotating the 5-brane web diagrams accompanied with Hanany-Witten transitions by moving 7-branes. The explicit realization of the dualities gives us the duality maps for the Coulomb branch moduli and parameters among these three theories.

 It is worth noting that in constructing 5-brane webs for an $SU(3)$ gauge theory with higher CS level, in particular, pure $SU(3)$ gauge theory with the CS level 7 which is dual to pure $G_2$ gauge theory, one may naively attempt to construct the theories with higher CS level by increasing the charge difference of the external 7-branes. In this way, the external 7-brane inevitably collide each other as moving each 7-brane to infinity. Up to the CS level 6, a suitable handling the monodromy cut of 7-branes would give a 5-brane web which the resulting 7-branes are no longer collide and thus can be taken to infinity. But for an $SU(3)$ theory with CS level 7 or higher, such procedure involving a 7-brane going across another 7-brane often leads to ill-defined 5-brane web such that a 7-brane after going across monodromy cut of other 7-brane bends toward the center of the 5-brane web due to 7-brane charge changes from the monodromy. Hence, though not conventional, a 5-brane construction for pure $SU(3)$ theory with CS level 7 seems best understood from the S-duality of pure $G_2$ gauge theory.

Starting from these 5-brane webs for the $G_2$  theories of various flavors or their dual $SU(3)$ or $Sp(2)$ theories, one can add different choices of hypermultiplets in an appropriate representation     and to build up the tree of  all theories connected by flavor decoupling or adding. In addition, a finer understanding of the 5-brane web for $G_2$ theory without flavor allows us to find the web for $SU(3)$ theory without flavor and CS level 9. These leads to the construction of the 5-brane webs for the whole family 
rank 2 gauge theories constructed in \cite{Jefferson:2018irk}.  A 5-brane web diagram for 
5d $SU(3)$ theory of the CS level $\frac32$ and with one hypermultiplet in the symmetric representation, can be constructed with O7$^+$-and O7$^-$-planes, which reveals a new 5-brane structure for this marginal theory.
From the constructed 5-brane web diagrams we can also obtain the duality map between the $SU(3)$ gauge theory with nine flavors and the CS level $\frac{3}{2}$ and the $Sp(2)$ gauge theory with eight flavors and single antisymmetric hypermultiplet.

The organization of this paper is as follows. In section \ref{sec:G2SU3Sp2}, we construct 5-brane web diagrams for the $G_2$ gauge theories with $0, 2$ and $6$ flavors and see the dualities to an $SU(3)$ or an $Sp(2)$ gauge theory with or without flavors. The duality map is obtained in each case. We also present a web diagram of the $Sp(2)$ gauge theories from the viewpoint of $SO(5)$. In section \ref{sec:SU3Sp2}, we consider deformations from the diagram in section \ref{sec:G2SU3Sp2} and obtain the other $SU(3)$ gauge theories and their dual $Sp(2)$ gauge theories with the duality map. Section \ref{sec:Sp23AS} is devoted for another deformation to the $Sp(2)$ gauge theory with three hypermultiplets in the antisymmetric representation from the viewpoint of $SO(5)$. In section \ref{sec:SU3CS9} we propose a 5-brane web diagram for the pure $SU(3)$ gauge theory with the CS level $9$. We will then summarize our results in section \ref{sec:concl}. Appendix \ref{sec:m0} explains some subtle identification of the inverse of the squared gauge coupling for $SO(5)$ gauge theories with spinors. In appendix \ref{sec:allwebs}, we summarize all the 5-brane webs for rank 2 theories which were constructed using geometries in \cite{Jefferson:2018irk}.

%% file: G2SU3Sp2.tex
\section{$G_2$-$SU(3)$-$Sp(2)$ sequence}\label{sec:G2SU3Sp2}
In this section, we first consider dualities involving $G_2$ gauge theories with flavors. A dual description of a $G_2$ gauge theory is given by an $SU(3)$ gauge theory and/or an $Sp(2)$ gauge theory depending on flavors \cite{Jefferson:2018irk}. We have constructed 5-brane web diagrams for $G_2$ gauge theories in \cite{Hayashi:2018bkd} and here generalize the construction to the case for the $G_2$ gauge theory with six flavors which may have a 6d UV completion. We will see the dualities from the viewpoint of 5-brane web diagrams.

\subsection{Without matter}
\label{sec:pureSU3CS7}

Before considering the duality involving $G_2$ gauge theory with flavors, we start from the case without matter. The pure $G_2$ gauge theory is dual to the pure $SU(3)$ gauge theory with the CS level $7$ \cite{Jefferson:2018irk}. We can also see the duality from the viewpoint of 5-brane webs.

Let us first review a 5-brane web diagram for the pure $G_2$ gauge theory. Two types of the 5-brane web diagram for the pure $G_2$ gauge theory have been proposed in \cite{Hayashi:2018bkd}. In order to see the duality to the pure $SU(3)$ gauge theory with the CS level $7$, it is useful to consider the pure $G_2$ diagram with an $\widetilde{\text{O5}}$-plane.

\begin{figure}
\centering
\subfigure[]{
\includegraphics[width=6cm]{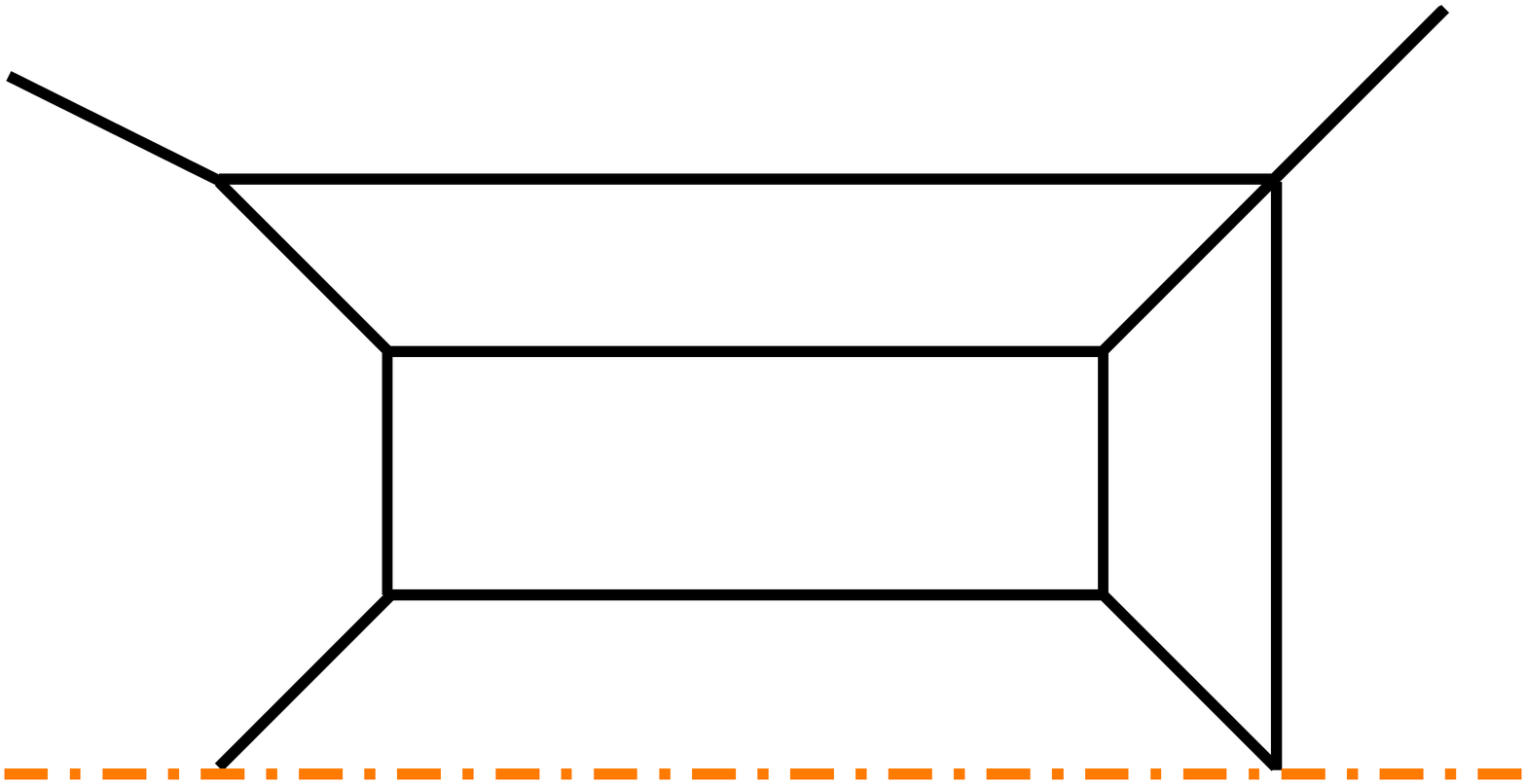} \label{fig:pureG2a}}
\subfigure[]{
\includegraphics[width=6cm]{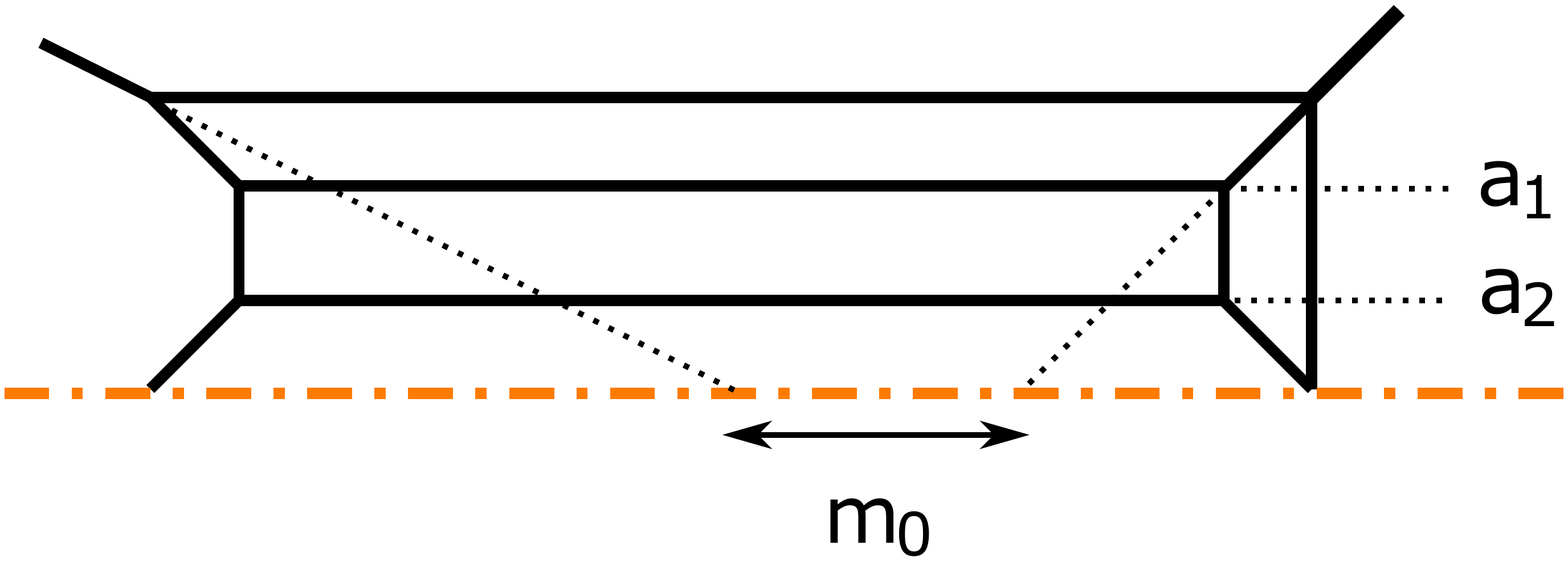} \label{fig:pureG2b}}
\caption{(a): A 5-brane web diagram for the pure $G_2$ gauge theory which is realized with an $\widetilde{\text{O5}}$-plane. The orange line represents the $\widetilde{\text{O5}}$-plane. The convention of the $\widetilde{\text{O5}}$-plane is different from the one used in \cite{Hayashi:2018bkd}. Here, the monodomry branch cut from a fractional D7-brane is extends from left to right whereas it extends from right to left in \cite{Hayashi:2018bkd}. (b): The gauge theory parameterization of the pure $G_2$ diagram.}
\label{fig:pureG2}
\end{figure}

The strategy to realize the 5-brane web diagram for the pure $G_2$ gauge theory was as follows. We first start from a 5-brane web diagram for the $SO(7)$ gauge theory with a hypermultiplet in the spinor representation. Then the Higgsing associated to the spinor matter yields the pure $G_2$ gauge theory at low energies. Hence if we apply the Higgsing procedure to the 5-brane web diagram for the 5-brane web of the $SO(7)$ gauge theory with one spinor, the resulting diagram should be a 5-brane web for the pure $G_2$ gauge theory. The diagram obtained in this way is depicted in Figure \ref{fig:pureG2a}. 
%%%%%%%%%%%%%%%%%%%%%%%%%%%%%%%%%
%%%%%%%%%%%%%%%%%%%%%%%%%%%%%%%%%%
Figure \ref{fig:pureG2b} shows the parameterization for the Coulomb branch moduli $a_1, a_2$ 
and the inverse of the squared gauge coupling $m_0$.

\begin{figure}
\centering
\includegraphics[width=8cm]{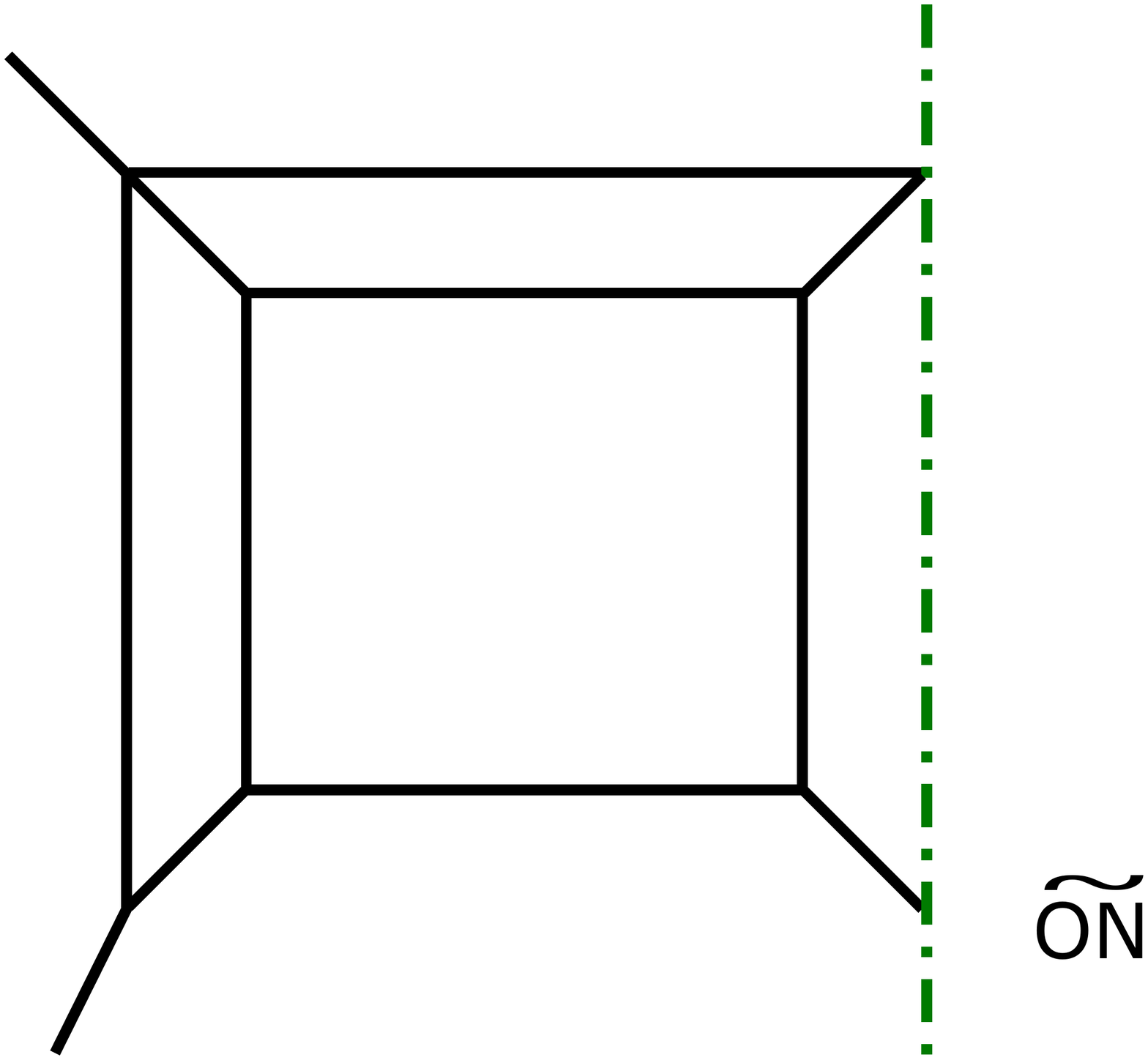}
\caption{A 5-brane web diagram for the pure $SU(3)$ gauge theory with the CS level $7$. The green line represents an $\widetilde{\text{ON}}$-plane.}
\label{fig:pureSU3CS7}
\end{figure}

From the 5-brane web for the pure $G_2$ gauge theory in Figure \ref{fig:pureG2a}, we can see the duality to the pure $SU(3)$ gauge theory with the CS level $7$. To see the duality, we first take  the S-duality which corresponds to the $\frac{\pi}{2}$ rotation for the diagram in Figure \ref{fig:pureG2a}. In terms of geometry it corresponds to the fiber-base duality \cite{Katz:1997eq, Aharony:1997bh, Bao:2011rc, Jefferson:2018irk}. Application of the $\frac{\pi}{2}$ rotation to the diagram in Figure \ref{fig:pureG2a} leads to a 5-brane web in Figure \ref{fig:pureSU3CS7}, where we have postulated the S-dual object of an $\widetilde{O5}^{\pm}$-plane as an $\widetilde{ON}^{\pm}$-plane  \cite{Kutasov:1995te, Sen:1996na, Sen:1998rg, Sen:1998ii, Kapustin:1998fa, Hanany:1999sj}. We claim that this 5-brane web in Figure \ref{fig:pureSU3CS7} represents pure $SU(3)$ gauge theory with Chern-Simons level 7. We justify this claim by comparing the area of the compact faces of the web diagram with the effective prepotential or the tension of monopole string. This claim can also be justified from the decoupling of two flavors from 5-brane description for the $SU(3)$ gauge theory with CS level 6 and two flavors, which has a clear 5-brane interpretation as a Higgsing of a quiver description $SU(2)-SU(3)_3-SU(2)$. We will discuss more detail in  section \ref{sec:G2wmatter}.   

%%%%%%%%%%%%%%%%%%%%%%%%%%%%%%%%%
%%%%%%%%%%%%%%%%%%%%%%%%%%%%%%%%%

\begin{figure}
\centering
\subfigure[]{
\includegraphics[width=6cm]{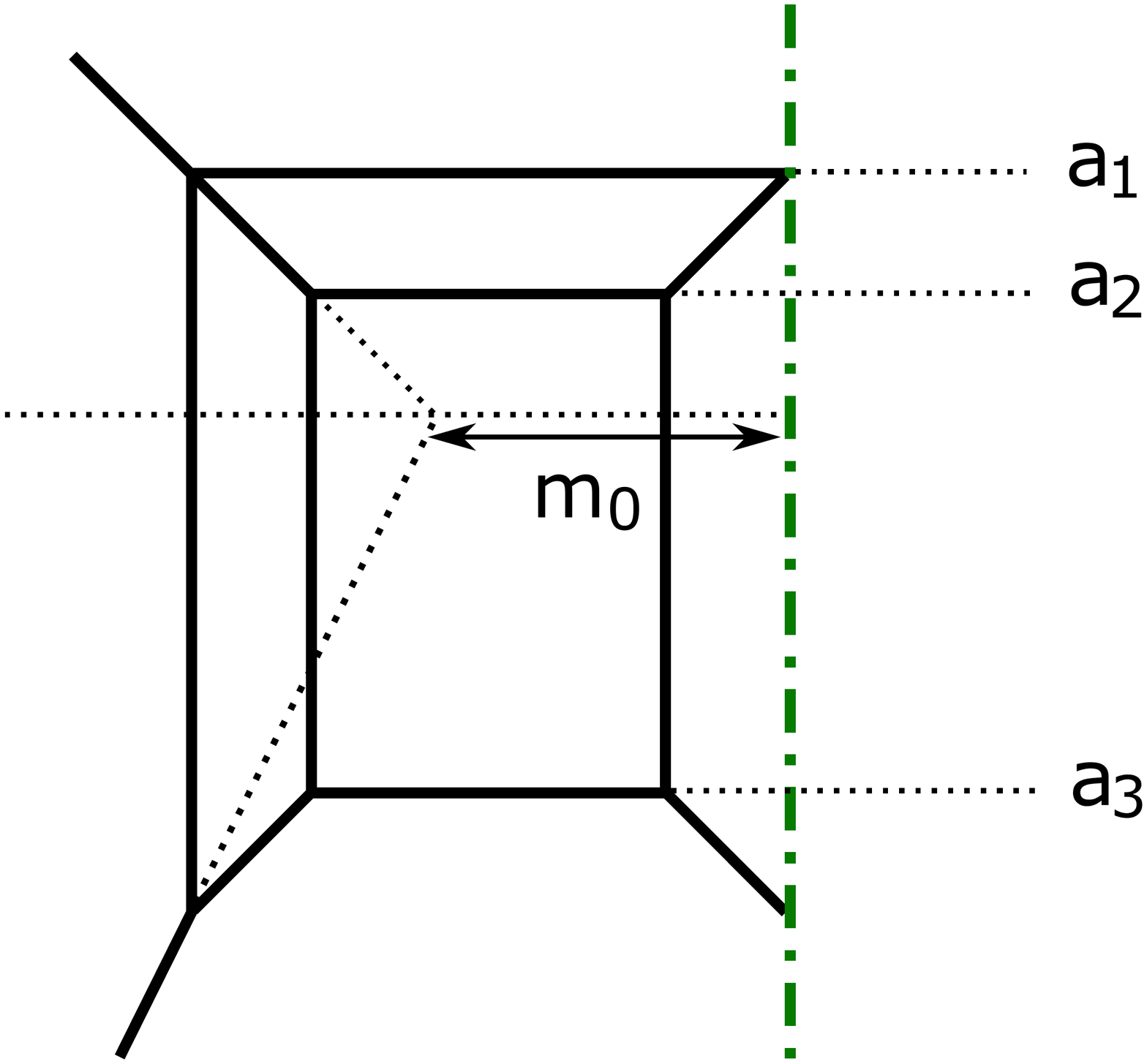} \label{fig:pureSU3CS7a}}
\subfigure[]{
\includegraphics[width=6cm]{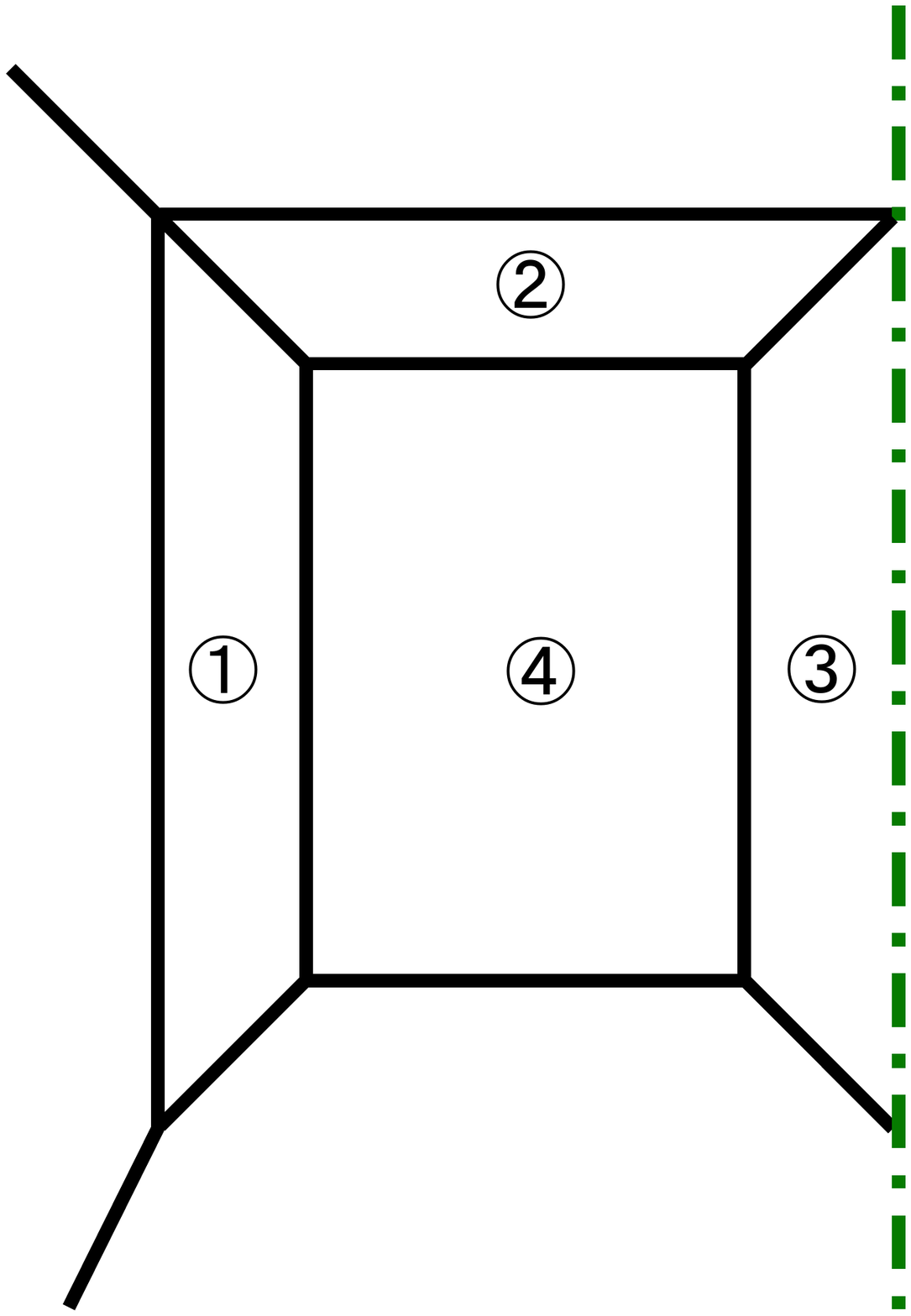} \label{fig:pureSU3CS7b}}
\caption{(a): The gauge theory parameterization for the 5-brane web diagram of the pure $SU(3)$ gauge theory in Figure \ref{fig:pureSU3CS7}. $a_1, a_2, a_3$ are the Coulomb branch moduli of $SU(3)$. The dotted line in the center is the location of the origin in the vertical direction. $m_0$ is the inverse of the squared gauge coupling. (b): A labeling for the faces in the 5-brane web in Figure \ref{fig:pureSU3CS7}.}
\label{fig:pureSU3CS7para}
\end{figure}

We now compute the area from the 5-brane web in Figure \ref{fig:pureSU3CS7} and compare it with 
 the effective prepotential or the tension of the monopole string for the pure $SU(3)$ gauge theory from the diagram. For that we first assign gauge theory parameters to the diagram as in Figure \ref{fig:pureSU3CS7a}. $a_1, a_2, a_3\; (a_1 + a_2 + a_3 = 0)$ are the Coulomb branch moduli and $m_0$ is the inverse of the squared gauge coupling. 
%%%%%%%%%%%%%%%%%%%%%%%%%%%%%%%%%
%%%%%%%%%%%%%%%%%%%%%%%%%%%%%%%%%
Then the area of the faces in Figure \ref{fig:pureSU3CS7b} becomes
\bea
\textcircled{\scriptsize 1} &=& (a_1 - a_2)(a_1 - a_3), \label{area1.pureSU3CS7}\\
\textcircled{\scriptsize 2} &=& (a_1-a_2)(m_0 + a_2), \label{area2.pureSU3CS7}\\
\textcircled{\scriptsize 3} &=&(a_1 - a_2)(a_1 - a_3), \label{area3.pureSU3CS7}\\
\textcircled{\scriptsize 4} &=& (a_2 - a_3)(m_0 - a_1 + 2a_2).\label{area4.pureSU3CS7}
\eea

We then compare the area \eqref{area1.pureSU3CS7}-\eqref{area4.pureSU3CS7} with the effective prepotential of the pure $SU(3)$ gauge theory with the CS level $7$. In general, the effective prepotential of a 5d gauge theory with a gauge group $G$ and matter is given by a cubic function of the Coulomb branch moduli $\phi_i$ \cite{Seiberg:1996qx, Morrison:1996xf, Intriligator:1997pq}
\bea
\mathcal{F}(\phi) = \frac{1}{2}m_0h_{ij}\phi_i\phi_j + \frac{\kappa}{6}d_{ijk}\phi_i\phi_j\phi_k + \frac{1}{12}\left(\sum_{r\in\text{roots}}\left|r \cdot \phi\right|^3 - \sum_f\sum_{w \in R_f}\left|w\cdot\phi - m_f\right|^3\right),\nn\\\label{prepotential}
\eea
where $m_0$ is the inverse of the squared gauge coupling, $\kappa$ is the classical Chern-Simons level and $m_f$ is the mass parameter for hypermultiplets in the representation $R_f$. We also defined $h_{ij} = \text{Tr}(T_iT_j)$ and $d_{ijk} = \frac{1}{2}\text{Tr}\left(T_i\{T_j, T_k\}\right)$ where $T_i$ are the Cartan generators of the Lie algebra associated to $G$. Then the effective prepotential for the pure $SU(3)$ gauge theory with the CS level $7$ becomes
\bea
\mathcal{F}_{SU(3)_7} &=& \frac{m_0}{2}(a_1^2 + a_2^2 + a_3^2) + \frac{1}{6}\left((a_1 - a_2)^3 + (a_1 - a_3)^3 + (a_2 - a_3)^3\right) + \frac{7}{6}\left(a_1^3 + a_2^3 + a_3^3\right)\nn\\
&=&m_0\left(\phi_1^2 -\phi_1\phi_2 + \phi_2^2\right) + \frac{4}{3}\phi_1^3 + 3\phi_1^2\phi_2 - 4\phi_1\phi_2^2 + \frac{4}{3}\phi_2^3, \label{prepot.SU3CS7}
\eea
where we changed the basis of the Coulomb branch moduli into the Dynkin basis by using the relation
\be
a_1 = \phi_1, \quad a_2 = -\phi_1 + \phi_2, \quad a_3 = -\phi_2, \label{SU3.Dynkinbasis}
\ee
in \eqref{prepot.SU3CS7}. The tension of the monopole string of the pure $SU(3)$ gauge theory with the CS level $7$ is given by taking a derivative of \eqref{prepot.SU3CS7} with respect to the Coulomb branch moduli $\phi_1, \phi_2$. Hence the tension is given by
\bea
\frac{\partial \mathcal{F}_{SU(3)_7}}{\partial \phi_1} &=& (2\phi_1 - \phi_2)(m_0 + 2\phi_1 + 4\phi_2), \label{tension1.pureSU3CS7}\\
\frac{\partial \mathcal{F}_{SU(3)_7}}{\partial \phi_2} &=& (-\phi_1 + 2\phi_2)(m_0 - 3\phi_1 + 2\phi_2). \label{tension2.pureSU3CS7}
\eea

We can compare the tension \eqref{tension1.pureSU3CS7}, \eqref{tension2.pureSU3CS7} with the area \eqref{area1.pureSU3CS7}-\eqref{area4.pureSU3CS7} to see the pure $SU(3)$ gauge theory realized by the web in Figure \ref{fig:pureSU3CS7} have the CS level $7$. It turns out that D3-brane does not cover each face of the diagram in Figure \ref{fig:pureSU3CS7b}. The comparison between the effective prepotential of the pure $G_2$ gauge theory and the area of the faces in \cite{Hayashi:2018bkd} revealed that one face which D3-brane is wrapped on is $\textcircled{\scriptsize 1} + \textcircled{\scriptsize 2} +2\textcircled{\scriptsize 3}$ and the other face is $\textcircled{\scriptsize 4}$. The coefficient $2$ in front of $\textcircled{\scriptsize 3}$ may be interpreted by the effect of including the mirror image. Then the explict comparison between \eqref{tension1.pureSU3CS7}, \eqref{tension2.pureSU3CS7} and \eqref{area1.pureSU3CS7}-\eqref{area4.pureSU3CS7} indeed gives the relations
\bea
\frac{\partial \mathcal{F}_{SU(3)_7}}{\partial \phi_1} &=& \textcircled{\scriptsize 1} + \textcircled{\scriptsize 2} + 2\times\textcircled{\scriptsize 3}, \label{relation1.pureSU3CS7}\\
\frac{\partial \mathcal{F}_{SU(3)_7}}{\partial \phi_2} &=& \textcircled{\scriptsize 4}. \label{relation2.pureSU3CS7}
\eea
Therefore, the equalities \eqref{relation1.pureSU3CS7} and \eqref{relation2.pureSU3CS7} imply that the pure $SU(3)$ gauge theory realized by the diagram in Figure \ref{fig:pureSU3CS7} has the CS level $7$.

Since we have a single 5-brane web diagram for the pure $G_2$ gauge theory and the pure $SU(3)$ gauge theory with the CS level $7$, it is also possible to obtain an explicit duality map between the parameters of the two theories. The length of a line in the diagram in Figure \ref{fig:pureSU3CS7} can be written by the two parameterizations. Since it is a single line the length written by the two parameterizations should be the same. Then we obtain the following duality map 
\bea
m_0^{SU(3)} &=& -\frac{m_0^{G_2}}{3},\\
\phi_1^{SU(3)} &=& \phi_2^{G_2} + \frac{1}{3}m_0^{G_2},\\
\phi_2^{SU(3)} &=& \phi_1^{G_2} + \frac{2}{3}m_0^{G_2},
\eea
where we used the Dynkin basis also for the Coulomb branch moduli of the pure $G_2$ gauge theory by using \eqref{G2.Dynkinbasis}.

\subsection{With matter}
\label{sec:G2wmatter}

In section \ref{sec:pureSU3CS7}, we saw that the 5-brane web diagram of the pure $G_2$ gauge theory in Figure \ref{fig:pureG2a} is S-dual to the 5-brane web diagram of the pure $SU(3)$ gauge theory with the CS level $7$ in Figure \ref{fig:pureSU3CS7}. In this section, we consider a 5-brane web diagram of the $G_2$ gauge theory with two hypermultiplets in the fundamental representation (flavors). The $G_2$ gauge theory with two flavors ($G_2+2\bF$) is dual not only to the $SU(3)$ gauge theory with two hypermultiplets in the fundamental representation and the CS level $6$ ($SU(3)_6+2\bF$) but also to the $Sp(2)$ gauge theory with two hypermultiplets in the antisymmetric representation and the non-trivial discrete theta angle ($Sp(2)_\pi+ 2\AS$) \cite{Jefferson:2018irk}. The two dualities are also indeed seen from the viewpoint of 5-brane webs. 

\begin{figure}
\centering
\subfigure[]{
\includegraphics[width=6cm]{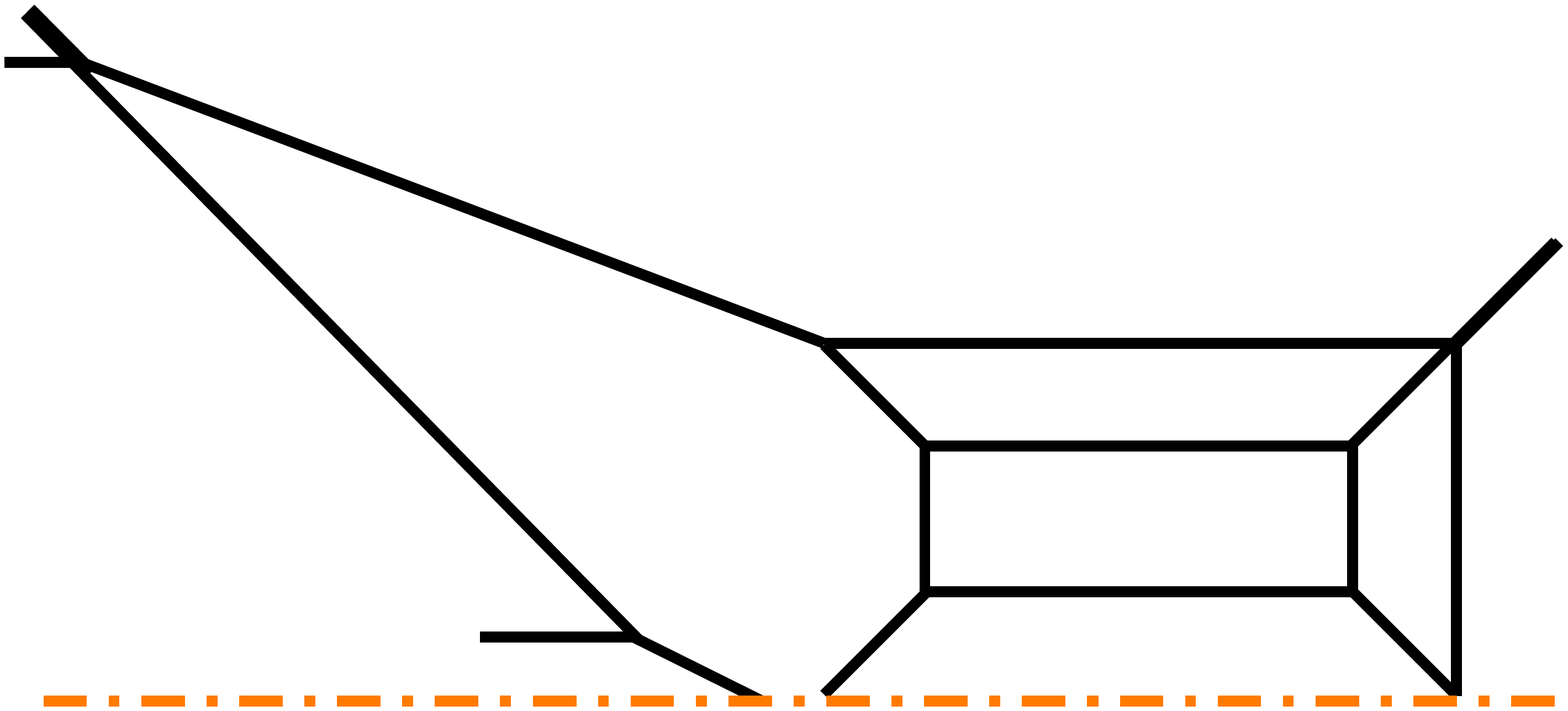} \label{fig:G2w2flvrsa}}
\subfigure[]{
\includegraphics[width=6cm]{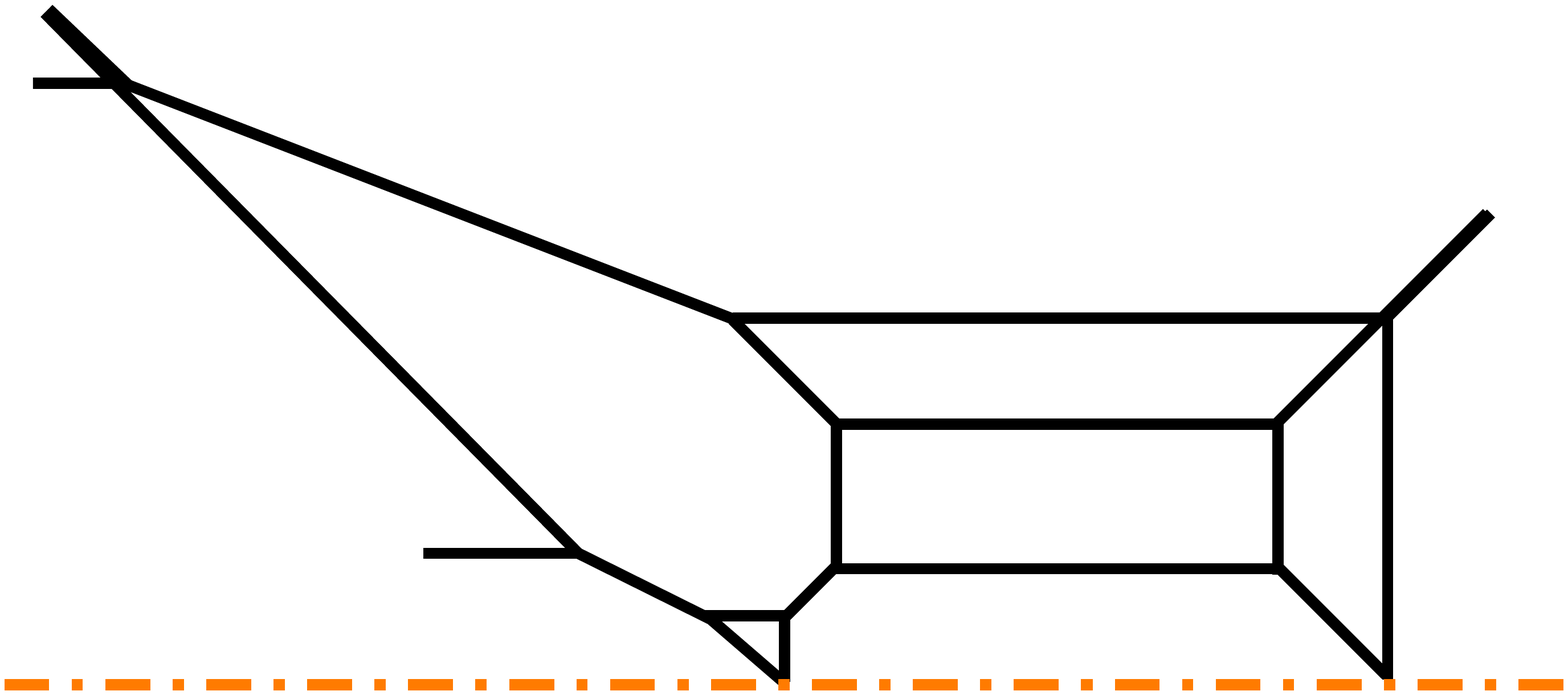} \label{fig:G2w2flvrsb}}
\caption{(a): A 5-brane web diagram for the $G_2$ gauge theory with two flavors plus two singlets. The matter is originated from two spinors of a $SO(7)$ gauge theory before the Higgsing. (b): Another 5-brane web diagram for the $G_2$ gauge theory with two flavors and two singlets, which is obtained by performing a flop transition to the diagram in Figure \ref{fig:G2w2flvrsa}.}
\label{fig:G2w2flvrs}
\end{figure}

The 5-brane web in Figure \ref{fig:pureG2a} for the pure $G_2$ gauge theory is obtained by Higgsing a 5-brane web for the $SO(7)$ gauge theory with one spinor. In order to introduce one flavor in the $G_2$ gauge theory we can consider a Higgsing of an $SO(7)$ gauge theory with a spinor or a flavor in addition to one spinor which we use for the Higgsing. A hypermultiplet in the vector representation of $SO(7)$ becomes one hypermultiplet in the fundamental representation of $G_2$ after the Higgsing, while a hypermultiplet in the spinor representation of $SO(7)$ becomes one flavor and a singlet of $G_2$. For later convenience, we introduce two flavors to the pure $G_2$ gauge theory by considering a Higgsing of the $SO(7)$ gauge theory with three spinors. The 5-brane web diagram obtained by the Higgsing is given in Figure \ref{fig:G2w2flvrsa}. One can perform a ``generalized flop transition'' \cite{Hayashi:2017btw, Hayashi:2018bkd} for the $Sp(0)$ part in Figure \ref{fig:G2w2flvrsa} and then the diagram becomes the one in Figure \ref{fig:G2w2flvrsb}.  
%%%%%%%%%%%%%%%%%%%%%%%%%%%%%%%%%
%%%%%%%%%%%%%%%%%%%%%%%%%%%%%%%%%

\begin{figure}
\centering
\subfigure[]{
\includegraphics[width=6cm]{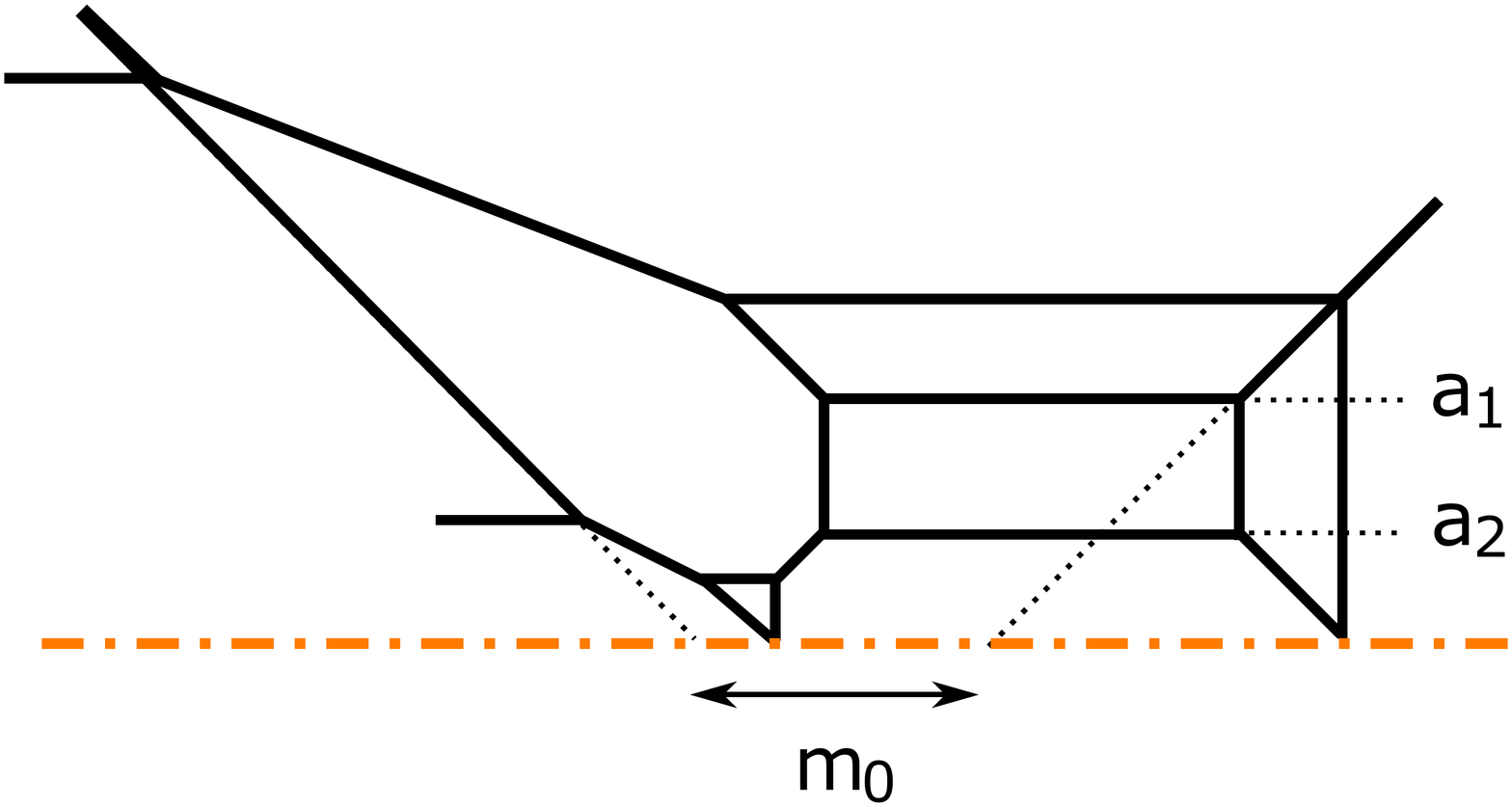} \label{fig:G2w2flvrspara1}}
\subfigure[]{
\includegraphics[width=6cm]{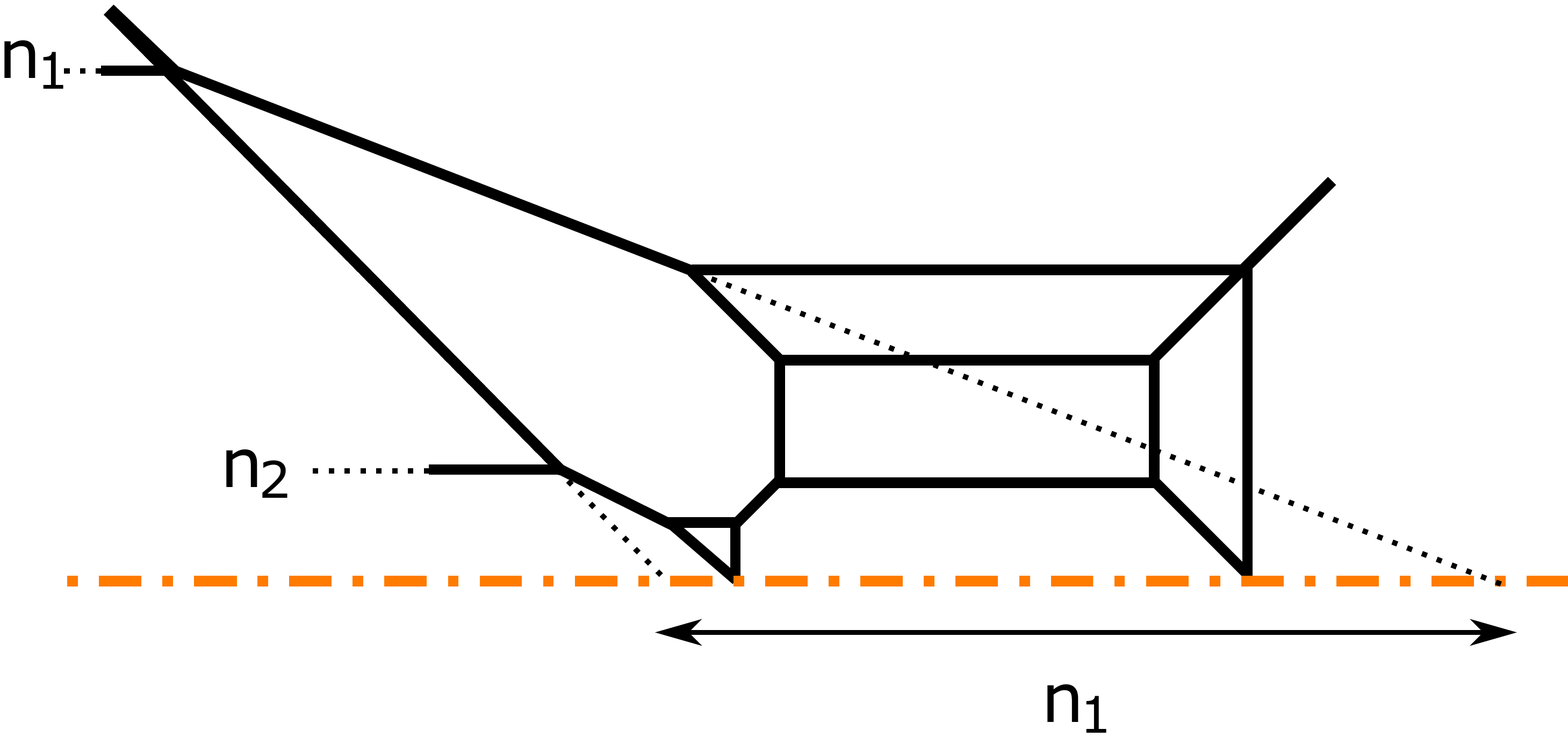} \label{fig:G2w2flvrspara2}}
\caption{(a): The parameterization for the inverse of the squared gauge coupling $m_0$ and the Coulomb branch moduli $a_1, a_2$. (b): Two parameters $n_1, n_2$ related to mass parameters for the two flavors.}
\label{fig:G2w2flvrspara}
\end{figure}

We can also assign the gauge theory parameters to the length of the 5-branes as in Figure \ref{fig:G2w2flvrspara1}. $a_1$ and $a_2$ are the Coulomb branch moduli of the $G_2$ gauge theory. 
%%%%%%%%%%%%%%%%%%%%%%%%%%%%%%%%%
%%%%%%%%%%%%%%%%%%%%%%%%%%%%%%%%%
Since the $G_2$ gauge theory with two flavors originates from the $SO(7)$ gauge theory with three spinors where two spinors are attached to the left side and one spinor, which is used for the Higgsing, is attached to the right side the diagram of the $SO(7)$ diagram before the Higgsing. Hence, the inverse of the squared gauge coupling of the $G_2$ gauge theory with two flavors can be read off similarly to the case of the $SO(7)$ gauge theory with two spinors. As explained in appendix \ref{sec:m0}, the inverse of the squared gauge coupling, $m_0$, in this case can be computed by extrapolating the leftmost $(1, -1)$ 5-brane and the rightmost $(1, 1)$ 5-brane. Hence $m_0$ of the $G_2$ gauge theory with two flavors can be chosen in Figure \ref{fig:G2w2flvrspara}.

The diagram contains two more parameters $n_1, n_2$ depicted in Figure \ref{fig:G2w2flvrspara2}, which are determined by the position of asymptotic 5-branes. The two parameters are related to the mass parameters for the two flavors. From the viewpoint of the $Sp(0)$, $n_1$ is the inverse of the squared gauge coupling of the $Sp(0)$ and $n_2$ is the mass parameter for one flavor of the $Sp(0)$. Hence, the two parameters are associated to the flavor symmetry $U(1) \times SO(3)$. The $SO(3)$ arises from the flavor D5-brane whose height is $n_2$ on top of an $\widetilde{\text{O5}}^-$-plane. On the other hand, the two flavors of the $G_2$ gauge theory yield an $Sp(2)$ flavor symmetry. Therefore, we need to change the basis given by the embedding $U(1) \times SO(3) \subset Sp(2)$ to obtain the mass parameters $m_1, m_2$ of $G_2+2\bF$ from $n_1, n_2$,~\footnote{We choose the normalization of the mass parameters $m_1, m_2$ so that they agree with the mass parameter in the expression \eqref{prepotential} of the effective prepotential.}
\be
m_1 =\frac12 (n_1+n_2), \qquad m_2= \frac12 (n_1-n_2).
\label{G2w2flvrsmass}
\ee

\begin{figure}
\centering
\includegraphics[width=8cm]{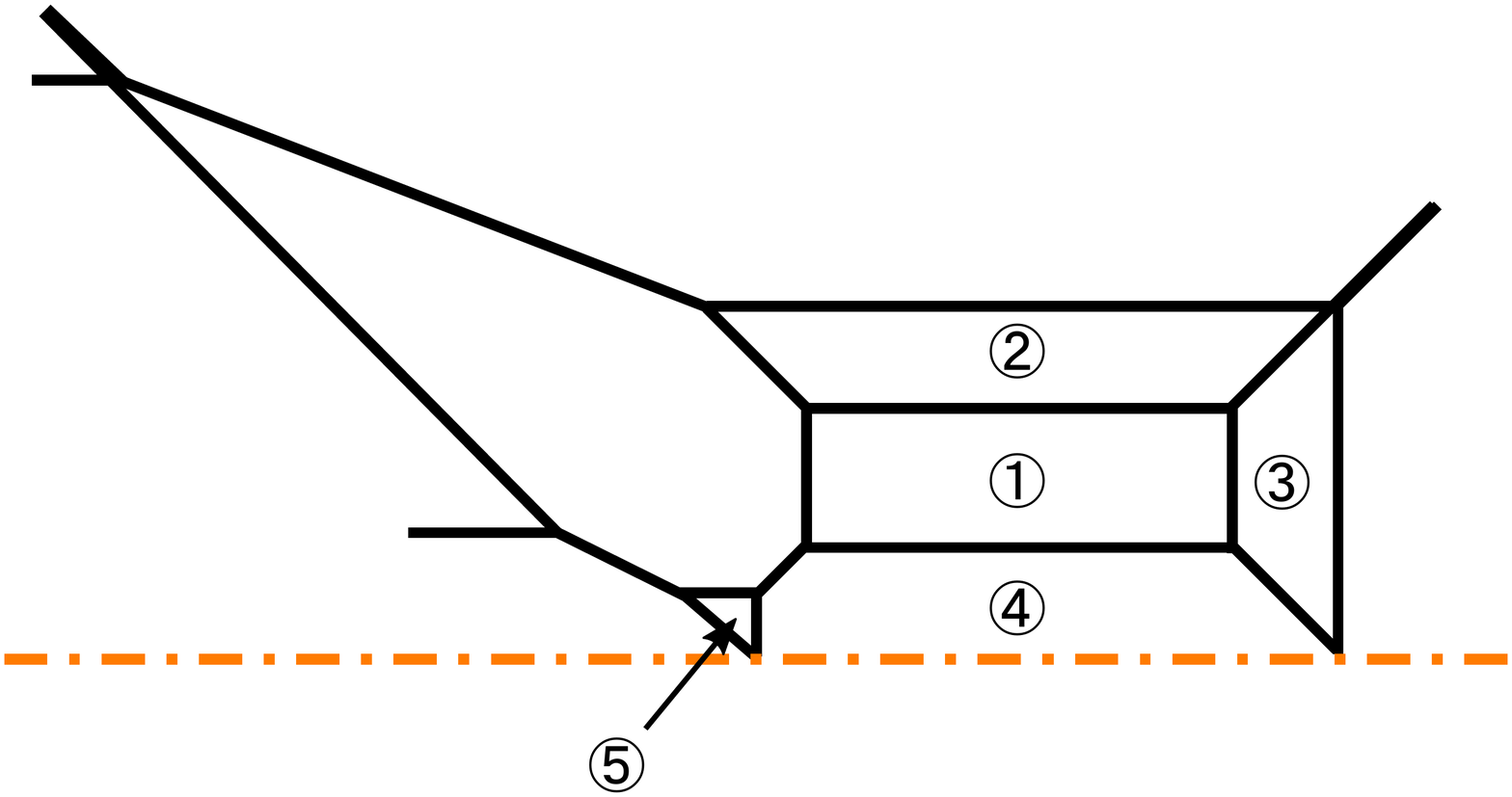}
\caption{A labeling for the faces in the diagram of the $G_2$ gauge theory with two flavors.}
\label{fig:G2w2flvrsface}
\end{figure}

One can check the validity of the parameterization by computing the prepotential of the $G_2$ gauge theory with two flavors. From the parameterization, we can compute the tension of the monopole string by the area of the faces in Figure \ref{fig:G2w2flvrsface}. The label of the faces are given in Figure \ref{fig:G2w2flvrsface}. 
%%%%%%%%%%%%%%%%%%%%%%%%%%%%%%%%%
%%%%%%%%%%%%%%%%%%%%%%%%%%%%%%%%%
The area of the faces is given by
\bea
\textcircled{\scriptsize 1} &=& (a_1 - a_2)(m_0-m_1-m_2+3a_1+a_2), \label{area1.G2w2flvrs}\\
\textcircled{\scriptsize 2} &=& a_2(m_0-m_1-m_2+3a_1+2a_2), \label{area2.G2w2flvrs}\\
\textcircled{\scriptsize 3} &=&a_1a_2, \label{area3.G2w2flvrs}\\
\textcircled{\scriptsize 4} &=& \frac{1}{2}\left(2m_0a_2 - 2m_1a_2 - m_2^2 + 2m_2a_1 - a_1^2  + 4a_1a_2+ 3a_2^2\right),\label{area4.G2w2flvrs}\\
\textcircled{\scriptsize 5} &=& \frac{1}{2}(-m_2 + a_1 + a_2)^2. \label{area5.G2w2flvrs}
\eea

We then compare the area \eqref{area1.G2w2flvrs}-\eqref{area5.G2w2flvrs} with the tension of the monopole string computed from the effective prepotential \eqref{prepotential}. In order to compute the prepotential of the $G_2$ gauge theory with two flavors, we need to determine the phase corresponding to the diagram in Figure \ref{fig:G2w2flvrsb}. The phase can be fixed from the requirement that the length of any 5-brane in the diagram is positive. Then the parameterization given in Figure \ref{fig:G2w2flvrspara1} and \eqref{G2w2flvrsmass} implies the phase 
\bea
&&a_1 + a_2 - m_1 < 0, \quad a_1 - m_1 < 0, \quad a_2 - m_1 < 0,\nn\\
&&-a_2 - m_1 < 0, \quad -a_1 - m_1 < 0, \quad -a_1 - a_2 - m_1 < 0,
\eea
for the flavor with the mass parameter $m_1$ and 
\bea
&&a_1 + a_2 - m_2 > 0, \quad a_1 - m_2 < 0, \quad a_2 - m_2 < 0,\nn\\
&&-a_2 - m_2 < 0, \quad -a_1 - m_2 < 0, \quad -a_1 - a_2 - m_2 < 0,
\eea
for the flavor with the mass parameter $m_2$. The effective prepotential of the $G_2$ gauge theory with two flavors on this phase becomes
\bea
\mathcal{F}_{G_2+2{\bf F}} &=& m_0(\phi_1^2 - 3\phi_1\phi_2 + 3\phi_2^3) +  \frac{1}{6}(-3m_1^3 -6m_1(\phi_1^2 - 3\phi_1\phi_2 + 3\phi_2^2)\\
&&-2m_2^3-3m_2^2\phi_2  -3m_2(2\phi_1^2 -6\phi_1\phi_2 + 5\phi_2^2) + 8\phi_1^3 - 24\phi_1^2\phi_2 + 18\phi_1\phi_2^2 + 7\phi_2^3),\nn
\label{prepot.G2w2flvrs}
\eea
where we used the Dynkin basis %\eqref{G2.Dynkinbasis} 
\bea
a_1 = \phi_1 - \phi_2, \quad a_2 = -\phi_1 + 2\phi_2. \label{G2.Dynkinbasis}
\eea
for the Coulomb branch moduli.

Then the tension of the monopole string of the $G_2$ gauge theory with two flavors is given by taking a derivative of the effective prepotential \eqref{prepot.G2w2flvrs} with respect to the Coulomb branch moduli $\phi_1, \phi_2$. Note that as in the case of the pure $SU(3)$ gauge theory with the CS level $7$, a D3-brane will not be wrapped on arbitrary five faces but needs to be wrapped on a particular linear combination. In particular we need to consider $\textcircled{\scriptsize 2} + \textcircled{\scriptsize 3} + 2\times\textcircled{\scriptsize 4} + \textcircled{\scriptsize 5}$ for the tension given by a derivative with respect to $\phi_2$. We indeed find the agreement between the linear combination of the area \eqref{area1.G2w2flvrs}-\eqref{area2.G2w2flvrs} and the tension of the monopole string,
\bea
\frac{\partial \mathcal{F}_{G_2+2{\bf F}}}{\partial \phi_1} &=&\textcircled{\scriptsize 1} , \label{tension1.G2w2flvrs}\\
\frac{\partial \mathcal{F}_{G_2+2{\bf F}}}{\partial \phi_2} &=&\textcircled{\scriptsize 2} + \textcircled{\scriptsize 3} + 2\times\textcircled{\scriptsize 4} + \textcircled{\scriptsize 5}. \label{tension2.G2w2flvrs}
\eea
The equalities \eqref{tension1.G2w2flvrs} and \eqref{tension2.G2w2flvrs} implies the correctness of the parameterization in Figure \ref{fig:G2w2flvrspara1} and \eqref{G2w2flvrsmass} and they also reconfirm that the diagram in Figure \ref{fig:G2w2flvrsb} gives rise to the $G_2$ gauge theory with two flavors. 

\subsubsection{Duality to $SU(3)_6 + 2{\bf F}$}
\label{sec:dualtoSU3w2flvrs}

\begin{figure}
\centering
\includegraphics[width=4.5cm]{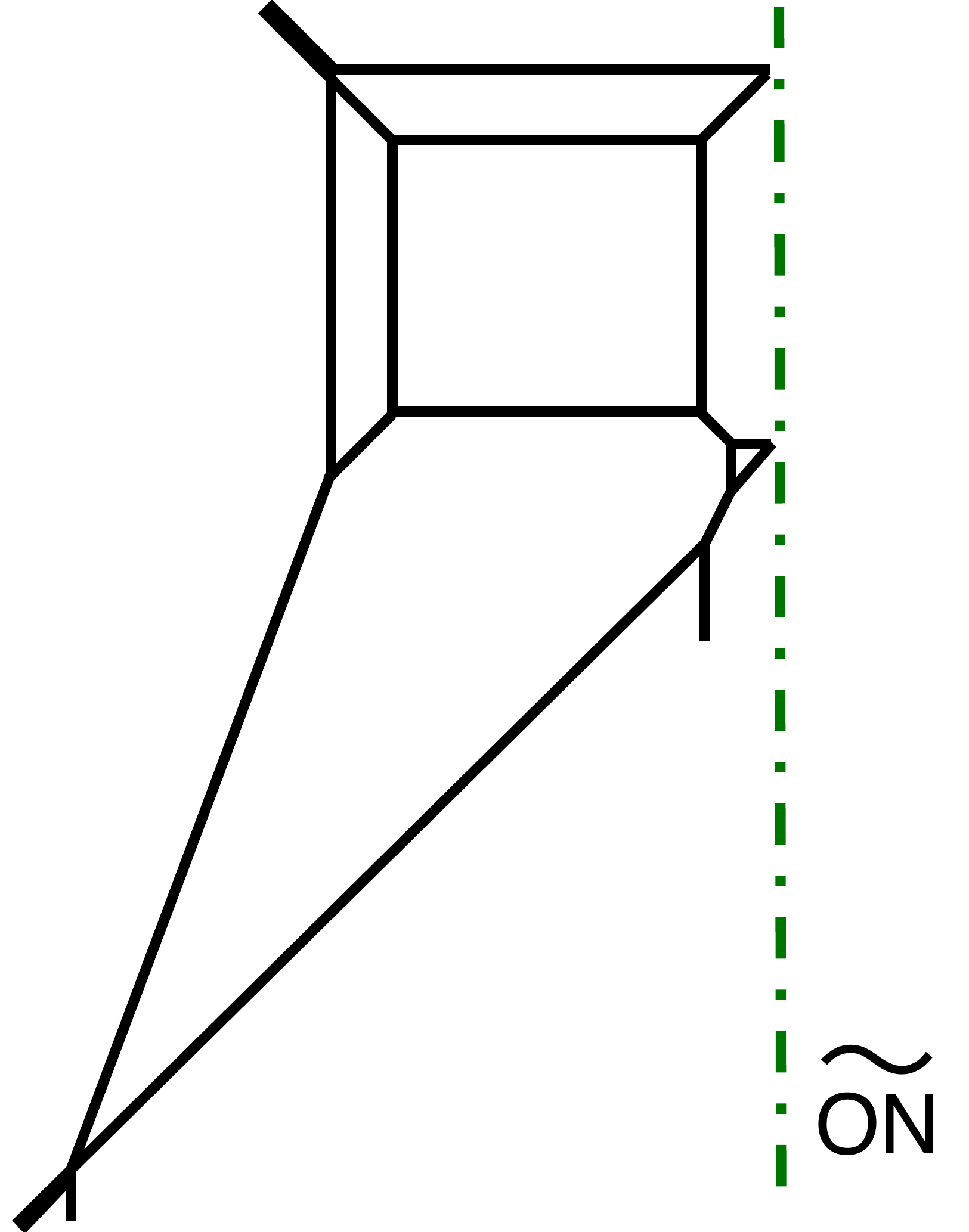}
\caption{The 5-brane web diagram with an $\widetilde{ON}$-plane obtained after performing the S-duality to the diagram in Figure \ref{fig:G2w2flvrsb}.}
\label{fig:SU3w2flvrsCS6}
\end{figure}

As we saw the duality between the pure $G_2$ gauge theory and the pure $SU(3)$ gauge theory with the CS level $7$ from the 5-brane webs in section \ref{sec:pureSU3CS7}, it is also possible to see the duality between the $G_2$ gauge theory with two flavors and the $SU(3)$ gauge theory with two flavors and the CS level $6$ from the 5-brane web in Figure \ref{fig:G2w2flvrsb}. Applying the S-duality to the diagram in Figure \ref{fig:G2w2flvrsb} yields a diagram in Figure \ref{fig:SU3w2flvrsCS6}. Since the diagram contains three color D5-branes, the diagram may be interpreted as an $SU(3)$ gauge theory. 
%%%%%%%%%%%%%%%%%%%%%%%%%%%%%%%%%
%%%%%%%%%%%%%%%%%%%%%%%%%%%%%%%%%

\begin{figure}
\centering
\subfigure[]{
\includegraphics[width=6cm]{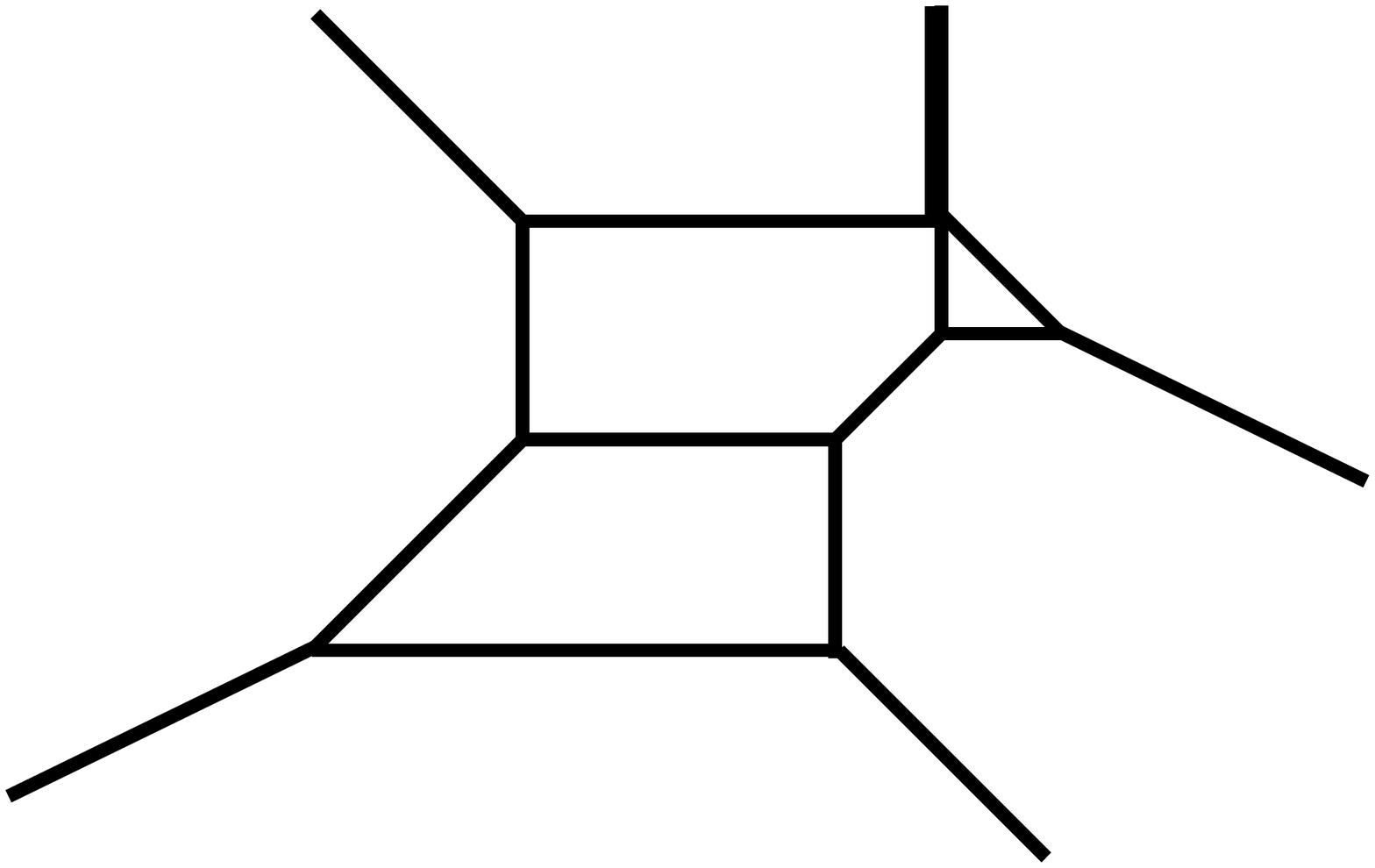} \label{fig:SU3w1AS}}
\subfigure[]{
\includegraphics[width=6cm]{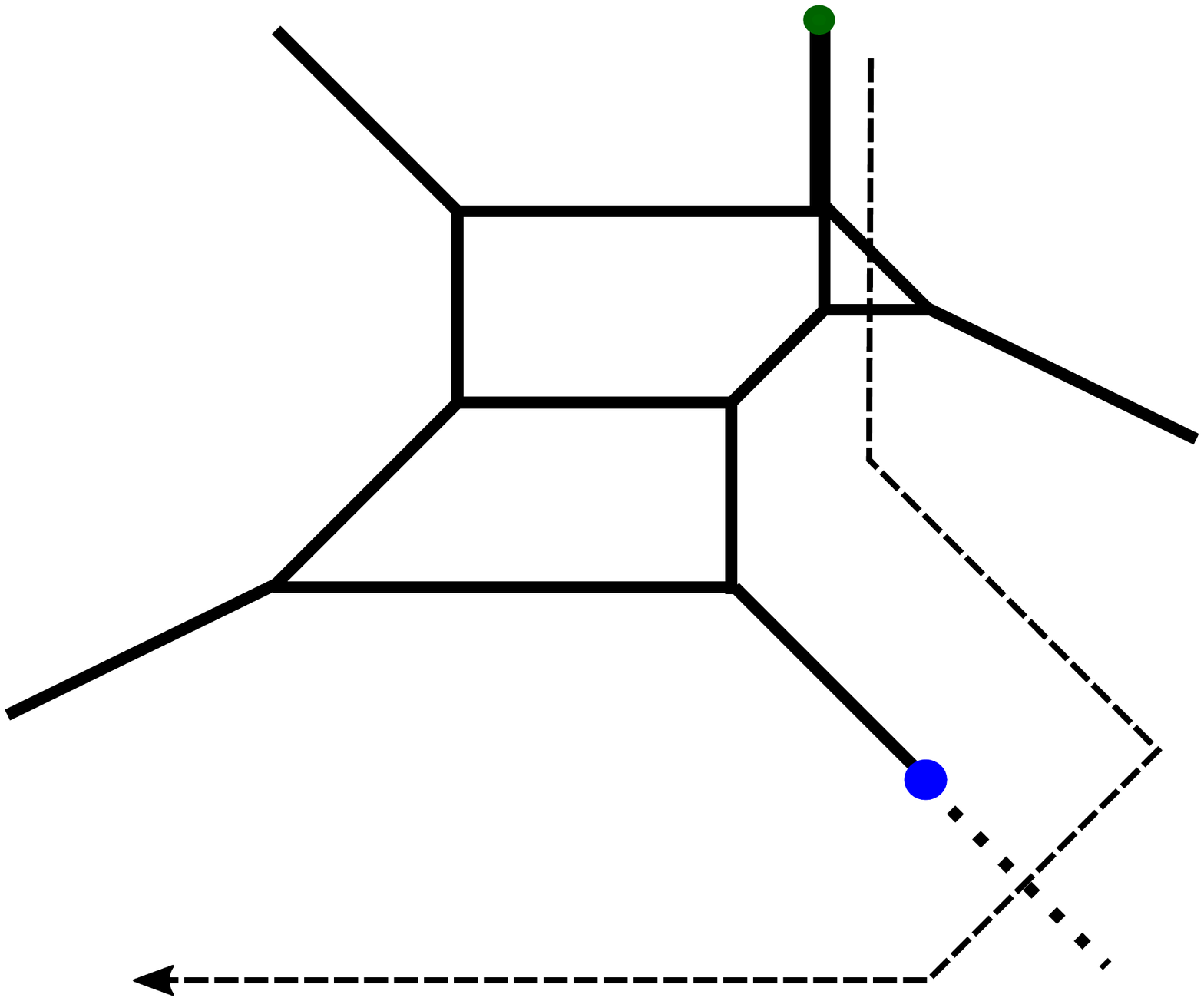} \label{fig:SU3w1ASdef1}}
\subfigure[]{
\includegraphics[width=6cm]{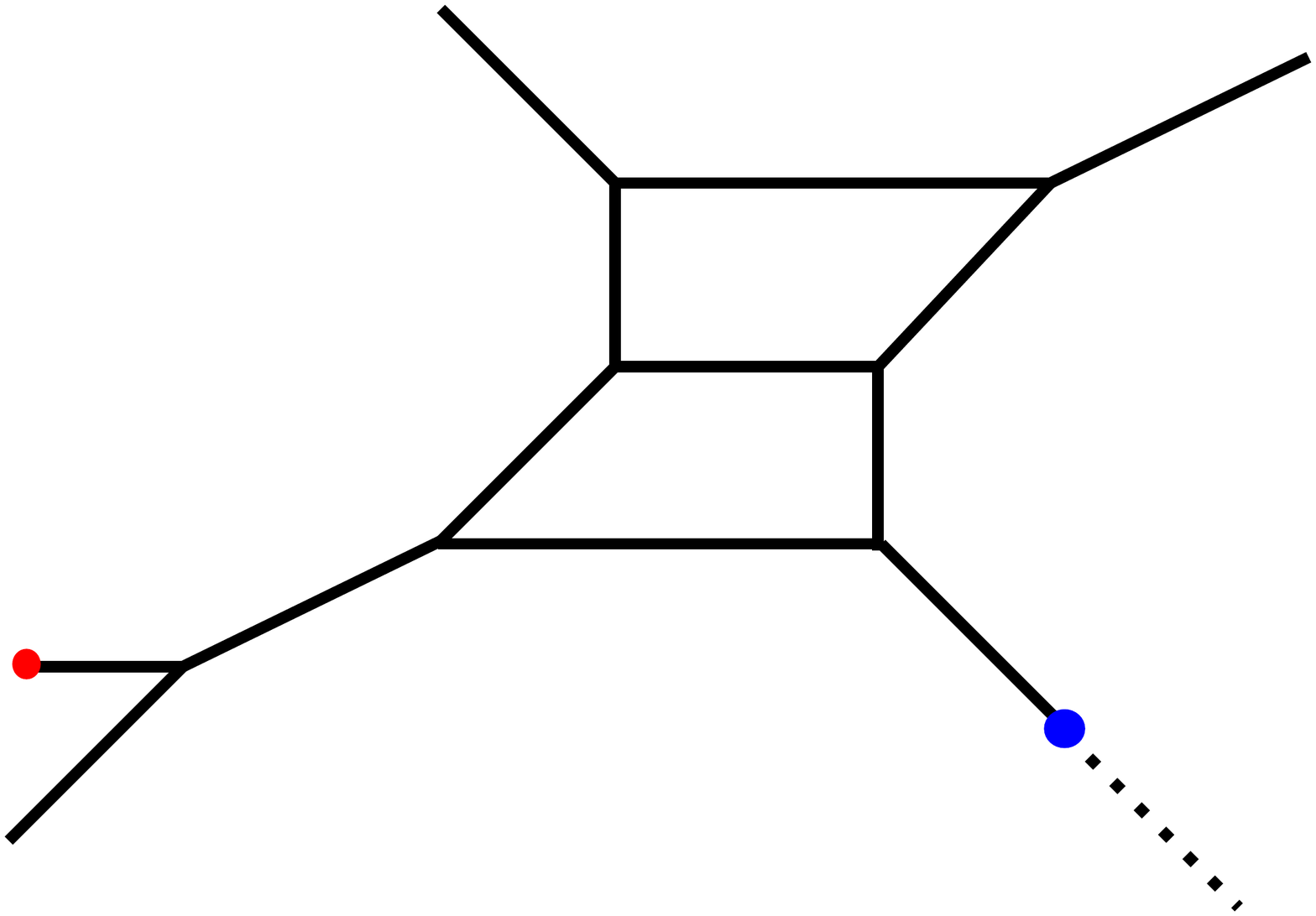} \label{fig:SU3w1ASdef2}}
\subfigure[]{
\includegraphics[width=6cm]{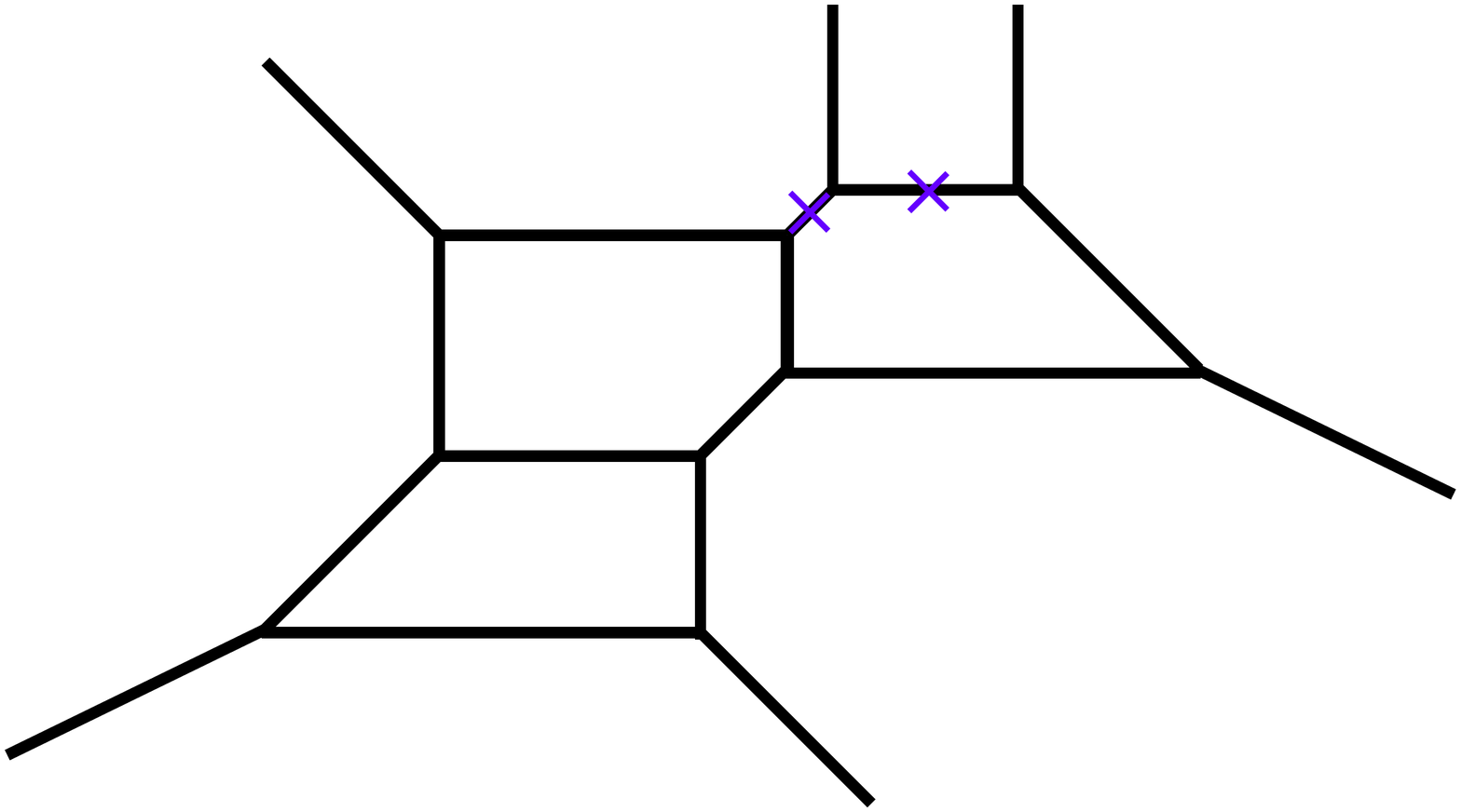} \label{fig:SU3SU2}}
\caption{(a): A diagram of an $SU(3)$ gauge theory one hypermultiplet in the antisymmetric representation and the CS level $\frac{1}{2}$. (b): The deformation which changes the diagram in Figure \ref{fig:SU3w1AS} into the one for an $SU(3)$ gauge theory with one flavor. (c): A diagram for the $SU(3)$ gauge theory with one flavor and the CS level $\frac{1}{2}$. (d): A diagram of the $SU(3)_{-1} - SU(2)$ quiver theory which can be Higgsed to the diagram in Figure \ref{fig:SU3w1AS}.}
\label{fig:SU3SU2Higgs}
\end{figure}

The diagram contains two more parameters except for the gauge coupling, which are determined by the position of asymptotic 5-branes. Hence the parameters are associated to mass parameters of some matter of the $SU(3)$ gauge theory. We argue that the matter is two hypermultiplets in the antisymmetric representation, which is equivalent to the antifundamental representation. Let us see a 5-brane web diagram for the $SU(3)$ gauge theory with one hypermultiplet in the antisymmetric representation and the CS level $\frac{1}{2}$ \cite{Bergman:2015dpa}, which is depicted in Figure \ref{fig:SU3w1AS}. 
%%%%%%%%%%%%%%%%%%%%%%%%%%%%%%%%%
%%%%%%%%%%%%%%%%%%%%%%%%%%%%%%%%%
Since the antisymmetric representation of $SU(3)$ is equivalent to the antifundamental representation, one can deform the diagram in Figure \ref{fig:SU3w1AS} into the one from which we can explicitly see the presence of one flavor. When we move the $(0, 1)$ 7-brane in Figure \ref{fig:SU3w1ASdef1} according to the indicated arrow, the HW transition gives a diagram in Figure \ref{fig:SU3w1ASdef2}. This is nothing but a diagram giving rise to the $SU(3)$ gauge theory with one flavor and the CS level $\frac{1}{2}$. 

This diagram can be obtained by a Higgsing from the $SU(3) - SU(2)$ quiver theory where the CS level of the $SU(3)$ is $-1$ \cite{Hayashi:2016jak}. The theory has an $SU(2)$ flavor symmetry associated to the two external NS5-branes extending in the upper direction in Figure \ref{fig:SU3SU2Higgs}. One can perform a Higgsing associated to the $SU(2)$ flavor symmetry, which corresponds to the tuning of the length of the 5-branes indicated by the purple $\times$ in Figure \ref{fig:SU3SU2}. The diagram exactly reduces to the one in Figure \ref{fig:SU3w1AS}. Namely, the Higgsing of the quiver $SU(3)_{-1}-SU(2)$ associated to the flavor symmetry $SU(2)$ can yield the $SU(3)$ gauge theory with a hypermultiplet in the antisymmetric representation and the CS level $\frac{1}{2}$. 

\begin{figure}
\centering
\subfigure[]{
\includegraphics[width=6cm]{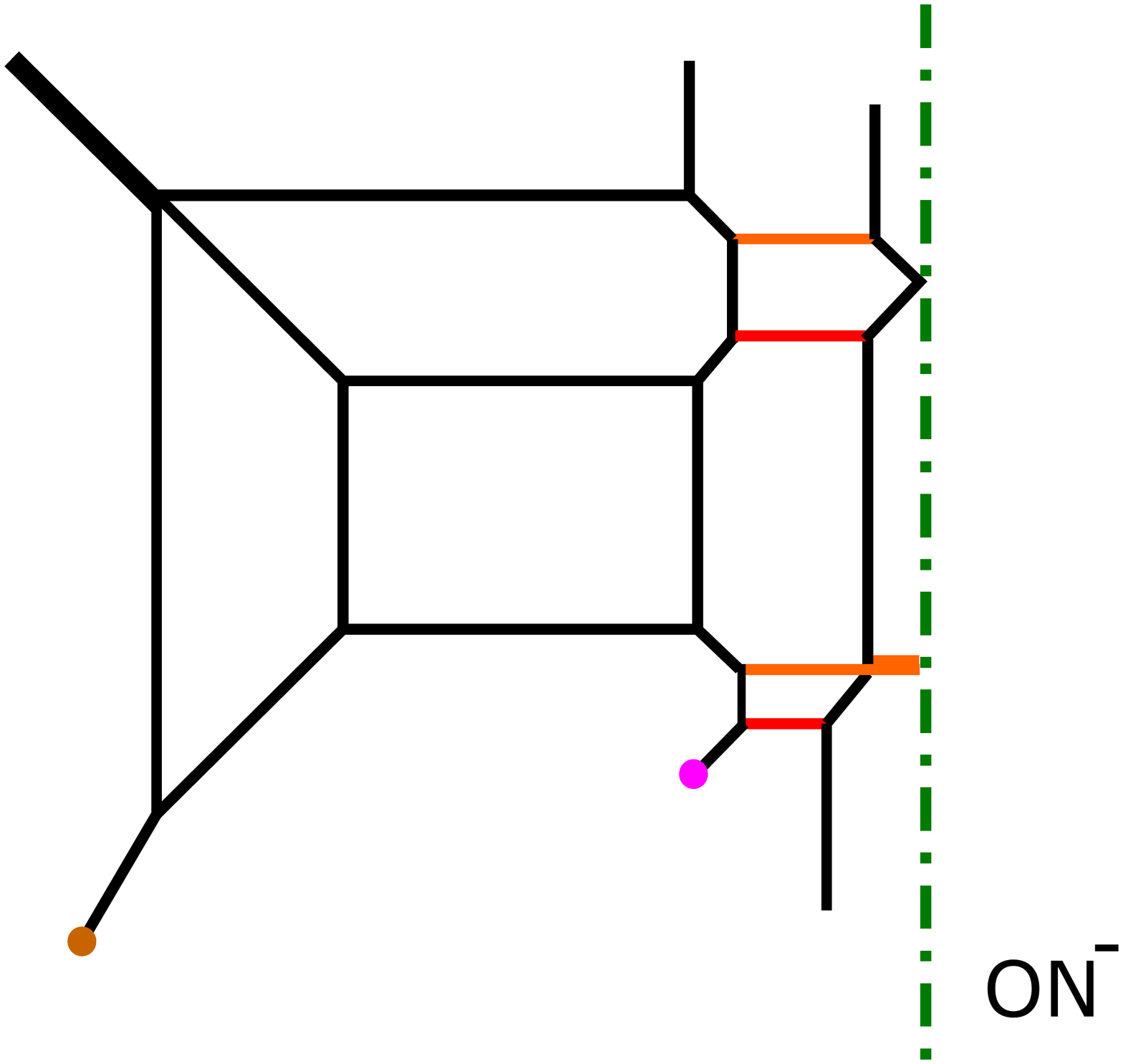} \label{fig:SU3SU2SU2}}
\subfigure[]{
\includegraphics[width=6cm]{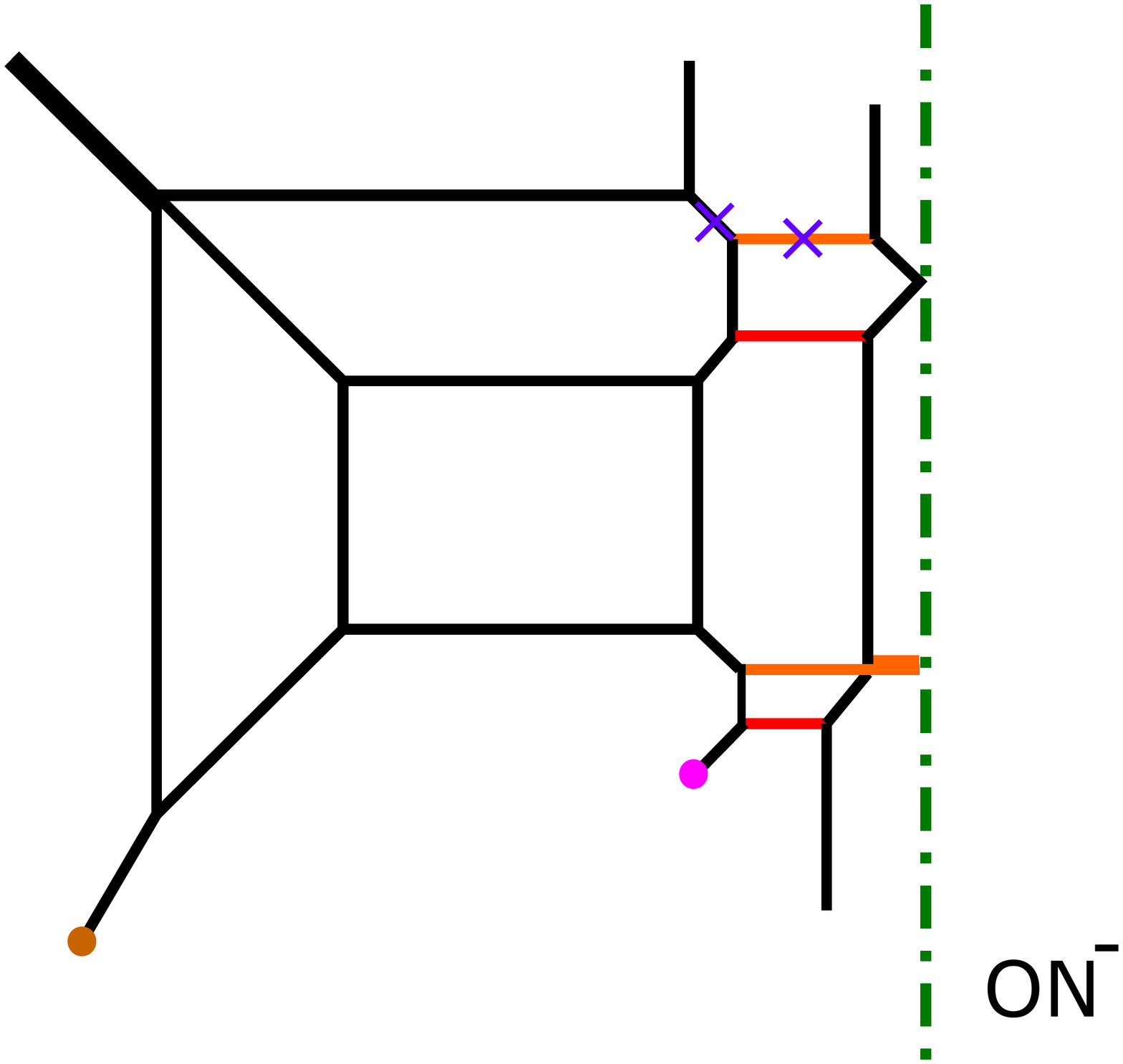} \label{fig:SU3SU2SU2Higgs1}}
\subfigure[]{
\includegraphics[width=6cm]{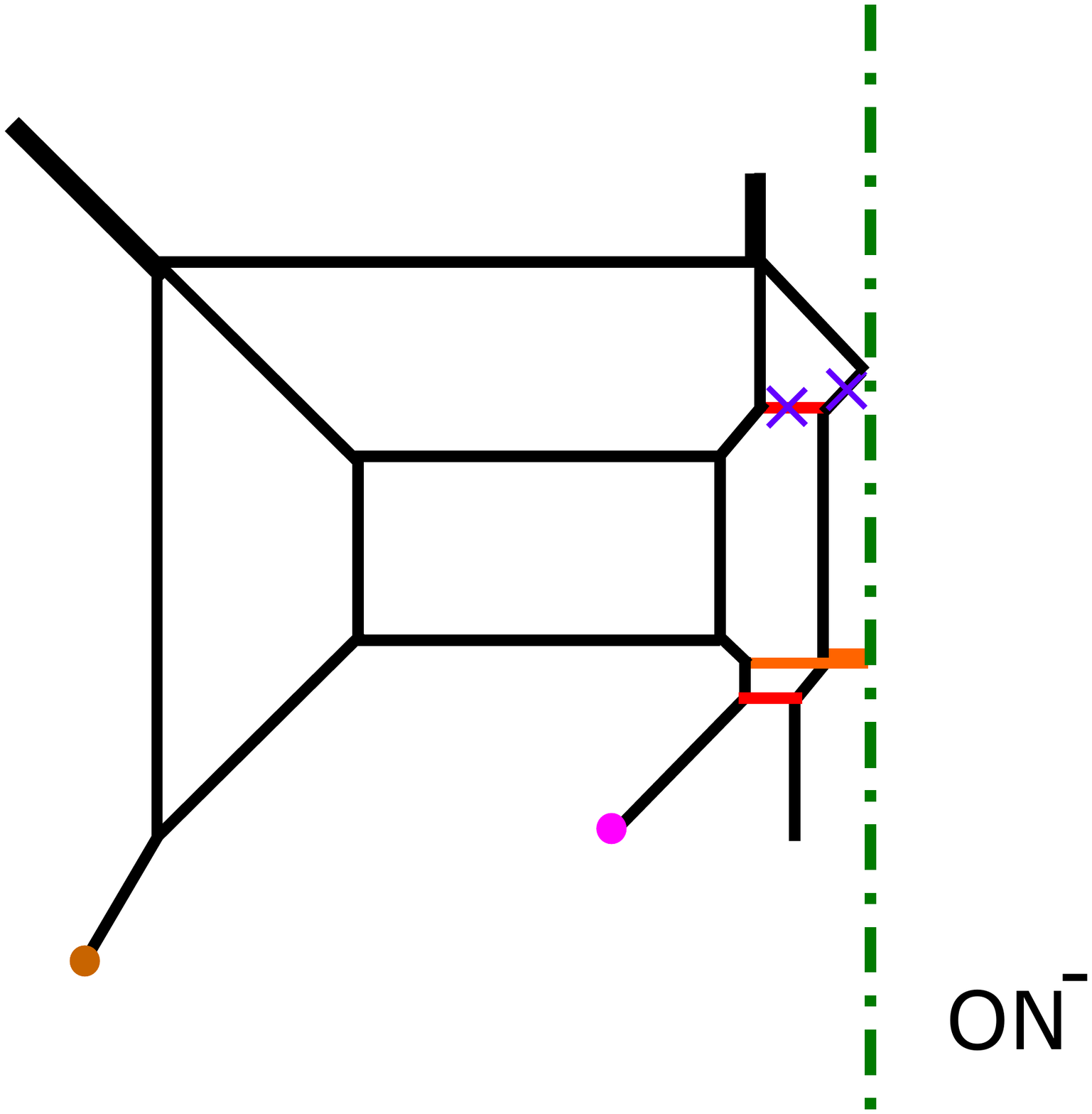} \label{fig:SU3SU2SU2Higgs2}}
\subfigure[]{
\includegraphics[width=6cm]{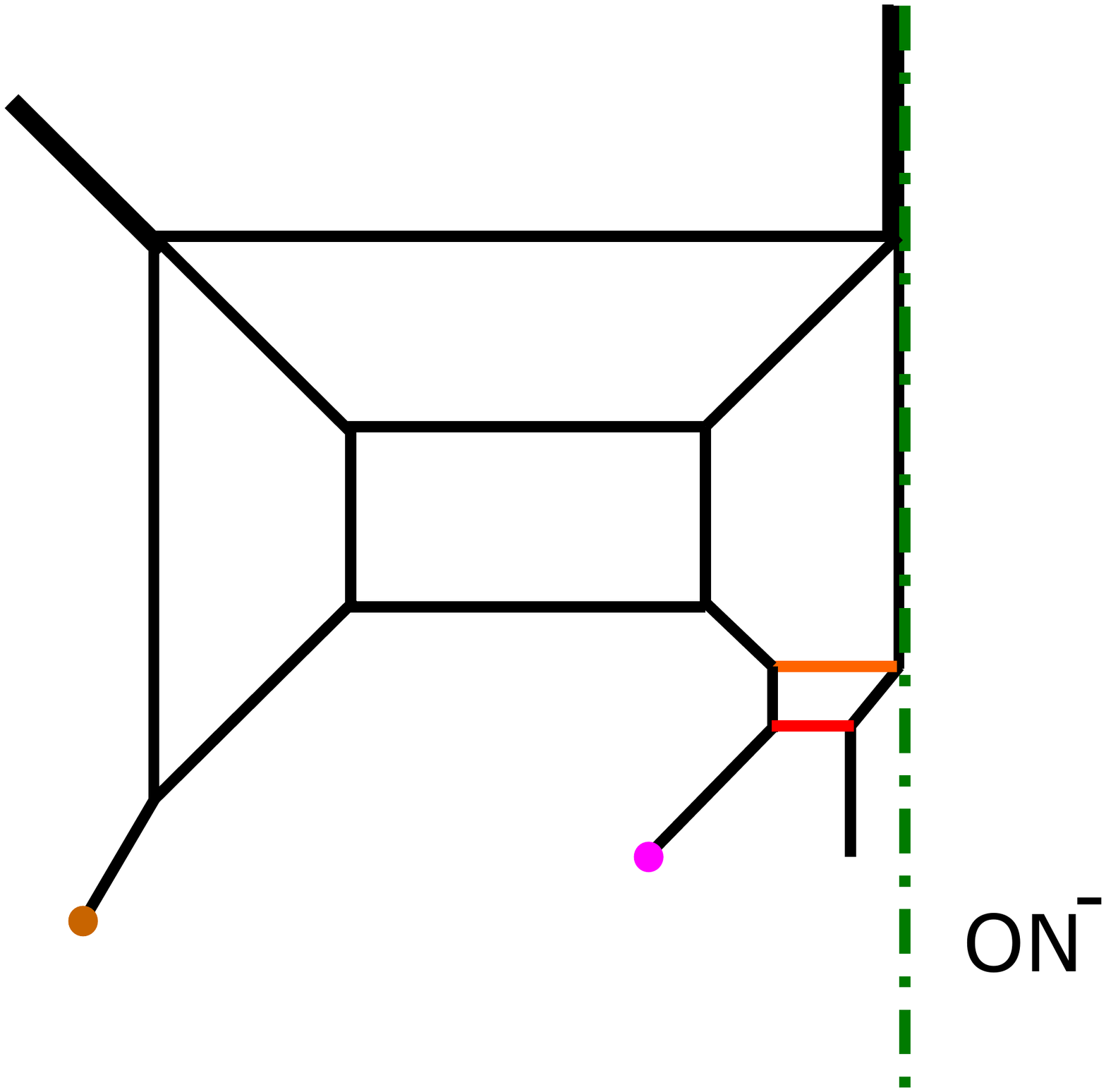} \label{fig:SU3SU2SU2Higgs3}}
\subfigure[]{
\includegraphics[width=6cm]{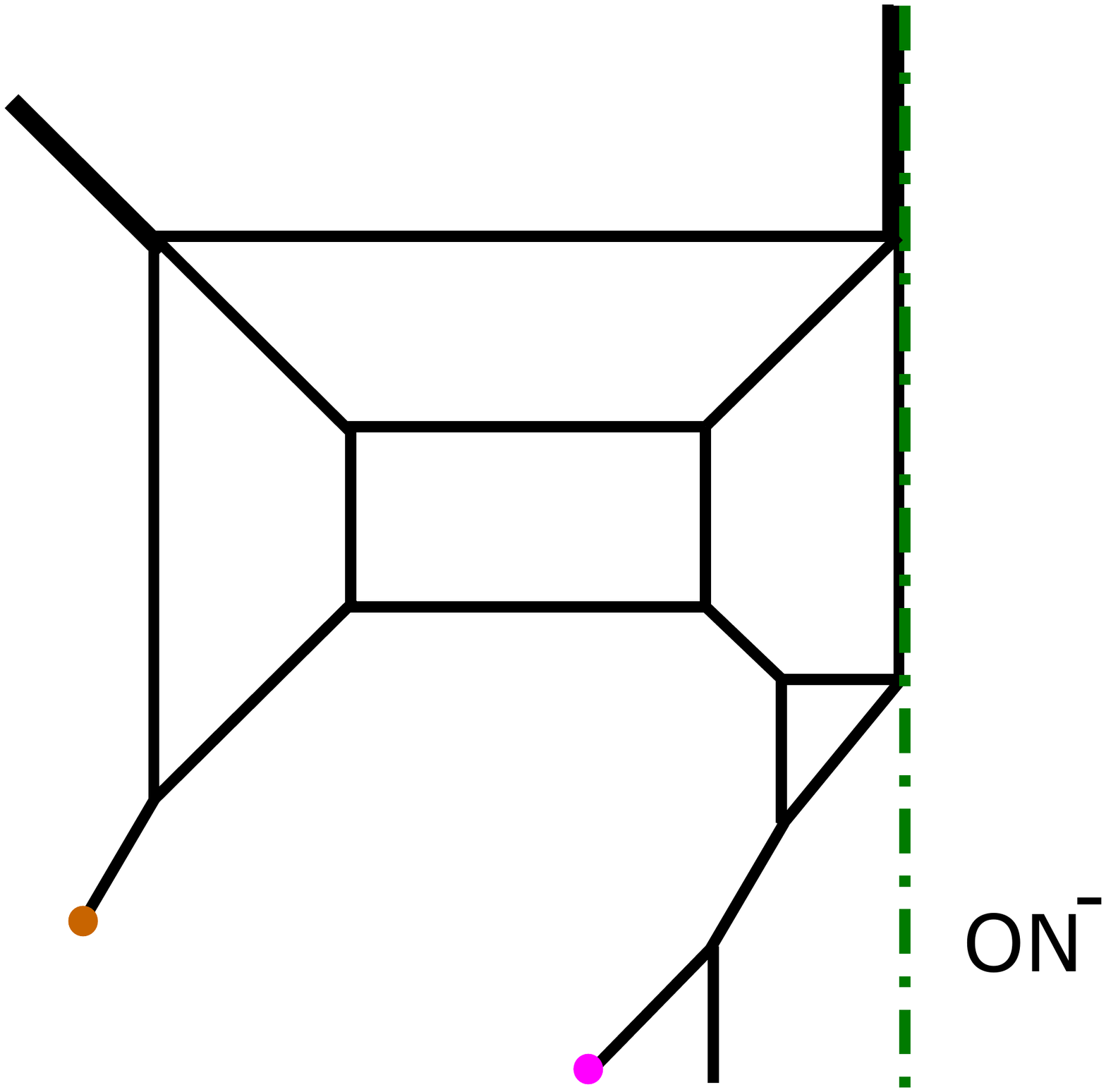} \label{fig:SU3SU2SU2Higgs4}}
\caption{(a): A diagram for the $SU(2) - SU(3)_3 - SU(2)$ quiver theory. (b): The first tuning. (c): The second tuning. (d): The diagram after the two Higgsings. (e): The diagram after applying a flop transition to the diagram in Figure \ref{fig:SU3SU2SU2Higgs3}. This diagram is equivalent to the one in Figure \ref{fig:SU3w2flvrsCS6}.}
\label{fig:SU3SU2SU2Higgs}
\end{figure}

Extending the idea of the Higgsing, an $SU(3)$ gauge theory with two antisymmetric hypermultiplets can originate from a Higgsing of the $SU(2) - SU(3)_3 - SU(2)$ quiver theory. Namely a Higgsing of the two $SU(2)$ gauge theories may yield two hypermultiplets in the antisymmetric representation for the $SU(3)$. The $SU(2) - SU(3)_3 - SU(2)$ quiver theory can be realized by using an ON$^-$-plane \cite{Hanany:1999sj, Hayashi:2015vhy} as in Figure \ref{fig:SU3SU2SU2}. 
%%%%%%%%%%%%%%%%%%%%%%%%%%%%%%%%%
%%%%%%%%%%%%%%%%%%%%%%%%%%%%%%%%%
The two color branes for one of the $SU(2)$ gauge group are given by the orange lines in Figure \ref{fig:SU3SU2SU2} and the two color branes for the other $SU(2)$ gauge group are represented by red segments in Figure \ref{fig:SU3SU2SU2}. The theory also has an $SO(4) \simeq SU(2) \times SU(2)$ flavor symmetry associated to the two external NS5-branes extending in the upper direction with an ON$^-$-plane in Figure \ref{fig:SU3SU2SU2}. We then perform two Higgsings which break the $SU(2) \times SU(2)$ flavor symmetry. The first Higgsing is realized by tuning the length of the 5-branes indicated by the purple $\times$ in Figure \ref{fig:SU3SU2SU2Higgs1}. Then the second Higgsing is achieved by tuning the length of the 5-branes indicated by the purple $\times$ in Figure \ref{fig:SU3SU2SU2Higgs2}. Then the resulting diagram becomes the one in Figure \ref{fig:SU3SU2SU2Higgs3}. After the two Higgsings, only one of the two color branes for each $SU(2)$ gauge group remains and the two $SU(2)$ gauge groups are broken. In order to connect to the diagram in Figure \ref{fig:SU3w2flvrsCS6}, we perform one flop transition and obtain the diagram in Figure \ref{fig:SU3SU2SU2Higgs4}. Although the diagram in Figure \ref{fig:SU3SU2SU2Higgs4} is written with an ON$^-$-plane. One can change the ON$^-$-plane into an $\widetilde{\text{ON}}^-$-plane by moving a fractional D7-brane which may be put at the end of an external NS5-brane on an ON$^-$-plane \cite{Zafrir:2015ftn, Hayashi:2018bkd}. Therefore, the Higgsings of the two $SU(2)$ in the quiver theory $SU(2) - SU(3)_3 - SU(2)$ yields the diagram in Figure \ref{fig:SU3w2flvrsCS6}, implying that the $SU(3)$ theory contain two hypermultiplets in the antisymmetric representation. Furthermore, the Higgsing of one $SU(2)$ in Figure \ref{fig:SU3SU2} increased the CS level by $\frac{3}{2}$. Hence it is natural to expect that the Higgsing of the two $SU(2)$ through the process \ref{fig:SU3SU2SU2Higgs1}-\ref{fig:SU3SU2SU2Higgs3} increases the CS level by $\frac{3}{2} \times 2 = 3$. Hence, the CS level after the Higgsing will be $6$.  In summary, the diagram in Figure \ref{fig:SU3w2flvrsCS6} may yield the $SU(3)$ gauge theory with two antisymmetric hypermultiplets, or equivalently two flavors, and the CS level $6$.

%%%%%%%%%%%%%%%%%%%%%%%%%%%%%%%%
\begin{figure}
\centering
\includegraphics[width=8cm]{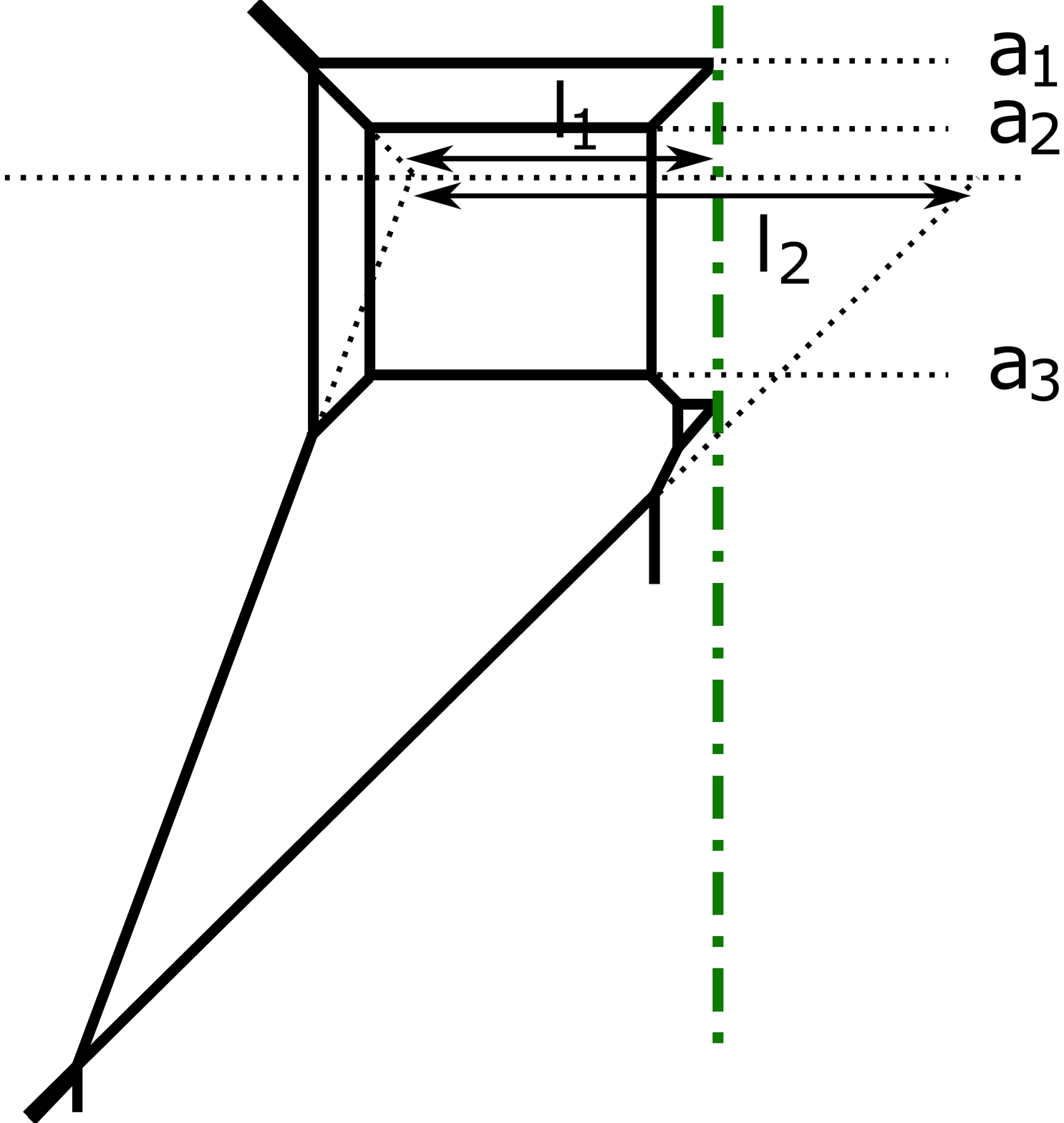} %\label{fig:SU3w2flCS6para1}
\caption{The gauge theory parameterization for the 5-brane web diagram of the $SU(3)$ gauge theory realized in Figure \ref{fig:SU3w2flvrsCS6}. $a_1, a_2, a_3$ are the Coulomb branch moduli of $SU(3)$. The dotted line in the center is the location of the origin in the vertical direction. $l_1$ and $l_2$ is related to the inverse of the squared gauge coupling $m_0$ by $l_1 + l_2 = 2m_0$.}
\label{fig:SU3w2flCS6para1}
%\label{fig:SU3w2flvrsCS6para}
\end{figure}
%%%%%%%%%%%%%%%%%%%%%%%%%%%%%%%%%
Let us confirm that the diagram in Figure \ref{fig:SU3w2flvrsCS6} gives rise to the $SU(3)$ gauge theory with two flavors and the CS level $6$ from the computation of the effective prepotential. For that we assign gauge theory parameters for the length of 5-branes in Figure \ref{fig:SU3w2flvrsCS6}. The inverse of the squared gauge coupling $m_0 = \frac{l_1 + l_2}{2}$ and the Coulomb branch moduli $a_1, a_2, a_3,\; (a_1 + a_2 + a_3 = 0)$ are given in Figure \ref{fig:SU3w2flCS6para1}, %\ref{fig:SU3w2flvrsCS6para1}, 
 which is the same parameterization as that for the pure $SU(3)$ gauge theory with the CS level $7$ in Figure \ref{fig:pureSU3CS7a}.

In order to see the parameterization of the mass parameters $m_1, m_2$, let us first recall how the mass parameter for antisymmetric hypermultiplet appears in Figure \ref{fig:SU3w1AS}.  
%%%%%%%%%%%%%%%%%%%%%%%%%%%%%%%%%
\begin{figure}
\centering
\subfigure[]{
\includegraphics[width=6cm]{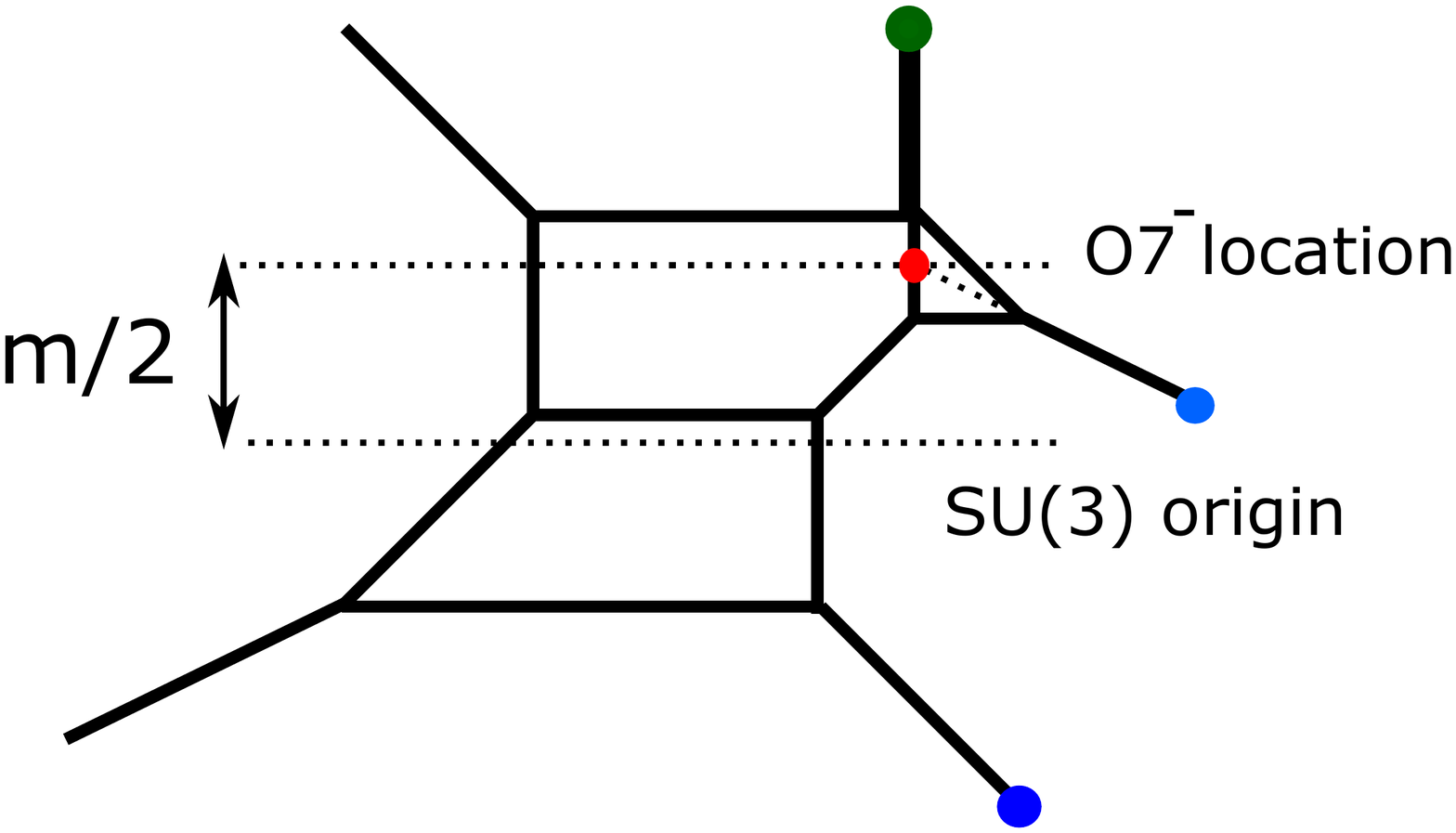} \label{fig:SU3w1ASpara}}
\subfigure[]{
\includegraphics[width=6cm]{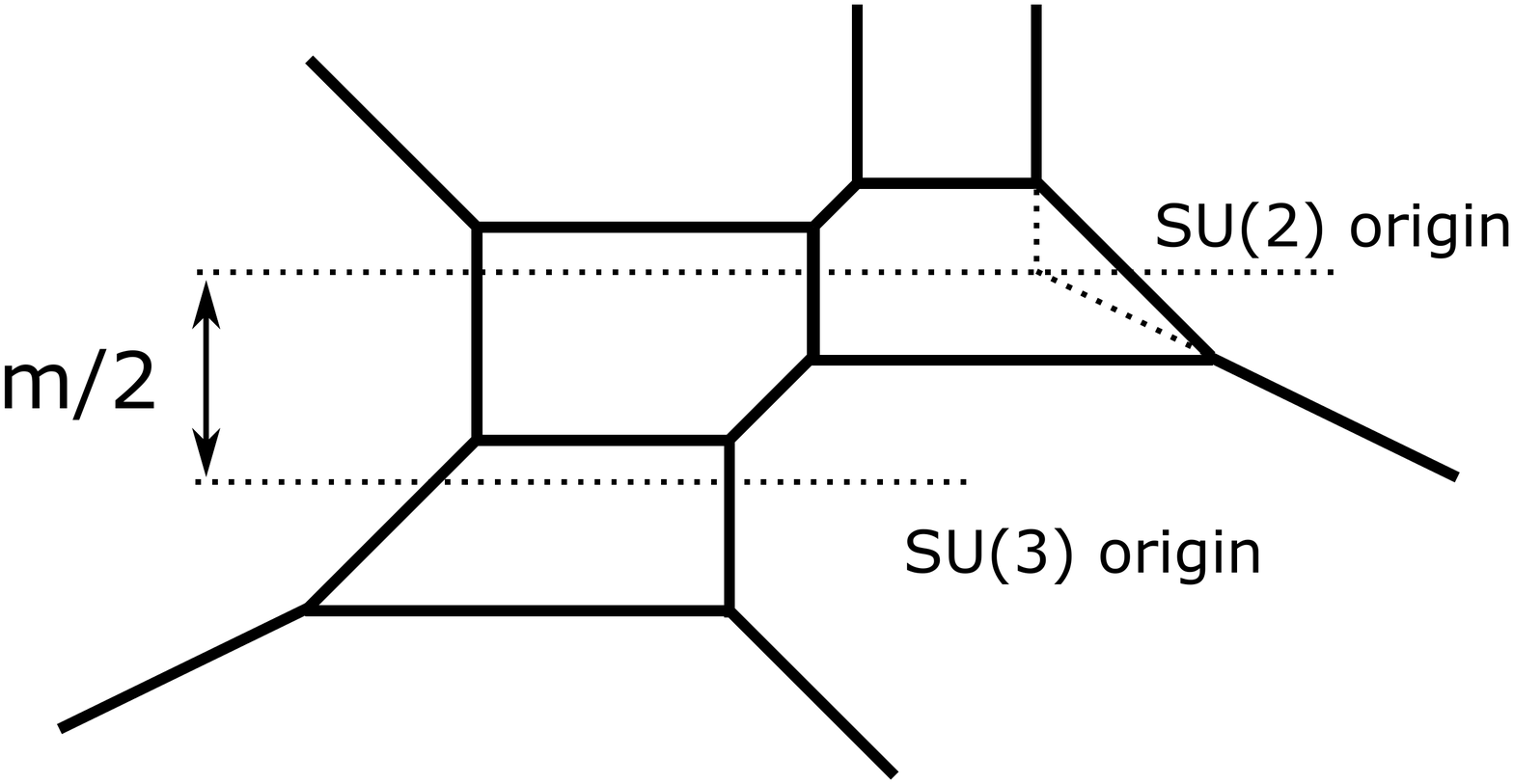} \label{fig:SU3SU2para}}
\subfigure[]{
\includegraphics[width=6cm]{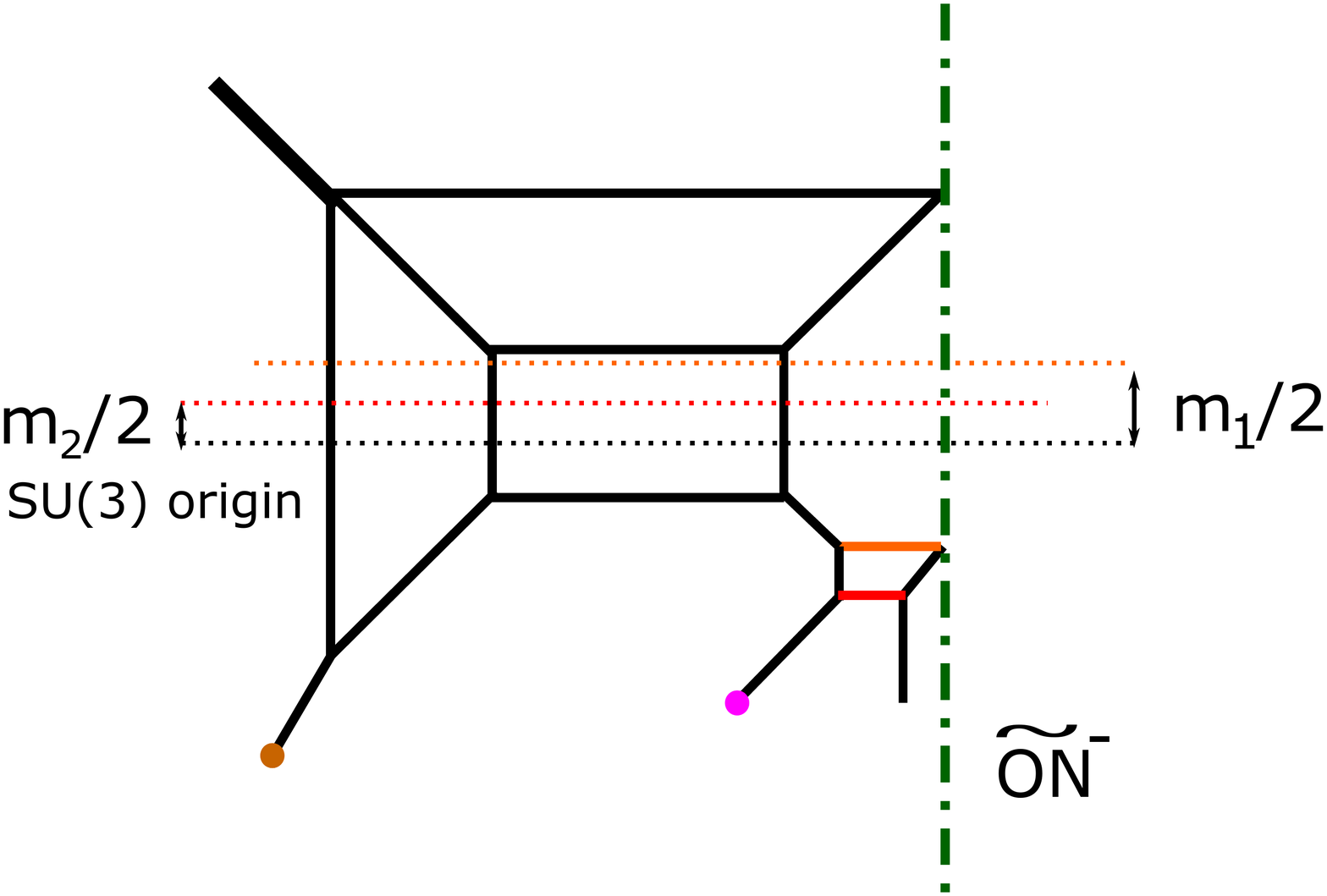} \label{fig:SU3SU2SU2para}}
\caption{(a): The mass parameter for a hypermultiplet in the antisymmetric representation in Figure \ref{fig:SU3w1AS} (b): How the mass parameter in Figure \ref{fig:SU3w1ASpara} is related to the length in the $SU(3)_{-1} - SU(2)_0$ quiver theory. (c): The mass parameters for two antisymmetric hypermultiplets. The origin for the $SU(3)$ color branes is denoted by the black dotted line. The origin for the $SU(2)$ realized by orange/red lines is depicted by the orange/red dotted line. }
\label{fig:antisymmass}
\end{figure}
%%%%%%%%%%%%%%%%%%%%%%%%%%%%%%%%%
The length associated to the mass parameter $m$ is depicted in Figure \ref{fig:SU3w1ASpara}. It is give by twice as long as the distance between the location of an O7$^-$-plane and the origin of the height for the $SU(3)$ color branes. The location of the O7$^-$-plane can be determined by the intersection point between the line of the $(0, 1)$ 5-brane and the line of the $(2, -1)$ 5-brane. In terms of the length in the diagram for the $SU(3)_{-1} - SU(2)$ quiver theory, the mass parameter is twice as long as the distance between the origin of the height for the $SU(3)$ color branes and the origin of the height for the $SU(2)$ color branes as in Figure \ref{fig:SU3SU2para}. The distance between the two origins of the $SU(3)$ and the $SU(2)$ is nothing but the mass parameter for the bifundamental matter. Therefore, the antisymmetric mass is originated from the bifundamental mass before the tuning. 

We can now generalize the discussion for the case with two antisymmetric hypermultiplets. Namely, a mass parameter is twice as long as the distance between the origin of the location of the  $SU(3)$ color branes and the origin of the location for the one of the $SU(2)$. The orange dotted line in Figure \ref{fig:SU3SU2SU2para} is originated from the origin for the $SU(2)$ color branes given by the orange lines in Figure \ref{fig:SU3SU2SU2}. One the other hand, the red dotted line in Figure \ref{fig:SU3SU2SU2para} is originated from the origin for the $SU(2)$ color branes given by the red lines in Figure \ref{fig:SU3SU2SU2}. Therefore, the two mass parameters are related to the distance between the origin of the $SU(3)$ and the orange dotted line or the red dotted line as in Figure \ref{fig:SU3SU2SU2para}.

For the comparison with the effective prepotential of the $SU(3)$ gauge theory with two flavors and the CS level $6$, we first compare the area of the faces of the diagram with the tension of the monopole string for the $SU(3)$ gauge theory with two flavors and the CS level $6$. We again use the labels for the faces in Figure \ref{fig:G2w2flvrsface}. Then the area of the five faces is given by 
\bea
\textcircled{\scriptsize 1} &=& \frac{1}{2}(-\phi_1 + 2\phi_2)(2m_0+m_1+m_2-6\phi_1+4\phi_2), \label{area1.SU3w2flvrsCS6}\\
\textcircled{\scriptsize 2} &=& (2\phi_1 - \phi_2)(\phi_1 + \phi_2), \label{area2.SU3w2flvrsCS6}\\
\textcircled{\scriptsize 3} &=&\frac{1}{2}(2\phi_1 - \phi_2)(2m_0 + m_1 + m_2 - 2\phi_1 + 2\phi_2), \label{area3.SU3w2flvrsCS6}\\
\textcircled{\scriptsize 4} &=& \frac{1}{2}\left(-m_1^2 - 2m_1\phi_1 + 3\phi_1^2 + 2\phi_1\phi_2 - 2\phi_2^2\right),\label{area4.SU3w2flvrsCS6}\\
\textcircled{\scriptsize 5} &=& \frac{1}{2}(m_1 + \phi_1)^2, \label{area5.SU3w2flvrsCS6}
\eea
where we used the Dynkin basis \eqref{SU3.Dynkinbasis}.

The area corresponds to the tension of the monopole string which can be computed by taking a derivative of the effective prepotential with respect to Coulomb branch moduli. The effective prepotential of the $SU(3)$ gauge theory with two flavors and the CS level $6$ may be calculated from the general formula \eqref{prepotential}. The condition that the length of the 5-branes in Figure \ref{fig:SU3w2flvrsCS6} is positive implies the following phase
\bea
-a_1 - m_1 < 0, \quad -a_2 - m_1 > 0, \quad -a_3 - m_1 > 0,
\eea
for one antisymmetric hypermultiplet with mass $m_1$ and 
\bea
-a_1 - m_2 >0, \quad -a_2 - m_2 > 0, \quad -a_3 - m_2 > 0,
\eea
for the other antisymmetric hypermultiplet with mass $m_2$. Here we used $a_1 + a_2 + a_3 = 0$ and expressed the weight of the antisymmetric representation by the weight of the antifundamental representation. Then the effective prepotential becomes
\bea
\mathcal{F}_{SU(3)_6+2{\bf F}} &=& m_0(\phi_1^2 - \phi_1\phi_2 + \phi_2^2) + \frac{1}{12}(m_1^3-6m_1^2\phi_1 + 6m_1\phi_2(-\phi_1 + \phi_2)\nn\\
&& + 3m_2^3+6m_2(\phi_1^2 - \phi_1\phi_2 + \phi_2^2) + 2(7\phi_1^3 + 18\phi_1^2\phi_2 - 24\phi_1\phi_2^2 + 8\phi_2^3)).\nn\\ \label{prepot.SU3w2flvrsCS6}
\eea
Then we find the expected relation between the area \eqref{area1.SU3w2flvrsCS6}-\eqref{area5.SU3w2flvrsCS6} and the tension of the monopole string,
\bea
\frac{\partial \mathcal{F}_{SU(3)_6+2{\bf F}}}{\partial \phi_1} &=&\textcircled{\scriptsize 2} + \textcircled{\scriptsize 3} + 2\times\textcircled{\scriptsize 4} + \textcircled{\scriptsize 5}, \label{tension1.SU3w2flvrsCS6}\\
\frac{\partial \mathcal{F}_{SU(3)_6+2{\bf F}}}{\partial \phi_2} &=&\textcircled{\scriptsize 1}. \label{tension2.SU3w2flvrsCS6}
\eea
The equalities \eqref{tension1.SU3w2flvrsCS6} and \eqref{tension2.SU3w2flvrsCS6} confirms that the $SU(3)$ gauge theory has the CS level $6$ and two flavors. 

By comparing the parameterization of the $G_2$ gauge theory and that of the $SU(3)$ gauge theory, we can also obtain the duality map between the parameters. The duality map is given by 
\bea
m_0^{SU(3)} &=& \frac{m_{\bF, 1}^{G_2} + m_{\bF, 2}^{G_2}}{2},\label{map1.SU3toG2w2flvrs}\\
m_{\AS, 1}^{SU(3)} &=& \frac{1}{3}\left(-m_0^{G_2} + m_{\bF, 1}^{G_2} - 2m_{\bF, 2}^{G_2}\right),\\
m_{\AS, 2}^{SU(3)} &=& \frac{1}{3}\left(-m_0^{G_2} - 2m_{\bF, 1}^{G_2} + m_{\bF, 2}^{G_2}\right),\\
\phi_1^{SU(3)} &=& \phi_2^{G_2} + \frac{1}{3}\left(m_0^{G_2} - m_{\bF, 1}^{G_2} - m_{\bF, 2}^{G_2}\right),\\
\phi_2^{SU(3)} &=& \phi_1^{G_2} + \frac{1}{3}\left(2m_0^{G_2} - 2m_{\bF, 1}^{G_2} -2m_{\bF, 2}^{G_2}\right), \label{map5.SU3toG2w2flvrs}
\eea
where we put the subindex standing for the representation for the matter. The labeling of the number for the masses is the same as before. We will use this convention for writing duality maps hereafter.

We note that if one decouples two hypermultiplets from the 5-brane web in Figure \ref{fig:SU3SU2SU2Higgs4} (or equivalently Figure \ref{fig:SU3w2flvrsCS6}), then the resulting diagram is same as the 5-brane web in Figure \ref{fig:pureSU3CS7}, which we claimed a 5-brane web for the pure $SU(3)$ theory with CS level 7. This hence provides a support of our construction of 5-brane web for the $SU(3)$ gauge theory with the CS level 7, discussed in section \ref{sec:pureSU3CS7}, as the decoupling of two flavors would increase the CS level by 1.

\subsubsection{Duality to $Sp(2)_{\pi} + 2{\bf AS}$}
\label{sec:dualtoSp2w2AS}

We have seen that the 5-brane diagram of the $G_2$ gauge theory with two flavors is S-dual to the diagram of the $SU(3)$ gauge theory with two flavors and the CS level $6$. In fact, the $G_2$ gauge theory with two flavors admits another dual description given by the $Sp(2)$ gauge theory with two hypermultiplets in the antisymmetric representation and the non-trivial discrete theta angle. We will argue that this duality can be also seen from the 5-brane web diagram. 

\begin{figure}
\centering
\subfigure[]{
\includegraphics[width=6cm]{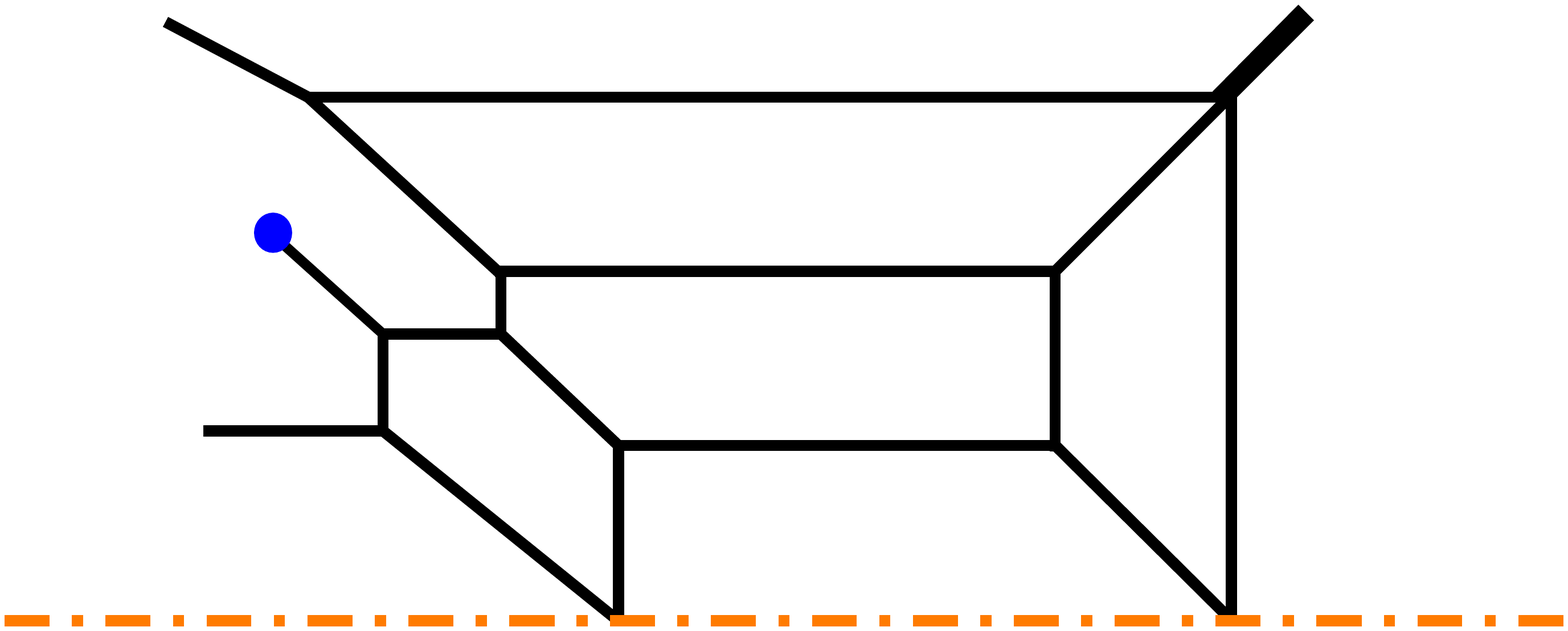} \label{fig:Sp2w2AS1}}
\subfigure[]{
\includegraphics[width=6cm]{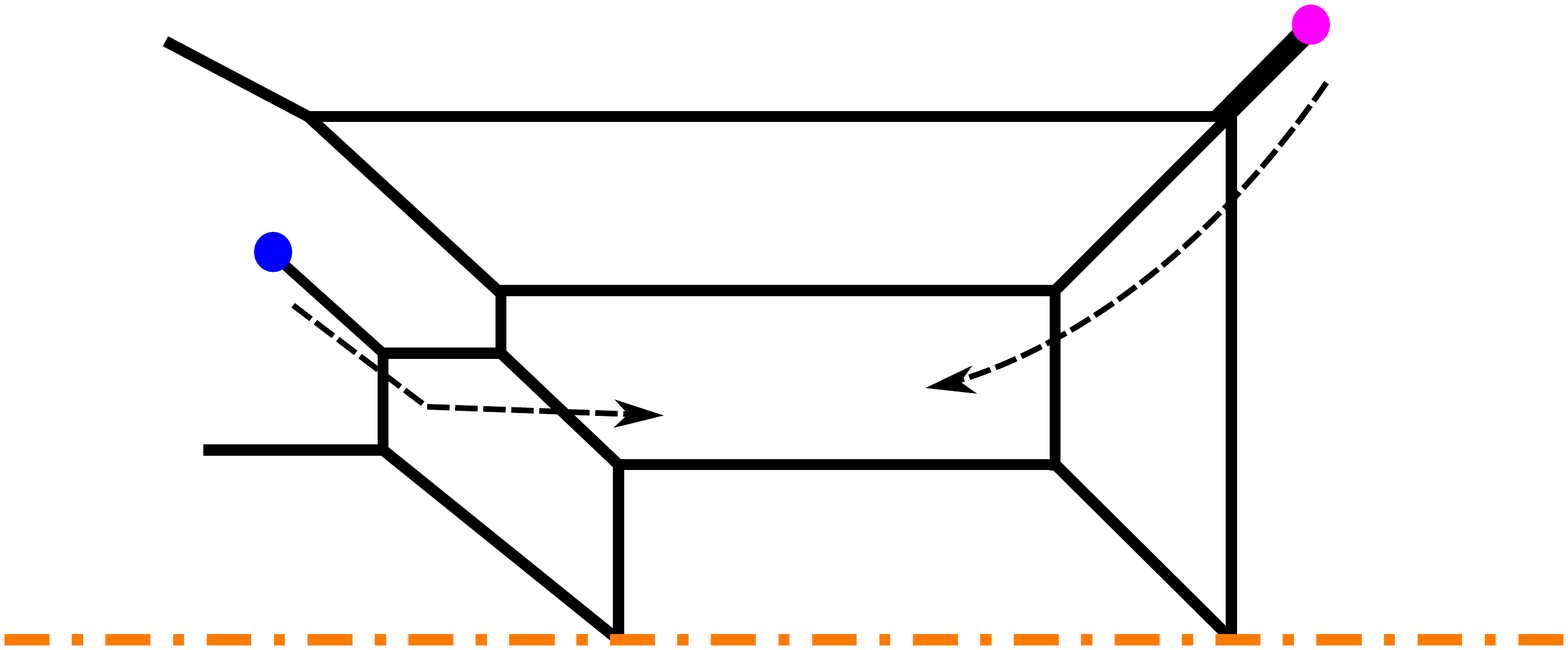} \label{fig:Sp2w2AS2}}
\caption{(a): The diagram obtained after applying three flop transitions to the diagram in Figure \ref{fig:G2w2flvrsb}. (b): Moving an external $(1, -1)$ 7-brane and an external $(1, 1)$ 7-brane.}
\label{fig:Sp2w2AS1to2}
\end{figure}

\begin{figure}
\centering
\subfigure[]{
\includegraphics[width=6cm]{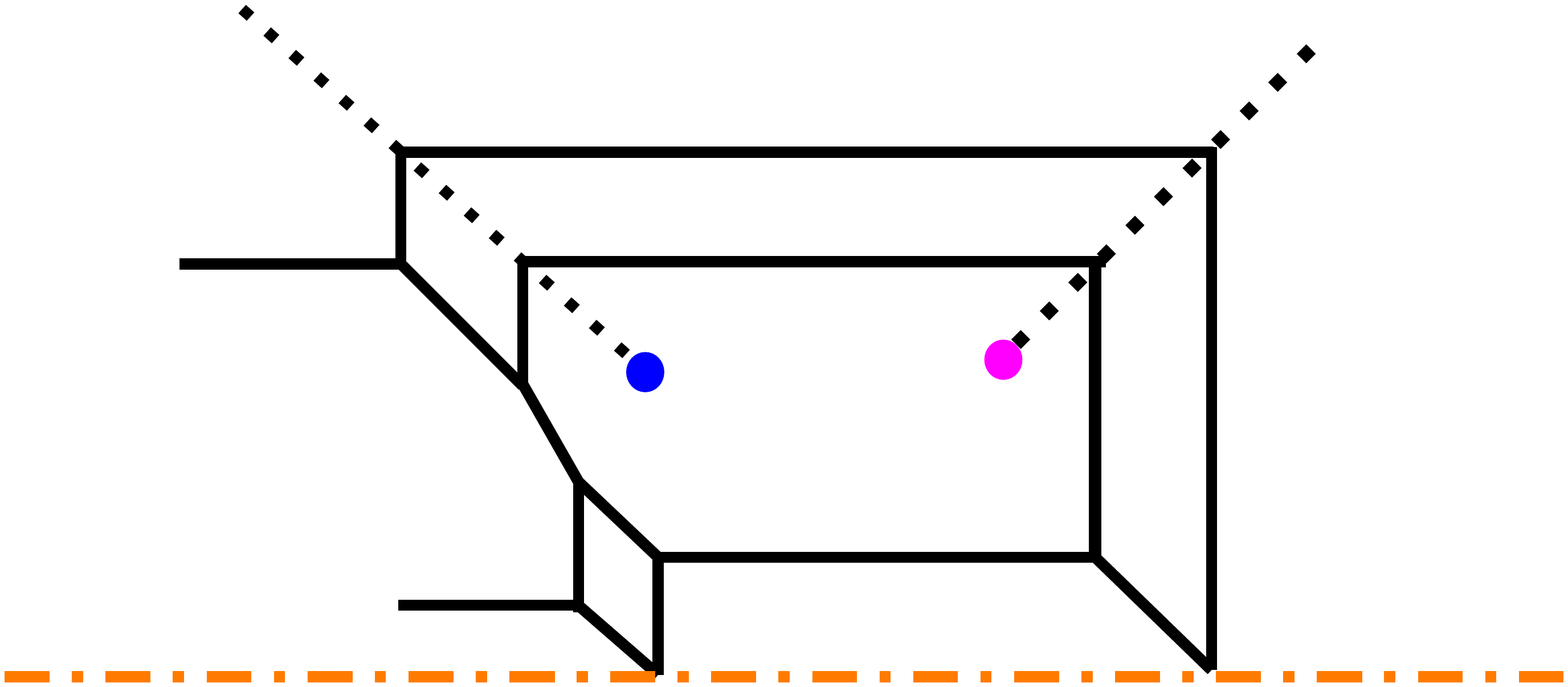} \label{fig:Sp2w2AS3}}
\subfigure[]{
\includegraphics[width=6cm]{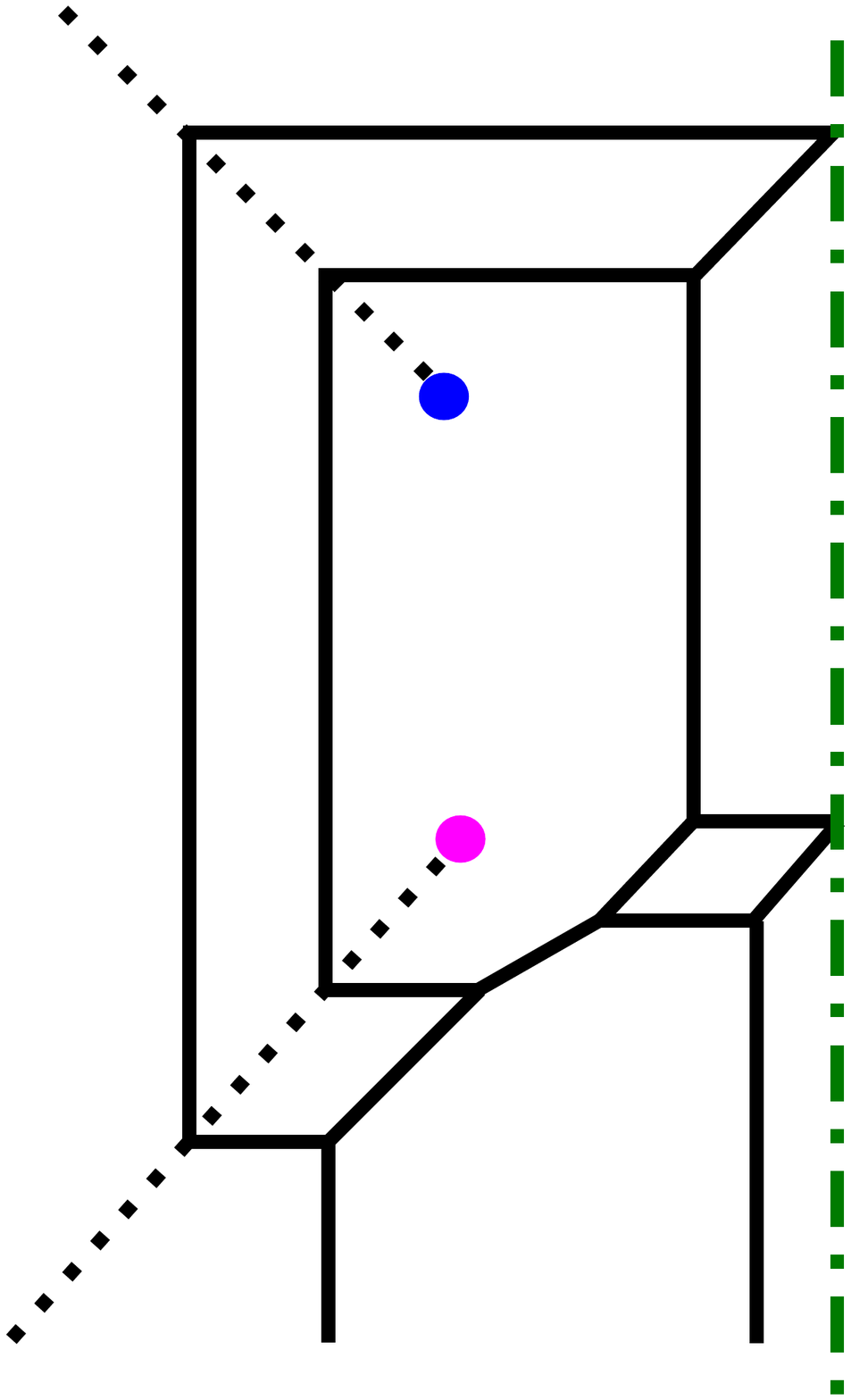} \label{fig:Sp2w2AS3a}}
\caption{(a): The diagram after moving the 7-branes in Figure \ref{fig:Sp2w2AS2}. (b): The diagram which is S-dual to the one in Figure \ref{fig:Sp2w2AS3}.}
\label{fig:Sp2w2AS3to3a}
\end{figure}

\begin{figure}
\centering
\subfigure[]{
\includegraphics[width=6cm]{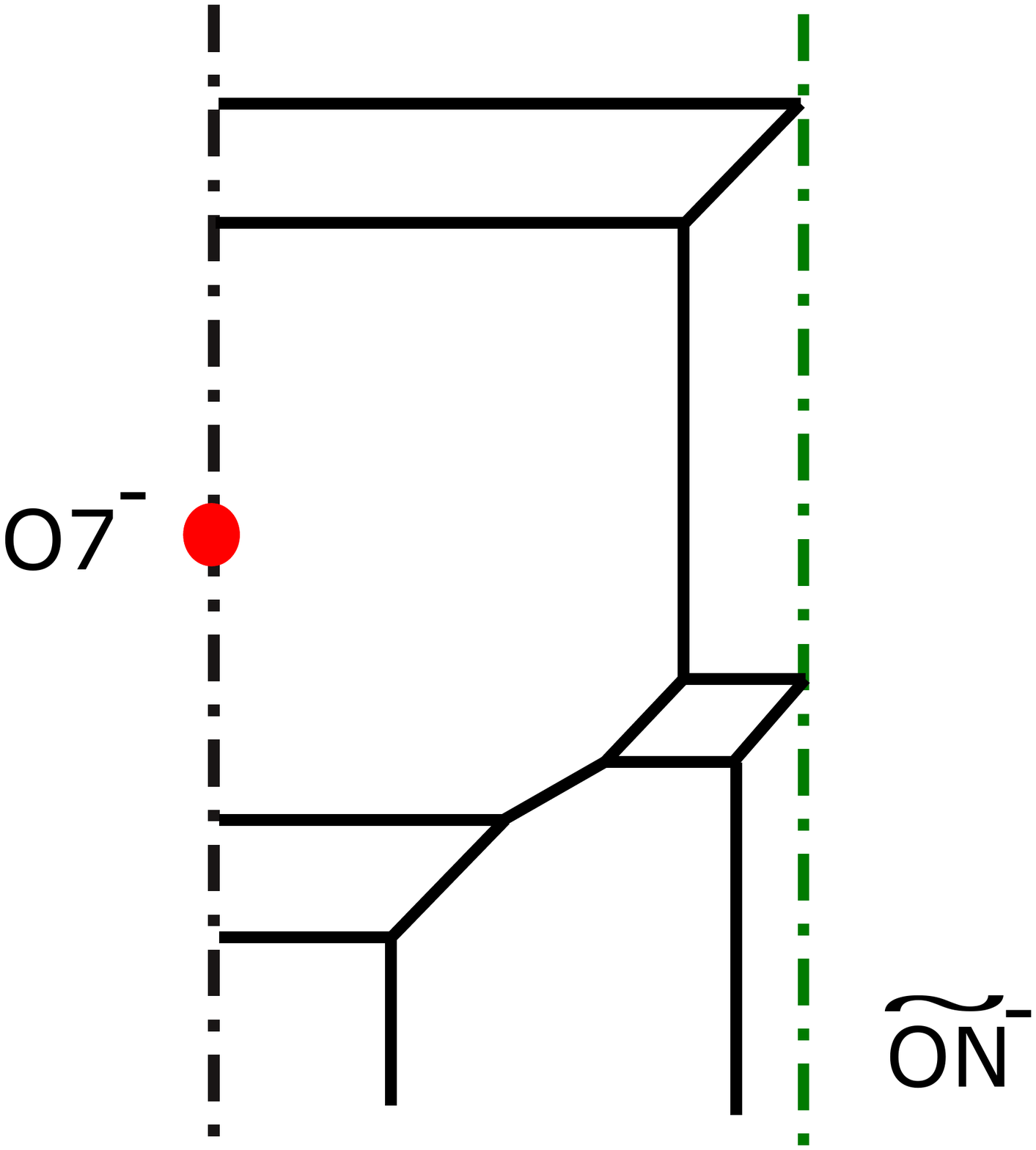} \label{fig:Sp2w2AS4}}
\subfigure[]{
\includegraphics[width=6cm]{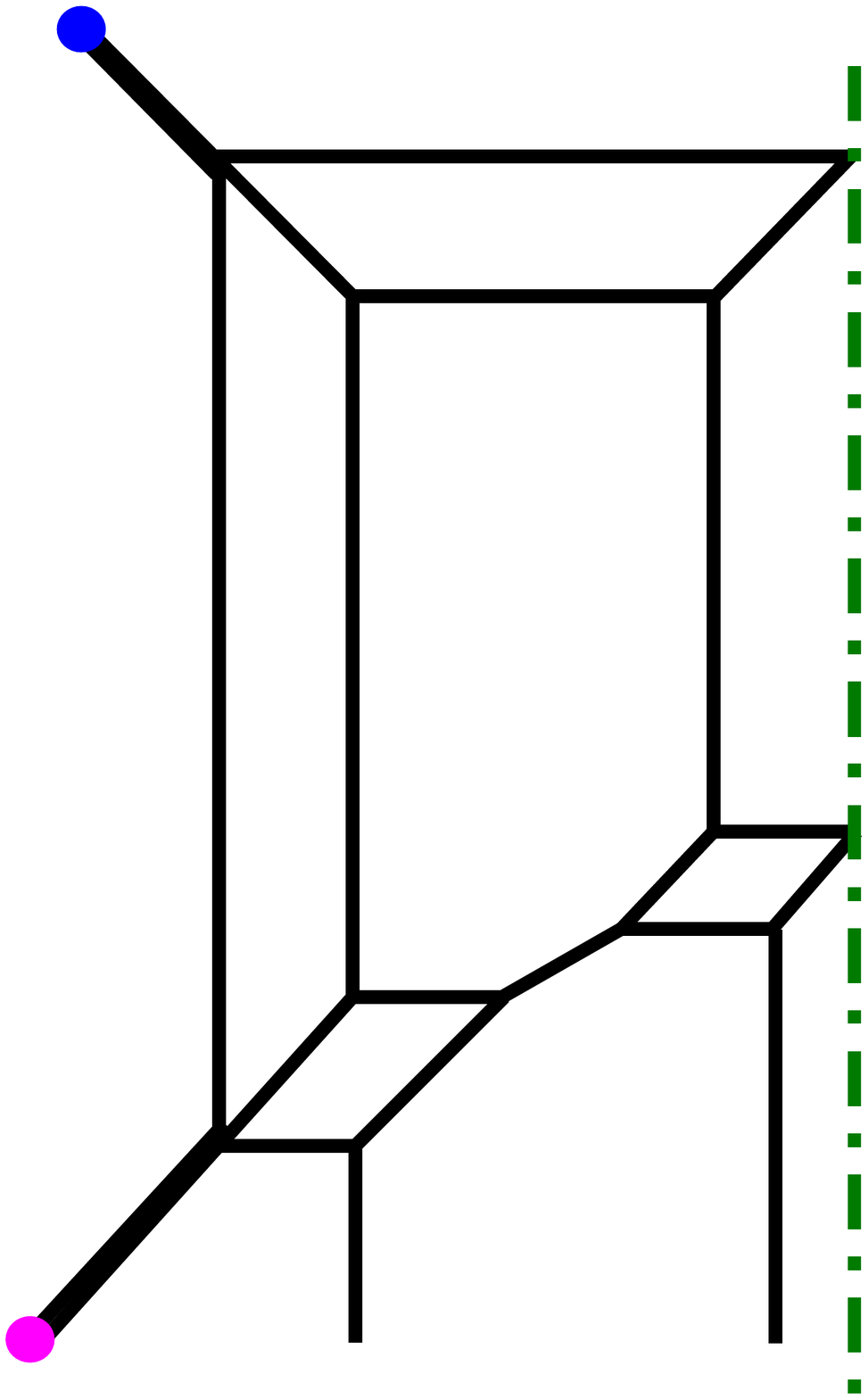} \label{fig:Sp2w2AS5}}
\caption{(a): Forming an O7$^-$-plane from the $(1, -1)$ 7-brane and the $(1, 1)$ 7-brane in Figure \ref{fig:Sp2w2AS3a}. (b): Moving the $(1, -1)$ 7-brane and the $(1. 1)$ 7-brane in Figure \ref{fig:Sp2w2AS3a} to infinitely far away.}
\label{fig:Sp2w2AS4to5}
\end{figure}

We first start from the 5-brane web for the $G_2$ gauge theory with two flavors in Figure \ref{fig:G2w2flvrsb}. In order to see the duality, we first perform two flop transitions and obtain a diagram in Figure \ref{fig:Sp2w2AS1}. We then move a $(1, -1)$ 7-brane and a $(1, 1)$ 7-brane according the arrows in Figure \ref{fig:Sp2w2AS2}. 
%%%%%%%%%%%%%%%%%%%%%%%%%%%%%%%%%
%%%%%%%%%%%%%%%%%%%%%%%%%%%%%%%%%
After moving the two 7-branes the diagram becomes the one in Figure \ref{fig:Sp2w2AS3}. At this stage, we apply the S-duality to the diagram in Figure \ref{fig:Sp2w2AS3}. Then the resulting configuration contains a pair of a $(1, -1)$ 7-brane and a $(1, 1)$ 7-brane in the same 5-brane chamber as in Figure \ref{fig:Sp2w2AS3a}. 
%%%%%%%%%%%%%%%%%%%%%%%%%%%%%%%%%
%%%%%%%%%%%%%%%%%%%%%%%%%%%%%%%%%
The $(1, -1)$ 7-brane and the $(1, 1)$ 7-brane may form an O7$^-$-plane \cite{Sen:1996vd} and we obtain the configuration in Figure \ref{fig:Sp2w2AS4}. Since we have four color D5-branes with an O7$^-$-plane, the theory realized by the diagram in Figure \ref{fig:Sp2w2AS4} may be an $Sp(2)$ gauge theory. 
%%%%%%%%%%%%%%%%%%%%%%%%%%%%%%%%%
%%%%%%%%%%%%%%%%%%%%%%%%%%%%%%%%%

\begin{figure}
\centering
\subfigure[]{
\includegraphics[width=6cm]{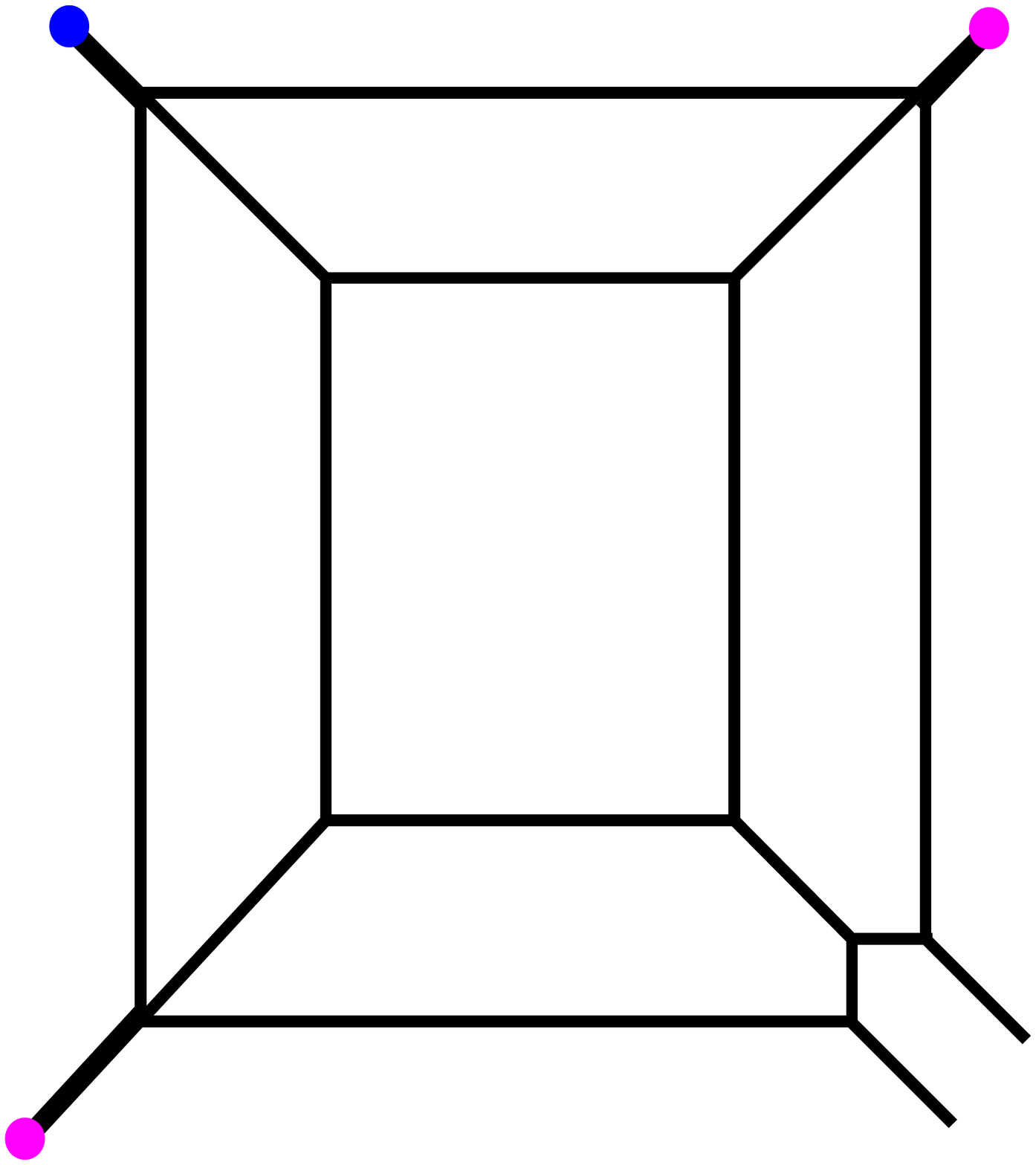} \label{fig:Sp2w1AS}}
\subfigure[]{
\includegraphics[width=6cm]{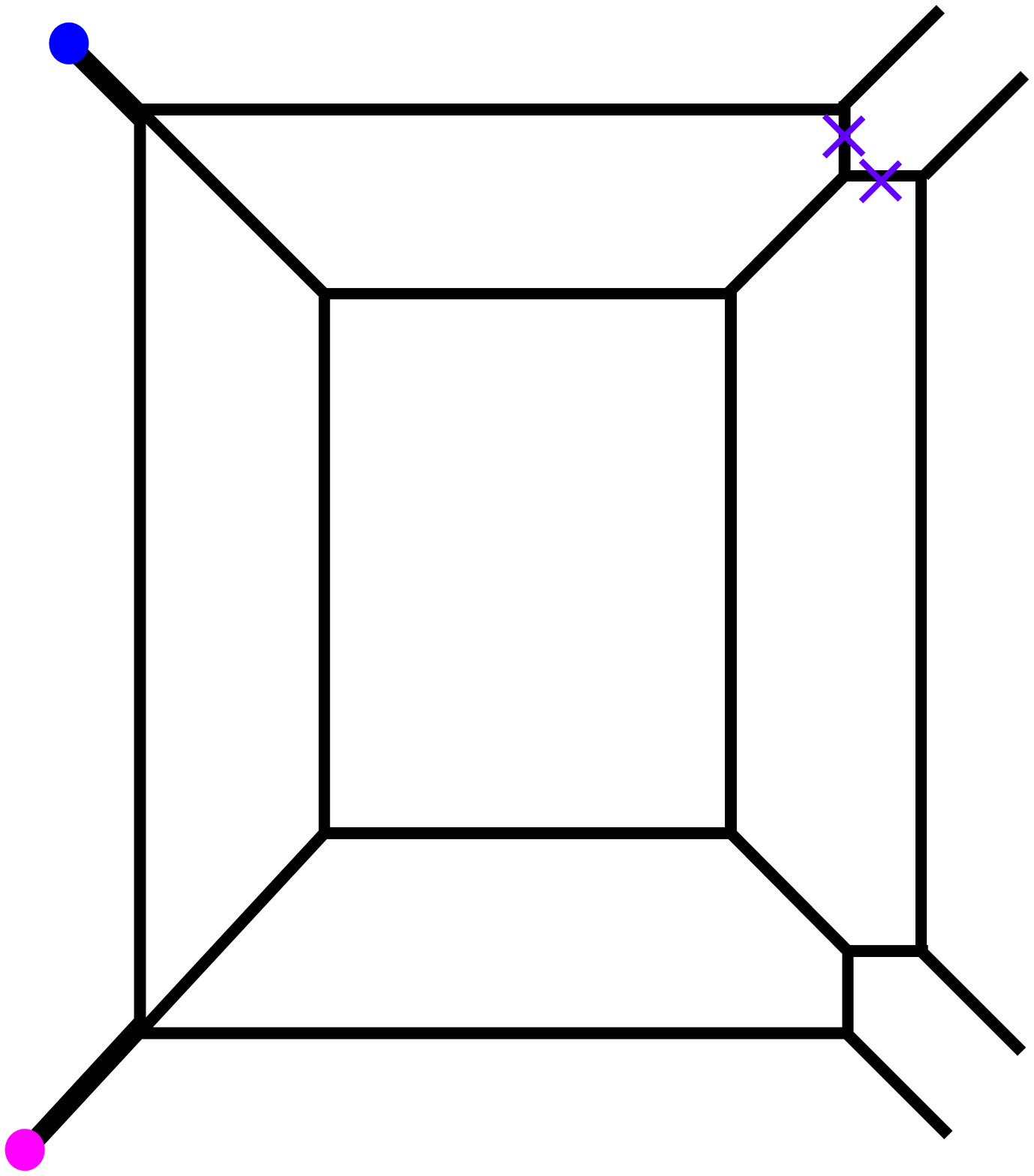} \label{fig:Sp2SU2}}
\caption{(a): A 5-brane web for the $Sp(2)$ gauge theory with one hypermultiplet in the antisymmetric representation. (b): A 5-brane web diagram for the $Sp(2)_0 - SU(2)_0$ quiver. Tuning the length of the 5-brane with the purple $\times$ yields the 5-brane in Figure \ref{fig:Sp2w1AS}.}
\label{fig:HiggsingtoSp2w1AS}
\end{figure}

A next question is whether the diagram in Figure \ref{fig:Sp2w2AS4} contains two hypermultiplets in the antisymmetric representation of the $Sp(2)$. As we saw in section \ref{sec:dualtoSU3w2flvrs}, the presence of the antisymmetric hypermultiplets can be understood from a Higgsing of an $SU(2)_0 - Sp(2) - SU(2)_0$ quiver theory also for an $Sp(2)$ gauge theory. For that let us first see how the antisymmetric hypermultiplet of an $Sp(2)$ gauge theory can appear from a 5-brane web. A 5-brane diagram for the $Sp(2)$ gauge theory with one antisymmetric hypermultiplet and the zero discrete theta angle is given in Figure \ref{fig:Sp2w1AS} \cite{Bergman:2015dpa}. The two external $(1, 1)$ 5-branes in Figure \ref{fig:Sp2w1AS} realizes an $SU(2)$ flavor symmetry from the one antisymmetric hypermultiplet. 
%%%%%%%%%%%%%%%%%%%%%%%%%%%%%%%%%
%%%%%%%%%%%%%%%%%%%%%%%%%%%%%%%%%
It is also possible to see that the diagram for the $Sp(2)_0$ gauge theory with one antisymmetric hypermultiplet in Figure \ref{fig:Sp2w1AS} can be obtained from a Higgsing of the $Sp(2)_0 - SU(2)_0$ quiver theory. A 5-brane web diagram of the $Sp(2) - SU(2)$ quiver theory with the zero discrete theta angle for both gauge groups is depicted in Figure \ref{fig:Sp2SU2}. The diagram shows an $SU(2) \times SU(2)$ flavor symmetry generated non-perturbatively from the viewpoint of the quiver theory. We can then perform a Higgsing associated to one of the $SU(2)$ flavor symmetry by tuning the length of 5-branes indicated by the purple $\times$ in Figure \ref{fig:Sp2SU2}. Then the Higgsing precisely yields the diagram in Figure \ref{fig:Sp2w1AS}. Therefore, the $Sp(2)$ gauge theory with a hypermultiplet in the antisymmetric representation and the zero discrete theta angle can be obtained from the Higgsing of the $Sp(2)_0 - SU(2)_0$ quiver theory. 

\begin{figure}
\centering
\subfigure[]{
\includegraphics[width=6cm]{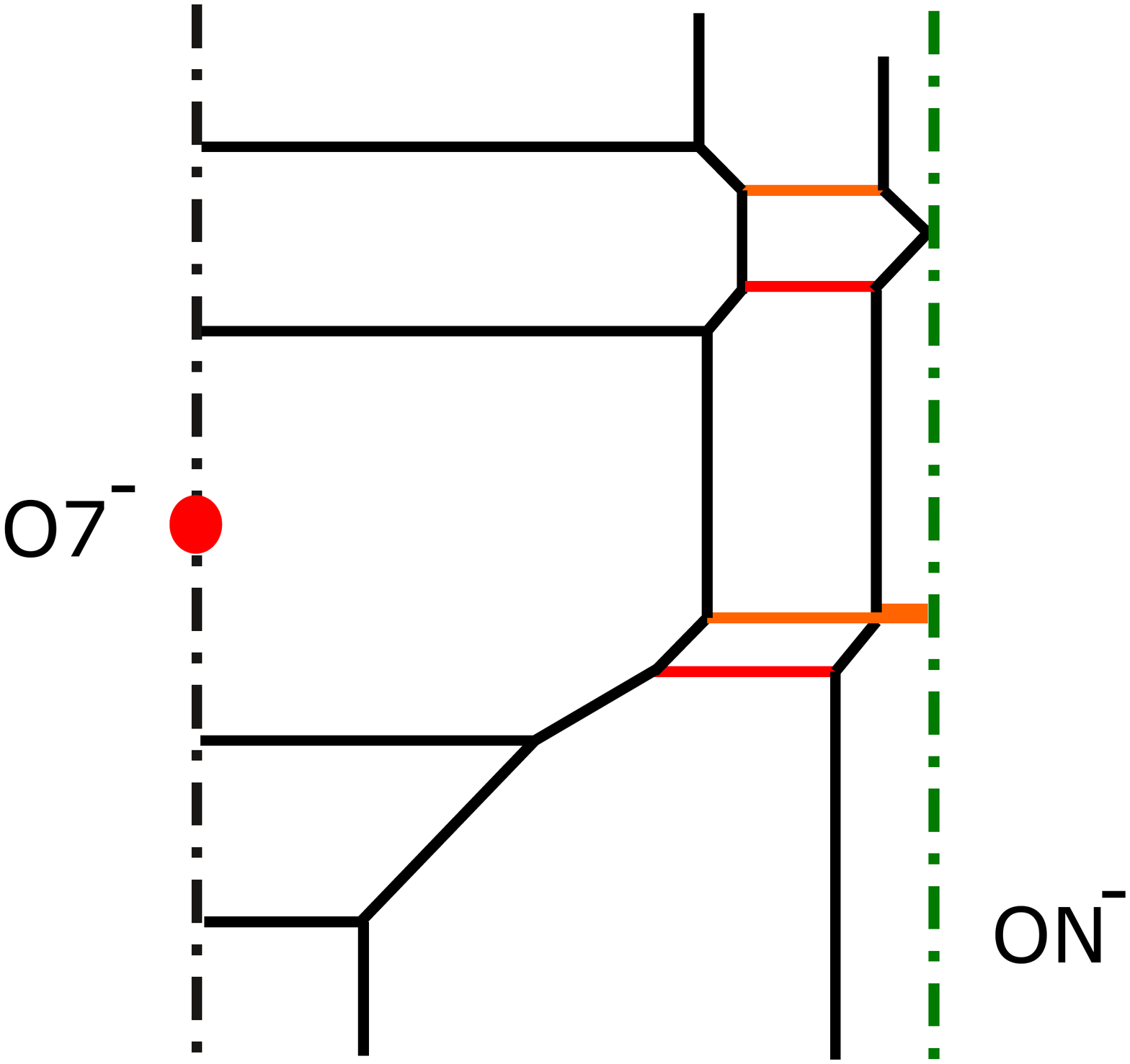} \label{fig:Sp2SU2SU2}}
\subfigure[]{
\includegraphics[width=6cm]{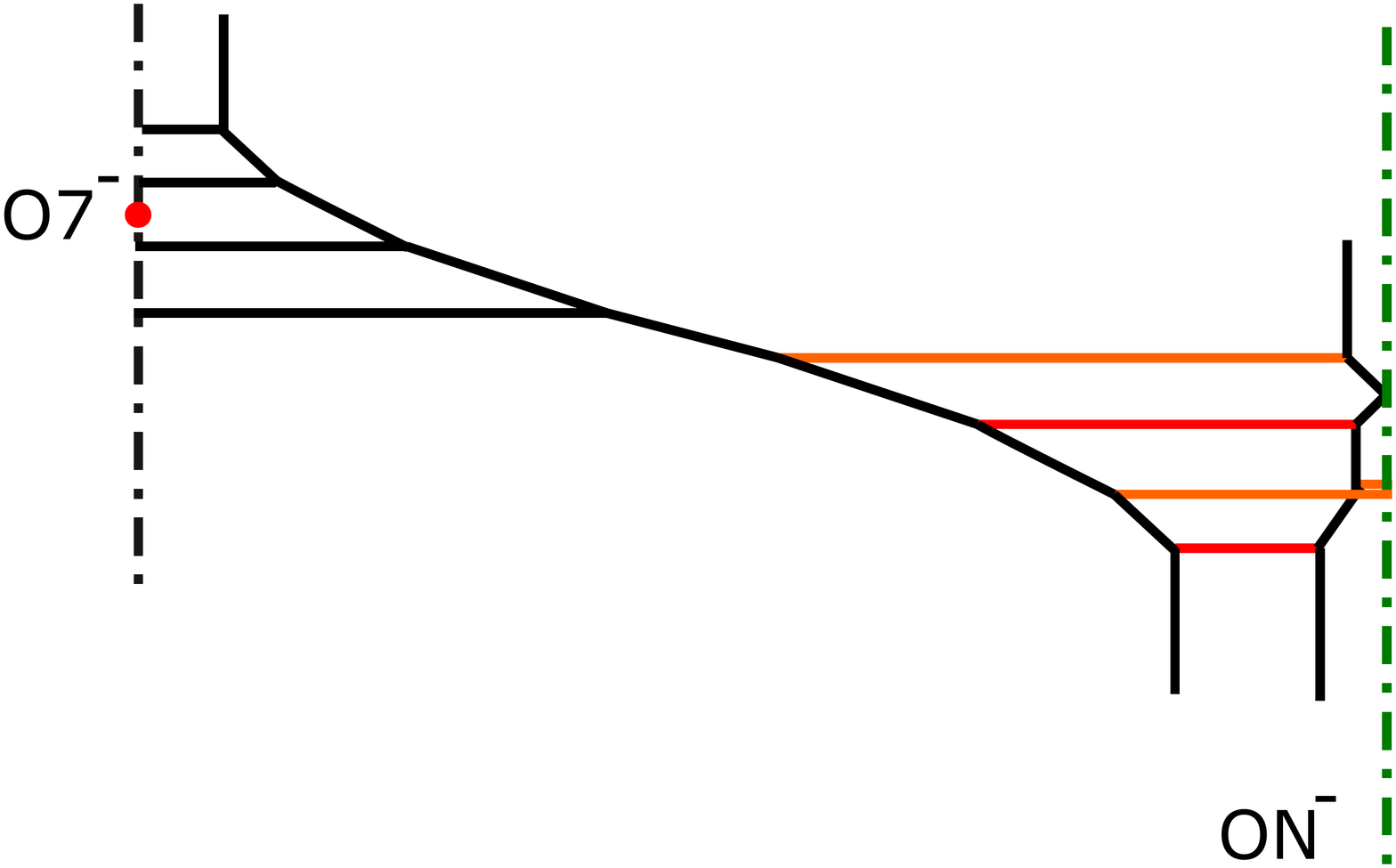} \label{fig:Sp2piSU2SU2}}
\caption{(a): A 5-brane diagram for an $SU(2)_0 - Sp(2)_{\pi} - SU(2)_0$ quiver gauge theory. The D5-branes in red yield color branes for one $SU(2)$ and the D5-branes in orange gives color branes for another $SU(2)$. (b): Another description for the diagram in Figure \ref{fig:Sp2SU2SU2} after performing several flop transitions. We can see that the $Sp(2)$ gauge theory has the non-trivial discrete theta angle.}
\label{fig:SP2SU2SU2}
\end{figure}

We now apply the Higgsing argument to a diagram with an ON$^-$-plane. Namely we consider a Higgsing of an $SU(2)-Sp(2)-SU(2)$ quiver theory to obtain a 5-brane diagram of an $Sp(2)$ gauge theory with two hypermultiplets in the antisymmetric representation. An $SU(2) - Sp(2) - SU(2)$ quiver theory is realized by a diagram in Figure \ref{fig:Sp2SU2SU2}. 
%%%%%%%%%%%%%%%%%%%%%%%%%%%%%%%%%
%%%%%%%%%%%%%%%%%%%%%%%%%%%%%%%%%
In Figure \ref{fig:Sp2SU2SU2}, the two $SU(2)$ gauge theories are realized from an ON$^-$-plane. The D5-branes in the red color represents one $SU(2)$ and the D5-branes in the orange color gives another $SU(2)$. From the parallel external $(0, 1)$ 5-branes, the diagram exhibits an $SO(4) \times SO(4)$ flavor symmetry. It turns out that we need to consider the $Sp(2)$ gauge theory with the discrete theta angle $\pi$ in order to connect to the diagram in Figure \ref{fig:Sp2w2AS4}. The discrete theta angle of the $Sp(2)$ gauge group can be more easily seen from the diagram in Figure \ref{fig:Sp2piSU2SU2}. When we take the length of the middle $(4, -1)$ 5-brane to be infinitely large, then the diagram is decomposed into two parts. The left part gives an $Sp(2)$ gauge theory and the right part yields the two disconnected $SU(2)$ gauge theories. Then the $Sp(2)$ gauge theory realized in the left part of the diagram in Figure \ref{fig:Sp2piSU2SU2} has the discrete theta angle $\pi$ \cite{Bergman:2015dpa, Hayashi:2016jak}. On the other hand, the diagram for the two disconnected $SU(2)$ gauge theories exhibits an $SO(4) \simeq SU(2) \times SU(2)$ flavor symmetry. Therefore the two $SU(2)$ gauge theories both have the zero discrete theta angle.

\begin{figure}
\centering
\subfigure[]{
\includegraphics[width=6cm]{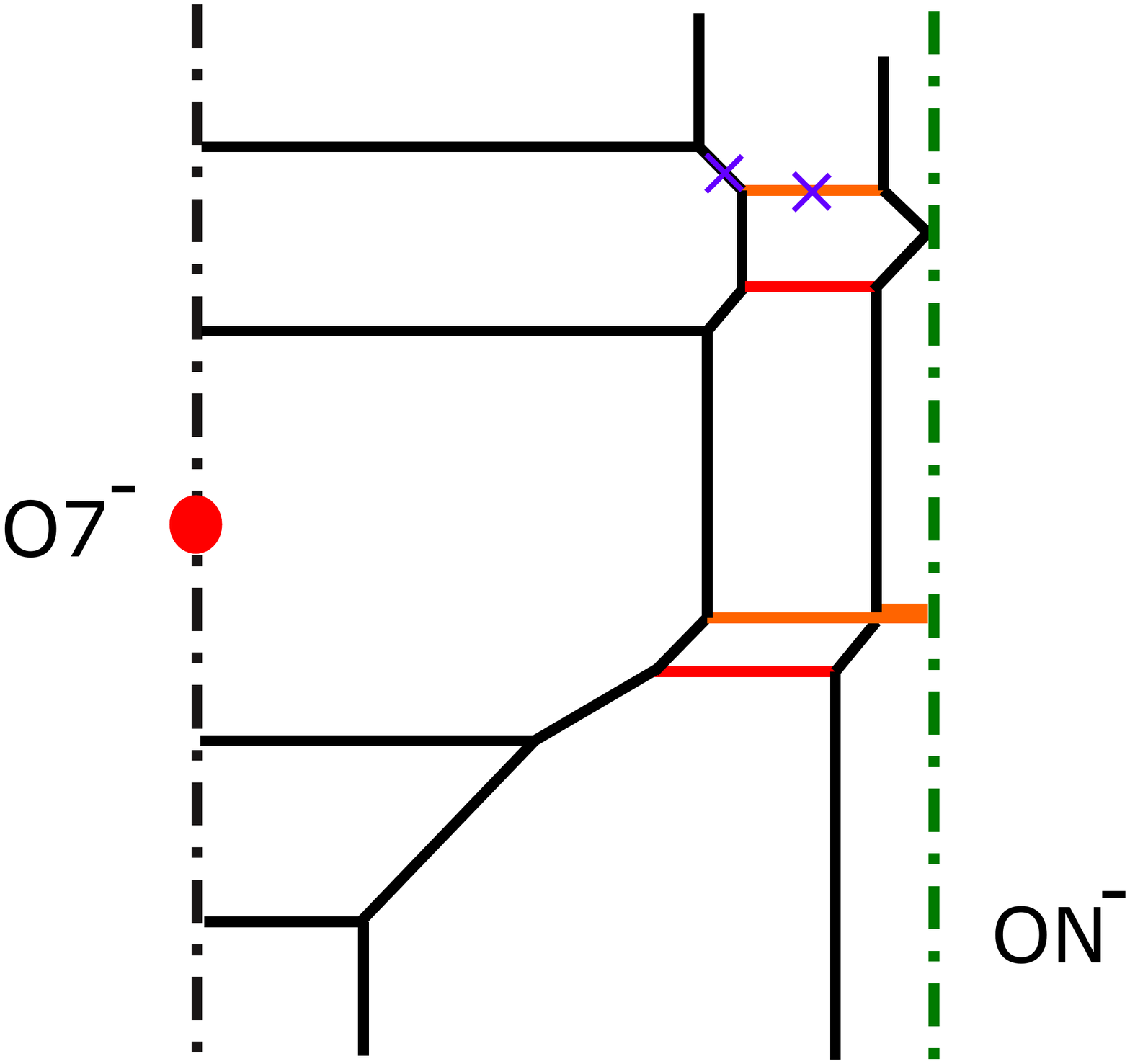} \label{fig:Sp2SU2SU2Higgs1}}
\subfigure[]{
\includegraphics[width=6cm]{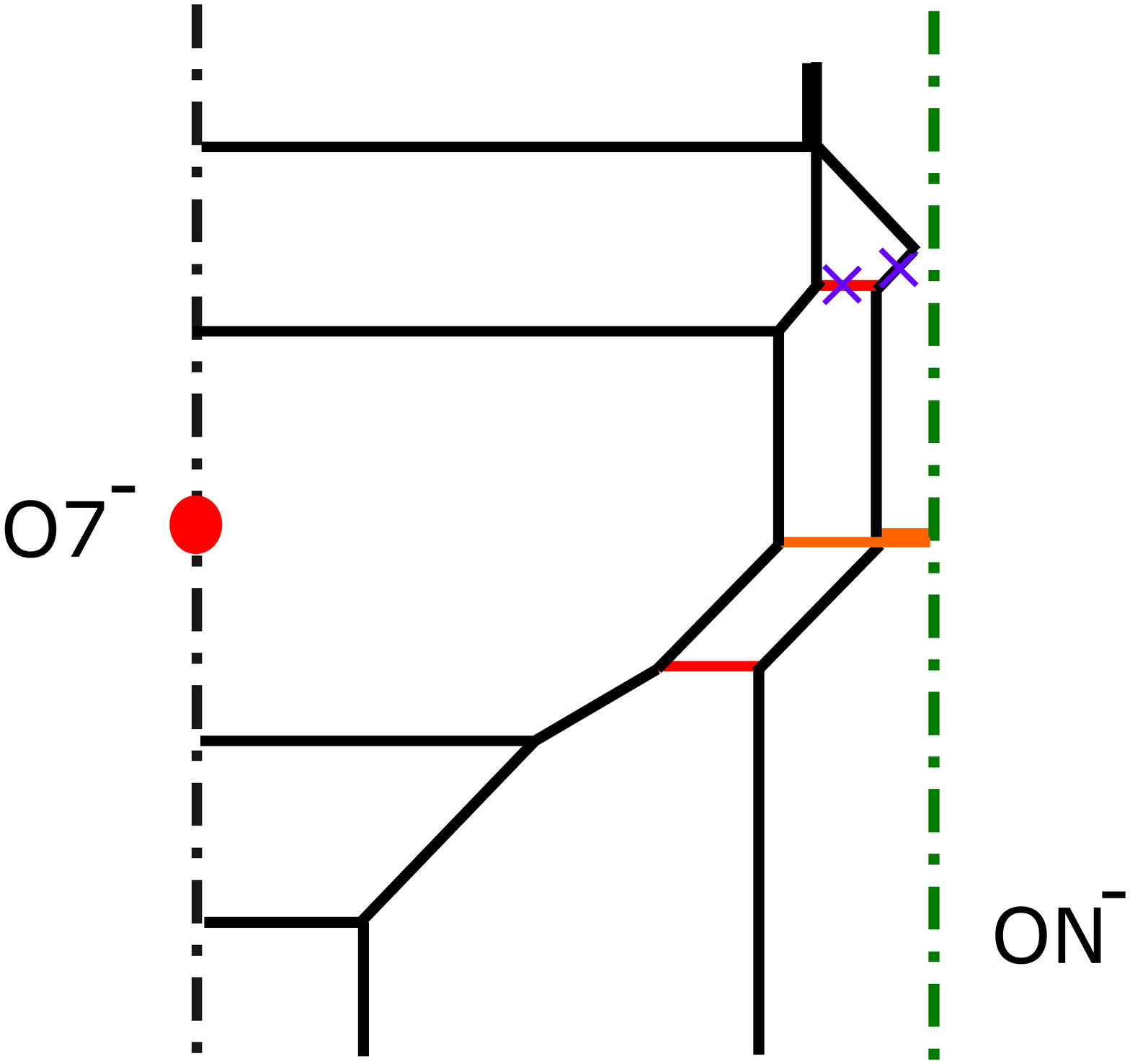} \label{fig:Sp2SU2SU2Higgs2}}
\subfigure[]{
\includegraphics[width=6cm]{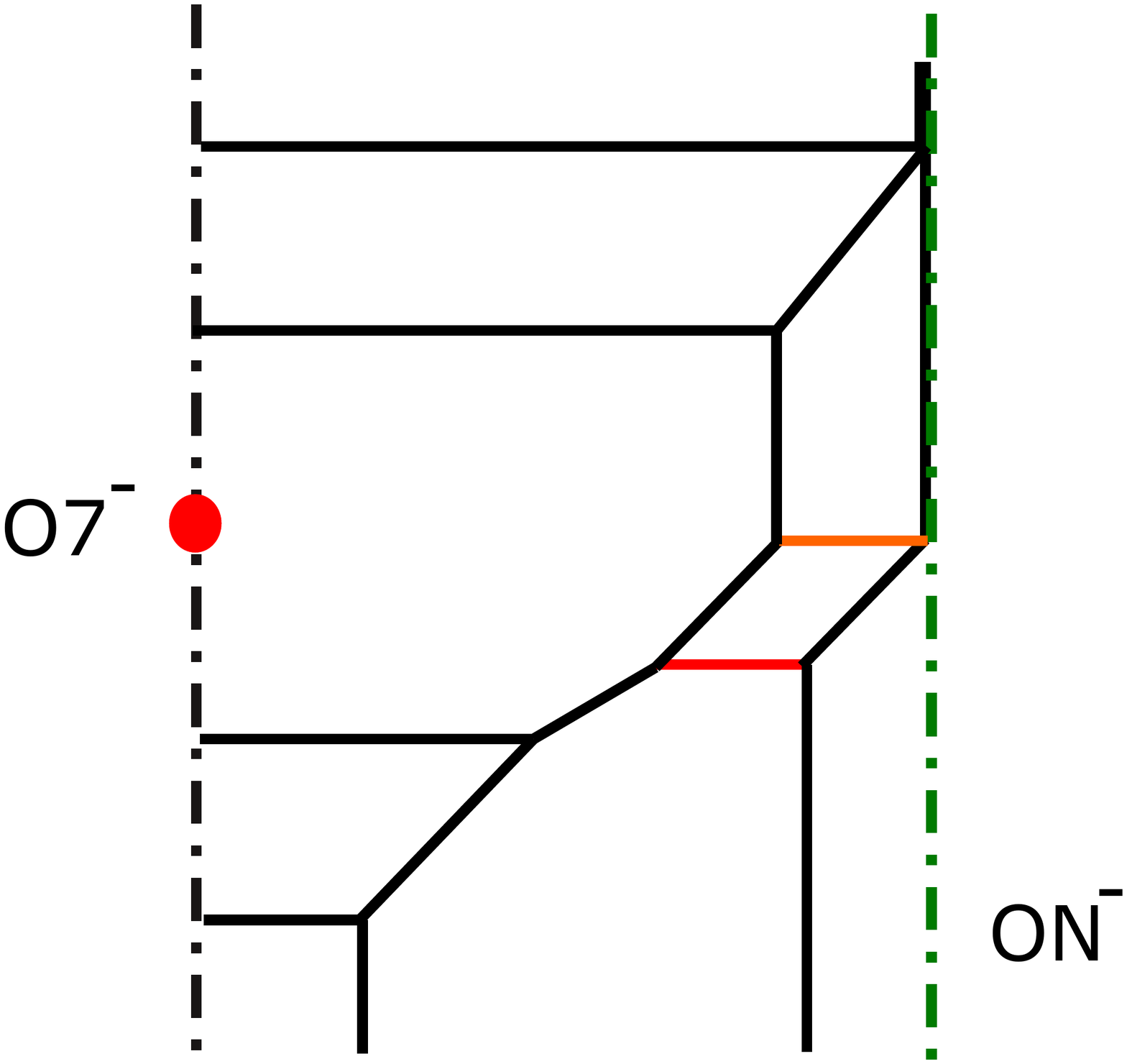} \label{fig:Sp2SU2SU2Higgs3}}
\caption{(a): Tuning for the first Higgsing. (b): Tuning for the second Higgsing. (c): The diagram after the two Higgsings.}
\label{fig:Sp2SU2SU2Higgs}
\end{figure}

From the diagram in Figure \ref{fig:Sp2SU2SU2}, we consider two Higgsings which break the two $SU(2)$ gauge groups. The Higgsing will also reduce the flavor group from $SO(4) \times SO(4) \to SO(4)$. We in particular consider a Higgsing which breaks the $SO(4)$ flavor symmetry associated to the parallel $(0, 1)$ 5-branes going in the upper direction. The Higgsing can be understood by two steps where each step is associated to each $SU(2)$ in the $SO(4)$. The first Higgsings is carried out by tuning the length of the 5-branes with the purple $\times$ in Figure \ref{fig:Sp2SU2SU2Higgs1} and the second Higgsing is done by setting the length of the 5-branes with the purple $\times$ in Figure \ref{fig:Sp2SU2SU2Higgs2} to be zero. 
%%%%%%%%%%%%%%%%%%%%%%%%%%%%%%%%%
%%%%%%%%%%%%%%%%%%%%%%%%%%%%%%%%%
The two Higgsings leave only one color brane for each $SU(2)$ gauge groups and the two $SU(2)$ gauge groups are broken. The resulting diagram after the Higgsing is depicted in Figure \ref{fig:Sp2SU2SU2Higgs3}. The diagram in Figure \ref{fig:Sp2SU2SU2Higgs3} is in fact equivalent to the diagram in Figure \ref{fig:Sp2w2AS4} by moving an $(0, 1)$ 7-brane which may be put at the end the external $(0, 1)$ 5-brane extending in the upper direction \cite{Zafrir:2015ftn, Hayashi:2018bkd}. Since the original $Sp(2)$ gauge theory has the non-trivial discrete theta angle, the theory after the Higgsing will also has the discrete theta angle $\pi$. Therefore, we can conclude that the diagram in Figure \ref{fig:Sp2w2AS4} may realize the $Sp(2)$ gauge theory with two hypermultiplets in the antisymmetric representation and the discrete theta angle $\pi$. Namely, the deformation of the diagram also shows that the $G_2$ gauge theory with two flavors is dual to the $Sp(2)$ gauge theory with two antisymmetric hypermultiplets and the discrete theta angle $\theta = \pi$.

\begin{figure}
\centering
\includegraphics[width=8cm]{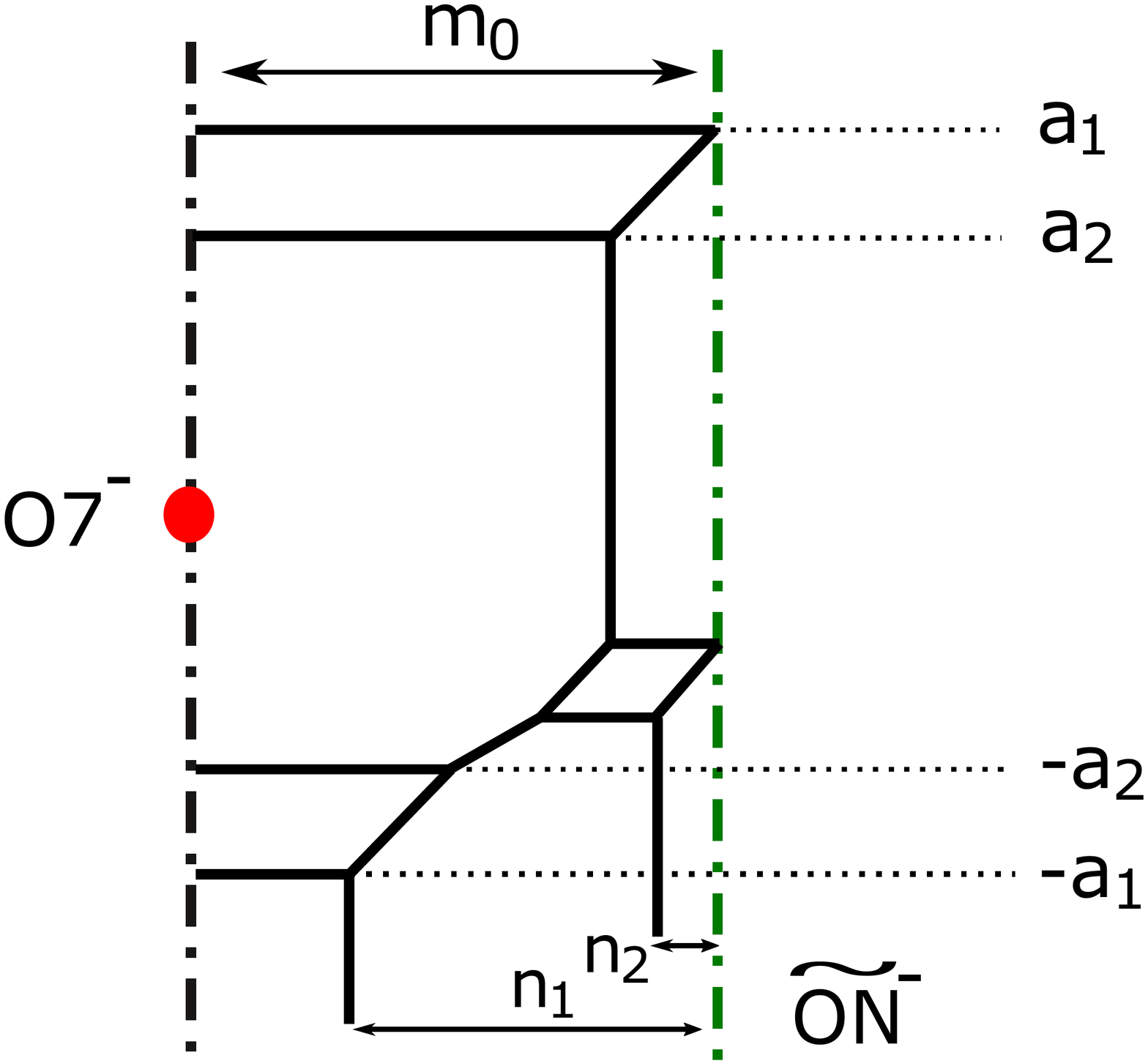}
\caption{The gauge theory parameterization for the $Sp(2)$ gauge theory with two antisymmetric hypermultiplets. }
\label{fig:Sp2w2ASpara}
\end{figure}

Let us also check if the diagram in Figure \ref{fig:Sp2w2AS4} or equivalently in Figure \ref{fig:Sp2w2AS5} yields the $Sp(2)_{\pi}$ gauge theory with two antisymmetric hypermultiplets from the calculation of the prepotential. The gauge theory parameterization for the $Sp(2)_{\pi}$ gauge theory with two antisymmetric hypermultiplets is given in Figure \ref{fig:Sp2w2ASpara}. $m_0$ is the inverse of the squared gauge coupling, $a_1, a_2$ are the Coulomb branch moduli of the $Sp(2)_{\pi}$ gauge theory. 
%%%%%%%%%%%%%%%%%%%%%%%%%%%%%%%%%
%%%%%%%%%%%%%%%%%%%%%%%%%%%%%%%%%
$n_1, n_2$ are related to the mass parameters for the two antisymmetric hypermultiplets. Note that $n_1, n_2$ are related to the chemical potentials for the $SO(5)$ flavor symmetry. On the other hand the mass parameters for the two antisymmetric hypermultiplets correspond to the chemical potentials for the $Sp(2)$ flavor symmetry. Hence the two mass parameters are given by
\be
n_1 = m_1 + m_2, \quad n_2 = m_1 - m_2. \label{Sp2w2ASmass}
\ee

\begin{figure}
\centering
\includegraphics[width=8cm]{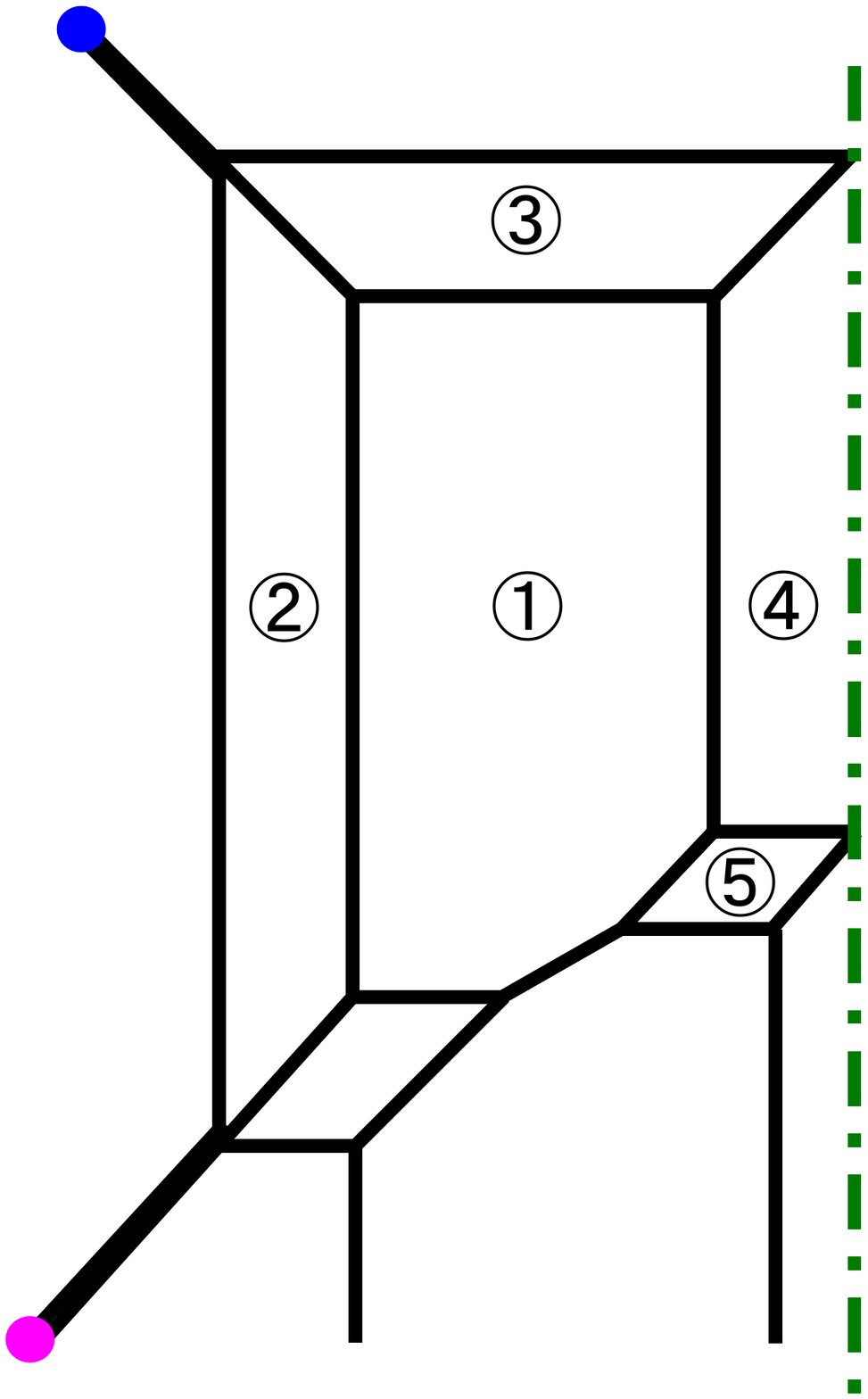}
\caption{A label for the faces in the diagram for the $Sp(2)_{\pi}$ gauge theory with two hypermultiplets in the antisymmetric representation.}
\label{fig:Sp2w2ASface}
\end{figure}

With the parameterization in Figure \ref{fig:Sp2w2ASpara} and \eqref{Sp2w2ASmass}, we can express the area of the faces whose label is depicted in Figure \ref{fig:Sp2w2ASface} in terms of the parameters of the $Sp(2)$ gauge theory with two hypermultiplets in the antisymmetric representation and the non-trivial discrete theta angle. 
%%%%%%%%%%%%%%%%%%%%%%%%%%%%%%%%%
%%%%%%%%%%%%%%%%%%%%%%%%%%%%%%%%%
The area of the five faces is given by
\bea
\textcircled{\scriptsize 1} &=& 2a_2m_0 - \frac{m_1^2}{2} + m_1(a_1 - a_2)  - \frac{m_2^2}{2} +m_2(a_1 - a_2) - a_1^2 + 3a_2^2, \label{area1.Sp2w2AS}\\
\textcircled{\scriptsize 2} &=& (a_1 - a_2)(m_0 - m_1 - m_2 + 2a_1 + a_2), \label{area2.Sp2w2AS}\\
\textcircled{\scriptsize 3} &=&(a_1 - a_2)(m_0 + a_2), \label{area3.Sp2w2AS}\\
\textcircled{\scriptsize 4} &=& \frac{1}{2}(a_1-a_2)(-2m_1 + 3a_1 + a_2),\label{area4.Sp2w2AS}\\
\textcircled{\scriptsize 5} &=& (a_1-a_2)(m_1-m_2). \label{area5.Sp2w2AS}
\eea

A linear combination of the area \eqref{area1.Sp2w2AS}-\eqref{area5.Sp2w2AS} corresponds to the tension of the monopole string and it can be computed also from a derivative of the effective prepotential with respect to a Coulomb branch modulus. The effective prepotential is computed from the general formula in \eqref{prepotential} and the diagram in Figure \ref{fig:Sp2w2AS4} corresponds to the phase 
\bea
a_1 + a_2 - m_1 > 0, \quad a_1 - a_2 - m_1 < 0, \quad -a_1 + a_2 - m_1 < 0, \quad -a_1 - a_2 - m_1 < 0,\nn\\
\eea
for one antisymmetric hypermultiplet with mass $m_1$ and 
\bea
a_1 + a_2 - m_2 > 0, \quad a_1 - a_2 - m_2 < 0, \quad -a_1 + a_2 - m_2 < 0, \quad -a_1 - a_2 - m_2 < 0,\nn\\
\eea
for the other antisymmetric hypermultiplet with mass $m_2$. Then the effective prepotential for the $Sp(2)_{\pi}$ gauge theory with the two antisymmetric hypermultiplets on the phase becomes
\bea
\mathcal{F}_{Sp(2)_{\pi}+2{\bf AS}} &=& m_0(2\phi_1^2 - 2\phi_1\phi_2 + \phi_2^2) + \frac{1}{6}(-m_1^3 - 3m_1^2\phi_2 + 3m_1(-4\phi_1^2+ 4\phi_1\phi_2- \phi_2^2)\nn\\
&&-m_2^3 -3m_2^2\phi_2 + 3m_2(-4\phi_1^2 + 4\phi_1\phi_2 -\phi_2^2) + 8\phi_1^3 + 12\phi_1^2\phi_2 - 18\phi_1\phi_2^2 +6\phi_2^3),\nn\\ \label{Sp2w2AS.prepot}
\eea
where we used the Dynkin basis for $Sp(2)$
\be
a_1 = \phi_1, \quad a_2 = -\phi_1 + \phi_2 \label{Sp2.Dynkinbasis}
\ee

Then taking the derivative of \eqref{Sp2w2AS.prepot} with respect to $\phi_1, \phi_2$ should correspond to the linear combination $\textcircled{\scriptsize 2} + \textcircled{\scriptsize 3} + 2\times\textcircled{\scriptsize 4} + \textcircled{\scriptsize 5}$ and $\textcircled{\scriptsize 1}$ respectively. Indeed the explict comparison between \eqref{area1.Sp2w2AS}-\eqref{area5.Sp2w2AS} and \eqref{Sp2w2AS.prepot} gives
\bea
\frac{\partial \mathcal{F}_{(Sp(2)_{\pi}+2{\bf AS})}}{\partial \phi_1} &=&\textcircled{\scriptsize 2} + \textcircled{\scriptsize 3} + 2\times\textcircled{\scriptsize 4} + \textcircled{\scriptsize 5}, \label{tension1.Sp2w2AS}\\
\frac{\partial \mathcal{F}_{(Sp(2)_{\pi}+2{\bf AS})}}{\partial \phi_2} &=&\textcircled{\scriptsize 1}. \label{tension2.Sp2w2AS}
\eea

\begin{figure}
\centering
\includegraphics[width=8cm]{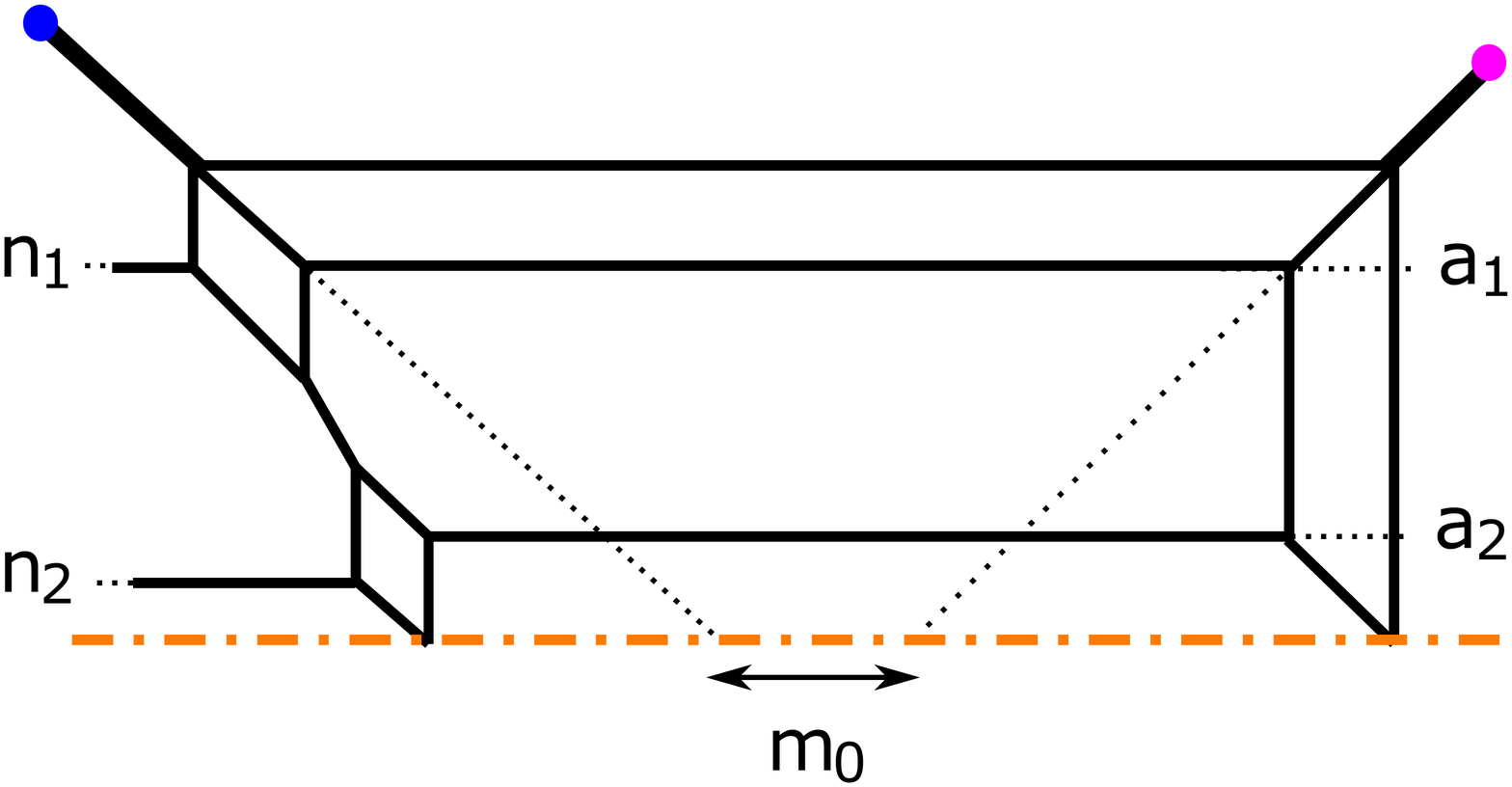}
\caption{The parameterization of the $G_2$ gauge theory with two flavors for the diagram which is S-dual to the one in Figure \ref{fig:Sp2w2AS5}.}
\label{fig:G2w2flvrspara3}
\end{figure}

By comparing the parameterization in Figure \ref{fig:Sp2w2ASpara} and \eqref{Sp2w2ASmass} for the $Sp(2)$ gauge theory with two antisymmetric hypermultiplets and the discrete theta angle $\pi$ with the parametrization of the $G_2$ gauge theory with two flavors in section \ref{sec:G2wmatter}, one can obtain the duality map between the $Sp(2)$ gauge theory and the $G_2$ gauge theory. Note that the S-dual of the diagram in Figure \ref{fig:Sp2w2AS5} yields a diagram of the $G_2$ gauge theory two flavors. The parameterization in section \ref{sec:G2wmatter} can be translated to the parameterization for the S-dual diagram as in Figure \ref{fig:G2w2flvrspara3}.
%%%%%%%%%%%%%%%%%%%%%%%%%%%%%%%%%
%%%%%%%%%%%%%%%%%%%%%%%%%%%%%%%%%
Again $n_1, n_2$ in Figure \ref{fig:G2w2flvrspara3} are related to the mass parameters by \eqref{G2w2flvrsmass}. Then, the comparison between the two parameterizations gives the duality map between the $Sp(2)$ gauge theory and the $G_2$ gauge theory
\bea
m_0^{Sp(2)} &=& -\frac{m_0^{G_2}}{2}, \label{map1.Sp2toG2w2flvrs}\\
m_{\AS, 1}^{Sp(2)} &=& m_{\bF, 1}^{G_2},\\
m_{\AS, 2}^{Sp(2)} &=& m_{\bF, 2}^{G_2},\\
\phi_1^{Sp(2)} &=& \phi_2^{G_2} + \frac{1}{2}m_0^{G_2},\\
\phi_2^{Sp(2)} &=& \phi_1^{G_2} + m_0^{G_2}. \label{map5.Sp2toG2w2flvrs}
\eea
Combining the map \eqref{map1.SU3toG2w2flvrs}-\eqref{map5.SU3toG2w2flvrs} with the map \eqref{map1.Sp2toG2w2flvrs}-\eqref{map5.Sp2toG2w2flvrs} yields the map between the $SU(3)$ gauge theory with two flavors and the CS level $6$ and the $Sp(2)$ gauge theory with two antisymmetric hypermultiplets and the non-trivial discrete theta angle 
\bea
m_0^{Sp(2)} &=& \frac{1}{4}(2m_0^{SU(3)} + 3m_{\AS, 1}^{SU(3)} + 3m_{\AS, 2}^{SU(3)}), \label{map1.Sp2toSU3w2flvrs}\\
m_{\AS, 1}^{Sp(2)} &=& \frac{1}{2}(2m_0^{SU(3)} + m_{\AS, 1}^{SU(3)} - m_{\AS, 2}^{SU(3)}),\\
m_{\AS, 2}^{Sp(2)} &=& \frac{1}{2}(2m_0^{SU(3)} - m_{\AS, 1}^{SU(3)} + m_{\AS, 2}^{SU(3)}),\\
\phi_1^{Sp(2)} &=& \phi_1^{SU(3)} + \frac{1}{4}(2m_0^{SU(3)} - m_{\AS, 1}^{SU(3)} - m_{\AS, 2}^{SU(3)}),\\
\phi_2^{Sp(2)} &=& \phi_2^{SU(3)} + \frac{1}{2}(2m_0^{SU(3)} - m_{\AS, 1}^{SU(3)} - m_{\AS, 2}^{SU(3)}). \label{map5.Sp2toSU3w2flvrs}
\eea

\subsection{Duality among marginal theories}
\label{sec:G2marginal}
In section \ref{sec:G2wmatter}, we started from the diagram of the $G_2$ gauge theory with two flavors and discussed that the diagram can be deformed into the one for the $SU(3)$ gauge theory with two flavors and the CS level $6$ and also into the one for the $Sp(2)$ gauge theory with two hypermultiplet in the antisymmetric representation and the discrete theta angle $\theta = \pi$. In order for the $G_2$ gauge theory to have a UV completion, we can add four more flavors to the $G_2$ gauge theory with two flavors. The UV completion of the $G_2$ gauge theory with six flavors is not a 5d SCFT but is supposed to be a 6d SCFT \cite{Zafrir:2015uaa, Jefferson:2017ahm}. It is in fact straightforward to extend the discussion of the dualities in section \ref{sec:G2wmatter} by adding four more flavors to the diagram for the $G_2$ gauge theory with four flavors. The $G_2$ gauge theory with six flavors is dual to the $SU(3)$ gauge theory with six flavors and the CS level $6$ and is also dual to the $Sp(2)$ gauge theory with four flavors and two hypermultiplets in the antisymmetric representation \cite{Jefferson:2018irk} and we will see the dualities from the 5-brane web diagram.

\begin{figure}
\centering
\subfigure[]{
\includegraphics[width=6cm]{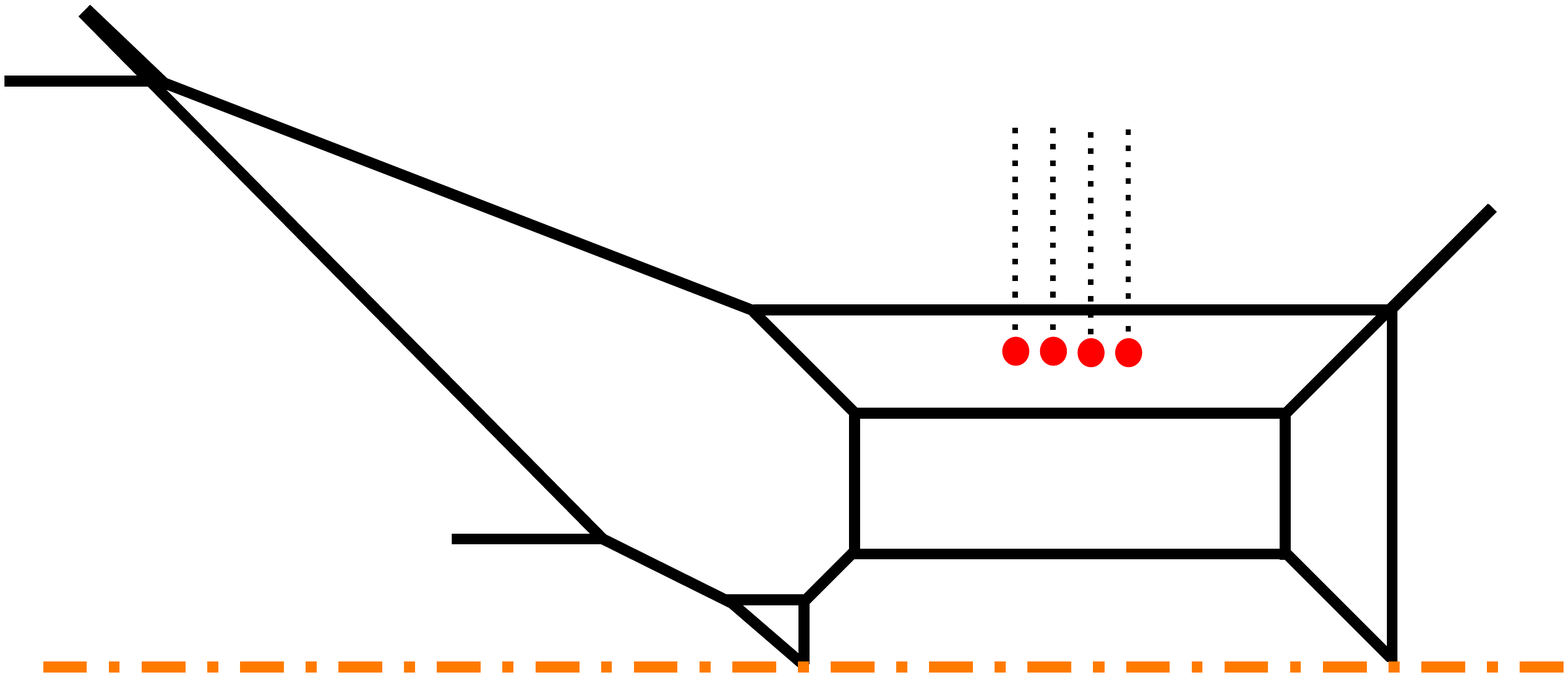} \label{fig:G2w6flvrs1}}
\subfigure[]{
\includegraphics[width=6cm]{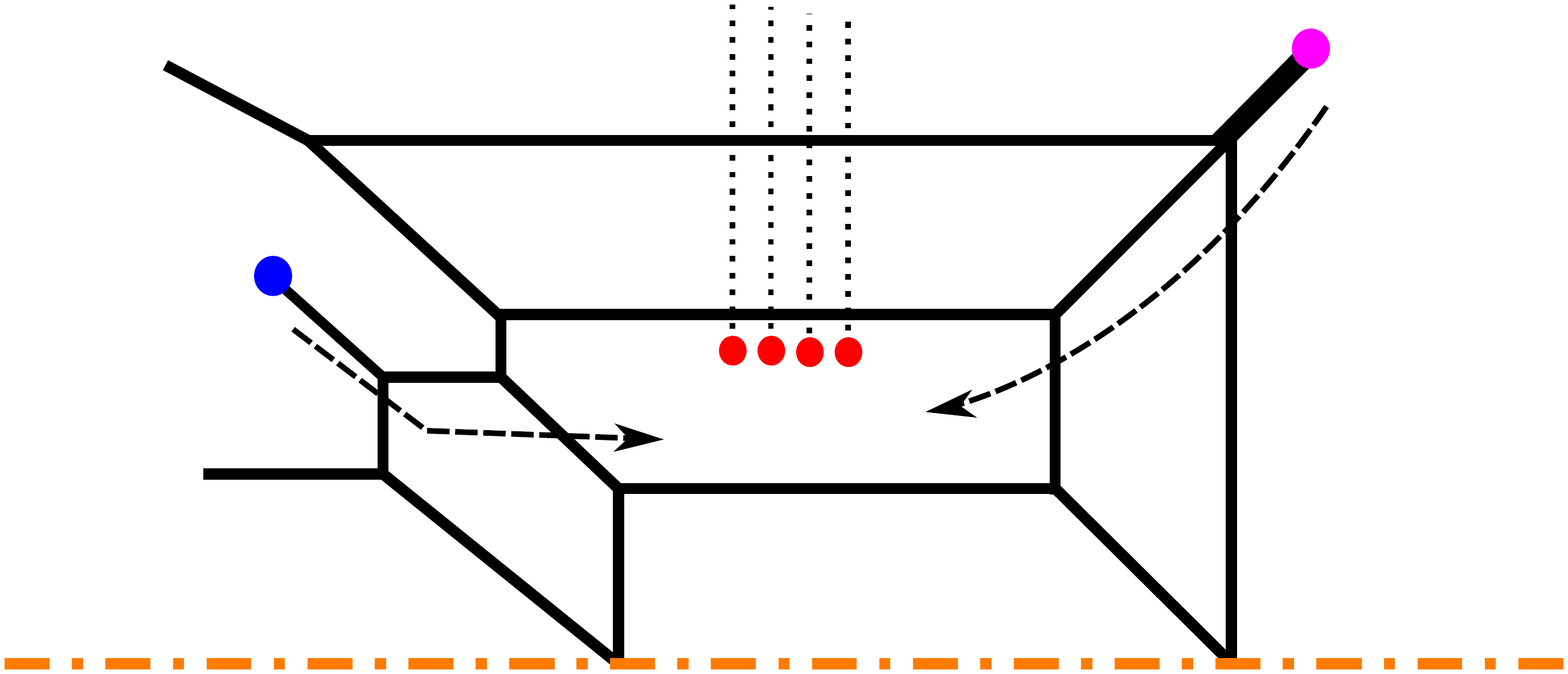} \label{fig:G2w6flvrs2}}
\subfigure[]{
\includegraphics[width=6cm]{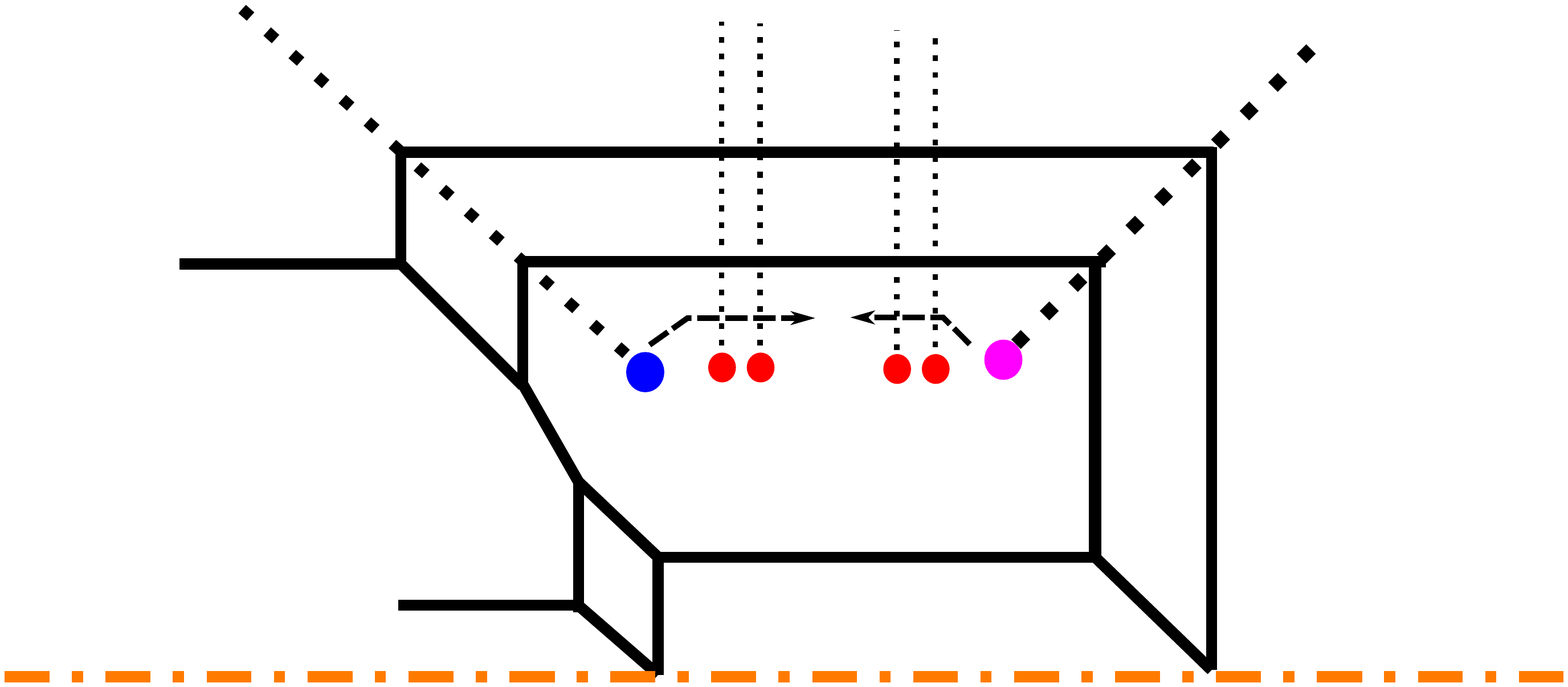} \label{fig:G2w6flvrs3}}
\subfigure[]{
\includegraphics[width=6cm]{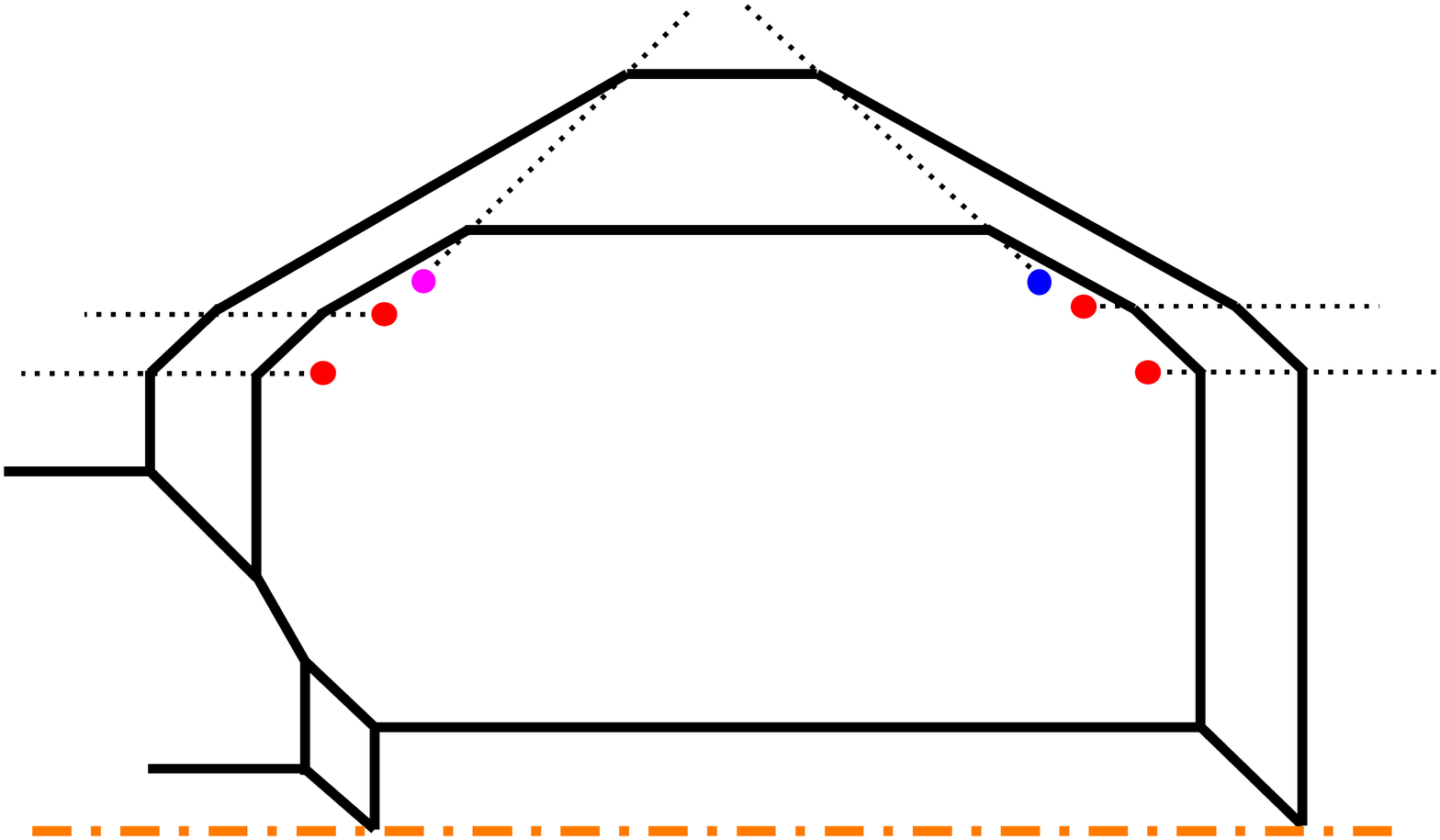} \label{fig:G2w6flvrs4}}
\subfigure[]{
\includegraphics[width=6cm]{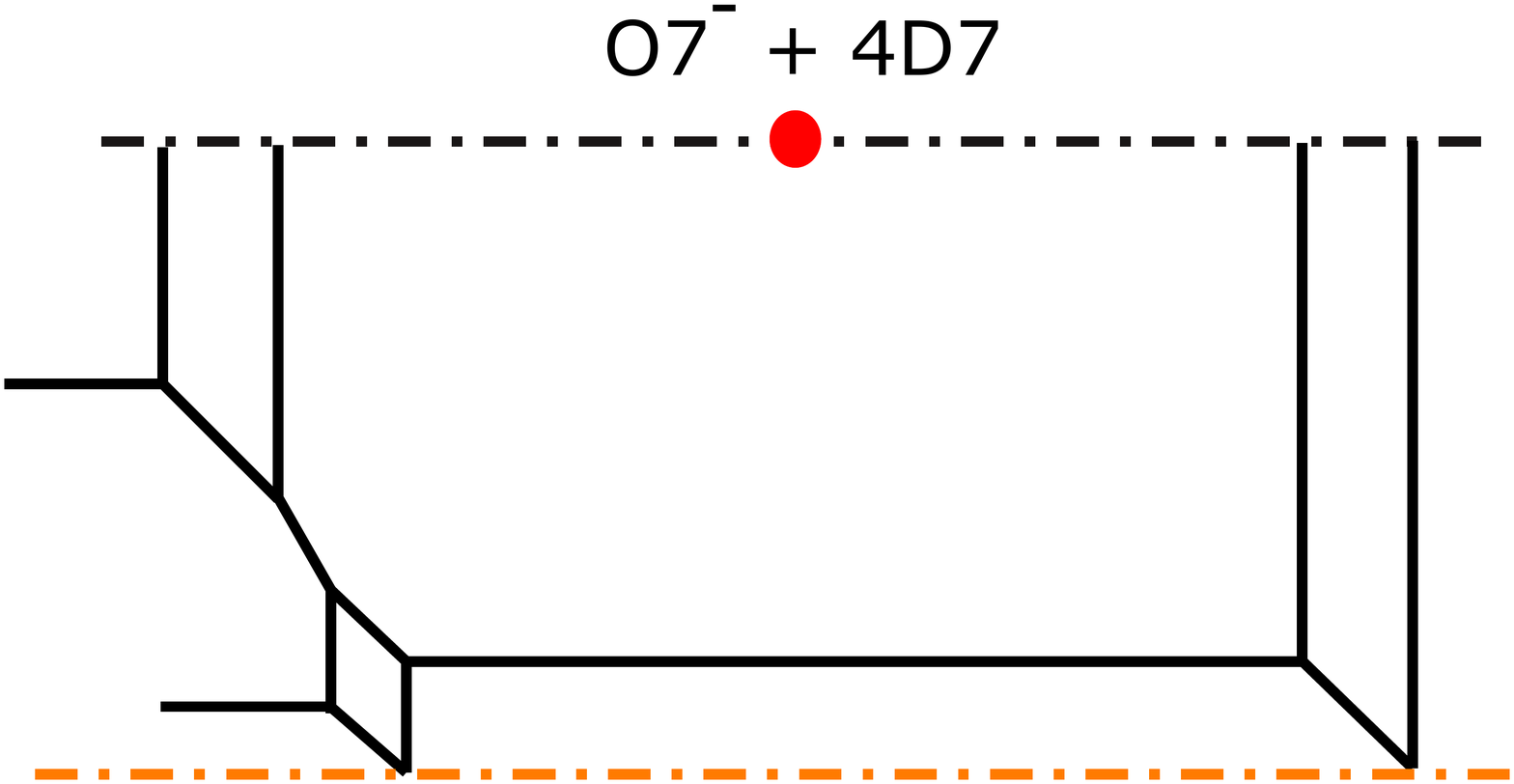} \label{fig:G2w6flvrs5}}
\caption{(a): Adding four D7-branes to the diagram in Figure \ref{fig:G2w2flvrsb}. (b): Applying the two flop transitions to the diagram in Figure \ref{fig:G2w6flvrs1}. (c): The diagram after moving the $(1, -1)$ 7-brane and the $(1, 1)$ 7-brane according to the arrows in Figure \ref{fig:G2w6flvrs2}. (d): The diagram after moving the $(1, -1)$ 7-brane and the $(1, 1)$ 7-brane according to the arrows in Figure \ref{fig:G2w6flvrs3}. (e): The diagram after forming an O7$^-$-plane from the $(1, -1)$ 7-brane and the $(1, 1)$ 7-brane in Figure \ref{fig:G2w6flvrs4}.}
\label{fig:G2w6flvrs}
\end{figure}

Let us first add four flavors to the diagram for the $G_2$ gauge theory with two flavors in Figure \ref{fig:G2w2flvrsb}. When we added two flavors to the diagram for the pure $G_2$ gauge theory, we introduced the flavors which originate from two spinors in the $SO(7)$ gauge theory before the Higgsing, This time, we introduce four flavors which originate from four hypermultiplets in the vector representation of $SO(7)$ before the Higgsing. The introduction of the flavors can be done by adding four D7-branes to the diagram for the $G_2$ gauge theory with two flavors as in Figure \ref{fig:G2w6flvrs1}.  
%%%%%%%%%%%%%%%%%%%%%%%%%%%%%%%%%
%%%%%%%%%%%%%%%%%%%%%%%%%%%%%%%%%
When we turn over the branch cuts of D7-branes in the horizontal directions, external 5-branes will cross each other. In order to obtain a consistent 5-brane web for the $G_2$ gauge theory with six flavors, we perform the flop transitions and move the $(1, -1)$ 7-brane and the $(1, 1)$ 7-brane as we did when we obtained the diagram for the $Sp(2)$ gauge theory. The diagram after the flop transitions and moving the 7-branes is given in Figure \ref{fig:G2w6flvrs3}. From this diagram, we let the $(1, -1)$ 7-brane and the $(1, 1)$ 7-brane cross the branch cuts of the D7-branes as in Figure \ref{fig:G2w6flvrs3}. The charge of the 7-branes changes and the diagram becomes the one in Figure \ref{fig:G2w6flvrs4}. At this stage, we have a pair of a $(1, 1)$ 7-brane and a $(1, -1)$ 7-brane, which can form an O7$^-$-plane. Therefore, the final diagram in Figure \ref{fig:G2w6flvrs5} contains a pair of an O7$^-$-plane and an $\widetilde{\text{O5}}$-plan in the vertical direction. The periodicity in the vertical direction implies that the theory has a 6d UV completion, which is consistent with the result in  \cite{Zafrir:2015uaa, Jefferson:2017ahm}.

\begin{figure}
\centering
\includegraphics[width=8cm]{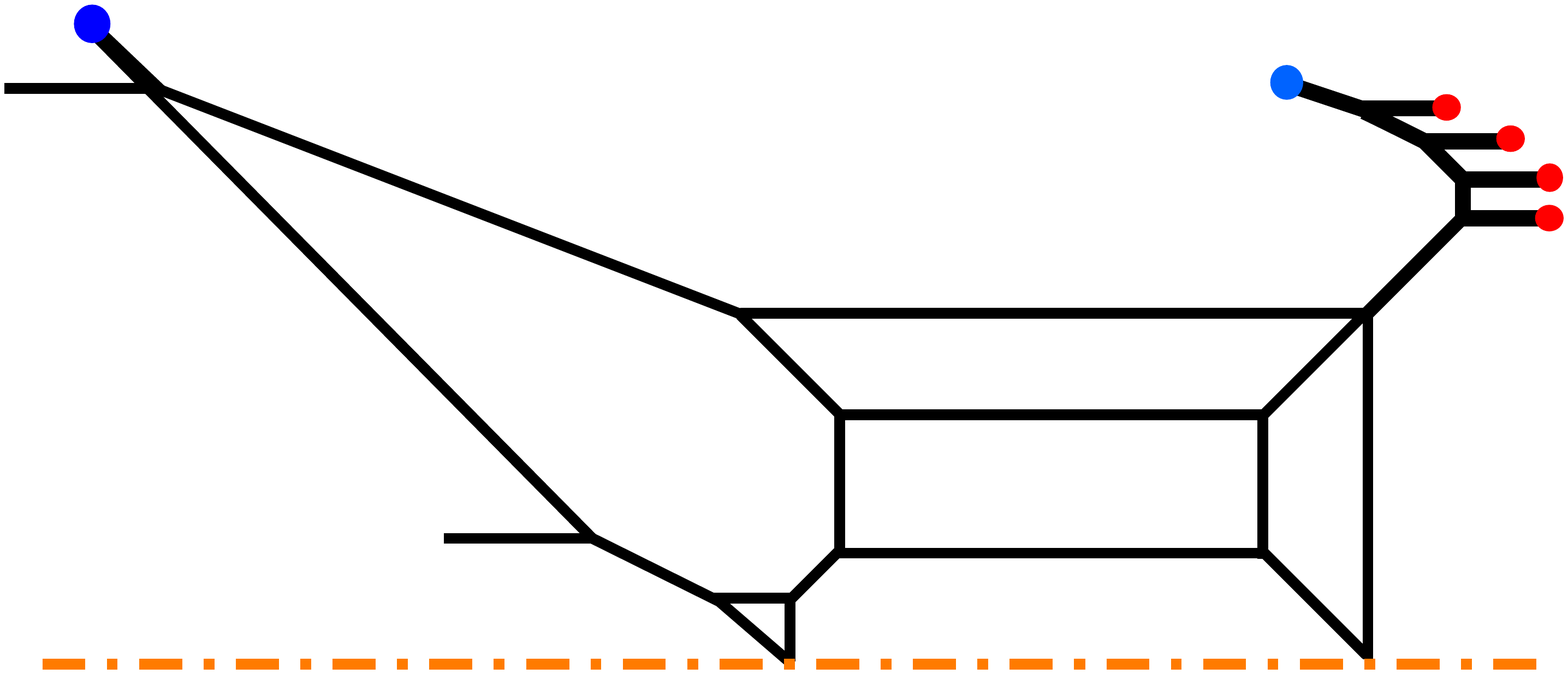}
\caption{A 5-brane web diagram for the $G_2$ gauge theory with six flavors after turning over the branch cuts of the D7-branes in the diagram in Figure \ref{fig:G2w6flvrs1}.}
\label{fig:G2w6flvrsdual1}
\end{figure}

\paragraph{Duality to $SU(3)$.}
From the diagram of the $G_2$ gauge theory with six flavors, it is also possible to get the duality to the $SU(3)$ gauge theory with four flavors and the CS level $4$, and to the $Sp(2)$ gauge theory with two antisymmetric hypermultiplets and four flavors. We first consider the duality to the $SU(3)$ gauge theory with six flavors and the CS level $4$. We first start from the diagram of the $G_2$ gauge theory with six flavors in Figure \ref{fig:G2w6flvrs1}. Turning over the branch cuts of the D7-branes in the horizontal direction yields a diagram in Figure \ref{fig:G2w6flvrsdual1}. 
%%%%%%%%%%%%%%%%%%%%%%%%%%%%%%%%%
%%%%%%%%%%%%%%%%%%%%%%%%%%%%%%%%%

\begin{figure}
\centering
\subfigure[]{
\includegraphics[width=6cm]{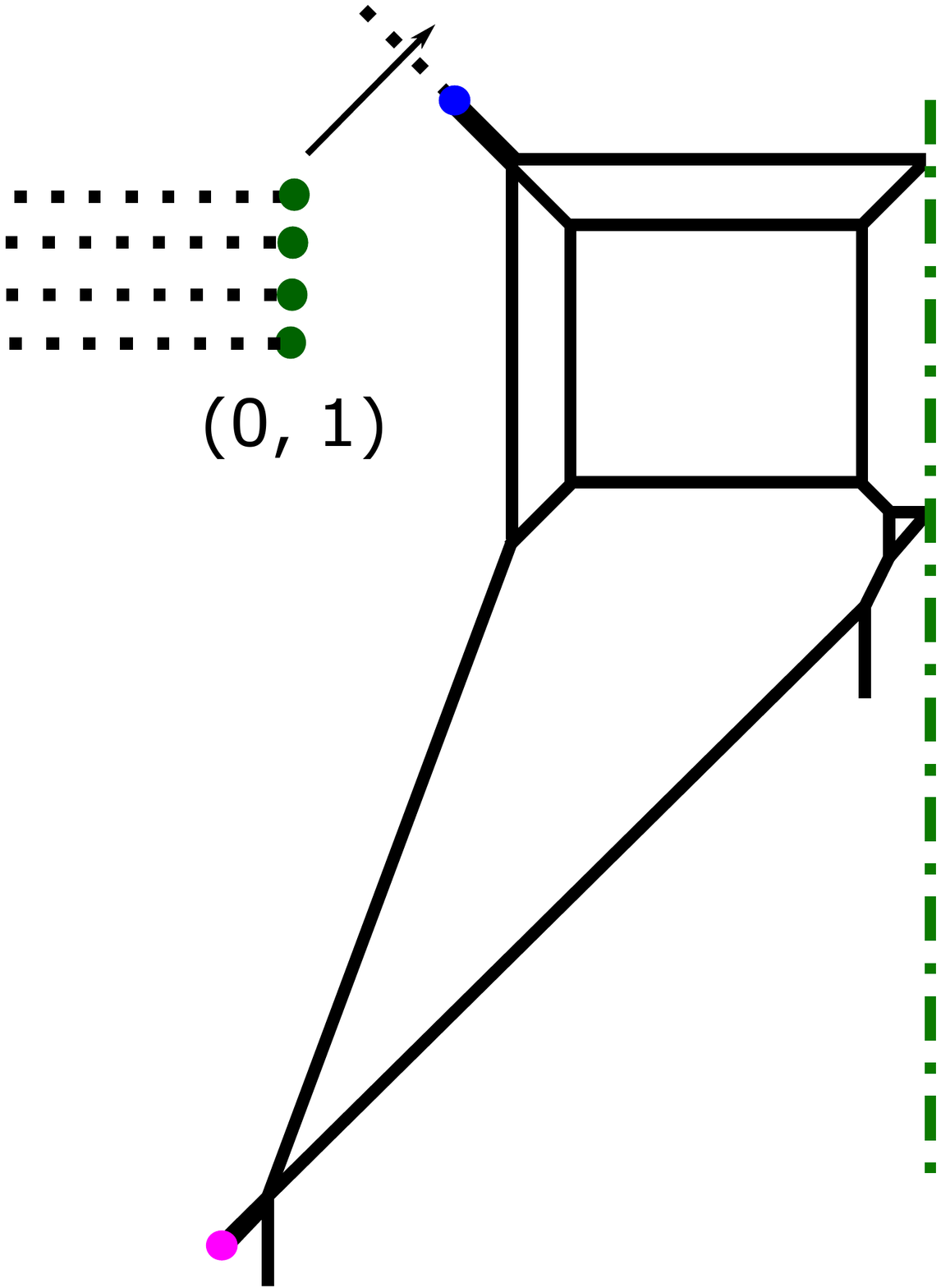} \label{fig:G2w6flvrsdual2}}
\subfigure[]{
\includegraphics[width=6cm]{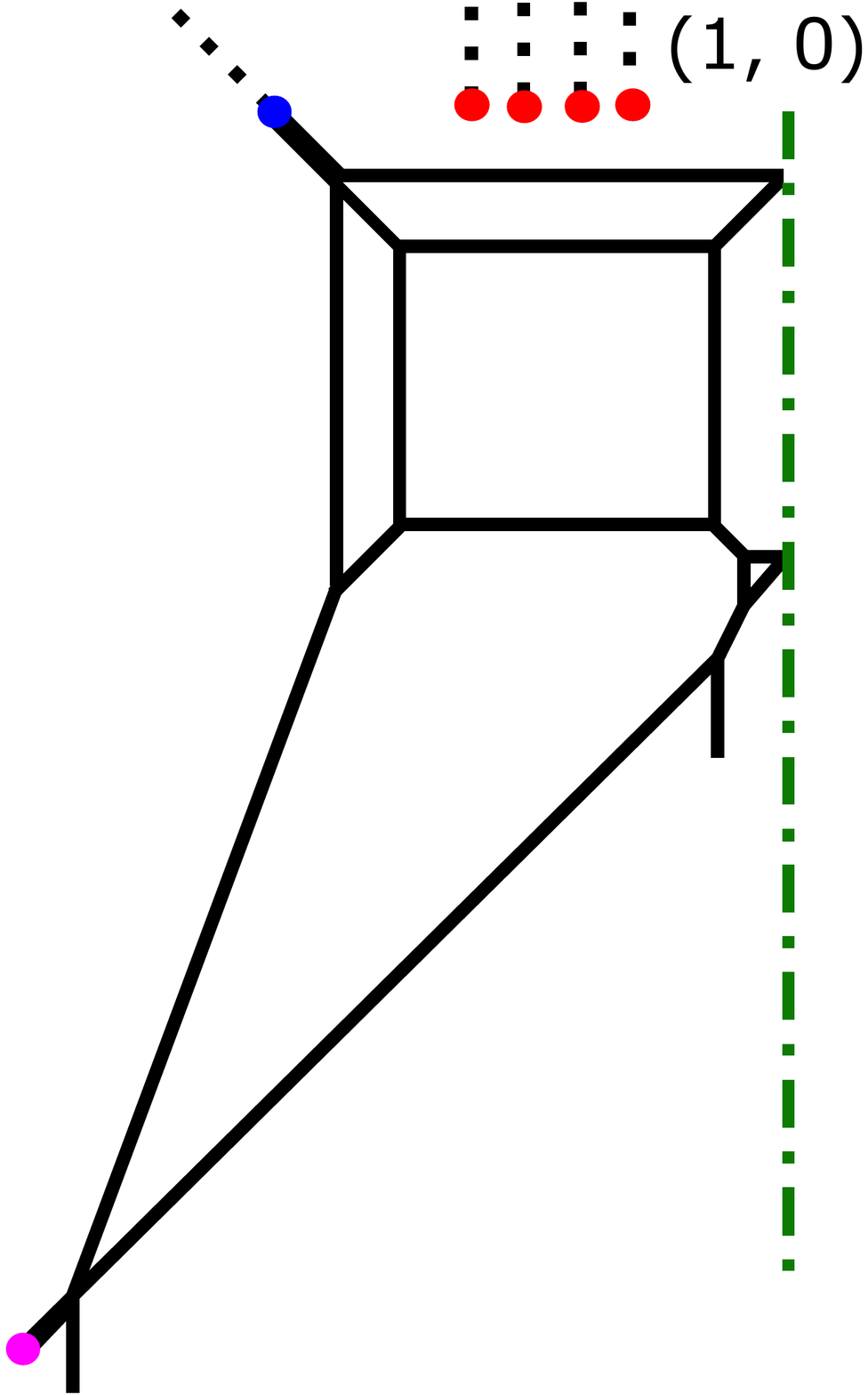} \label{fig:G2w6flvrsdual3}}
\subfigure[]{
\includegraphics[width=6cm]{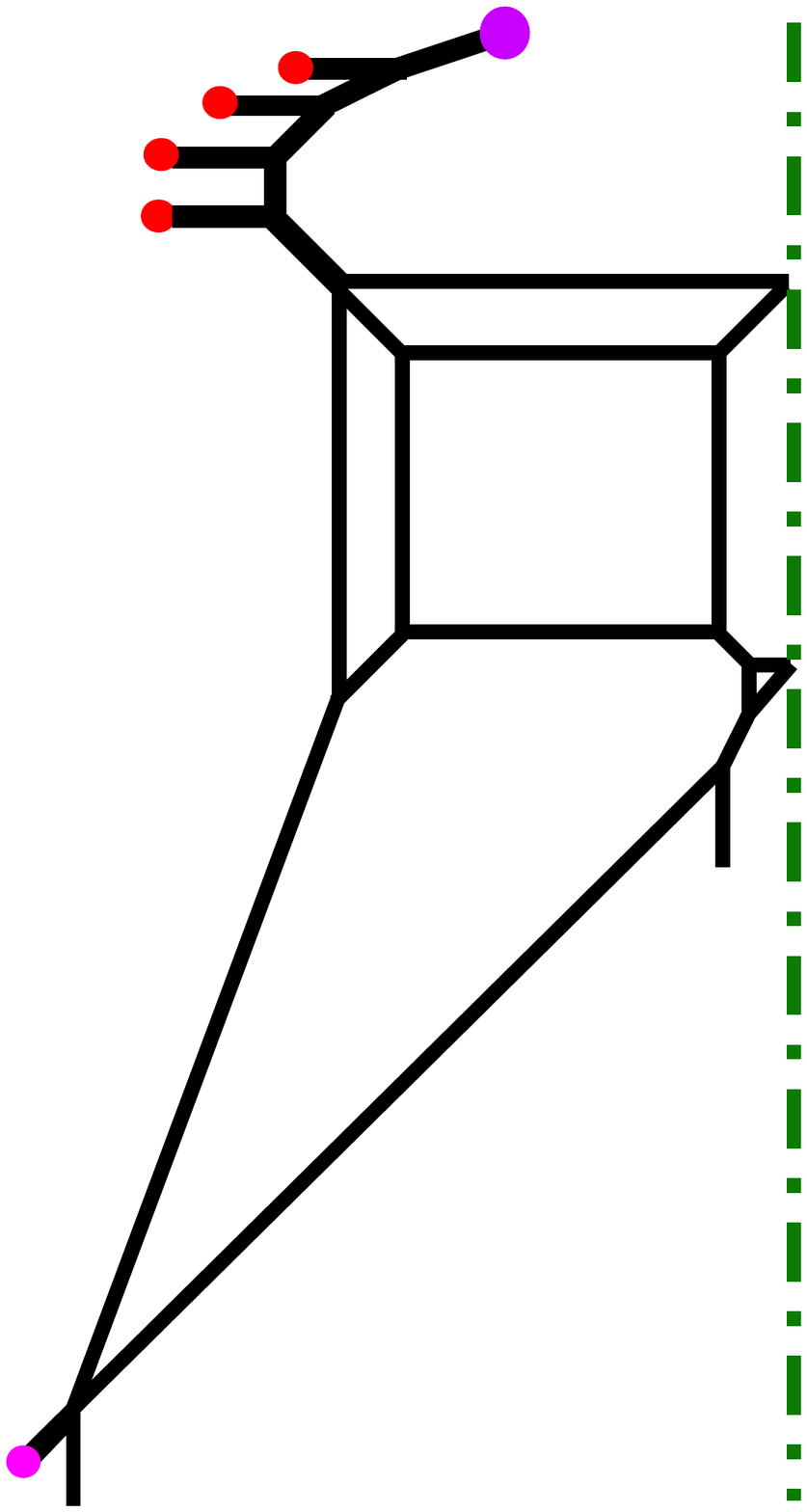} \label{fig:SU3w6flvrsCS4}}
\caption{(a): The S-dual to the diagram in Figure \ref{fig:G2w6flvrs1}. (b): After crossing the branch cut of the $(1, -1)$ 7-brane and four $(0, 1)$ 7-branes change into four $(1, 0)$ 7-branes. (c): A diagram which can be interpreted as the $SU(3)$ gauge theory with six flavors and the CS level $4$.}
\label{fig:G2w6flvrsdualtoSU3}
\end{figure}

One the other hand, the S-dual to the diagram in Figure \ref{fig:G2w6flvrs1} gives rise to a diagram in Figure \ref{fig:G2w6flvrsdual2}. 
%%%%%%%%%%%%%%%%%%%%%%%%%%%%%%%%%
%%%%%%%%%%%%%%%%%%%%%%%%%%%%%%%%%
As in the case for the $G_2$ gauge theory with two flavors in section \ref{sec:dualtoSU3w2flvrs}, the existence of the three color D5-branes in Figure \ref{fig:G2w6flvrsdual2} implies that the diagram yields an $SU(3)$ gauge theory. Furthermore, we can change the four $(0, 1)$ 7-branes into the four $(1, 0)$ 7-branes when the $(0, 1)$ 7-branes cross the branch cut of the $(1, -1)$ 7-brane. Then the four D7-branes can create flavor D5-branes as in Figure \ref{fig:SU3w6flvrsCS4}. Compared to the diagram in Figure \ref{fig:SU3w2flvrsCS6}, the diagram contains four more flavors. Since decoupling the flavors in the same direction yields the diagram for the $SU(3)$ gauge theory with two flavors and the CS level $6$, the CS level for the $SU(3)$ gauge theory in Figure \ref{fig:SU3w6flvrsCS4} should be $4$. Hence, we can conclude that the diagram gives the $SU(3)$ gauge theory with six flavors and the CS level $4$.

\begin{figure}
\centering
\includegraphics[width=8cm]{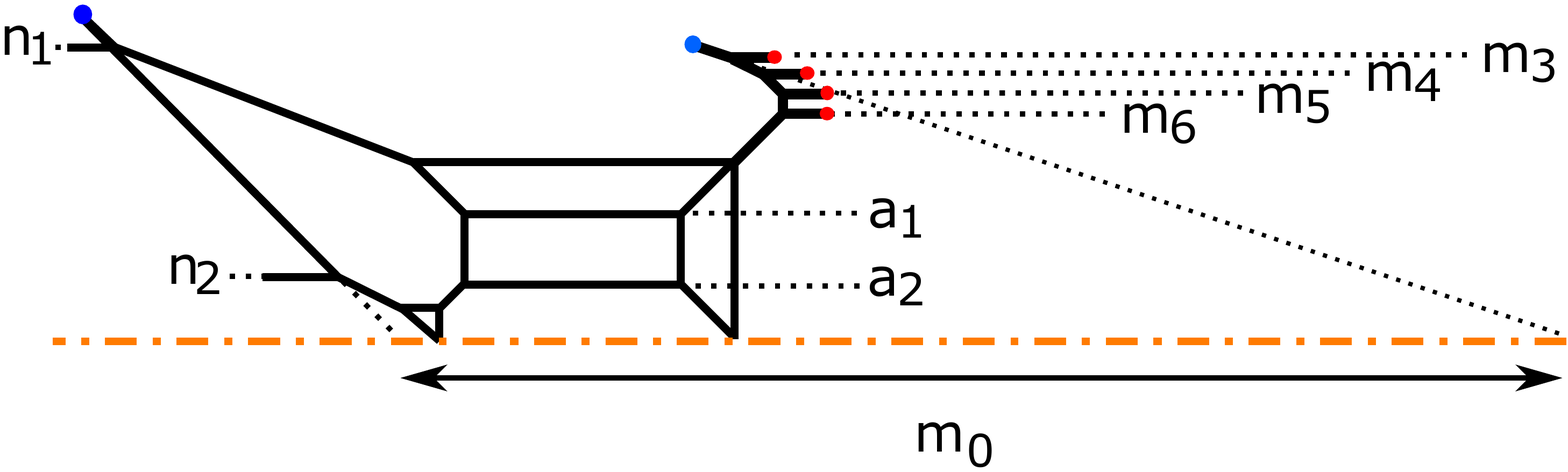}
\caption{The $G_2$ gauge theory parameterization for the diagram in Figure \ref{fig:G2w6flvrsdual1}.}
\label{fig:G2w6flvrspara}
\end{figure}

\begin{figure}
\centering
\includegraphics[width=8cm]{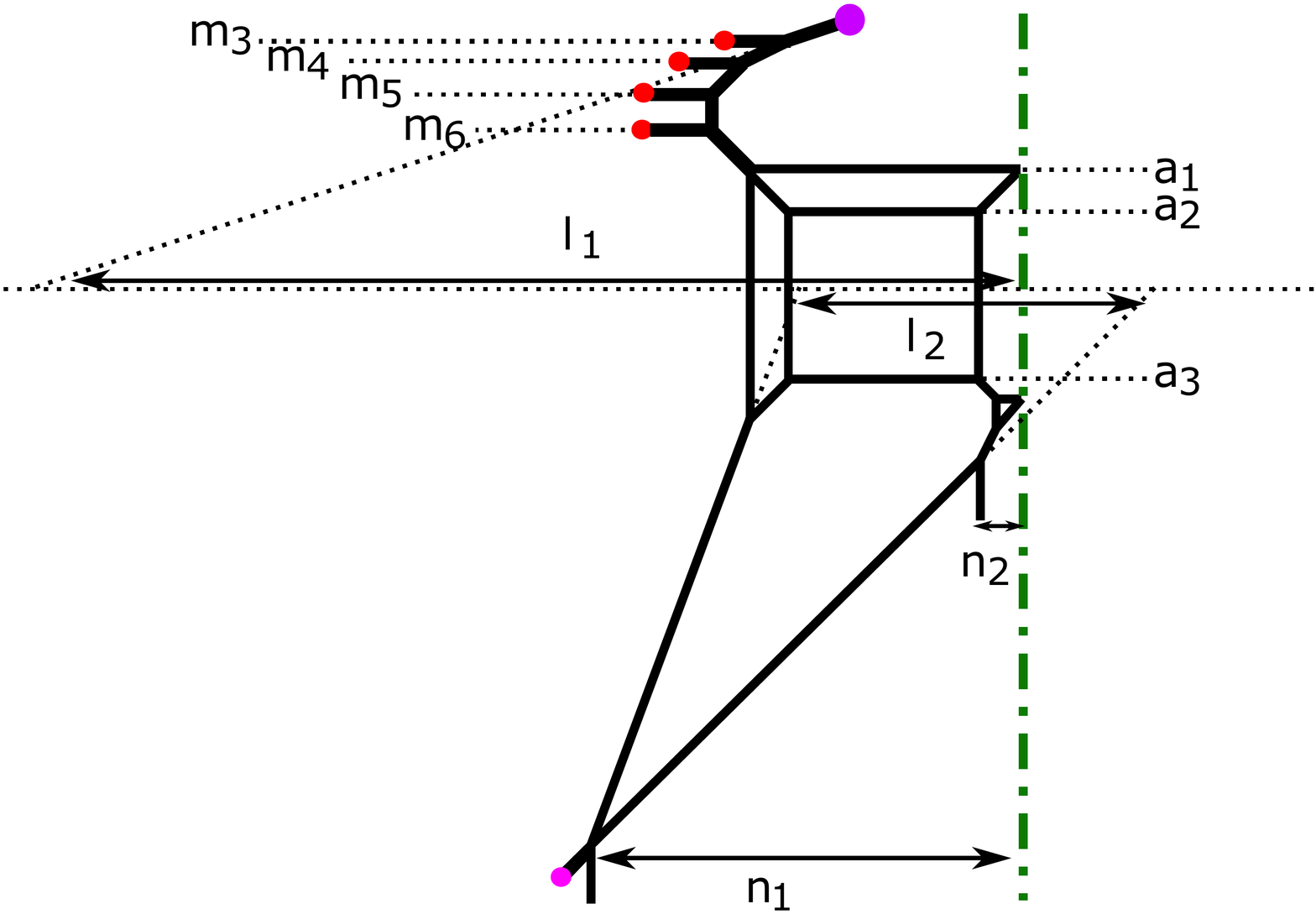}
\caption{The $SU(3)$ gauge theory parameterization for the diagram in Figure \ref{fig:SU3w6flvrsCS4}.}
\label{fig:SU3w6flvrsCS4para}
\end{figure}

Here let us see the parameterization of the $G_2$ gauge theory with six flavors and the $SU(3)$ gauge theory with six flavors and the CS level $4$ and obtain the duality map between the two theories. The gauge theory parameterization for the $G_2$ gauge theory with six flavors is given in Figure \ref{fig:G2w6flvrspara}. 
%%%%%%%%%%%%%%%%%%%%%%%%%%%%%%%%%
%%%%%%%%%%%%%%%%%%%%%%%%%%%%%%%%%
$m_0$ is the inverse of the squared gauge coupling, $a_1, a_2$ are the Coulomb branch moduli and $m_3, m_4, m_5, m_6$ are the mass parameters. $n_1, n_2$ are also related to the two mass parameters by \eqref{G2w2flvrsmass}. It is also possible to obtain the gauge theory parameterization for the $SU(3)$ gauge theory with six flavors and the CS level $4$ by extending the parametrization in section \ref{sec:dualtoSU3w2flvrs}. The parameterization for the $SU(3)$ gauge theory is given in Figure \ref{fig:SU3w6flvrsCS4para}.
%%%%%%%%%%%%%%%%%%%%%%%%%%%%%%%%%
%%%%%%%%%%%%%%%%%%%%%%%%%%%%%%%%%
The inverse of the squared gauge couping $m_0$ is given by $m_0 = \frac{l_1 + l_2}{2}$. $a_1, a_2, a_3,\; (a_1 + a_2 + a_3=0)$ are the Coulomb branch moduli and $m_3, m_4, m_5, m_6$ are the mass parameters for the additional four flavors. The two other mass parameters $m_1, m_2$ enter in $l_1, l_2$ and $n_2$ by\footnote{On the other hand, $n_1$ is given by $n_1 = 2m_0 - (m_3 + m_4 + m_5 + m_6)$.}
\bea
l_1 &=&m_0 + \frac{1}{2}\left(m_1 + m_2 + m_3 + m_4 + m_5 + m_6\right),\\
l_2 &=&m_0  - \frac{1}{2}\left(m_1 + m_2 + m_3 + m_4 + m_5 + m_6\right),\\
n_2 &=& m_1 - m_2. 
\eea

The comparison between the two parameterizations in Figure \ref{fig:G2w6flvrspara} and in Figure \ref{fig:SU3w6flvrsCS4para} gives rise to the duality map
\bea
m_0^{SU(3)} &=& \frac{1}{2}m_0^{G_2} - \frac{1}{2}\lambda_1, \label{map1.SU3toG2w6flvrs}\\
m_{\AS, 1}^{SU(3)} &=& -m_{\bF, 2}^{G_2} + \lambda_1,\\
m_{\AS, 2}^{SU(3)} &=& -m_{\bF, 1}^{G_2} + \lambda_1,\\
m_{\bF, i}^{SU(3)} &=& m_{\bF, i}^{G_2} - \lambda_1,\quad (i=3, \cdots, 6),\\
\phi_1^{SU(3)} &=&  \phi_2^{G_2} - \lambda_1,\\
\phi_2^{SU(3)} &=& \phi_1^{G_2} - 2\lambda_1, \label{map9.SU3toG2w6flvrs}
\eea
where 
\be
\lambda_1 = -\frac{1}{3}m_0^{G_2} + \frac{1}{3}\sum_{i=1}^6m^{G_2}_{\bF, i}.
\ee

\begin{figure}
\centering
\includegraphics[width=8cm]{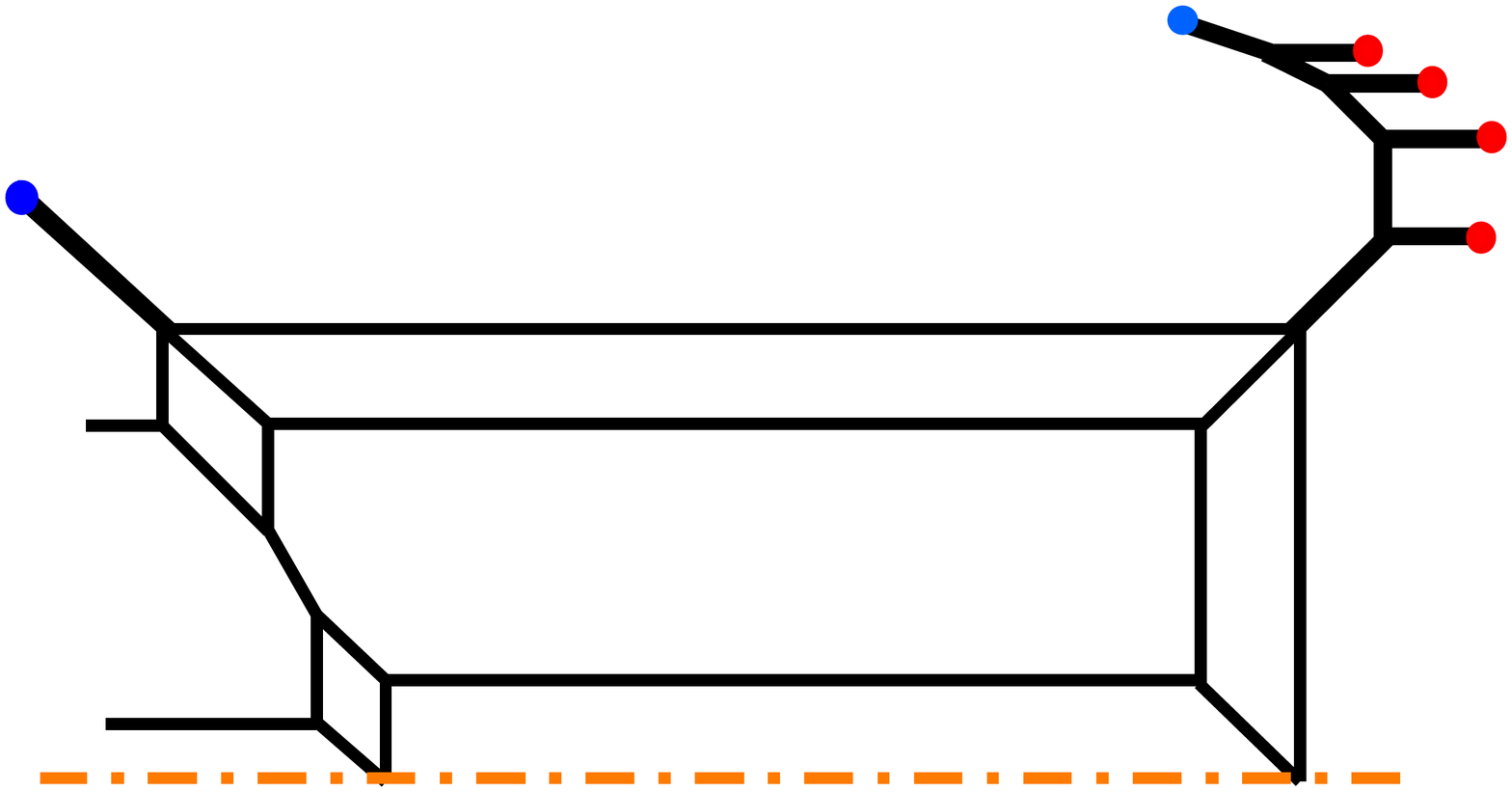}
\caption{A 5-brane web diagram for the $G_2$ gauge theory with six flavors after moving the 7-branes in the diagram in Figure \ref{fig:G2w6flvrs3}.}
\label{fig:G2w6flvrsdualtoSp2}
\end{figure}

\begin{figure}
\centering
\subfigure[]{
\includegraphics[width=6cm]{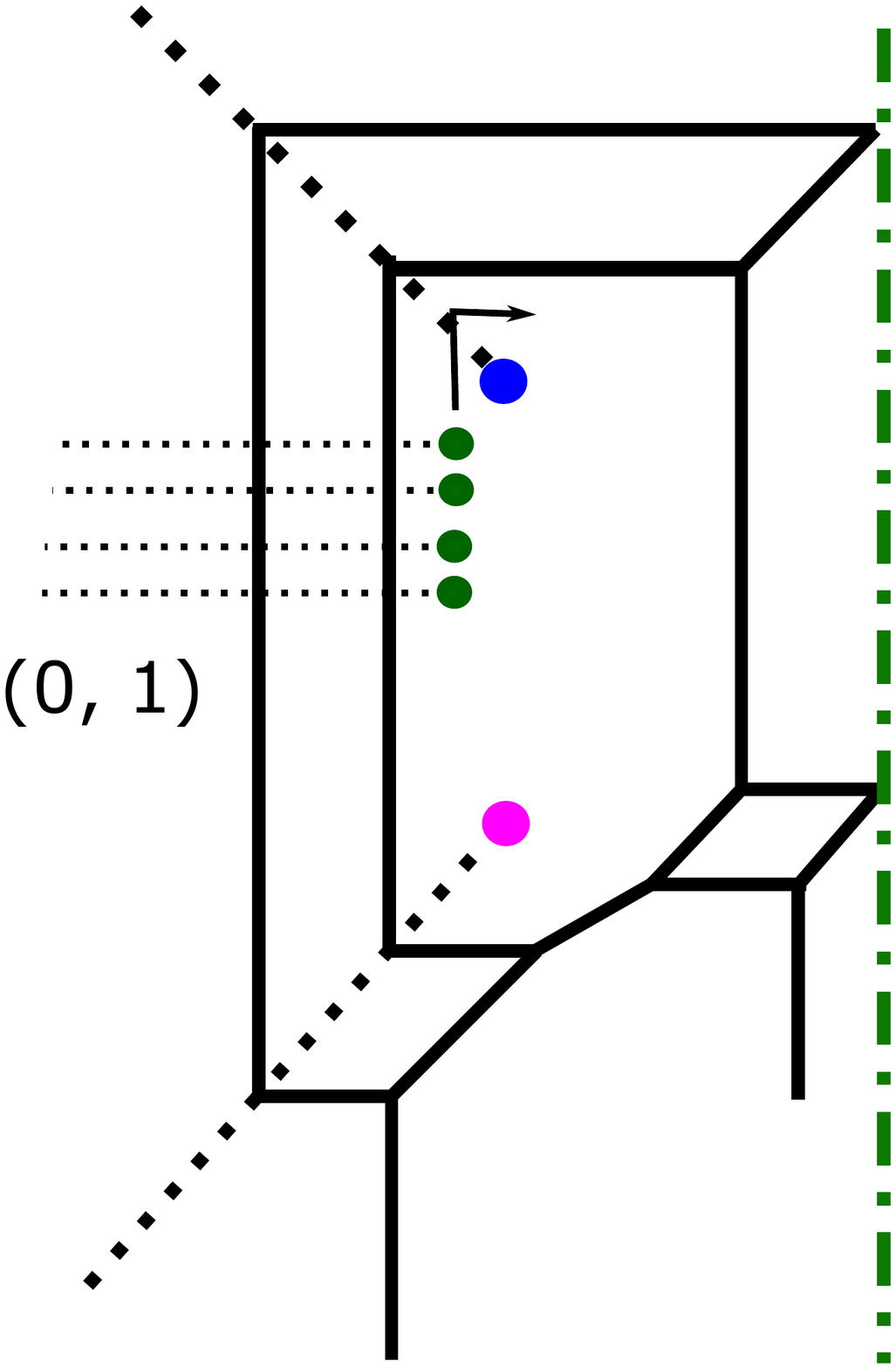} \label{fig:Sp2w2AS4F1}}
\subfigure[]{
\includegraphics[width=6cm]{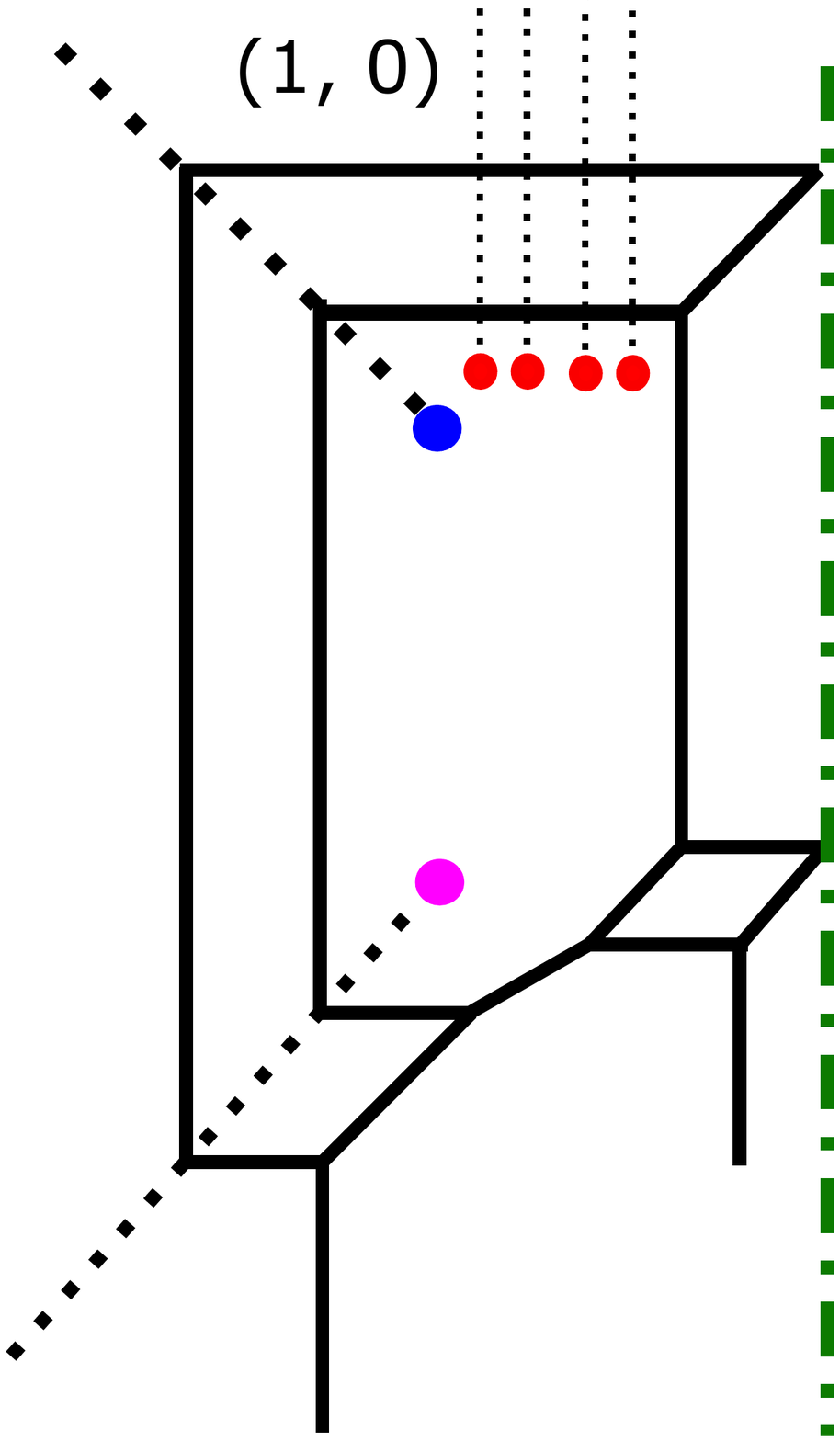} \label{fig:Sp2w2AS4F2}}
\subfigure[]{
\includegraphics[width=6cm]{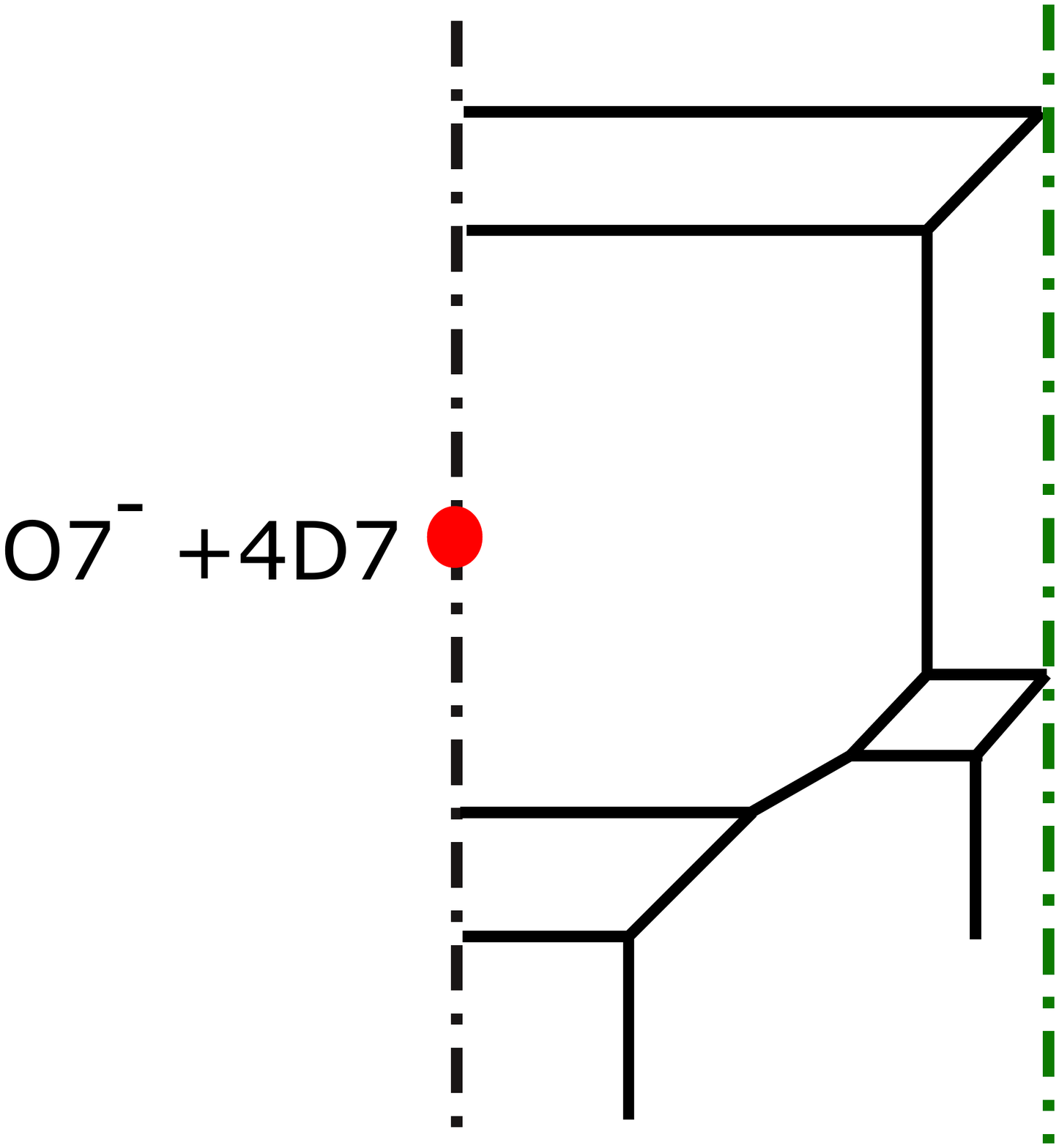} \label{fig:Sp2w2AS4F3a}}
\subfigure[]{
\includegraphics[width=6cm]{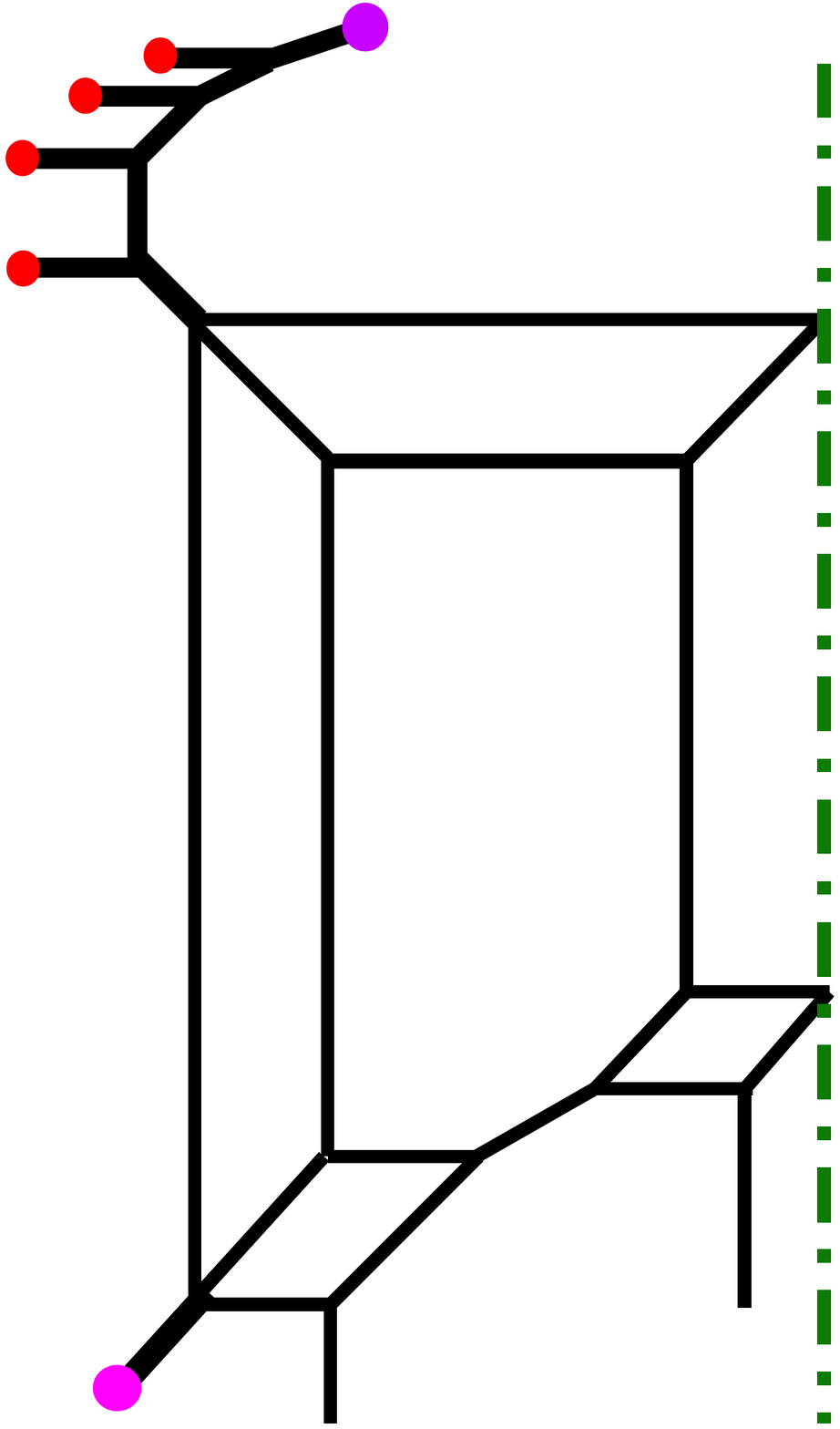} \label{fig:Sp2w2AS4F3}}
\caption{(a): The S-dual to the diagram in Figure \ref{fig:G2w6flvrs3}. (b): After crossing the branch cut of the $(1, -1)$ 7-brane and four $(0, 1)$ 7-branes change into four $(1, 0)$ 7-branes. (c): The $(1, -1)$ 7-brane and the $(1, 1)$ 7-brane forms an O7$^-$-plane and the diagram which can be interpreted as the $Sp(2)$ gauge theory with two antisymmetric hypermultiplets and four flavors. (d): The diagram when we send the 7-branes in Figure \ref{fig:Sp2w2AS4F2} to infinitely far. }
\label{fig:Sp2w2AS4F}
\end{figure}

\paragraph{Duality to $Sp(2)$.}
We can also see the duality to the $Sp(2)$ gauge theory with two antisymmetric hypermultiplets and four flavors. For that we start from the diagram in Figure \ref{fig:G2w6flvrs3} which gives the $G_2$ gauge theory with six flavors. When we send the 7-branes to infinitely far, the diagram becomes the one in Figure \ref{fig:G2w6flvrsdualtoSp2}. 
%%%%%%%%%%%%%%%%%%%%%%%%%%%%%%%%%
%%%%%%%%%%%%%%%%%%%%%%%%%%%%%%%%%
Then we consider applying the S-duality to the diagram in Figure \ref{fig:G2w6flvrs3}. Then the resulting configuration becomes the one in Figure \ref{fig:Sp2w2AS4F1}. 
%%%%%%%%%%%%%%%%%%%%%%%%%%%%%%%%%
%%%%%%%%%%%%%%%%%%%%%%%%%%%%%%%%%
Then we can move the four $(0, 1)$ 7-branes according the arrow in Figure \ref{fig:Sp2w2AS4F1} and then the four $(0, 1)$ 7-branes change into four $(1, 0)$ 7-branes as in Figure \ref{fig:Sp2w2AS4F2}. %Then pulling the four $(1, 0)$ 7-branes in the horizontal direction yields the diagram in Figure \ref{fig:Sp2w2AS4F3}. 
Then the $(1, -1)$ 7-brane and the $(1, 1)$ 7-brane in Figure \ref{fig:Sp2w2AS4F2} can form an O7$^-$-plane and the diagram becomes the one in Figure \ref{fig:Sp2w2AS4F3a}. The presence of the four flavor D7-branes in Figure \ref{fig:Sp2w2AS4F3} implies the existence of four hypermultiplets in the fundamental representation of $Sp(2)$. Hence, the diagram in Figure \ref{fig:Sp2w2AS4F3a} yields the $Sp(2)$ gauge theory with two hypermultiplets in the antisymmetric representation and the four hypermultiplets in the fundamental representation. An equivalent diagram when we send the 7-branes in Figure \ref{fig:Sp2w2AS4F2} to infinitely far is also depicted in Figure \ref{fig:Sp2w2AS4F3}. 

\begin{figure}
\centering
\includegraphics[width=8cm]{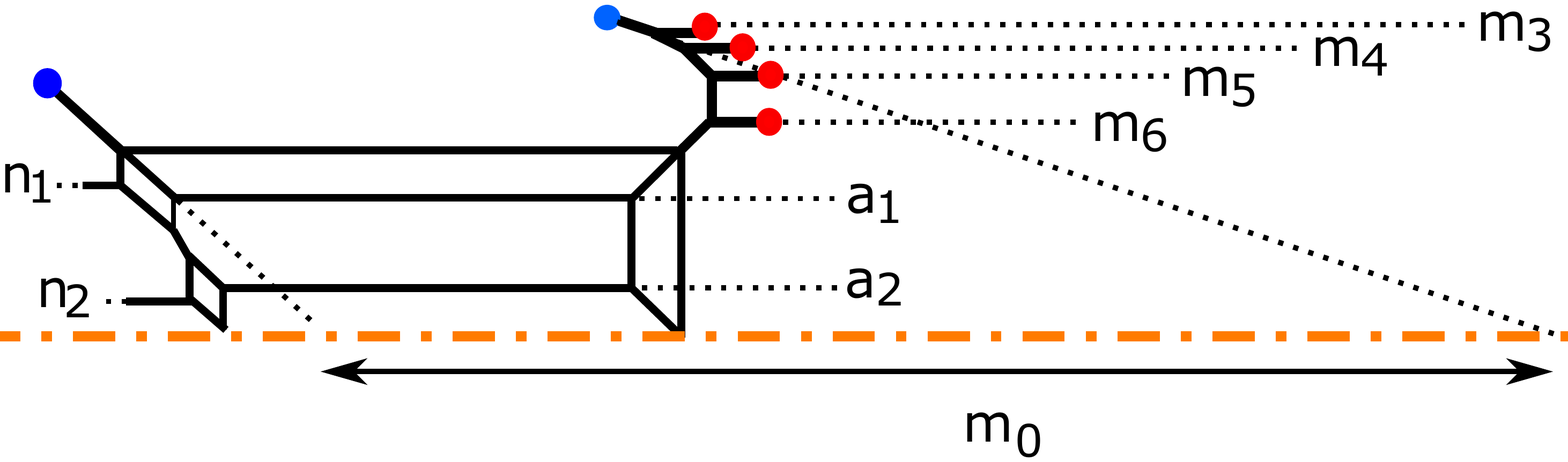}
\caption{The $G_2$ gauge theory parameterization for the diagram in Figure \ref{fig:G2w6flvrsdualtoSp2}.}
\label{fig:G2w6flvrspara2}
\end{figure}

\begin{figure}
\centering
\includegraphics[width=8cm]{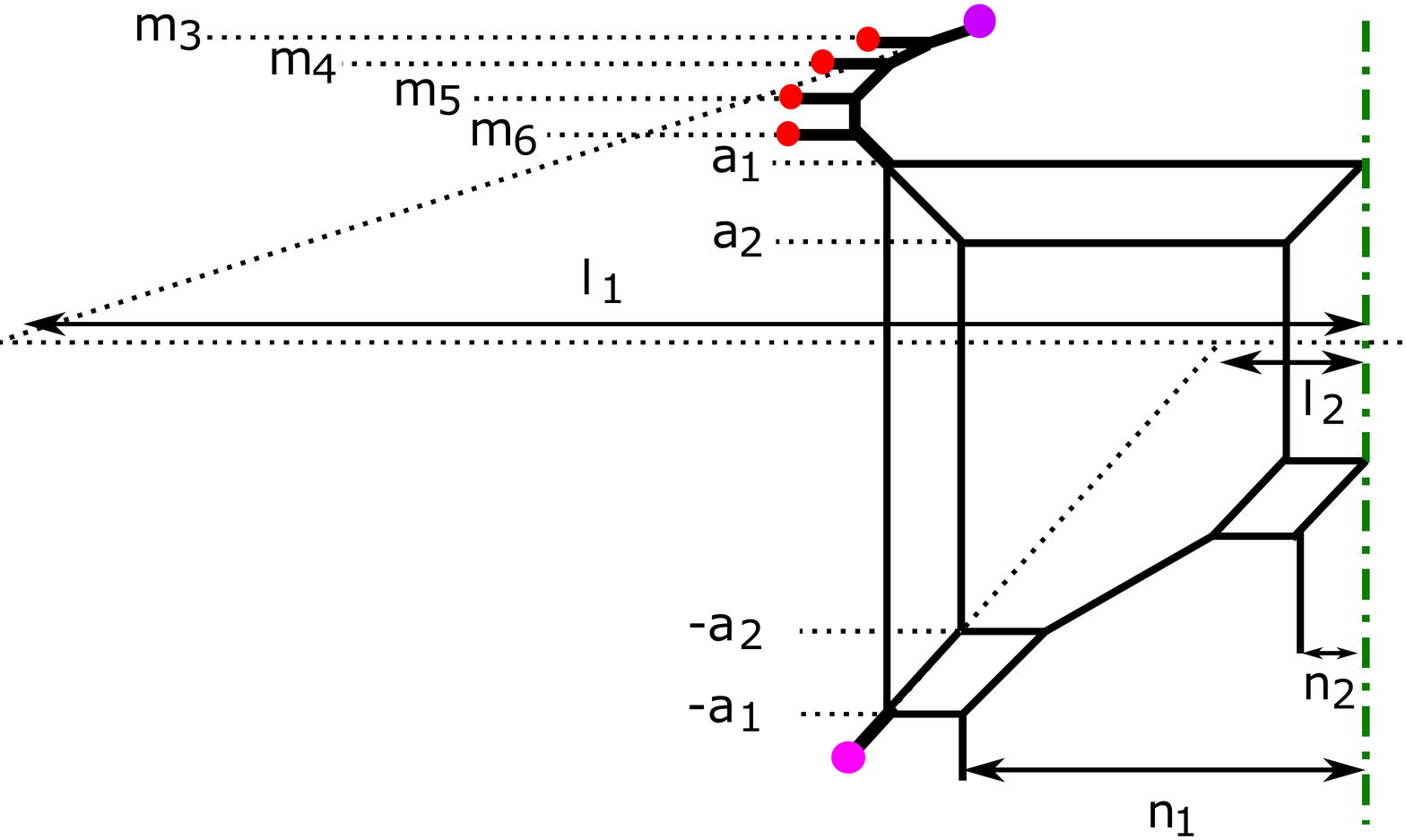}
\caption{The $Sp(2)$ gauge theory parameterization for the diagram in Figure \ref{fig:Sp2w2AS4F3}.}
\label{fig:Sp2w2AS4Fpara}
\end{figure}

We can also see the parameterization of the both two theories by generalizing the parameterization in section \ref{sec:dualtoSp2w2AS} and also obtain the duality map between the two theories. The gauge theory parameterization for the $G_2$ gauge theory realized in Figure \ref{fig:G2w6flvrsdualtoSp2} is given in Figure \ref{fig:G2w6flvrspara2}. 
%%%%%%%%%%%%%%%%%%%%%%%%%%%%%%%%%
%%%%%%%%%%%%%%%%%%%%%%%%%%%%%%%%%
$m_0$ is the inverse of the squared gauge coupling, $a_1, a_2$ are the Coulomb branch moduli and $m_3, m_4, m_5, m_6$ are the mass parameters. $n_1, n_2$ are related to the two other mass parameters by \eqref{G2w2flvrsmass}. On the other hand, the gauge theory parametrization for the $Sp(2)$ gauge theory realized in Figure \ref{fig:Sp2w2AS4F3} is depicted in Figure \ref{fig:Sp2w2AS4Fpara}.
%%%%%%%%%%%%%%%%%%%%%%%%%%%%%%%%%
%%%%%%%%%%%%%%%%%%%%%%%%%%%%%%%%%
The inverse of the squared gauge coupling is given by $m_0 = \frac{l_1 + l_2}{2}$, $a_1, a_2$ are the Coulomb branch moduli and $m_3, m_4, m_5, m_6$ are the mass parameters for the four flavors. $n_1, n_2$ are related to the mass parameters for two antisymmetric hypermultiplets and the relation is given by \eqref{Sp2w2ASmass}.  

The comparison between the two parameterizations in Figure \ref{fig:G2w6flvrspara2} and in Figure \ref{fig:Sp2w2AS4Fpara} yields the duality map between the $G_2$ gauge theory and the $Sp(2)$ gauge theory,
\bea
m_0^{Sp(2)} &=& \frac{m_0^{G_2}}{2}, \label{map1.Sp2toG2w6flvrs}\\
m_{\AS, 1}^{Sp(2)} &=& m_{\bF, 1}^{G_2},\\
m_{\AS, 2}^{Sp(2)} &=& m_{\bF, 2}^{G_2},\\
m_{\bF, i}^{Sp(2)} &=& m_{\bF, i}^{G_2} -\lambda_2, \quad (i= 3, \cdots, 6), \\
\phi_1^{Sp(2)} &=& \phi_2^{G_2} - \lambda_2,\\
\phi_2^{Sp(2)} &=& \phi_1^{G_2} - 2\lambda_2, \label{map9.Sp2toG2w6flvrs}
\eea
where we defined
\be
\lambda_2 = -\frac{1}{2}m_0^{G_2} + \frac{1}{2}\sum_{i=3}^6m_{\bF, i}^{G_2}.
\ee
Combining the map \eqref{map1.SU3toG2w6flvrs}-\eqref{map9.SU3toG2w6flvrs} between the $SU(3)$ gauge theory and the $G_2$ gauge theory with the map \eqref{map1.Sp2toG2w6flvrs}-\eqref{map9.Sp2toG2w6flvrs} between the $Sp(2)$ gauge theory and the $G_2$ gauge theory, we can also obtain the map between the $Sp(2)$ gauge theory and the $SU(3)$ gauge theory, 
\bea
m_0^{Sp(2)} &=& m_0^{SU(3)} + \frac{1}{2}\sum_{i=1}^2m_{\AS, i}^{SU(3)} -\lambda_3,\label{map1.Sp2toSU3w6flvrs}\\
m_{\AS, i}^{Sp(2)} &=& m_{\AS, i}^{SU(3)} - 2\lambda_3,\quad (i=1, 2)\\
m_{\bF, j}^{Sp(2)} &=&m_{\bF, j}^{SU(3)} -  \lambda_3, \quad (j = 3, \cdots, 6),\\
\phi_1^{Sp(2)} &=& \phi_1^{SU(3)} - \lambda_3, \\
\phi_2^{Sp(2)} &=&\phi_2^{SU(3)} - 2\lambda_3, \label{map9.Sp2toSU3w6flvrs}
\eea
where
\be
\lambda_3 = -\frac{1}{2}m_0^{SU(3)} + \frac{1}{4}\sum_{i=1}^2m_{\AS, i}^{SU(3)} + \frac{1}{4}\sum_{j=3}^6m_{\bF, j}^{SU(3)}.
\ee

\subsection{Realization as $SO(5) + 2{\bf V} + 4{\bf S}$}\label{subsec:SO52F4S}

In section \ref{sec:dualtoSp2w2AS} and section \ref{sec:G2marginal}, we have seen the realization of the $Sp(2)$ gauge group from four D5-branes with an O7$^-$-plane. In fact, the diagram may be deformed to a diagram which can be interpreted as an $SO(5)$ gauge theory. This is consistent with the fact that there is an isomorphism $\mathfrak{so}(5) \simeq \mathfrak{sp}(2)$ at the level of the Lie algebra. 

To see the deformation, we start from the diagram in Figure \ref{fig:G2w6flvrs1} for the $G_2$ gauge theory with six flavors. In section \ref{sec:G2marginal} we have seen this diagram can be deformed into the one in Figure \ref{fig:Sp2w2AS4F3a}, yielding the $Sp(2)$ gauge theory with two antisymmetric hypermultiplets and four flavors. Here we consider a different deformation. 
%%%%%%%%%%%%%%%%%%%%%%%%%%%%%%%%%
\begin{figure}
\centering
\subfigure[]{
\includegraphics[width=6cm]{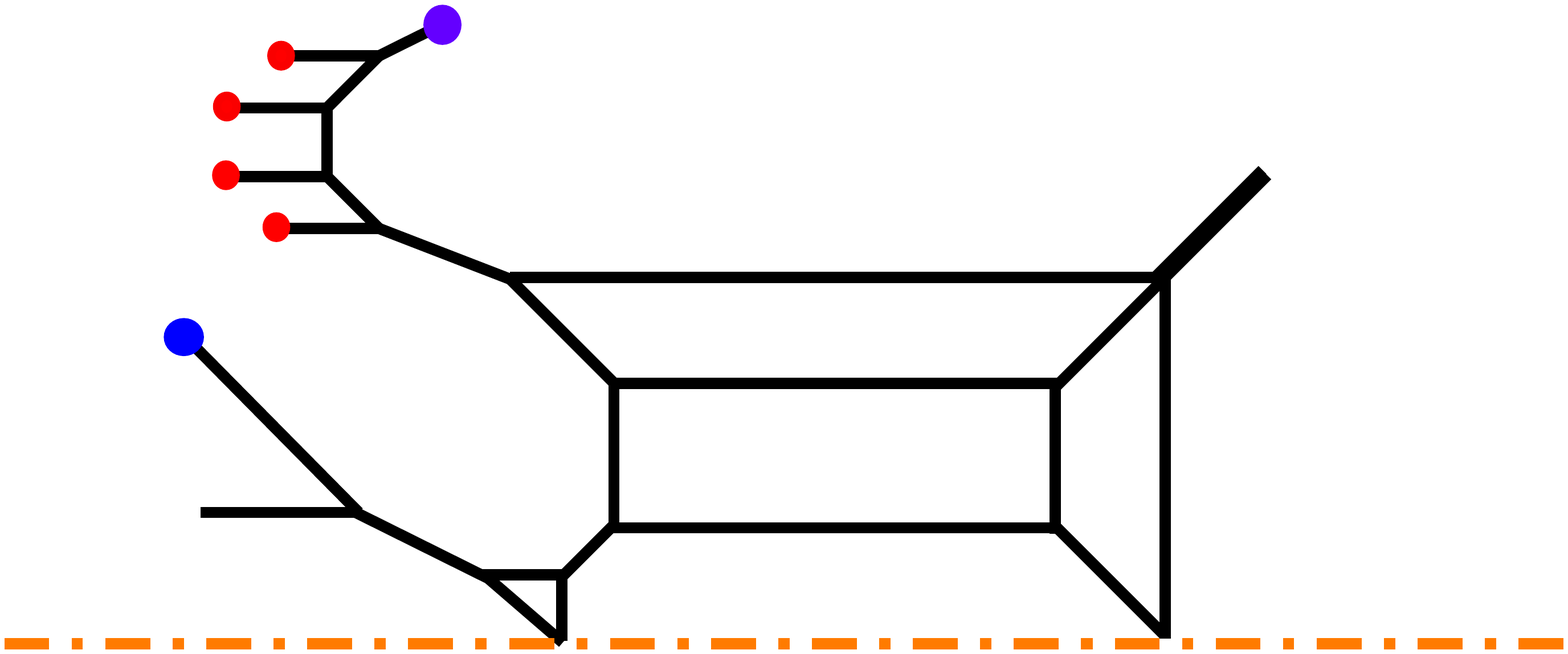} \label{fig:G2toSO5def1}}
\subfigure[]{
\includegraphics[width=6cm]{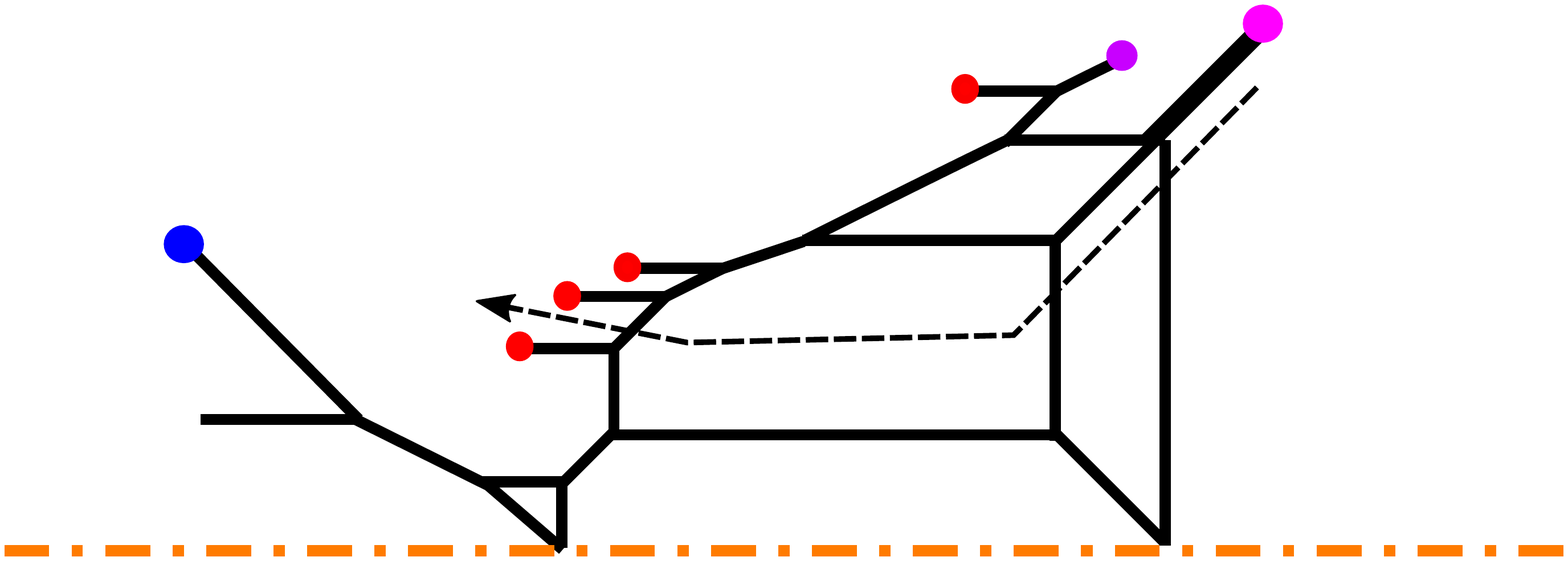} \label{fig:G2toSO5def2}}
\subfigure[]{
\includegraphics[width=6cm]{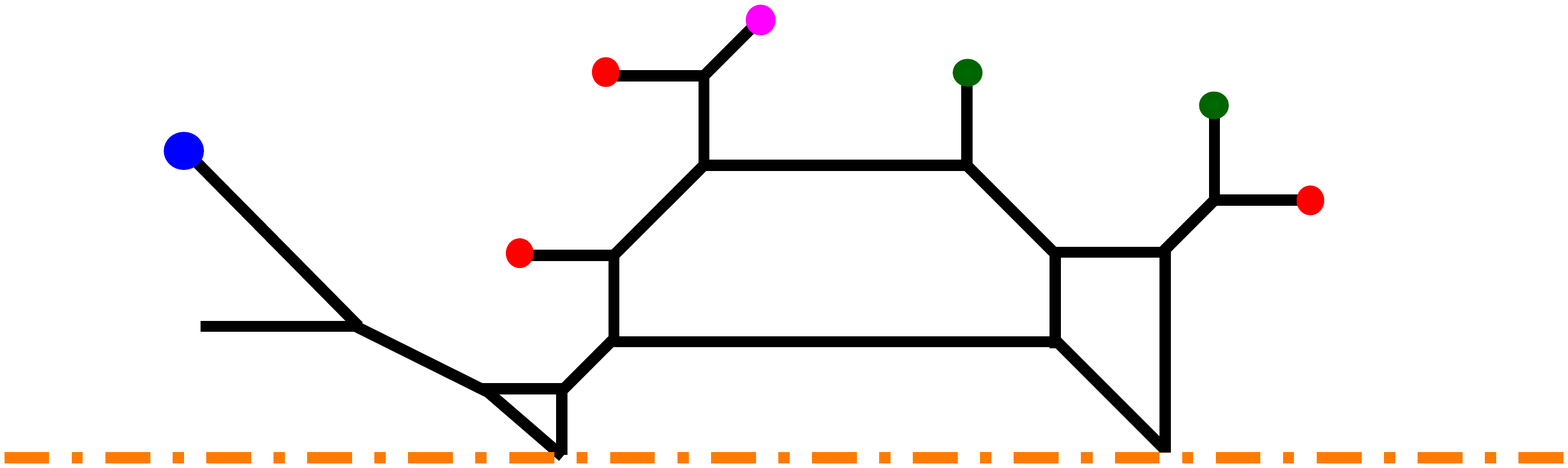} \label{fig:G2toSO5def3}}
\subfigure[]{
\includegraphics[width=6cm]{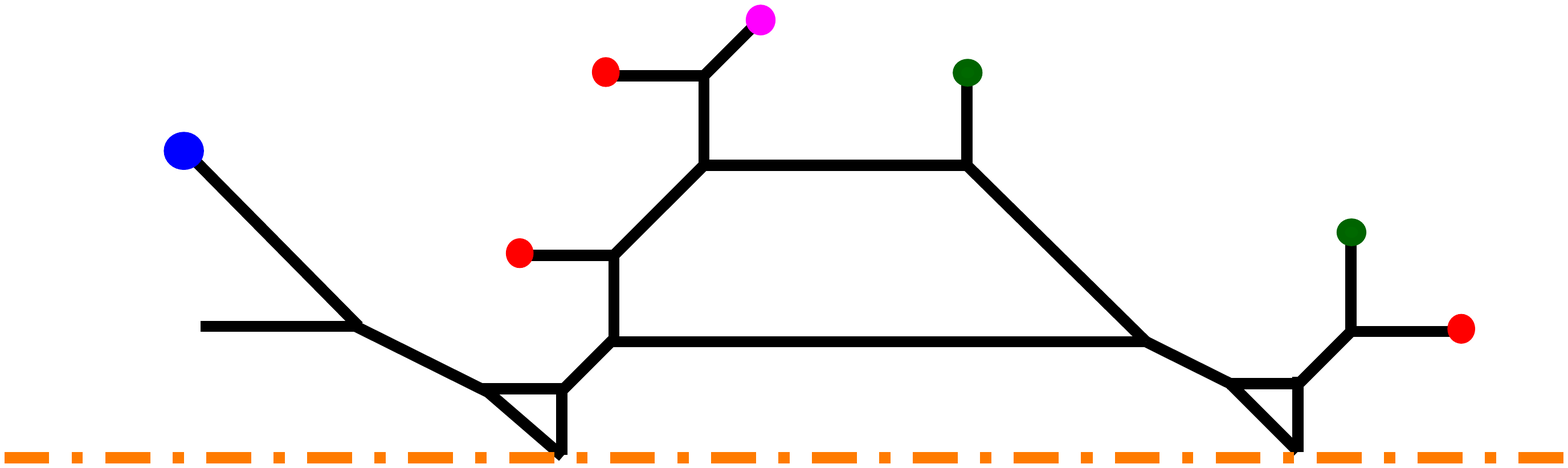} \label{fig:G2toSO5def4}}
\subfigure[]{
\includegraphics[width=6cm]{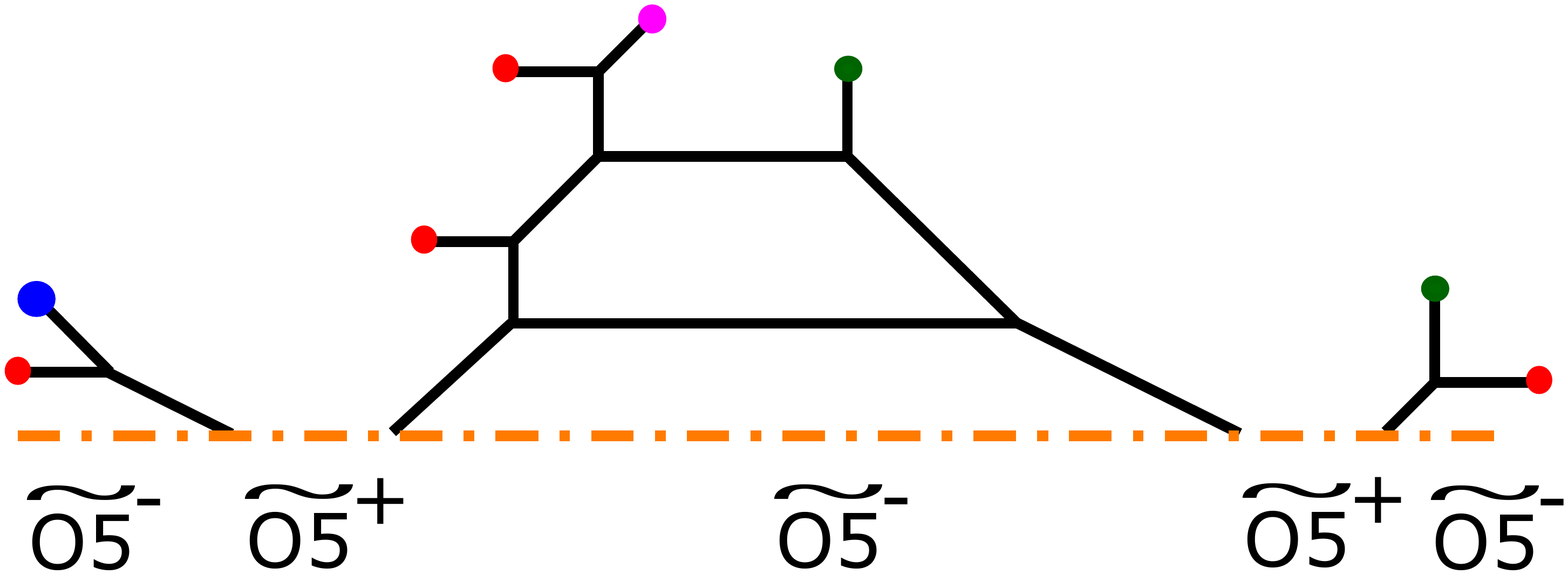} \label{fig:G2toSO5def5}}
\caption{(a): A diagram for the $G_2$ gauge theory with six flavors. (b): Applying flop transitions to the diagram in Figure \ref{fig:G2toSO5def1}. (c): The diagram after moving the $(1, 1)$ 7-brane according to the arrow in Figure \ref{fig:G2toSO5def3}. (d): The diagram obtained after applying a flop transition to the diagram in Figure \ref{fig:G2toSO5def4}. (e): The diagram obtained after applying generalized flop transitions to the diagram in Figure \ref{fig:G2toSO5def4}.}
\label{fig:G2toSO5def}
\end{figure}
%%%%%%%%%%%%%%%%%%%%%%%%%%%%%%%%%
First, the diagram in Figure \ref{fig:G2w6flvrs1} can be written as the one in Figure \ref{fig:G2toSO5def1}. Applying flop transitions yields the diagram in Figure \ref{fig:G2toSO5def2}. From the diagram in Figure \ref{fig:G2toSO5def2}, we move the $(1, 1)$ 7-brane according to the arrow in Figure \ref{fig:G2toSO5def3}, giving rise to the diagram in Figure \ref{fig:G2toSO5def4}. Then, performing further flop transitions changes the diagram finally into the one in Figure \ref{fig:G2toSO5def5}. The diagram in Figure \ref{fig:G2toSO5def5} is exactly the diagram for the $SO(5)$ gauge theory with two hypermultiplets in the vector representation and four hypermultiplets in the spinor representation, which is equivalent to the $Sp(2)$ gauge theory with two hypermultiplets in the antisymmetric representation and four hypermultiplets in the fundamental representation when we do not see the global structure. Therefore, one can also consider a sequence of decoupling hypermultiplets in terms of the $SO(5)$ viewpoint. 

\begin{figure}
\centering
\includegraphics[width=8cm]{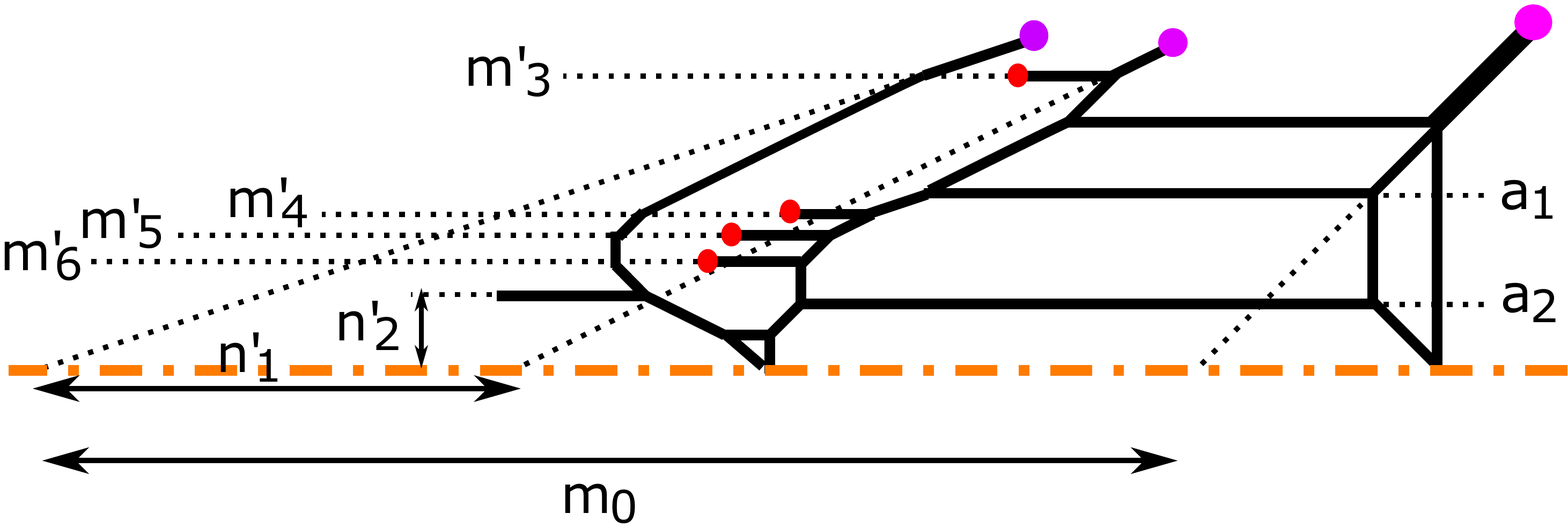}
\caption{The parameterization for the diagram for the $G_2$ gauge theory with six flavors in Figure \ref{fig:G2toSO5def2} .}
\label{fig:G2toSO5para}
\end{figure}

\begin{figure}
\centering
\includegraphics[width=8cm]{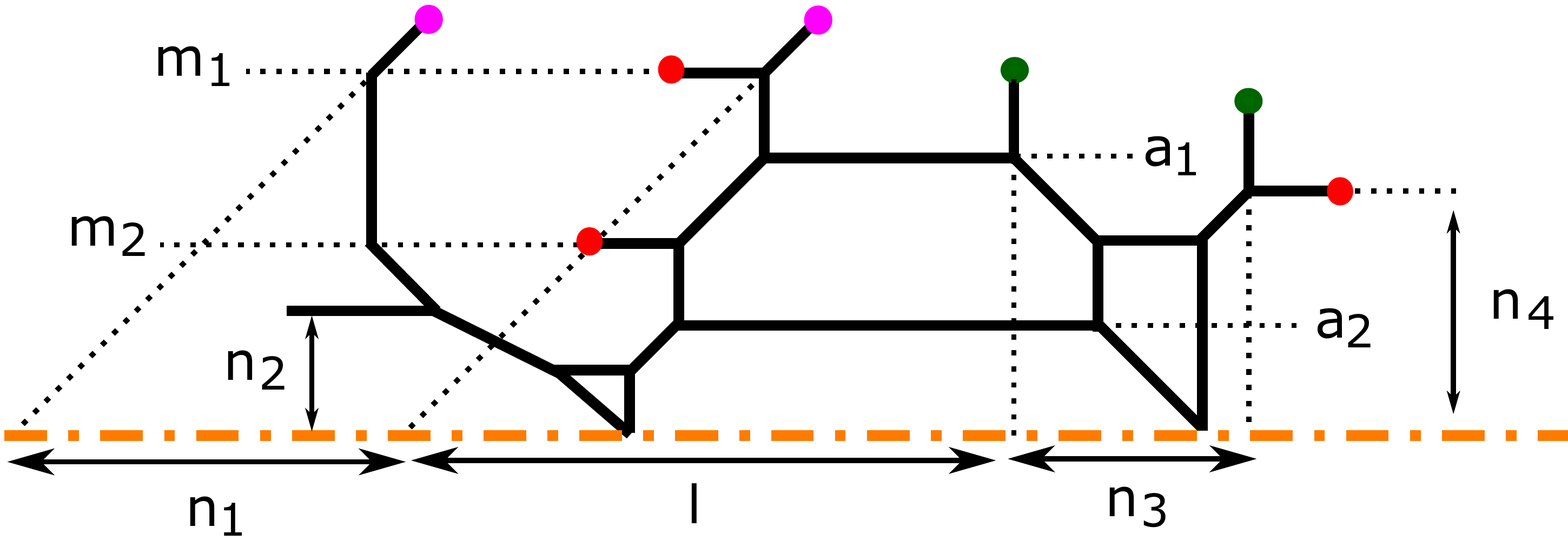}
\caption{The parameterization for the diagram for the $SO(5)$ gauge theory with two vectors and four spinors in Figure \ref{fig:G2toSO5def3} .}
\label{fig:SO5toG2para}
\end{figure}

From the deformation from Figure \ref{fig:G2toSO5def1} to Figure \ref{fig:G2toSO5def5}, one can determine the duality map between the $G_2$ gauge theory with six flavors and the $SO(5)$ gauge theory with two hypermultiplets in the vector representation and four hypermultiplets in the spinor representation. To determine the duality map we compare the diagram in Figure \ref{fig:G2toSO5def2} with the diagram in Figure \ref{fig:G2toSO5def3}. The parameterization for the $G_2$ gauge theory with six flavors is given in Figure \ref{fig:G2toSO5para}. 
%%%%%%%%%%%%%%%%%%%%%%%%%%%%%%%%%
%%%%%%%%%%%%%%%%%%%%%%%%%%%%%%%%%
$a_1, a_2$ are the Coulomb branch moduli and $m_0$ is the inverse of the squared gauge coupling. $n'_1, n'_2$ are related to two mass parameters by \eqref{G2w2flvrsmass}, namely
\be
n'_1 = m'_1 + m'_2, \quad n'_2 = m'_1 - m'_2.
\ee
The other mass parameters $m'_3, m'_4, m'_5, m'_6$ appear directly in the diagram in Figure \ref{fig:G2toSO5para}. On the other hand, the parameterization for the $SO(5)$ gauge theory with two vectors and four spinors is given in Figure \ref{fig:SO5toG2para}. 
%%%%%%%%%%%%%%%%%%%%%%%%%%%%%%%%%
%%%%%%%%%%%%%%%%%%%%%%%%%%%%%%%%%
$a_1, a_2$ are the Coulomb branch moduli and $m_1, m_2$ are the mass parameters for the two hypermultiplets in the vector representation. $n_1, n_2, n_3, n_4$ are the mass parameters for the four hypermultiplets in the spinor representation. Since the mass parameters for the two flavors in section \ref{sec:G2wmatter} originate from the mass parameters of the two spinors of the $SO(7)$ gauge theory before the Higgsing, we choose the mass parameters for the four spinors similarly to \eqref{G2w2flvrsmass}, namely
\bea
n_1 = m_3 + m_4, \quad n_2 = m_3 - m_4, \quad n_3 = m_5 + m_6, \quad n_4 = m_5 - m_6. 
\eea
On the other hand, the length corresponding to $m_0$, the inverse of the squared gauge coupling, is different from that for the $G_2$ gauge theory with two flavors or the $SO(7)$ gauge theory with two spinors. As explained in appendix \ref{sec:m0}, $m_0$ for the $SO(5)$ gauge theory with four spinors, attaching two spinors on the both sides of the diagram, is given by
\be
m_0 = l + \frac{1}{2}n_1 + \frac{1}{2}n_3. \label{m0.SO5w2V4S}
\ee

By comparing the parameterization in Figure \ref{fig:G2toSO5para} with the parameterization in Figure \ref{fig:SO5toG2para} with the deformation depicted in Figure \ref{fig:G2toSO5def2}, we can determine the duality map between the $G_2$ gauge theory with six flavors and the $SO(5)$ gauge theory with two vectors and the four spinors. The duality map is given by 
\bea
m_0^{SO(5)} &=& \frac{m_0^{G_2}}{2}, \label{map1.SO5toG2w6flvrs}\\
m_{{\bf V}, i}^{SO(5)} &=& m_{\bF, 7-i}'^{G_2}, \quad (i=1, 2)\\
m_{{\bf S}, j+2}^{SO(5)} &=& m'^{G_2}_{\bF, j} - \lambda_4, \quad (j=1, \cdots, 4),\\
\phi_1^{SO(5)} &=& \phi_1^{G_2} - 2\lambda_4, \\
\phi_2^{SO(5)} &=& \phi_2^{G_2} - \lambda_4,\label{map9.SO5toG2w6flvrs}
\eea
where 
\bea
\lambda_4 = -\frac{1}{2}m_0^{G_2} + \frac{1}{2}\sum_{i=1}^4m_{\bF, i}^{G_2}
\eea
and we used the Coulomb branch moduli in the Dynkin basis of $SO(5)$
\be
\phi_1 = a_1, \quad \phi_2 = \frac{1}{2}(a_1 + a_2).  \label{SO5.Dynkinbasis}
\ee

We can also see the map between the $Sp(2)$ gauge theory with two antisymmetric hypermultiplets and four flavors discussed in section \ref{sec:G2marginal} with the $SO(5)$ gauge theory from the comparison of the duality map \eqref{map1.Sp2toG2w6flvrs}-\eqref{map9.Sp2toG2w6flvrs} with \eqref{map1.SO5toG2w6flvrs}-\eqref{map9.SO5toG2w6flvrs}. In order to obtain a simple map, we first rename the mass parameters for the $G_2$ gauge theory by
\bea
m_1'^{G_2} = m_3^{G_2}, \quad m_2'^{G_2} = m_4^{G_2}, \quad m_3'^{G_2} = m_5^{G_2}, \quad m_4'^{G_2} = m_6^{G_2}, \quad m_5'^{G_2} = m_1^{G_2}, \quad m_6'^{G_2} = m_2^{G_2}.\nn\\
\eea
Then the map between the $Sp(2)$ gauge theory and the $SO(5)$ gauge theory becomes
\bea
m_0^{SO(5)} &=& m_0^{Sp(2)},\label{map1.SO5toSp2}\\
m_{{\bf V}, i}^{SO(5)} &=& m_{\AS, i}^{Sp(2)}, \quad (i=1, 2),\\
m_{{\bf S}, j}^{SO(5)} &=& m_{\bF, j}^{Sp(2)}, \quad (j=3, 4, 5), \\
m_{{\bf S}, 6}^{SO(5)} &=& -m_{{\bf F}, 6}^{Sp(2)}, \label{map7.SO5toSp2}\\
\phi_k^{SO(5)} &=& \phi_{3-k}^{Sp(2)}, \quad (k=1, 2). \label{map9.SO5toSp2}
\eea
The map \eqref{map1.SO5toSp2}-\eqref{map9.SO5toSp2} is reasonable since the $SO(5)$ gauge theory is equivalent to the $Sp(2)$ gauge theory when we ignore the global structure. Note that there is a minus sign for the map \eqref{map7.SO5toSp2}. This is because the diagram for the $Sp(2)$ gauge theory with two antisymmetric hypermultiplets and four flavors in section \ref{sec:G2marginal} has the discrete theta angle $\pi$ as it was obtained by adding four flavors to the $Sp(2)_{\pi}$ gauge theory with two antisymmetric hypermultiplets. On the other hand, the $SO(5)$ gauge theory in the diagram in Figure \ref{fig:G2toSO5def3} has zero discrete theta angle. Hence the minus sign in \eqref{map7.SO5toSp2} is necessary to change the discrete theta angle. 

%% file: SU3Sp2.tex
\section{$SU(3)$-$Sp(2)$-$SU(2)\times SU(2)$ sequences}\label{sec:SU3Sp2}

In this section, we consider deformations that lead to theories of gauge groups $SU(3)$ and $Sp(2)$  which are dual to each other without involving $G_2$. We start with the marginal theory $Sp(2)+2\AS+4\bF$ in the $G_2-SU(3)-Sp(2)$ sequence and decouple hypermultiplets in the antisymmetric representation. After decoupling one antisymmetric hypermultiplet, we obtain $Sp(2)+1\AS+4\bF$, and then adding more flavors yields another marginal theory $Sp(2) +1 \AS + 8\bF$, which is dual to $SU(3)_{\frac32} + 9 \bF$, and it will be discussed in section \ref{subsec:Sp21AS8F}. We also discuss yet another deformation by decoupling the remaining antisymmetric hypermultiplet and then obtain $Sp(2) + 10\bF$, which is dual to $SU(3)_{0} + 10 \bF$, and it will be discussed in \ref{subsec:SU3CS010F}.

%--------
\begin{figure}
\centering
\subfigure[]{
\includegraphics[width=4cm]{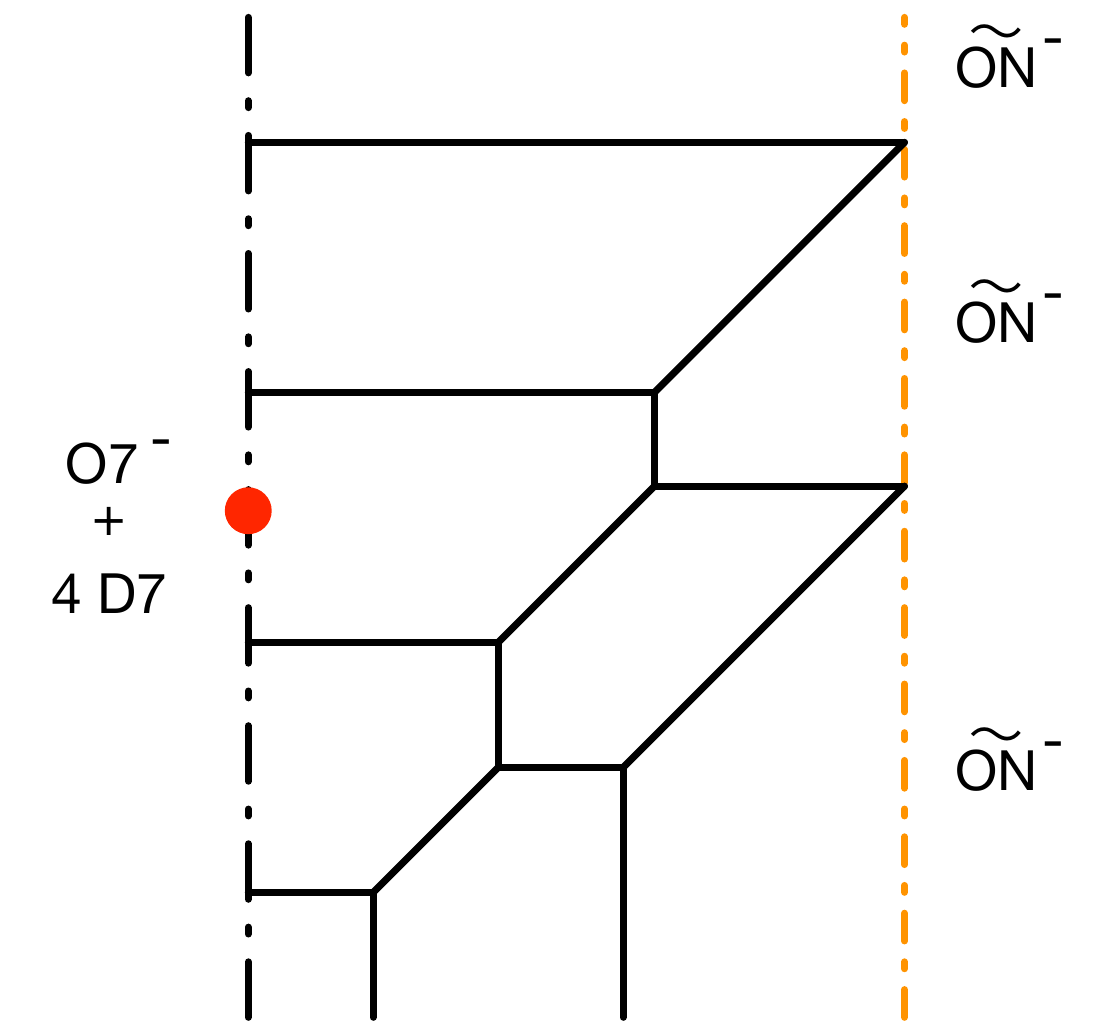} \label{fig:Sp22AS4F-1}}\qquad\qquad\qquad\qquad
\subfigure[]{
\includegraphics[width=3cm]{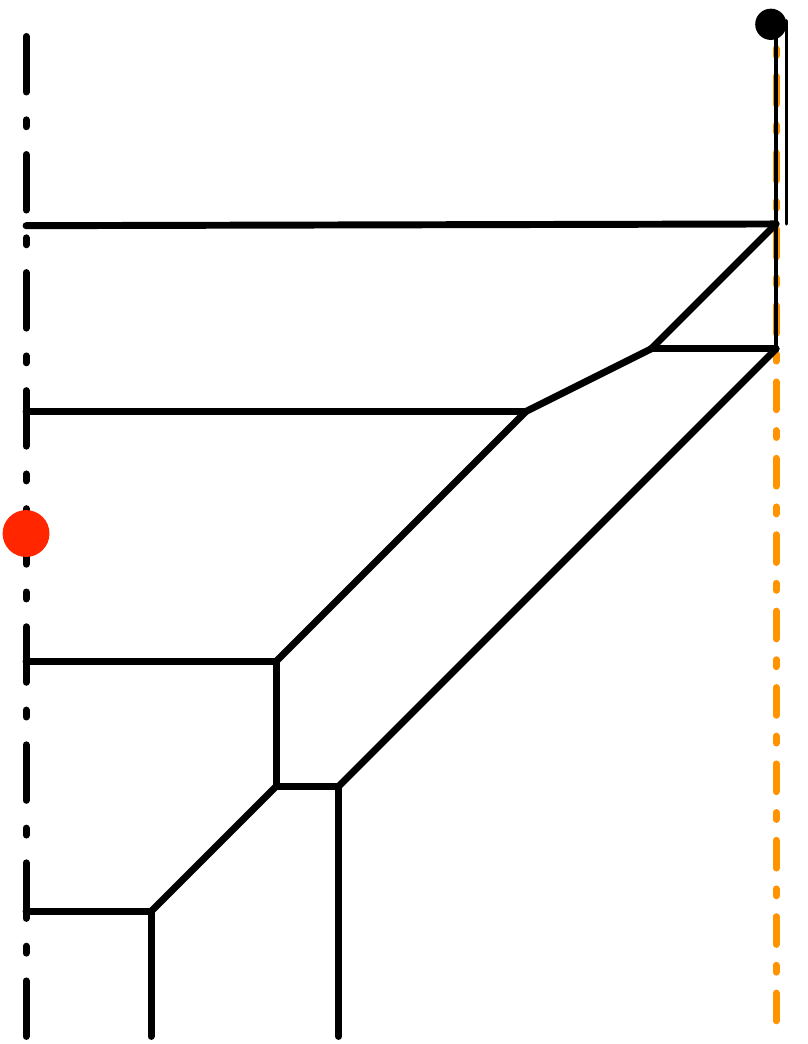} \label{fig:Sp22AS4F-2}}\\
\hspace{1.4cm}\subfigure[]{
\includegraphics[width=4cm]{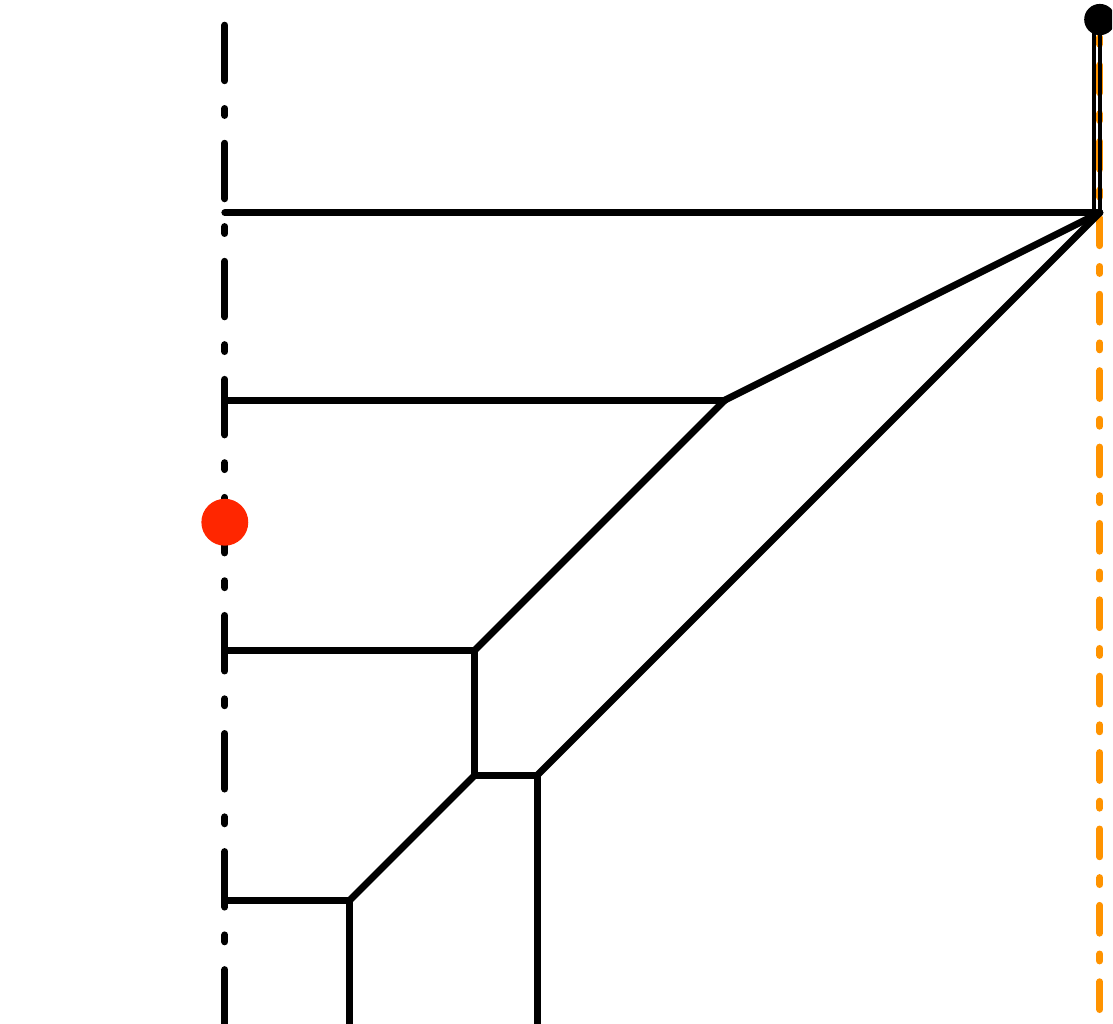} \label{fig:Sp22AS4F-3}}\qquad\qquad\qquad\qquad
\subfigure[]{
\includegraphics[width=4.2cm]{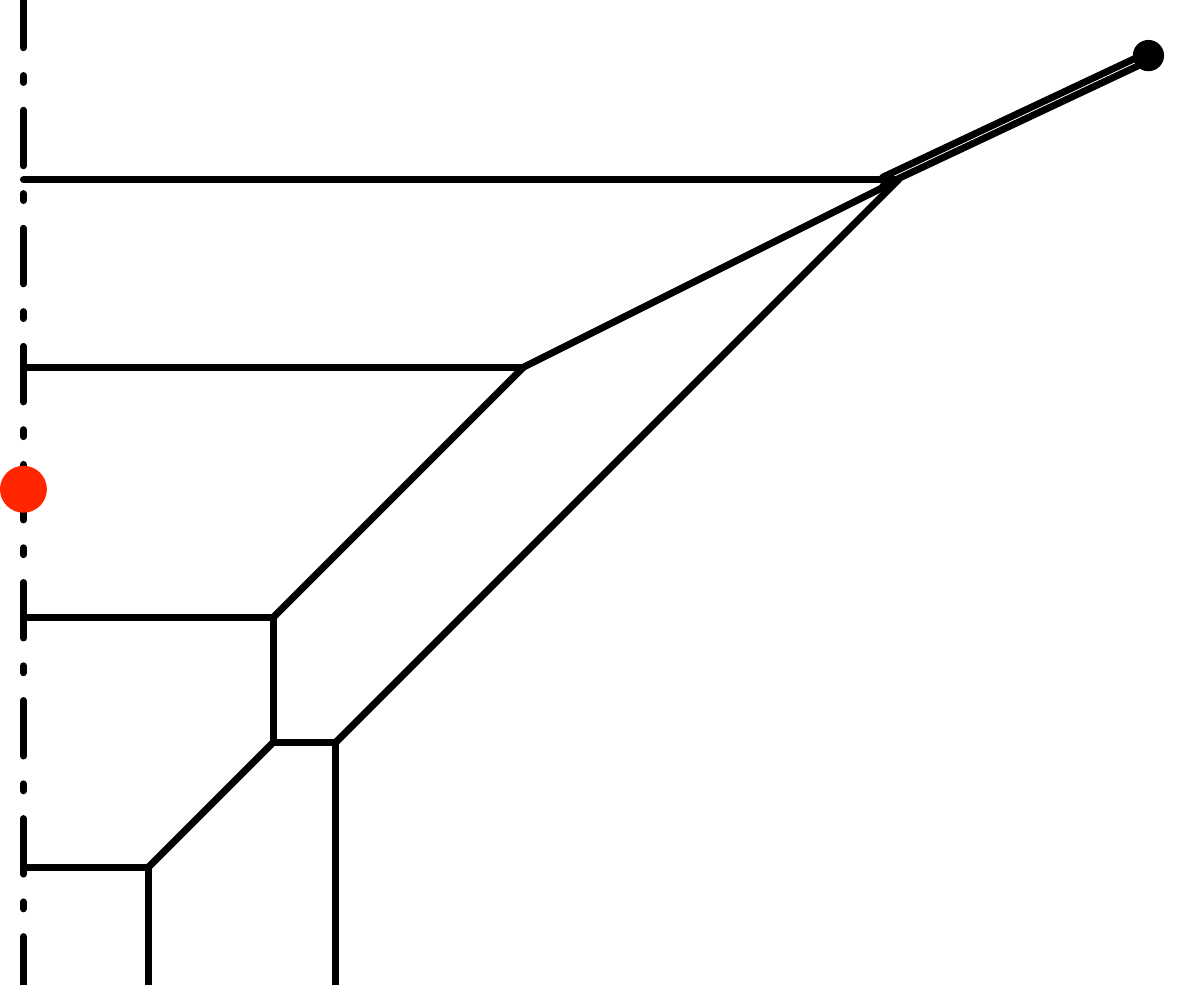} \label{fig:Sp22AS4F-4}} 
\caption{(a): A brane configuration for $Sp(2)+ 2\AS+4\bF$ with an O7$^-$- and an $\widetilde{\text{ON}}$-planes.  (b): A diagram after a flop transition together with relocating a half $D7$-brane to the upper side, which turns the $\widetilde{\text{ON}}$-plane to an ON-plane. (c): An intermediate process of a deformation from Figure \ref{fig:Sp22AS4F-2} to Figure \ref{fig:Sp22AS4F-3} by sending $m_1$ to $\infty$ with $m_2$ kept fixed.  (d): The resulting brane configuration for $Sp(2)+ 1\AS+4\bF$ with an O7$^-$-plane but without an ON-plane.}
\label{fig:Sp22AS4F}
\end{figure}
%--------------------

\subsection{Deformation to 
5-brane web of $SU(3)_{\frac{3}{2}} + 9 {\bf F}$,  $Sp(2) +1 {\bf AS} + 8{\bf F}$ and $[5\bF+SU(2)]\times[SU(2)+2\bF]$}\label{subsec:Sp21AS8F}
As explained in the previous section \ref{sec:G2marginal}, a 5-brane configuration for $G_2+ 6\bF$ can be deformed to display a 5-brane configuration for $Sp(2)+ 2\AS+4\bF$. For example, Figure \ref{fig:G2w6flvrs} shows a deformation from the diagram of $G_2 + 6\bF$ and the last diagram is S-dual to the diagram for $Sp(2) + 2\AS + 4\bF$ given in Figure \ref{fig:Sp2w2AS4F3a}. In this section, we first discuss decoupling of a hypermultiplet in the antisymmetric representation ($\AS$) by starting from a 5-brane web, for instance, Figure \ref{fig:G2w6flvrs5} or equivalently Figure \ref{fig:Sp22AS4F-1}. There are four flavor D7-branes stacked on top of an O7$^-$-plane, and there are two external NS5-branes in Figure \ref{fig:Sp22AS4F-1}. The height of the four D7-branes gives the mass of flavors while the position of the two NS5-branes along the horizontal axis is related to the mass of two antisymmetric hypermultiplets. The precise relation between the mass of antisymmetric hypermultiplets and the length in the diagram in Figure \ref{fig:Sp22AS4F-1} was obtained in \eqref{Sp2w2ASmass}. In particular, the distance between the two NS5-branes parameterizes two times of the mass of one antisymmetric hypermultiplet, $2m_2$. The distance from the $\widetilde{\text{ON}}$-plane to the center of mass position of the two NS5-branes parametrizes the mass of the other antisymmetric hypermultiplet, $m_1$. Let us consider a case where we decouple one $\AS$ by taking $m_1\to\infty$. To this end, we first need to perform a flop transition in such a way as depicted from Figures \ref{fig:Sp22AS4F-1} to \ref{fig:Sp22AS4F-2}.  When we move from the diagram in Figure \ref{fig:Sp22AS4F-1} to the one in Figure \ref{fig:Sp22AS4F-2}, we also transform the $\widetilde{\text{ON}}$-plane into an ON-plane by moving a fractional D7-brane, where the precise process can be found in \cite{Zafrir:2015ftn}. Then we can take the limit $m_1 \to \infty$ with $m_2$ fixed, which can be also realized by sending the horizontal position of the ON-plane to infinitely right. The process is depicted from Figure \ref{fig:Sp22AS4F-2} to Figure \ref{fig:Sp22AS4F-4}. When $m_1$ becomes larger compared to the diagram in Figure \ref{fig:Sp22AS4F-3}, we conjecture that the upper right configuration may involve two $(2, 1)$ 5-branes off from the ON-plane\footnote{In \cite{Hayashi:2017btw}, a similar decoupling limit is discussed, that is the decoupling process from the rank 1 $\tilde{E}_1$ theory to the $E_0$ theory from the perspective of a brane configuration in the presence of an $O5$-plane. Due to the generalized flop transitions, the 5-brane configuration for the pure $Sp(1)_{\pi}$ gauge theory was deformed to have two long NS5-branes near the O5-plane, and taking a limit where the length of the NS5-branes become infinitely long may effectively yields a brane configuration with the long NS5-branes which end on an $[0,1]$ 7-brane.} which preserve the charge conservation, and then eventually the diagram may be effectively described without the ON-plane as in Figure \ref{fig:Sp22AS4F-4} in the limit $m_1 \to \infty$. 
This leads to a brane configuration for $Sp(2)+ 1\AS+4\bF$ in Figure \ref{fig:Sp22AS4F-4}. We note that as shown in \cite{Hayashi:2017btw}, some of the transitions in the deformation shown in Figure \ref{fig:Sp22AS4F}  corresponds to different phases of the Seiberg-Witten curve which can be obtained from the diagrams in Figure \ref{fig:Sp22AS4F}.

%---------------
\begin{figure}
\centering
\subfigure[]{
\includegraphics[width=4.5cm]{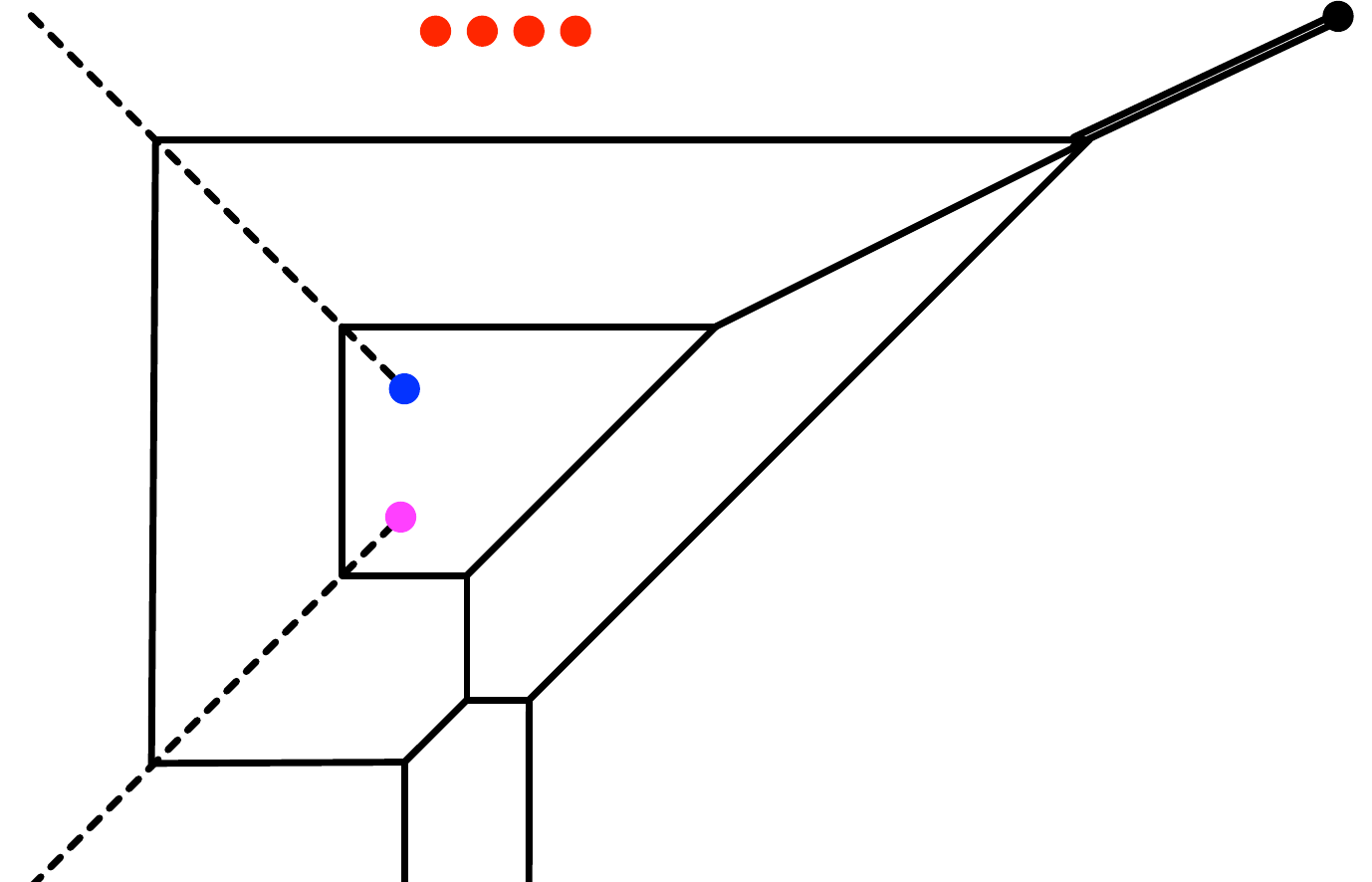} \label{fig:Sp21AS4Fs2}}
\subfigure[]{
\includegraphics[width=4cm]{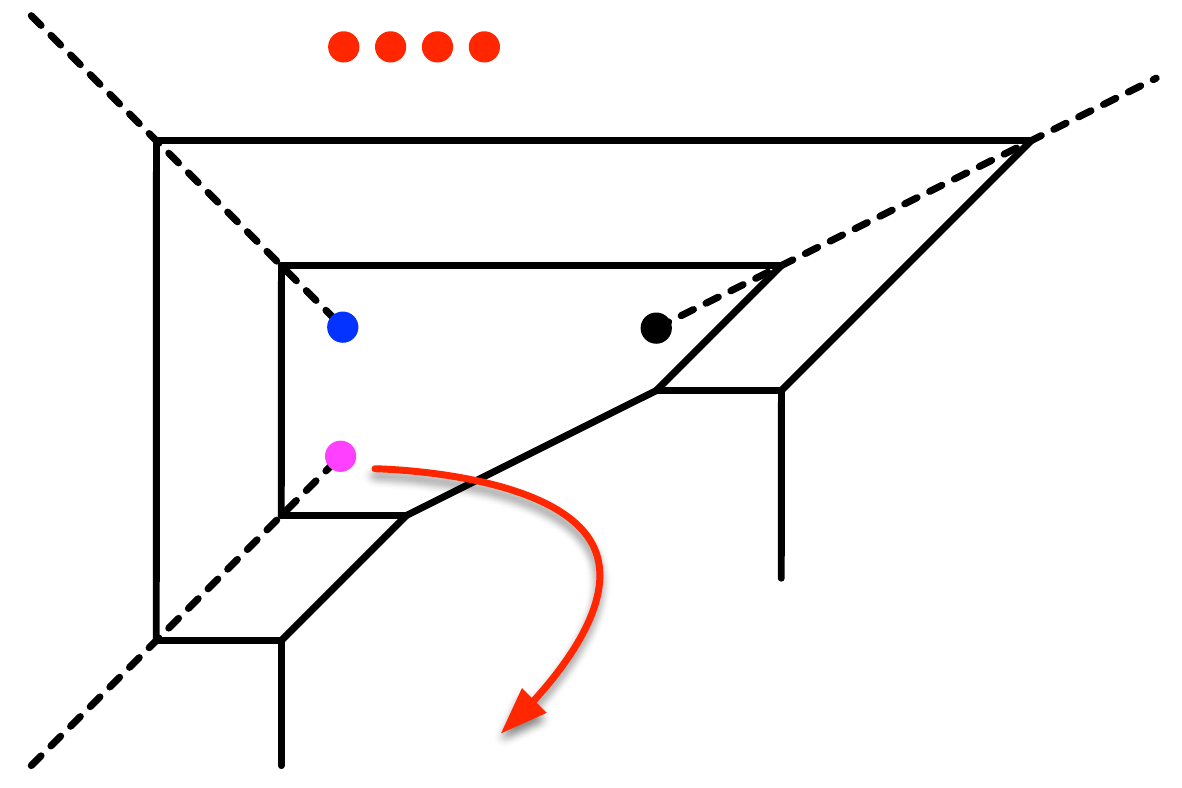} \label{fig:Sp21AS4Fs3}}\quad
\subfigure[]{
\includegraphics[width=4cm]{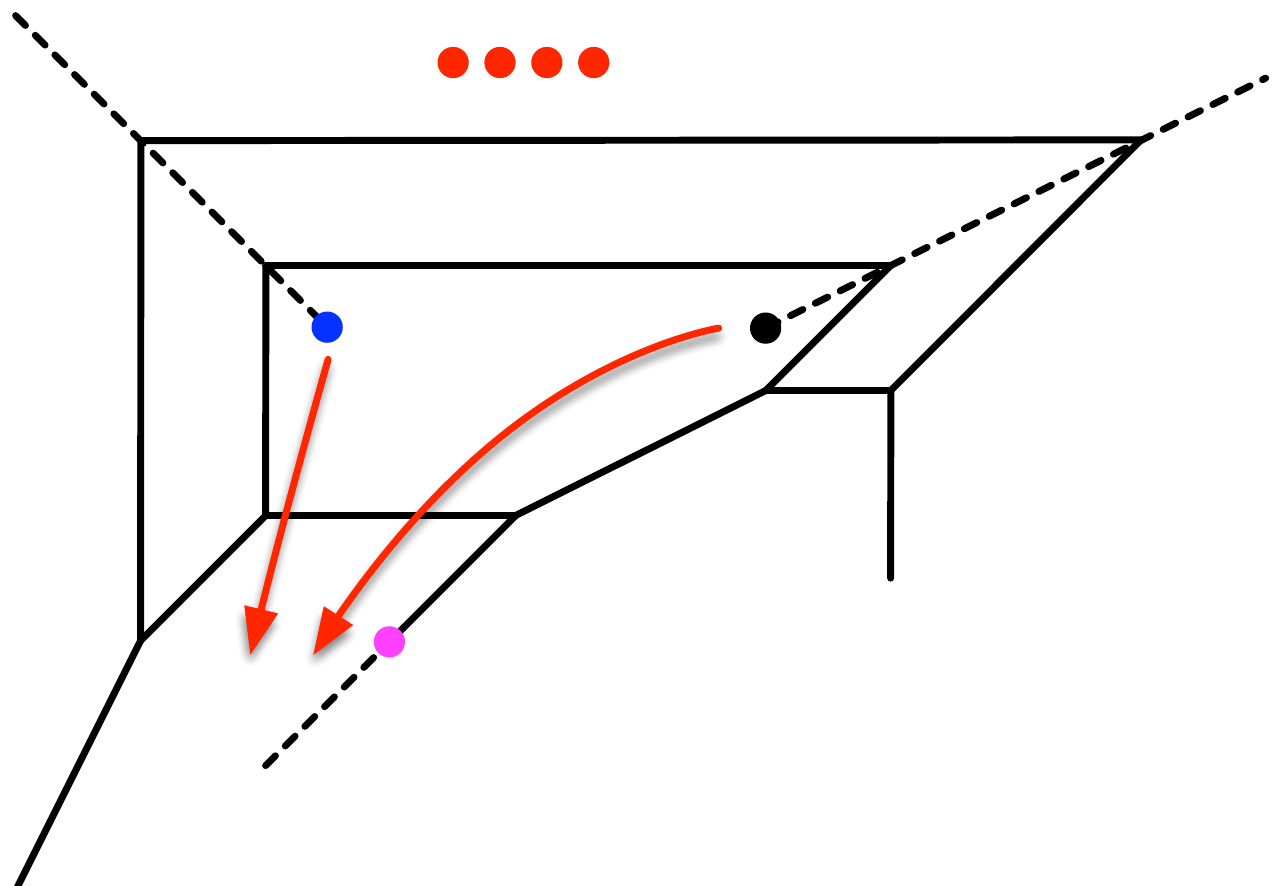} \label{fig:Sp21AS4Fs4}}
\subfigure[]{
\includegraphics[width=9cm]{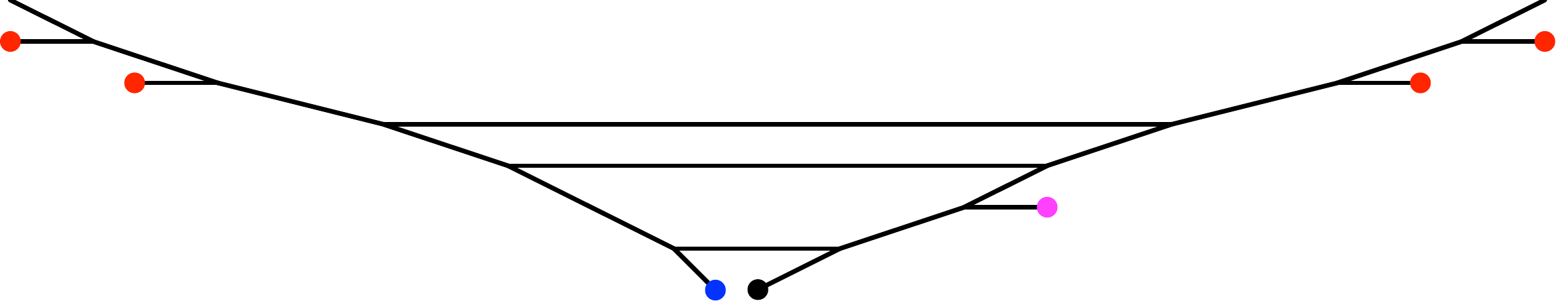}\label{fig:SU35FCS32}}
\caption{(a): A 5-brane web for $Sp(2)+ 1\AS+4\bF$ with the resolution of an O7$^-$-plane into a $[1,-1]$ (blue dot) and a $[1,1]$ (pink dot) 7-branes, where 4 D7-branes (four red dots) are allocated in the upper part. (b): Taking the $[1,1]$ 7-brane outside the 5-brane loop after a flop transition and moving the $[2,1]$ 7-brane (black dot) inside the 5-brane loops. (c): Moving the $[1,-1]$ and $[2,1]$ 7-branes outside the 5-brane loop through the lower D5-brane to make a configuration where an $SU(3)$ gauge theory description is manifest. (d): The resulting 5-brane web for $SU(3)_{\frac72}+ 5\bF$.
}
\label{fig:Sp21AS4Fdual}
\end{figure}
%---------------
\paragraph{Duality between $Sp(2)+ 1\AS+8\bF$ and $SU(3)_{\frac32}+ 9\bF$}
We first see a duality between $Sp(2)+ 1\AS+4\bF$ and $SU(3)_{\frac72}+ 5\bF$ starting from the configuration in Figure \ref{fig:Sp22AS4F-4}. From the perspective of the 5-brane web, this duality can be seen as re-arrangement of 7-branes as depicted in Figure \ref{fig:Sp21AS4Fdual}. By resolving the O7$^-$ into two 7-branes of the charge $[1,-1]$ (blue dot) and $[1,1]$ (pink dot), one finds that the resulting diagram is given by Figure \ref{fig:Sp21AS4Fs2}, where $4$ D7-branes (four red dots) are allocated in the upper part for convenience. After applying a flop transition, the brane configure becomes Figure \ref{fig:Sp21AS4Fs3}, where we also take the $[2,1]$ 7-brane (black dot) inside the 5-brane loops. From this configuration it is possible to move 7-branes around to obtain a brane configuration where the presence of an $SU(3)$ gauge group is manifest. Firstly, we take the $[1,1]$ 7-brane (pink dot) outside along the arrow in Figure \ref{fig:Sp21AS4Fs3}, resulting in the web diagram in Figure \ref{fig:Sp21AS4Fs4}. We then take out the remaining two 7-branes of the charge $[1,-1]$ and $[2, 1]$  across the lower $D5$-brane in Figure \ref{fig:Sp21AS4Fs4}. After moving the 4 D7-branes to the left and the right, we reach a diagram in Figure \ref{fig:SU35FCS32}, which manifestly realizes $SU(3)_{\frac72}+ 5\bF$.

We now consider deformation to the theories with higher flavors from $Sp(2)+1\AS+4\bF$ or $SU(3)_{\frac72}+ 5\bF$. From Figure \ref{fig:Sp21AS4Fs2}, adding more flavors to $Sp(2)+1\AS+4\bF$ is straightforward as one can introduce more D7-branes (red dots). Since adding the D7-branes in the same way for the diagram of $SU(3)_{\frac72}+5\bF$ in Figure \ref{fig:SU35FCS32} should give an equivalent theory, one readily expects that  $Sp(2)+1\AS+(4+n)\bF$ is dual to $SU(3)_{\frac72-\frac{n}{2}}+ (5+n)\bF$ with $n \geq 0$. 
From the point of view of 5-brane web, one can add up to four more flavors to Figure \ref{fig:Sp21AS4Fs2}, and the brane configuration can at most possesses 8 D7-branes which corresponds to $Sp(2)+1\AS+8\bF$, whose UV fixed point exists in six dimensions \cite{Hayashi:2015zka}. Namely the upper bound for the $n$ is four and the marginal theories which are dual to each other are given by
$SU(3)_{\frac32}+ 9\bF$ \cite{Zafrir:2015rga} 
and $Sp(2)+1\AS+8\bF$. 

%---------------
\begin{figure}
\centering
\subfigure[]{
\includegraphics[width=3.5cm]{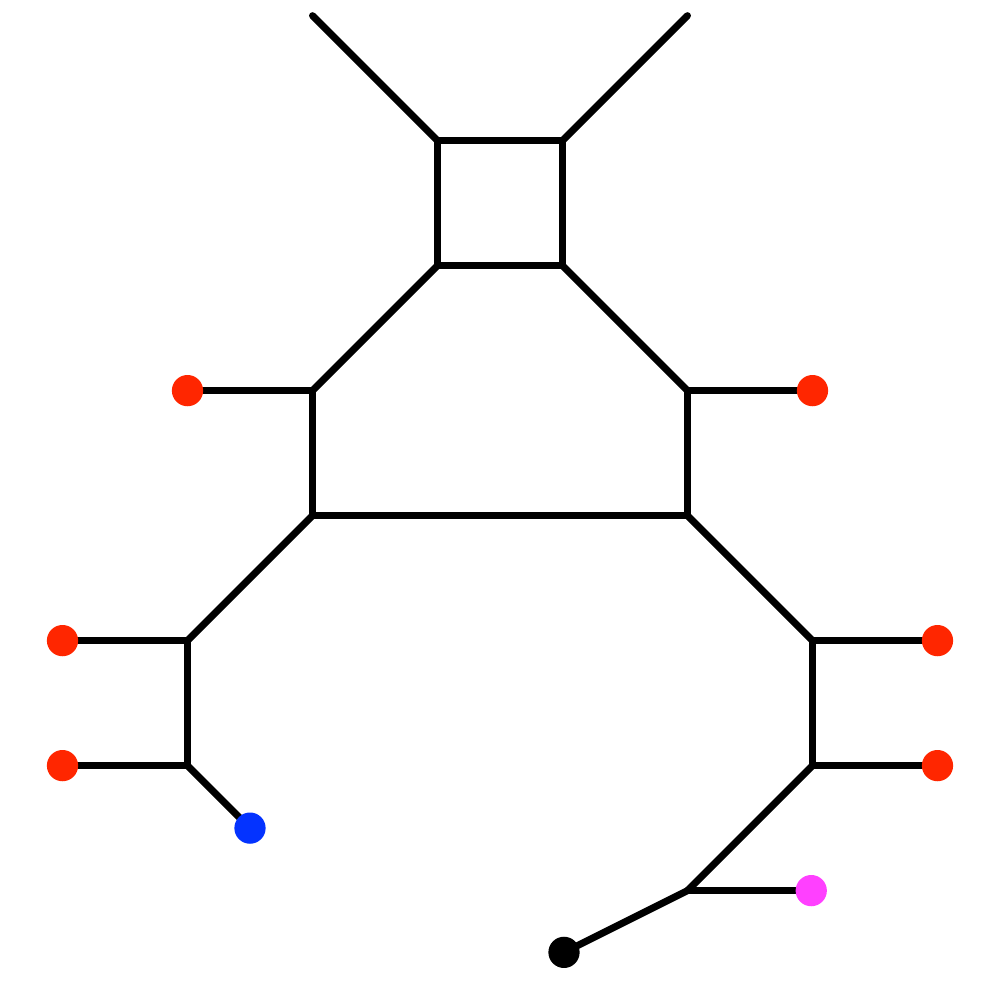} \label{fig:SU2xSU2+5F-1}}\quad
\subfigure[]{
\includegraphics[width=3.5cm]{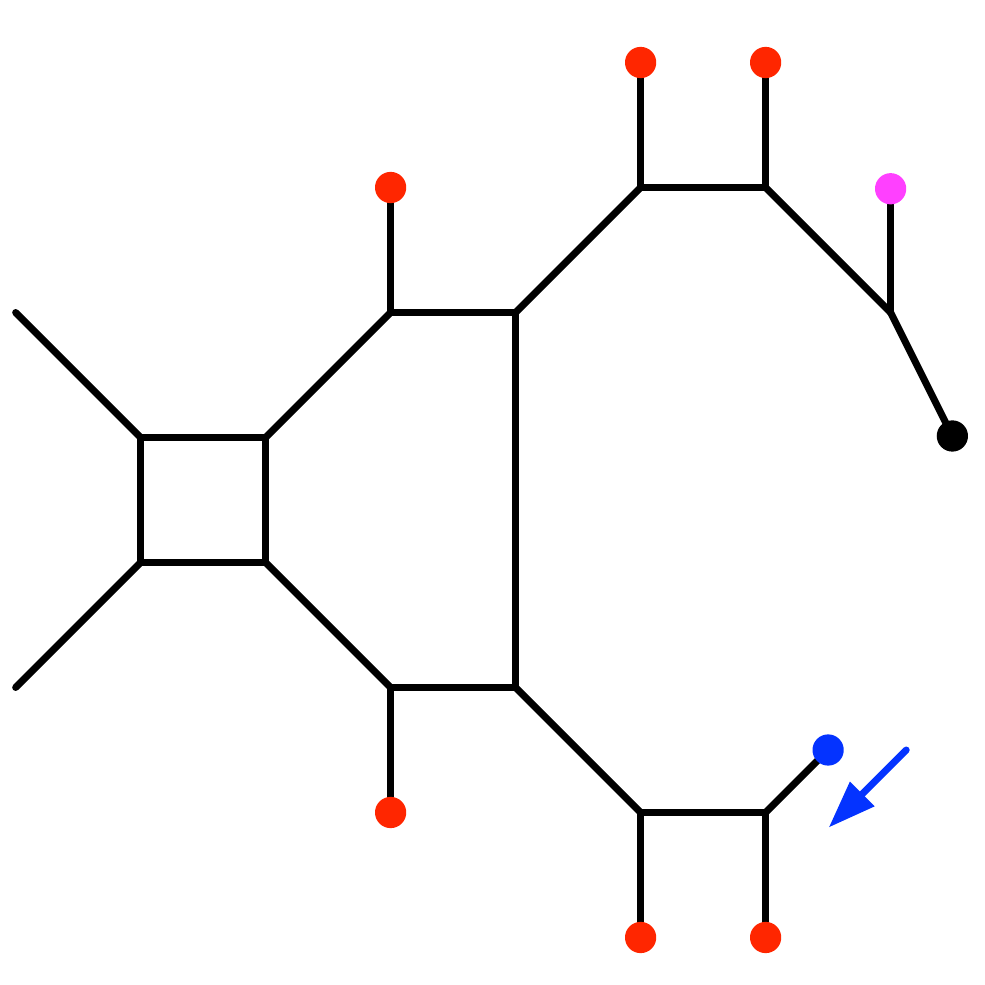} \label{fig:SU2xSU2+5F-2}}\quad
\subfigure[]{
\includegraphics[width=3.5cm]{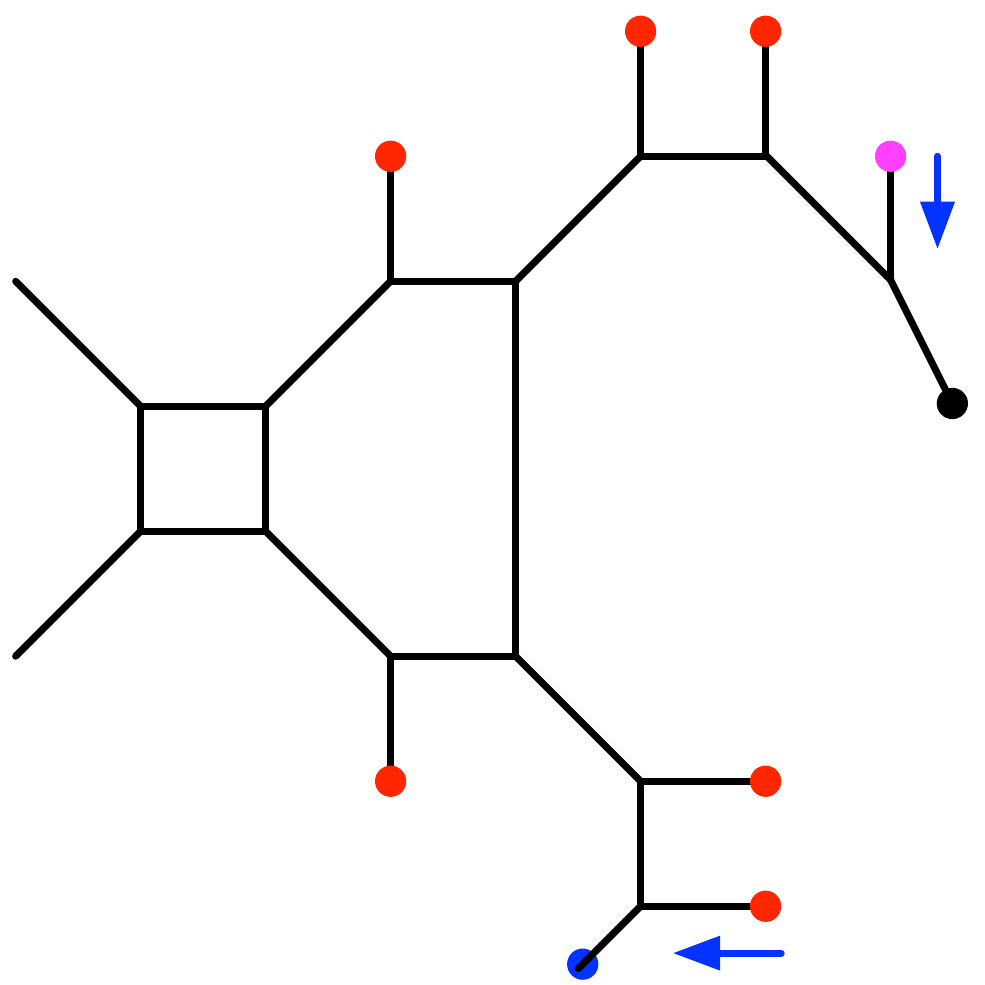} \label{fig:SU2xSU2+5F-3}}\quad
\subfigure[]{
\includegraphics[width=3.5cm]{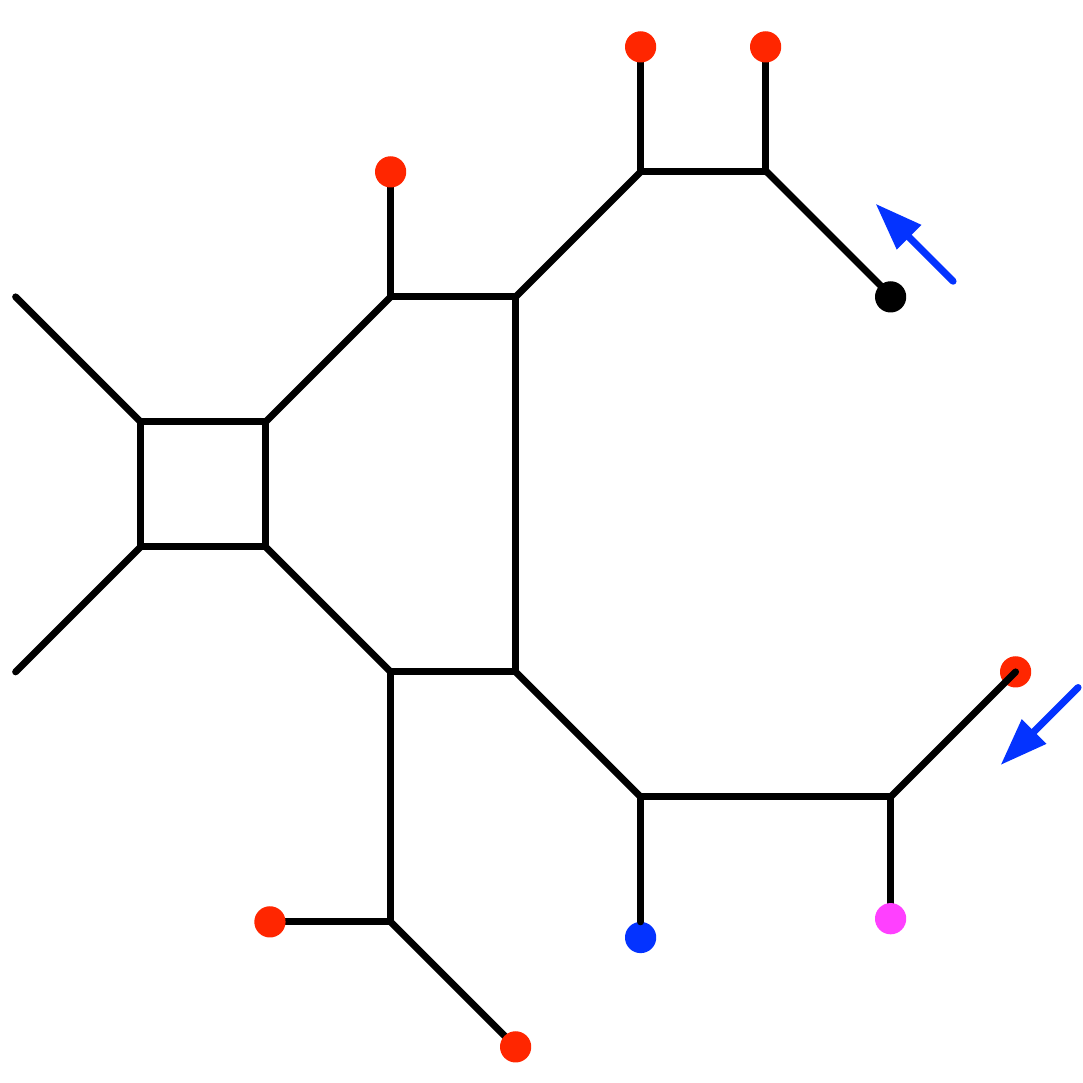}\label{fig:SU2xSU2+5F-4}}\quad
\subfigure[]{
\includegraphics[width=3.5cm]{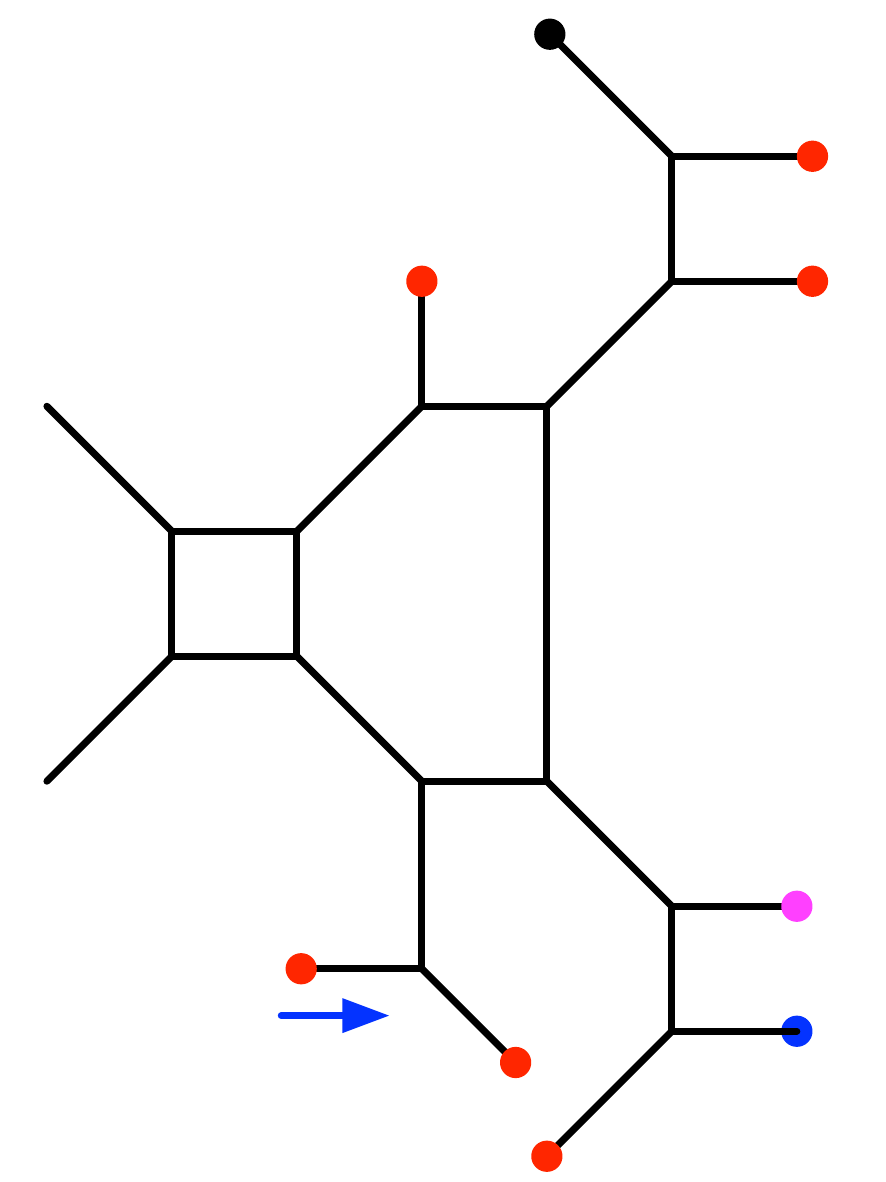}\label{fig:SU2xSU2+5F-5}}\quad
\subfigure[]{
\includegraphics[width=4.5cm]{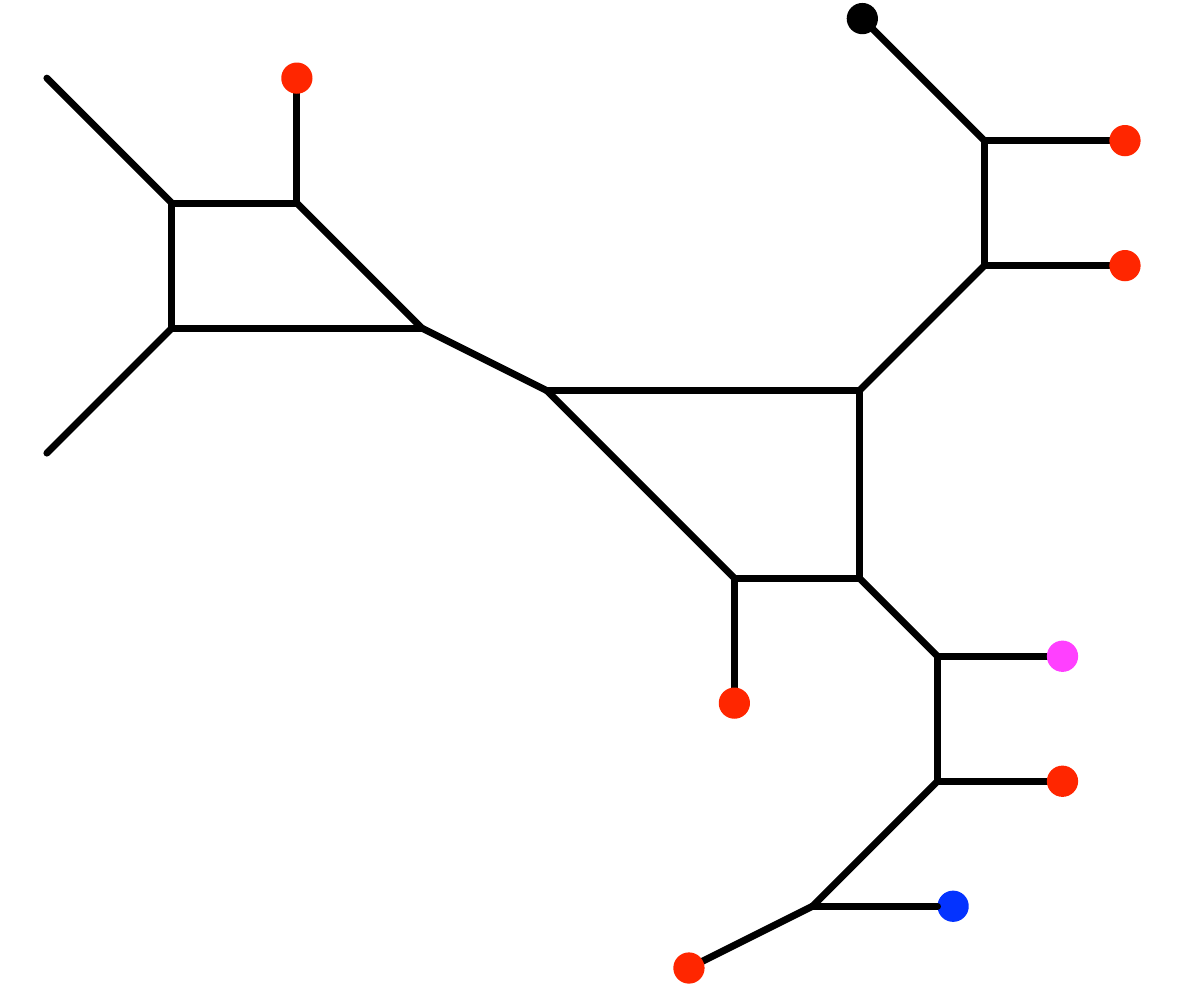}\label{fig:SU2xSU2+5F-6}}
\caption{(a): A mass deformed web-diagram for $SU(3)_{\frac52}+7\bF$. (b): The S-dual version of Figure \ref{fig:SU2xSU2+5F-1}. (c) and (d): various 7-brane motions along the directions of the arrows. (e) and (f): 5-brane webs for $SU(2)_\pi\times [SU(2) +5\bF]$.
}
\label{fig:SU2xSU2-5F}
\end{figure}
%---------------
\paragraph{ $SU(2)\times SU(2)$ quiver description.}

$Sp(2)+1\AS+N_f\bF$ with $6\le N_f\le 8$ has yet another dual description as a quiver theory $[SU(2) +(N_f-6)\bF]\times [SU(2) +5\bF]$ \footnote{It was discussed in \cite{Jefferson:2017ahm, Jefferson:2018irk} that there is some subtlety in the CFT limit of some $SU(2)\times SU(2)$ quiver descriptions.}, which is the quiver consisting of the $SU(2)$ gauge theory with $(N_f-6)$ flavors and the $SU(2)$ gauge theory with five flavors and a bi-fundamental hypermultiplet that transforms as $\bf(2, 2)$ of $SU(2)\times SU(2)$. The duality can be understood from $SU(3)_{\frac{11-N_f}{2}}+(N_f+1)\bF$. $Sp(2)+1\AS+N_f\bF$ is dual to  $SU(3)_{\frac{11-N_f}{2}}+(N_f+1)\bF$ and $SU(3)_{\frac{11-N_f}{2}}+(N_f+1)\bF$ is in fact S-dual to the quiver theory $[SU(2) +(N_f-6)\bF]\times [SU(2) +5\bF]$. 

This can be also explicitly seen from 5-brane webs. As a representative example, we consider $SU(3)_\frac52+7\bF$ (or $Sp(2)+1\AS+6\bF$). A 5-brane web diagram for a mass deformed configuration of $SU(3)_\frac52+7\bF$ is given in Figure \ref{fig:SU2xSU2+5F-1}. Its S-dual transformed web is given in Figure \ref{fig:SU2xSU2+5F-2}. After various 7-brane motions depicted from Figure \ref{fig:SU2xSU2+5F-2} to Figure \ref{fig:SU2xSU2+5F-5}, we find that the resulting 5-brane configuration shows the quiver theory of $\left[SU(2)\right] \times \left[SU(2)+5\bF\right]$ as in Figure \ref{fig:SU2xSU2+5F-5}. In order see the discrete theta angle for the pure $SU(2)$ part, we consider a flop transition from Figure \ref{fig:SU2xSU2+5F-5} to Figure \ref{fig:SU2xSU2+5F-6}. Then we can see that the pure $SU(2)$ part in Figure \ref{fig:SU2xSU2+5F-6} implies the non-trivial discrete theta angle and the quiver theory more precisely is given by $\left[SU(2)_{\pi}\right] \times \left[SU(2)+5\bF\right]$. 
Our finding is also consistent with the claim \cite{Jefferson:2018irk} that a dual of $SU(3)_\frac52+7\bF$ (or $Sp(2)+1\AS+6\bF$) is $SU(2)_\pi\times [SU(2) +5\bF]$.

\begin{figure}
\centering
\subfigure[]{
\includegraphics[width=3.5cm]{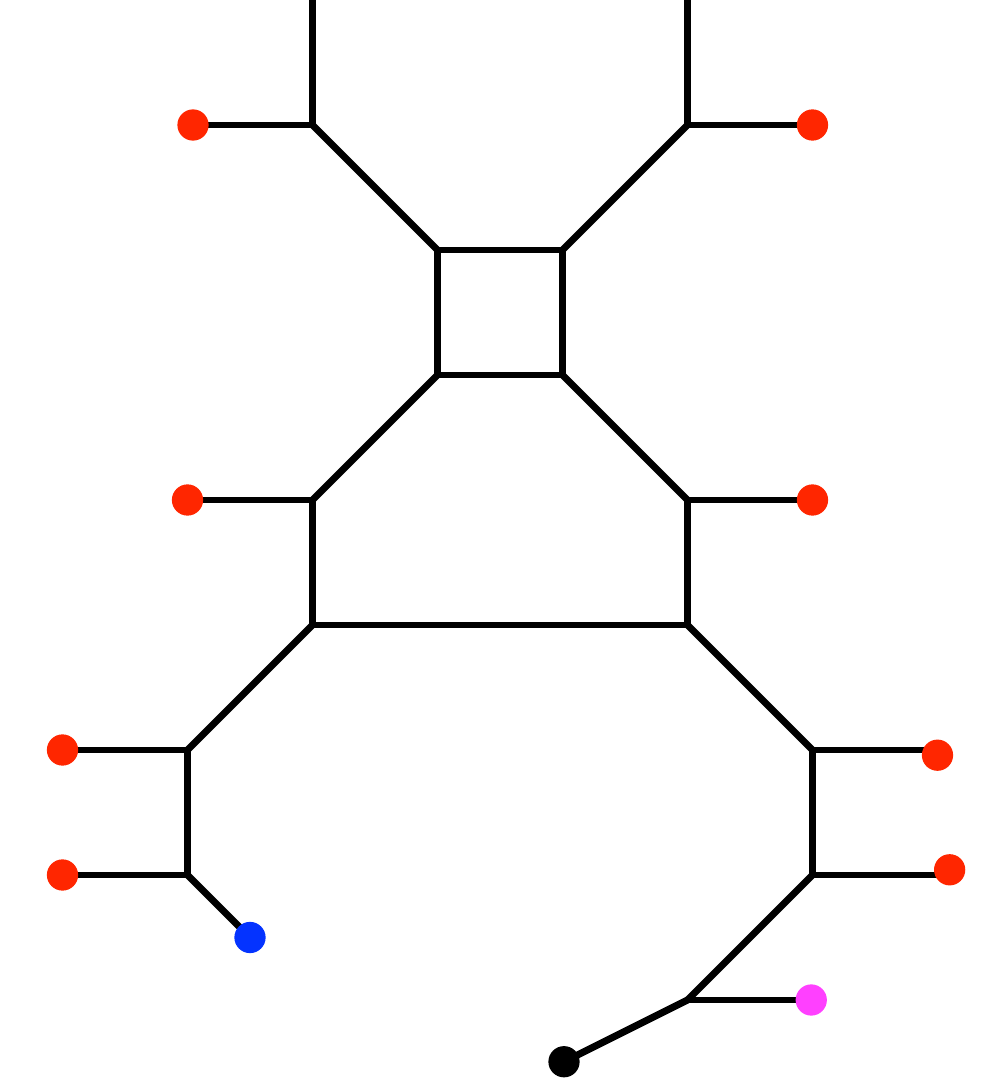} \label{fig:SU3+9F}}\qquad\qquad\qquad
\subfigure[]{
\includegraphics[width=3.5cm]{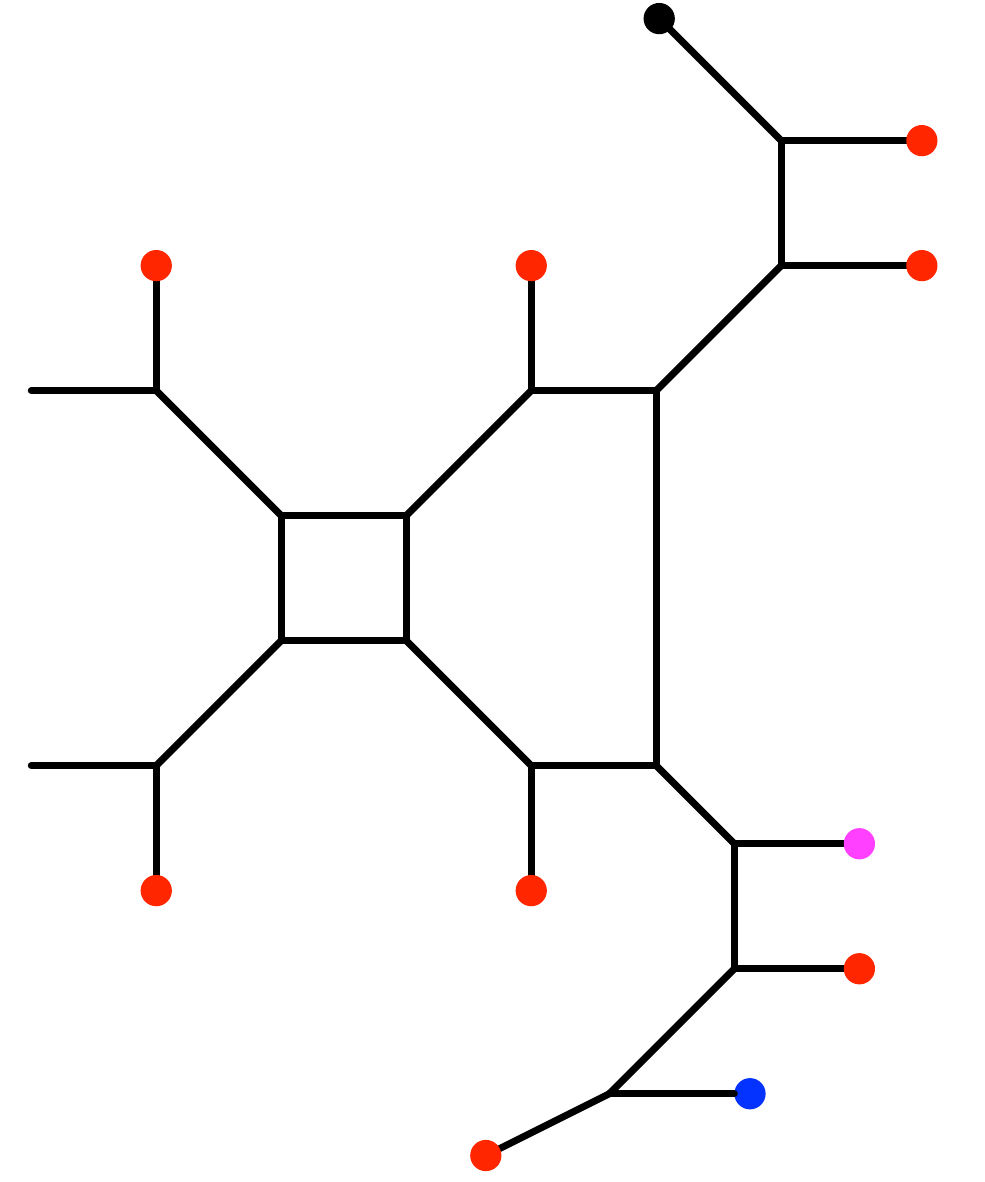} \label{fig:S2F+SU2xSU2+5F}}
\caption{(a) $SU(3)_\frac32+9\bF$. (b) $[SU(2)+2\bF]\times [SU(2) +5\bF]$.}
\label{fig:SU2xSU2+7F}
\end{figure}

It is straightforward to add more flavors to duality relation between $SU(3)_\frac52+7\bF$ and $SU(2)_\pi\times [SU(2) +5\bF]$. With more flavors, the following theories are S-dual to each other:   
\begin{align}
	&SU(3)_2+8\bF~\leftrightarrow~ [SU(2)+1\bF]\times [SU(2) +5\bF], ~{\rm and}\\
	&SU(3)_\frac32+9\bF~\leftrightarrow~[SU(2)+2\bF]\times [SU(2) +5\bF].	
\end{align}
The  corresponding web diagrams for the marginal case are given in Figure \ref{fig:SU2xSU2+7F}.

We note that there is another decoupling from the quiver theory which yields different dualities. 
For example, for the $[SU(2)+2\bF]\times [SU(2) +5\bF]$, there are two possible decoupling of a flavor. One is the decoupling of a flavor from the first $SU(2)$, which was already discussed, and it gives $[SU(2)+1\bF]\times [SU(2) +5\bF]$ dual to 
$SU(3)_2+8\bF$. The other one is the decoupling of a flavor from the second $SU(2)$, which gives $[SU(2)+2\bF]\times [SU(2) +4\bF]$. This theory turns out to be dual to $SU(3)_1+8\bF$ and also to $Sp(2)+8\bF$, which we will discuss in more detail in section \ref{subsec:SU3CS010F}. 

The decoupling of a flavor from the quiver theory $[SU(2)+1\bF]\times [SU(2) +5\bF]$ is, in particular, interesting as it allows three different ways of decoupling of a flavor. Recall that an $SU(2)$ theory with a flavor can lead to the pure $SU(2)$ theory with 
different discrete theta angles, $SU(2)_0$ and $SU(2)_\pi$, depending on taking the mass of the flavor to be $\pm \infty$. By decoupling a flavor in the first $SU(2)$, one hence finds two quiver gauge theories,  
$SU(2)_0\times [SU(2) +5\bF]$ and $SU(2)_\pi\times [SU(2) +5\bF]$. Here the latter theory $SU(2)_\pi\times [SU(2) +5\bF]$ is dual to $SU(3)_\frac32+7\bF$ and also to $Sp(2)+1\AS + 6\bF$ as we discussed before.  %------------------------------

%-----------------------------
\subsection{Duality map between $SU(3)_{\frac{3}{2}} + 9 {\bf F}$ and $Sp(2) +1 {\bf AS} + 8{\bf F}$}
\label{subsec:SU3CS329Fmap}
%------------------------------

In the previous subsection, we saw the duality between the $SU(3)$ gauge theory with nine flavors and the CS level $\frac{3}{2}$ and the $Sp(2)$ gauge theory with one antisymmetric hypermultiplet and eight fundamental hypermultiplets. Since we have the diagrams for the two theories, it is possible to obtain the duality map between the parameters of the two theories. Before obtaining the duality map between the marginal theories, let us start from an easier example by decouple eight flavors from the both theories. We decouple the eight flavors by sending the mass of the flavors to $+\infty$ and redefine the gauge coupling. Then the decoupling yields a duality between the $SU(3)$ gauge theory with one flavor and the CS level $\frac{11}{2}$ and the $Sp(2)$ gauge theory with one antisymmetric hypermultiplet and the non-trivial discrete theta angle. We first obtain the duality map between them. 

%-----------------------------
\begin{figure}
\centering
%\subfigure[]{
\includegraphics[width=8cm]{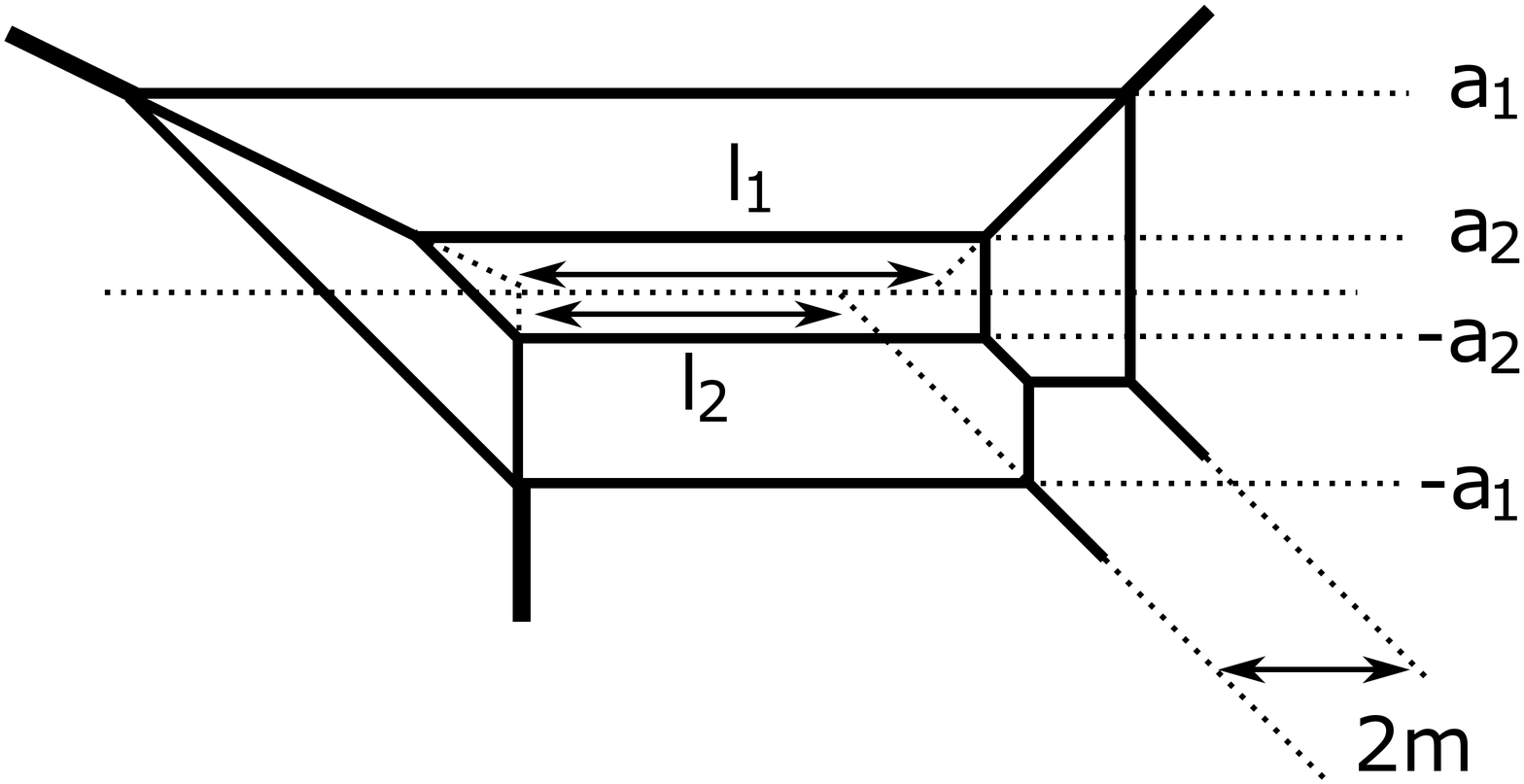} 
%}\qquad\qquad\qquad
%\subfigure[]{
%\includegraphics[width=3.5cm]{2F+SU2xSU2+5F.pdf} \label{fig:S2F+SU2xSU2+5F}}
\caption{The gauge theory parameterization for the $Sp(2)_{\pi}$ gauge theory with a hypermultiplet in the antisymmetric representation.}
\label{fig:Sp2pi1}
\end{figure}
%-----------------------------
The 5-brane web diagram and the gauge theory parameterization for the $Sp(2)_{\pi}$ gauge theory with one antisymmetric hypermultiplet is given in Figure \ref{fig:Sp2pi1}. $a_1, a_2$ are the Coulomb branch moduli of the $Sp(2)$ gauge theory and $m$ is the mass parameter for the antisymmetric hypermutliplet. It turns out that the inverse of the squared gauge coupling is given by
\be
m_0 = \frac{1}{2}(l_1 + l_2 + m). \label{m0.Sp2w1AS}
\ee

Let us confirm the choice of the parameters by comparing the area of the faces in the diagram \ref{fig:Sp2pi1} and the tension of a monopole string computed from the effective prepotential of the $Sp(2)$ gauge theory.
%-----------------------------
\begin{figure}
\centering
\includegraphics[width=8cm]{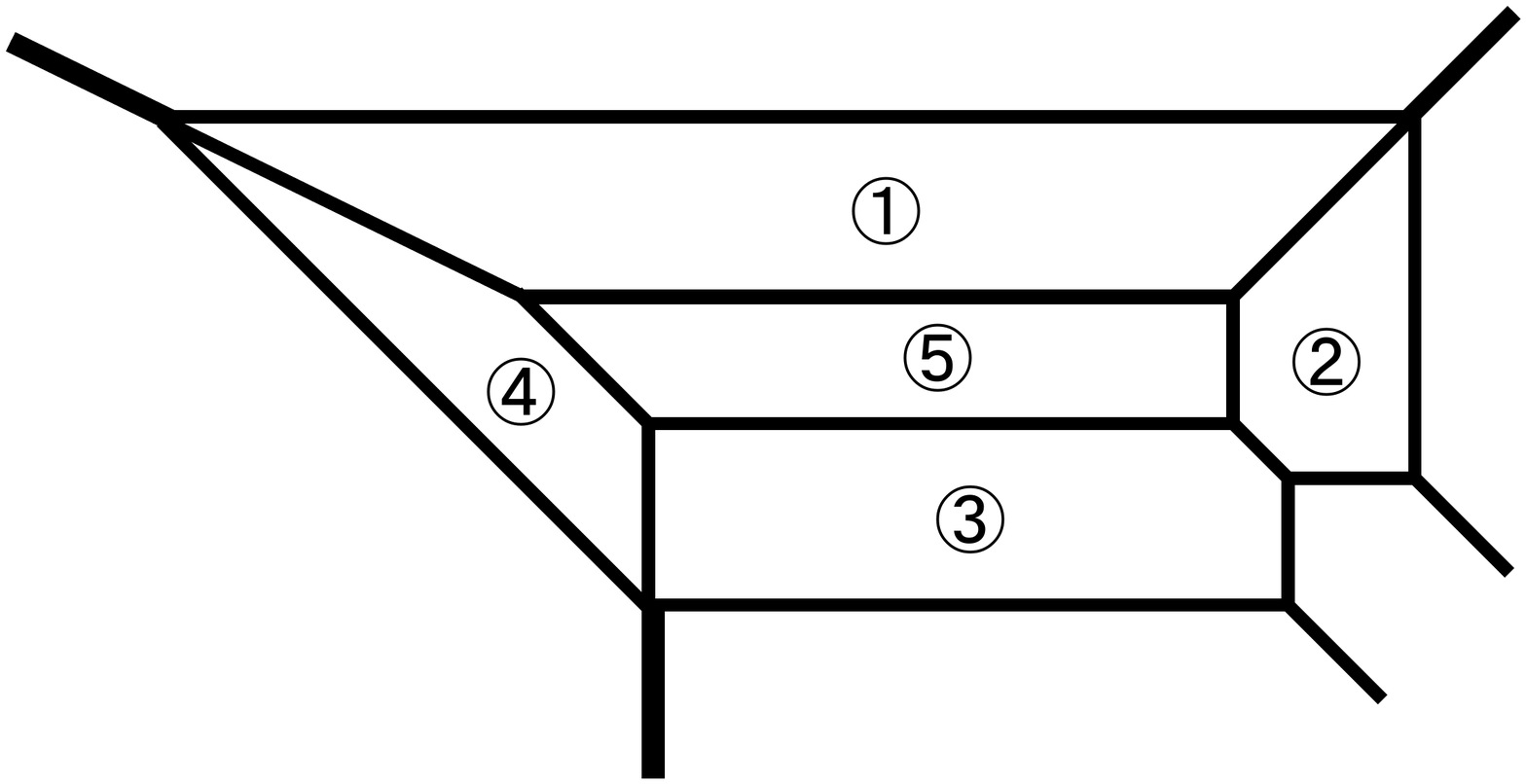} 
\caption{A labeling for five faces in the diagram for the $Sp(2)_{\pi}$ gauge theory with one antisymmetric hypermultiplet.}
\label{fig:Sp2w1ASface}
\end{figure}
%-----------------------------
The area for the faces labeled in Figure \ref{fig:Sp2w1ASface} becomes
\bea
\textcircled{\scriptsize 1} &=& \frac{1}{2}(a_1 - a_2)(2m_0 + 3a_1 + 3a_2), \label{area1.Sp2w1AS}\\
\textcircled{\scriptsize 2} &=& a_1^2 - a_2^2 - \frac{m^2}{2}, \label{area2.Sp2w1AS}\\
\textcircled{\scriptsize 3} &=& \frac{1}{2}(2m_0(a_1 - a_2) + a_1^2 - a_2^2 - m^2), \label{area3.Sp2w1AS}\\
\textcircled{\scriptsize 4} &=& a_1^2 - a_2^2,\label{area4.Sp2w1AS}\\
\textcircled{\scriptsize 5} &=& 2a_2(m_0 + 2a_2). \label{area5.Sp2w1AS}
\eea

On the other hand, the effective prepotential of the $Sp(2)$ gauge theory with one antisymmetric hypermultiplet can be calculated by using \eqref{prepotential}. The phase associated to the antisymmetric hypermultiplet from the diagram in Figure \ref{fig:Sp2pi1} is 
\be
a_1 + a_2 - m > 0, \quad a_1 - a_2 - m > 0, \quad -a_1 + a_2 - m > 0, \quad -a_1 - a_2 - m > 0. \label{phase.Sp2w1AS}
\ee
Hence the effective prepotential of the $Sp(2)$ gauge theory becomes 
\be
\mathcal{F}_{Sp(2)_{\pi} + 1 {\bf AS}} = m_0(2\phi_1^2 - 2\phi_1\phi_2 + \phi_2^2) - m^2\phi_1 + \frac{4}{3}\phi_2(3\phi_1^2 - 3\phi_1\phi_2 + \phi_2^2), \label{F.Sp2w1AS}
\ee
where we used the Coulomb branch moduli in the Dynkin basis of $Sp(2)$ given by \eqref{Sp2.Dynkinbasis}. 

Then the derivative of the prepotential \eqref{F.Sp2w1AS} with repsect to the Coulomb branch moduli yield the tension of a monopole string. A D3-brane can be wrapped on a face $\textcircled{\scriptsize 1}  + \textcircled{\scriptsize 2} + \textcircled{\scriptsize 3} + \textcircled{\scriptsize 4}$ or on a face $\textcircled{\scriptsize 5}$, and the explict comparison between the area \eqref{area1.Sp2w1AS}-\eqref{area5.Sp2w1AS} and the tension calculated from \eqref{F.Sp2w1AS} indeed yields 
\bea
\frac{\partial \mathcal{F}_{Sp(2)_{\pi} + 1{\bf AS}}}{\partial \phi_1} &=& \textcircled{\scriptsize 1}  + \textcircled{\scriptsize 2} + \textcircled{\scriptsize 3} + \textcircled{\scriptsize 4},\label{tension1.Sp2w1AS}\\
\frac{\partial \mathcal{F}_{Sp(2)_{\pi} + 1{\bf AS}}}{\partial \phi_2} &=& \textcircled{\scriptsize 5}, \label{tension2.Sp2w1AS} 
\eea
which confirms the gauge theory parameterization in the diagram in Figure \ref{fig:Sp2pi1} and \eqref{m0.Sp2w1AS}.

\begin{figure}
\centering
%\subfigure[]{
\includegraphics[width=8cm]{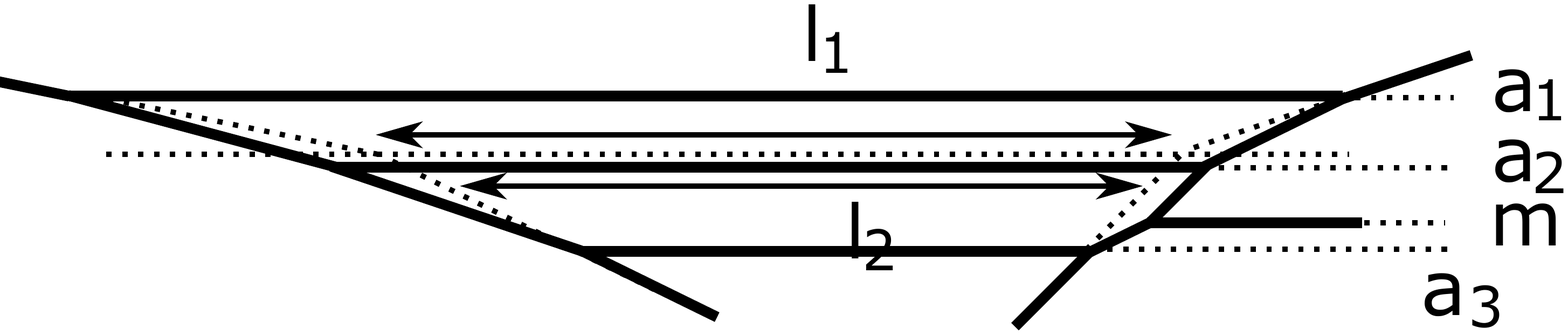} 
%}\qquad\qquad\qquad
%\subfigure[]{
%\includegraphics[width=3.5cm]{2F+SU2xSU2+5F.pdf} \label{fig:S2F+SU2xSU2+5F}}
\caption{The gauge theory parameterization for the $SU(3)$ gauge theory with one flavor and the CS level $\frac{11}{2}$. }
\label{fig:SU3w1F}
\end{figure}

The diagram in Figure \ref{fig:Sp2pi1} can be deformed into the one for the $SU(3)$ gauge theory with one flavor and the CS level $\frac{11}{2}$, The deformation is essentially given in Figure \ref{fig:Sp2pi1} and the resulting web and also the gauge theory parameterization for the $SU(3)$ gauge theory are depicted in Figure \ref{fig:SU3w1F}. 
%------------------------------
%-----------------------------
$a_1, a_2, a_3$ with $a_1 + a_2 + a_3 = 0$ are the Coulomb branch moduli and $m$ is the mass parameter for the one flavor. The inverse of the squared gauge coupling $m_0$ is given by 
\be
m_0 = \frac{1}{2}(l_1 + l_2). \label{m0.SU3w1F}
\ee

\begin{figure}
\centering
\includegraphics[width=8cm]{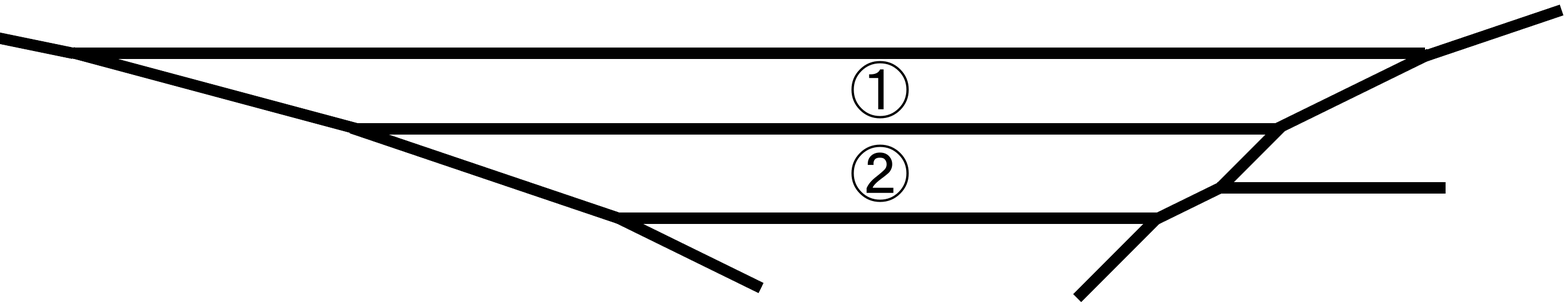} 
\caption{A labeling for two faces in the diagram for the $SU(3)_{\frac{11}{2}}$ gauge theory with one flavor.}
\label{fig:SU3w1Fface}
\end{figure}

Let us also compare the area of the faces in Figure \ref{fig:SU3w1F} with the tension of a monopole string for completeness. The faces labeled in Figure \ref{fig:SU3w1Fface}
%-----------------------------
%-----------------------------
yield the area
\bea
\textcircled{\scriptsize 1} &=& \frac{1}{2}(a_1 - a_2)(2m_0 + m + 6a_1 + 2a_2 - 4a_3),\label{area1.SU3w1F}\\
\textcircled{\scriptsize 2} &=& \frac{1}{2}(2m_0(a_2 - a_3) - m^2 + (a_2 + a_3) m + 4a_2^2 - 4a_2a_3 - a_3^2). \label{area2.SU3w1F}
\eea
On the other hand, the phase for the $SU(3)_{\frac{11}{2}}$ gauge theory with one flavor is given by
\be
a_1 - m > 0, \quad a_2 - m > 0, \quad a_3 - m> 0,
\ee
and the effective prepotential becomes
\bea
\mathcal{F}_{SU(3)_{\frac{11}{2}}+1{\bf F}} &=& \frac{1}{2}m_0(\phi_1^2 - \phi_1\phi_2 + \phi_2^2) + \frac{1}{12}m^3 + \frac{1}{2}m\phi_1(\phi_1 - \phi_2) - \frac{1}{2}m^2\phi_2\nn\\
&& + \frac{4}{3}\phi_1^3 + 2\phi_1^2\phi_2 - 3\phi_1\phi_2^2 + \frac{7}{6}\phi_2^3, \label{F.SU3w1F}
\eea
where we used the Coulomb branch moduli in the Dynkin basis \eqref{SU3.Dynkinbasis}. The area of the faces \eqref{area1.SU3w1F} and \eqref{area2.SU3w1F} agrees with the derivative of the prepotential \eqref{F.SU3w1F} with respect to the Coulomb branch moduli $\phi_1, \phi_2$ by
\bea
\frac{\partial \mathcal{F}_{SU(3)_{\frac{11}{2}} + 1{\bf F}}}{\partial \phi_1} &=& \textcircled{\scriptsize 1},\label{tension1.SU3w1F}\\
\frac{\partial \mathcal{F}_{SU(3)_{\frac{11}{2}} + 1{\bf F}}}{\partial \phi_2} &=& \textcircled{\scriptsize 2}. \label{tension2.SU3w1F} 
\eea

Since we know the deformation between the diagrams for the $Sp(2)_{\pi}$ gauge theory with one antisymmetric hypermultiplet and the $SU(3)_{\frac{11}{2}}$ gauge theory with one flavor as well as their parameterization, comparing the two diagrams may give the duality map between the two parameterization. The duality map then is given by 
\begin{align}
	m_0^{SU(3)} &= -\frac13 m_0^{Sp(2)}+\frac56 m^{Sp(2)}_{\AS},\\
		m_{\bF}^{SU(3)} & =- m_{\AS}^{Sp(2)}+2\lambda,\\
		\phi_1^{SU(3)} &= \phi_1^{Sp(2)} - \lambda, \\
         \phi_2^{SU(3)} &= \phi_2^{Sp(2)} - 2\lambda,
	\end{align}
where
\begin{align}
	\lambda = -\frac13 m_0^{Sp(2)}+ \frac13 m_{\AS}^{Sp(2)}.
\end{align}

It is now straightforward to obtain the duality map for the $SU(3)_{\frac{3}{2}}$ gauge theory with nine flavors and the $Sp(2)$ gauge theory with one antisymmetirc hypermultiplet an eight flavors. Adding eight flavors in both theories can be accomplished by introducing eight D7-branes in the two diagrams. The height of the eight D7-branes in the two diagrams are equal to each other and the definition of the inverse of the gauge coupling also changes according to the change of the slope of the external 5-branes which we used in the diagrams in Figure \ref{fig:Sp2pi1} and Figure \ref{fig:SU3w1F}. Then the duality map between $SU(3)_{-\frac{3}{2}}$ with nine flavors and $Sp(2)$ gauge theory with one antisymmetric hypermultiplet and eight flavors is given by
\begin{align}
	m_0^{SU(3)} &= m_0^{Sp(2)}-\frac12 m^{Sp(2)}_{\AS},\label{eq:m0su3m0Sp29F}\\
	m_{\bF}^{SU(3)} & =- m_{\AS}^{Sp(2)}+2\lambda,\\
	m_{\bF, i}^{SU(3)} & = m_{\bF, i}^{Sp(2)} -\lambda\qquad (i=1,\cdots, 8),\\
	\phi_1^{SU(3)} &= \phi_1^{Sp(2)} -\lambda,\\
	\phi_2^{SU(3)} &= \phi_2^{Sp(2)} - 2\lambda,
\end{align}
where
\begin{align}
	\lambda = -\frac13 m_0^{Sp(2)}+ \frac13 m_{\AS}^{Sp(2)}+ \frac16 \sum_{i=1}^{8}m_{\bF, i}^{Sp(2)}.
\end{align}
%-----------------------------
Or if we express the $Sp(2)$ gauge theory parameters in terms of the $SU(3)$ gauge theory parameters, the map becomes 
\begin{align}
	m_0^{Sp(2)} &= m_0^{SU(3)}-\frac12 m^{SU(3)}_{\bF} - \lambda',\label{eq:m0su3m0Sp29Fv2}\\
	m_{\AS}^{Sp(2)} & =- m_{\bF}^{SU(3)} - 2\lambda',\\
	m_{\bF, i}^{Sp(2)} & = m_{\bF, i}^{SU(3)} -\lambda' \qquad (i=1,\cdots, 8),\\
	\phi_1^{Sp(2)} &= \phi_1^{SU(3)} -\lambda',\\
	\phi_2^{Sp(2)} &= \phi_2^{SU(3)} - 2\lambda',\label{eq:phi2su3phi2sp29F}
\end{align}
where
\begin{align}
	\lambda' = -\frac12 m_0^{SU(3)} - \frac14 m_{\bF}^{SU(3)}+ \frac14 \sum_{i=1}^{8}m_{\bF, i}^{SU(3)}.
\end{align}

%%%%%%%%%%%%%%%%%%%%%%%%%%%%

\subsection{Periodicity for the diagrams of $Sp(2) +1 {\bf AS} + 8{\bf F}$ and $SU(3)_{\frac{3}{2}} + 9 {\bf F}$}
\label{subsec:taoDiagram}
For a marginal theory, which can be viewed as a 6d theory on a circle, it may be natural to assume that the bare coupling of the marginal theory would be the radius of the compactification circle. It is then expected that bare couplings of each dual theory are equal to each other as they would correspond to the same radius. For instance, the bare couplings of $SU(3)_0 + 10 {\bf F}$ and $Sp(2) + 10 {\bf F}$ are equal to each other, which can be also explicitly seen from their 5-brane webs as done in \cite{Hayashi:2016abm}. However, from the duality maps we obtained, some dual theories which are marginal have had different $m_0$. For instance, the bare coupling of $SU(3)_4+6\bF$ is different from that of  $Sp(2) + 2\AS + 4\bF$ as in \eqref{map1.Sp2toSU3w6flvrs}. Another example is the bare couplings of $SU(3)_{\frac{3}{2}} + 9 {\bf F}$ and $Sp(2) +1 {\bf AS} + 8{\bf F}$ as shown in \eqref{eq:m0su3m0Sp29F}. It is then natural to ask which $m_0$ is related to the period of a circle associated to the circle compactification of a 6d theory. In this subsection, we consider 5-brane configurations of $SU(3)_{\frac{3}{2}} + 9 {\bf F}$ and $Sp(2) +1 {\bf AS} + 8{\bf F}$ and compare their bare couplings with the period of the diagrams.
%---------------
\begin{figure}
\centering
\includegraphics[width=14cm]{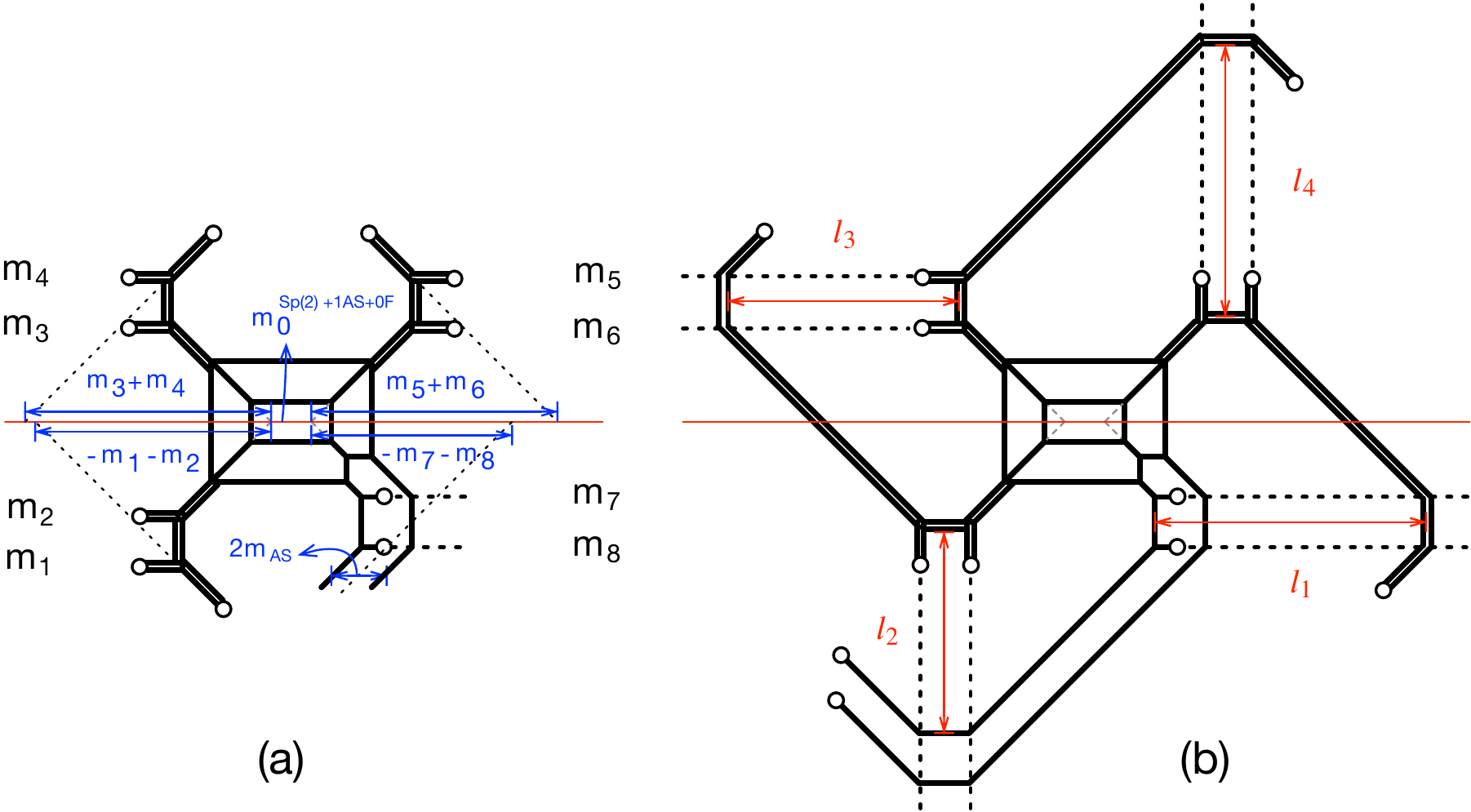} 
\caption{(a): The parameterization for $Sp(2)+1\AS+8\bF$. $m_0$ for $N_f = 8$ is given by $m_0^{N_f=8} = m_0^{N_f=0}-\frac12\big(m_1+m_2-m_3-m_4-m_5-m_6+m_7+m_8 \big)$.  (b): A Tao diagram for $Sp(2)+1\AS+8\bF$.}
\label{fig:Sp21A8FTao}
\end{figure}
%---------------

Marginal theories whose 5-brane configuration can be constructed without an orientifold are described by a particular 5-brane configuration of special properties: a shape of an infinite rotating spiral with constant period, call it a Tao web diagram \cite{Kim:2015jba,Hayashi:2015fsa}. Since a Tao diagram is periodic, the period associated to the diagram can be read off from the configuration of a Tao diagram. 

Consider a 5-brane web for $Sp(2) +1 {\bf AS} + 8{\bf F}$. For instance, Figure \ref{fig:Sp21A8FTao}(a) is an example of a 5-brane web configuration for $Sp(2) +1 {\bf AS} + 8{\bf F}$. Applying the Hanay-Witten transitions explained in \cite{Kim:2015jba}, one can readily get a Tao web diagram for $Sp(2) +1 {\bf AS} + 8{\bf F}$ \cite{Hayashi:2015zka} depicted in Figure \ref{fig:Sp21A8FTao}(b). It follows from Figure \ref{fig:Sp21A8FTao}(a) that the inverse of the squared gauge coupling of $Sp(2) +1 {\bf AS} + 8{\bf F}$ can be diagrammatically computed by taking the average of the asymptotic distances on the center of the Coulomb branch moduli from two pairs of the external 5-branes: 
\begin{align}
	m_0^{Sp(2) +1 {\bf AS} + 8{\bf F}} = &~m_0^{Sp(2) +1 {\bf AS} + 0{\bf F}}\nn\\
	&-\frac12\big(m_1^{Sp(2)}+m_2^{Sp(2)}-m_3^{Sp(2)}-m_4^{Sp(2)}-m_5^{Sp(2)}-m_6^{Sp(2)}+m_7^{Sp(2)}+m_8^{Sp(2)} \big).\nonumber\\
\end{align}
The length $l_i$ in Figure \ref{fig:Sp21A8FTao}(b) can be expressed by the gauge theory parameters as
\begin{align}
	l_1&=m_5^{Sp(2)}+m_6^{Sp(2)}+m_{\AS},  &l_2&=m_0^{Sp(2) +1 {\bf AS} + 0{\bf F}}-m_7^{Sp(2)}-m_8^{Sp(2)}-m_\AS,\cr
	l_3&=-m_1^{Sp(2)}-m_2^{Sp(2)}, &l_4&=m_0^{Sp(2) +1 {\bf AS} + 0{\bf F}}+m_3^{Sp(2)}+m_4^{Sp(2)}.
\end{align}
Then the period of the Tao diagram in Figure \ref{fig:Sp21A8FTao} is given by the sum of the length $l_i, (i=1, 2, 3, 4)$ and it turns out to be equal to $2m_0^{Sp(2) + 1\AS + 8\bF}$:
\begin{align}
\tau_{Sp(2) +1 {\bf AS} + 8{\bf F}}=
\sum_{i=1}^{4} \, l_i = 2 \,m_0^{Sp(2) +1 {\bf AS} + 8{\bf F}}.
\end{align}
Namely, the inverse of the squared gauge coupling of $Sp(2) + 1\AS + 8\bF$ is directly related to the period of the Tao diagram.

%------------
\begin{figure}
\centering
\includegraphics[width=9cm]{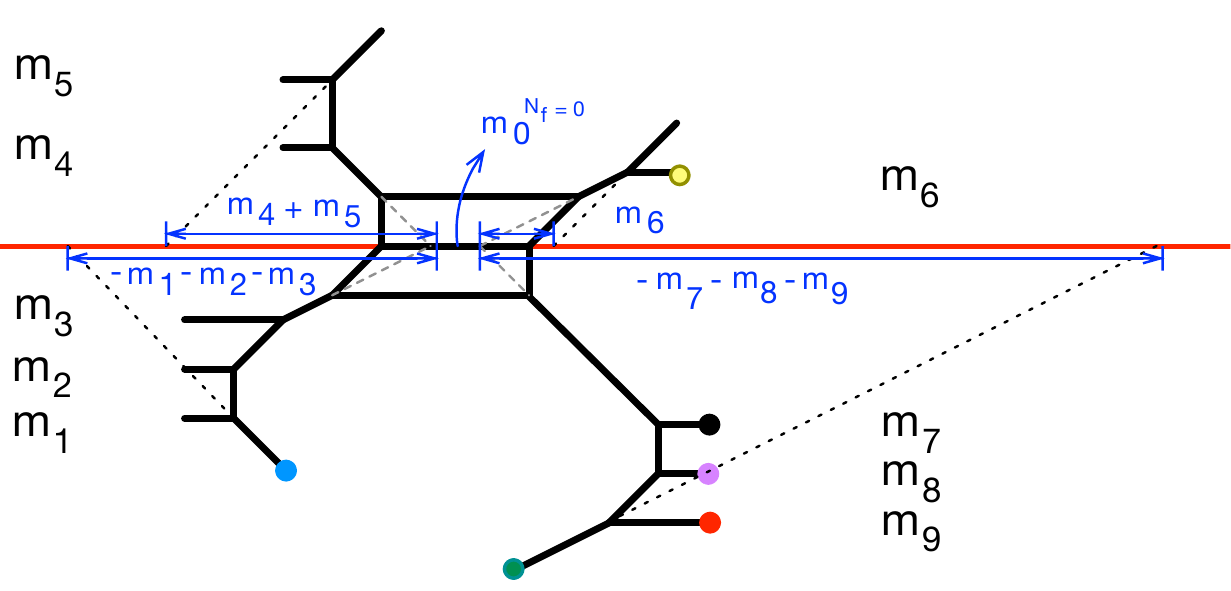} 
\caption{A 5-brane web for $SU(3)_\frac{3}{2}+9\bF$ and the parameterization. $m_0$ for $N_f = 9$ is given by $m_0^{N_f=9} = m_0^{N_f=0}-\frac12\big(m_1+m_2+m_3-m_4-m_5-m_6+m_7+m_8+m_9 \big)$.}
\label{fig:SU3CS32Nf9m0}
\end{figure}
%------------
We now consider a Tao web diagram for $SU(3)_{\frac{3}{2}} + 9 {\bf F}$ which is a bit involving. From a 5-brane configuration for $SU(3)_{\frac{3}{2}} + 9 {\bf F}$ given in Figure \ref{fig:SU3CS32Nf9m0}, $m_0$ for $SU(3)_{\frac32} + 9\bF$ is expressed as a linear combination of the mass parameters $m_i$ and $m_0$ for the pure $SU(3)_0$ gauge theory 
\begin{align}
	m_0^{SU(3)_{\frac{3}{2}} + 9 {\bf F}} = m_0^{SU(3)_0 + 0 {\bf F}}-\frac12\big(m_1+m_2+m_3-m_4-m_5-m_6+m_7+m_8+m_9 \big).
\end{align}

\begin{figure}
\centering
\includegraphics[width=11cm]{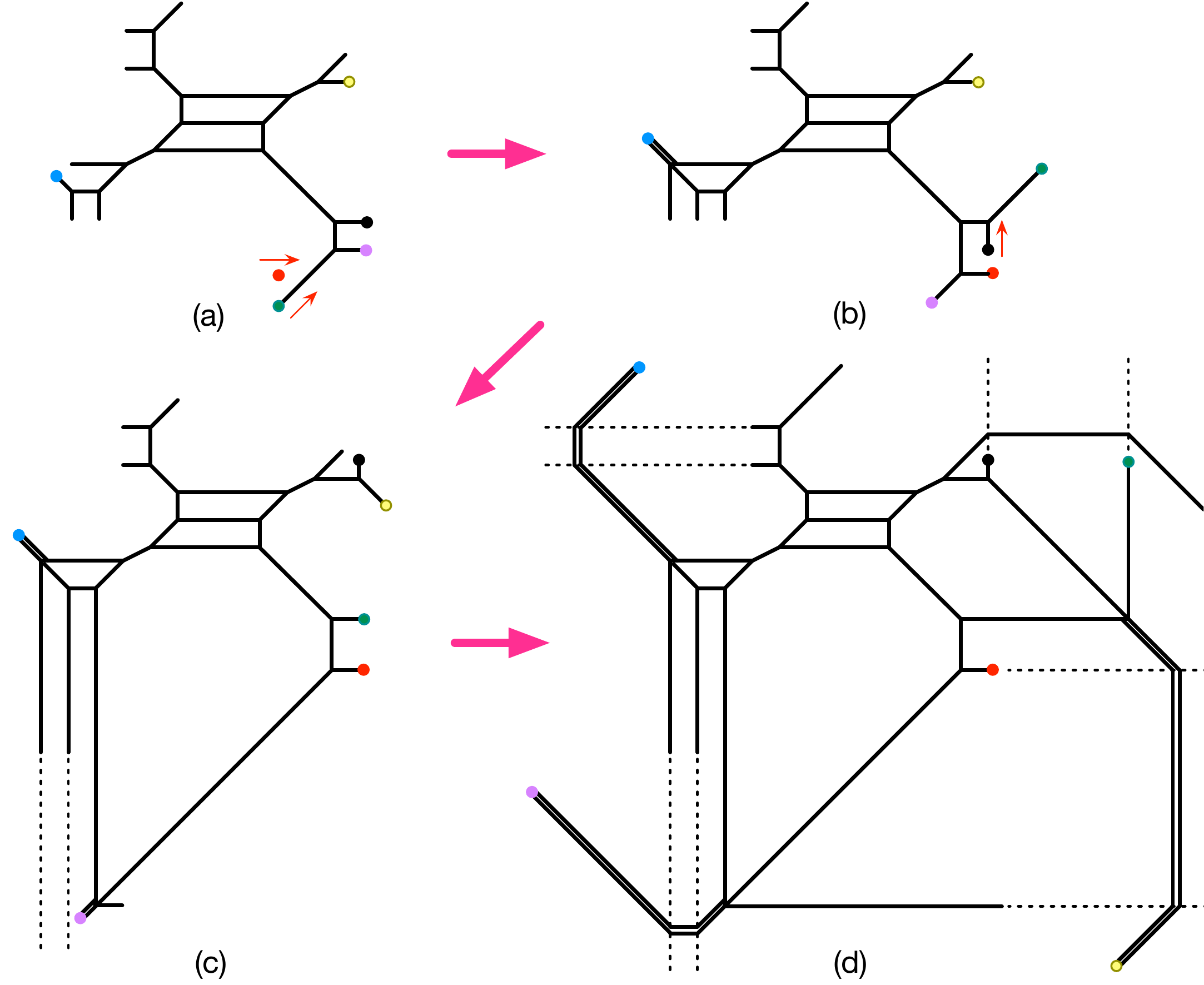} 
\caption{The deformation from a 5-brane web of $SU(3)_\frac{3}{2}+9\bF$ to its Tao web diagram. The 5-brane configuration in Figure \ref{fig:SU3CS32Nf9m0} can be deformed to the 5-brane web (a). Applying a particular successive Hanany-Witten transitions by moving 7-branes, we arrive at a Tao web diagram given in (d).}
\label{fig:SU3CS32Nf9Tao}
\end{figure}

A Tao web diagram can be obtained by a successive application of Hanany-Witten transition with  a particular 7-brane motion explained in Figure \ref{fig:SU3CS32Nf9Tao}. 
%------------
%------------
For example, one can start with Figure \ref{fig:SU3CS32Nf9m0} and perform Hanany-Witten transitions associated with the red and blue 7-branes to get Figure \ref{fig:SU3CS32Nf9Tao}(a). And further performing Hanany-Witten transitions in a particular order described in Figure \ref{fig:SU3CS32Nf9Tao} yields the diagram in Figure \ref{fig:SU3CS32Nf9Tao}(d). In Figure \ref{fig:SU3CS32Nf9Tao}(d), we denoted the dotted lines for the monodromy cuts of some 7-branes. By letting all other 7-branes go through these monondromy cuts, 7-brane charges for those 7-branes are changed and in fact, these particular monodromy cuts are chosen so that all other 7-branes keep passing through the cuts and they form a spiral shape with a constant period. The periodic structure can be more explicitly seen in Figure \ref{fig:SU3CS32Nf9Tao}. 
%------------
\begin{figure}
\centering
\includegraphics[width=13cm]{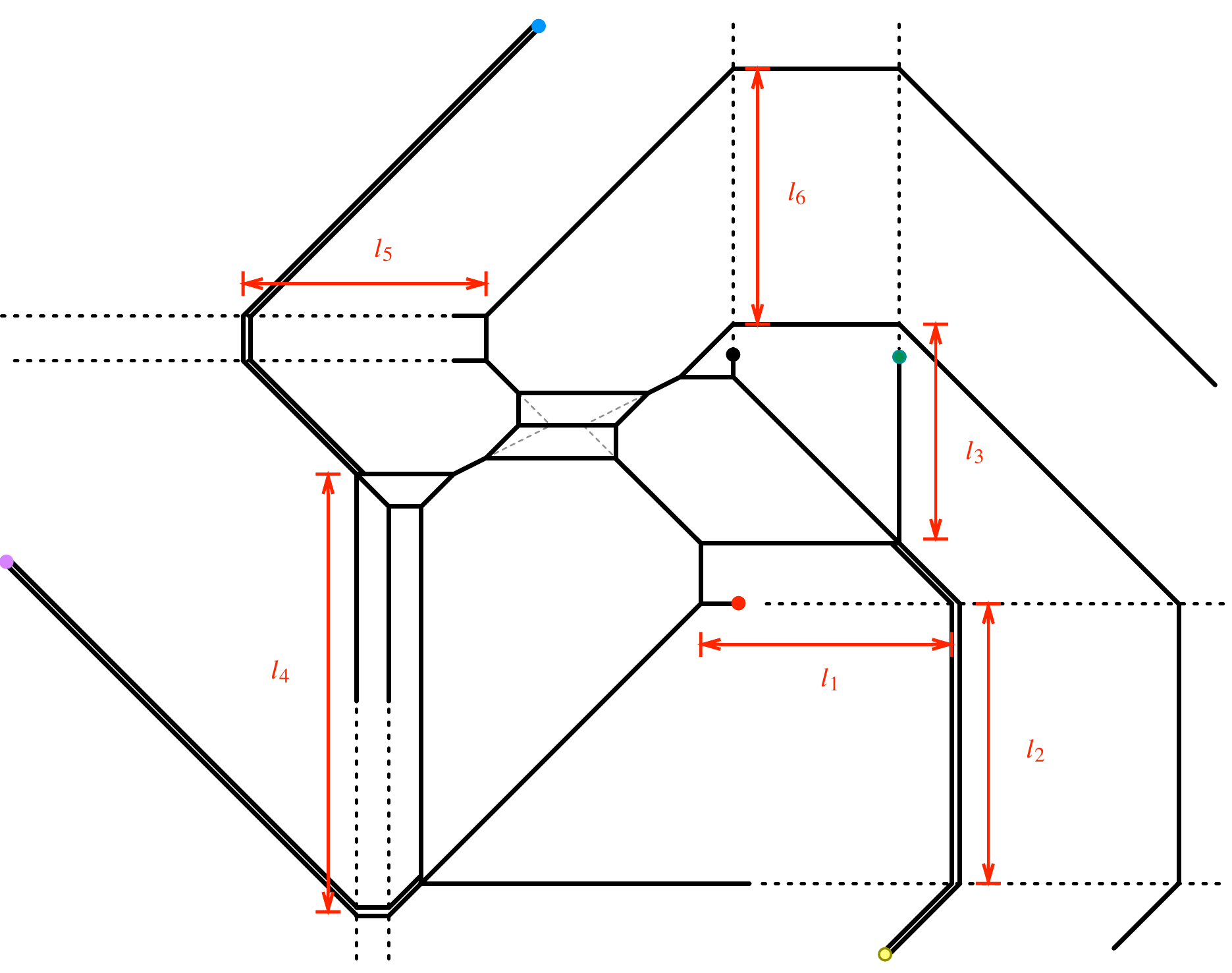} 
\caption{A Tao web diagram for $SU(3)_\frac{3}{2}+9\bF$ and the period is given by $\tau_{SU(3)_\frac{3}{2}+9\bF}={l_1+l_2+l_3+l_4+l_5+l_6}$.}
\label{fig:SU3Nf9Taom0}
\end{figure}
%------------
The length $l_i, (i=1, \cdots, 6)$ in Figure \ref{fig:SU3CS32Nf9Tao} are related to the mass parameters $m_i, (i=1, \cdots, 9)$ and $m_0$ for the pure $SU(3)_0$ gauge theory as
where 
\begin{align}
	\l_1 &=m_6+m_7-m_8-m_9,\qquad
	&{\l_2} &=m_0^{N_f=0}-m_2-m_3-m_7,\cr
	{\l_3} &=-m_6-m_7-m_8,\qquad
	&{\l_4} &=m_0^{N_f=0}-m_1-m_7-m_9,\cr
	{\l_5} &=-m_1-m_2-m_3,\qquad 
	&{\l_6} &=m_0^{N_f=0}+m_4+m_5+m_6.
\end{align}
Then the diagram in Figure \ref{fig:SU3CS32Nf9Tao} implies that the period is the sum of $l_i, (i=1, \cdots, 9)$ and it yields 
\begin{align}
\tau_{SU(3)_{\frac{3}{2}} + 9 {\bf F}}\,	= \sum_{i=1}^6 \,l_i= 3 m_0^{SU(3)_\frac{3}{2} + 9 {\bf F}}-\frac12 \sum_{i=1}^{9} m_i. \label{period.SU3Tao}
\end{align}
Note that, unlike the $Sp(2) +1 {\bf AS} + 8{\bf F}$ case, the period $\tau_{SU(3)_{\frac{3}{2}} + 9 {\bf F}}$ is not 
given by $2\,m_0^{SU(3)_\frac{3}{2}+ 9 {\bf F}}$. It is however easy to see that, by applying the duality map between the two theories \eqref{eq:m0su3m0Sp29Fv2}, the period of $SU(3)_{\frac{3}{2}} + 9 {\bf F}$  
is equal to $2m_0^{Sp(2) + 1\AS + 8\bF}$ and hence it is equivalent to the period of $Sp(2) +1 {\bf AS} + 8{\bf F}$,
\begin{align}
	\tau_{SU(3)_{\frac{3}{2}} + 9 {\bf F}} 
	= 2 \,m_0^{Sp(2) +1 {\bf AS} + 8{\bf F}}
	=\tau_{Sp(2) +1 {\bf AS} + 8{\bf F}} .
\end{align}
Since two marginal theories are dual to each other, it is expected that they have the same period as the UV completion 6d theory on a circle whose radius directly related to the period of two different 5d descriptions. Namely, only the $2m_0$ of $Sp(2) + 1\AS + 8\bF$ is directly equal to the period of the Tao diagram but the $3m_0$ for $SU(3)_{\frac32} + 9\bF$ needs a shift depending on the mass parameters as in \eqref{period.SU3Tao} in order to form the period.

%%%%%%%%%%%%%%%%%%%%%%%%%%%%%

%-----------------------------
\subsection{Further deformation to 
5-brane webs of $SU(3)_0 + 10 {\bf F}$, $Sp(2) + 10 {\bf F}$, and $[4\bF+ SU(2)]\times[SU(2)+4\bF]$}
\label{subsec:SU3CS010F}
%---------------
\begin{figure}
\centering
\subfigure[]{
\includegraphics[width=4cm]{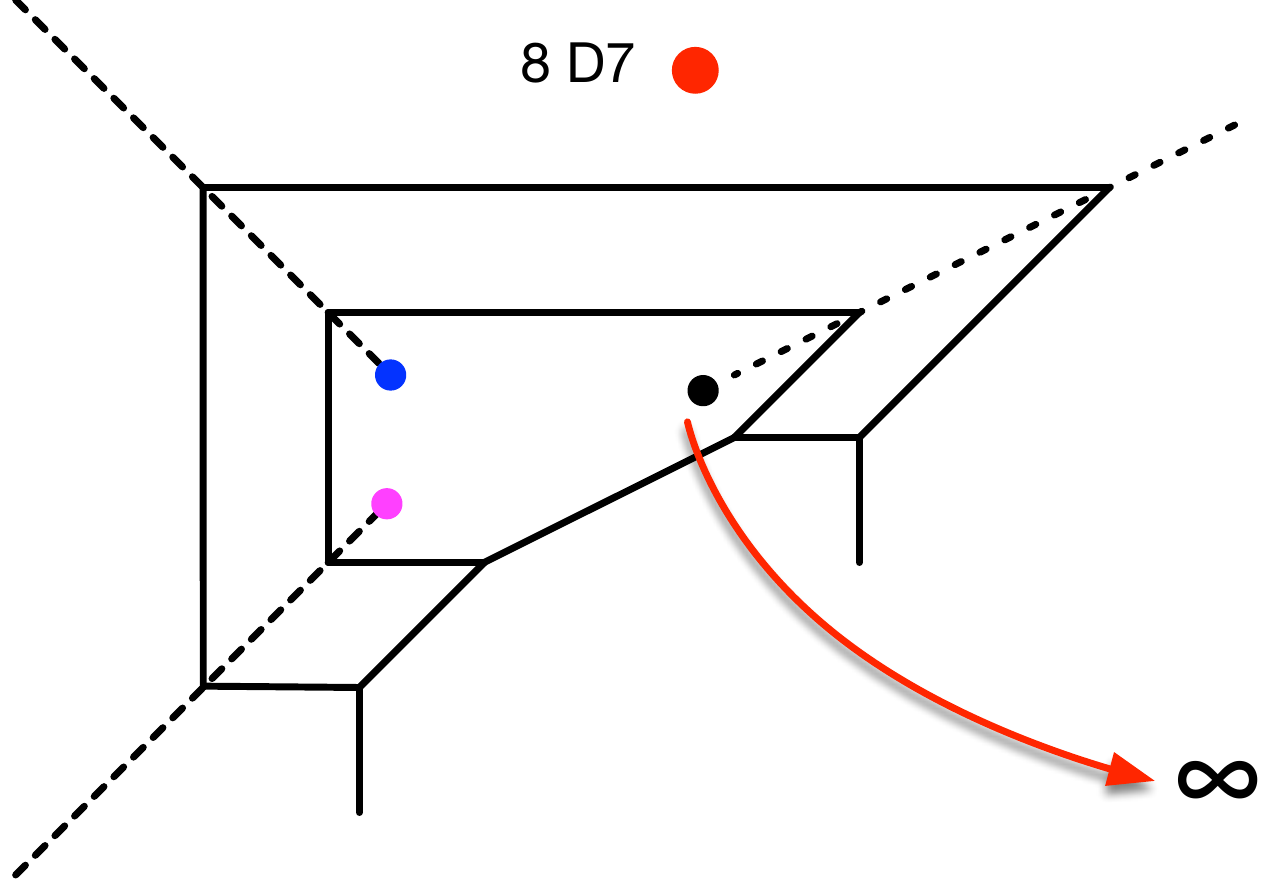} \label{fig:Sp21AS8F2flop}}\quad\qquad
\subfigure[]{
\includegraphics[width=6cm]{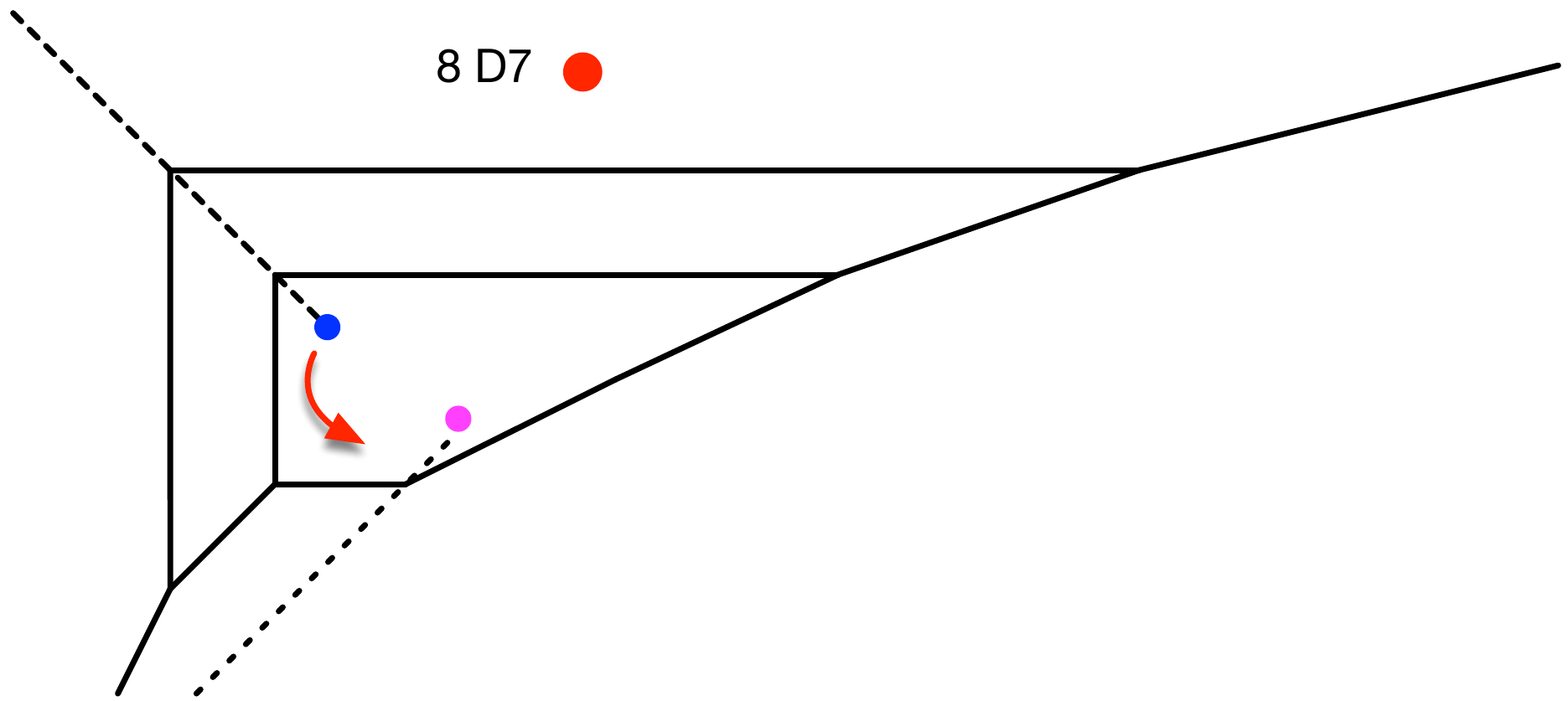} \label{fig:Sp28F1}}\quad
\subfigure[]{
\includegraphics[width=9cm]{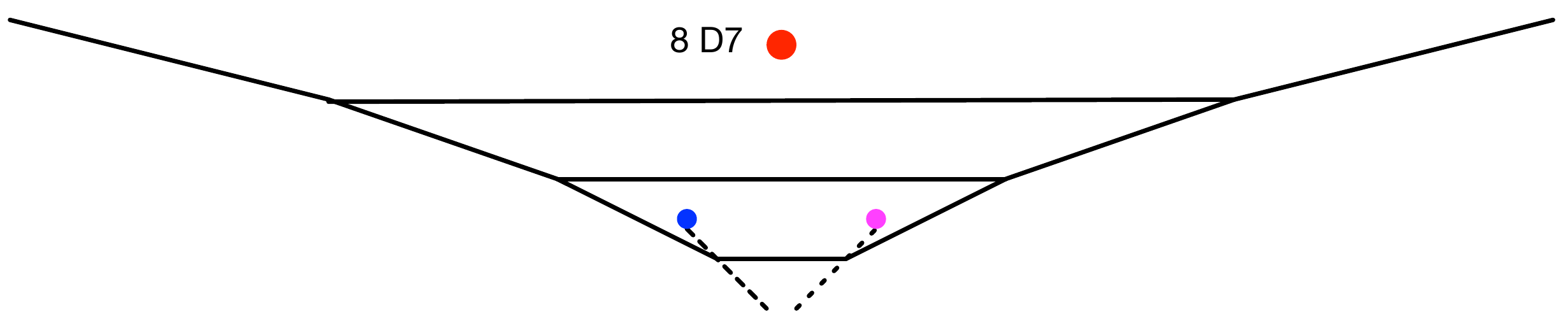} \label{fig:Sp28F2}}
\subfigure[]{
\includegraphics[width=9cm]{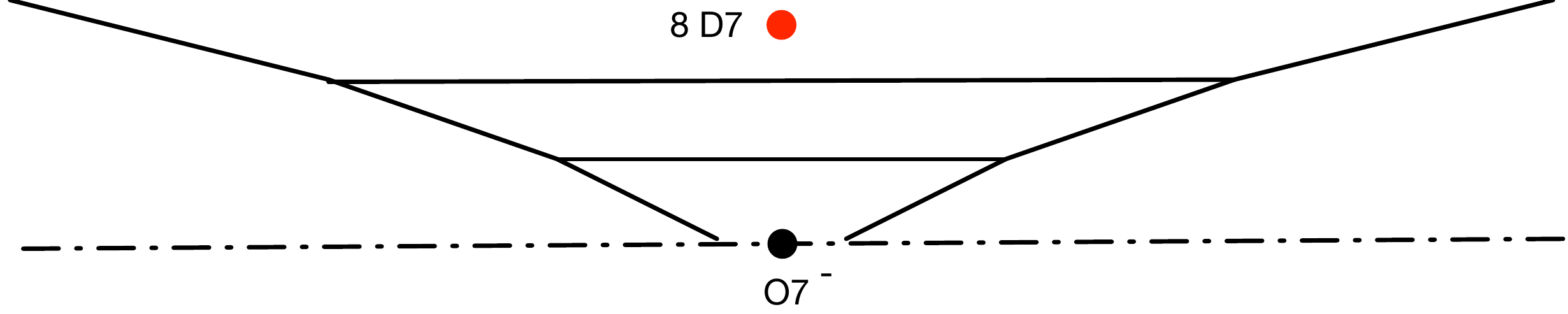} \label{fig:Sp28F3}}
\subfigure[]{
\includegraphics[width=9cm]{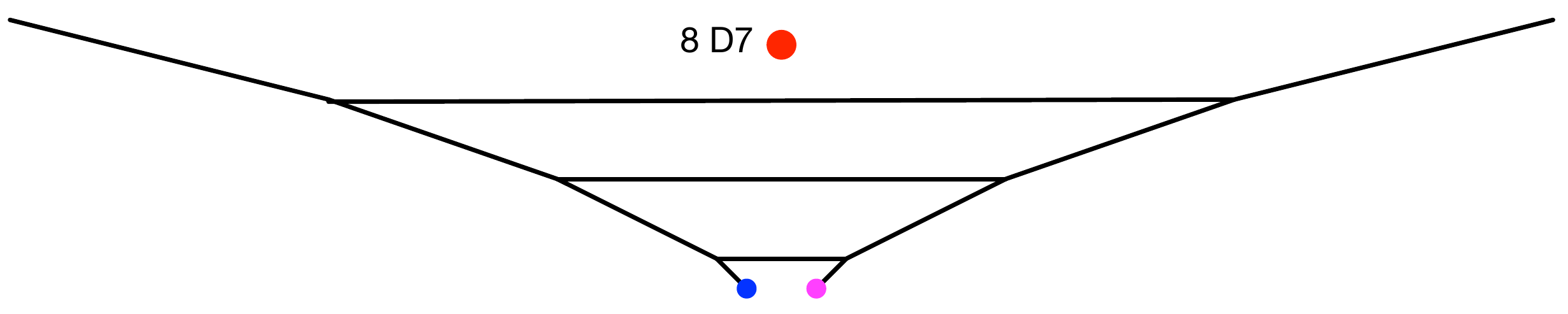} \label{fig:SU38F}}
\caption{(a): Decoupling an antisymmetric hypermultiplet by taking the 7-brane of the charge $[2,1]$.~  
(b) and (c): Moving a 7-brane of charge $[1,-1]$ to put together with a 7-brane of charge $[1,1]$ in a 5-brane loop.~ (d): Recombining the two 7-branes to make an O7$^-$-plane leading to a brane configuration for $Sp(2)+8\bF$.~ (e): After resolving O7$^-$back  into the two 7-branes, the resulting 5-brane configuration becomes equivalent to that of $SU(3)_1+8\bF$.}
\label{fig:Sp21A8Fdecoupling}
\end{figure}
%---------------

In section \ref{subsec:Sp21AS8F}, we considered a deformation of a marginal theory $Sp(2)+2\AS+4\bF$ by decoupling of a hypermultiplet in the antisymmetric representation. We then introduced four more hypermultiplets in the fundamental representation which led to another marginal theory $Sp(2)+1\AS+8\bF$.  In this subsection, in a similar manner, we consider a deformation of $Sp(2)+1\AS+8\bF$ by decoupling the  hypermultiplet in the antisymmetric representation. The mass of the antisymmetric hypermultiplet is proportional to the distance between the two external NS5-brane in Figure \ref{fig:Sp21AS8F2flop}. To decouple this antisymmetric hypermultiplet, we take this mass to infinity, or equivalently we take the distance be infinite. To this end, as depicted in Figure \ref{fig:Sp21AS8F2flop}, we bring out the $[2, 1]$ 7-brane outside the 5-brane loops and move it to infinitely far away from the diagram. The resulting  web diagram is given in Figure \ref{fig:Sp28F1}, where we also moved the $[1,1]$ 7-brane to the right. Then we move the $[1,-1]$  7-brane next to the $[1,1]$ 7-brane by rotating the cut of the $[1,-1]$ 7-brane 
so that it extends in the lower direction as depicted in Figure \ref{fig:Sp28F2}. It is then readily seen that one can recombine the two 7-branes of the charge $[1,-1]$ and $[1,1]$ to make an O7$^-$-plane and thus the resulting 5-brane configuration is a familiar configuration for $Sp(2)+8\bF$ as given in Figure \ref{fig:Sp28F3}. Instead of forming an O7$^-$-plane, one can resolve the O7$^-$-plane back into the two 7-branes. The resulting 5-brane configuration then shows $SU(3)+8\bF$ where the corresponding CS level for this $SU(3)$ theory is $1$. Hence the diagram gives $Sp(2)+8\bF$ and $SU(3)_1+ 8\bF$, and they are dual to each other \cite{Gaiotto:2015una,Hayashi:2015zka}.

From the perspective of S-duality, the decoupling corresponds to decoupling a flavor from the flavors associated with the $SU(2)+5\bF$ of the quiver $[SU(2)+2\bF]\times [SU(2) +5\bF]$. For instance, $SU(3)_1+ 8\bF$ can be obtained by taking one of the lower D7-branes in Figure \ref{fig:S2F+SU2xSU2+5F} is taken to $-\infty$. Then the S-dual of the diagram yields $[SU(2)+2\bF]\times [SU(2) +4\bF]$. 

As discussed, we can consider adding more flavors to $Sp(2)+8\bF$ and $SU(3)_1 + 8\bF$ in the same way. The marginal theory one can obtain in this way is $Sp(2)+10\bF$ and $SU(3)_0+ 10\bF$, which are dual to each other. $SU(3)_0+ 10\bF$ is S-dual to $[SU(2)+4\bF]\times [SU(2) +4\bF]$. The duality map between $Sp(2)+10\bF$ and $SU(3)_0+10\bF$ has been already obtained in \cite{Hayashi:2016abm}. For book-keeping purpose, we summarize the map here. For convenience, we label the $Sp(2)$ parameters with a prime ($'$) and the $SU(3)$ parameters without a prime. For each instanton factor ($m_0', m_0$), the Coulomb moduli parameters ($a'_i, a_i$) and the mass parameters ($m'_i,m_i$), the duality map between $Sp(2)+10\bF$ and $SU(3)+10\bF$ is given as follows: 
%------------
\begin{figure}
\centering
\subfigure[]{
\includegraphics[width=5cm]{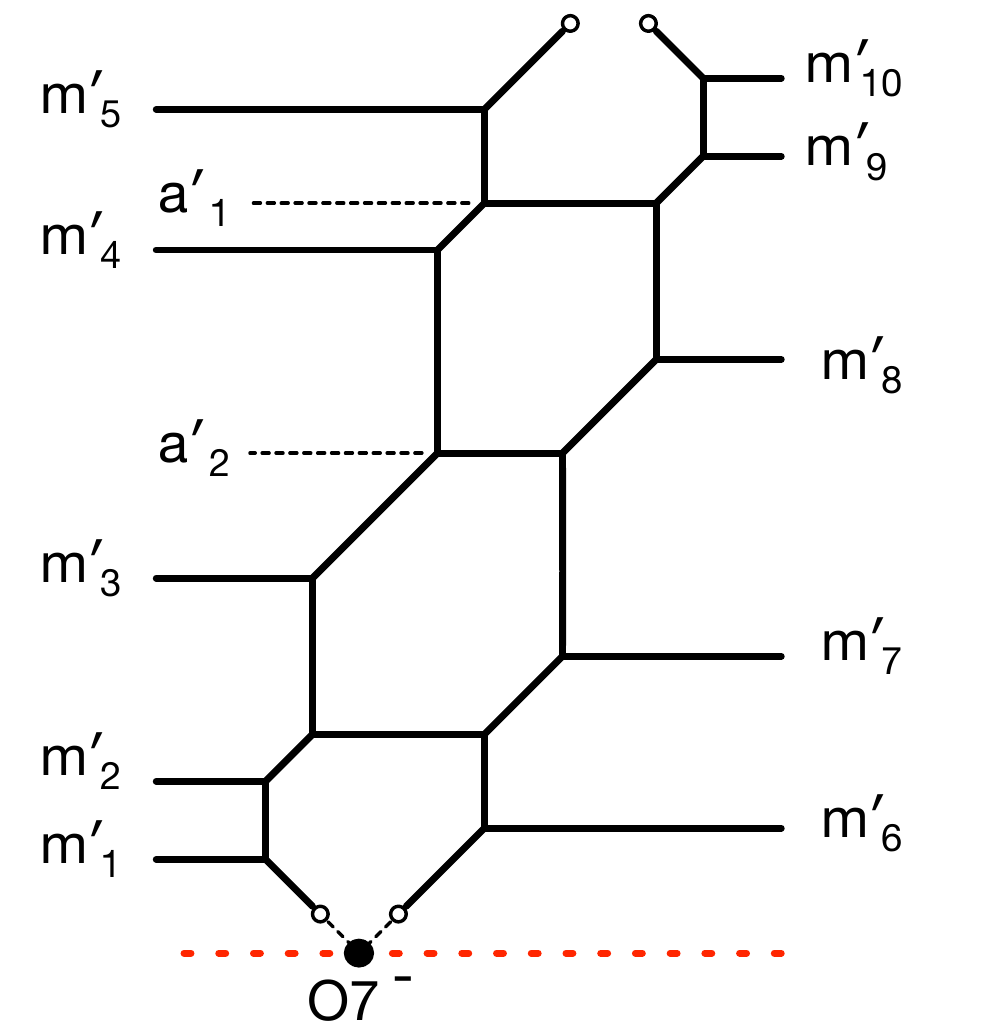} \label{fig:5dSU3Sp2}}\qquad\qquad\qquad
\subfigure[]{
\includegraphics[width=5cm]{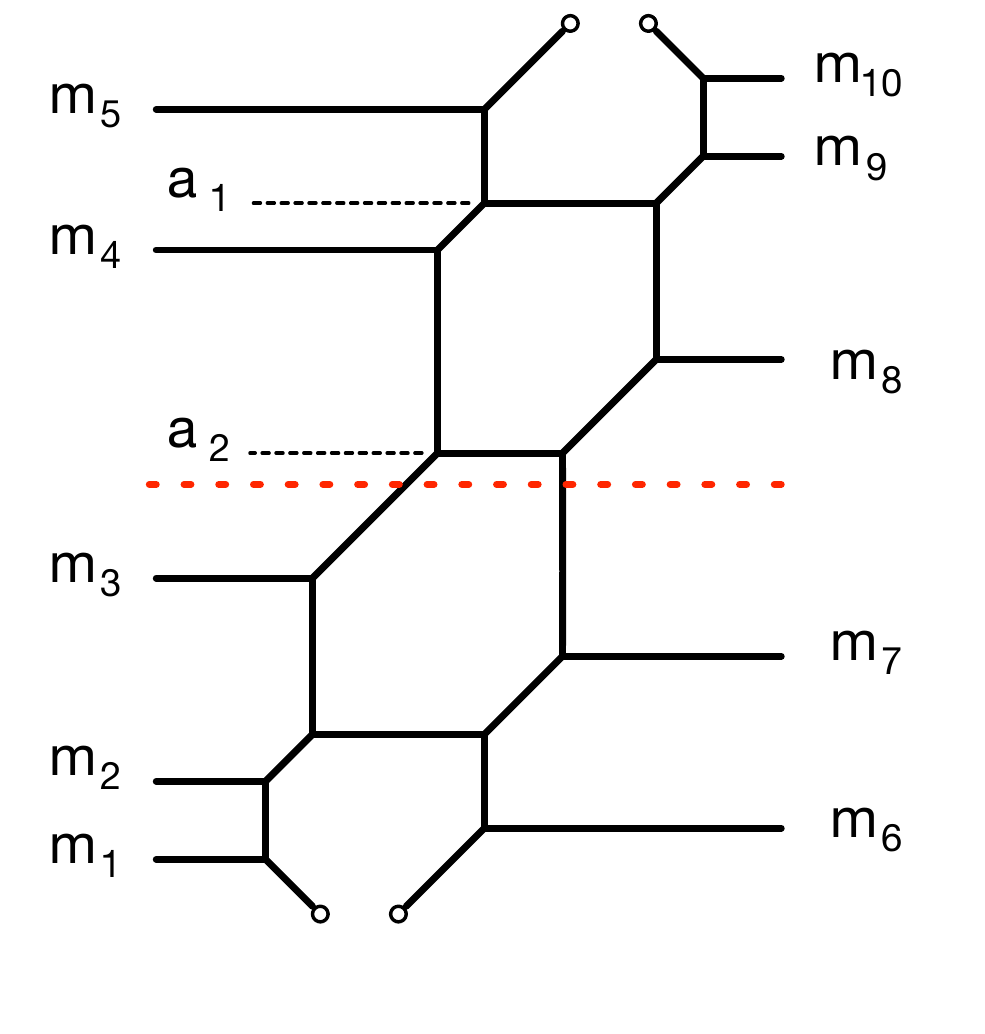} \label{fig:5dSU310F}}
\caption{(a): A 5-brane web diagram with an O7$^-$-plane for $Sp(2)+10\bF$. (b): A 5-brane web diagram for $SU(3)_0+10\bF$ where the red dotted line indicates the origin of the Coulomb branch. 
}
\label{fig:SU3+10F}
\end{figure}
%------------
\begin{align}
m_0'& = m_0;~\, \quad\qquad\qquad a'_j = a_j+  \frac12\ell,	\qquad (j=1,2);\cr
m'_i &= m_i+\frac12\ell\, ,\qquad  m'_{i+5} = - \,m_{i+5} -\frac12\ell, \qquad (i=1, \cdots, 5).
\end{align}
where $\ell = m_0-\displaystyle\frac12\sum_{i=1}^{10}m_i$ and the relation of the Coulomb branch moduli and the mass parameters with the length in the diagrams is summarized in Figure \ref{fig:SU3+10F}.

\subsection{5-brane web for $SU(3)_0+1\bF+ 1\mathbf{Sym}$}\label{sec:SU31Sym1F}
%------------
\begin{figure}
\centering
\subfigure[$SU(3)_{0}+1\bF+1\mathbf{Sym}$]{
\includegraphics[width=7cm]{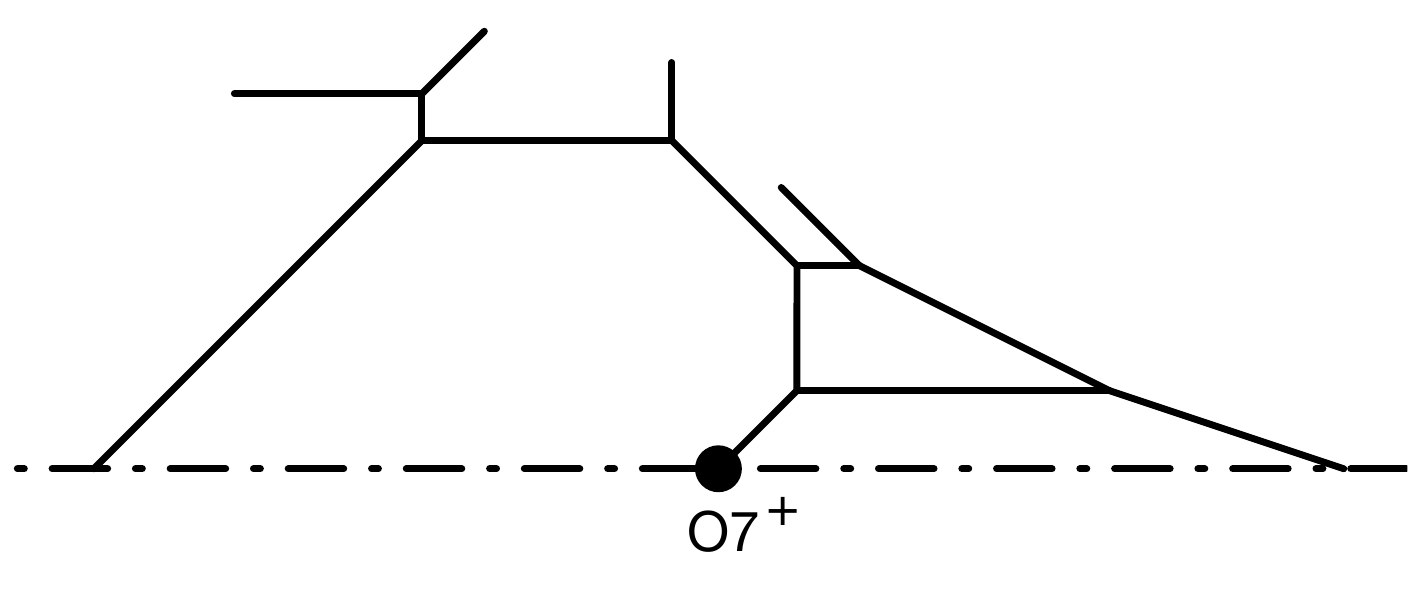} \label{fig:SU31F1Sym}}
\quad
\subfigure[$SU(3)_{-\frac12}+1\mathbf{Sym}$]{
\includegraphics[width=7cm]{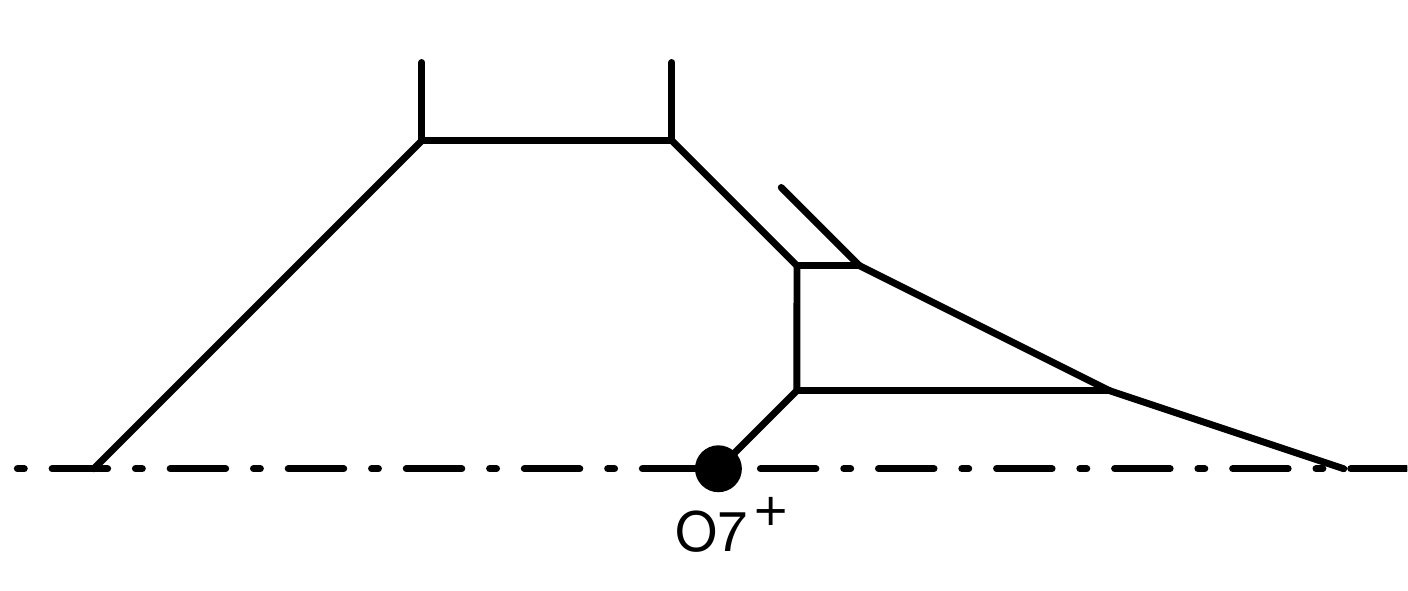} \label{fig:SU31Sym}
}
\caption{5-brane web diagrams for $SU(3)$ gauge theories with a symmetric hypermultiplet.}
\label{fig:SU31SymGeneric}
\end{figure}
%------------
There is another deformation from an $SU(3)$ theory with a flavor. It is to add a hypermultiplet in the symmetric representation ($\mathbf{Sym}$), which may yield $SU(3)_0+1\bF + 1\mathbf{Sym}$. We note that this theory is a marginal theory as the prepotential contribution of a symmetric hypermultiplet can be effectively ``equivalent'' to that of eight hypermultiplets in the fundamental representation and a hypermultiplet in the antisymmetric representation ($1\mathbf{Sym}\sim 8\bF+ 1\AS$). It follows that $SU(3)_0+1\bF + 1\mathbf{Sym}$ would give an equivalent prepotential as that of $SU(3)_0+9\bF+1\AS$ or $SU(3)_0+10\bF$, which has a 6d UV fixed point. However a 5-brane configuration for $SU(3)_0+1\bF +1\mathbf{Sym}$ \cite{Hayashi:2015vhy} is quite distinct from the brane configuration for $SU(3)_0+10\bF$. The $SU(N)$ theory with a symmetric hypermultiplet is described by the introduction of an O7$^+$-plane on which an NS5-bane ends \cite{Bergman:2015dpa}. For instance, see Figure \ref{fig:SU31SymGeneric}, which shows a 5-brane web for $SU(3)_0+1\bF+ 1\mathbf{Sym}$ in Figure \ref{fig:SU31F1Sym} and a 5-brane web for $SU(3)_{-\frac{1}{2}} + 1\mathbf{Sym}$ in Figure \ref{fig:SU31Sym}. Using this 5-brane web diagram for $SU(3)_0+1\bF + 1\mathbf{Sym}$, we will show that the areas of the compact faces of the web diagram agree with the monopole tension from the effective prepotential. 

%------------
\begin{figure}
\centering
\includegraphics[width=11cm]{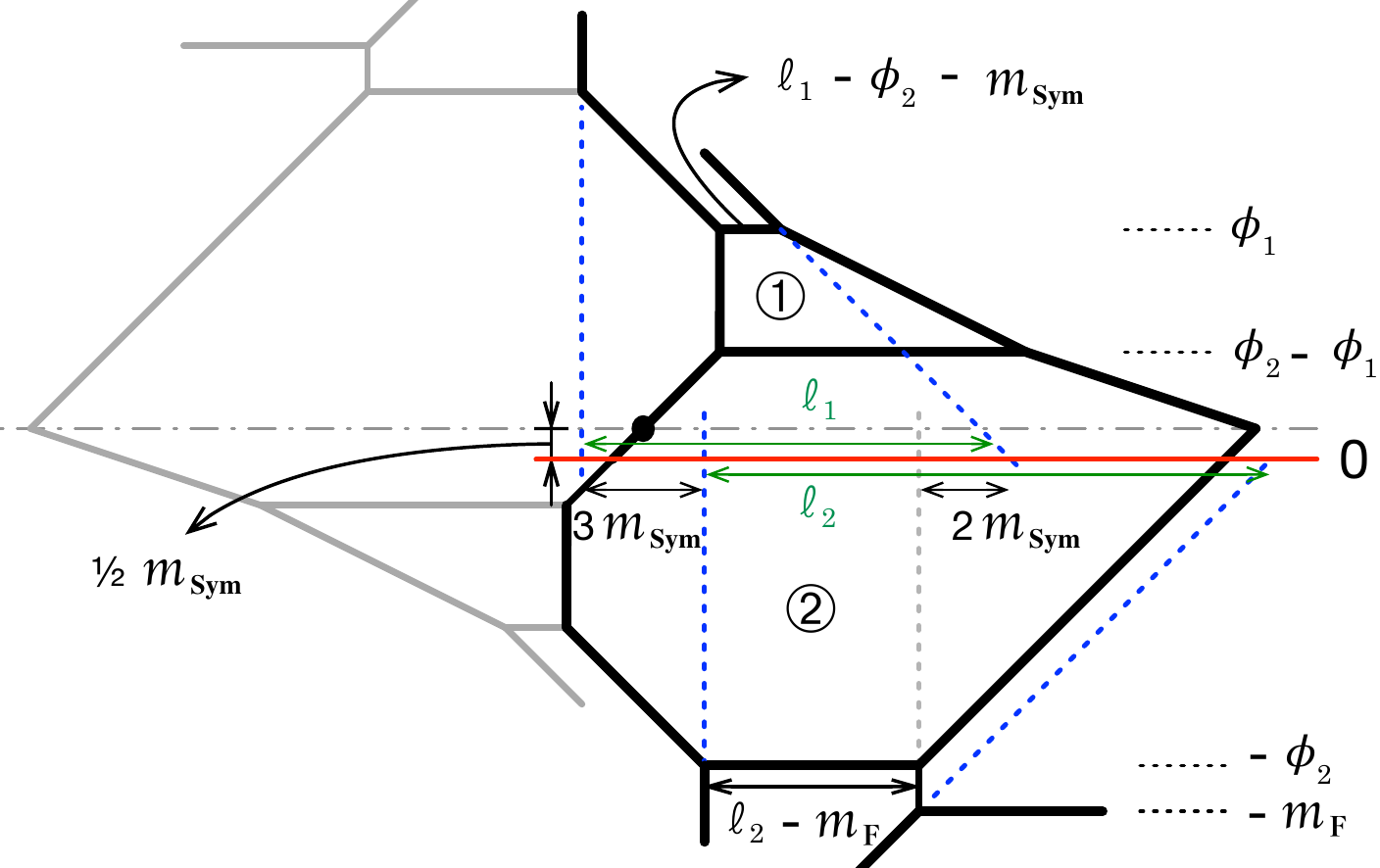} \label{}
\caption{$SU(3)_0+1\bF+1\mathbf{Sym}$ with mass parameters of the hypermultiplets. The 5-brane web below the monodromy cut of an O7$^+$-plane is the reflected 
image due to the O7$^+$-plane. The center of the Coulomb branch is denoted by the red line in the middle.}
\label{fig:SU31F1SymArea}
\end{figure}
%------------

In Figure \ref{fig:SU31F1SymArea} which is a 5-brane web diagram describing $SU(3)_0+1\bF + 1\mathbf{Sym}$ with the mass parameters $m_{\bF}$ and $m_{\bf Sym}$. We reflected 5-brane webs on the left against the O7$^+$-plane to the right below, and chose the right part as the fundamental region which looks similar to that of an $SU(3)$ theory (it is the bold faced 5-brane web in the figure). In this way, one can readily compute to the area of the compact faces. The parameters in this 5-brane web are measured from the center of the Coulomb branch which is the horizontal line in red. The distance between the O7$^+$-plane cut and the center of the Coulomb branch moduli corresponds to the half mass of the hypermultiplet in the symmetric representation, $\frac12 m_{\bf Sym}$, which is a natural generalization of the definition of the mass for an antisymmetric hypermultiplet discussed in section \ref{sec:dualtoSU3w2flvrs}. The bare coupling $m_0$ is defined as usual by the average of two extrapolated external 5-branes which are expressed as blue dotted lines in Figure \ref{fig:SU31F1SymArea}. The blue dotted lines intersecting with the center line for the Coulomb branch moduli give rise to two distances $\ell_1$ and $\ell_2$. It is not difficult to see that the two distances are related by $\ell_1=\ell_2+ 5 m_{\bf Sym}-m_{\bF}$. The bare coupling is defined by the average of the two asymptotic distances of external 5-branes 
\begin{align}
	m_0~& = \frac12 (l_1+l_2),
\end{align}
and hence $\ell_1, \ell_2$ can be expressed as
\begin{align}
	l_1  = m_0+\frac52\,m_{\bf Sym}-\frac12\,m_{\bF},\qquad  l_2 = m_0-\frac52\,m_{\bf Sym}+\frac12\,m_{\bF} .
\end{align}
A little bit of algebra then yields that the area of the compact faces  $\textcircled{\scriptsize 1}$ and $\textcircled{\scriptsize 2}$ in Figure \ref{fig:SU31F1SymArea} are given by
\begin{align}
\textcircled{\scriptsize 1}&=~(2\phi_1-\phi_2) \Big(m_0+ 2\,\phi_1-2\,\phi_2-\frac12 m_{\bF}+\frac32 m_{\bf Sym}\Big),\label{eq:MT4su31S1Fa}\\
\textcircled{\scriptsize 2} &=~
	m_0\, (-\phi_1+2\,\phi_2)- 3\,\phi_1^2 + 4\,\phi_1\phi_2 -\frac{\phi_2^2}{2}+ m_\bF\Big(\frac{\phi_1}{2}-\phi_2\Big) -\frac32 m^2_{\bf Sym} - \frac{3}{2}m_{\bf Sym}\phi_1.\label{eq:MT4su31S1Fb}
\end{align}
%---------------
\begin{figure}
\centering
\includegraphics[width=11cm]{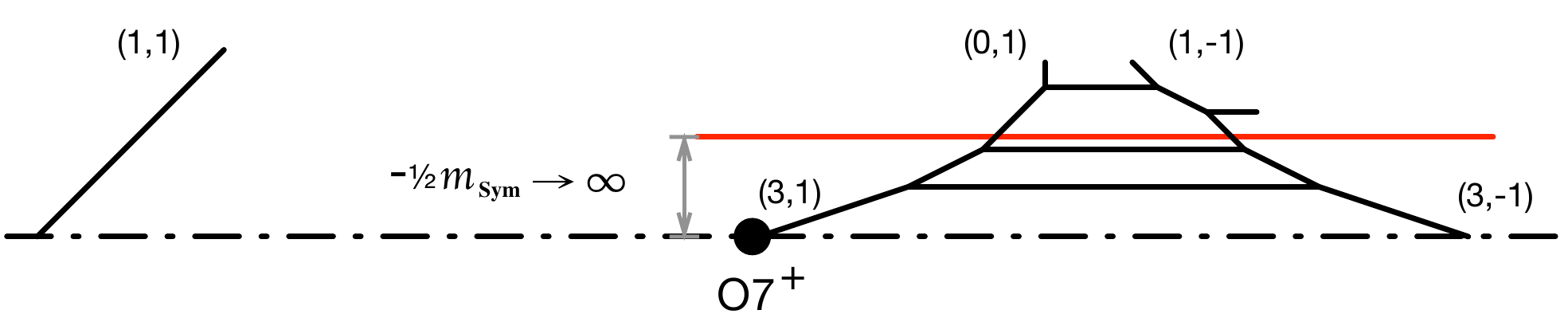}
\caption{A deformed web diagram for $SU(3)_0+1\bF + 1\mathbf{Sym}$ which enables one to decouple a hypermultiplet in the symmetric representation by taking its mass to $-\infty$.
}
 \label{fig:O7w1Fdecoupling0}
\end{figure}
%---------------
%---------------
\begin{figure}
\centering
\includegraphics[width=6cm]{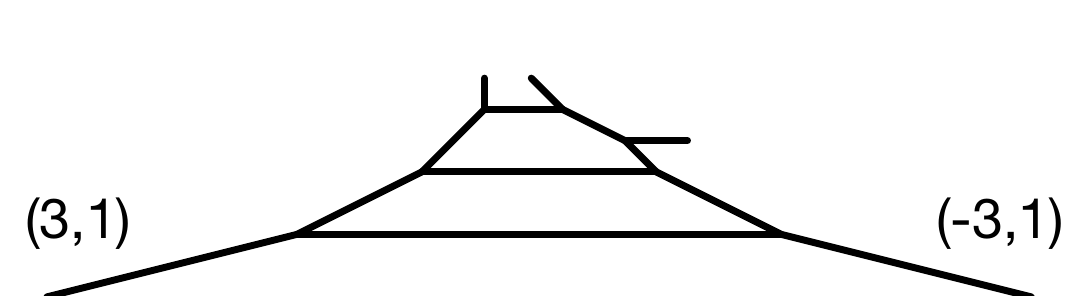} 
\caption{
An effective description of the diagram in Figure \ref{fig:O7w1Fdecoupling0} when one sends $m_{\bf Sym} \to -\infty$. The diagram exhibits $SU(3)_{-\frac72}+1\bF$. }
\label{fig:O7w1Fdecoupling}
\end{figure}
%---------------

The effective prepotential is computed from \eqref{prepotential}. The phase of the parameters corresponding to the configuration of Figure \ref{fig:SU31F1SymArea} is  
\begin{align}
	m_{\bF}\, \geq\, \phi_2 \,\geq \,\phi_1\,\geq\, 
\frac12 \,\phi_2 \,\geq\, m_{\bf Sym} \,\geq\, 0.
\end{align}
The prepotential reads
\begin{align}
{\mathcal F}_{SU(3)_0+1\bF + 1\mathbf{Sym}}=& ~ 
m_0\, (\phi_1^2 -\phi_1\phi_2 +\phi_2^2)+\frac43\phi_1^3- 3\,\phi_1^2\phi_2+2\,\phi_1\phi_2^2-\frac16\phi_2^3-\frac14m_{\bF}^3
\cr
&-\frac12m_\bF \,(\phi_1^2 -\phi_1\phi_2+ \phi_2^2)+\frac32m_{\bf Sym}\, \phi_1(\phi_1 -\phi_2) - \frac{3}{2}m_{\bf Sym}^2\phi_2.
\end{align}
It is straightforward to see that the monopole string tension agrees with the area from the 5-brane web for $SU(3)_0+1\bF + 1\mathbf{Sym}$:
\begin{align}
	\frac{\partial{\mathcal F}_{SU(3)_0+1\bF + 1\mathbf{Sym}}}{\partial \phi_1} &~=\textcircled{\scriptsize 1},\label{eq:MT4su31S1Fa}\\
	\frac{\partial{\mathcal F}_{SU(3)_0+1\bF + 1\mathbf{Sym}}}{\partial \phi_2} &~=\textcircled{\scriptsize 2}.\label{eq:MT4su31S1Fb}
\end{align}

We close the subsection with a comment on decoupling of hypermultiplets. First we can take $m_{\bF}\to\infty$ to decouple a flavor, which leads to $SU(3)_{-\frac12} + 1\mathbf{Sym}$ as shown in Figure \ref{fig:SU31Sym}\footnote{From Figure \ref{fig:SU31F1SymArea}, we take $-m_{\bF} \to -\infty$ and hence the CS level decreases by a half.}. We can see that the area after the flavor decoupling reproduces the monopole tensions of $SU(3)_{-\frac12} + 1\mathbf{Sym}$ from the corresponding prepotential. It is also possible to take the mass of a symmetric hypermultiplet to $-\infty$ in order to decouple the hypermultiplet in the symmetric representation. For that, consider a deformed web diagram for $SU(3)_0+1\bF + 1\mathbf{Sym}$ depicted in Figure \ref{fig:O7w1Fdecoupling0}, where three color D5-branes are put in on the right. On the left, there is a $(1, 1)$ 5-brane coming from the reflection of $(-3,1)$ 5-brane due to the O7$^+$-plane. The mass of the symmetric matter $m_{\bf Sym}$ is given by the distance between O7$^+$-plane and the center of the Coulomb branch (denoted as a red line). Since the origin of the Coulomb branch moduli is above the location of the O7$^+$-plane, the distance between them is given by $-\frac{1}{2}m_{\bf Sym}$. By taking $m_{\bf Sym}\to-\infty$, one gets a web digram given in Figure \ref{fig:O7w1Fdecoupling}.

%%%%%%%%%%%%%%%%%%%%%%%%%%%%%%%%%
\subsection{5-brane web for $SU(3)_\frac32+ 1\mathbf{Sym}$}\label{sec:SU31SymCS32}
%---------------
\begin{figure}
\centering
\includegraphics[width=13cm]{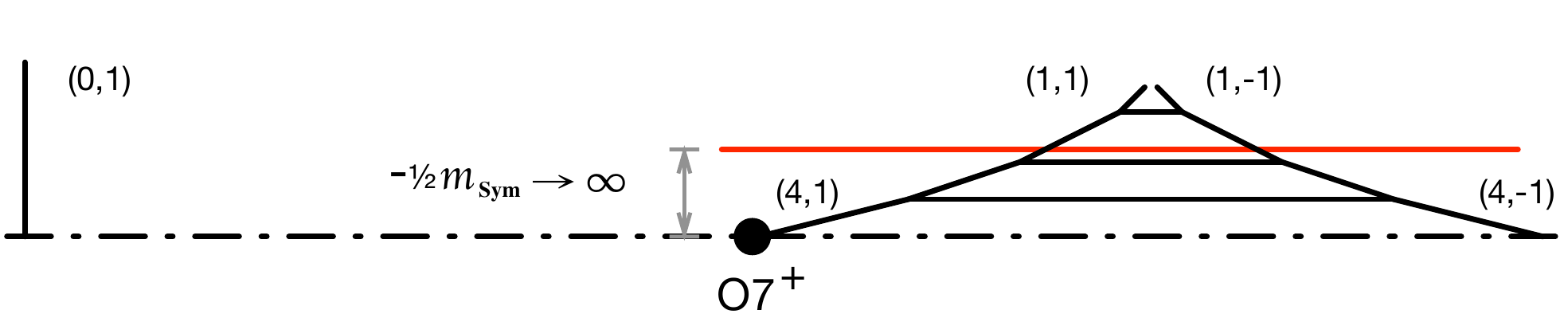}
\caption{A web diagram for $SU(3)_{-\frac32} + 1\mathbf{Sym}$ from which one can decouple a hypermultiplet in the symmetric representation by taking its mass to $-\infty$, leading a web diagram for pure $SU(3)_{-5}$ theory. 
}
 \label{fig:O7+1Sym+CS32}
\end{figure}
%---------------
%---------------
\begin{figure}
\centering
\includegraphics[width=15cm]{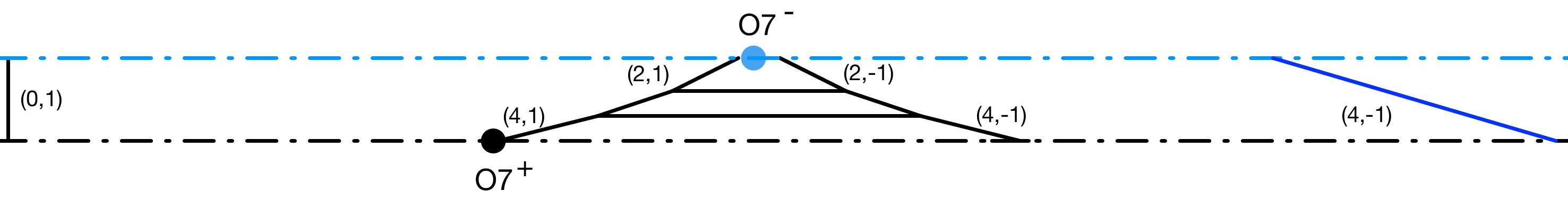}
\caption{A web diagram for $SU(3)_{-\frac32} + 1\mathbf{Sym}$ with two O7-planes by recombining a pair of 7-branes of charges $[1,1]$ and $[1,-1]$ into an O7$^-$-plane in Figure \ref{fig:O7+1Sym+CS32}. As a result, an NS5-brane one the left goes through the branch cut of the O7$^-$-plane reappear as a $(4, -1)$ 5-brane on the right, which is the solid blue line. This goes through again to the cut of O7$^+$-plane reappear as an NS5-branes on the left of the first NS5-brane. This makes a 5-brane configuration with infinitely many NS5-branes on the left and infinitely many $(4,-1)$ 5-branes on the right.
}
 \label{fig:O7+O7-SU3Sym}
\end{figure}
%---------------
Here, we consider yet another marginal theory: $SU(3)_{\frac32}+1\mathbf{Sym}$. Similar to the 5-brane web configuration of Figure \ref{fig:O7w1Fdecoupling0}, one has a 5-brane web for $SU(3)_{-\frac32}+1\mathbf{Sym}$ depicted in Figure \ref{fig:O7+1Sym+CS32}. As discussed in the previous section, the mass of a symmetric hypermultiplet parameterizes the distance between O7$^+$-plane and the center of the Coulomb branch, as shown in Figure \ref{fig:O7+1Sym+CS32}. It is then straightforward to see that taking it mass to $-\infty$ which shifts the CS level by $-\frac72$ gives rise to $SU(3)_{-5}$. 

This 5-brane configuration has an intriguing aspect which is quite different from 5-brane web for $SU(3)_\frac12+ 1\mathbf{Sym}+1F$. As discussed in \cite{Hayashi:2015vhy}, one can recombine three 7-branes of the charges $[1,-1]$, $[0,1]$ and $[1,1]$ in Figure \ref{fig:O7w1Fdecoupling0} to deform the 5-brane configuration to be a 5-brane configuration with an O7$^+$-plane and an O7$^-$-plane, connected by an NS5-brane. This hence makes the theory manifestly marginal. It is, in fact, a twisted compactification of a 6d theory \cite{Hayashi:2015vhy}. One can attempt to recombine 7-branes in a 5-brane web diagram for $SU(3)_\frac32+ 1\mathbf{Sym}$. For example, see Figure \ref{fig:O7+O7-SU3Sym}. It is 
a 5-brane configuration with two different O7-planes but an NS5-brane is not connected to two O7-plane, rather the NS5-brane is left away. As there are two O7-planes, this NS5-brane goes through the branch cut of an O7$^-$-plane reappears as a $(4,-1)$ 5-brane on the other side of O7$^-$-plane, as shown in Figure \ref{fig:O7+O7-SU3Sym}. In fact, this configuration does not stop here. The $(4,-1)$ 5-brane (the blue solid line in the figure) again goes through the branch cut of an O7$^+$-plane, and comes out as an NS5-brane on the left side of the first NS5-brane, which again reappear on the right side of the first $(4,-1)$ 5-brane, and this pattern is repeated. This 5-brane configuration for $SU(3)_\frac32+ 1\mathbf{Sym}$ with an O7$^+$-and O7$^-$-planes separated apart along the vertical direction of 5-brane plane, gives rise to a new kind of 5-brane configuration representing a twisted compactification of a 6d theory with an infinitely repeated 5-branes on the left and right sides of two O7-planes.

%%%%%%%%%%%%%%%%%%
\paragraph{New 5-brane web diagram for 5d $SU(N)_{{N}/{2}}+1\mathbf{Sym}$.} 
%---------------
\begin{figure}
\centering
\includegraphics[width=13cm]{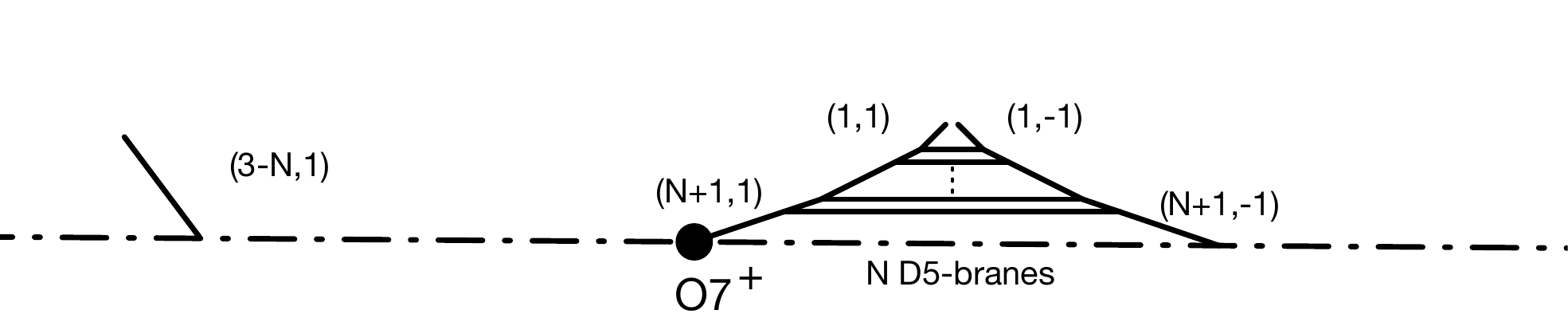}
\caption{A web diagram for $SU(N)_{-\frac{N}2} + 1\mathbf{Sym}$ from which one can decouple a hypermultiplet in the symmetric representation by taking its mass to $-\infty$, leading a web diagram for pure $SU(N)_{-N-2}$ theory.
}
 \label{fig:O7+SUN1SymCSN2}
\end{figure}
%---------------
%---------------
\begin{figure}
\centering
\includegraphics[width=14.5cm]{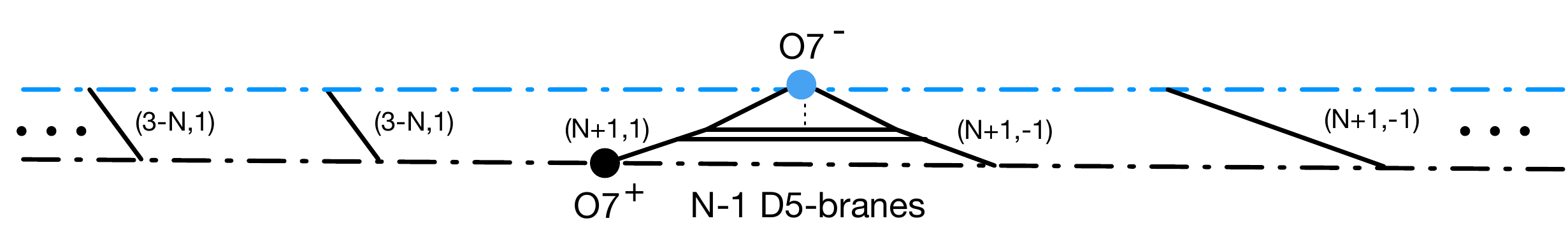}
\caption{A web diagram for $SU(N)_{-\frac{N}2} + 1\mathbf{Sym}$ with two O7-planes. Due to two O7-planes, this 5-brane configuration has infinitely many $(3-N,1)$ 5-branes on the left and infinitely many $(N+1,-1)$ 5-branes on the right.
}
 \label{fig:O7+-SUN1SymCSN2}
\end{figure}
%---------------
It is straightforward to construct a 5-brane configuration for $SU(N)_{\frac{N}{2}}+1\mathbf{Sym}$ as depicted in Figure \ref{fig:O7+SUN1SymCSN2}. We note that $N\ge 2$. As before, 5d $SU(N)_{\frac{N}{2}}+1\mathbf{Sym}$ has a 5-brane configuration with two O7-planes with two sets of infinitely repeated 5-branes of the charges $(3-N,1)$ and $(N+1,1)$, on the left and right sides of the O7-planes, which is depicted in Figure \ref{fig:O7+-SUN1SymCSN2}. As the positions of these infinitely repeated 5-branes depend on the separation between two O7-planes, one can express the periodicity of the infinitely repeated 5-branes as a linear function of the vertical separation between two O7-planes.

We note that for $SU(N)_{\frac{N}{2}}+1\mathbf{Sym}$, it is easy to see that decoupling of a symmetric hypermultiplet shifts the CS level $\kappa$ by 
\begin{align}
\kappa ~\rightarrow~\kappa+\frac{N+4}{2}.
\end{align}
It follows that decoupling of a symmetric hypermultiplet from 5d $SU(N)_{\frac{N}{2}}+1\mathbf{Sym}$ gives either $SU(N)_{N+2}$ or $SU(N)_{2}$ (modular the sign of the CS level).

%% file: Sp23AS.tex
\section{$Sp(2)$ gauge theory with $3{\bf AS}$}\label{sec:Sp23AS}
%---------------
\begin{figure}
\centering
\subfigure[]{
\includegraphics[width=4cm]{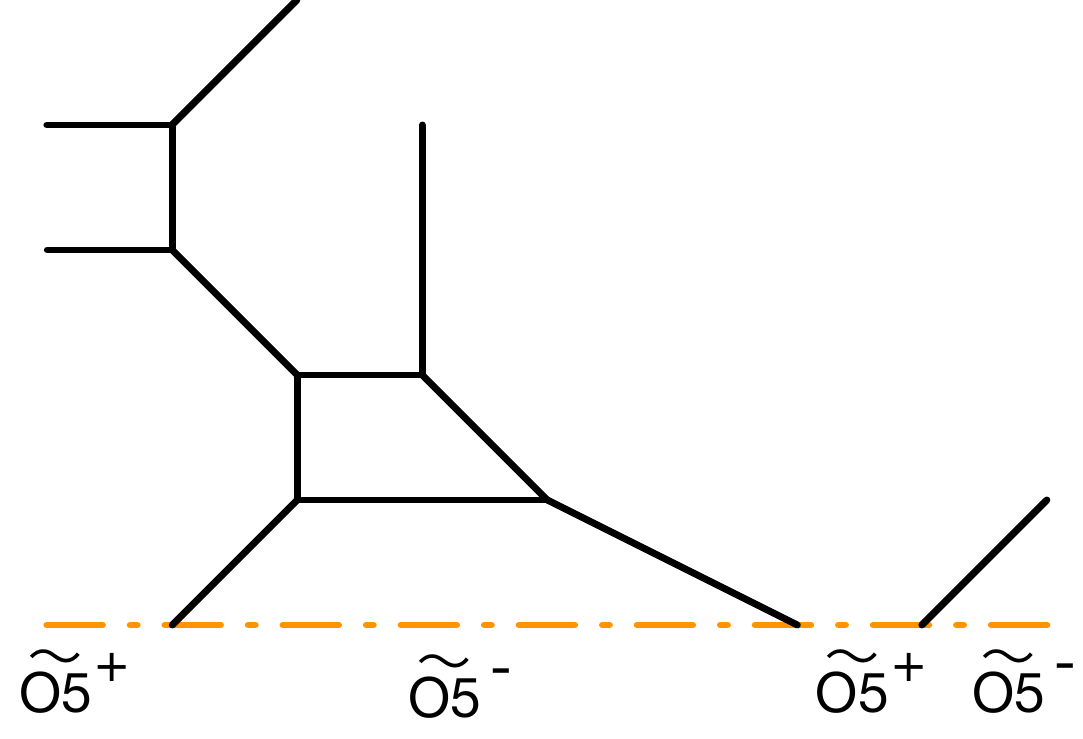} \label{fig:SO52F1S}}\\
\subfigure[]{
\includegraphics[width=4cm]{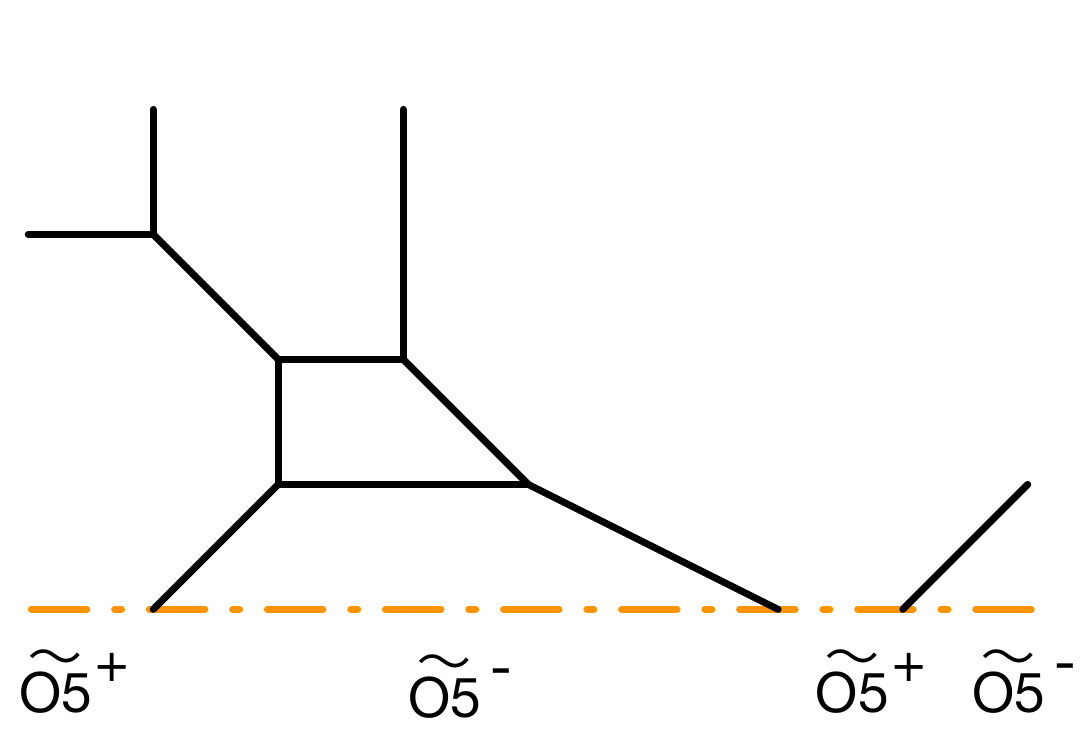} \label{fig:SO51F1S}}\quad
\subfigure[]{
\includegraphics[width=4cm]{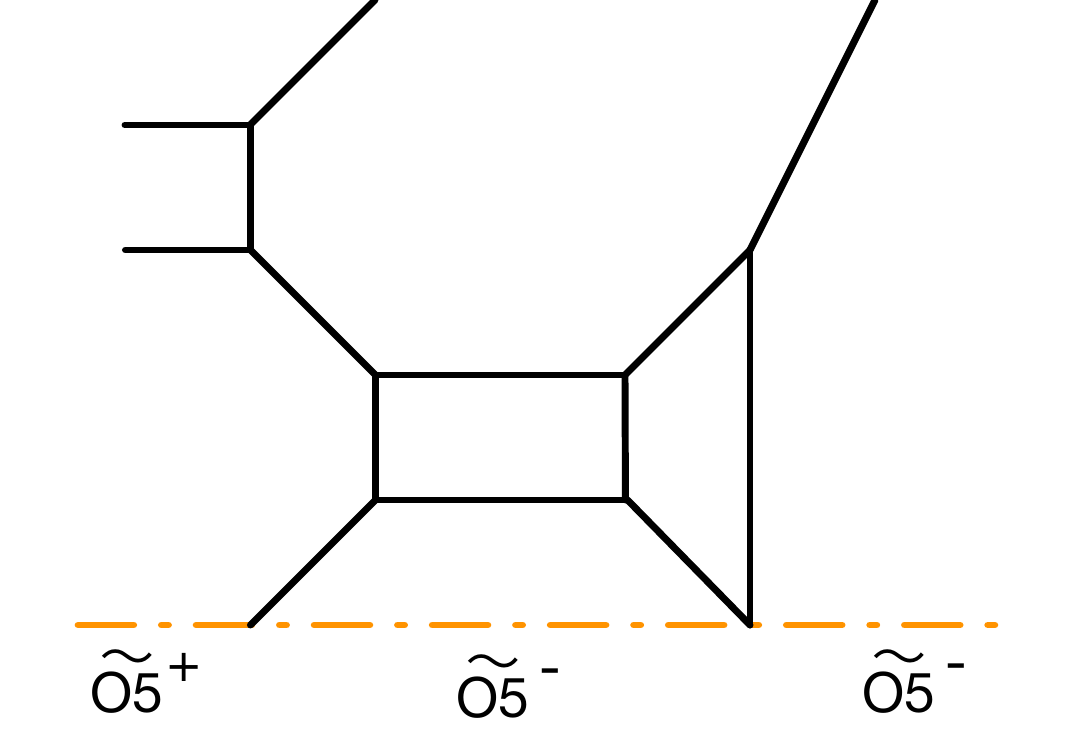} \label{fig:SO52FthetaPi}}\quad
\subfigure[]{
\includegraphics[width=4cm]{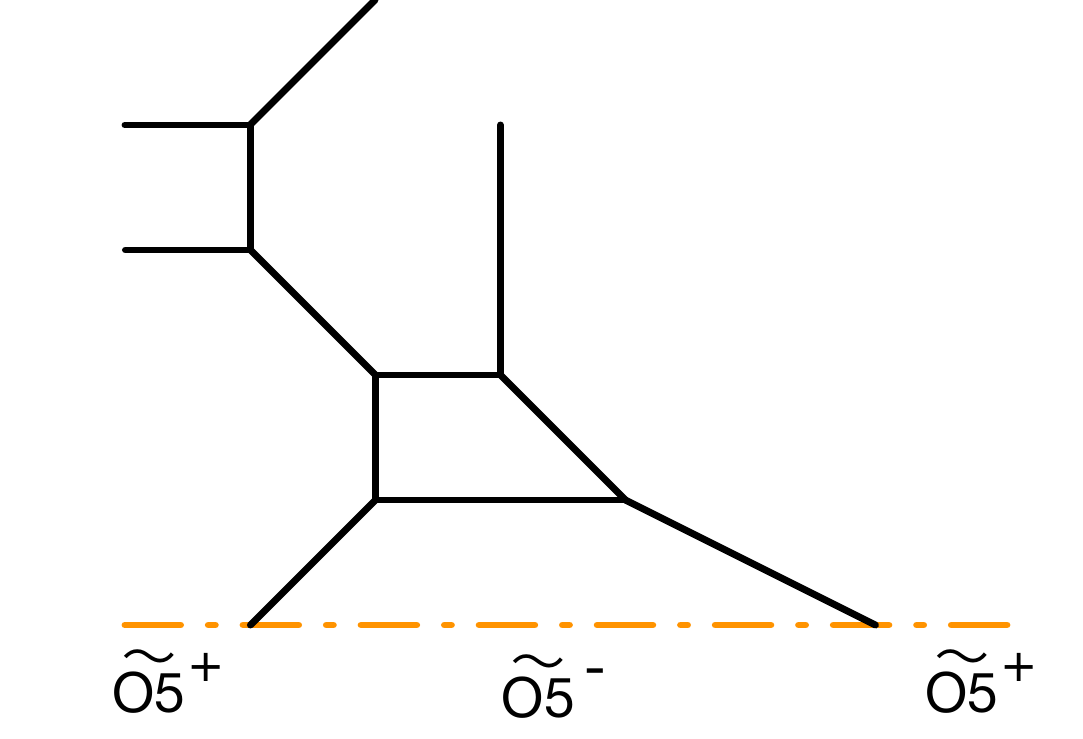} \label{fig:SO52Ftheta0}}
\caption{(a): A 5-brane web diagram for $SO(5)+2\bF+1\bS$. Different ways of decoupling a hypermultiplet yield the following three different theories.  (b): A 5-brane web for $SO(5)+1\bF+1\bS$, obtained by decoupling a vector from (a). (c): A 5-brane web for $SO(5)_\pi+2\bF$, obtained by decoupling a spinor taking negative infinite mass from (a). (d): A 5-brane web for $SO(5)_0+2\bF$, obtained by decoupling a spinor taking positive infinite mass from (a).
}
\label{fig:SO52F1Sdecoupling}
\end{figure}
%---------------
In the $G_2-SU(3)-Sp(2)$ sequence, $G_2+3\bF$ is dual to $Sp(2)+2\AS+1\bF$ and it can be also understood as $SO(5)+2\bF+1\bS$. Its decoupling is in particular interesting because depending on how we decouple the fundamental hypermultiplet for $Sp(2)$, it leads to two different discrete $\theta$-angle for the $Sp(2)$ theory. Moreover, it allows us to deform the theory by adding another hypermultiplet in the antisymmetric representation. Namely, we can properly decouple the flavor from $Sp(2)+2\AS+1\bF$ to obtain $Sp(2)_0+2\AS$ and then add one more hypermultiplet in the antisymmetric representation, which gives rise to another marginal theory $Sp(2)_0+3\AS$ or equivalently $SO(5)_0+3\bF$. 

In this section, we consider the deformation leading to the $Sp(2)_0+3\AS$ marginal theory. Introducing three hypermultiplets in the antisymmetric representation in a 5-brane web is not yet clear, so it is better to change  the brane configuration for $Sp(2)$ to that for $SO(5)_0$ as adding an antisymmetric hypermultiplet to an $Sp(2)$ theory is equivalent to adding a vector to an $SO(5)$ theory.  

Now we start with a brane configuration for $SO(5)+2\bF+1\bS$  given in Figure \ref{fig:SO52F1S}. There are three different possible decouplings as depicted in Figure \ref{fig:SO52F1Sdecoupling}. By decoupling a vector, we get $SO(5)+1\bF+1\bS$ (Figure \ref{fig:SO51F1S}). By decoupling a spinor taking its mass to negative infinity, 
 we get $SO(5)_\pi+2\bF$ (Figure \ref{fig:SO52FthetaPi})
Decoupling a spinor by taking the mass of the spinor matter infinite, we get $SO(5)_0+2\bF$ (Figure \ref{fig:SO52Ftheta0}).
%------------------------------------------
\begin{figure}%[H]
	\centering
\includegraphics[width=8cm]{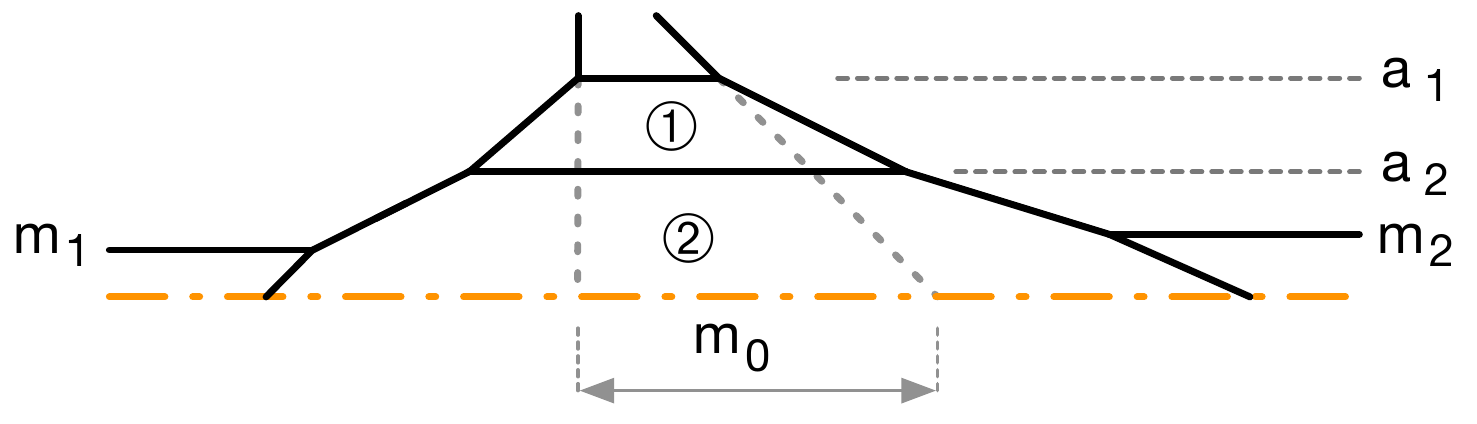}
\caption{A 5-brane web diagram with an $O5$-plane for $SO(5)+2{\bf F}$}
\label{fig:so5+2F}
\end{figure}
%------------------------------------------

For completeness, let us compare the area with the monopole tension from the effective prepotential of $SO(5)$ gauge theories with antisymmetric hypermultiplets. We start from $SO(5) + 2\bF$ in Figure \ref{fig:SO52Ftheta0}. By mass deformations, we can get a web diagram for $SO(5)_0+2\bF$ given in Figure \ref{fig:so5+2F}.  One can then read off the monopole tension of the theory from the areas in the brane configuration: 
\begin{align}\label{eq:so5Ta}
\textcircled{\scriptsize 1}&=\frac{1}{2} \left(a_1-a_2\right) \left(a_1-3 a_2+2 m_0\right), 
	\\
\textcircled{\scriptsize 2}&=	\frac12 \Big(2 a_2 m_0-a_2^2+4 a_1 a_2-m_1^2-m_2^2\Big),\label{eq:so5Tb}
\end{align}
where the range of the parameters are given as
\begin{align}
	a_1 \geq a_2\geq m_i\qquad\qquad (i=1,2). \label{eq:SO5Weyl1}
\end{align}
We now compare the area of the two faces with the monopole string tension from the prepotential for $SO(5)$+2{\bf F} in the %Weyl 
chamber.  
The effective prepotential in the phase \eqref{eq:SO5Weyl1} is then given by  
\begin{align}
	\mathcal{F}_{\,SO(5)+2{\bf F}} ~=& \, m_0 \big( \phi_1^2 - 2\,\phi_1\phi_2 + 2\,\phi_2^2\big)+ 
\frac13 \Big(4\, \phi_1^3 - 9\, \phi_1^2\phi_2 +6\, \phi_1\phi_2^2 +4 \,\phi_2^3 \Big)\crcr
&-\frac{1}{12} \sum_{i=1}^{2} \Big(12 m_i^2 \phi _2+m_i^3+16 \phi _2^3-24 \phi_1 \phi
   _2^2+12 \phi _1^2 \phi _2 \Big), \label{F.SO5w2F}
\end{align}
where we used the Dynkin basis \eqref{SO5.Dynkinbasis}.
One can see explicitly that the monopole tension computed from \eqref{F.SO5w2F} is related to the area \eqref{eq:so5Ta} and \eqref{eq:so5Tb} by 
\begin{align}\label{eq:so5Tphi}
\frac{\partial{\mathcal{F}_{\,SO(5)+2{\bf F}}}}{\partial \phi_1} &=\textcircled{\scriptsize 1},
	\qquad\qquad
\frac{\partial{\mathcal{F}_{\,SO(5)+2{\bf F}}}}{\partial \phi_2}   =2\times\textcircled{\scriptsize 2}.
\end{align}

\paragraph{$SO(5)+3\bF$ case.} 
%---------------
\begin{figure}
\centering
\subfigure[]{
\includegraphics[width=8.5cm]{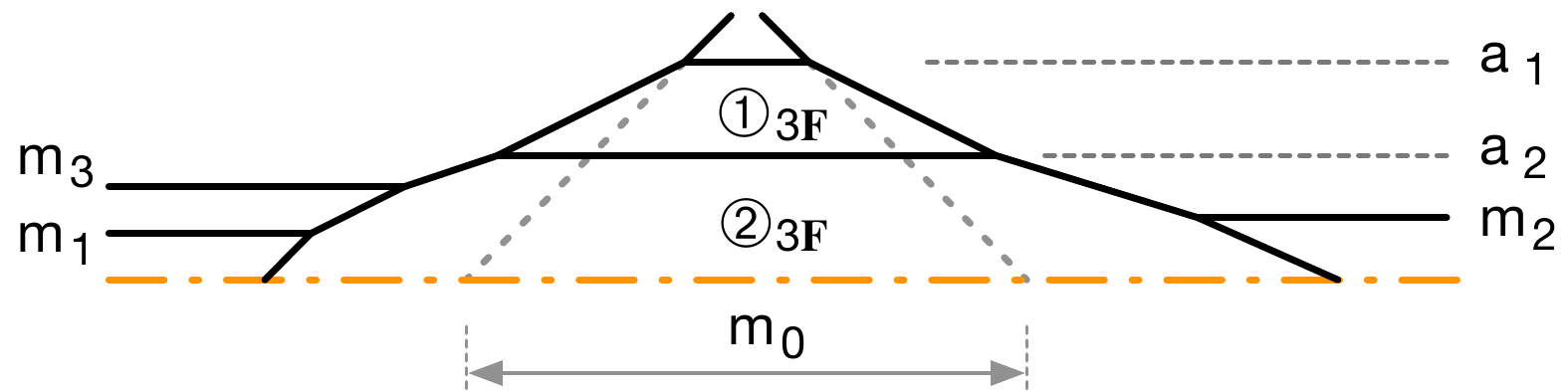} \label{fig:SO53Ftheta0-area}
}
\subfigure[]{
\includegraphics[width=8.5cm]{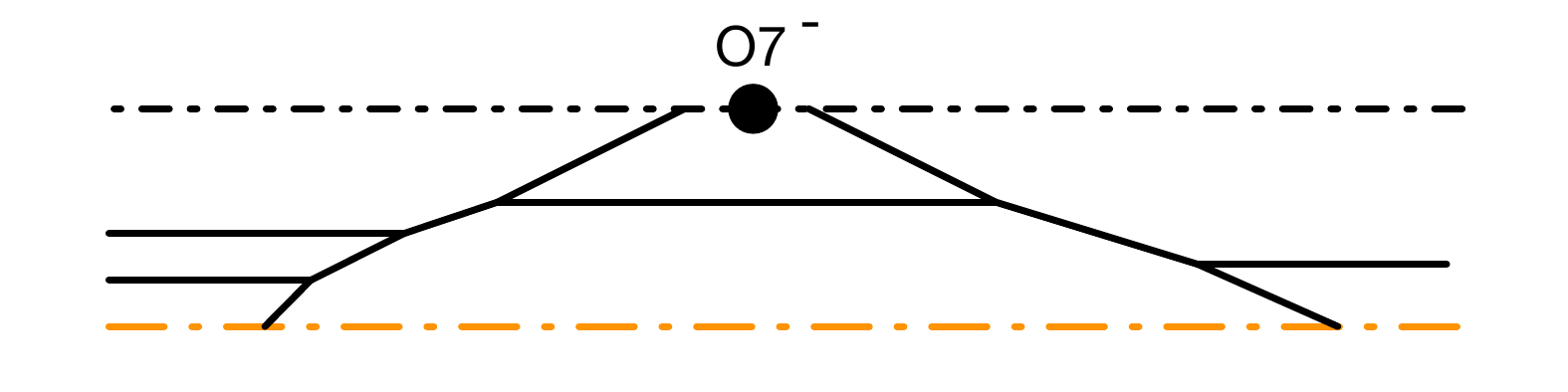} \label{fig:SO53Fth0-O7-}
}
\caption{A 5-brane web diagram with an O5-plane for $SO(5)+3{\bf F}$.
}
\label{fig:SO53Fth0}
\end{figure}
%---------------
For the $SO(5)$ theory with three hypermultiplets in the fundamental representation, a 5-brane web diagram with an O5-plane is depicted in Figure \ref{fig:SO53Fth0}. Note that the two external 5-branes in Figure \ref{fig:SO53Ftheta0-area} are of the charges $(1,1)$ and $(1,-1)$. Then a $(1, 1)$ 7-brane and a $(1, -1)$ 7-brane can end on the external 5-branes respectively and they can be combined to be an O7$^-$-plane as shown in Figure \ref{fig:SO53Fth0-O7-}. It hence has a periodic direction in the 5-brane plane, and therefore it is a marginal theory, which can be understood as a twisted compactification \cite{Hayashi:2015vhy}.

The area of the compact faces in the web diagram in Figure \ref{fig:SO53Ftheta0-area} are then given by
\begin{align}
\textcircled{\scriptsize 1}_{3\bF}&=\left(a_1-a_2\right) \left({m_0}-2 a_2\right),\label{eq:so5Ta3F1}
	\\
\textcircled{\scriptsize 2}_{3\bF}&=\frac12\Big(2a_2m_0- 2a_2^2 + 4a_1a_2-m_1^2-m_2^2-m_3^2\Big).\label{eq:so5Ta3F2}
\end{align}
We now compare these area with the monopole string tension from the prepotential. 
The diagram in Figure \ref{fig:SO53Ftheta0-area} is in the phase 
\be
a_1 \geq a_2 \geq m_i, \quad i=1, 2, 3. \label{phase.SO5w3F}
\ee 
Then the effective prepotential for $SO(5)+3\bF$ in this phase is given by  
\begin{align}
	\mathcal{F}_{\,SO(5)+3{\bf F}} ~=& \, m_0 \big( \phi_1^2 - 2\,\phi_1\phi_2 + 2\,\phi_2^2\big)+ 
\frac13 \Big(4\, \phi_1^3 - 9\, \phi_1^2\phi_2 +6\, \phi_1\phi_2^2 +4 \,\phi_2^3 \Big)\crcr
&-\frac{1}{12} \sum_{i=1}^{3} \Big(12 m_i^2 \phi _2+m_i^3+16 \phi _2^3-24 \phi_1 \phi
   _2^2+12 \phi _1^2 \phi _2 \Big)
.
\end{align}
As expected, 
one can readily see that the monopole string tension agree with the area of the faces of the web diagram in Figure \ref{fig:SO53Fth0},
\begin{align}\label{eq:so5Tphi3Fcheck}
%	T_1
\frac{\partial{\mathcal{F}_{SO(5)+3\bF}}}{\partial \phi_1} &
   =\textcircled{\scriptsize 1}_{3\bF},
	\qquad\qquad
%T_2&
\frac{\partial{\mathcal{F}_{SO(5)+3{\bf F}}}}{\partial \phi_2}  =
2\times\textcircled{\scriptsize 2}_{3\bF}. 
\end{align}

%% file: SU3CS9.tex
\section{5-brane web for pure $SU(3)_9$ gauge theory}\label{sec:SU3CS9}

\begin{figure}
\centering
\includegraphics[width=8cm]{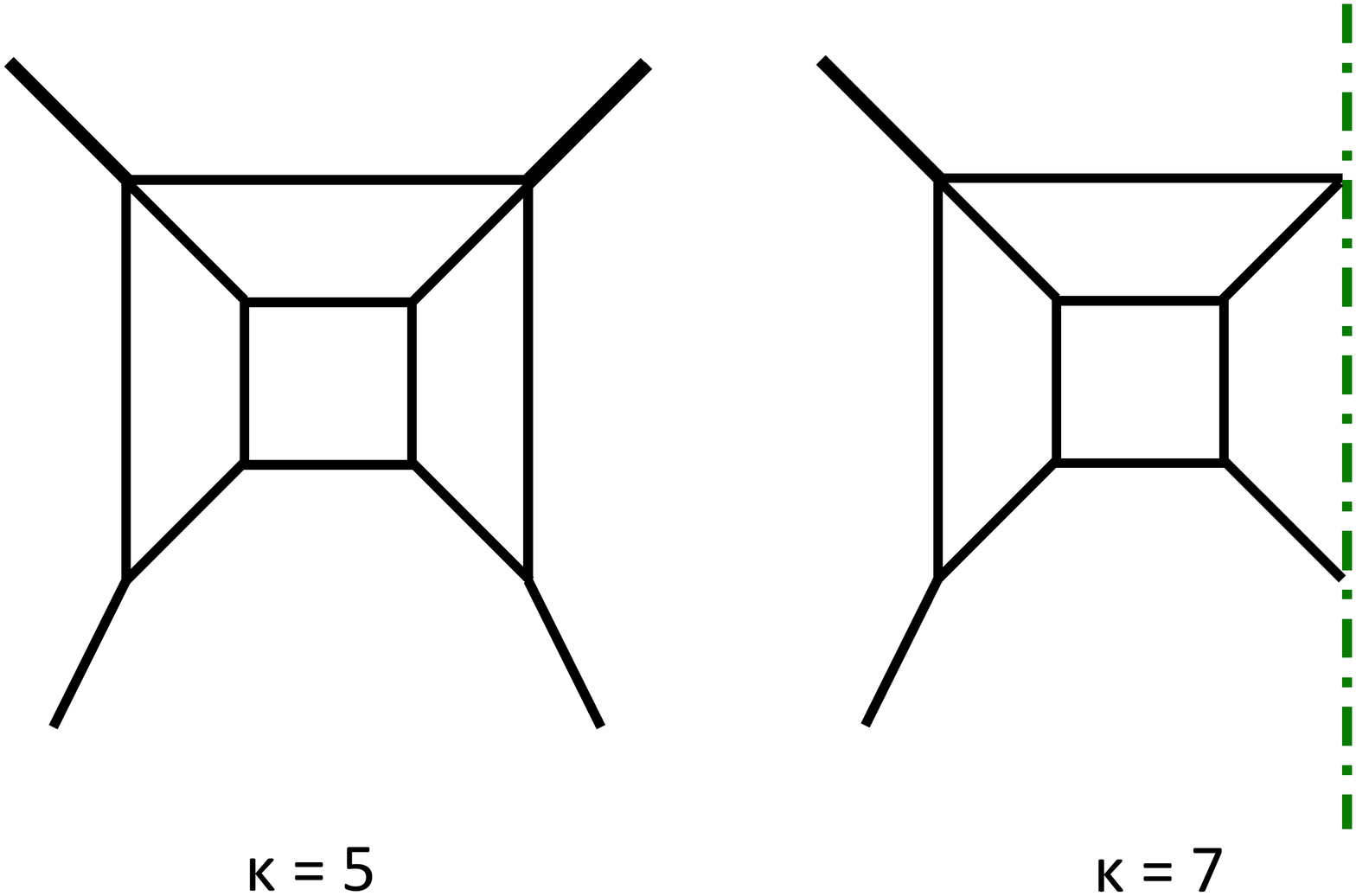}
\caption{The left diagram realizes the pure $SU(3)$ gauge theory with the CS level $5$ whereas the right diagram realizes the pure $SU(3)$ gauge theory with the CS level $7$.}
\label{fig:pureSU3CS5vs7}
\end{figure}

\begin{figure}
\centering
\includegraphics[width=8cm]{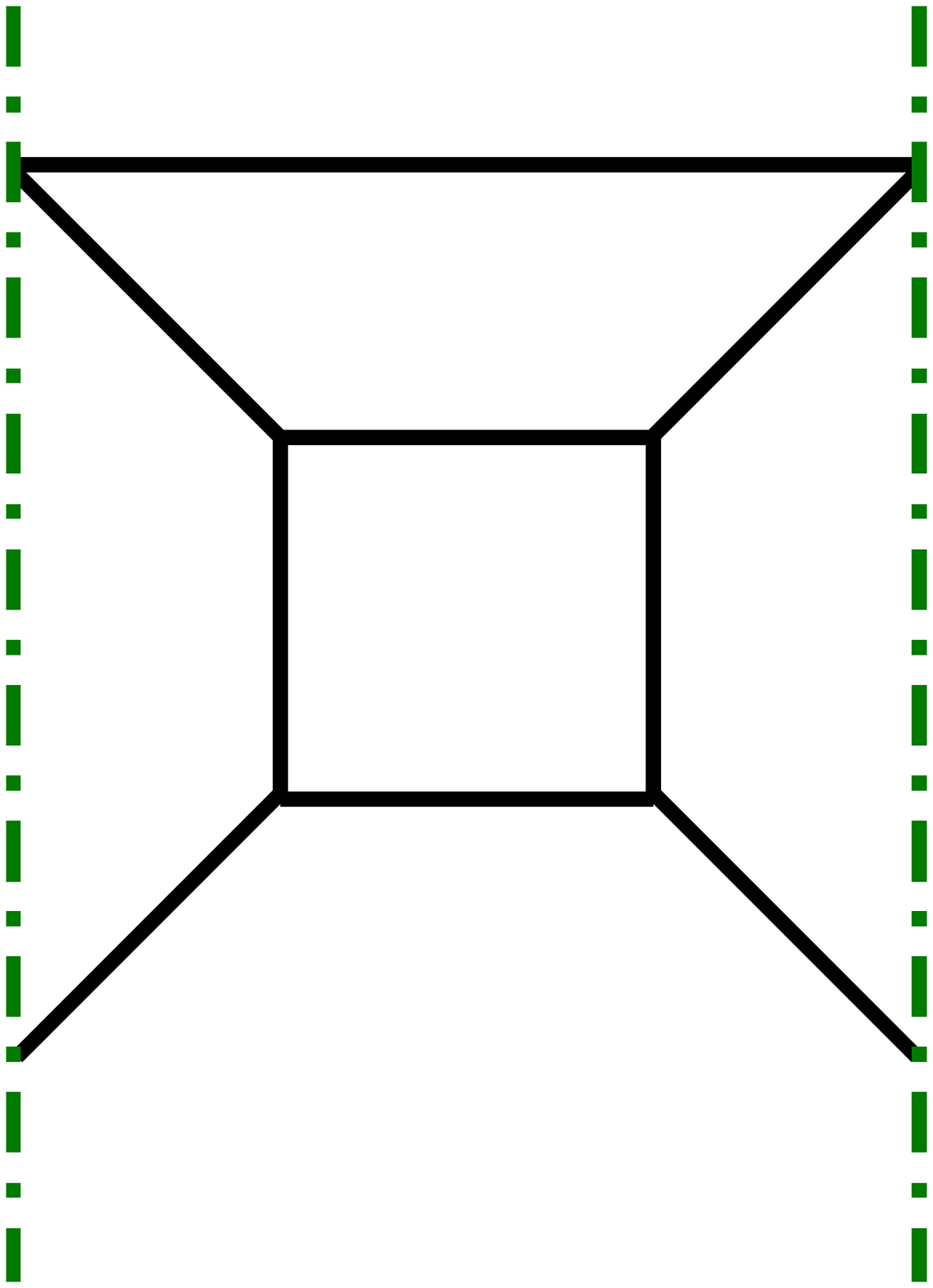}
\caption{A 5-brane web diagram for the pure $SU(3)$ gauge theory with the CS level $9$.}
\label{fig:pureSU3CS9}
\end{figure}

In section \ref{sec:pureSU3CS7}, we realized a 5-brane web diagram which yields the pure $SU(3)$ gauge theory with the CS level $7$. In fact, it turns out that an extension of the diagram gives a 5-brane diagram of the pure $SU(3)$ gauge theory with the CS level $9$. In order to see the extension, it is useful to compare a 5-brane web diagram for the pure $SU(3)$ gauge theory with the CS level $5$ with the 5-brane web diagram for the pure $SU(3)$ gauge theory with the CS level $7$. The two diagrams are depicted in Figure \ref{fig:pureSU3CS5vs7}. 
%%%%%%%%%%%%%%%%%%%%%%%%%%%%%%%%%
%%%%%%%%%%%%%%%%%%%%%%%%%%%%%%%%%
The increase of the CS level by $2$ is implemented by replacing one side of the diagram of the pure $SU(3)$ gauge theory with the CS level $5$ with an ON-plane. Hence, it is natural to guess that replacing another side of the diagram of the pure $SU(3)$ gauge theory with the CS level $7$ with an ON-plane may give rise to a diagram of the pure $SU(3)$ gauge theory with the CS level $9$. We then propose that the diagram in Figure \ref{fig:pureSU3CS9} gives rise to a 5-brane web diagram for the pure $SU(3)$ gauge theory with the CS level $9$. 
%%%%%%%%%%%%%%%%%%%%%%%%%%%%%%%%%
%%%%%%%%%%%%%%%%%%%%%%%%%%%%%%%%%

\begin{figure}
\centering
\includegraphics[width=8cm]{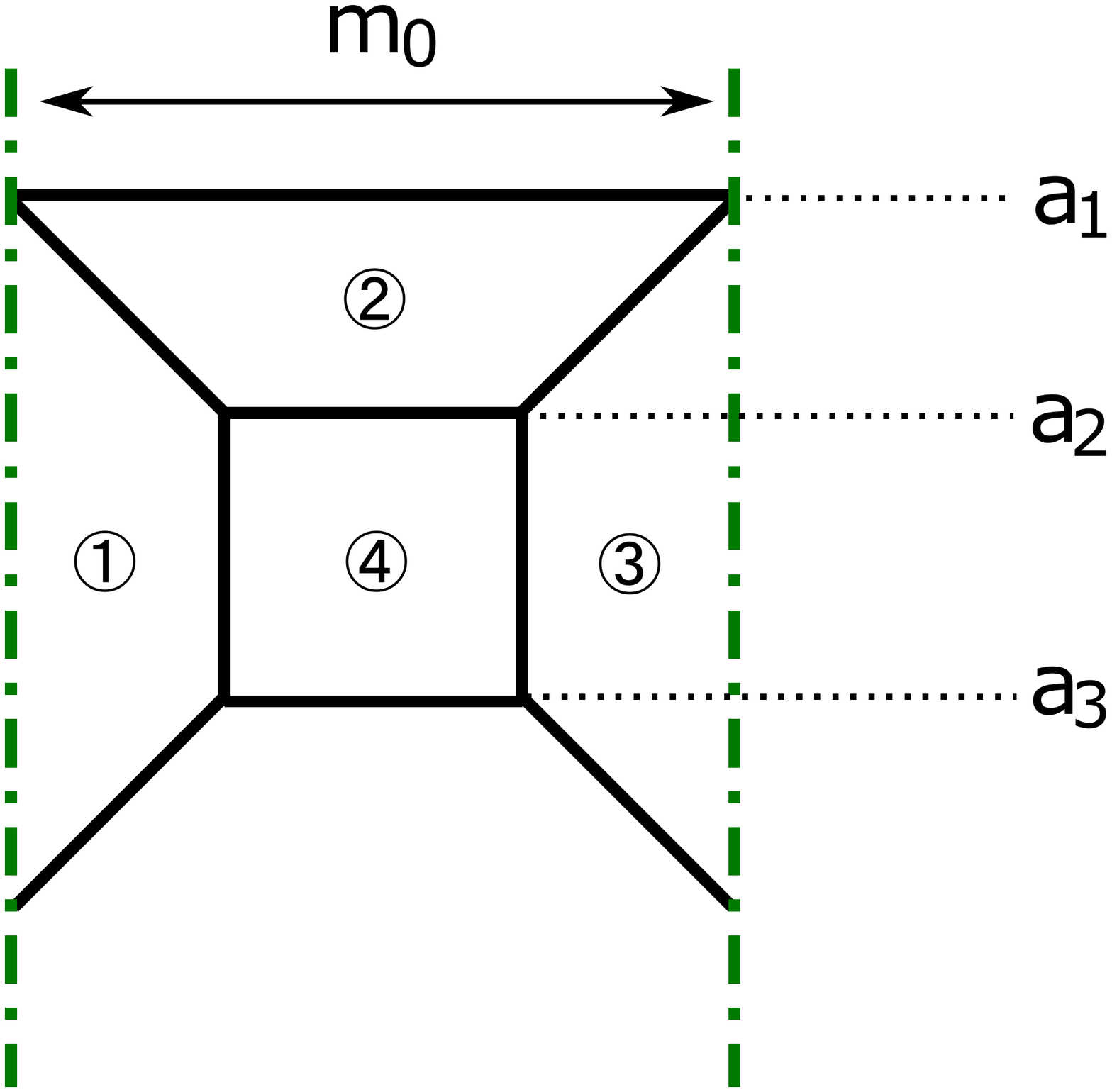}
\caption{The parameterization for the pure $SU(3)$ gauge theory with the CS level $9$. $a_1, a_2, a_3$ are the Coulomb branch moduli and $m_0$ is the inverse of the squared gauge coupling.}
\label{fig:pureSU3CS9para}
\end{figure}

One can check the claim by computing the tension of the monopole string from the 5-brane web diagram in Figure \ref{fig:pureSU3CS9}. The tension is given by the area and we can compare the area with the result expected from the field theory. In order to write the area by the gauge theory parameters of the $SU(3)$ gauge theory, we assign the Coulomb branch moduli $a_1, a_2, a_3, \; (a_1 + a_2 + a_3 = 0)$ and the inverse of the squared gauge coupling $m_0$ as in Figure \ref{fig:pureSU3CS9para}. 
%%%%%%%%%%%%%%%%%%%%%%%%%%%%%%%%%
%%%%%%%%%%%%%%%%%%%%%%%%%%%%%%%%%
Then the area of the four faces in Figure \ref{fig:pureSU3CS9para} becomes 
\bea
\textcircled{\scriptsize 1} &=& (a_1 - a_2)(a_1 - a_3), \label{area1.SU3CS9}\\
\textcircled{\scriptsize 2} &=& (a_1 - a_2)(m_0 - a_1 + a_2),  \label{area2.SU3CS9}\\
\textcircled{\scriptsize 3} &=& (a_1 - a_2)(a_1 - a_3), \label{area3.SU3CS9}\\
\textcircled{\scriptsize 4} &=& (a_2 - a_3)(m_0 - 2a_1 + 2a_2). \label{area4.SU3CS9}
\eea

Let us then compare the area with the tension of the monopole string of the pure $SU(3)$ gauge theory with the CS level $9$. Since we do not have matter, the theory have only one phase and the effective prepotential can be computed from \eqref{prepotential} and it becomes
\bea
\mathcal{F}_{SU(3)_9} &=& \frac{m_0}{2}(a_1^2 + a_2^2 + a_3^2) + \frac{1}{6}\left((a_1 - a_2)^3 + (a_1 - a_3)^3 + (a_2 - a_3)^3\right) + \frac{9}{6}\left(a_1^3 + a_2^3 + a_3^3\right)\nn\\
&=&m_0\left(\phi_1^2 -\phi_1\phi_2 + \phi_2^2\right) + \frac{4}{3}\phi_1^3 + 4\phi_1^2\phi_2 - 5\phi_1\phi_2^2 + \frac{4}{3}\phi_2^3, \label{prepot.SU3CS9}
\eea
where we changed the basis for the Coulomb branch moduli into the Dynkin basis in \eqref{prepot.SU3CS9} by using \eqref{SU3.Dynkinbasis}. Then the tension of the monopole string is given by taking a derivative of the prepotential with respect to $\phi_1$ and $\phi_2$. Hence the tension from \eqref{prepot.SU3CS9} is 
\bea
\frac{\partial \mathcal{F}_{SU(3)_9}}{\partial \phi_1} &=& (2\phi_1 - \phi_2)(m_0 + 2\phi_1 + 5\phi_2),\label{tension1.SU3CS9}\\ 
\frac{\partial \mathcal{F}_{SU(3)_9}}{\partial \phi_2} &=& (-\phi_1 + 2\phi_2)(m_0 - 4\phi_1 + 2\phi_2). \label{tension2.SU3CS9}
\eea

Now we can compare the tension \eqref{tension1.SU3CS9} and \eqref{tension2.SU3CS9} with the area \eqref{area1.SU3CS9}-\eqref{area4.SU3CS9}. As in the case of the comparison between the area and the monopole string tension for the pure $G_2$ gauge theory, we need to consider a linear combination among \eqref{area1.SU3CS9}-\eqref{area4.SU3CS9} to obtain the area of a face where D3-brane covers \cite{Hayashi:2018bkd}. More specifically, the area corresponding to the tension \eqref{tension1.SU3CS9} should be $2\textcircled{\scriptsize 1} + \textcircled{\scriptsize 2} + 2\textcircled{\scriptsize 3}$ while the area corresponding to the tension \eqref{tension2.SU3CS9} is simply given by $\textcircled{\scriptsize 4}$. Indeed, it is straightforward to check the equalities 
\bea
2\textcircled{\scriptsize 1} + \textcircled{\scriptsize 2} + 2\textcircled{\scriptsize 3} &=& \frac{\partial \mathcal{F}_{SU(3)_9}}{\partial \phi_1},\\
\textcircled{\scriptsize 4} &=&\frac{\partial \mathcal{F}_{SU(3)_9}}{\partial \phi_2}.
\eea
from the explicit expressions of \eqref{area1.SU3CS9}-\eqref{area4.SU3CS9} and \eqref{tension1.SU3CS9}-\eqref{tension2.SU3CS9}. This gives an evidence that the diagram in Figure \ref{fig:pureSU3CS9} gives rise to the pure $SU(3)$ gauge theory with the CS level $9$. 

%% file: concl.tex
\section{Conclusion}\label{sec:concl}
In the paper, we proposed all the 5-brane webs of rank 2 superconformal theories classified via geometries in \cite{Jefferson:2018irk}, and discussed their mutual dualities from the perspective of S-duality and the Hanany-Witten transitions arising by moving 7-branes. As many of 5-brane webs for such rank 2 theories are already known, our focus has been those theories newly proposed in \cite{Jefferson:2018irk}, which did not have 5-brane descriptions. 

We explicitly constructed 5-brane webs for all the marginal theories. We compared the area of the web diagram for each theory with the monopole string tension calculated from the effective prepotential, which showed the perfect agreement. We also found the duality map among dual theories.  
For instance, explicit 5-brane webs are presented in section \ref{sec:G2SU3Sp2} for the $G_2$ gauge theories with six flavors ($G_2+6\bF$) and its dual theories, the $Sp(2)$ gauge theory with four flavors and two hypermultiplets in the antisymmetric representation ($Sp(2)+2\AS+4\bF$) and the $SU(3)$ gauge theory with six flavors and Chern-Simons level 4 ($SU(3)_4+6\bF$). We also present 5-brane web from the viewpoint of the $SO(5)$ theory with two hypermultiplets in the fundamental representation and four hypermultiplets in the spinor representation ($SO(5)+2{\bf V}+ 4{\bf S}$). The duality map among the theories in the $G_2$-$SU(3)$-$Sp(2)$ sequences are also discussed in sections \ref{sec:G2marginal} and \ref{subsec:SO52F4S}.

From the $G_2-SU(3)-Sp(2)$ sequence, we also discussed various deformations: (i) One can deform the theory by decoupling one antisymmetric hypermultiplet from $Sp(2)+2\AS+4\bF$ and then by adding flavors which leads to the $Sp(2)$ gauge theory with one antisymmetric and eight flavors ($Sp(2)+ 1\AS+ 8\bF$) which is dual to the $SU(3)$ gauge theory with nine flavors and the CS level $\frac{3}{2}$ ($SU(3)_{3/2}+ 9\bF$). (ii) One can further decouple the antisymmetric hypermultiplet and add more flavors to get the $Sp(2)$ gauge theory with ten flavors ($Sp(2) + 10\bF$) which is dual to the $SU(3)$ gauge theory with ten flavors and the CS level $0$ ($SU(3)_0 + 10\bF$). (iii) One can also deform the theory to the $SU(3)$ gauge theory with one symmetric hypermultiplet and one flavor with the zero CS level ($SU(3)_0 + 1{\bf Sym} + 1\bF$) and also to the $SU(3)$ gauge theory with only one symmetric hypermultiplet with the CS level $3/2$ ($SU(3)_{\frac32} + 1{\bf Sym}$). (iv) Another possible deformation is to deform the theory to the $Sp(2)_0$ theory with three hypermultiplets in the antisymmetric representation and the discrete theta angle zero ($Sp(2)_0+3\AS$) or equivalently the $SO(5)_0$ theory with three vectors and the discrete theta angle zero ($SO(5)_0+3\bF$).

We note that the 5-brane web diagrams for the marginal $SU(3)$ theories, $SU(3)_0+1{\bf Sym}+ 1\bF$ and $SU(3)_\frac32+1{\bf Sym}$ 
are obtained with an O7$^+$-plane. 
In particular, a 5-brane web diagram for $SU(3)_{\frac32} + 1{\bf Sym}$ can be constructed with O7$^+$- and O7$^-$-planes with 5-branes appearing repeatedly on both sides of O7-planes with a periodic structure. This can be straightforwardly generalized to rank $N$, exhibiting a new 5-brane structure for 5d $SU(N)_{\frac{N}{2}}+1{\bf Sym}$ theory. We also note that, inspired by the brane web for the pure $G_2$ gauge theory with an O5-plane, 
we constructed a 5-brane web for the marginal $SU(3)$ gauge theory with the CS level $9$, which requires two $\widetilde{ON}$-planes\footnote{
We found that the $SU(3)$ gauge theories with the CS level from 3 to 6 also have a 5-brane web description with an $\widetilde{ON}$-plane, while the $SU(3)$ gauge theory with the CS level 7 is only possible with an $\widetilde{ON}$-plane. See Appendix \ref{sec:allwebs}.}. 
An $\widetilde{ON}$-plane appears not only as the S-dual object of an $\widetilde{O5}$-plane, but also naturally as Higgsing and decoupling of the D-type quiver theory with an $ON^0$-plane.
Thus, a 5-brane web with an $\widetilde{ON}$-plane sometimes can allow a field theory description.

The duality between $Sp(2)+ 1\AS+ 8\bF$ and $SU(3)_{3/2}+ 9\bF$ can be understood as a particular sequence of Hanany-Witten transitions by moving 7-branes. From the duality map \eqref{eq:m0su3m0Sp29F} - \eqref{eq:phi2su3phi2sp29F} and their brane configurations (Tao web diagrams in Figures \ref{fig:Sp21A8FTao} and \ref{fig:SU3Nf9Taom0}), these two marginal theories have the same period which is expressed as two times the inverse of the bare gauge coupling of ${Sp(2)+1\AS+8\bF}$ squared, denoted by $2\, m_0^{Sp(2)}$. 
We note that, when expressed in terms of the parameters of $SU(3)_{3/2}+ 9\bF$, the period is not $2\, m_0^{SU(3)}$ 
but more complicated, which is, however, equivalent to $2\, m_0^{Sp(2)}$  
under the duality map.

By decoupling hypermultiplets from the marginal theories, 
 we can obtain 5-brane webs for 5d superconformal theories with various hypermultiplets. Various 5d theories with less number of hypermultiplet can be obtained from decoupling of hypermultiplets from another marginal theory. For example, we decoupled the symmetric matter to obtain a brane web for $SU(3)_{-\frac72} + 1{\bf F}$. 
This decoupling generates various RG flows among rank 2 theories. Following a summary figure presented in \cite{Jefferson:2018irk}, we also summarize the 5-brane webs for rank 2 theories, their duality relations, and RG flows in Figure \ref{fig:listoffigures}.

It would be interesting to study the 6d origin of the marginal theories discussed in the paper. 
It seems generic that for a marginal $SU$ gauge theory with non-zero CS level has the property that the compactification radius (or the period) is composed of a linear combination of the bare coupling and the mass parameters of the hypermultiplets, which may indicate some intriguing interplay between the compactification radius and the mass parameters. It would also be interesting to further study  such relation from the perspective of its 6d origin. 
Another interesting future direction would be further confirm the duality relation from BPS operator counting from the gauge theories. For instance, one may compute superconformal indices for dual theories and confirm the duality map, which would be another consistency check for the duality maps that we obtained from 5-brane webs. 
Finally, pursuing 
5-brane webs for superconformal theories of higher rank greater than 2 which may lead to new 5-brane perspective on higher Chern-Simons levels and hypermultiplet in other representations than what was discussed. For instance, the hypermultiplet in the rank 3 antisymmetric representation can be constructed \cite{HKLY2018}.

%% file: readoffm0.tex
\section{The gauge coupling for $SO(2N+1)$ gauge theory with spinors}
\label{sec:m0}

In section \ref{subsec:SO52F4S}, the inverse of the squared gauge coupling $m_0$ for the $SO(5)$ gauge theory with two vector hypermultiplets and four spinor hypermultiplets was defined by \eqref{m0.SO5w2V4S}, using the parameters in Figure \ref{fig:SO5toG2para}. The definition of $m_0$ in a web was in fact different from that of the $G_2$ gauge theory with two flavors given in Figure \ref{fig:G2w2flvrspara}. The $G_2$ gauge theory with two flavors was obtained from the Higgsing of the $SO(7)$ gauge theory with three spinors. Therefore, how to read off $m_0$ from the diagram was different between the $SO(5)$ gauge theory and the $SO(7)$ gauge theory.  In this appendix, we give an explanation of the difference by using the effective prepotential of an $SO(2N+1)$ gauge theory with spinors.

\subsection{Decoupling of a spinor}

We first discuss how decoupling one spinor of an $SO(2N+1)$ gauge theory affects the inverse of the squared gauge coupling by using the effective prepotential. The effective prepotential for the $SO(2N+1)$ gauge theory with $N_f$ vectors and $N_s$ spinors can be calculated from the general expression \eqref{prepotential} and it is given by
\begin{align}\label{eq:prep}
\mathcal{F}_{SO(2N+1)+N_f{\bf V} + N_s{\bf S}} = 
&
\frac{1}{2} m_0 \sum_{i=1}^N a_i{}^2 
+ \frac{1}{6} \left( 
\sum_{1 \le i < j  \le  N} \left[ |a_i-a_j|^3 + |a_i+a_j|^3 \right]
+ \sum_{i=1}^N |a_i|{}^3
\right)
\cr
& 
- 
\frac{1}{12} \sum_{i=1}^{N}  \sum_{j=1}^{N_f} 
\left( 
\left| a_i  - m_j \right|^3
+ \left| - a_i  - m_j \right|^3
\right)
\cr
& 
- \frac{1}{12} \sum_{k=1}^{N_s} 
%\sum_{ \{ s_i =\pm  \} } 
\sum_{s_1 = \pm 1} \sum_{s_2 = \pm 1} \cdots \sum_{s_N = \pm 1}
\left| \frac{1}{2} \left(  \sum_{i=1}^{N} s_i a_i \right) - m_k \right|^3.
\end{align}
We then consider decoupling one spinor by sending $m_{N_s} \to +\infty$. Then the terms involving the $N_s$th spinor in the last line of \eqref{eq:prep} become
\begin{align}
&
- \frac{1}{12} %\sum_{k=1}^{N_s} 
%\sum_{ \{ s_i =\pm  \} } 
\sum_{s_1 = \pm 1} \sum_{s_2 = \pm 1} \cdots \sum_{s_N = \pm 1}
\left| \frac{1}{2} \left(  \sum_{i=1}^{N} s_i a_i \right) - m_{N_s} \right|^3 
%\cr
= 
&
%- \frac{1}{12} 
%\sum_{s_2' = \pm 1} \cdots \sum_{s_N' = \pm 1}
%\left(
%\left( \frac{1}{2} \left( \sum_{i=1}^{N} s'_i a_i \right) + m_{N_s} \right)^3
%+ \left( - \frac{1}{2} \left( \sum_{i=1}^{N} s'_i a_i \right) + m_{N_s} \right)^3
%\right)
%\cr
%= 
%&
%- \frac{1}{6} 
%\sum_{s_2' = \pm 1} \cdots \sum_{s_N' = \pm 1}
%\left( 
%\frac{3}{4} \left( \sum_{i=1}^{N} s'_i a_i \right)^2 m_{N_s}
%+ m_{N_s} {}^3
%\right)
%\cr
%= 
%&
%- \frac{1}{8} m_{N_s}
%\sum_{s_2' = \pm 1} \cdots \sum_{s_N' = \pm 1}
%\left( \sum_{i=1}^{N} s'_i a_i \right)^2 
%- \frac{1}{6}m_{N_s} {}^3 
%\sum_{s_2' = \pm 1} \cdots \sum_{s_N' = \pm 1} 1
%\cr
%= 
%&
%- \frac{1}{8} m_{N_s}
%\left(
%\sum_{s_2' = \pm 1} \cdots \sum_{s_N' = \pm 1}
%\sum_{i=1}^{N} \sum_{j=1}^{N} s'_i s'_j  a_i  a_j
%\right)
%- \frac{1}{6}m_{N_s} {}^3 \times 2^{N-1}
%\cr
%= 
%&
%- \frac{1}{8} m_{N_s}
%\sum_{s_2' = \pm 1} \cdots \sum_{s_N' = \pm 1}
%\left(
%\sum_{i=1}^{N} a_i{}^2 
%+ 
%2 \sum_{1 \le i < j \le N} s'_i s'_j  a_i  a_j
%\right)
%- \frac{2^{N-2}}{3}m_{N_s} {}^3
%\cr
%= 
%&
%- \frac{1}{8} m_{N_s} \sum_{i=1}^{N} a_i{}^2 
%\left(
%\sum_{s_2' = \pm 1} \cdots \sum_{s_N' = \pm 1} 1
%\right)
%- \frac{1}{4} m_{N_s} \sum_{1 \le i < j \le N} a_i  a_j
%\left(
%\sum_{s_2' = \pm 1} \cdots \sum_{s_N' = \pm 1} s'_i s'_j  
%\right)
%- \frac{2^{N-2}}{3}m_{N_s} {}^3
%\cr
%= 
%&
%- \frac{1}{8} m_{N_s} \sum_{i=1}^{N} a_i{}^2 \times 2^{N-1}
%- \frac{1}{4} m_{N_s} \sum_{1 \le i < j \le N} a_i  a_j \times 0
%- \frac{2^{N-2}}{3}m_{N_s} {}^3 
%\cr
%= 
%&
- 2^{N-4} m_{N_s} \sum_{i=1}^{N} a_i{}^2 
- \frac{2^{N-2}}{3}m_{N_s} {}^3.
\end{align}
Therefore, in the limit where $m_{N_s} \to \infty$, the effective prepotential \eqref{eq:prep} becomes
\begin{align}
\mathcal{F}_{SO(2N+1)+N_f{\bf V} + (N_s-1){\bf S}} &= 
%&
\frac{1}{2} \left( m_0 - 2^{N-3} m_{N_s} \right) \sum_{i=1}^N a_i{}^2 
\cr
&+ \frac{1}{6} \left( 
\sum_{1 \le i < j  \le  N} \left[ |a_i-a_j|^3 + |a_i+a_j|^3 \right]
+ \sum_{i=1}^N |a_i|{}^3
\right)
\cr
& 
- 
\frac{1}{12} \sum_{i=1}^{N}  \sum_{k=1}^{N_f} 
\left( 
\left| a_i  - m_j \right|^3
+ \left| - a_i  - m_j \right|^3
\right)
\cr
& 
- 
\frac{1}{12} \sum_{k=1}^{N_s - 1} 
%\sum_{ \{ s_i =\pm  \} } 
\sum_{s_1 = \pm 1} \sum_{s_2 = \pm 1} \cdots \sum_{s_N = \pm 1}
\left| \frac{1}{2} \left(  \sum_{i=1}^{N} s_i a_i \right) - m_k \right|^3,
\end{align}
up to the constant term $- \frac{2^{N-2}}{3} m_{N_s}{}^3$ which we can discard.
In order to obtain the effective prepotential for the theory after decoupling one spinor with mass $m_{N_s}$,  we need to identify the new (inverse of the squared) gauge coupling constant $m_0^{\rm new}$ as
\begin{align}\label{eq:newm0}
m_0^{\rm new} = m_0 - 2^{N-3} m_{N_s}.
\end{align}
Unlike the case of decoupling a vector, the shift for $m_0$ depends on $N$.

%%%%%%%%%%%%%%%%%%%%%%%%%%%%%%%%%%%%%%%%%%%%%%%%%%%%%
%%%%%%%%%%%%%%%%%%%%%%%%%%%%%%%%%%%%%%%%%%%%%%%%%%%%%
\subsection{Reading off gauge coupling from web}
Using the general formula for the shift of $m_0$ after decoupling a spinor, we identify the length corresponding to $m_0$ from a 5-brane web diagram for the $SO(2N+1)$ gauge theory with spinors. 
\subsubsection{One spinor case}
\begin{figure}
\centering
\includegraphics[width=5cm]{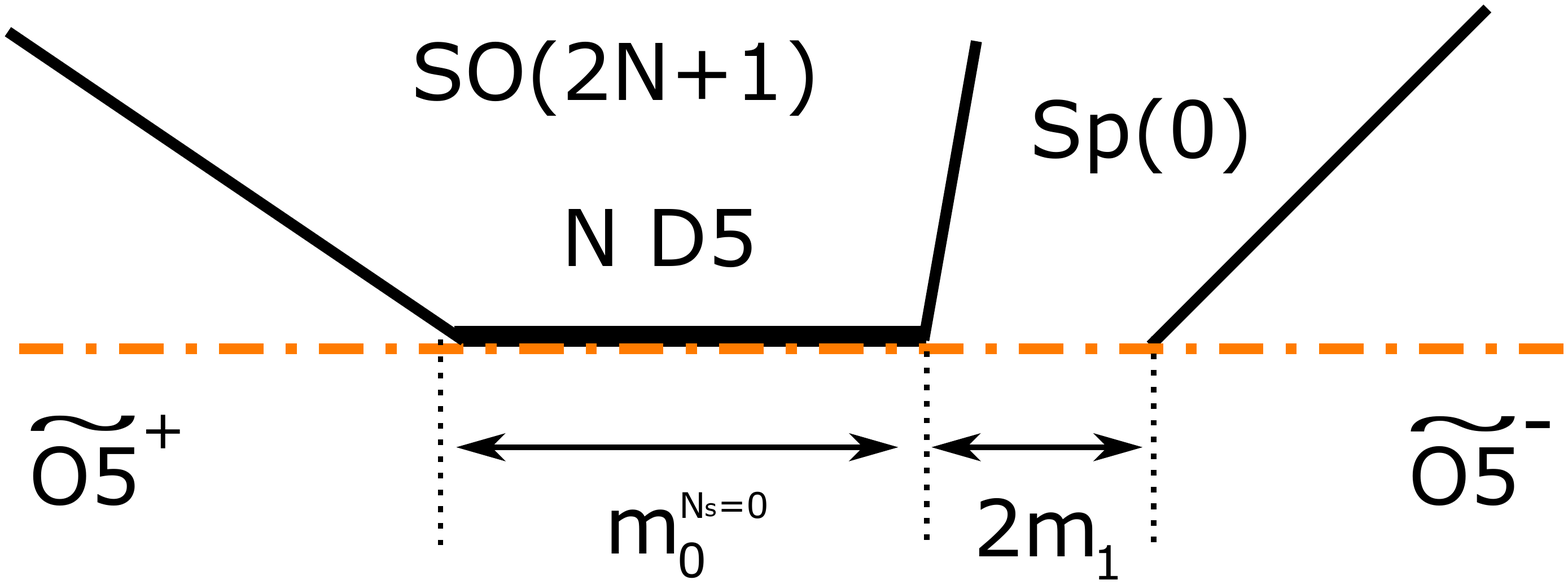}
\caption{The parameters for the $SO(2N+1)$ gauge theory with one spinor.}
\label{fig:SO2N+1w1S}
\end{figure}

For simplicity, we consider the $SO(2N+1)$ gauge theory with one spinor.
We denote the inverse of the squared gauge coupling by $m_0^{N_s=1}$ and the mass of the spinor by $m_1$.
The inverse of the squared gauge coupling after decoupling one spinor is denoted by $m_0^{N_s=0}$.
A 5-brane web for the $SO(2N+1)$ gauge theory with one spinor is depicted in Figure \ref{fig:SO2N+1w1S}.
%%%%%%%%%%%%%%%%%%%%%%%%
%%%%%%%%%%%%%%%%%%%%%%%%
The mass parameter $m_1$ is related to a half of the length between the $(N-2, 1)$ 5-brane and the $(1, 1)$ 5-brane on the $\widetilde{\text{O5}}$-plane in Figure \ref{fig:SO2N+1w1S}, which can be interpreted as a half of the inverse of the squared gauge coupling of the ``$Sp(0)$'' part. Hence a 5-brane web for the pure $SO(2N+1)$ gauge theory after decoupling the spinor can be realized by moving $(1, 1)$ 5-brane at the right hand side to infinitely right.
Then, it is straightforward to read of the inverse of the squared gauge coupling $m_0^{N_s=0}$ of the pure $SO(2N+1)$ gauge theory in a symmetric phase (vanishing Coulomb branch parameter) since it is simply the distance between the $(N-1, -1)$ 5-brane on the left and the $(N-2, 1)$ 5-brane on the $\widetilde{\text{O5}}$-plane in Figure \ref{fig:SO2N+1w1S}.

\begin{figure}
\centering
%\begin{minipage}{0.28\hsize}
%\centering
\subfigure[]{\includegraphics[width=4.5cm]{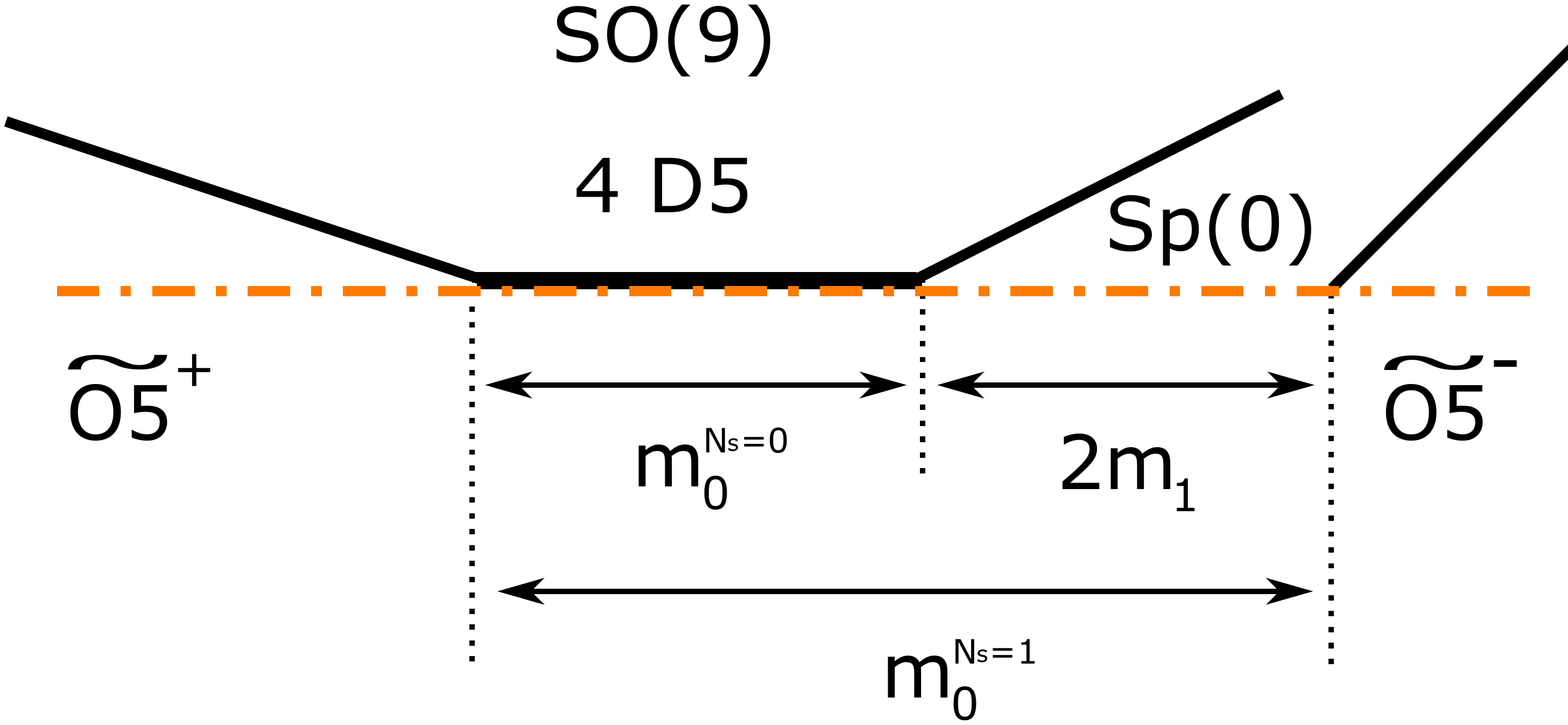}\label{fig:SO9}}
%\caption{Parametrization for SO(9) gauge theory with spinor.}
%\end{minipage}
%\hspace{0.01\hsize}
%\begin{minipage}{0.28\hsize}
%\centering
\subfigure[]{\includegraphics[width=4.5cm]{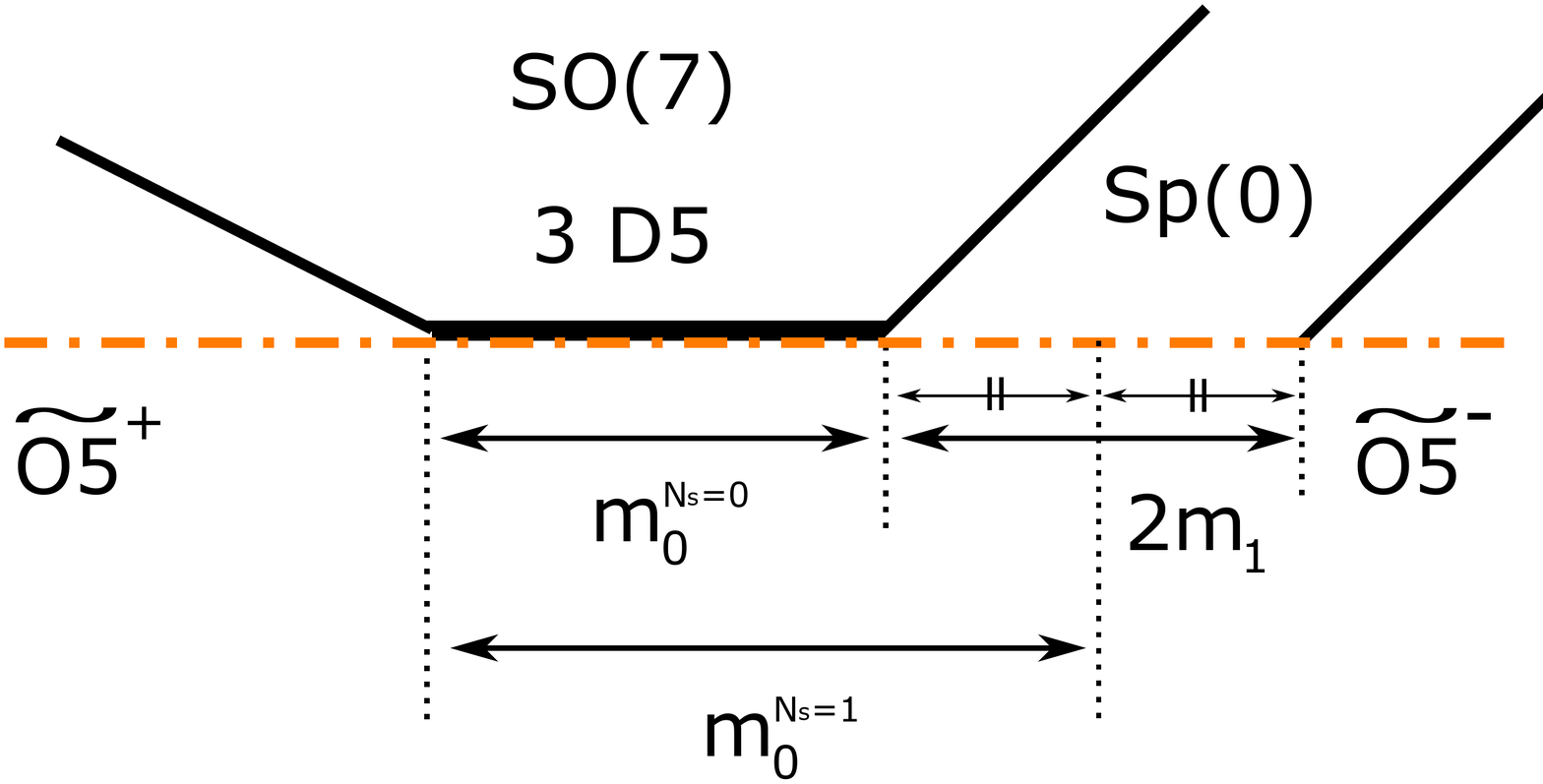}\label{fig:SO7}}
%\caption{Parametrization for SO(7) gauge theory with spinor.}
%\end{minipage}
%\hspace{0.01\hsize}
%\begin{minipage}{0.28\hsize}
%\centering
\subfigure[]{\includegraphics[width=4.5cm]{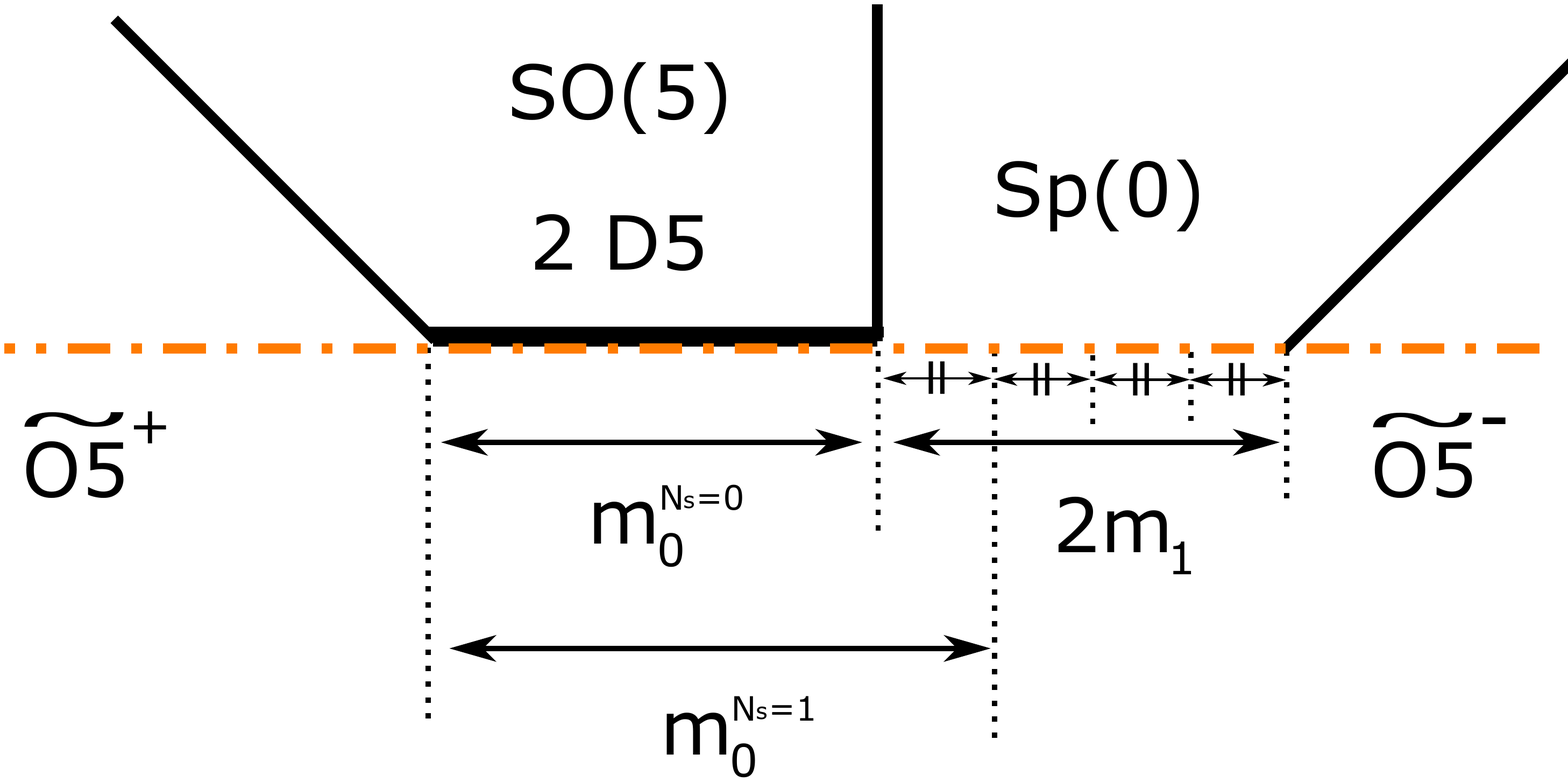}\label{fig:SO5}}
\caption{(a): $m_0^{N_s=1}$ for the $SO(9)$ gauge theory with a spinor. (b): $m_0^{N_s=1}$ for the $SO(7)$ gauge theory with a spinor. (c): $m_0^{N_s=1}$ for the $SO(5)$ gauge theory with a spinor.}
%\end{minipage}
\end{figure}

Since we can identify $m_0^{N_s = 0}$ in the web in Figure \ref{fig:SO2N+1w1S}, the relation \eqref{eq:newm0} tells us how to read off $m_0^{N_s=1}$ from the web. The relation is given by
\begin{align}
m_0^{N_s=1} = m_0^{N_s=0} + 2^{N-3} m_{1}. \label{m0.1spinor}
\end{align}
For example, Eq.~\eqref{m0.1spinor} yields $m_0^{N_s=1} =m_0^{N_s=0} + 2 m_{1}$ for the $SO(9)$ gauge theory. Then the length corresponding to $m_0^{N_s=1} = m_0^{N_s=0} + 2m_1$ is depicted in Figure \ref{fig:SO9}. Namely, we should use the ``outside'' point where the $(1, 1)$ 5-brane ends as in Figure \ref{fig:SO9}.
For the $SO(7)$ gauge theory, the relation becomes $m_0^{N_s=1} = m_0^{N_s=0} + m_{1}$.
The diagram for the $SO(7)$ gauge theory is depicted in Figure \ref{fig:SO7}. Unlike the case for the $SO(9)$ gauge theory, we use the ``middle'' point between the NS5-brane and the $(1, 1)$ 5-brane to define $m_0^{N_s = 1}$ as in Figure \ref{fig:SO7}.
Finally, for the $SO(5)$ gauge theory, Eq.~\eqref{m0.1spinor} gives $m_0^{N_s=1} = m_0^{N_s=0} + \frac{m_{1}}{2}$. Therefore, the ``quarter'' point between NS5-brane and the $(1, 1)$ 5-brane the needs to be used to define $m_0^{N_s=1}$ as in Figure \ref{fig:SO5}.

\subsubsection{Two spinor case}

\begin{figure}
\centering
\includegraphics[width=5cm]{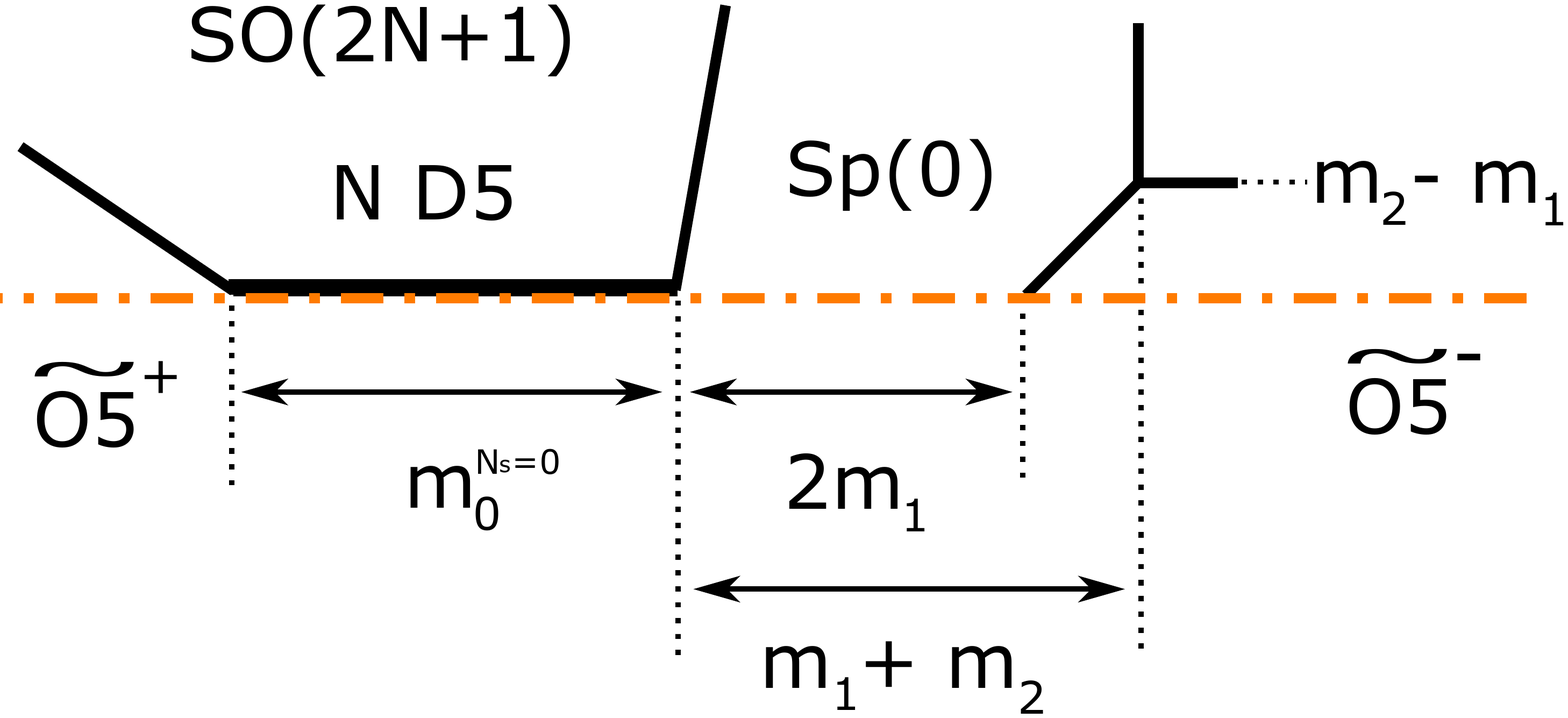}
\caption{The parameters for the $SO(2N+1)$ gauge theory with two spinors.}
\label{fig:SOw2S}
\end{figure}

Next, we consider a case for the $SO(2N+1)$ gauge theory with two spinors.
The two spinors are realized at one side as depicted in Figure \ref{fig:SOw2S}. 
%%%%%%%%%%%%%%%%%%%%%%%
%%%%%%%%%%%%%%%%%%%%%%%
We denote the inverse of the squared gauge coupling for the $SO(2N+1)$ gauge theory with two spinors by $m_0^{N_s=2}$ and we write the masses of the two spinors by $m_1$ and $m_2$.
These parameters can be read off from the web in Figure \ref{fig:SOw2S}.
Note that this parametrization is consistent with decoupling of the spinor matter with mass $m_2$,
which corresponds to sending the height of the flavor D5-brane extending in the left direction to infinitely far, keeping the length between the $(N-2, 1)$ 5-brane and the $(1, 1)$ 5-brane on the $\widetilde{\text{O5}}$-plane. 
It turns out that the inverse of the squared gauge coupling for the ``$Sp(0)$'' part is identified with $m_1+m_2$ rather than $2m_1+2m_2$, which makes the way of reading off $m_0^{N_s=2}$ different from the one spinor cases.

\begin{figure}
\centering
%\begin{minipage}{0.28\hsize}
%\centering
\subfigure[]{\includegraphics[width=4.5cm]{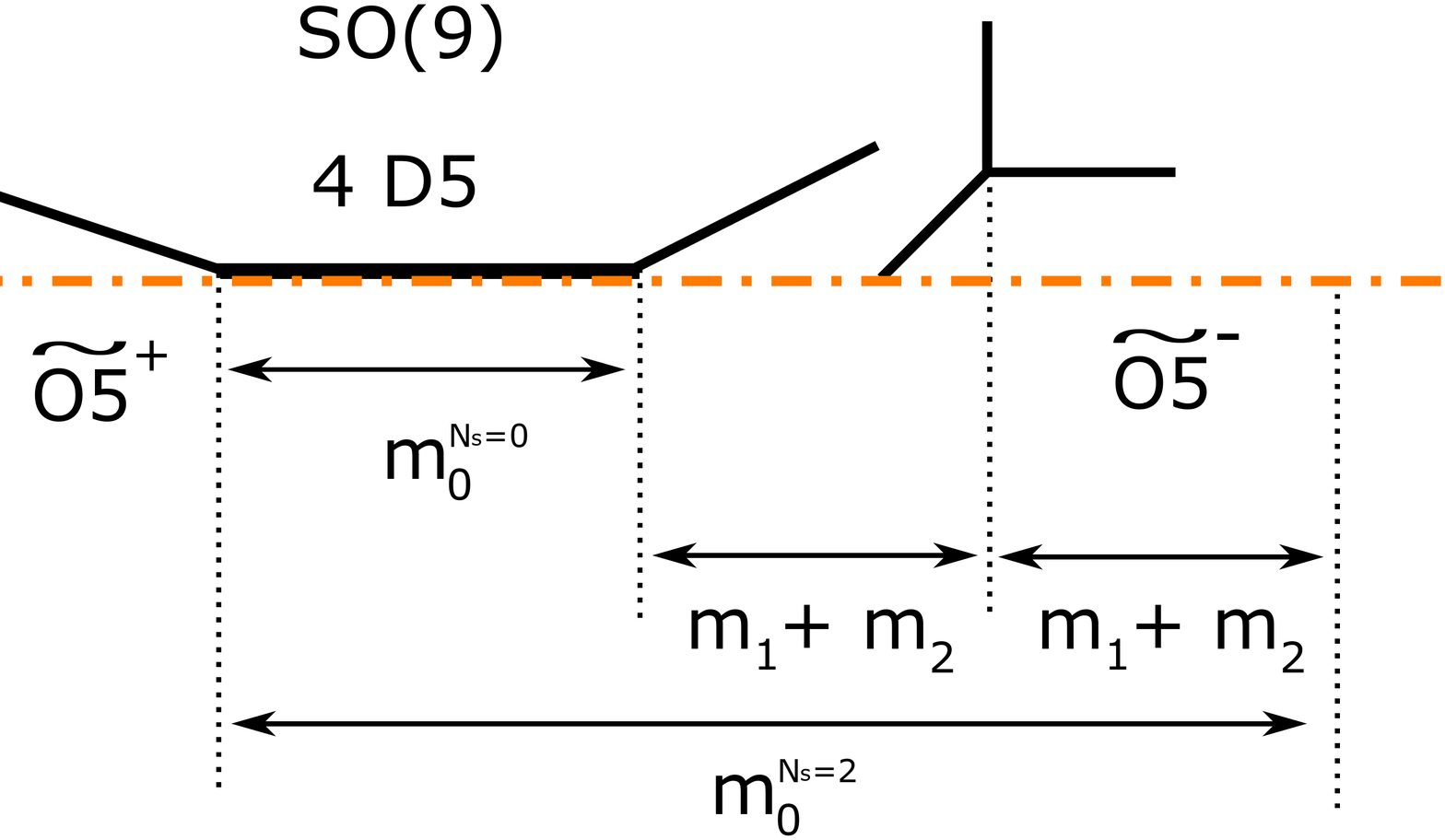}\label{fig:SO9w2S}}
%\caption{Parametrization for SO(9) gauge theory with spinor.}
%\end{minipage}
%\hspace{0.01\hsize}
%\begin{minipage}{0.28\hsize}
%\centering
\subfigure[]{\includegraphics[width=4.5cm]{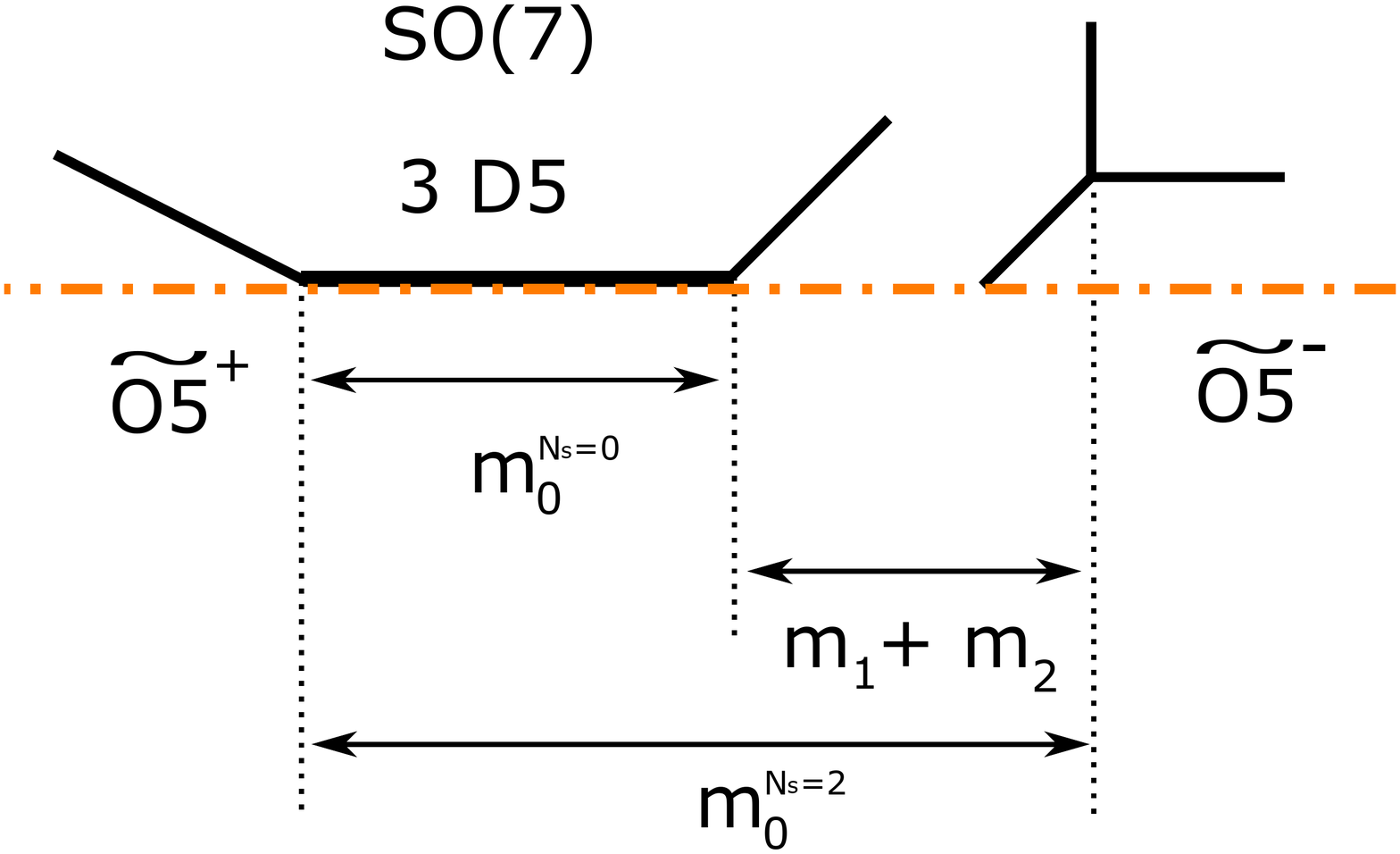}\label{fig:SO7w2S}}
%\caption{Parametrization for SO(7) gauge theory with spinor.}
%\end{minipage}
%\hspace{0.01\hsize}
%\begin{minipage}{0.28\hsize}
%\centering
\subfigure[]{\includegraphics[width=4.5cm]{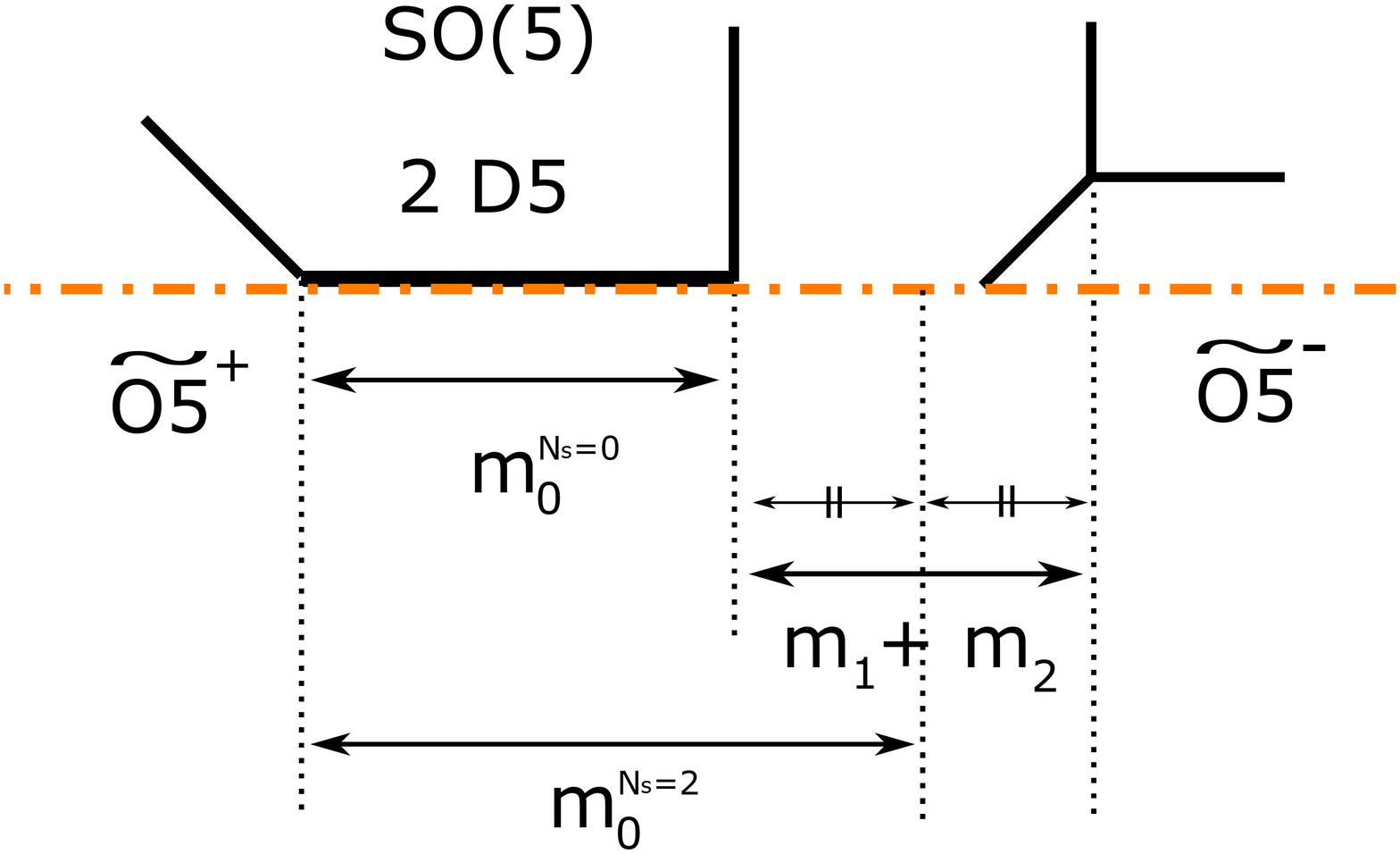}\label{fig:SO5w2S}}
\caption{(a): The parametrization for the $SO(9)$ gauge theory with two spinors. (b): The parametrization for the $SO(7)$ gauge theory with two spinors. (c): The parametrization for the $SO(5)$ gauge theory with two spinors.}
%\end{minipage}
\end{figure}

Using the relation \eqref{eq:newm0} twice gives the relation between $m_0^{N_s=0}$ and $m_0^{N_s=2}$
\begin{align}
m_0^{N_s=2} = m_0^{N_s=0} + 2^{N-3} (m_{1}+m_{2}).\label{m0.2spinors}
\end{align}
Since $m_0^{N_s=0}$ is given as depicted in Figure \ref{fig:SOw2S}, we can determine the length corresponding to $m0^{N_s=2}$ by utilizing the relation \eqref{m0.2spinors}. 
For example, a diagram for the $SO(9)$ gauge theory with two spinors in the symmetric phase is depicted in Figure \ref{fig:SO9w2S}. In this case, the relation \eqref{m0.2spinors} is expressed as $m_0^{N_s=2} =m_0^{N_s=0} + 2 (m_{1}+m_2)$.
Therefore, the length corresponding to $m_0^{N_s=2}$ is given by adding the distance between the $(2, 1)$ 5-brane and the $(1, 1)$ 5-brane on the $\widetilde{\text{O5}}$-plane twice to $m_0^{N_s=0}$ as in Figure \ref{fig:SO9w2S}.
For the $SO(7)$ gauge theory with two spinors, we have $m_0^{N_s=2} = m_0^{N_s=0} + m_{1}+m_2$ from \eqref{m0.2spinors}. Then, to define $m_0^{N_s=2}$, we need to use the ``outside'' point which is given by extrapolating the NS5-brane as depicted in Figure \ref{fig:SO7w2S}.
Finally we consider the $SO(5)$ gauge theory with two spinors. The relation \eqref{m0.2spinors} yields $m_0^{N_s=2} = m_0^{N_s=0} + \frac{m_{1}+m_2}{2}$. 
Therefore, we should use the ``middle'' point between the NS5-brane on the left and the NS5-brane on the right in order to define $m_0^{N_s=2}$. The explicit length corresponding to $m_0^{N_s=2}$ is drawn in Figure \ref{fig:SO5w2S}.

%% file: app2-1.tex
\section{The web diagrams for rank two SCFTs}
\label{sec:allwebs}

In this appendix, we give the 5-brane web diagrams for all the 5d $\mathcal{N}=1$ SCFTs with rank 2 classified in \cite{Jefferson:2018irk}. In Figure \ref{fig:listoffigures},  all the diagrams for such theories are listed. The theory in the box with gray color has 6d UV fixed point. In other words, certain 6d $(1,0)$ SCFTs compactified on $S^1$ give the 5d gauge theories inside the gray box. All the other 5d theories are obtained by RG flows triggered by relevant deformation, which is described by the arrows. 
The gauge theories inside the same box are the dual theories, which have identical SCFTs at their UV fixed point.\footnote{Some quiver theories may have subtlety as their parameter regions may not be directly connected to the UV fixed point as pointed out in \cite{Jefferson:2018irk}.}
In some boxes, non-Lagrangian theories are given, which are specified by the Calabi-Yau geometry. We exclude 5-brane webs which can be obtained by an S-duality transformation or trivial Hanany-Witten transitions, unless they have manifest Lagrangian descriptions.

In each box in Figure \ref{fig:listoffigures}, a Figure number from \ref{Fig:SU3-6F-4} to \ref{Fig:SU3-9} is given, in which the corresponding web diagrams are 
depicted, where the external 5-branes are attached to 7-branes which should be understood as being taken to infinity. Note that, for all the diagrams we list in the appendix, we can move all the 7-branes to infinity by finite steps. 
In this way, a conventional attempt to construct 5-brane for pure $SU(3)_7$ theory is not very useful. 
One may consider a naive diagram for $SU(3)_7$ like the left diagram in Figure \ref{Fig:SU3-6}(b), but it does not lead to a ``finite" diagram like the right diagram in (b). In other words, one cannot move the 7-branes to infinity in finite steps, and hence a 5-brane web for describing the pure $SU(3)_7$ theory may need to rely on unconventional 5-brane constructions. As presented in Figure \ref{Fig:SU3-7}, it is described by introducing an $\widetilde{ON}$-plane.

In some of the 
Figures, web diagrams are classified into several groups labelled by (a), (b), (c), etc. The web diagrams in the same group can be transformed to each other diagrammatically by using flop transitions, Hanany-Witten transitions, $SL(2,\mathbb{Z})$ S-duality transformation as well as reflection.  The ``flop transitions'' include the generalized ones found in \cite{Hayashi:2017btw}.
The web diagrams with different groups do not have such obvious diagrammatical transformations from one to the other even though they correspond to an identical SCFT. If the group label (a), (b), (c), ... are not written, it means that all the diagrams in the Figure can be transformed to each other diagrammatically. 
The RG flows of the SCFTs are understood diagrammatically as a certain limit often combined with the flop transitions. 

In some cases, more than one diagrams are depicted for one gauge theory. In order to emphasize the difference of such diagrams, we put brief explanations in the bracket after the name of the theory. For example, the second diagram in Figure \ref{Fig:SU3-5F-9/2} denoted as $SU(3)_{9/2} + 5 \mathbf{F}$ $(=1\mathbf{AS} + 4\mathbf{F})$. This means that one out of 5 flavors are actually realized as the antisymmetric tensor representation using $\widetilde{ON}^-$-plane while the other 4 flavors are conventionally realized as fundamental representation using D7-branes. 
Although antisymmetric tensor and fundamental are identical for $SU(3)$ gauge group, they nicely characterize the difference of the web diagrams. 

Also, for the SO(5) gauge theories with hypermultiplets in spinor representation, we often obtain several diagrams due to the fact that the spinor representation can be realized both at the left hand side and the right hand side of the $SO(5)$ gauge theory in the web diagram. For example, in Figure \ref{Fig:SU3-4F-5}, the last diagram in the group (b) is denoted as ``$SO(5)+2\mathbf{V}+2\mathbf{S} (1\mathbf{S}+1\mathbf{S})$''. This means that one spinor is realized at the left hand side while the other one is at the right hand side. On the contrary, the diagrams in the group (a) and group (c) are explained as ``$2 \mathbf{S} \,\, \& \,\, \theta =0$'' and ``$2 \mathbf{S} \,\, \& \,\, \theta =\pi$'', respectively. This means that the two spinors are realized at the left side for both cases. The remaining information $\theta=0$ or $\theta=\pi$ denote the discrete theta angle.
Although discrete theta angle is not really defined for the $SO(5)$ theory with spinors, it is used just to briefly specify the configuration at the right hand side of this diagram.
The configuration for ``$\theta=0$'' is the same as the right half of the diagram in Figure \ref{fig:SO52Ftheta0} while ``$\theta=\pi$'' is the one in Figure \ref{fig:SO52FthetaPi}. 

The explanations for the $G_2$ gauge theories are a little bit tricky.
For example, in Figure \ref{Fig:SU3-5F-9/2}, the first diagram is denoted as ``$G_2 + 5 \mathbf{F}$  $(=1\mathbf{S} + 4 \mathbf{V})$''. This actually means that this diagram for the $G_2$ gauge theory with 5 flavor is obtained by Higgsing one hypermultiplet in spinor representation of $SO(7)$ gauge theory with two hypermultiplets in spinor representation and four hypermultiplets in vector representation. Since one out of two hypermultiplets in spinor representation disappear in the process of Higgsing, what remains in the $G_2$ gauge theory is 1 flavor coming from the spinor representation and the 4 flavors coming from vector representation. This information of the origins are briefly explained as ``$1\mathbf{S} + 4 \mathbf{V}$''.  
Analogously, ``$G_2 + 5 \mathbf{F}$  $(=2\mathbf{S} + 3 \mathbf{V})$'' for the diagram below means that the 2 flavor originates from spinor representation while the 3 flavor originates from the vector representation of the parent $SO(7)$ gauge theory. 

All the external $(p,q)$ 5-branes are terminated by $(p,q)$ 7-branes in all the diagrams in the following Figures. In some cases, we can straightforwardly move $(p,q)$ 7-branes to $(p,q)$ direction and obtain the web diagrams consists only of $(p,q)$ 5-branes. 
For other cases, $(p,q)$ 7-branes go across the branch cut created by other $(p,q)$ 7-branes. In this case, we need to properly move $(p,q)$ 7-branes taking into account this monodromy as well as Hanany-Witten transition in order to move all the $(p,q)$ 7-branes to infinity. 
Especially in the diagrams for the 6d theories, it is not possible to move all the $(p,q)$ 7-branes to infinity at least in finite step unless they does not exist from the beginning as in Figure \ref{Fig:SU3-9}.

%% file: app2-2.tex
%\newpage
\begin{figure}[]
\hspace{-0.3cm}
\begin{tikzpicture}
\tikzset{block/.style={draw, align=center, inner sep=0.3mm}};
%
%%%%%%%%%%%%%%%%%%%%
%%% [Going down from ninth level] %%
%%%%%%%%%%%%%%%%%%%%
%
%%%%%%%%%%%%%%%%%%%%
%%%%%%% Ninth Level %%%%%%%
%%%%%%%%%%%%%%%%%%%%
%
\node (SU2-SU2-00)[block] {
\scalebox{0.7}{Fig. \ref{Fig:SU2-SU2-00}} \\ \vspace{-6.5mm}\\
\scalebox{0.6}{$SU(2)_0$-$SU(2)_0$}
};
\node (SU2-SU2-pi0) [block,right = 0.1cm of SU2-SU2-00] {
\scalebox{0.7}{Fig. \ref{Fig:SU2-SU2-pi0}} \\ \vspace{-6.5mm}\\
\scalebox{0.6}{$SU(2)_{\pi}$-$SU(2)_{0}$}
};
\node (SU3-2F-0) [block, right = 0.1cm of SU2-SU2-pi0] {
\scalebox{0.7}{Fig. \ref{Fig:SU3-2F-0}} \\ \vspace{-6.5mm} \\
\scalebox{0.6}{$SU(3)_{0} + 2\mathbf{F}$} 
\vspace{-3mm}\\
\scalebox{0.6}{$SU(2)_{\pi}$-$SU(2)_{\pi}$}
};

\node (SU3-2F-1) [block, right = 0.1cm of SU3-2F-0] {
\scalebox{0.7}{Fig. \ref{Fig:SU3-2F-1}} \\ \vspace{-6.5mm} \\ 
\scalebox{0.6}{$SU(3)_{1}  \!+\!  2\mathbf{F}$}
};
\node (SU3-2F-2) [block, right = 0.1cm of SU3-2F-1] {
\scalebox{0.7}{Fig. \ref{Fig:SU3-2F-2}} \\ \vspace{-6.5mm} \\ 
\scalebox{0.6}{$SU(3)_{2}  \!+\!  2\mathbf{F}$}
};
\node (SU3-2F-3) [block, right = 0.1cm of SU3-2F-2] {
\scalebox{0.7}{Fig. \ref{Fig:SU3-2F-3}} \\ \vspace{-6.5mm} \\ 
\scalebox{0.6}{$SU(3)_{3}  \!+\!  2\mathbf{F}$}
};
\node (SU3-2F-4) [block, right = 0.1cm of SU3-2F-3] {
\scalebox{0.7}{Fig. \ref{Fig:SU3-2F-4}} \\ \vspace{-6.5mm} \\ 
\scalebox{0.6}{$SU(3)_{4}  \!+\!  2\mathbf{F}$}
\hspace{-1mm}\vspace{-3mm} \\
\scalebox{0.6}{$Sp(2)  \!+\!  2\mathbf{F}$}
};
\node (SU3-2F-5) [block, right = 0.1cm of SU3-2F-4] {
\scalebox{0.7}{Fig. \ref{Fig:SU3-2F-5}} \\ \vspace{-6.5mm} \\ 
\scalebox{0.6}{$SU(3)_{5}  \!+\!  2\mathbf{F}$}
\hspace{-1mm} \vspace{-3mm} \\
\scalebox{0.6}{$Sp(2)  \!+\!  1\mathbf{AS}  \!+\!  1\mathbf{F}$}
};
\node (SU3-2F-6) [block, right = 0.1cm of SU3-2F-5] {
\scalebox{0.7}{Fig. \ref{Fig:SU3-2F-6}} \\ \vspace{-6.5mm} \\ 
\scalebox{0.6}{$SU(3)_{6}  \!+\!  2\mathbf{F}$}
\vspace{-3mm}\\
\scalebox{0.6}{$Sp(2)_{\pi} \!+\!  2\mathbf{AS}$}
\hspace{-1mm}\vspace{-3mm}\\
\scalebox{0.6}{$G_2 \!+\! 2 \mathbf{F}$}
};
\node (Sp2-2AS-0) [block, right = 0.1cm of SU3-2F-6] {
\scalebox{0.7}{Fig. \ref{Fig:Sp2-2AS-0}} \\ \vspace{-6.5mm} \\ 
\scalebox{0.6}{$Sp(2)_0  \!+\!  2\mathbf{AS} $}
};
%
%%%%%%%%%%%%%%%%%%%%
%%%%%%% Tenth Level %%%%%%%
%%%%%%%%%%%%%%%%%%%%
%
\node (Sp2-1AS-0) [block, below = 14mm of Sp2-2AS-0] {
\scalebox{0.7}{Fig. \ref{Fig:Sp2-1AS-0}} \\ \vspace{-6.5mm} \\ 
\scalebox{0.6}{$Sp(2)_0  \!+\!  1\mathbf{AS}$}
};
\node (SU3-1F-13/2) [block, left = 1mm of Sp2-1AS-0] {
\scalebox{0.7}{Fig. \ref{Fig:SU3-1F-13/2}} \\ \vspace{-6.5mm} \\ 
\scalebox{0.6}{$SU(3)_{\frac{13}{2}}  \!+\!  1\mathbf{F}$}
\hspace{-1mm}\vspace{-3mm}\\
\scalebox{0.6}{$G_2 \!+\! 1 \mathbf{F}$}
};
\node (SU3-1F-11/2) [block, left = 1mm of SU3-1F-13/2] {
\scalebox{0.7}{Fig. \ref{Fig:SU3-1F-11/2}} \\ \vspace{-6.5mm} \\ 
\scalebox{0.6}{$SU(3)_{\frac{11}{2}}  \!+\!  1\mathbf{F}$}
\hspace{-1mm}\vspace{-3mm}\\
\scalebox{0.6}{$Sp(2)_{\pi} \!+\! 1 \mathbf{AS}$}
};
\node (SU3-1F-9/2) [block, left = 1mm of SU3-1F-11/2] {
\scalebox{0.7}{Fig. \ref{Fig:SU3-1F-9/2}} \\ \vspace{-6.5mm} \\ 
\scalebox{0.6}{$SU(3)_{\frac{9}{2}}  \!+\!  1\mathbf{F}$}
\hspace{-1mm}\vspace{-3mm}\\
\scalebox{0.6}{$Sp(2) \!+\! 1 \mathbf{F}$}
};
\node (SU3-1F-7/2) [block, left = 1mm of SU3-1F-9/2] {
\scalebox{0.7}{Fig. \ref{Fig:SU3-1F-7/2}} \\ \vspace{-6.5mm} \\ 
\scalebox{0.6}{$SU(3)_{\frac{7}{2}}  \!+\!  1\mathbf{F}$}
};
\node (SU3-1F-5/2) [block, left = 1mm of SU3-1F-7/2] {
\scalebox{0.7}{Fig. \ref{Fig:SU3-1F-5/2}} \\ \vspace{-6.5mm} \\ 
\scalebox{0.6}{$SU(3)_{\frac{5}{2}}  \!+\!  1\mathbf{F}$}
};
\node (SU3-1F-3/2) [block, left = 1mm of SU3-1F-5/2] {
\scalebox{0.7}{Fig. \ref{Fig:SU3-1F-3/2}} \\ \vspace{-6.5mm} \\ 
\scalebox{0.6}{$SU(3)_{\frac{3}{2}}  \!+\!  1\mathbf{F}$}
};
\node (SU3-1F-1/2) [block, left = 1mm of SU3-1F-3/2] {
\scalebox{0.7}{Fig. \ref{Fig:SU3-1F-1/2}} \\ \vspace{-6.5mm} \\ 
\scalebox{0.6}{$SU(3)_{\frac{1}{2}}  \!+\!  1\mathbf{F}$}
};
\node (F1-dP2) [block, left = 1mm of SU3-1F-1/2] {
\scalebox{0.7}{Fig. \ref{Fig:F1-dP2}} \\ \vspace{-6.5mm} \\ 
\scalebox{0.6}{``$\mathbb{F}_1 \cup$ dP$_2$''}
};
%
%%%%%%%%%%%%%%%%%%%%
%%%%%% Eleventh Level %%%%%%
%%%%%%%%%%%%%%%%%%%%
%
\node (Sp2-0) [block, below = 15mm of Sp2-1AS-0] {
\scalebox{0.7}{Fig. \ref{Fig:Sp2-0}} \\ \vspace{-6.5mm} \\ 
\scalebox{0.6}{$Sp(2)_0$}
};
\node (SU3-7) [block, left = 4mm of Sp2-0] {
\scalebox{0.7}{Fig. \ref{Fig:SU3-7}} \\ \vspace{-6.5mm} \\ 
\scalebox{0.6}{$SU(3)_{7}$}
\hspace{-1mm}\vspace{-3mm}\\
\scalebox{0.6}{$G_2$}
};
\node (SU3-6) [block, left = 4mm of SU3-7] {
\scalebox{0.7}{Fig. \ref{Fig:SU3-6}} \\ \vspace{-6.5mm} \\ 
\scalebox{0.6}{$SU(3)_{6}$}
};
\node (SU3-5) [block, left = 4mm of SU3-6] {
\scalebox{0.7}{Fig. \ref{Fig:SU3-5}} \\ \vspace{-6.5mm} \\ 
\scalebox{0.6}{$SU(3)_{5}$}
\vspace{-3mm} \\
\scalebox{0.6}{$Sp(2)_{\pi}$}
};
\node (SU3-4) [block, left = 4mm of SU3-5] {
\scalebox{0.7}{Fig. \ref{Fig:SU3-4}} \\ \vspace{-6.5mm} \\ 
\scalebox{0.6}{$SU(3)_{4}$}
};
\node (SU3-3) [block, left = 2mm of SU3-4] {
\scalebox{0.7}{Fig. \ref{Fig:SU3-3}} \\ \vspace{-6.5mm} \\ 
\scalebox{0.6}{$SU(3)_{3}$}
};
\node (SU3-2) [block, left = 2mm of SU3-3] {
\scalebox{0.7}{Fig. \ref{Fig:SU3-2}} \\ \vspace{-6.5mm} \\ 
\scalebox{0.6}{$SU(3)_{2}$}
};
\node (SU3-1) [block, left = 2mm of SU3-2] {
\scalebox{0.7}{Fig. \ref{Fig:SU3-1}} \\ \vspace{-6.5mm} \\ 
\scalebox{0.6}{$SU(3)_{1}$}
};
\node (SU3-0) [block, left = 2mm of SU3-1] {
\scalebox{0.7}{Fig. \ref{Fig:SU3-0}} \\ \vspace{-6.5mm} \\ 
\scalebox{0.6}{$SU(3)_{0}$}
};
\node (F2-dP1) [block, left = 4mm of SU3-0] {
\scalebox{0.7}{Fig. \ref{Fig:F2-dP1}} \\ \vspace{-6.5mm} \\ 
\scalebox{0.6}{``$\mathbb{F}_2 \cup$ dP$_1$''}
};
%
%%%%%%%%%%%%%%%%%%%%
%%%%%% Twelfth Level %%%%%%
%%%%%%%%%%%%%%%%%%%%
%
%
\node (F3-P2) [block, below = 3mm of F2-dP1] {
\scalebox{0.7}{Fig. \ref{Fig:F3-P2}} \\ \vspace{-6.5mm} \\ 
\scalebox{0.6}{``$\mathbb{F}_3 \cup \mathbb{P}^2 $''}
};
\node (F6-P2) [block, below = 3mm of Sp2-0] {
\scalebox{0.7}{Fig. \ref{Fig:F6-P2}} \\ \vspace{-6.5mm} \\ 
\scalebox{0.6}{``$\mathbb{F}_6 \cup \mathbb{P}^2 $''}
};
%%%%%%%%%%%%%%%%%%%%
%%  Going up from eighth level  %%%
%%%%%%%%%%%%%%%%%%%%
%
%%%%%%%%%%%%%%%%%%%%
%%%%%% Eighth level %%%%%%%
%%%%%%%%%%%%%%%%%%%%
%
\node (Sp2-3AS-0) [block, fill=gray!15, above = 9mm of Sp2-2AS-0] {
\scalebox{0.7}{Fig. \ref{Fig:Sp2-3AS-0}} \\ \vspace{-6.5mm} \\ 
\scalebox{0.6}{$Sp(2)_{0} \!+\! 3\mathbf{AS}$}
};
\node (SU3-3F-11/2) [block, left = 1mm of Sp2-3AS-0] {
\scalebox{0.7}{Fig. \ref{Fig:SU3-3F-11/2}} \\ \vspace{-6.5mm} \\ 
\scalebox{0.6}{$SU(3)_{\frac{11}{2}} \!+\! 3\mathbf{F}$}
\hspace{-1mm} \vspace{-3mm}\\
\scalebox{0.6}{$Sp(2) \!+\! 2 \mathbf{AS} \!+\!1 \mathbf{F}$}
\hspace{-2mm} \vspace{-3mm} \\
\scalebox{0.6}{$G_2 \!+\! 3 \mathbf{F}$}
};
\node (SU3-3F-9/2) [block, left = 1mm of SU3-3F-11/2] {
\scalebox{0.7}{Fig. \ref{Fig:SU3-3F-9/2}} \\ \vspace{-6.5mm} \\ 
\scalebox{0.6}{$SU(3)_{\frac{9}{2}} \!+\! 3\mathbf{F}$}
\hspace{-1mm} \vspace{-3mm}\\
\scalebox{0.6}{$Sp(2) \!+\! 1 \mathbf{AS} \!+\!2 \mathbf{F}$}
};
\node (SU3-3F-7/2) [block, left = 1mm of SU3-3F-9/2] {
\scalebox{0.7}{Fig. \ref{Fig:SU3-3F-7/2}} \\ \vspace{-6.5mm} \\ 
\scalebox{0.6}{$SU(3)_{\frac{7}{2}} \!+\! 3\mathbf{F}$}
\hspace{-1mm} \vspace{-3mm}\\
\scalebox{0.6}{$Sp(2) \!+\! 3 \mathbf{F}$}
};
\node (SU3-3F-5/2) [block, left = 1mm of SU3-3F-7/2] {
\scalebox{0.7}{Fig. \ref{Fig:SU3-3F-5/2}} \\ \vspace{-6.5mm} \\ 
\scalebox{0.6}{$SU(3)_{\frac{5}{2}} \!+\! 3\mathbf{F}$}
};
\node (SU3-3F-3/2) [block, left = 1mm of SU3-3F-5/2] {
\scalebox{0.7}{Fig. \ref{Fig:SU3-3F-3/2}} \\ \vspace{-6.5mm} \\ 
\scalebox{0.6}{$SU(3)_{\frac{3}{2}} \!+\! 3\mathbf{F}$}
};
\node (SU3-3F-1/2) [block, left = 1mm of SU3-3F-3/2] {
\scalebox{0.7}{Fig. \ref{Fig:SU3-3F-1/2}} \\ \vspace{-6.5mm} \\ 
\scalebox{0.6}{$SU(3)_{\frac{1}{2}} \!+\! 3\mathbf{F}$}
\hspace{-1mm} \vspace{-3mm}\\
\scalebox{0.6}{$SU(2)_{\pi}$-$SU(2)$-$[1]$}
};
\node (SU2-SU2-1F-0) [block,  left = 5mm of SU3-3F-1/2] {
\scalebox{0.7}{Fig. \ref{Fig:SU2-SU2-1F-0}} \\ \vspace{-6.5mm} \\ 
\scalebox{0.6}{$[1]$-$SU(2)$-$SU(2)_0$}
};
%
%%%%%%%%%%%%%%%%%%%%
%%%%%% Seventh level %%%%%%%
%%%%%%%%%%%%%%%%%%%%
%
\node (SU3-4F-5) [block, above = 3mm of SU3-3F-11/2] {
\scalebox{0.7}{Fig. \ref{Fig:SU3-4F-5}} \\ \vspace{-6.5mm} \\ 
\scalebox{0.6}{$SU(3)_{5} \!+\! 4\mathbf{F}$}
\hspace{-1mm} \vspace{-3mm}\\
\scalebox{0.6}{$Sp(2) \!+\! 2 \mathbf{AS} \!+\!2 \mathbf{F}$}
\hspace{-2mm} \vspace{-3mm} \\
\scalebox{0.6}{$G_2 \!+\! 4 \mathbf{F}$}
};
\node (SU3-4F-4) [block, left = 1mm of SU3-4F-5] {
\scalebox{0.7}{Fig. \ref{Fig:SU3-4F-4}} \\ \vspace{-6.5mm} \\ 
\scalebox{0.6}{$SU(3)_{4} \!+\! 4\mathbf{F}$}
\hspace{-1mm} \vspace{-3mm}\\
\scalebox{0.6}{$Sp(2) \!+\! 1 \mathbf{AS} \!+\!3 \mathbf{F}$}
};
\node (SU3-4F-3) [block, left = 1mm of SU3-4F-4] {
\scalebox{0.7}{Fig. \ref{Fig:SU3-4F-3}} \\ \vspace{-6.5mm} \\ 
\scalebox{0.6}{$SU(3)_{3} \!+\! 4\mathbf{F}$}
\hspace{-1mm} \vspace{-3mm}\\
\scalebox{0.6}{$Sp(2) \!+\! 4 \mathbf{F}$}
};
\node (SU3-4F-2) [block, left = 1mm of SU3-4F-3] {
\scalebox{0.7}{Fig. \ref{Fig:SU3-4F-2}} \\ \vspace{-6.5mm} \\ 
\scalebox{0.6}{$SU(3)_{2} \!+\! 4\mathbf{F}$}
};
\node (SU3-4F-1) [block, left = 1mm of SU3-4F-2] {
\scalebox{0.7}{Fig. \ref{Fig:SU3-4F-1}} \\ \vspace{-6.5mm} \\ 
\scalebox{0.6}{$SU(3)_{1} \!+\! 4\mathbf{F}$}
\hspace{-1mm} \vspace{-3mm}\\
\scalebox{0.6}{$SU(2)_{\pi}$-$SU(2)$-$[2]$}
};
\node (SU3-4F-0) [block, left = 1mm of SU3-4F-1] {
\scalebox{0.7}{Fig. \ref{Fig:SU3-4F-0}} \\ \vspace{-6.5mm} \\ 
\scalebox{0.6}{$SU(3)_{0} \!+\! 4\mathbf{F}$}
\hspace{-1mm} \vspace{-3mm}\\
\scalebox{0.6}{$[1]$-$SU(2)$-$SU(2)$-$[1]$}
};
\node (SU2-SU2-2F-0) [block, left = 1mm of SU3-4F-0] {
\scalebox{0.7}{Fig. \ref{Fig:SU2-SU2-2F-0}} \\ \vspace{-6.5mm} \\ 
\scalebox{0.6}{$[2]$-$SU(2)$-$SU(2)_0$}
};
%
%%%%%%%%%%%%%%%%%%%%
%%%%%%% Sixth level %%%%%%%
%%%%%%%%%%%%%%%%%%%%
%
\node (SU3-5F-9/2) [block, above = 3mm of SU3-4F-5] {
\scalebox{0.7}{Fig. \ref{Fig:SU3-5F-9/2}} \\ \vspace{-6.5mm} \\ 
\scalebox{0.6}{$SU(3)_{\frac{9}{2}} \!+\! 5\mathbf{F}$}
\hspace{-1mm} \vspace{-3mm}\\
\scalebox{0.6}{$Sp(2) \!+\! 2 \mathbf{AS} \!+\!3 \mathbf{F}$}
\hspace{-2mm} \vspace{-3mm} \\
\scalebox{0.6}{$G_2 \!+\! 5 \mathbf{F}$}
};
\node (SU3-5F-7/2) [block, left = 1mm of SU3-5F-9/2] {
\scalebox{0.7}{Fig. \ref{Fig:SU3-5F-7/2}} \\ \vspace{-6.5mm} \\ 
\scalebox{0.6}{$SU(3)_{\frac{7}{2}} \!+\! 5\mathbf{F}$}
\hspace{-1mm} \vspace{-3mm}\\
\scalebox{0.6}{$Sp(2) \!+\! 1 \mathbf{AS} \!+\!4 \mathbf{F}$}
};
\node (SU3-5F-5/2) [block, left = 2mm of SU3-5F-7/2] {
\scalebox{0.7}{Fig. \ref{Fig:SU3-5F-5/2}} \\ \vspace{-6.5mm} \\ 
\scalebox{0.6}{$SU(3)_{\frac{5}{2}} \!+\! 5\mathbf{F}$}
\hspace{-1mm} \vspace{-3mm}\\
\scalebox{0.6}{$Sp(2) \!+\! 5 \mathbf{F}$}
};
\node (SU3-5F-3/2) [block, left = 2mm of SU3-5F-5/2] {
\scalebox{0.7}{Fig. \ref{Fig:SU3-5F-3/2}} \\ \vspace{-6.5mm} \\ 
\scalebox{0.6}{$SU(3)_{\frac{3}{2}} \!+\! 5\mathbf{F}$}
\hspace{-1mm} \vspace{-3mm}\\
\scalebox{0.6}{$SU(2)_{\pi}$-$SU(2)$-$[3]$}
};
\node (SU3-5F-1/2) [block, left = 1mm of SU3-5F-3/2] {
\scalebox{0.7}{Fig. \ref{Fig:SU3-5F-1/2}} \\ \vspace{-6.5mm} \\ 
\scalebox{0.6}{$SU(3)_{\frac{1}{2}} \!+\! 5\mathbf{F}$}
\hspace{-1mm} \vspace{-3mm}\\
\scalebox{0.6}{$[1]$-$SU(2)$-$SU(2)$-$[2]$}
};
\node (SU2-SU2-3F-0) [block, left = 1mm of SU3-5F-1/2] {
\scalebox{0.7}{Fig. \ref{Fig:SU2-SU2-3F-0}} \\ \vspace{-6.5mm} \\ 
\scalebox{0.6}{$[3]$-$SU(2)$-$SU(2)_0$}
};
%
%
%%%%%%%%%%%%%%%%%%%%
%%%%%%% Fifth level %%%%%%%
%%%%%%%%%%%%%%%%%%%%
%
\node (SU3-6F-4) [block, fill=gray!15, above = 3mm of SU3-5F-9/2] {
\scalebox{0.7}{Fig. \ref{Fig:SU3-6F-4}} \\ \vspace{-6.5mm} \\ 
\scalebox{0.6}{$SU(3)_{4} \!+\! 6\mathbf{F}$}
\hspace{-1mm} \vspace{-3mm}\\
\scalebox{0.6}{$Sp(2) \!+\! 2 \mathbf{AS} \!+\!4 \mathbf{F}$}
\hspace{-2mm} \vspace{-3mm} \\
\scalebox{0.6}{$G_2 \!+\! 6 \mathbf{F}$}
};
\node (SU3-6F-3) [block, left = 1mm of SU3-6F-4] {
\scalebox{0.7}{Fig. \ref{Fig:SU3-6F-3}} \\ \vspace{-6.5mm} \\ 
\scalebox{0.6}{$SU(3)_{3} \!+\! 6\mathbf{F}$}
\hspace{-1mm} \vspace{-3mm}\\
\scalebox{0.6}{$Sp(2) \!+\! 1 \mathbf{AS} \!+\!5 \mathbf{F}$}
};
\node (SU3-6F-2) [block, left = 1mm of SU3-6F-3] {
\scalebox{0.7}{Fig. \ref{Fig:SU3-6F-2}} \\ \vspace{-6.5mm} \\ 
\scalebox{0.6}{$SU(3)_{2} \!+\! 6\mathbf{F}$}
\hspace{-1mm} \vspace{-3mm}\\
\scalebox{0.6}{$Sp(2) \!+\! 6 \mathbf{F}$}
\hspace{-1mm} \vspace{-3mm}\\
\scalebox{0.6}{$SU(2)_{\pi}$-$SU(2)$-$[4]$}
};
\node (SU3-6F-1) [block, left = 2mm of SU3-6F-2] {
\scalebox{0.7}{Fig. \ref{Fig:SU3-6F-1}} \\ \vspace{-6.5mm} \\ 
\scalebox{0.6}{$SU(3)_{1} \!+\! 6\mathbf{F}$}
\hspace{-1mm} \vspace{-3mm}\\
\scalebox{0.6}{$[1]$-$SU(2)$-$SU(2)$-$[3]$}
};
\node (SU3-6F-0) [block, left = 4mm of SU3-6F-1] {
\scalebox{0.7}{Fig. \ref{Fig:SU3-6F-0}} \\ \vspace{-6.5mm} \\ 
\scalebox{0.6}{$SU(3)_{0} \!+\! 6\mathbf{F}$}
\hspace{-1mm} \vspace{-3mm}\\
\scalebox{0.6}{$[2]$-$SU(2)$-$SU(2)$-$[2]$}
};
\node (SU2-SU2-4F-0) [block, left = 1mm of SU3-6F-0] {
\scalebox{0.7}{Fig. \ref{Fig:SU2-SU2-4F-0}} \\ \vspace{-6.5mm} \\ 
\scalebox{0.6}{$[4]$-$SU(2)$-$SU(2)_0$}
};
%
%
%%%%%%%%%%%%%%%%%%%%
%%%%%%% Fourth level %%%%%%%
%%%%%%%%%%%%%%%%%%%%
%
\node (SU3-7F-5/2) [block, above = 5mm of SU3-6F-3] {
\scalebox{0.7}{Fig. \ref{Fig:SU3-7F-5/2}} \\ \vspace{-6.5mm} \\ 
\scalebox{0.6}{$SU(3)_{\frac{5}{2}} \!+\! 7\mathbf{F}$}
\hspace{-1mm} \vspace{-3mm}\\
\scalebox{0.6}{$Sp(2) \!+\! 1 \mathbf{AS} \!+\!6 \mathbf{F}$}
\hspace{-1mm} \vspace{-3mm}\\
\scalebox{0.6}{$SU(2)_{\pi}$-$SU(2)$-$[5]$}
};
\node (SU3-7F-3/2) [block, left = 1mm of SU3-7F-5/2] {
\scalebox{0.7}{Fig. \ref{Fig:SU3-7F-3/2}} \\ \vspace{-6.5mm} \\ 
\scalebox{0.6}{$SU(3)_{\frac{3}{2}} \!+\! 7\mathbf{F}$}
\hspace{-1mm} \vspace{-3mm}\\
\scalebox{0.6}{$Sp(2) \!+\! 7 \mathbf{F}$}
\hspace{-1mm} \vspace{-3mm}\\
\scalebox{0.6}{$[1]$-$SU(2)$-$SU(2)$-$[4]$}
};
\node (SU3-7F-1/2) [block, left = 1mm of SU3-7F-3/2] {
\scalebox{0.7}{Fig. \ref{Fig:SU3-7F-1/2}} \\ \vspace{-6.5mm} \\ 
\scalebox{0.6}{$SU(3)_{\frac{1}{2}} \!+\! 7\mathbf{F}$}
\hspace{-1mm} \vspace{-3mm}\\
\scalebox{0.6}{$[2]$-$SU(2)$-$SU(2)$-$[3]$}
};
\node (SU2-SU2-5F-0) [block, left = 1mm of SU3-7F-1/2] {
\scalebox{0.7}{Fig. \ref{Fig:SU2-SU2-5F-0}} \\ \vspace{-6.5mm} \\ 
\scalebox{0.6}{$SU(2)_0$-$SU(2)$-$[5]$}
};
%
%%%%%%%%%%%%%%%%%%%%
%%%%%%% Third level %%%%%%%
%%%%%%%%%%%%%%%%%%%%
%
\node (SU3-8F-2) [block, above = 5mm of SU3-7F-5/2] {
\scalebox{0.7}{Fig. \ref{Fig:SU3-8F-2}} \\ \vspace{-6.5mm} \\ 
\scalebox{0.6}{$SU(3)_{2} \!+\! 8\mathbf{F}$}
\hspace{-1mm} \vspace{-3mm}\\
\scalebox{0.6}{$Sp(2) \!+\! 1 \mathbf{AS} \!+\!7 \mathbf{F}$}
\hspace{-1mm} \vspace{-3mm}\\
\scalebox{0.6}{$[1]$-$SU(2)$-$SU(2)$-$[5]$}
};
\node (SU3-8F-1) [block, left = 1mm of SU3-8F-2] {
\scalebox{0.7}{Fig. \ref{Fig:SU3-8F-1}} \\ \vspace{-6.5mm} \\ 
\scalebox{0.6}{$SU(3)_{1} \!+\! 8\mathbf{F}$}
\hspace{-1mm} \vspace{-3mm}\\
\scalebox{0.6}{$Sp(2) \!+\! 8 \mathbf{F}$}
\hspace{-1mm} \vspace{-3mm}\\
\scalebox{0.6}{$[2]$-$SU(2)$-$SU(2)$-$[4]$}
};
\node (SU3-8F-0) [block, left = 1mm of SU3-8F-1] {
\scalebox{0.7}{Fig. \ref{Fig:SU3-8F-0}} \\ \vspace{-6.5mm} \\ 
\scalebox{0.6}{$SU(3)_{0} \!+\! 8\mathbf{F}$}
\hspace{-1mm} \vspace{-3mm}\\
%\scalebox{0.6}{$Sp(2) \!+\! 8 \mathbf{F}$}
%\hspace{-1mm} \vspace{-3mm}\\
\scalebox{0.6}{$[3]$-$SU(2)$-$SU(2)$-$[3]$}
};
%
%%%%%%%%%%%%%%%%%%%%
%%%%%%% Second level %%%%%%%
%%%%%%%%%%%%%%%%%%%%
%
\node (SU3-9F-3/2) [block, fill=gray!15, above = 5mm of SU3-8F-2] {
\scalebox{0.7}{Fig. \ref{Fig:SU3-9F-3/2}} \\ \vspace{-6.5mm} \\ 
\scalebox{0.6}{$SU(3)_{\frac{3}{2}} \!+\! 9\mathbf{F}$}
\hspace{-1mm} \vspace{-3mm}\\
\scalebox{0.6}{$Sp(2) \!+\! 1 \mathbf{AS} \!+\!8 \mathbf{F}$}
\hspace{-1mm} \vspace{-3mm}\\
\scalebox{0.6}{$[2]$-$SU(2)$-$SU(2)$-$[5]$}
};
\node (SU3-9F-1/2) [block, left = 1mm of SU3-9F-3/2] {
\scalebox{0.7}{Fig. \ref{Fig:SU3-9F-1/2}} \\ \vspace{-6.5mm} \\ 
\scalebox{0.6}{$SU(3)_{\frac{1}{2}} \!+\! 9\mathbf{F}$}
\hspace{-1mm} \vspace{-3mm}\\
\scalebox{0.6}{$Sp(2) \!+\!9 \mathbf{F}$}
\hspace{-1mm} \vspace{-3mm}\\
\scalebox{0.6}{$[3]$-$SU(2)$-$SU(2)$-$[4]$}
};
%
%%%%%%%%%%%%%%%%%%%%
%%%%%%% First level %%%%%%%
%%%%%%%%%%%%%%%%%%%%
%
\node (SU3-10F-0) [block, fill=gray!15, above = 3mm of SU3-9F-1/2] {
\scalebox{0.7}{Fig. \ref{Fig:SU3-10F-0}} \\ \vspace{-6.5mm}\\ 
\scalebox{0.6}{$SU(3)_{0} \!+\! 10\mathbf{F}$}
\hspace{-1mm} \vspace{-3mm}\\
\scalebox{0.6}{$Sp(2) \!+\!10 \mathbf{F}$}
\hspace{-1mm} \vspace{-3mm}\\
\scalebox{0.6}{$[4]$-$SU(2)$-$SU(2)$-$[4]$}
};
%
%
%
%%%%%%%%%%%%%%%%%%%%%%%
%%%%%%% Isolated SCFTs  %%%%%%%
%%%%%%%%%%%%%%%%%%%%%%%
%
\node (SU3-1Sym) [block, left = 2mm of F1-dP2] {
\scalebox{0.7}{Fig. \ref{Fig:SU3-1Sym}} \\ \vspace{-6.5mm}\\ 
\scalebox{0.6}{$SU(3)_{\frac{1}{2}} \!+\! 1\mathbf{Sym}$}
};
\node (SU3-1Sym-1F) [block, fill=gray!15, above = 3mm of SU3-1Sym] {
\scalebox{0.7}{Fig. \ref{Fig:SU3-1Sym-1F}} \\ \vspace{-6.5mm}\\ 
\scalebox{0.6}{$SU(3)_{0} \!+\! 1\mathbf{Sym} \!+\! 1\mathbf{F}$}
};
\node (SU3-9) [block, fill=gray!15, left= 50mm of SU3-10F-0] {
\scalebox{0.7}{Fig. \ref{Fig:SU3-9}} \\ \vspace{-6.5mm}\\ 
\scalebox{0.6}{$SU(3)_{9}$}
};
%
%
%%%%%%%%%%     Added in v2  %%%%%%%%%%%%%%%%%%
\node (SU3-1Sym-3/2) [block, fill=gray!15, left = 2mm of SU3-1Sym] {
\scalebox{0.7}{Fig. \ref{Fig:SU3-1Sym-3/2}} \\ \vspace{-6.5mm}\\ 
\scalebox{0.6}{$SU(3)_{\frac{3}{2}} \!+\! 1\mathbf{Sym} $}
};
%%%%%%%%%%%%%%%%%%%%%%%%%%%%%%%%%%%%%
%
%%%%%%%%%%%%%%%%%%%%
%%%%%%%%%%%%%%%%%%%%
%%%%%%%%%%%%%%%%%%%%
% %%%%%   Arrows %%%%%%%%%
%%%%%%%%%%%%%%%%%%%%
%%%%%%%%%%%%%%%%%%%%
%%%%%%%%%%%%%%%%%%%%
%
%
%
% First level -- Second Level 
%
\draw[-latex] (SU3-10F-0)--(SU3-9F-1/2);
%
% Second level -- Third Level 
%
%
\draw[-latex] (SU3-9F-1/2)--(SU3-8F-0);
\draw[-latex] (SU3-9F-1/2)--(SU3-8F-1);
\draw[-latex] (SU3-9F-3/2)--(SU3-8F-1);
\draw[-latex] (SU3-9F-3/2)--(SU3-8F-2);
%
%
% Third level -- Fourth Level 
%
%
%
\node(SU3-8F-2b)[below = -1mm of SU3-8F-2]{};
\node(SU2-SU2-5F-0a)[above = -1mm of SU2-SU2-5F-0]{};
\draw[-latex] (SU3-8F-0)--(SU3-7F-1/2);
\draw[-latex] (SU3-8F-1)--(SU3-7F-1/2);
\draw[-latex] (SU3-8F-1)--(SU3-7F-3/2);
\draw[-latex] (SU3-8F-2)--(SU3-7F-3/2);
\draw[-latex] (SU3-8F-2)--(SU3-7F-5/2);
\draw[-latex] (SU3-8F-2b)--(SU2-SU2-5F-0a);
%
%
% Fourth level -- Fifth Level 
%
%
%
\node(SU3-7F-3/2b)[below left= -4mm of SU3-7F-3/2]{};
\node(SU2-SU2-4F-0a)[above = -1mm of SU2-SU2-4F-0]{};
\draw[-latex] (SU2-SU2-5F-0)--(SU2-SU2-4F-0);
\draw[-latex] (SU3-7F-1/2)--(SU3-6F-0);
\draw[-latex] (SU3-7F-1/2)--(SU3-6F-1);
\draw[-latex] (SU3-7F-3/2)--(SU3-6F-1);
\draw[-latex] (SU3-7F-3/2)--(SU3-6F-2);
\draw[-latex] (SU3-7F-5/2)--(SU3-6F-2);
\draw[-latex] (SU3-7F-5/2)--(SU3-6F-3);
\draw[-latex] (SU3-7F-3/2b)--(SU2-SU2-4F-0a);
%
%
% Fifth level -- Six Level 
%
%
%
\draw[-latex] (SU2-SU2-4F-0)--(SU2-SU2-3F-0);
\draw[-latex] (SU3-6F-0)--(SU3-5F-1/2);
\draw[-latex] (SU3-6F-1)--(SU3-5F-1/2);
\draw[-latex] (SU3-6F-1)--(SU3-5F-3/2);
\draw[-latex] (SU3-6F-2)--(SU3-5F-3/2);
\draw[-latex] (SU3-6F-2)--(SU3-5F-5/2);
\draw[-latex] (SU3-6F-3)--(SU3-5F-5/2);
\draw[-latex] (SU3-6F-3)--(SU3-5F-7/2);
\draw[-latex] (SU3-6F-4)--(SU3-5F-7/2);
\draw[-latex] (SU3-6F-4)--(SU3-5F-9/2);
\node(SU3-6F-1b)[below left= -4mm of SU3-6F-1]{};
\node(SU2-SU2-3F-0a)[above = -1mm of SU2-SU2-3F-0]{};
\draw[-latex] (SU3-6F-1b)--(SU2-SU2-3F-0a);
%
%
%
% Sixth level -- Seventh Level 
%
%
%
\draw[-latex] (SU2-SU2-3F-0)--(SU2-SU2-2F-0);
\draw[-latex] (SU3-5F-1/2)--(SU2-SU2-2F-0);
\draw[-latex] (SU3-5F-1/2)--(SU3-4F-0);
\draw[-latex] (SU3-5F-1/2)--(SU3-4F-1);
\draw[-latex] (SU3-5F-3/2)--(SU3-4F-1);
\draw[-latex] (SU3-5F-3/2)--(SU3-4F-2);
\draw[-latex] (SU3-5F-5/2)--(SU3-4F-2);
\draw[-latex] (SU3-5F-5/2)--(SU3-4F-3);
\draw[-latex] (SU3-5F-7/2)--(SU3-4F-3);
\draw[-latex] (SU3-5F-7/2)--(SU3-4F-4);
\draw[-latex] (SU3-5F-9/2)--(SU3-4F-4);
\draw[-latex] (SU3-5F-9/2)--(SU3-4F-5);
%
%
% Seventh level -- Eighth Level 
%
%
%
\draw[-latex] (SU2-SU2-2F-0)--(SU2-SU2-1F-0);
\draw[-latex] (SU3-4F-0)--(SU2-SU2-1F-0);
\draw[-latex] (SU3-4F-0)--(SU3-3F-1/2);
\draw[-latex] (SU3-4F-1)--(SU3-3F-1/2);
\draw[-latex] (SU3-4F-1)--(SU3-3F-3/2);
\draw[-latex] (SU3-4F-2)--(SU3-3F-3/2);
\draw[-latex] (SU3-4F-2)--(SU3-3F-5/2);
\draw[-latex] (SU3-4F-3)--(SU3-3F-5/2);
\draw[-latex] (SU3-4F-3)--(SU3-3F-7/2);
\draw[-latex] (SU3-4F-4)--(SU3-3F-7/2);
\draw[-latex] (SU3-4F-4)--(SU3-3F-9/2);
\draw[-latex] (SU3-4F-5)--(SU3-3F-9/2);
\draw[-latex] (SU3-4F-5)--(SU3-3F-11/2);
%
%
%
% Eighth level -- Ninth Level 
%
%
%
\draw[-latex] (SU2-SU2-1F-0)--(SU2-SU2-00);
\draw[-latex] (SU2-SU2-1F-0)--(SU2-SU2-pi0);
\draw[-latex] (SU3-3F-1/2)--(SU3-2F-0);
\draw[-latex] (SU3-3F-1/2)--(SU3-2F-1);
\draw[-latex] (SU3-3F-3/2)--(SU3-2F-1);
\draw[-latex] (SU3-3F-3/2)--(SU3-2F-2);
\draw[-latex] (SU3-3F-5/2)--(SU3-2F-2);
\draw[-latex] (SU3-3F-5/2)--(SU3-2F-3);
\draw[-latex] (SU3-3F-7/2)--(SU3-2F-3);
\draw[-latex] (SU3-3F-7/2)--(SU3-2F-4);
\draw[-latex] (SU3-3F-9/2)--(SU3-2F-4);
\draw[-latex] (SU3-3F-9/2)--(SU3-2F-5);
\draw[-latex] (SU3-3F-11/2)--(SU3-2F-5);
\draw[-latex] (SU3-3F-11/2)--(SU3-2F-6);
\draw[-latex] (SU3-3F-11/2)--(Sp2-2AS-0);
\draw[-latex] (Sp2-3AS-0)--(Sp2-2AS-0);
\draw[-latex] (SU3-3F-1/2)--(SU2-SU2-pi0);
%
%
%
%
% Ninth level -- Tenth Level 
%
%
%
\draw[-latex] (SU2-SU2-pi0) -- (F1-dP2);
\draw[-latex] (SU3-2F-0) -- (SU3-1F-1/2);
\draw[-latex] (SU3-2F-1) -- (SU3-1F-1/2);
\draw[-latex] (SU3-2F-1) -- (SU3-1F-3/2);
\draw[-latex] (SU3-2F-2) -- (SU3-1F-3/2);
\draw[-latex] (SU3-2F-2) -- (SU3-1F-5/2);
\draw[-latex] (SU3-2F-3) -- (SU3-1F-5/2);
\draw[-latex] (SU3-2F-3) -- (SU3-1F-7/2);
\draw[-latex] (SU3-2F-4) -- (SU3-1F-7/2);
\draw[-latex] (SU3-2F-4) -- (SU3-1F-9/2);
\draw[-latex] (SU3-2F-5) -- (SU3-1F-9/2);
\draw[-latex] (SU3-2F-5) -- (SU3-1F-11/2);
\draw[-latex] (SU3-2F-6) -- (SU3-1F-11/2);
\draw[-latex] (SU3-2F-6) -- (SU3-1F-13/2);
\draw[-latex] (Sp2-2AS-0) -- (Sp2-1AS-0);
\node(SU3-2F-5b)[below = -1.5mm of SU3-2F-5]{};
\node(Sp2-1AS-0a)[above = -1mm of Sp2-1AS-0]{};
\draw[-latex] (SU3-2F-5b) -- (Sp2-1AS-0a);
%
%
% Tenth level -- Eleventh Level 
%
%
%
\draw[-latex] (F1-dP2) -- (F2-dP1);
\draw[-latex] (SU3-1F-1/2) -- (SU3-0);
\draw[-latex] (SU3-1F-1/2) -- (SU3-1);
\draw[-latex] (SU3-1F-3/2) -- (SU3-1);
\draw[-latex] (SU3-1F-3/2) -- (SU3-2);
\draw[-latex] (SU3-1F-5/2) -- (SU3-2);
\draw[-latex] (SU3-1F-5/2) -- (SU3-3);
\draw[-latex] (SU3-1F-7/2) -- (SU3-3);
\draw[-latex] (SU3-1F-7/2) -- (SU3-4);
\draw[-latex] (SU3-1F-9/2) -- (SU3-4);
\draw[-latex] (SU3-1F-9/2) -- (SU3-5);
\draw[-latex] (SU3-1F-11/2) -- (SU3-5);
\draw[-latex] (SU3-1F-11/2) -- (SU3-6);
\draw[-latex] (SU3-1F-13/2) -- (SU3-6);
\draw[-latex] (SU3-1F-13/2) -- (SU3-7);
\draw[-latex] (Sp2-1AS-0) -- (Sp2-0);
\node(SU3-1F-9/2b)[below=-1.5mm of SU3-1F-9/2]{};
\node(Sp2-0a)[above=-1mm of Sp2-0]{};
\draw[-latex] (SU3-1F-9/2b) -- (Sp2-0a);
%
%
% Eleventh level -- Twelfth Level 
%
%
%
\node(SU3-4b)[below = -1mm of SU3-4]{};
\node(SU3-2b)[below = -1mm of SU3-2]{};
\draw[-latex] (SU3-4b) -- (F6-P2);
\draw[-latex] (Sp2-0) -- (F6-P2);
\draw[-latex] (F2-dP1) -- (F3-P2);
\draw[-latex] (SU3-2b) -- (F3-P2);
%
%
%
% Remaining Isolated
%
%
%
\node(SU3-1F-7/2a)[above = -1mm of SU3-1F-7/2]{};
\node(SU3-3a)[above= -1mm of SU3-3]{};
\node(SU3-3al)[left= 0mm of SU3-3a]{};
\node(SU3-4a)[above= -1mm of SU3-4]{};
\node(SU3-4al)[left= 0mm of SU3-4a]{};
\node(SU3-1Symb)[below = -1.5mm of SU3-1Sym]{};
\node(SU3-1Symbr)[right = 1mm of SU3-1Symb]{};
\draw[-latex](SU3-1Sym-1F) -- (SU3-1Sym);
\draw[-latex](SU3-1Sym-1F) -- (SU3-1F-7/2a);
\draw[-latex,color=blue](SU3-1Symbr) -- (SU3-3al);
\draw[-latex,color=blue](SU3-1Symbr) -- (SU3-4al);
%
%
%
%
%%%%%%%%%%     Added in v2  %%%%%%%%%%%%%%%%%%
\node(SU3-1Sym-3/2b)[below = -1.5mm of SU3-1Sym-3/2]{};
\node(SU3-1Sym-3/2br)[right = 1mm of SU3-1Sym-3/2b]{};
\node(SU3-2a)[above= -1mm of SU3-2]{};
\node(SU3-2al)[left= 0mm of SU3-2a]{};
\node(SU3-5a)[above= -1mm of SU3-5]{};
\node(SU3-5al)[left= 0mm of SU3-5a]{};
\draw[-latex,color=red](SU3-1Sym-3/2br) -- (SU3-2al);
\draw[-latex,color=red](SU3-1Sym-3/2br) -- (SU3-5al);
%%%%%%%%%%%%%%%%%%%%%%%%%%%%%%%%%%%%
%
%
%
%
\end{tikzpicture}
\caption{List of 5d $\mathcal{N}=1$ theories with rank 2 and their relations through RG flows, where the marginal theories are shaded in gray. %It was discussed in \cite{Jefferson:2017ahm, Jefferson:2018irk} that there is some subtlety in the CFT limit of the  $SU(2)\times SU(2)$ quiver descriptions associated with Figures 74, 75, and 121. 
}
\label{fig:listoffigures} 
\end{figure}
%\end{tiny}

\begin{figure}
\centering
\includegraphics[width=15cm]{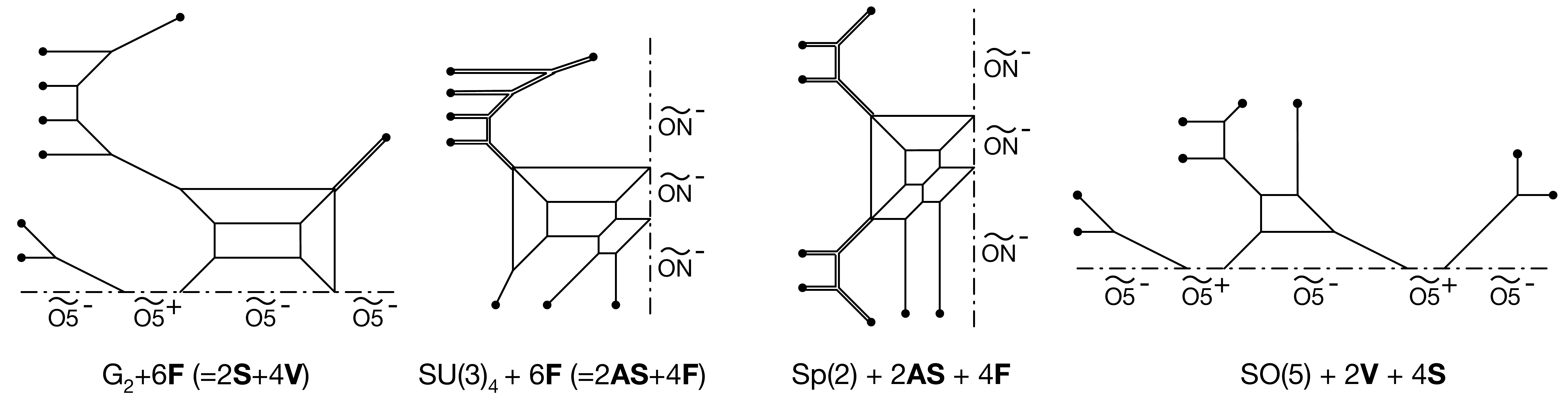}
\caption{Various diagrams representing $G_2+6\mathbf{F}$,  $SU(3)_{4}+6\mathbf{F}$, $Sp(2)+2\mathbf{AS}+4\mathbf{F}$, and $SO(5) + 2\mathbf{V} + 4\mathbf{S}$. All the diagrams are related by HW transitions, generalized flop transitions, and $SL(2,\mathbb{Z})$ transformation and reflection.}
\label{Fig:SU3-6F-4}
\end{figure}
\begin{figure}
\centering
\includegraphics[width=15cm]{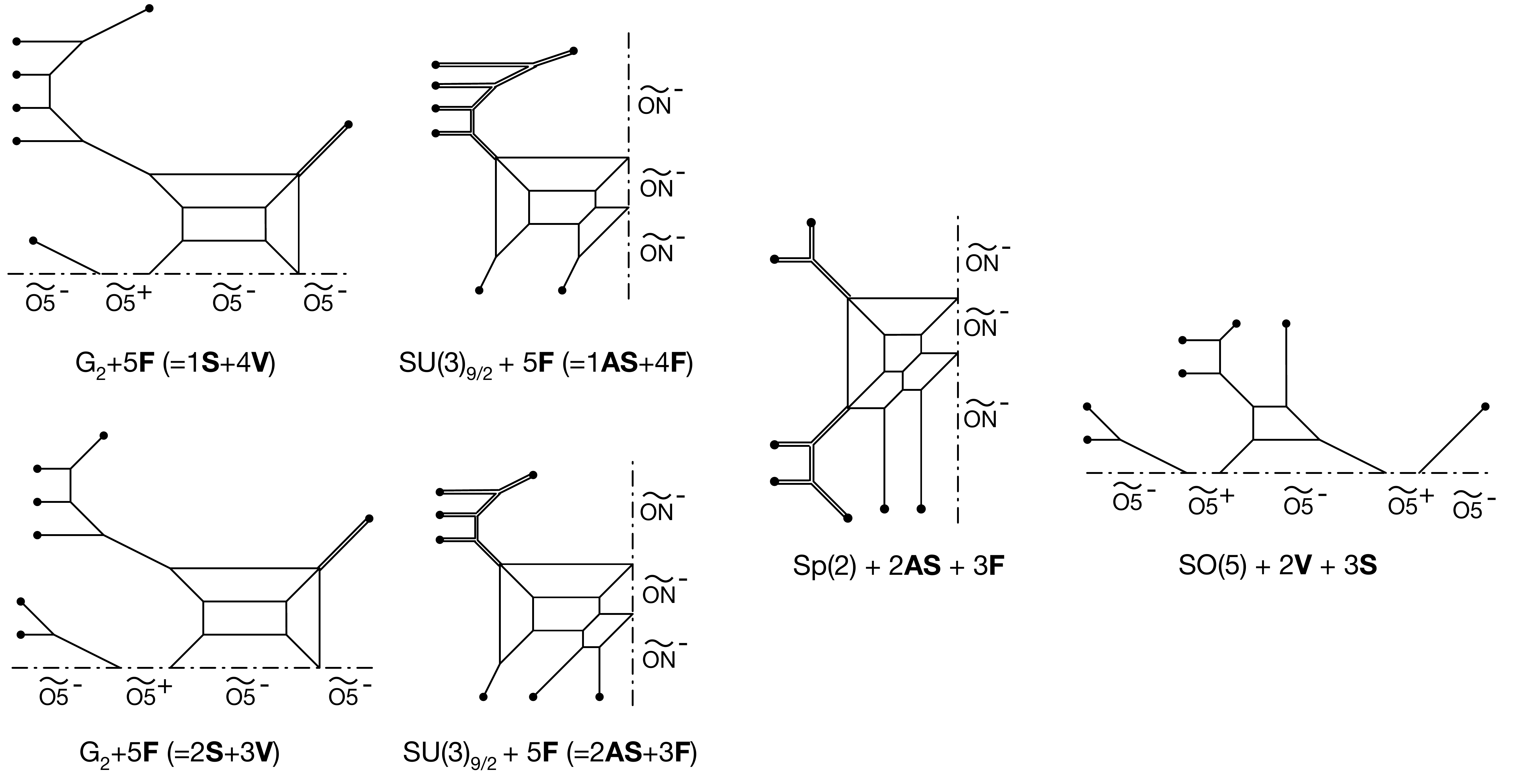}
\caption{Various diagrams representing $G_2+5\mathbf{F}$,  $SU(3)_{\frac{9}{2}}+5\mathbf{F}$, $Sp(2)+2\mathbf{AS}+3\mathbf{F}$, $SO(5) + 2\mathbf{V} + 3\mathbf{S}$. These diagrams are obtained from any of the diagrams in Figure \ref{Fig:SU3-6F-4} by a certain limit corresponding to a relevant deformation.}
\label{Fig:SU3-5F-9/2}
\end{figure}
\begin{figure}
\centering
\includegraphics[width=15cm]{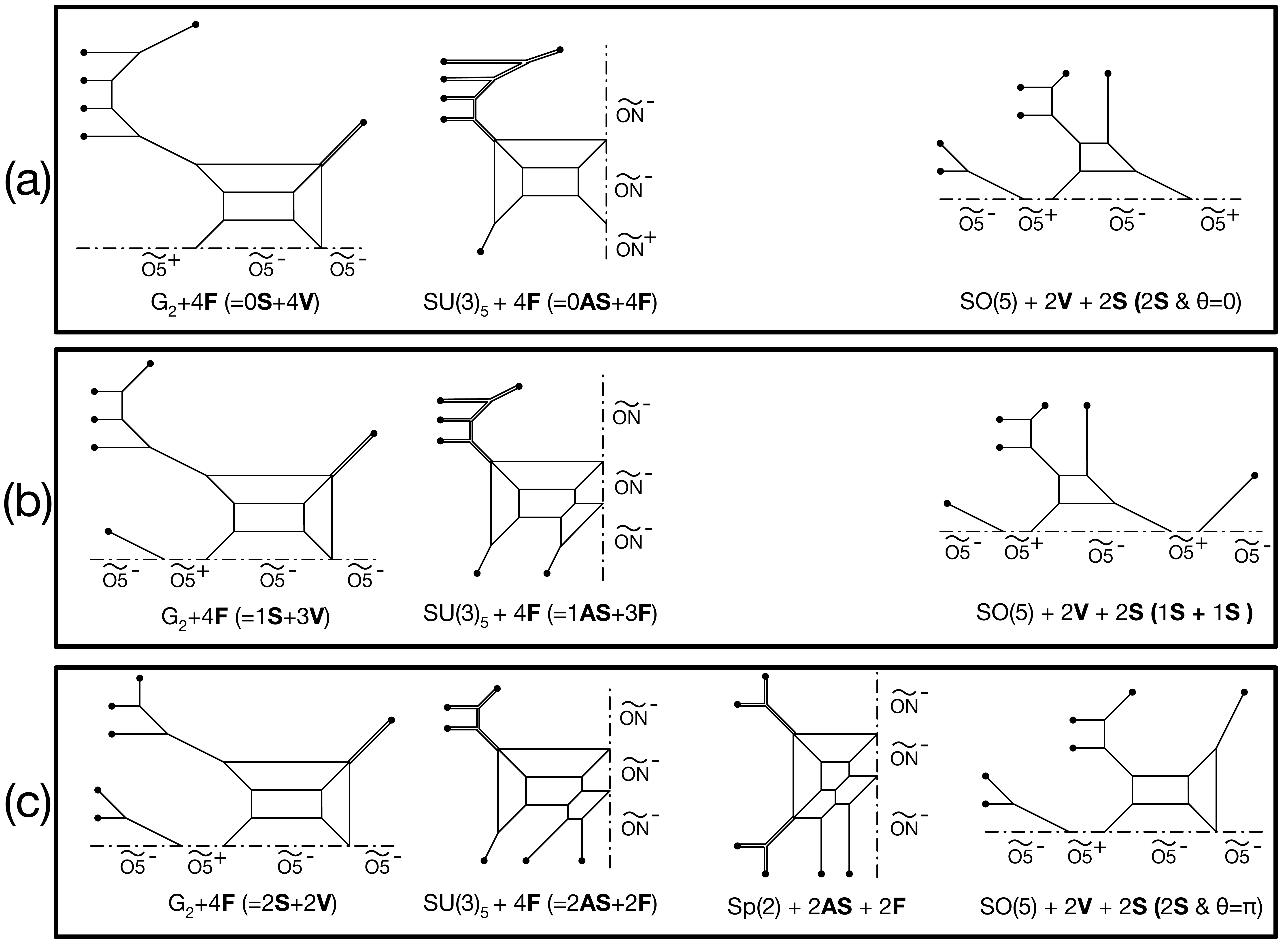}
\caption{Various diagrams representing $G_2+4\mathbf{F}$,  $SU(3)_{5}+4\mathbf{F}$, $2\mathbf{AS}+2\mathbf{F}$, $SO(5) + 2\mathbf{V} + 2\mathbf{S}$. The diagrams in the same group are related by Hanany-Witten transitions, generalized flop transitions, and $SL(2,\mathbb{Z})$ transformation and reflection. Although the diagrams in the different groups are not related in such a way, all the diagrams are obtained from the one in Figure \ref{Fig:SU3-5F-9/2} with different diagrammatic limit.}
\label{Fig:SU3-4F-5}
\end{figure}
\begin{figure}
\centering
\includegraphics[width=15cm]{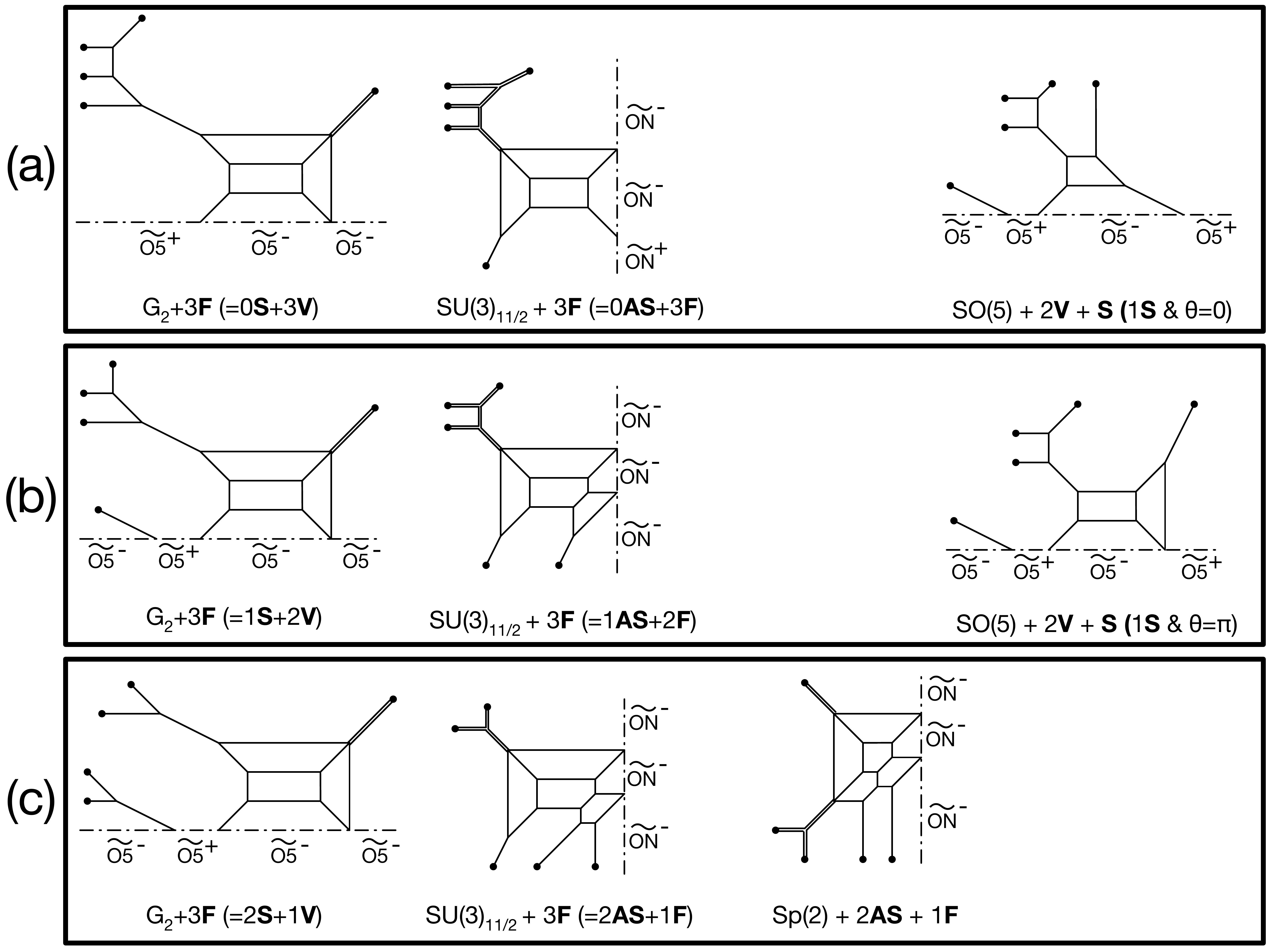}
\caption{Various diagrams representing $G_2+3\mathbf{F}$,  $SU(3)_{\frac{11}{2}}+3\mathbf{F}$, $1\mathbf{AS}+2\mathbf{F}$, $2\mathbf{AS}+1\mathbf{F}$, $SO(5) + 2\mathbf{V} + 1\mathbf{S}$. The diagrams in group (a) are obtained from group (a) and group (b) in Figure \ref{Fig:SU3-4F-5}. Group (b) is obtained from group (b) and group (c) in Figure \ref{Fig:SU3-4F-5}. Group (c) is obtained from group (c) in Figure \ref{Fig:SU3-4F-5}.}
\label{Fig:SU3-3F-11/2}
\end{figure}
\begin{figure}
\centering
\includegraphics[width=15cm]{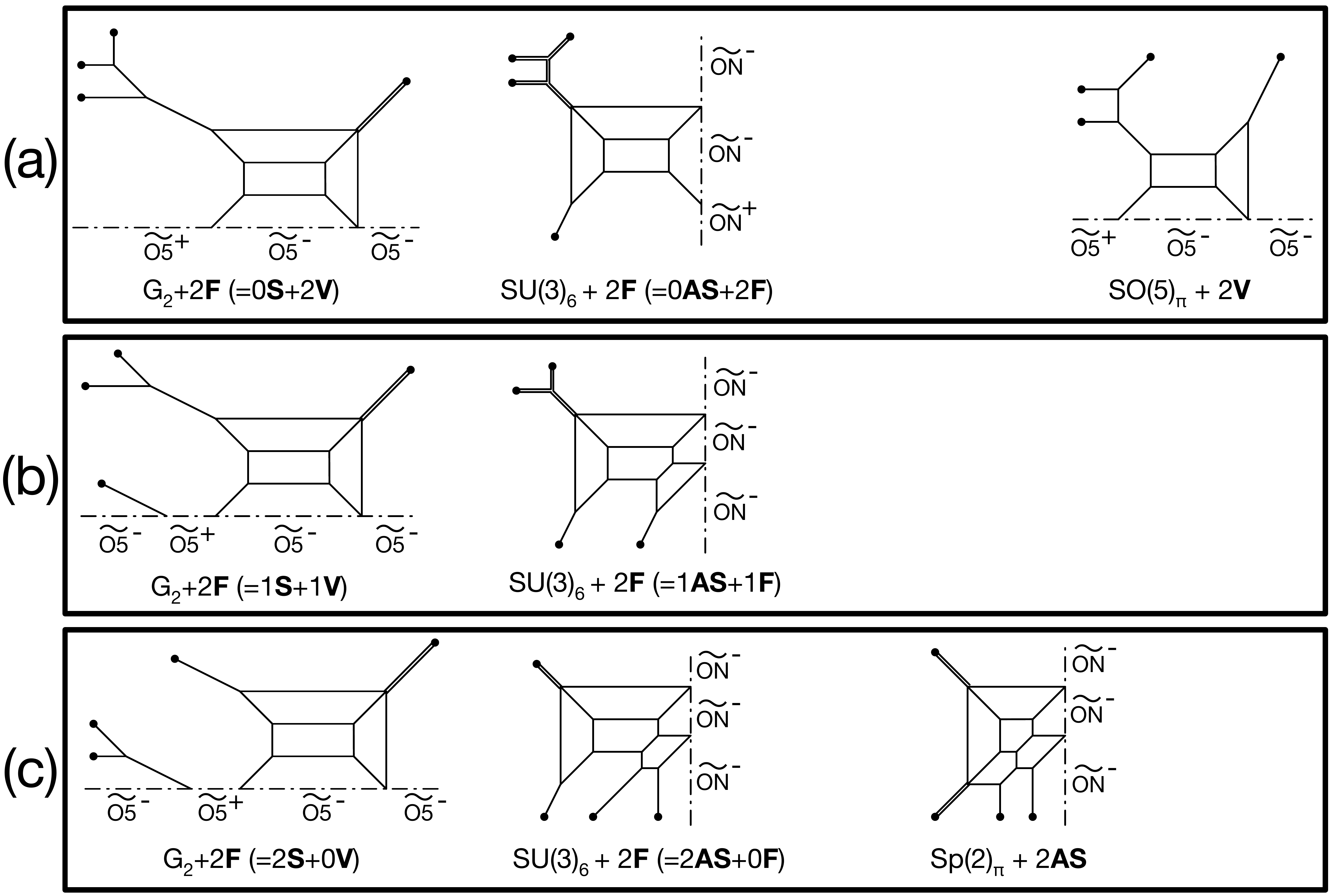}
\caption{Various diagrams representing $G_2+2\mathbf{F}$,  $SU(3)_{6}+2\mathbf{F}$, $1\mathbf{AS}+1\mathbf{F}$, $2\AS$, $SO(5)_\pi + 2\mathbf{V}$, and $Sp(2)_{\pi}+2\AS$ which are obtained from Figure \ref{Fig:SU3-3F-11/2}.}
\label{Fig:SU3-2F-6}
\end{figure}
\begin{figure}
\centering
\includegraphics[width=15cm]{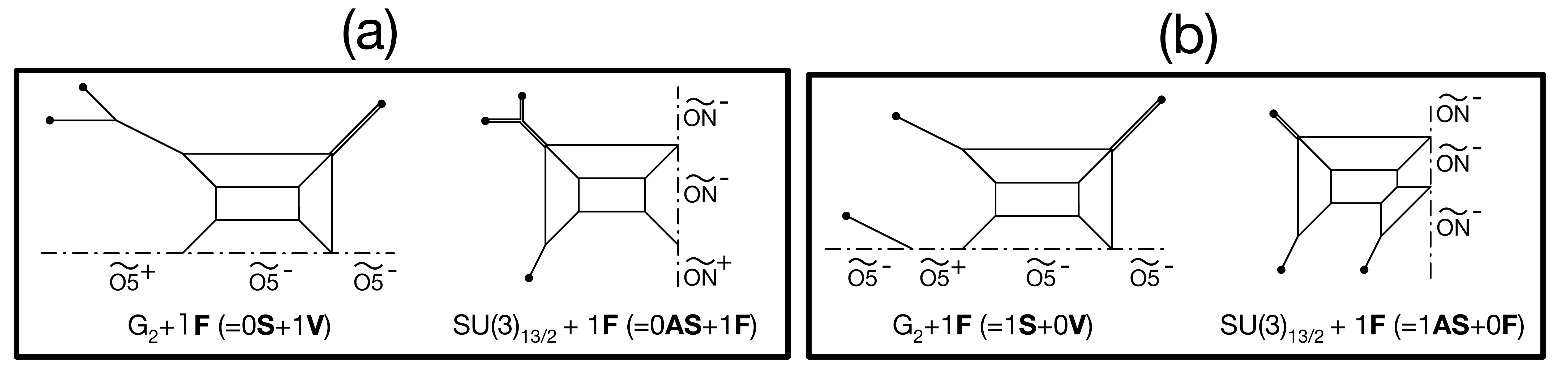}
\caption{Various diagrams representing $G_2+1\mathbf{F}$,  $SU(3)_{\frac{13}{2}}+1\mathbf{F}$, $SO(5)_{\pi} + 2\mathbf{V}$, and $Sp(2)_{\pi}+2\AS$ which are obtained from Figure \ref{Fig:SU3-3F-11/2}.}
\label{Fig:SU3-1F-13/2}
\end{figure}
\begin{figure}
\centering
\includegraphics[width=8cm]{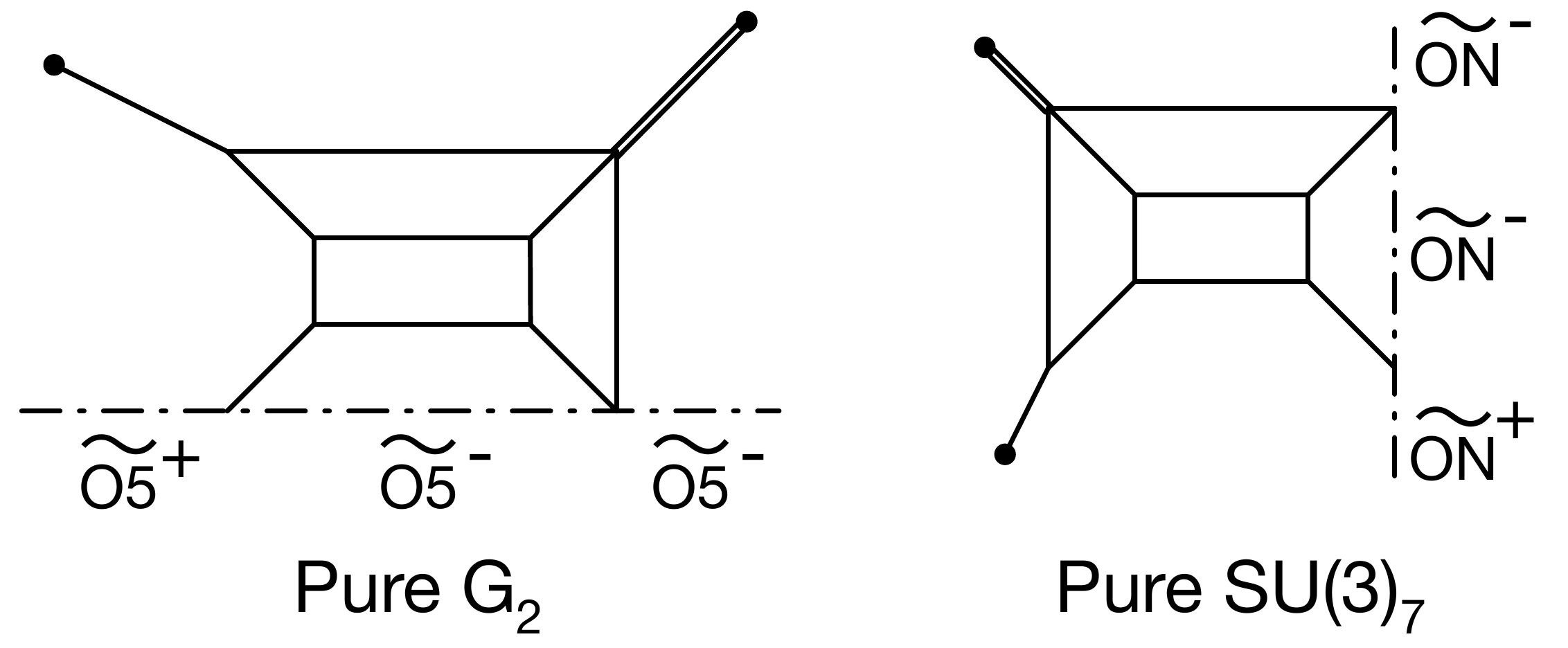}
\caption{5-brane webs for pure $G_2$ and pure $SU(3)_7$, which are obtained Figure \ref{Fig:SU3-1F-13/2}.}
\label{Fig:SU3-7}
\end{figure}
\begin{figure}
\centering
\includegraphics[width=4cm]{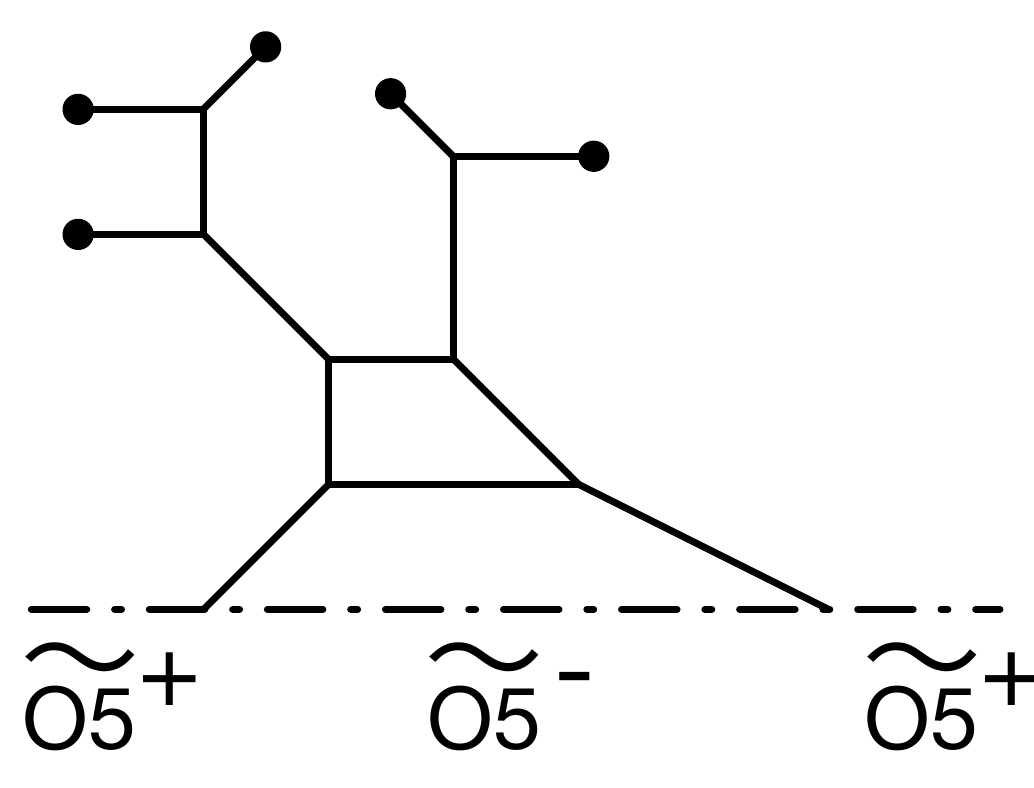}
\caption{A 5-brane web for $Sp(2)_0 + 3 \mathbf{AS}$ realized as $SO(5)_0 + 3 \mathbf{V}$}
\label{Fig:Sp2-3AS-0}
\end{figure}
\begin{figure}
\centering
\includegraphics[width=12cm]{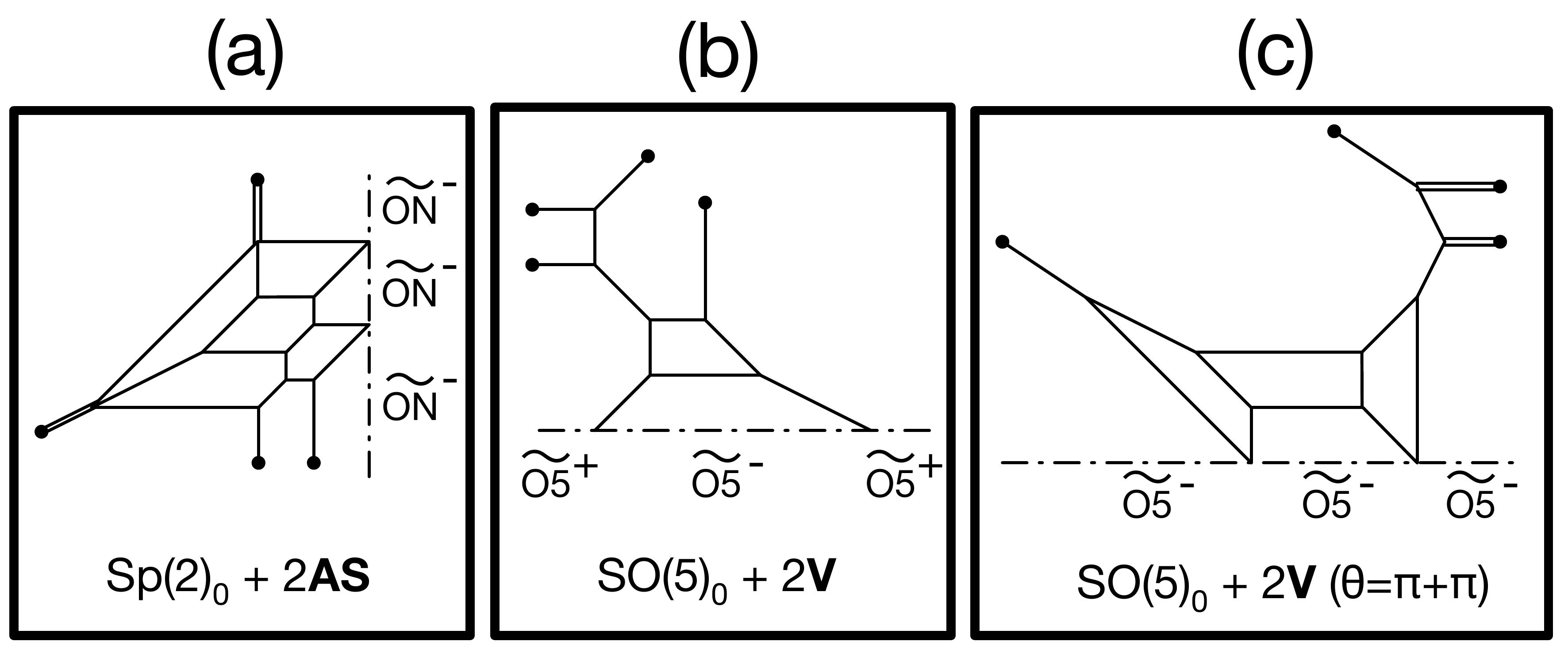}
\caption{5-brane webs for (a) $Sp(2)_0+2\AS$, (b)$SO(5)_0+2\bF$, and (c) $SO(5)_{2\pi}+2\bF$.}
\label{Fig:Sp2-2AS-0}
\end{figure}
\begin{figure}
\centering
\includegraphics[width=12cm]{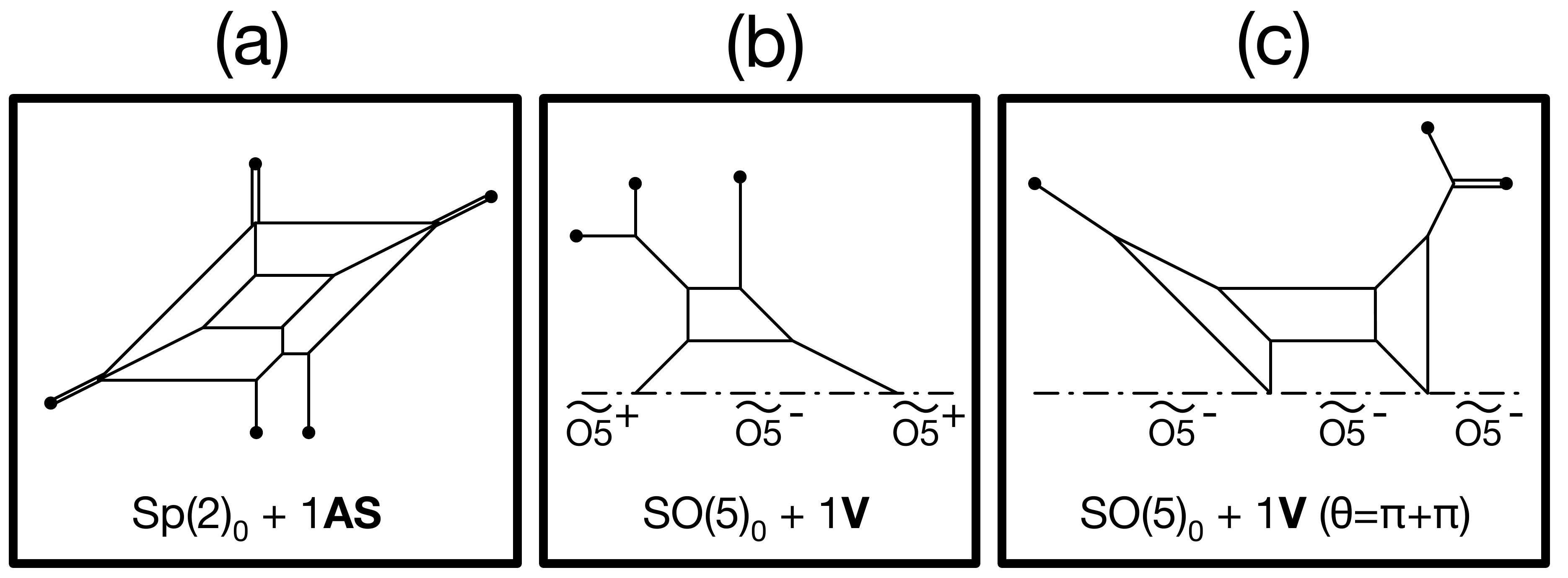}
\caption{5-brane webs for (a) $Sp(2)_0+1\AS$, (b) $SO(5)_0+1\bF$, and (c) $SO(5)_{2\pi}+1\bF$.}
\label{Fig:Sp2-1AS-0}
\end{figure}
\begin{figure}
\centering
\includegraphics[width=12cm]{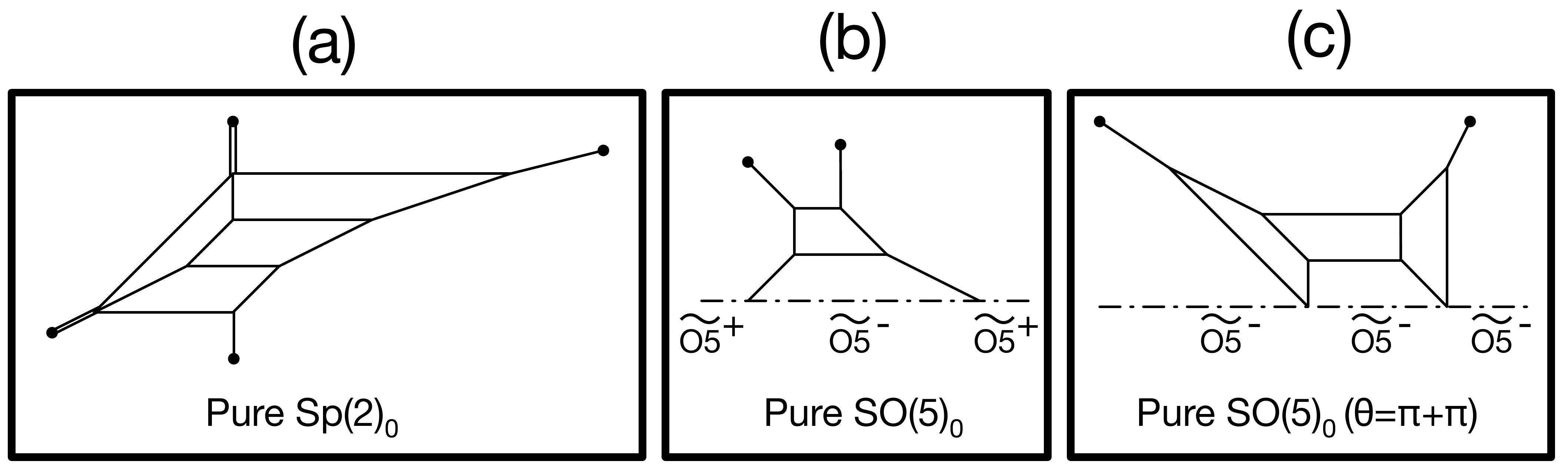}
\caption{5-brane webs for (a) $Sp(2)_0$, (b) $SO(5)_0$, and (c) $SO(5)_{2\pi}$.}
\label{Fig:Sp2-0}
\end{figure}
\begin{figure}
\centering
\includegraphics[width=7cm]{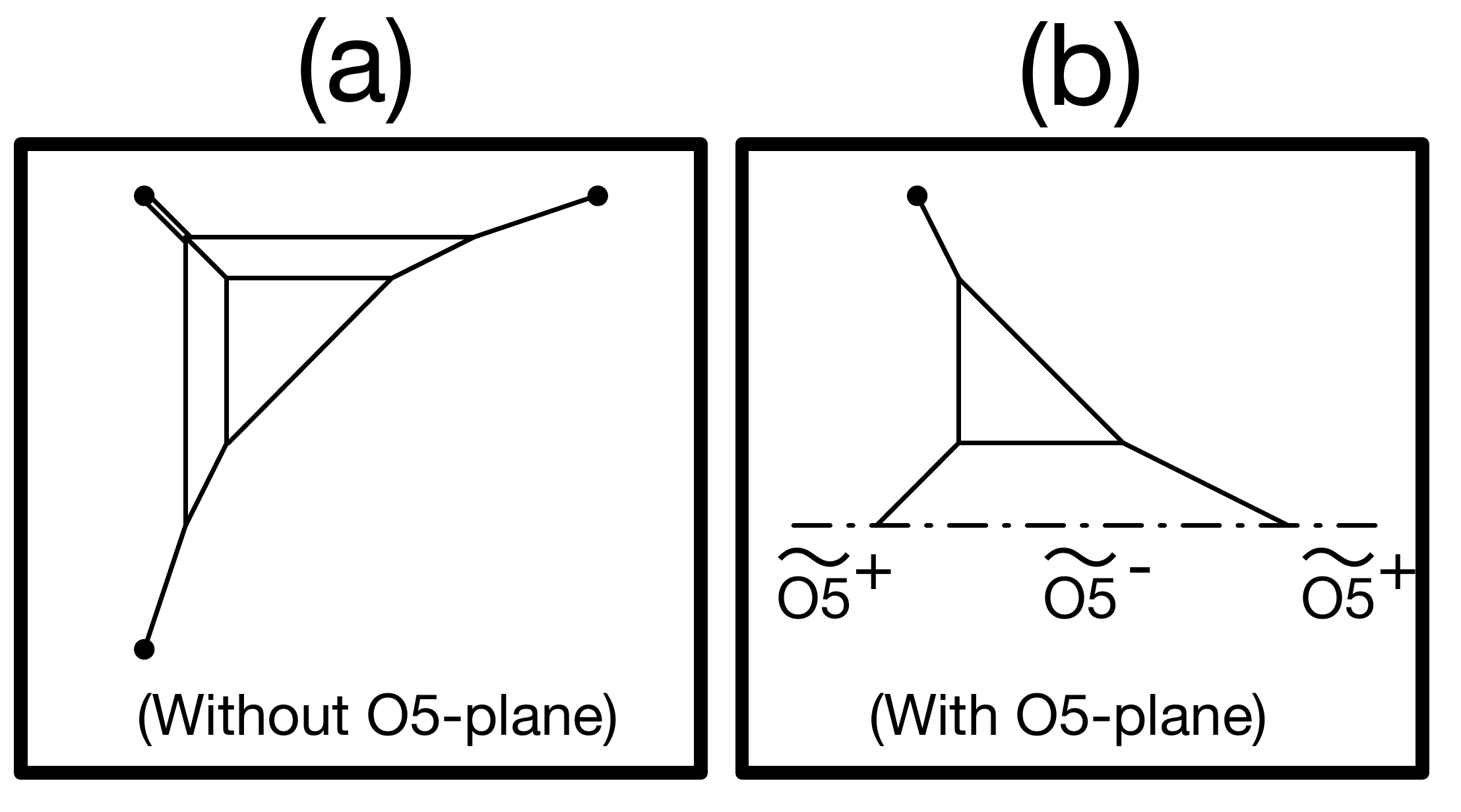}
\caption{A non-Lagrangian theories denoted by $\mathbb{F}_6 \cup\mathbb{P}^2$ that are obtained the RG flows on \ref{Fig:Sp2-0}.}
\label{Fig:F6-P2}
\end{figure}
\begin{figure}
\centering
\includegraphics[width=12cm]{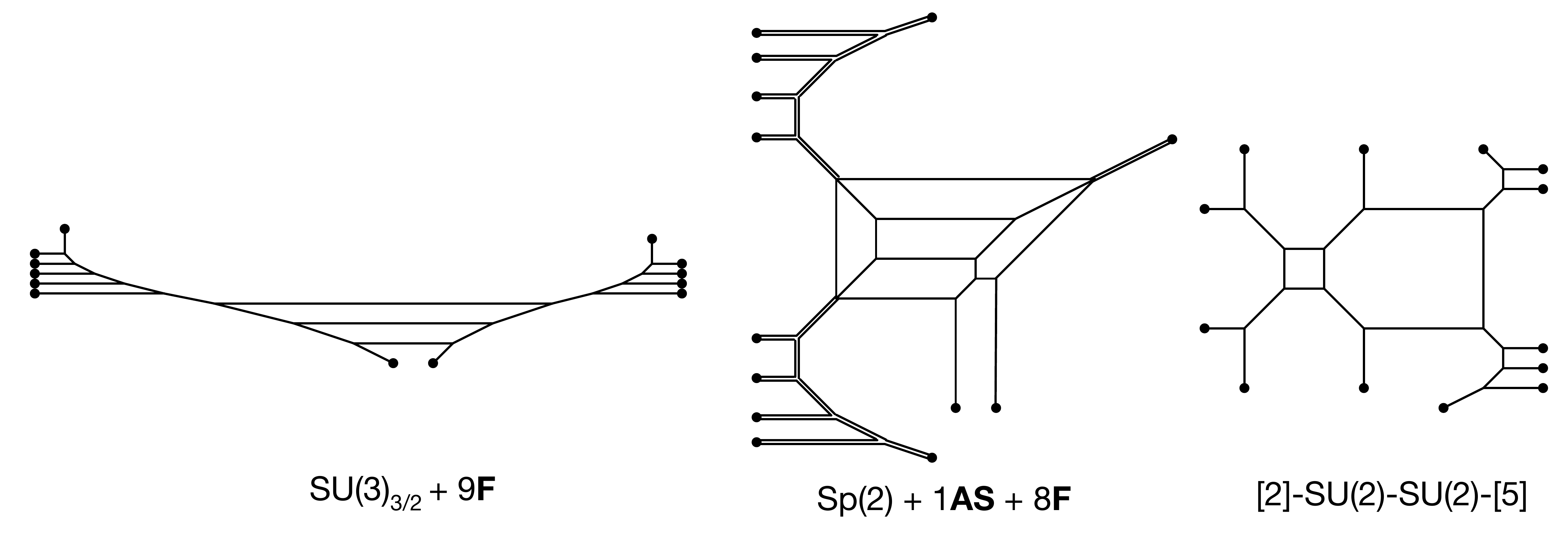}
\caption{Various 5-brane webs for $SU(3)_\frac32+9\bF$, $Sp(2)+1\AS+8\bF$, and $[SU(2)+2\bF]\times [SU(2)+5\bF]$.}
\label{Fig:SU3-9F-3/2}
\end{figure}
\begin{figure}
\centering
\includegraphics[width=12cm]{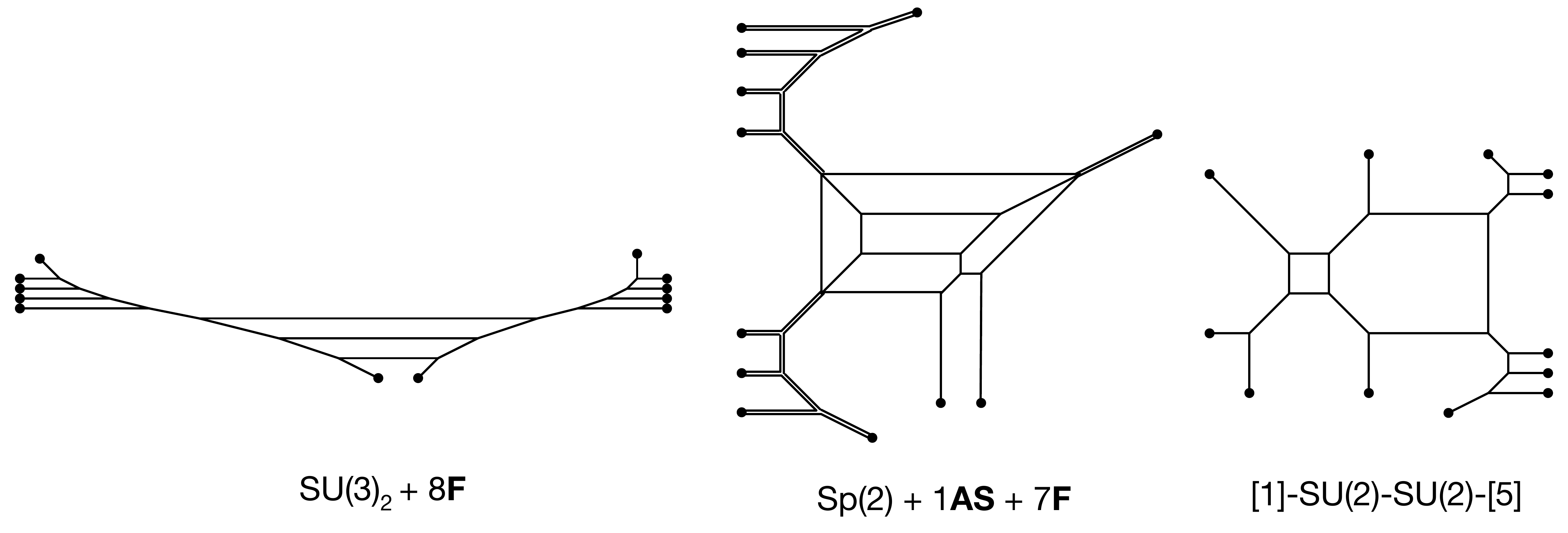}
\caption{Various 5-brane webs for $SU(3)_2+8\bF$, $Sp(2)+1\AS+7\bF$, and $[SU(2)+1\bF]\times [SU(2)+5\bF]$.}
\label{Fig:SU3-8F-2}
\end{figure}
\begin{figure}
\centering
\includegraphics[width=12cm]{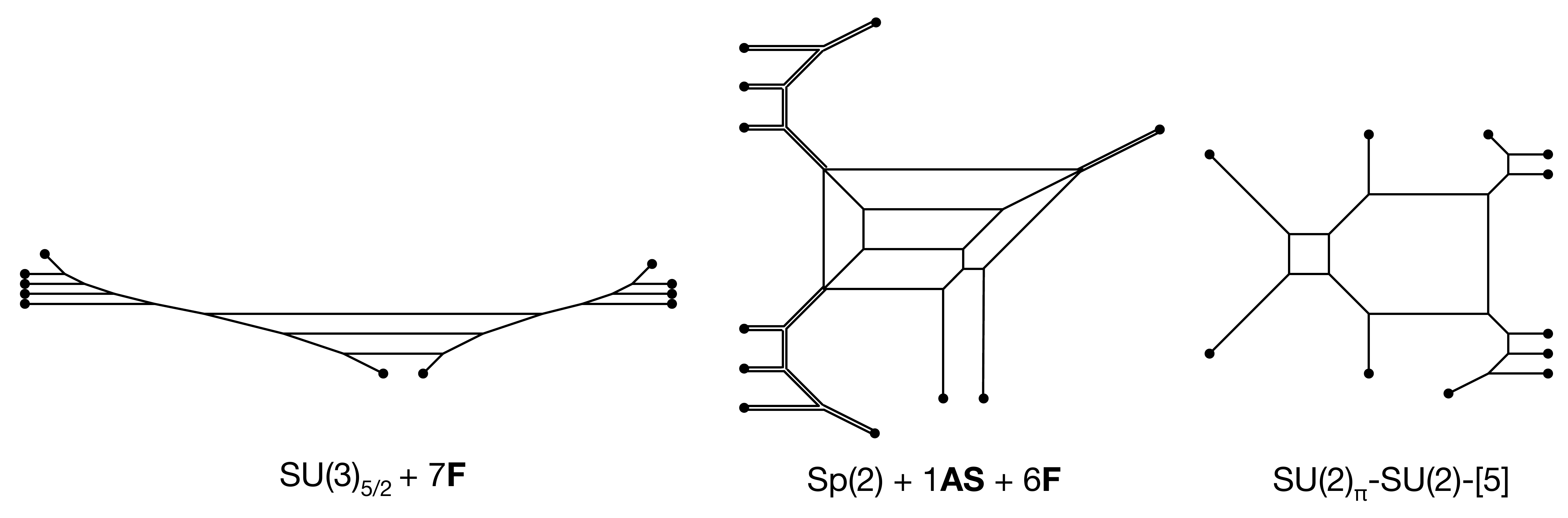}
\caption{Various 5-brane webs for $SU(3)_\frac52+7\bF$, $Sp(2)+1\AS+6\bF$, and $SU(2)_\pi\times [SU(2)+5\bF]$.}
\label{Fig:SU3-7F-5/2}
\end{figure}
\begin{figure}
\centering
\includegraphics[width=8cm]{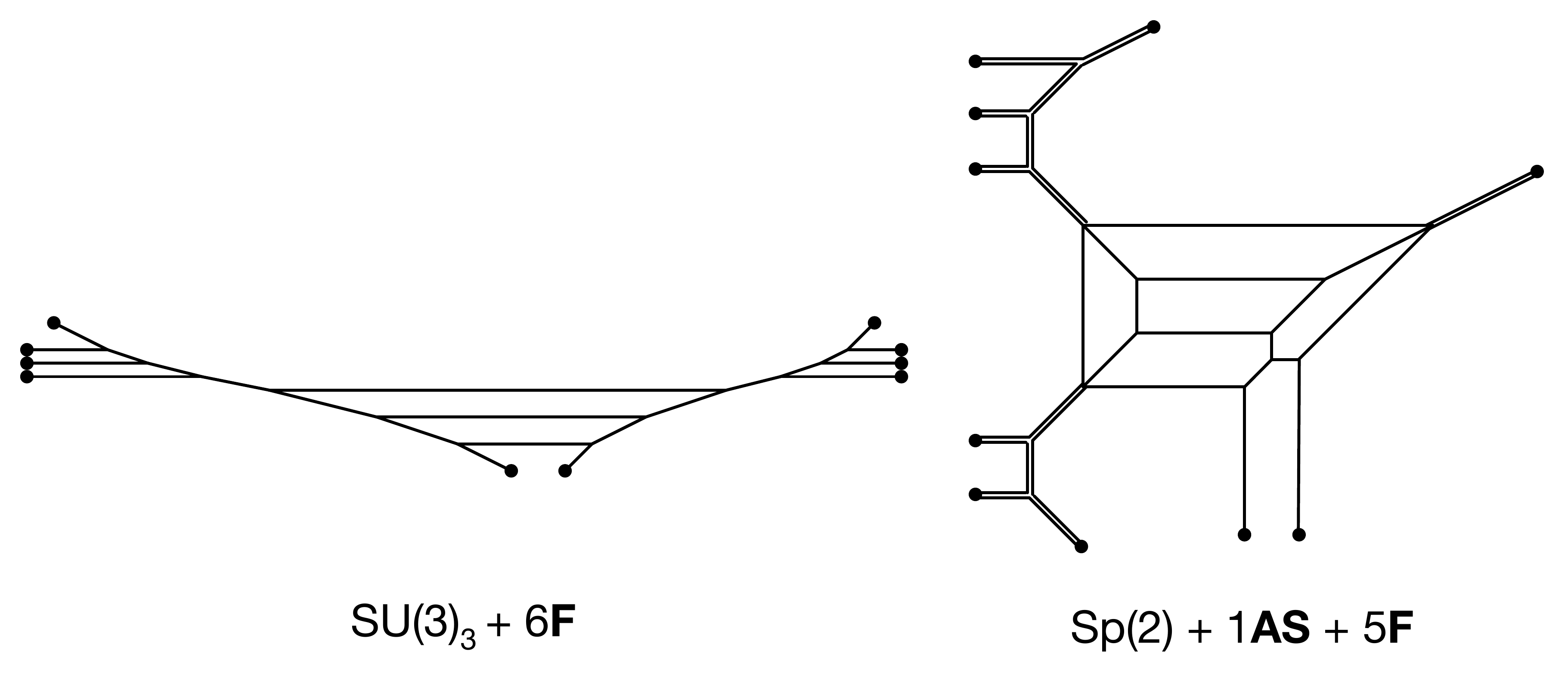}
\caption{Various 5-brane webs for $SU(3)_3+6\bF$ and $Sp(2)+1\AS+5\bF$.}
\label{Fig:SU3-6F-3}
\end{figure}
\begin{figure}
\centering
\includegraphics[width=15cm]{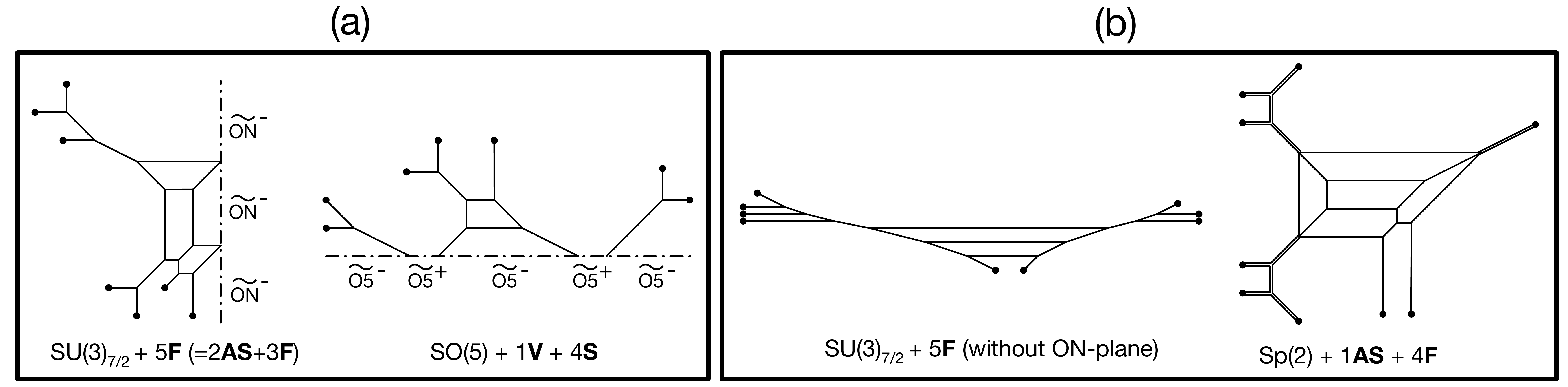}
\caption{5-brane webs for $SU(3)_\frac 72+5\bF$ and $SO(5)+1\bF+4{\bf S}$ (or $Sp(2)+1\AS+4\bF$) (a) with an orientifold  and (b) without an orientifold.}
\label{Fig:SU3-5F-7/2}
\end{figure}
\begin{figure}
\centering
\includegraphics[width=12cm]{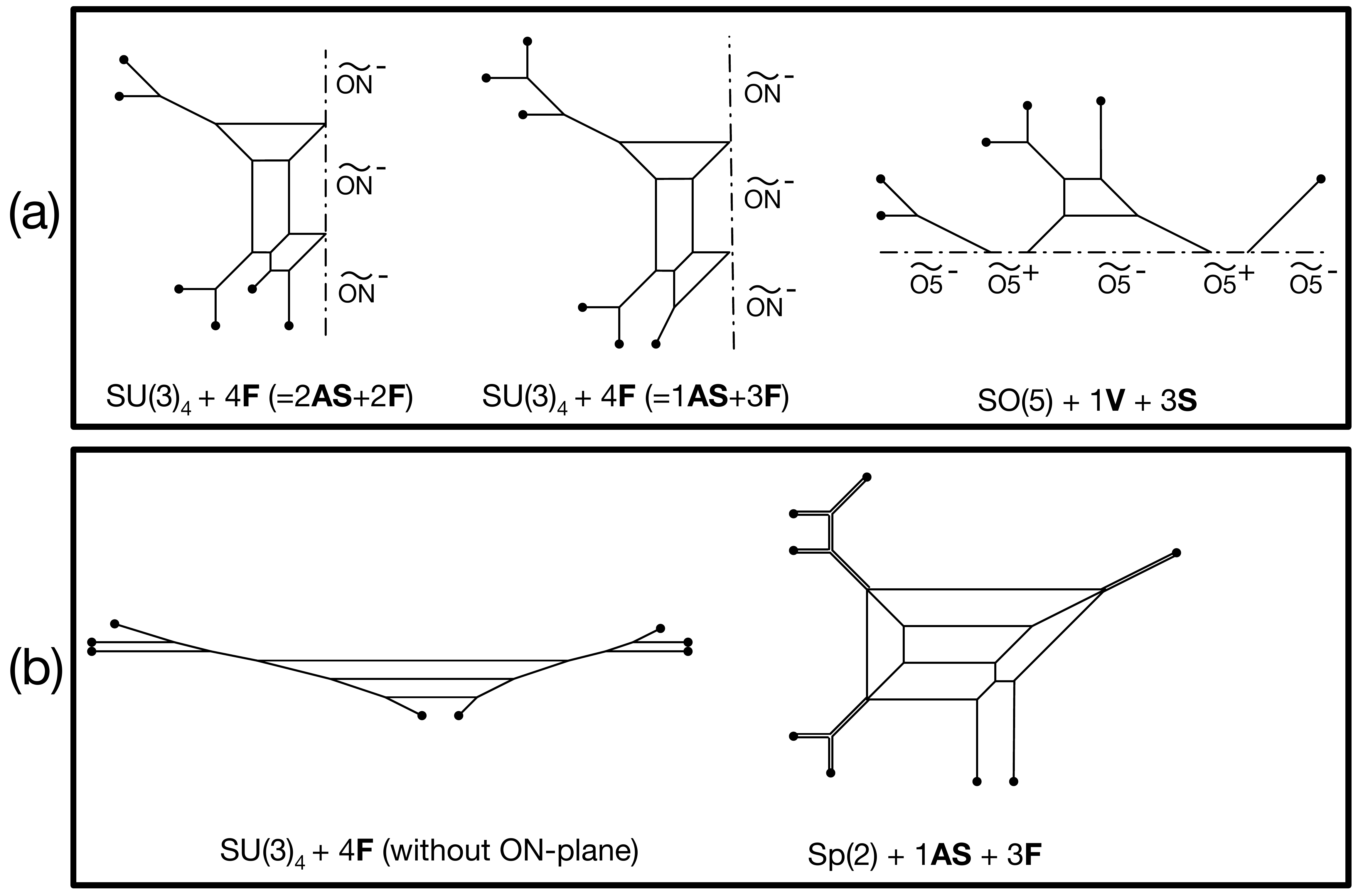}
\caption{5-brane webs for $SU(3)_4+4\bF$ and $SO(5)+1\bF+3{\bf S}$ (or $Sp(2)+1\AS+3\bF$) (a) with an orientifold  and (b) without an orientifold.}
\label{Fig:SU3-4F-4}
\end{figure}
\begin{figure}
\centering
\includegraphics[width=15cm]{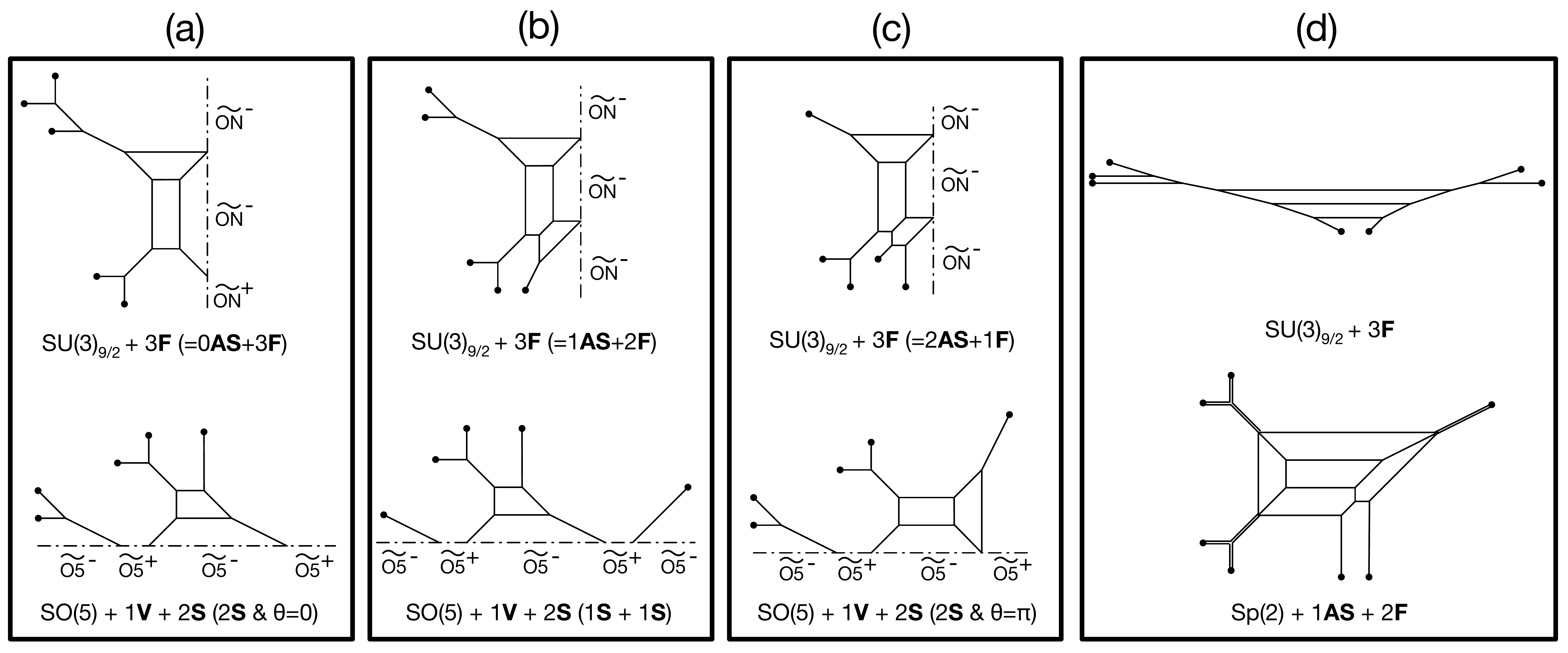}
\caption{Various 5-brane webs for $SU(3)_\frac92+3\bF$ and its dual $Sp(2) +1\AS +2\bF$ or $SO(5)+1\bF+2{\bf S}$. }
\label{Fig:SU3-3F-9/2}
\end{figure}
\begin{figure}
\centering
\includegraphics[width=15cm]{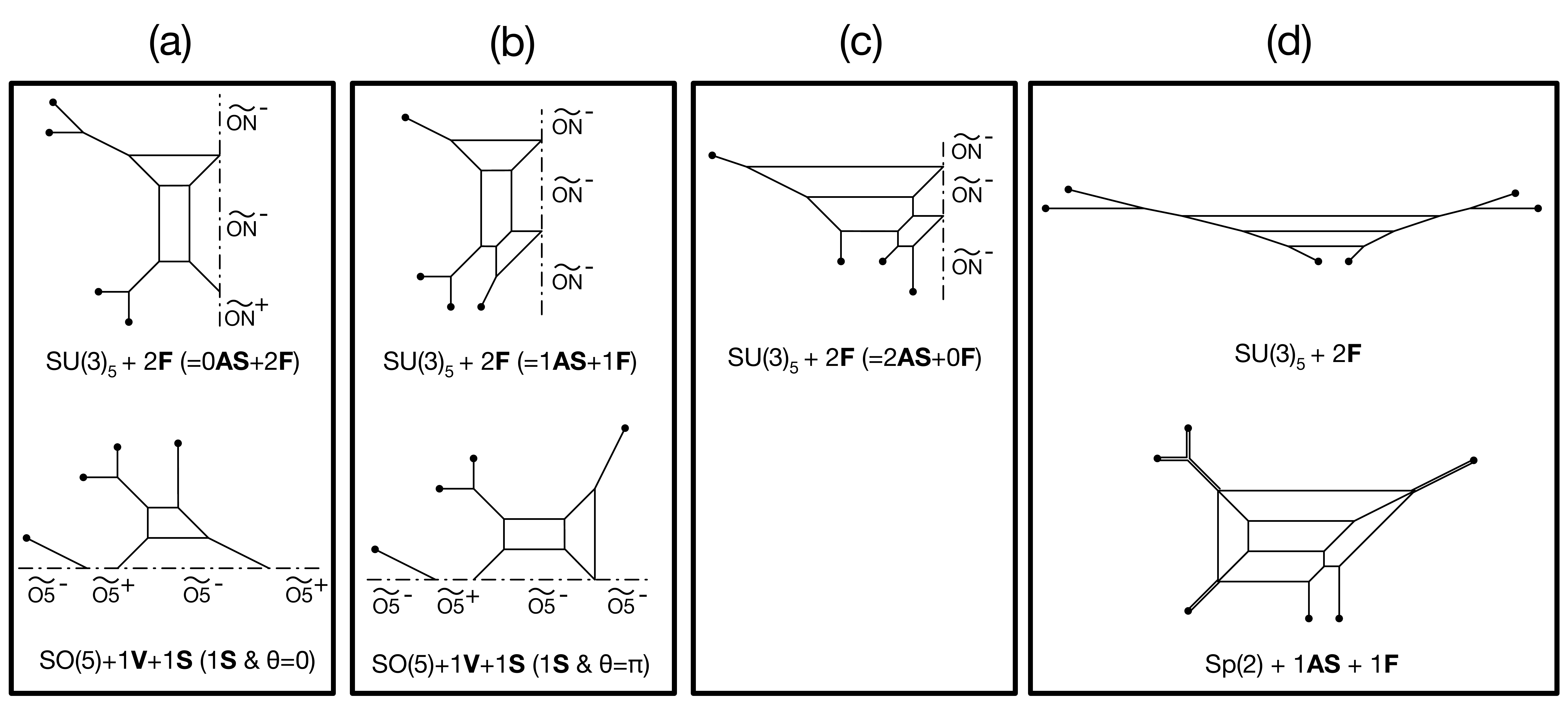}
\caption{Various 5-brane webs for $SU(3)_5+2\bF$ and its dual $Sp(2)+1\AS +1\bF$ or $SO(5)+1\bF+1{\bf S}$.}
\label{Fig:SU3-2F-5}
\end{figure}
\begin{figure}
\centering
\includegraphics[width=15cm]{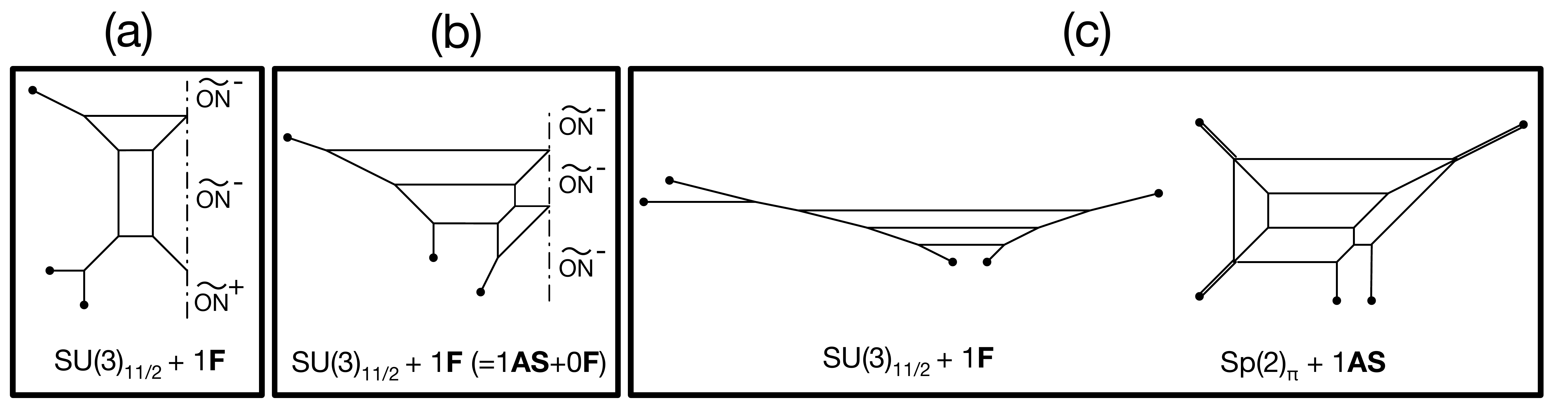}
\caption{Various 5-brane webs for $SU(3)_\frac{11}{2}+1\bF$ and its dual $Sp(2)_\pi +1\AS $.}
\label{Fig:SU3-1F-11/2}
\end{figure}
\begin{figure}
\centering
\includegraphics[width=12cm]{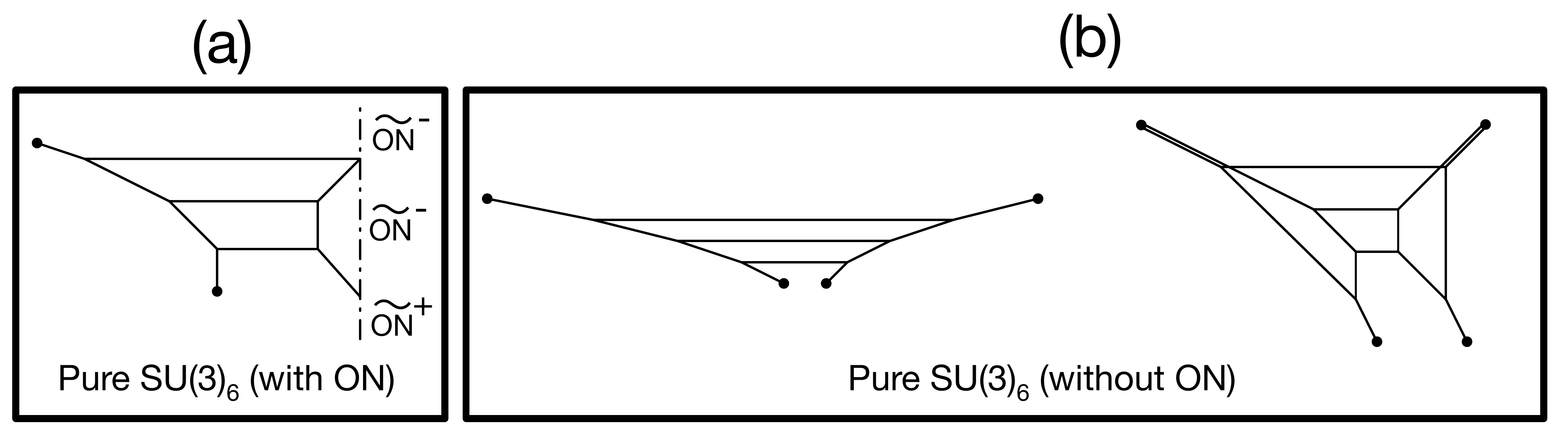}
\caption{%\color{red}
5-brane webs for pure $SU(3)_6$. (a) A 5-brane web with an ON-plane. (b) A 5-brane web without an ON-plane. In (b), the left one is a diagram naively showing $SU(3)_6$, while the right one is a 5-brane web when two 7-branes in the bottom part of the left figure are pulled out upward along the direction of their charges. %One may consider a naive diagram for pure $SU(3)_7$ like the left diagram in (b), but it does not lead to a ``finite" diagram like the right diagram in (b). 
}
\label{Fig:SU3-6}
\end{figure}
\begin{figure}
\centering
\includegraphics[width=12cm]{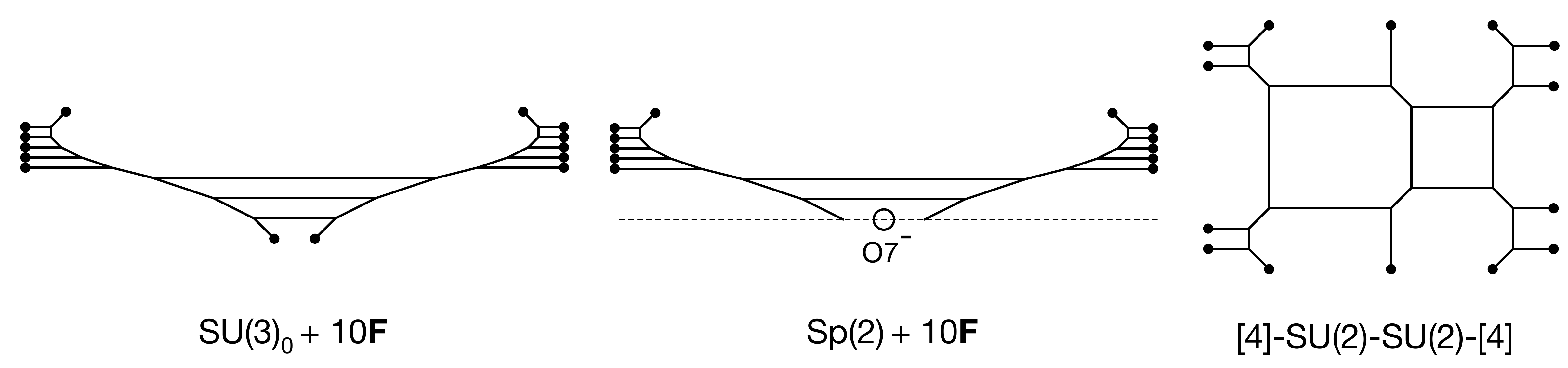}
\caption{5-brane webs for $SU(3)_0+10\bF$, $Sp(2)+10\bF$, and $[SU(2)+4\bF]\times [SU(2)+4\bF]$, which are dual to one anther.}
\label{Fig:SU3-10F-0}
\end{figure}
\begin{figure}
\centering
\includegraphics[width=12cm]{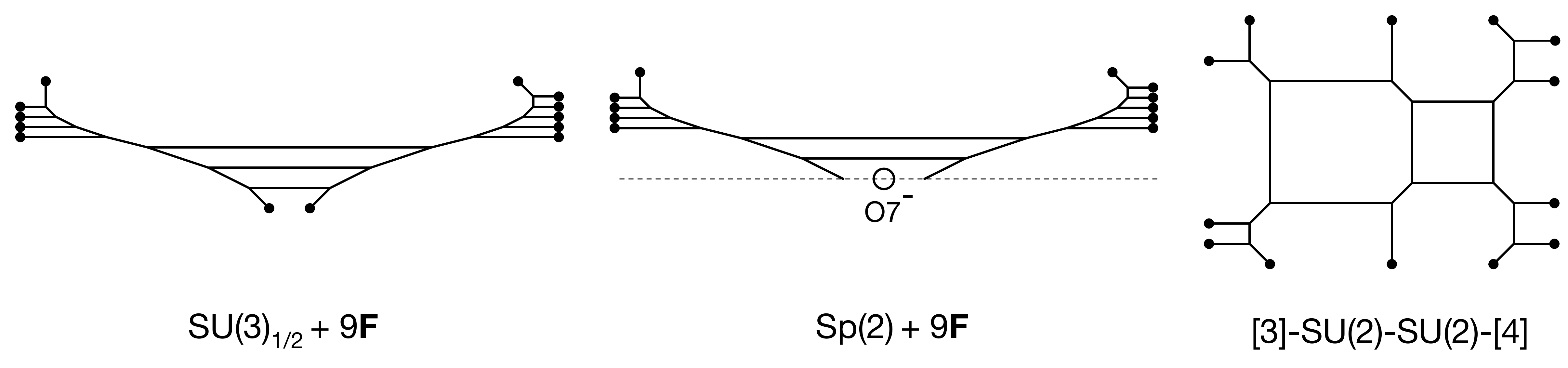}
\caption{5-brane webs for $SU(3)_\frac12+9\bF$, $Sp(2)+9\bF$, and $[SU(2)+3\bF]\times [SU(2)+4\bF]$, which are dual to one anther.}
\label{Fig:SU3-9F-1/2}
\end{figure}
\begin{figure}
\centering
\includegraphics[width=12cm]{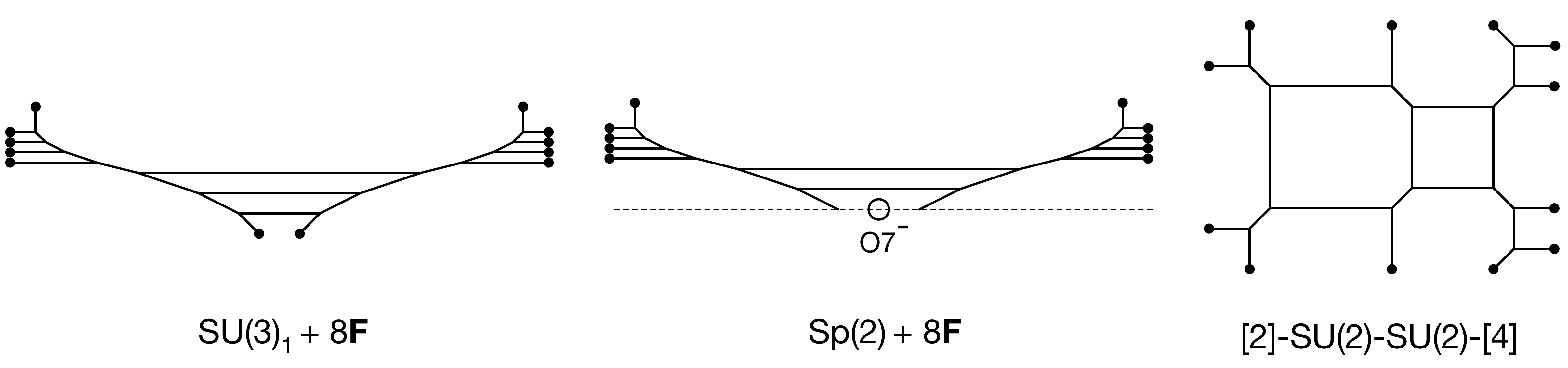}
\caption{5-brane webs for $SU(3)_1+8\bF$, $Sp(2)+8\bF$, and $[SU(2)+2\bF]\times [SU(2)+4\bF]$, which are dual to one anther.}
\label{Fig:SU3-8F-1}
\end{figure}
\begin{figure}
\centering
\includegraphics[width=12cm]{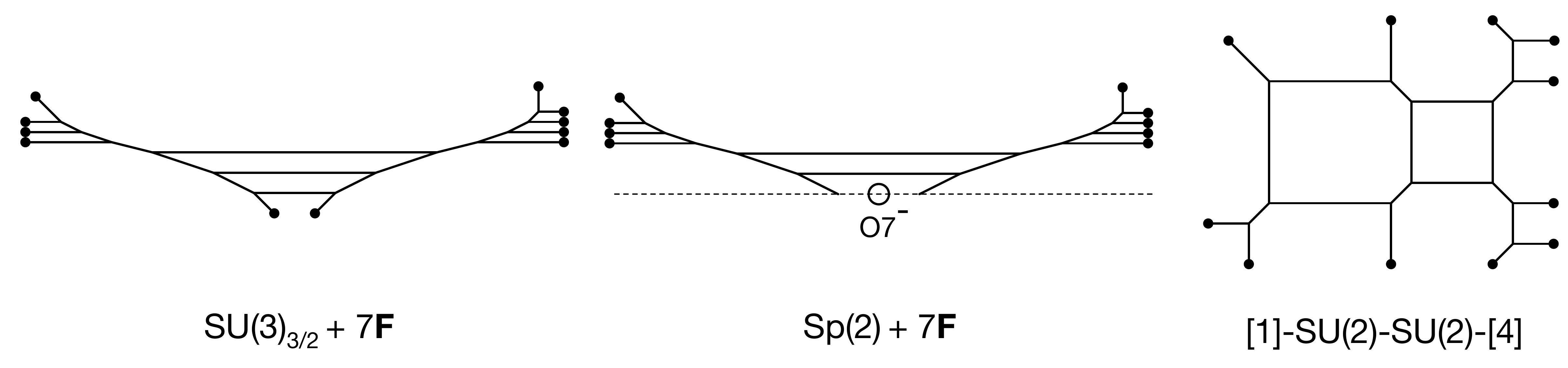}
\caption{5-brane webs for $SU(3)_\frac32+7\bF$, $Sp(2)+7\bF$, and $[SU(2)+1\bF]\times [SU(2)+4\bF]$, which are dual to one anther.}
\label{Fig:SU3-7F-3/2}
\end{figure}
\begin{figure}
\centering
\includegraphics[width=12cm]{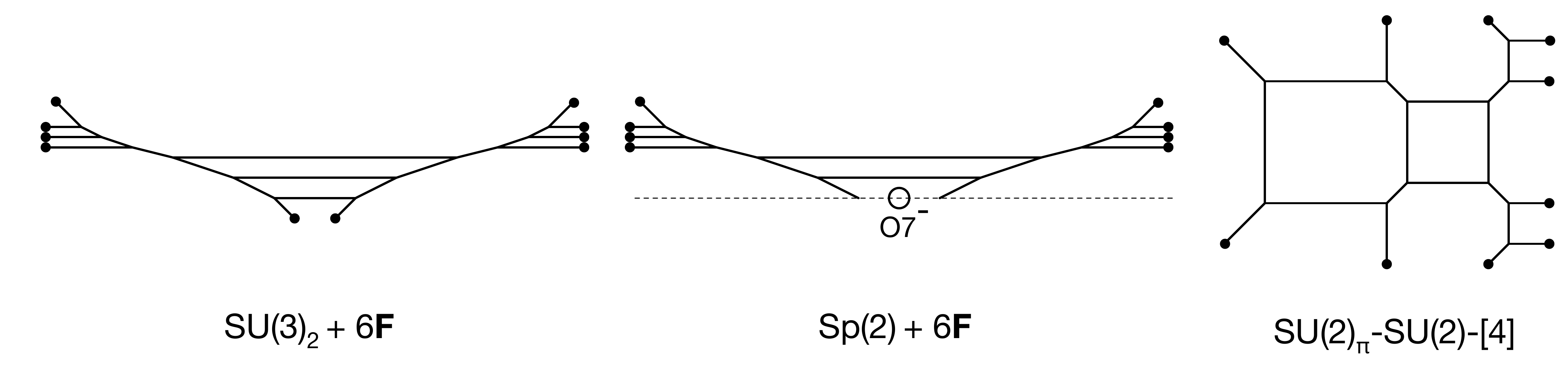}
\caption{5-brane webs for $SU(3)_2+6\bF$, $Sp(2)+6\bF$, and $SU(2)_\pi\times [SU(2)+4\bF]$, which are dual to one anther.}
\label{Fig:SU3-6F-2}
\end{figure}
\begin{figure}
\centering
\includegraphics[width=8cm]{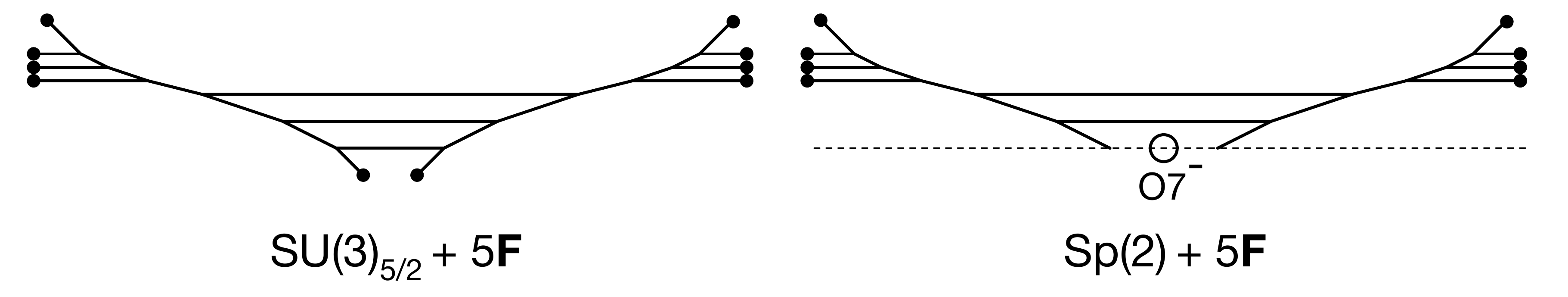}
\caption{5-brane webs for $SU(3)_\frac52+5\bF$ and its dual $Sp(2)+5\bF$.}
\label{Fig:SU3-5F-5/2}
\end{figure}
\begin{figure}
\centering
\includegraphics[width=8cm]{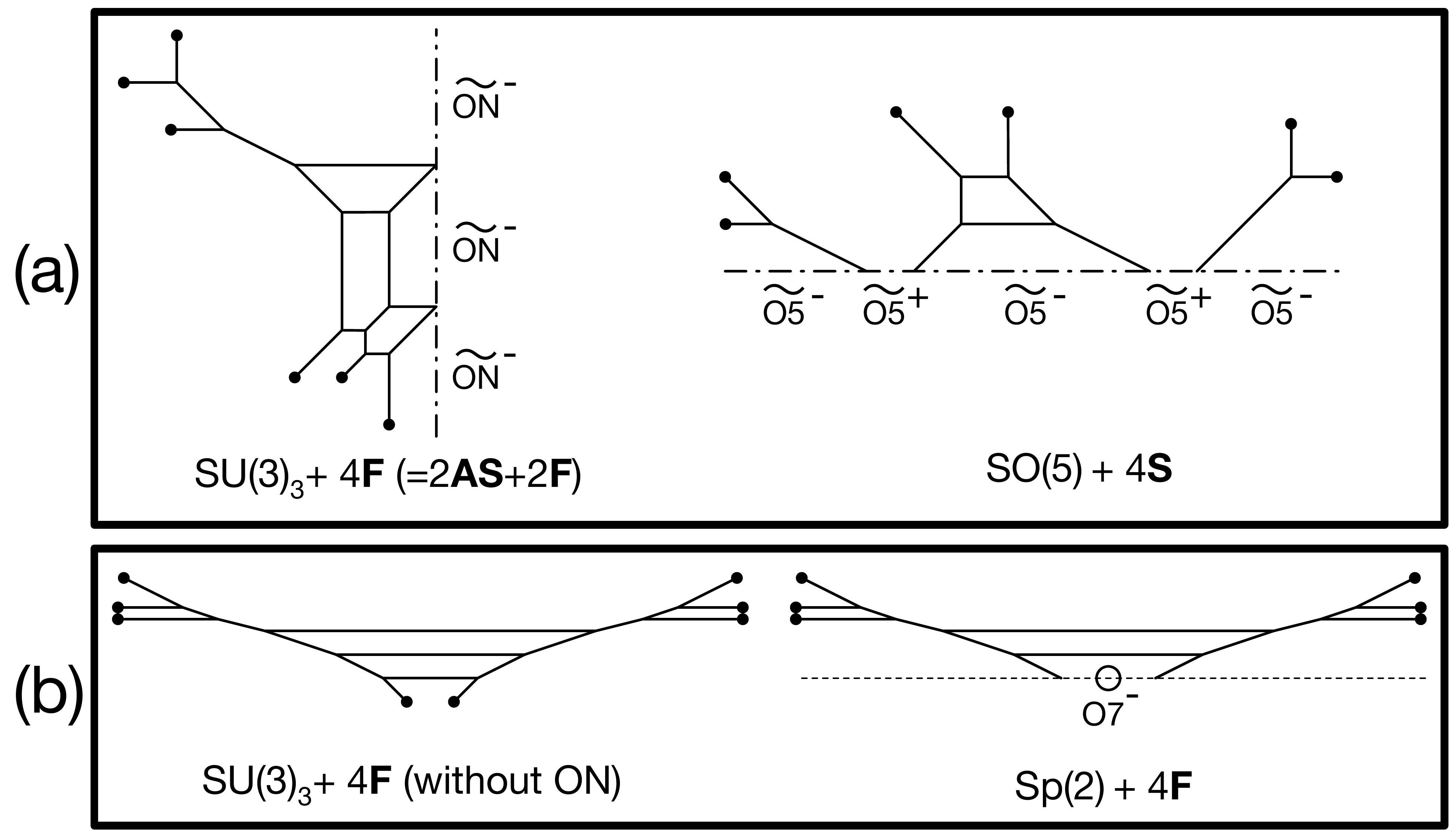}
\caption{5-brane webs for $SU(3)_3+4\bF$ and its dual $Sp(2)+4\bF$ or $SO(5)+4{\bf S}$.}
\label{Fig:SU3-4F-3}
\end{figure}
\begin{figure}
\centering
\includegraphics[width=12cm]{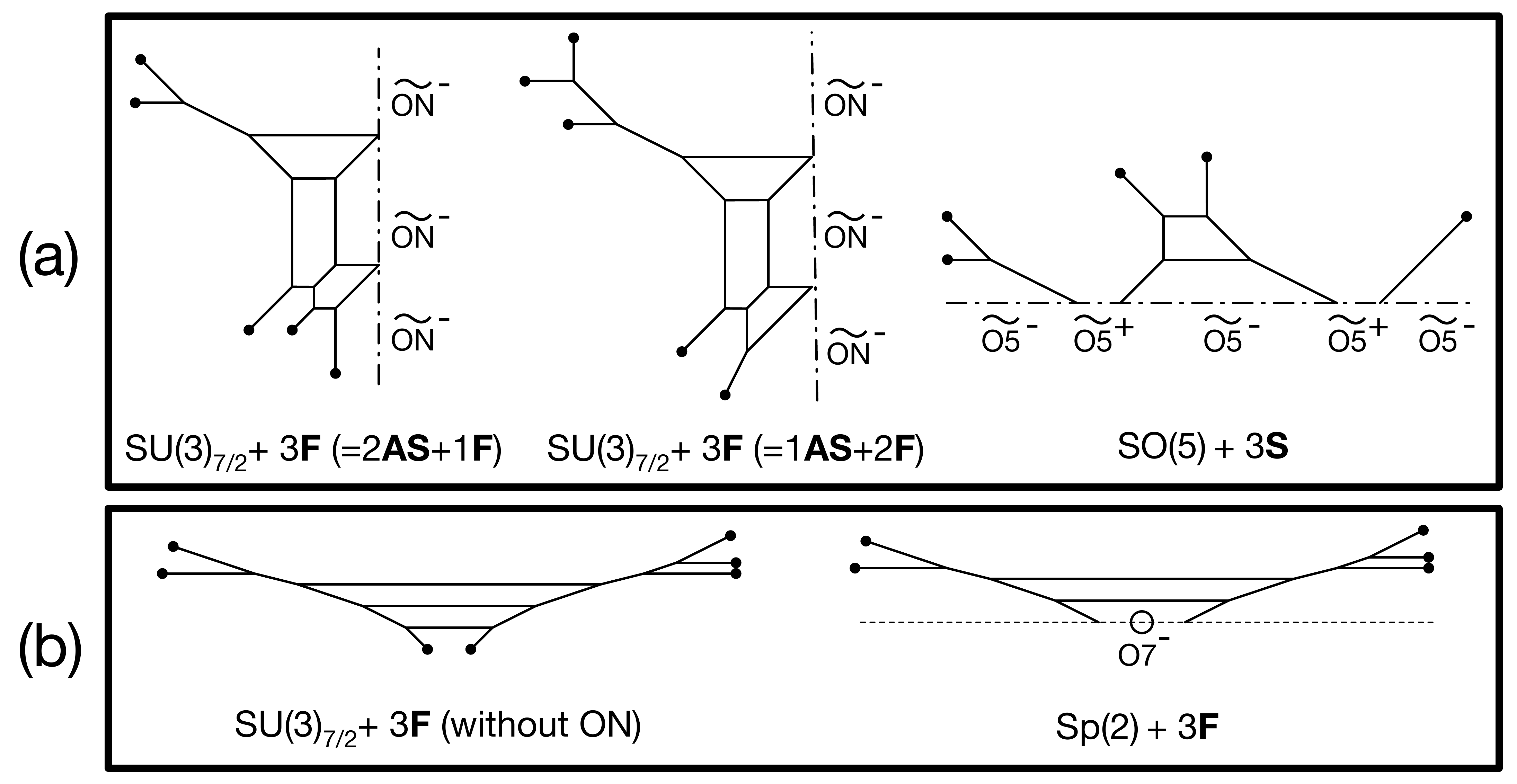}
\caption{Various 5-brane webs for $SU(3)_
\frac72+3\bF$ and its dual $Sp(2)+3\bF$ or $SO(5)+3{\bf S}$.}
\label{Fig:SU3-3F-7/2}
\end{figure}
\begin{figure}
\centering
\includegraphics[width=15cm]{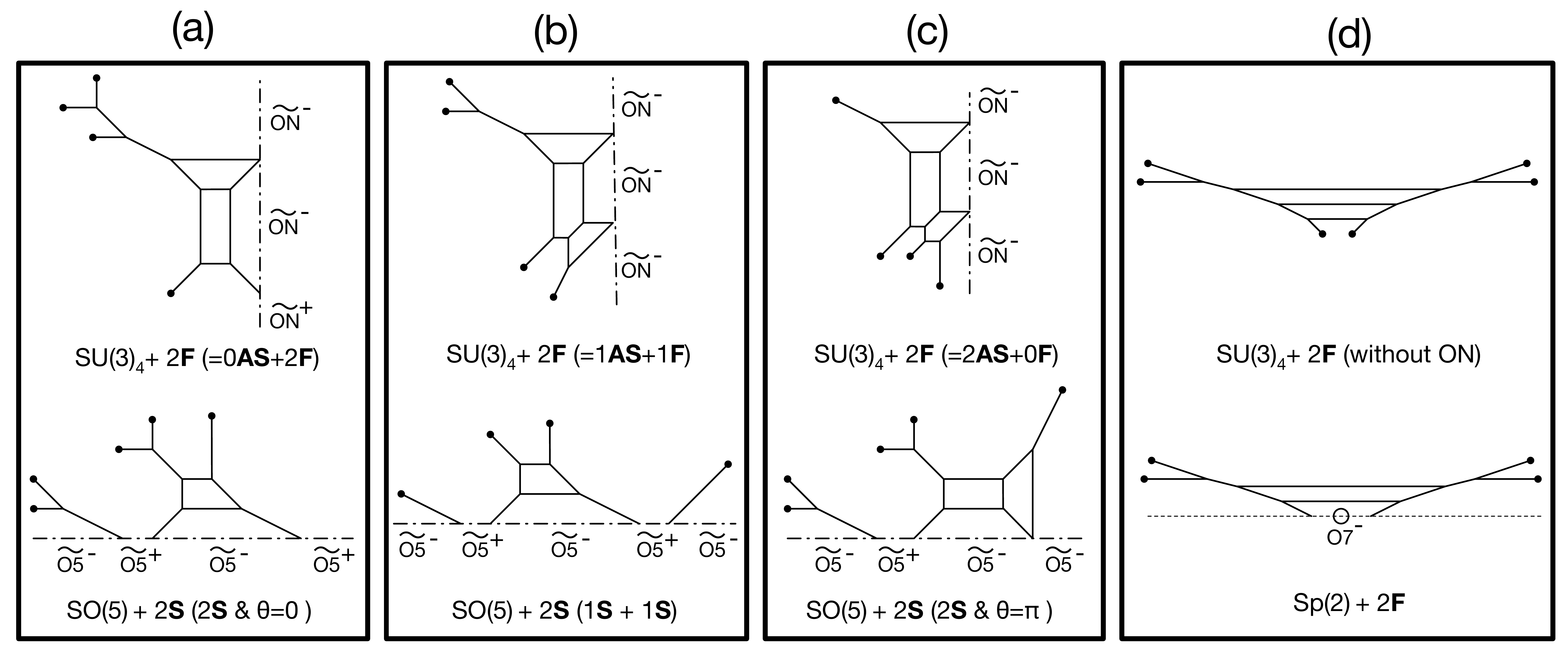}
\caption{Various 5-brane webs for $SU(3)_
4+2\bF$ and its dual $Sp(2)+2\bF$ or $SO(5)+2{\bf S}$.}
\label{Fig:SU3-2F-4}
\end{figure}
\begin{figure}
\centering
\includegraphics[width=12cm]{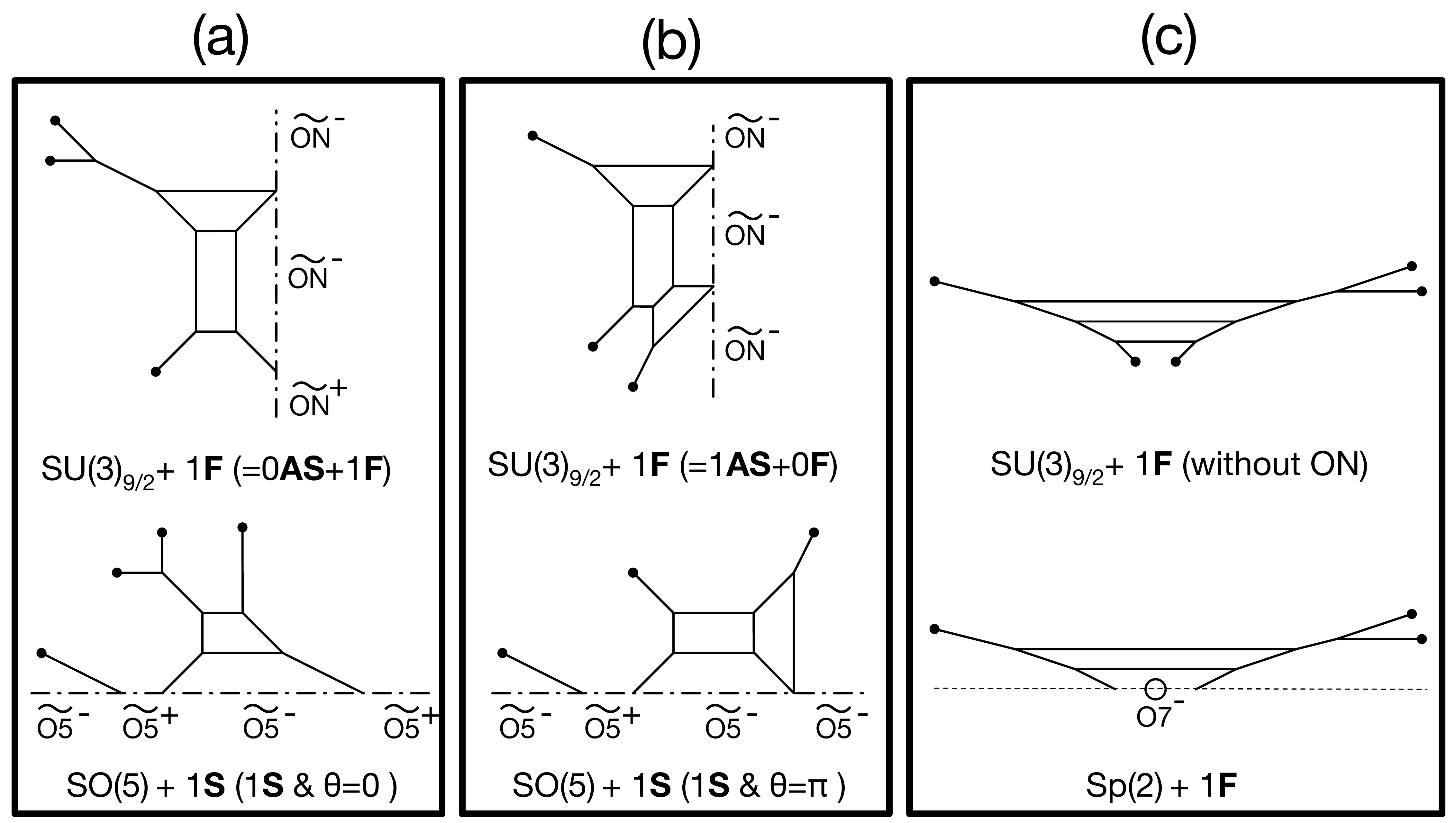}
\caption{Various 5-brane webs for $SU(3)_
\frac92+1\bF$ and its dual $Sp(2)+1\bF$ or $SO(5)+1{\bf S}$.}
\label{Fig:SU3-1F-9/2}
\end{figure}
\begin{figure}
\centering
\includegraphics[width=8cm]{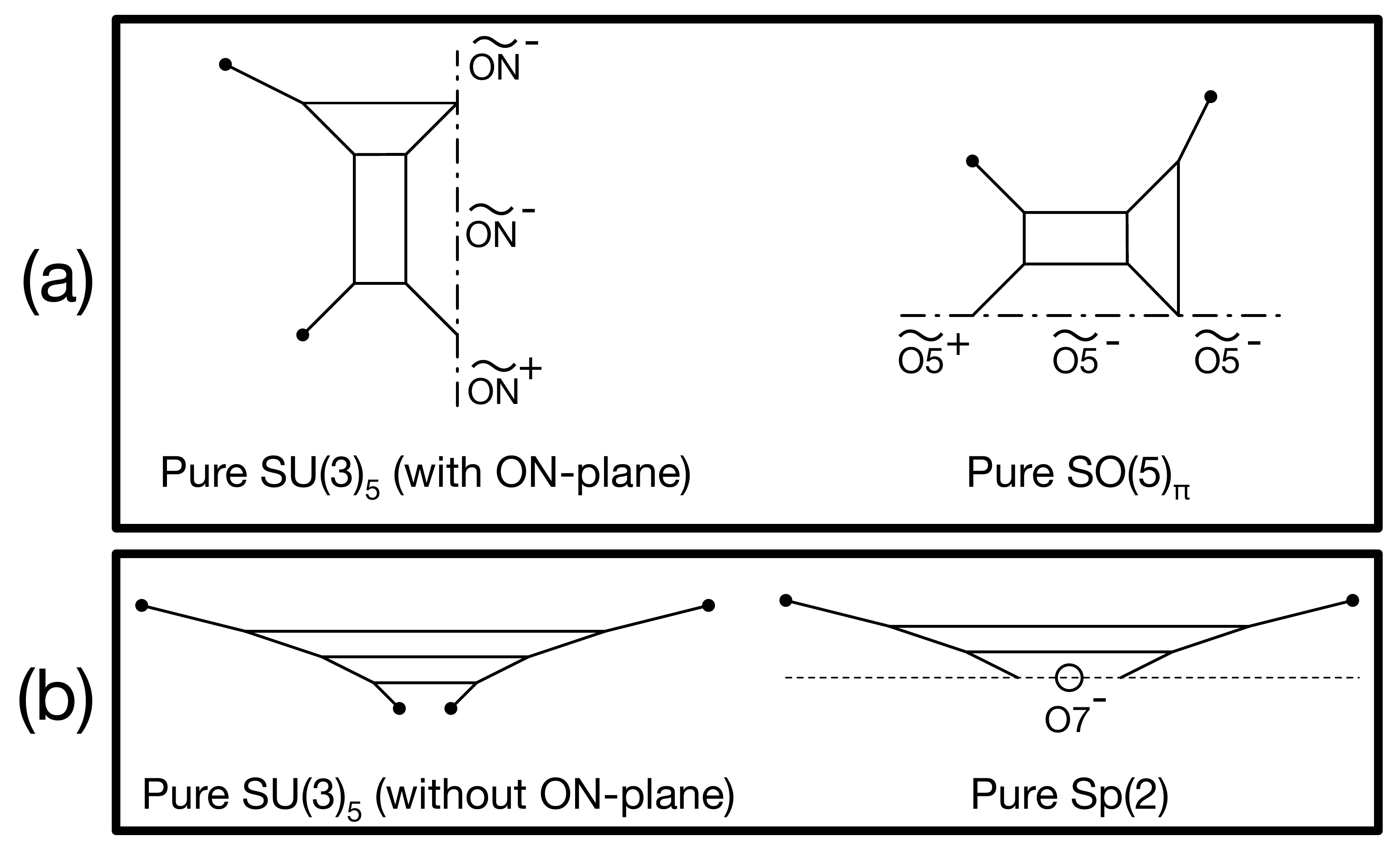}
\caption{Various 5-brane webs for $SU(3)_
5$ and its dual $Sp(2)_\pi$ or $SO(5)_\pi$. 
For figure (b), the discrete theta angle for the pure $Sp(2)$ theory is determined by how an $O7^-$-plane is resolved. For instance, the resolution of an $O7^-$-plane into a pair of [1,-1] and [1,1] 7-branes corresponds to the discrete theta angle $\theta=\pi$ for $Sp(2)$, while the resolution of an $O7^-$-plane into a pair of [0,-1] and [2,1] 7-branes corresponds to $\theta=0$ for $Sp(2)$ 
%Note that it is not obvious to read off what the discrete theta angle of the pure $Sp(2)$ theory from a web diagram with an $O7^-$-plane depicted in figure (b), 
%but the flop transition distinguishes $\theta=0$ and $\theta=\pi$ 
\cite{Bergman:2015dpa}.
}
\label{Fig:SU3-5}
\end{figure}
\begin{figure}
\centering
\includegraphics[width=8cm]{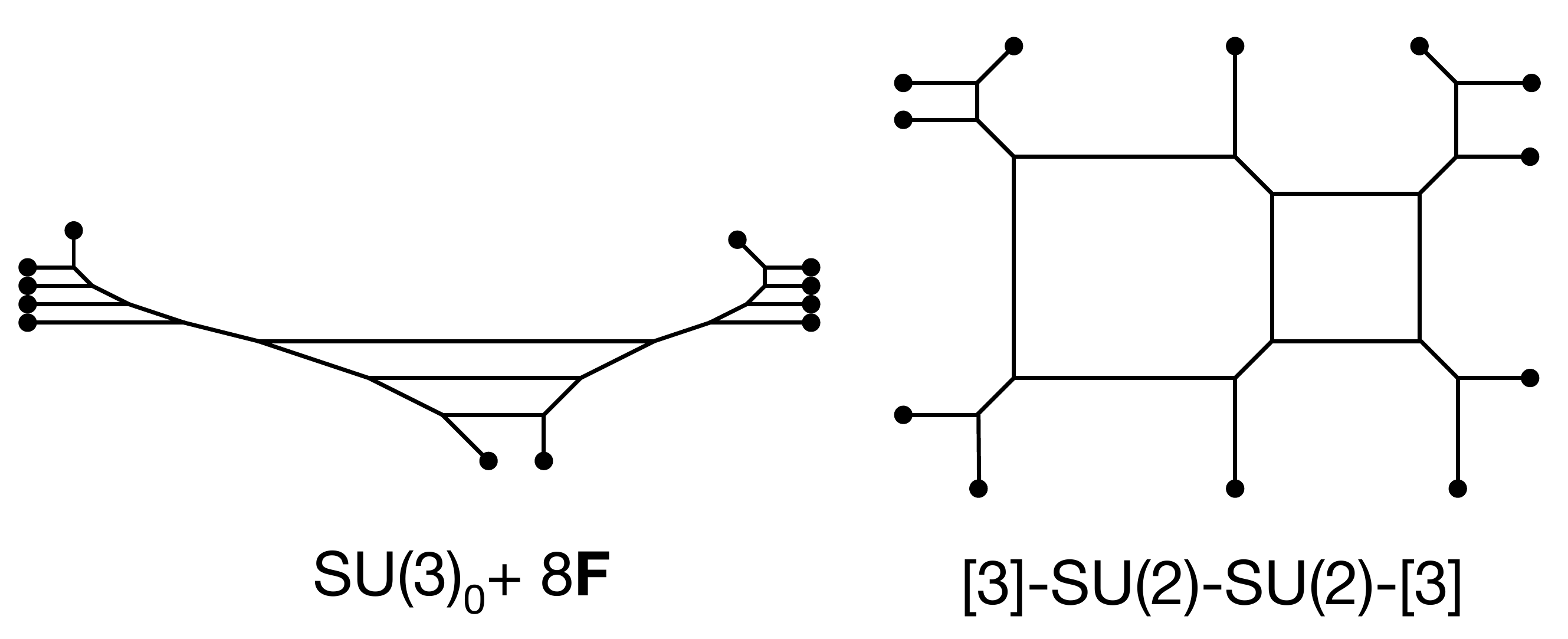}
\caption{5-brane webs for $SU(3)_0+8\bF$ and its dual quiver $[SU(2)+3\bF]\times [SU(2)+3\bF]$.}
\label{Fig:SU3-8F-0}
\end{figure}
\begin{figure}
\centering
\includegraphics[width=8cm]{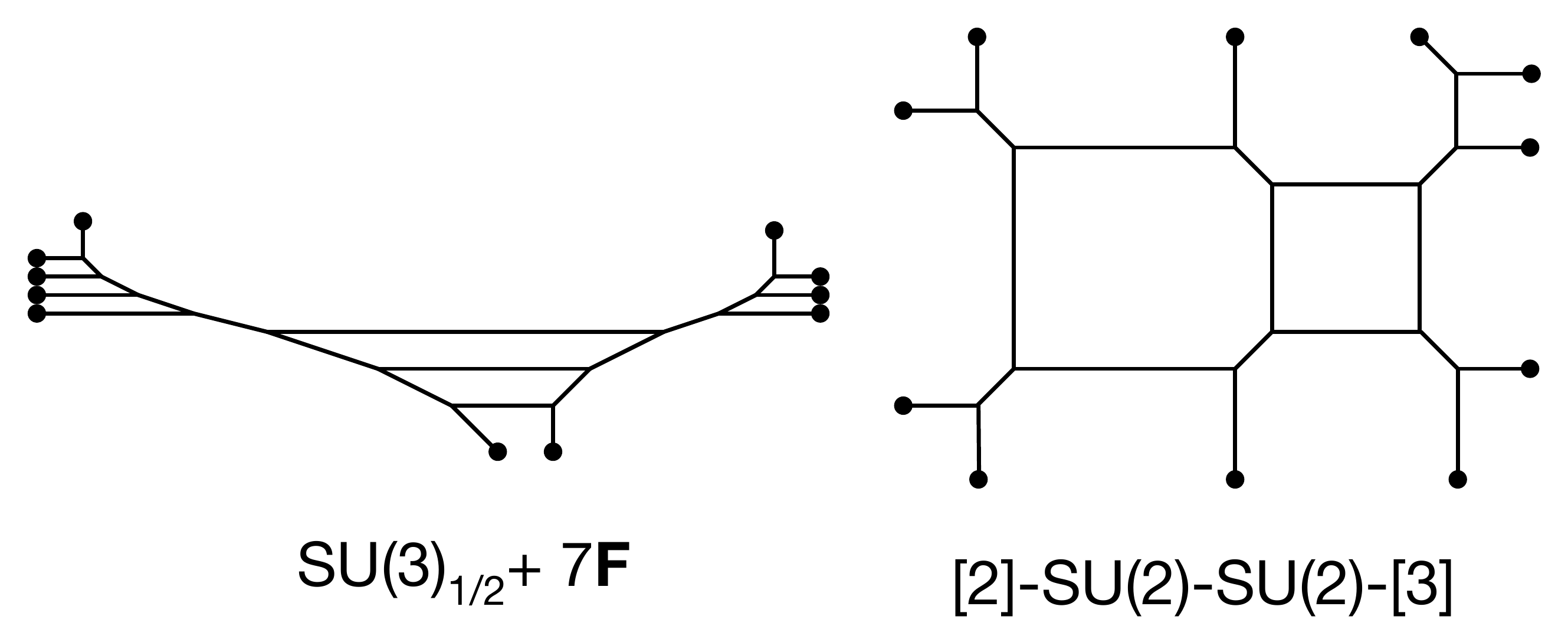}
\caption{5-brane webs for $SU(3)_\frac12+7\bF$ and its dual quiver $[SU(2)+2\bF]\times [SU(2)+3\bF]$.}
\label{Fig:SU3-7F-1/2}
\end{figure}
\begin{figure}
\centering
\includegraphics[width=8cm]{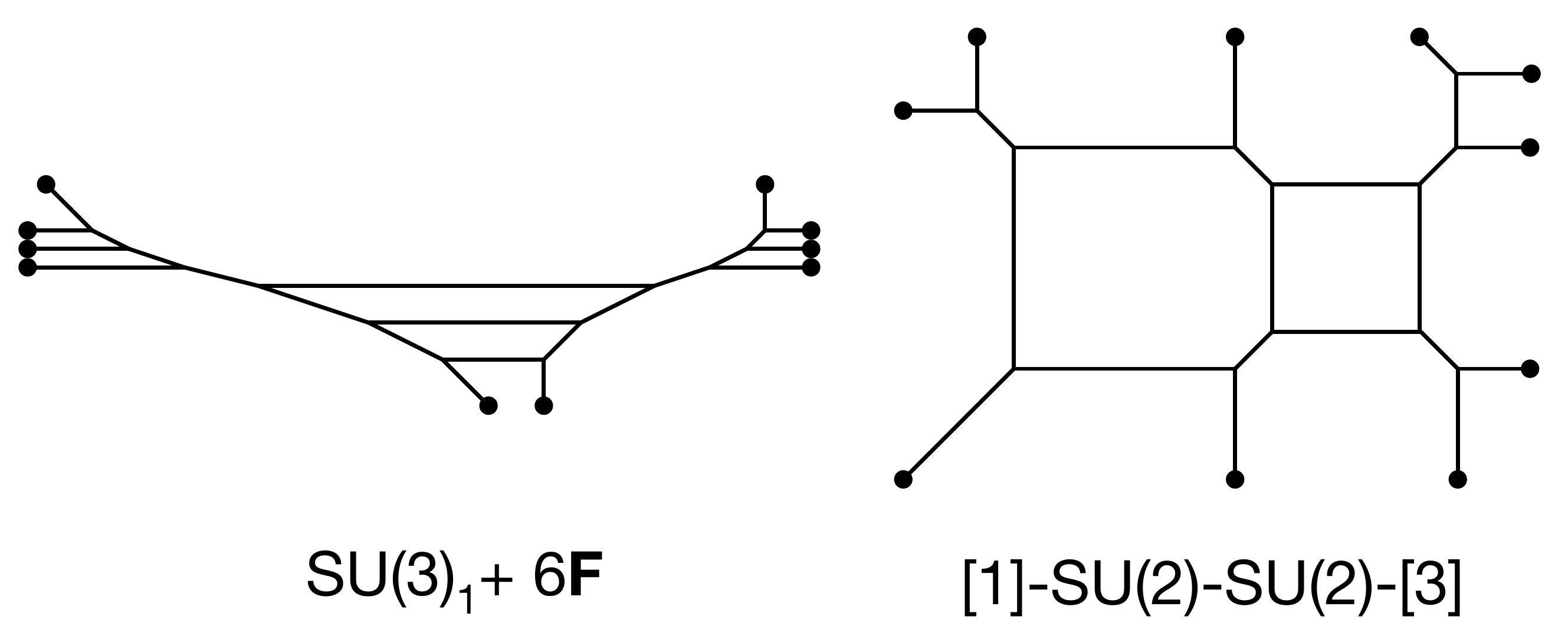}
\caption{5-brane webs for $SU(3)_1+6\bF$ and its dual quiver $[SU(2)+1\bF]\times [SU(2)+3\bF]$.}
\label{Fig:SU3-6F-1}
\end{figure}
\begin{figure}
\centering
\includegraphics[width=8cm]{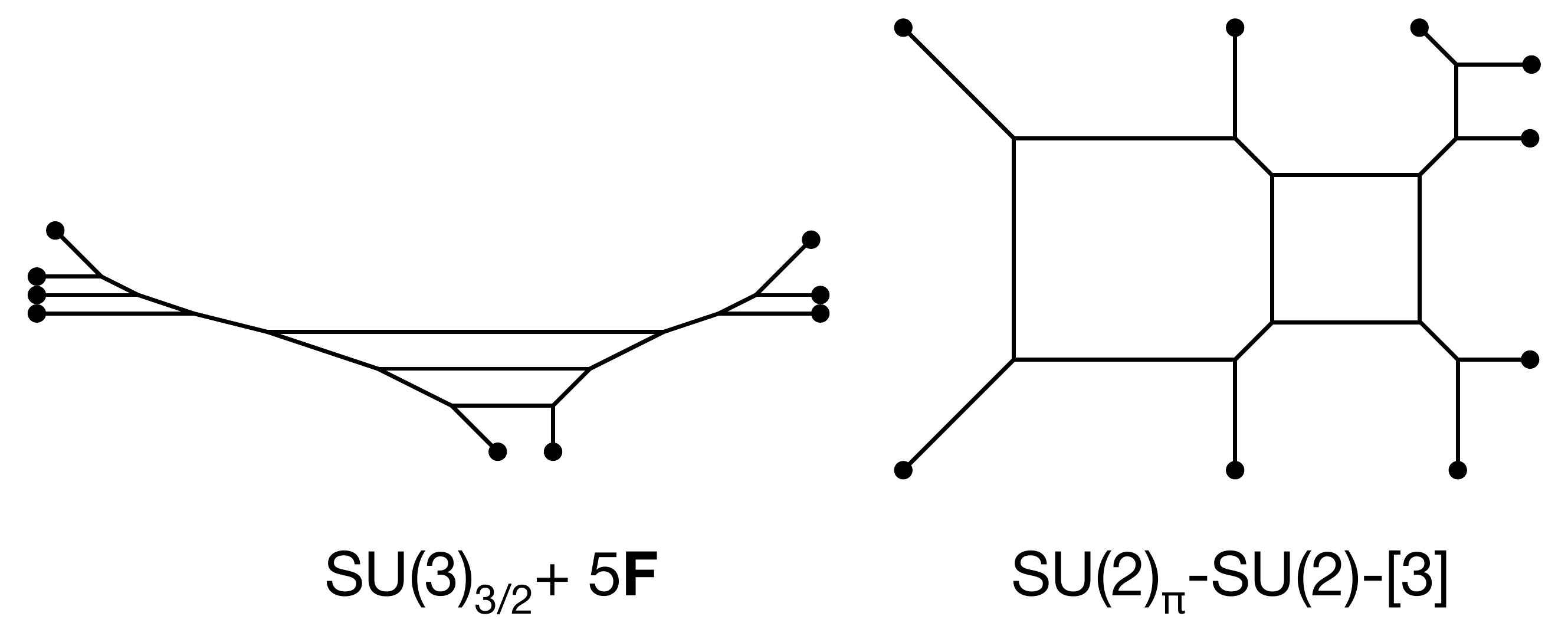}
\caption{5-brane webs for $SU(3)_\frac32+5\bF$ and its dual quiver $SU(2)_\pi\times [SU(2)+3\bF]$.}
\label{Fig:SU3-5F-3/2}
\end{figure}
\begin{figure}
\centering
\includegraphics[width=4cm]{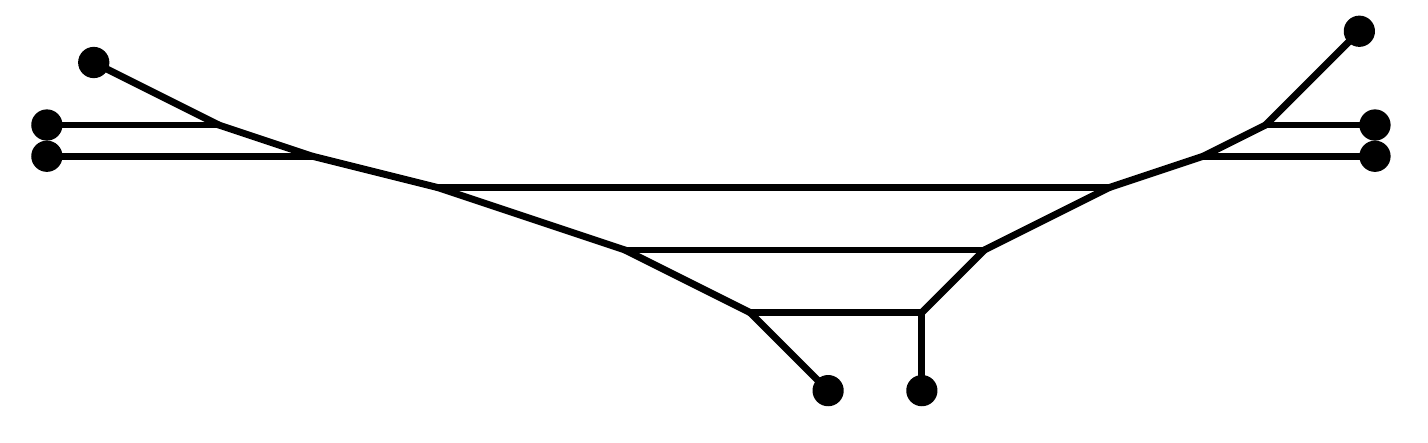}
\caption{A 5-brane web for $SU(3)_{2} + 4 \mathbf{F}$}
\label{Fig:SU3-4F-2}
\end{figure}
\begin{figure}
\centering
\includegraphics[width=8cm]{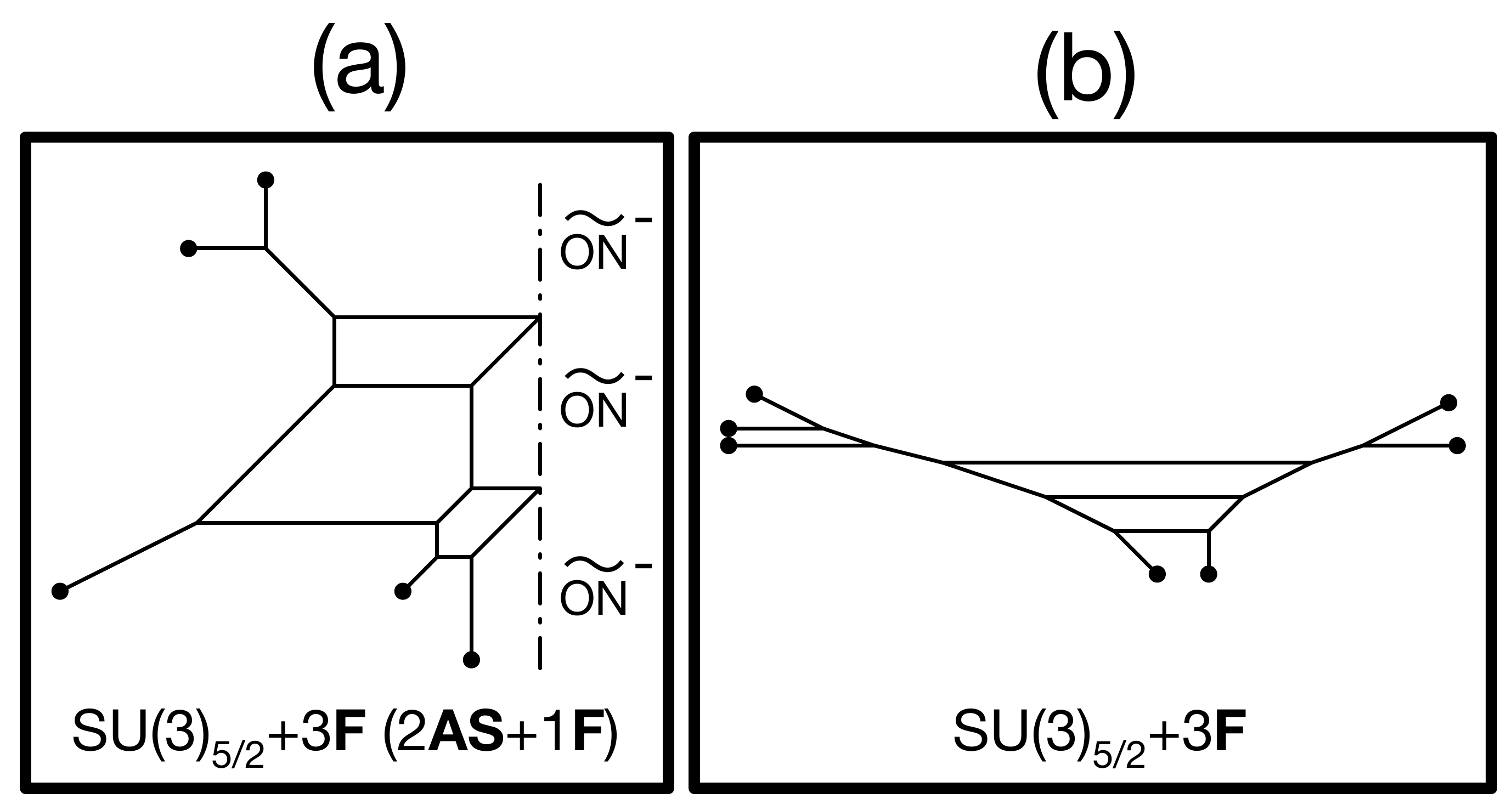}
\caption{5-brane webs for $SU(3)_\frac52+3\bF$ with and without an $\widetilde{ON}$-plane.}
\label{Fig:SU3-3F-5/2}
\end{figure}
\begin{figure}
\centering
\includegraphics[width=12cm]{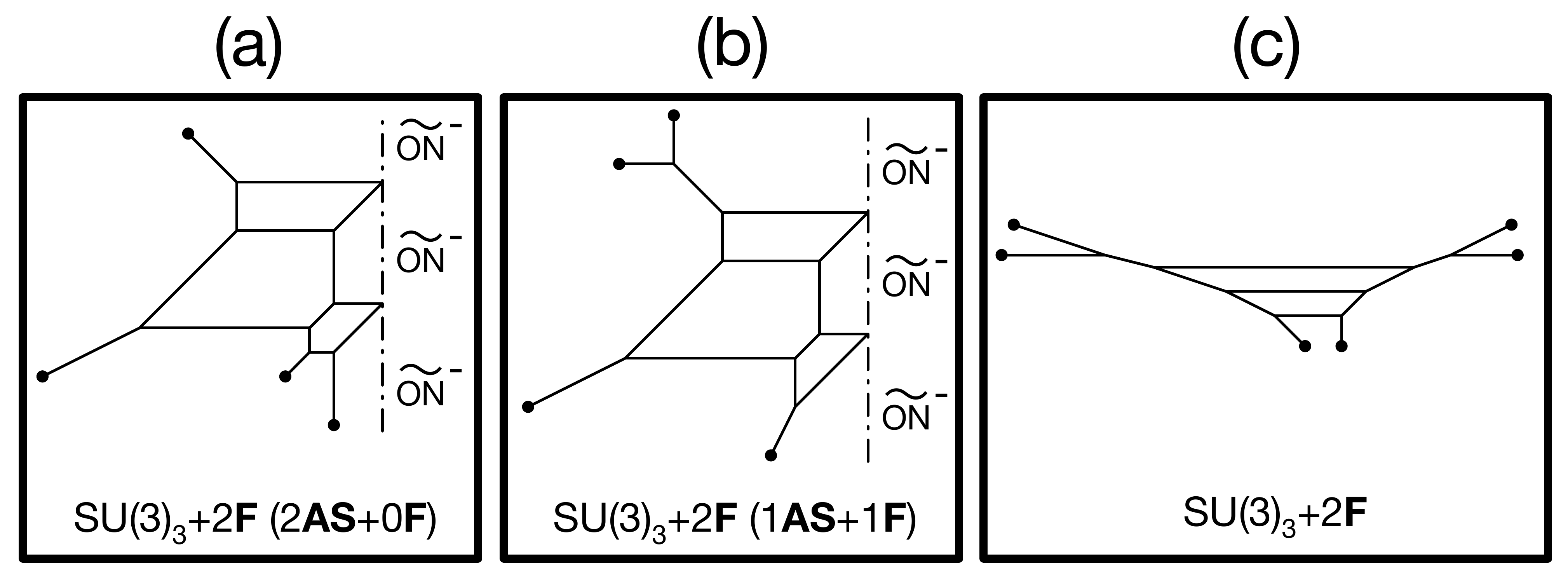}
\caption{Various 5-brane webs for $SU(3)_3+2\bF$ with and without an $\widetilde{ON}$-plane.}
\label{Fig:SU3-2F-3}
\end{figure}
\begin{figure}
\centering
\includegraphics[width=12cm]{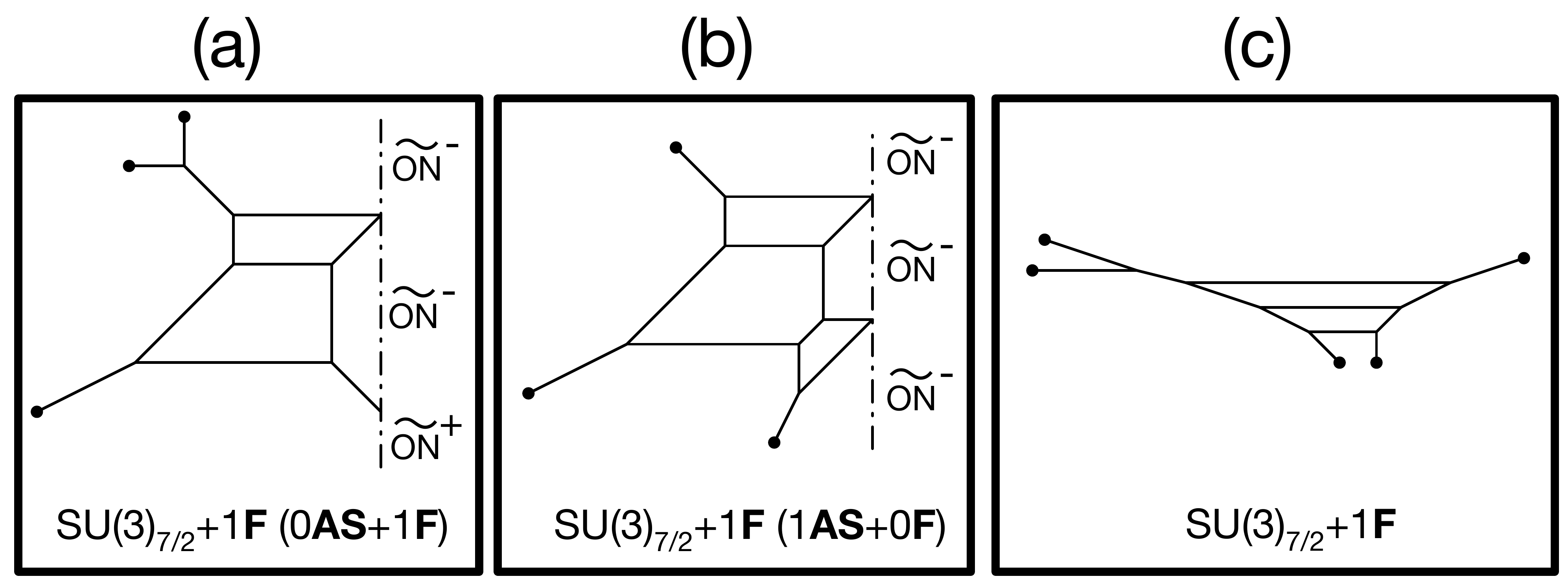}
\caption{Three different 5-brane configurations: (a) $SU(3)_\frac72+1\bF$ with an $\widetilde{ON}$-plane. (b) $SU(3)_\frac72+1\AS$ with an $\widetilde{ON}$-plane. (c) $SU(3)_\frac72+1\bF$ without an $\widetilde{ON}$-plane.}
\label{Fig:SU3-1F-7/2}
\end{figure}
\clearpage
\begin{figure}
\centering
\includegraphics[width=8cm]{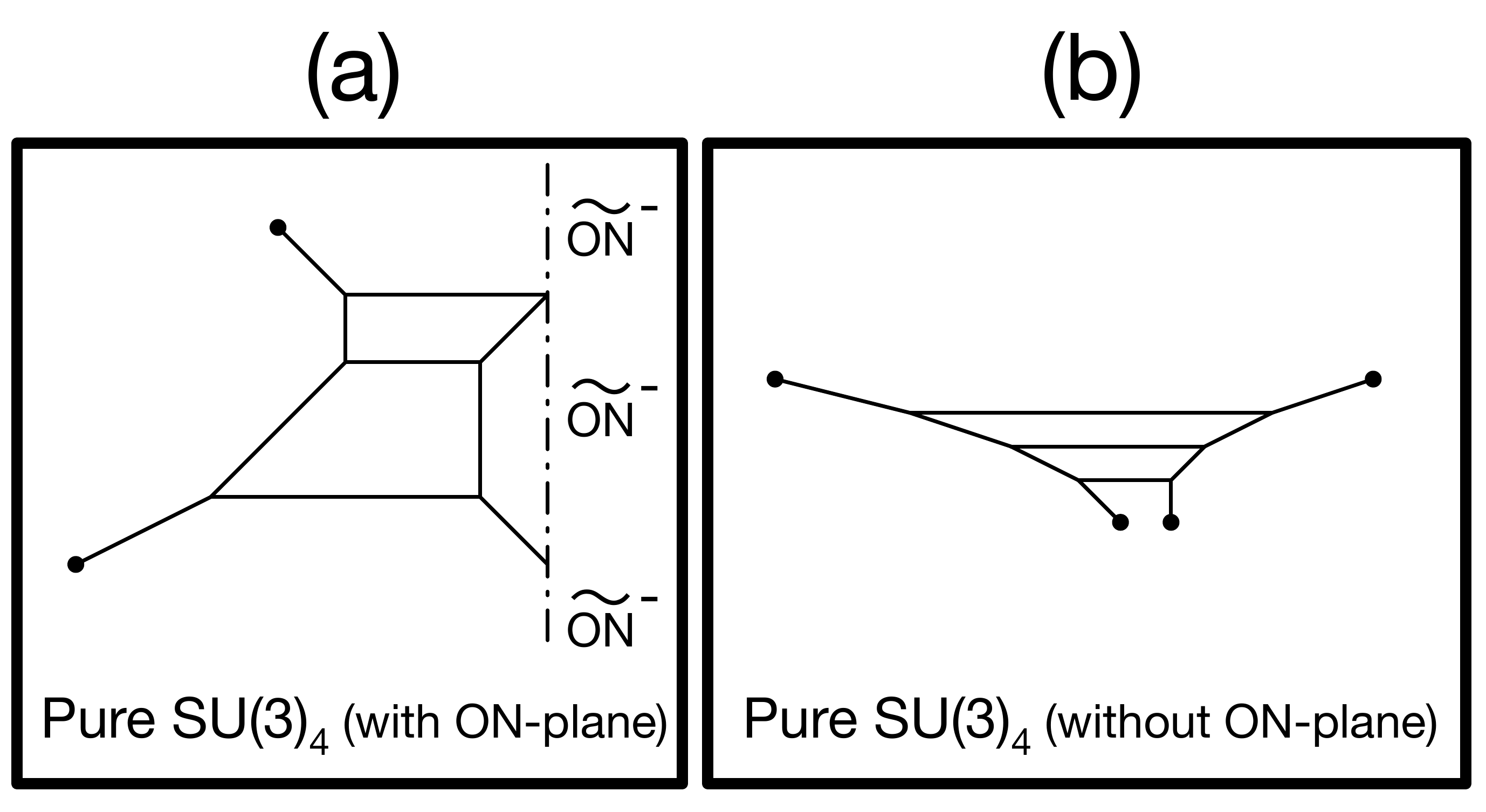}
\caption{(a) A web diagram for $SU(3)_4$ with an $\widetilde{ON}$-plane. (b) A conventional brane web diagram for $SU(3)_4$.}
\label{Fig:SU3-4}
\end{figure}
\begin{figure}
\centering
\includegraphics[width=8cm]{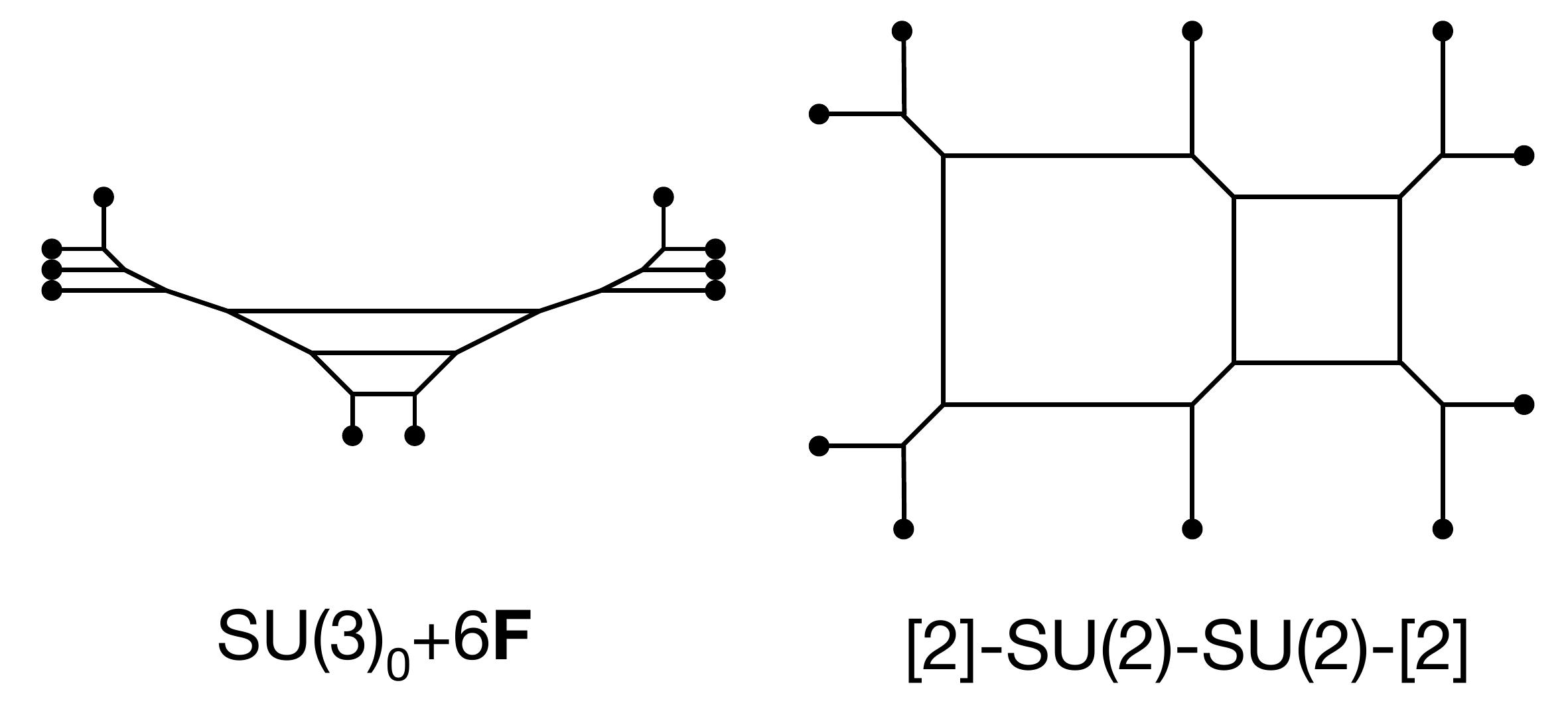}
\caption{5-brane webs for $SU(3)_{0} + 6 \mathbf{F}$ and its dual quiver theory $[SU(2)+2\bF]\times [SU(2)+2\bF]$.}
\label{Fig:SU3-6F-0}
\end{figure}
\begin{figure}
\centering
\includegraphics[width=8cm]{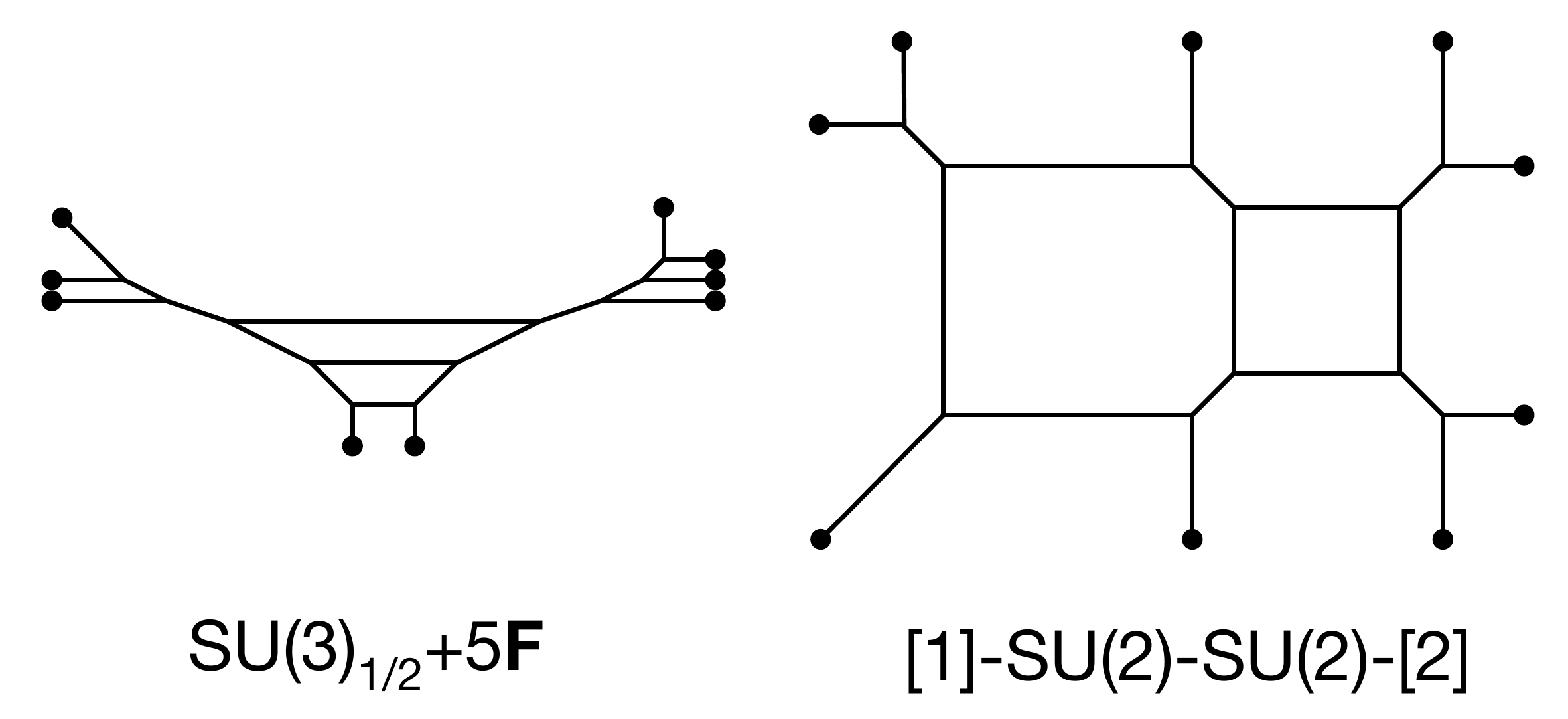}
\caption{5-brane webs for $SU(3)_{\frac12} + 5 \mathbf{F}$ and its dual quiver theory $[SU(2)+1\bF]\times [SU(2)+2\bF]$.}
\label{Fig:SU3-5F-1/2}
\end{figure}
\begin{figure}
\centering
\includegraphics[width=8cm]{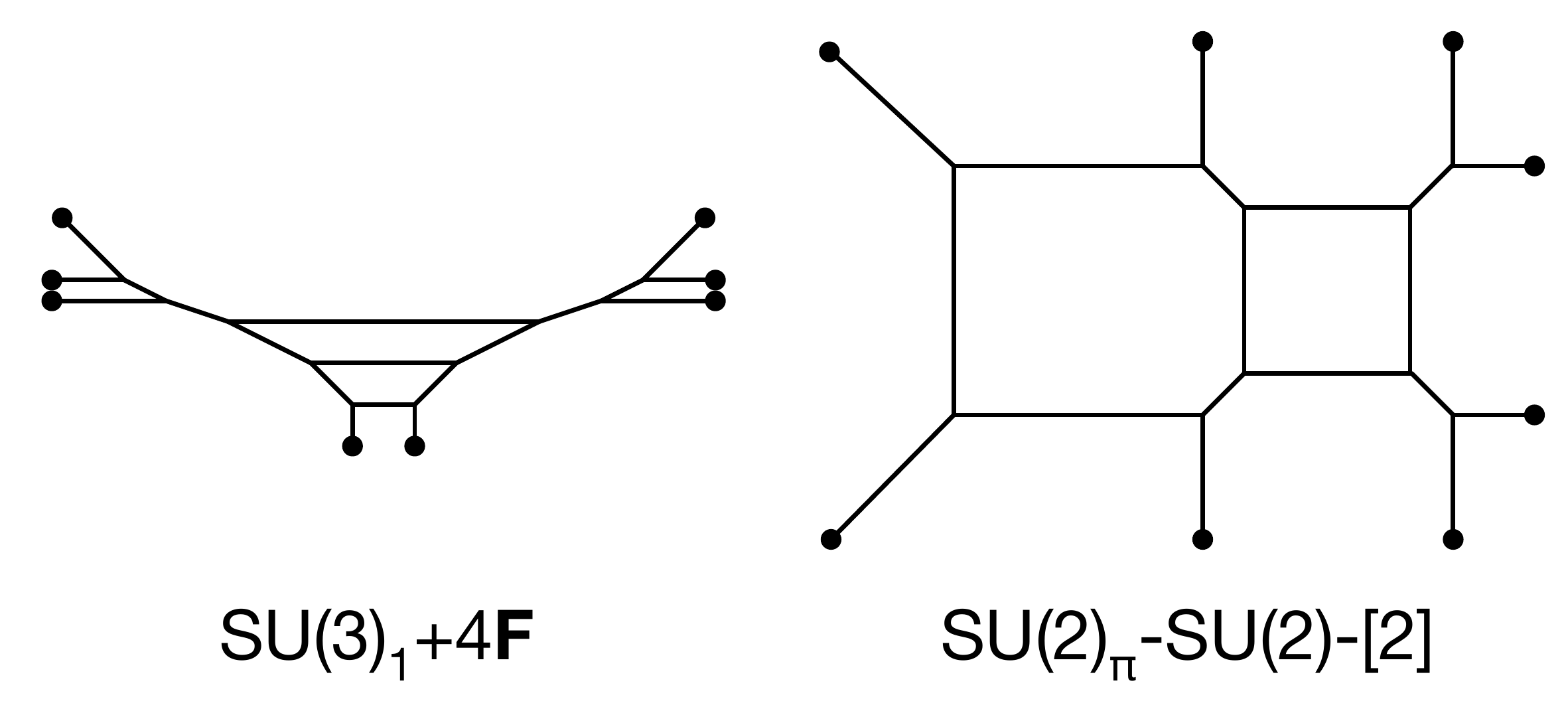}
\caption{5-brane webs for $SU(3)_{1} + 4 \mathbf{F}$ and its dual quiver theory $SU(2)_\pi\times [SU(2)+2\bF]$.}
\label{Fig:SU3-4F-1}
\end{figure}
\begin{figure}%[H]
\centering
\includegraphics[width=4cm]{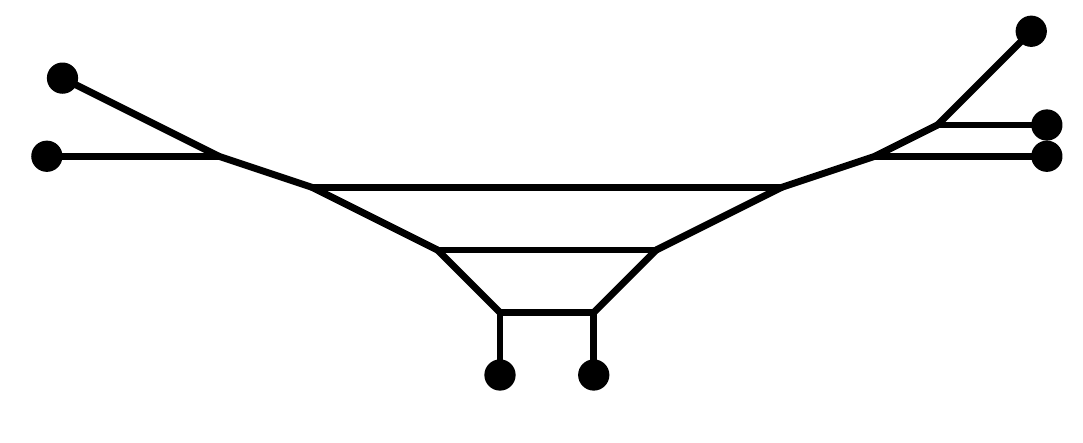}
\caption{A 5-brane web diagram for $SU(3)_{\frac{3}{2}} + 3 \mathbf{F}$.}
\label{Fig:SU3-3F-3/2}
\end{figure}
\begin{figure}%[H]
\centering
\includegraphics[width=8cm]{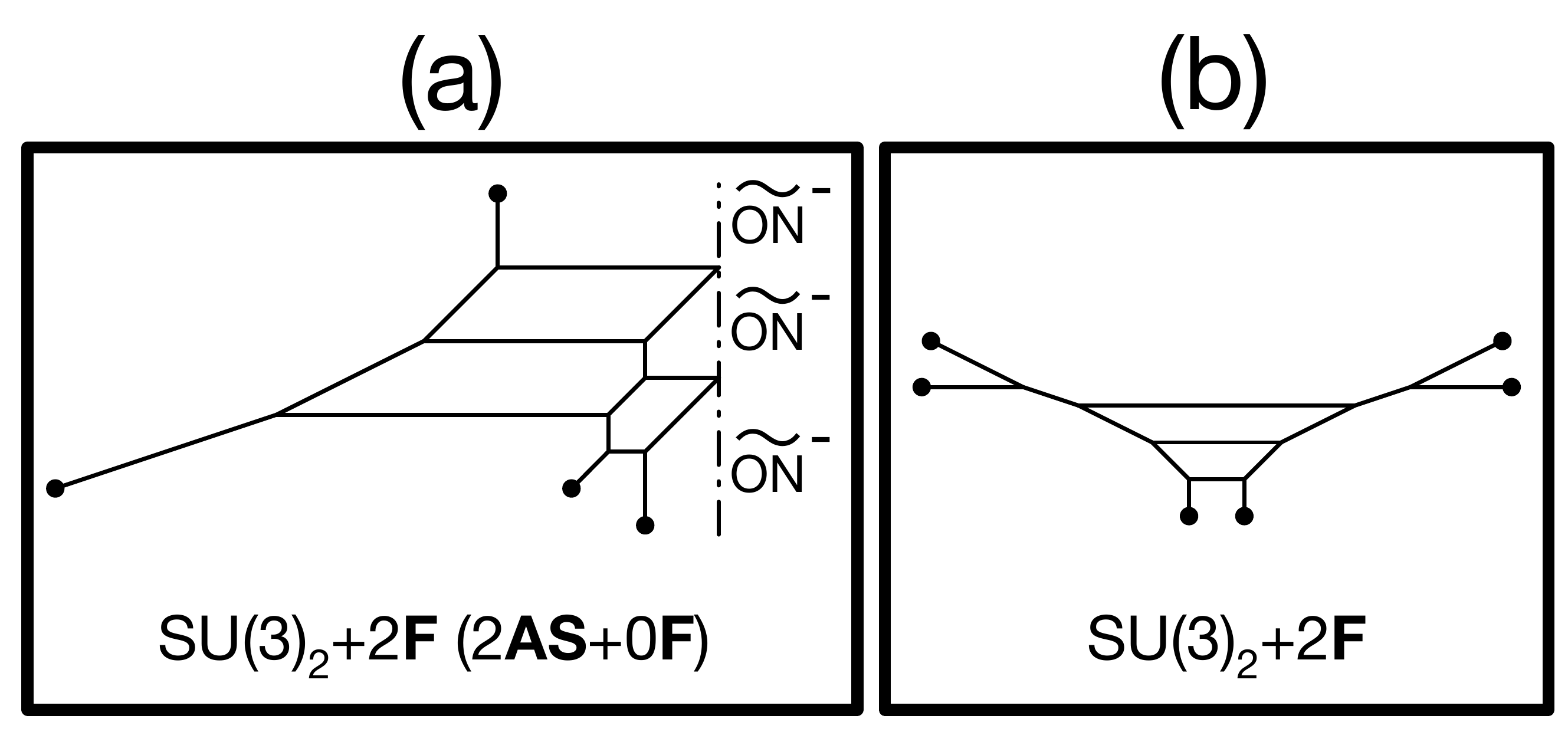}
\caption{(a) A web diagram for $SU(3)_2+2\bF$ with an $\widetilde{ON}$-plane which can be understood as $SU(3)_2+2\AS$. (b) A conventional brane web diagram for $SU(3)_2+2\bF$.}
\label{Fig:SU3-2F-2}
\end{figure}
\begin{figure}
\centering
\includegraphics[width=8cm]{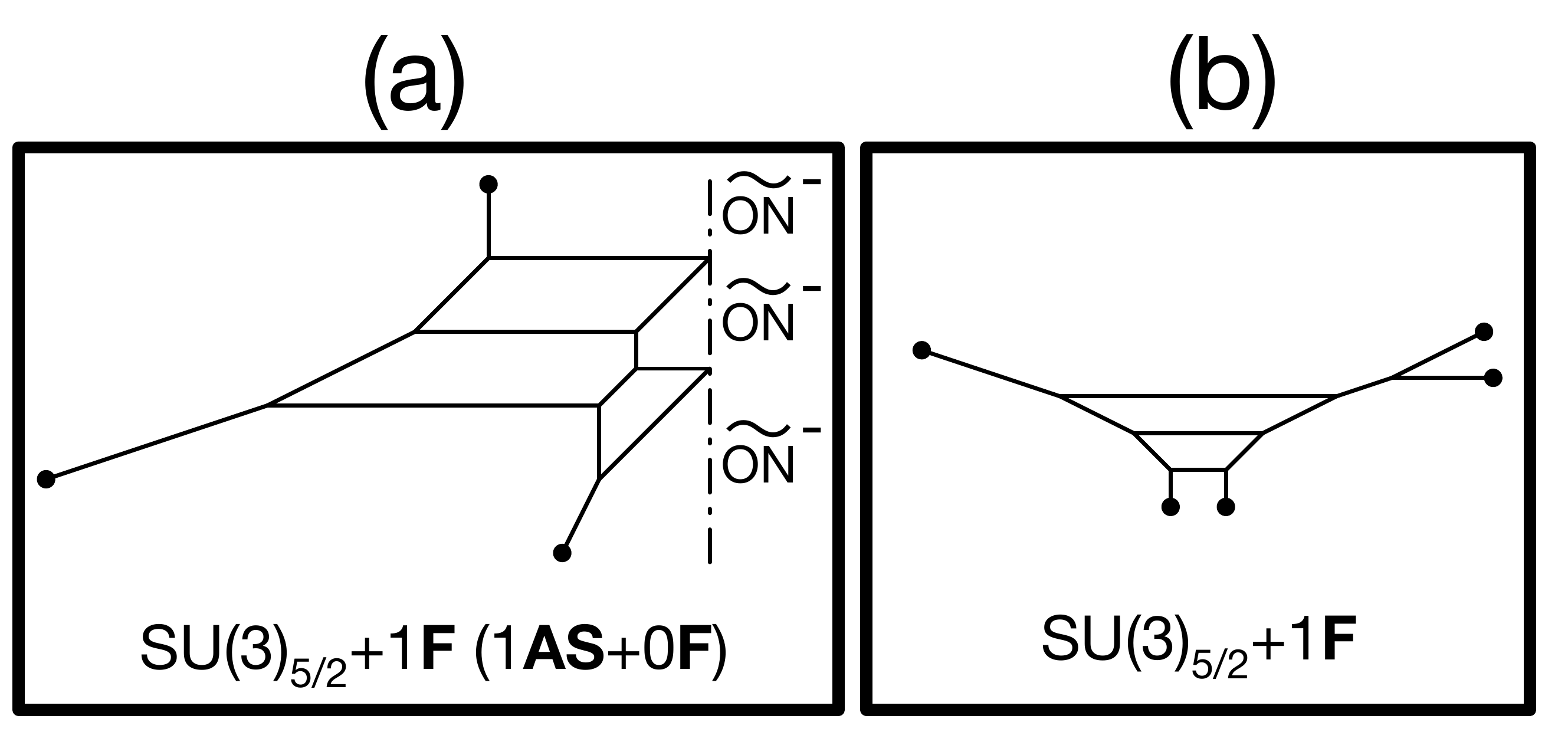}
\caption{(a) A web diagram for $SU(3)_\frac52+1\bF$ theory with an $\widetilde{ON}$-plane.~(b) a conventional web diagram for $SU(3)_\frac52+1\bF$ without an $\widetilde{ON}$-plane.}
\label{Fig:SU3-1F-5/2}
\end{figure}
\begin{figure}
\centering
\includegraphics[width=8cm]{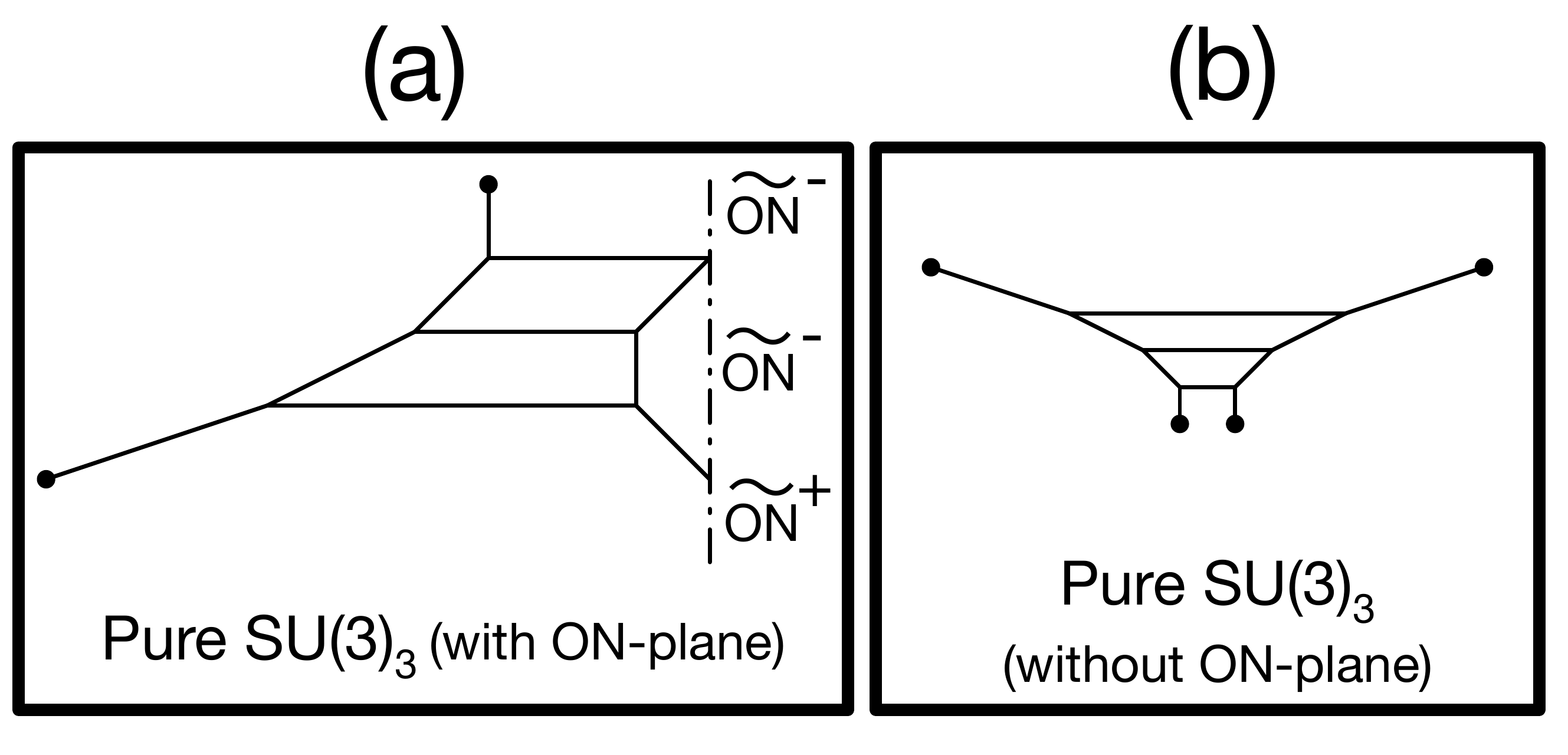}
\caption{(a) A web diagram for the pure $SU(3)_3$ theory with an $\widetilde{ON}$-plane.~(b) a conventional web diagram for the pure $SU(3)_3$ theory without an $\widetilde{ON}$-plane. }
\label{Fig:SU3-3}
\end{figure}
\begin{figure}
\centering
\includegraphics[width=8cm]{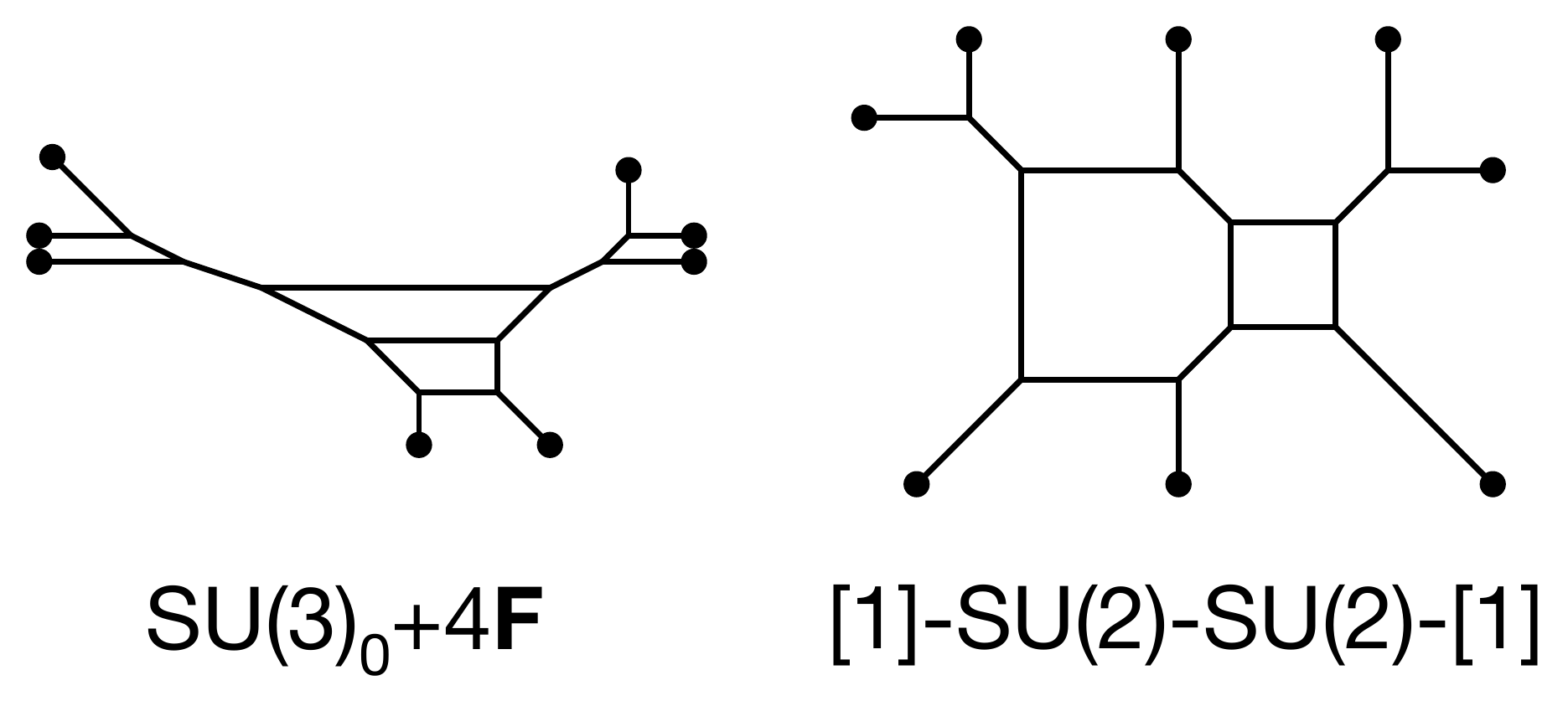}
\caption{5-brane webs for $SU(3)_0+4\bF$ and its S-dual quiver theory $[SU(2)+1\bF]\times [SU(2)+1\bF]$. The duality can be seen by an S-duality and Hanany-Witten transitions.}
\label{Fig:SU3-4F-0}
\end{figure}
\begin{figure}
\centering
\includegraphics[width=8cm]{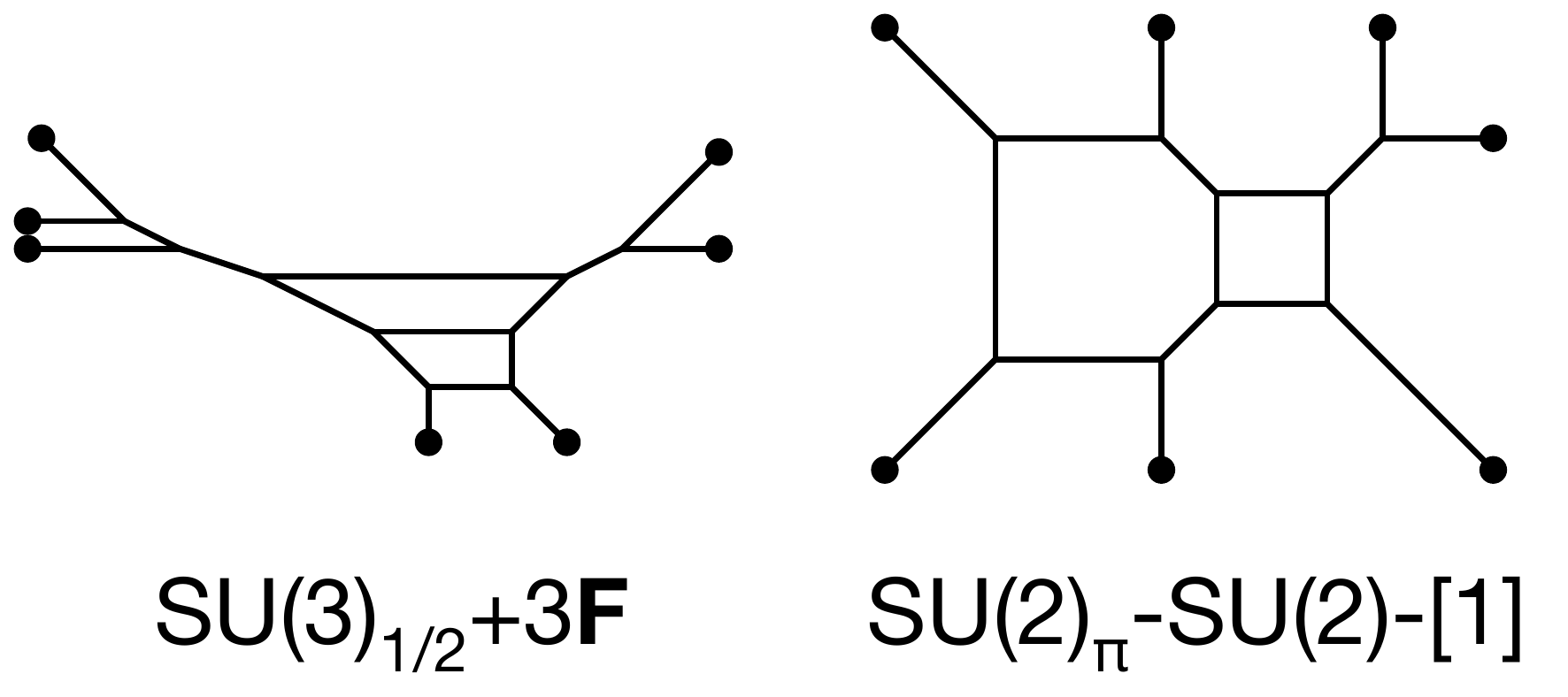}
\caption{5-brane webs for $SU(3)_\frac12+3\bF$ and $SU(2)_\pi\times [SU(2)+1\bF]$ which are dual to each other. }
\label{Fig:SU3-3F-1/2}
\end{figure}
\begin{figure}
\begin{minipage}{0.45\hsize}
\centering
\includegraphics[width=4cm]{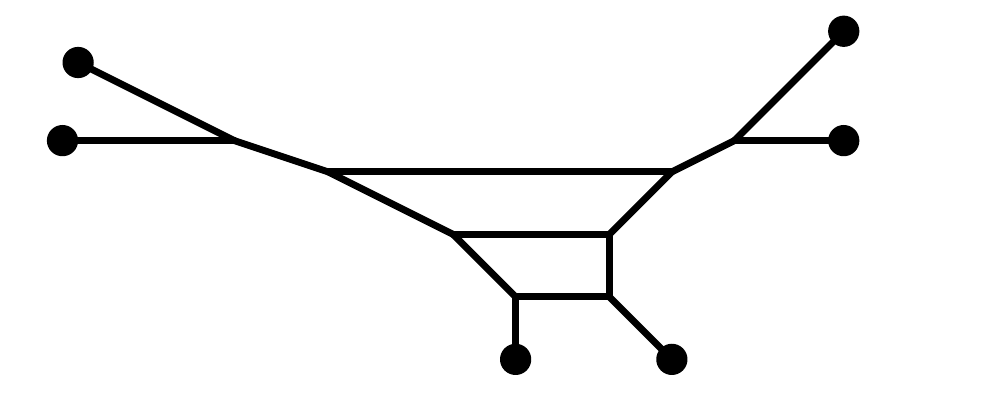}
\caption{$SU(3)_{1} + 2 \mathbf{F}$}
\label{Fig:SU3-2F-1}
\end{minipage}
%\end{figure}
%%
%\begin{figure}
\begin{minipage}{0.45\hsize}
\centering
\includegraphics[width=4cm]{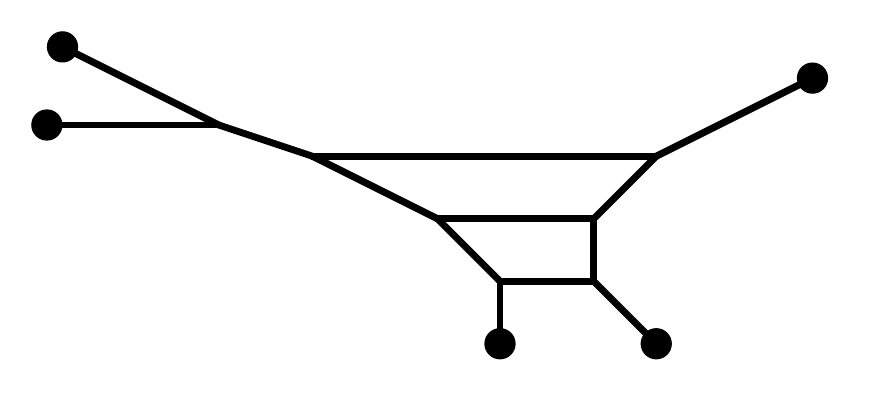}
\caption{$SU(3)_{\frac{3}{2}} + 1 \mathbf{F}$}
\label{Fig:SU3-1F-3/2}
\end{minipage}
\end{figure}
\begin{figure}%[H]
\begin{minipage}{0.45\hsize}
\centering
\includegraphics[width=4cm]{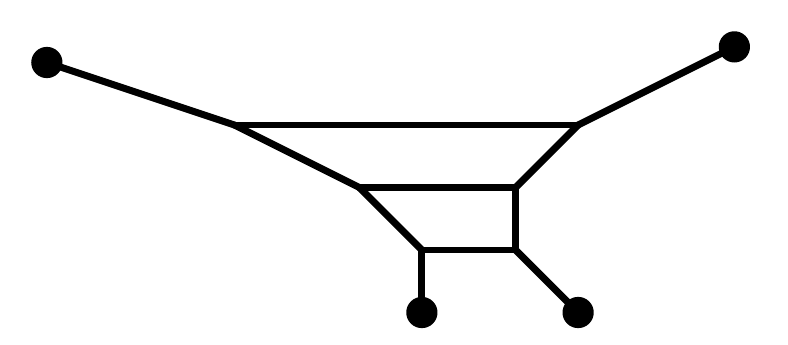}
\caption{Pure $SU(3)_2$}
\label{Fig:SU3-2}
\end{minipage}
%\end{figure}
%%
%\begin{figure}
\begin{minipage}{0.45\hsize}
\centering
\includegraphics[width=4cm]{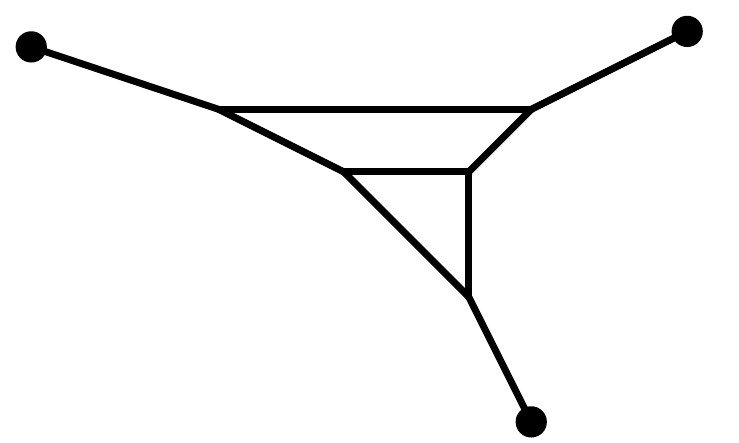}
\caption{$\mathbb{F}_3 \cup \mathbb{P}^2$}
\label{Fig:F3-P2}
\end{minipage}
\end{figure}
\begin{figure}%[H]
\begin{minipage}{0.45\hsize}
\centering
\includegraphics[width=4cm]{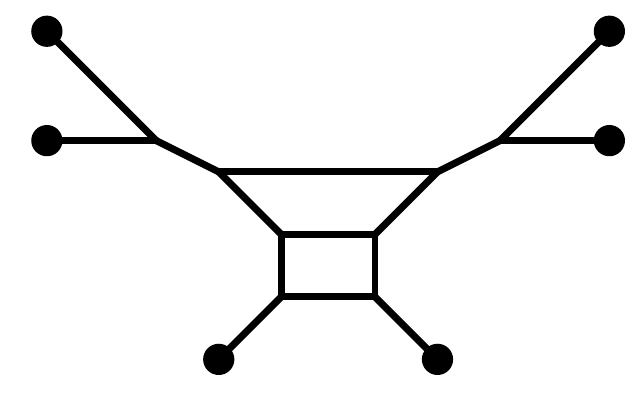}
\caption{$SU(3)_{1} + 2 \mathbf{F}$}
\label{Fig:SU3-2F-0}
\end{minipage}
%\end{figure}
%%
%\begin{figure}
\begin{minipage}{0.45\hsize}
\centering
\includegraphics[width=4cm]{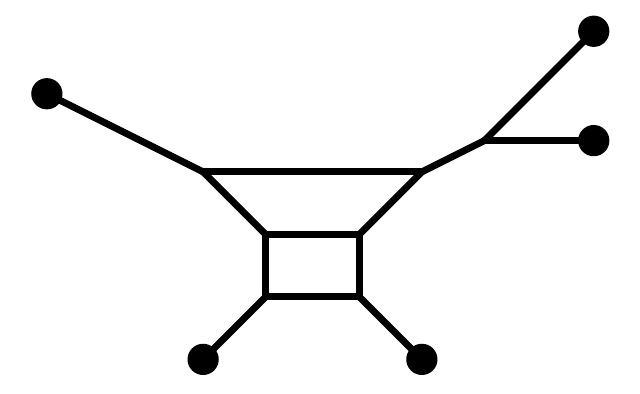}
\caption{$SU(3)_{\frac{1}{2}} + 1 \mathbf{F}$}
\label{Fig:SU3-1F-1/2}
\end{minipage}
\end{figure}
\begin{figure}%[H]
\begin{minipage}{0.45\hsize}
\centering
\includegraphics[width=4cm]{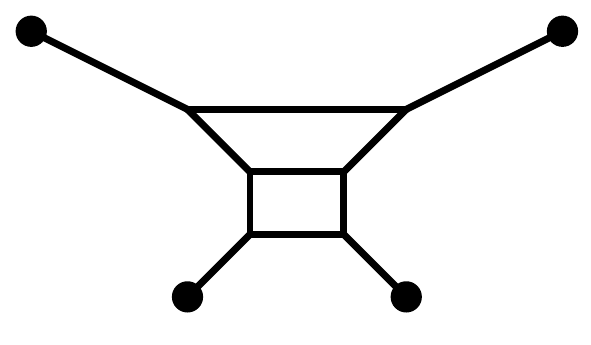}
\caption{Pure $SU(3)_1$}
\label{Fig:SU3-1}
\end{minipage}
%\end{figure}
%%
%\begin{figure}
\begin{minipage}{0.45\hsize}
\centering
\includegraphics[width=4cm]{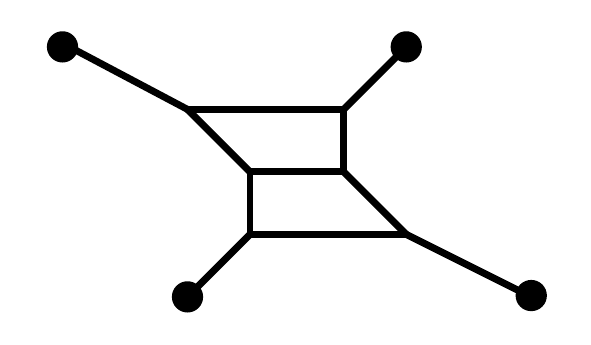}
\caption{Pure $SU(3)_0$}
\label{Fig:SU3-0}
\end{minipage}
\end{figure}
\begin{figure}
\begin{minipage}{0.45\hsize}
\centering
\includegraphics[width=4cm]{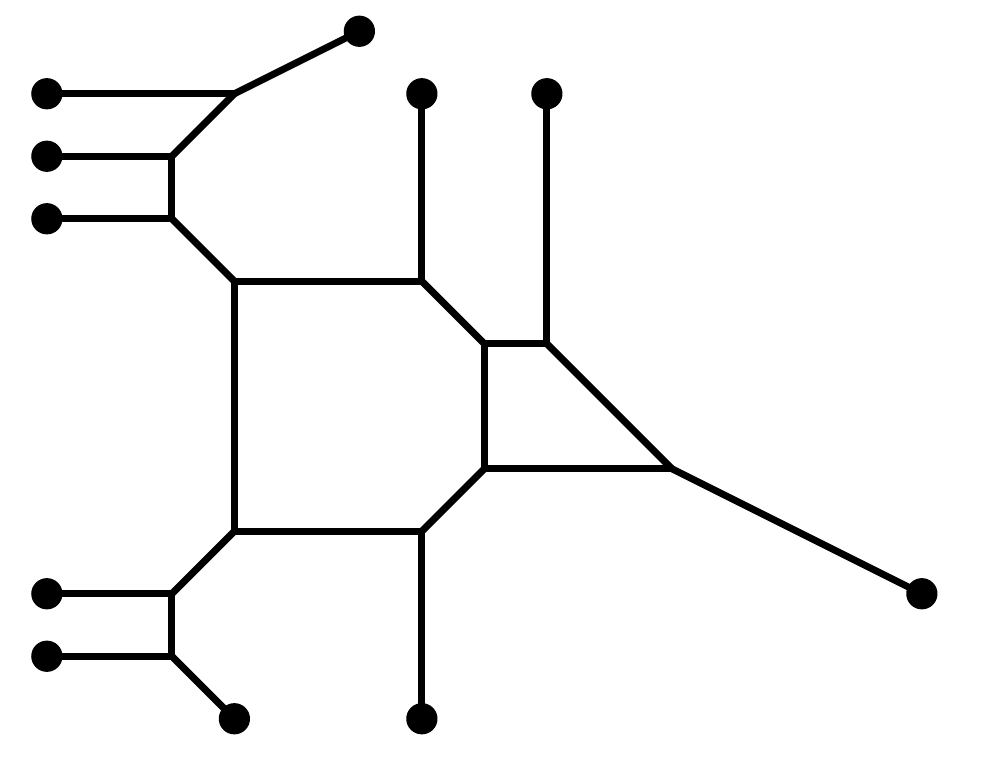}
\caption{$[SU(2)+5\bF]-SU(2)_0$}
\label{Fig:SU2-SU2-5F-0}
\end{minipage}
%\end{figure}
%%
%\begin{figure}
\begin{minipage}{0.45\hsize}
\centering
\includegraphics[width=4cm]{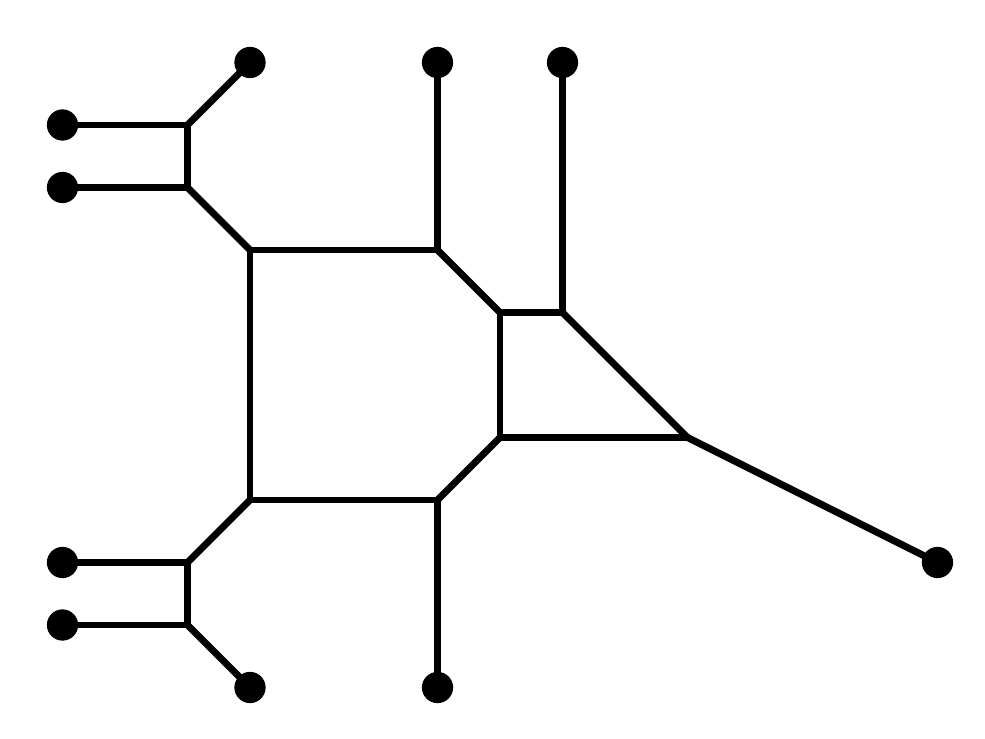}
\caption{$[SU(2)+4\bF]-SU(2)_0$}
\label{Fig:SU2-SU2-4F-0}
\end{minipage}
\end{figure}
\begin{figure}
\begin{minipage}{0.45\hsize}
\centering
\includegraphics[width=4cm]{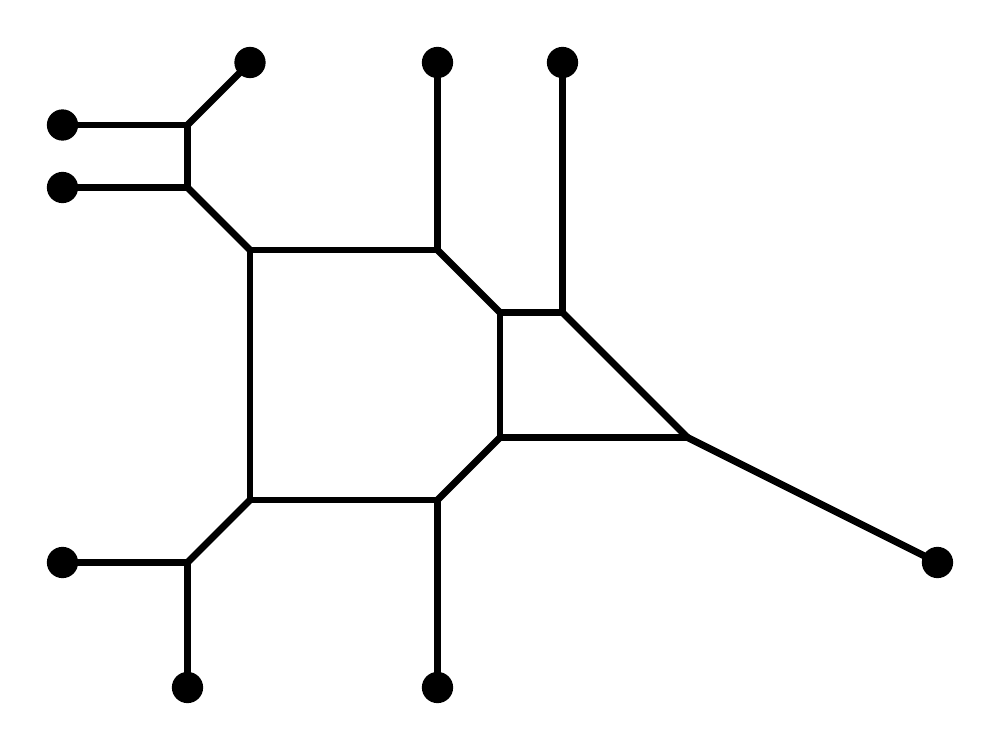}
\caption{$[SU(2)+3\bF]-SU(2)_0$}
\label{Fig:SU2-SU2-3F-0}
\end{minipage}
%\end{figure}
%%
%\begin{figure}
\begin{minipage}{0.45\hsize}
\centering
\includegraphics[width=4cm]{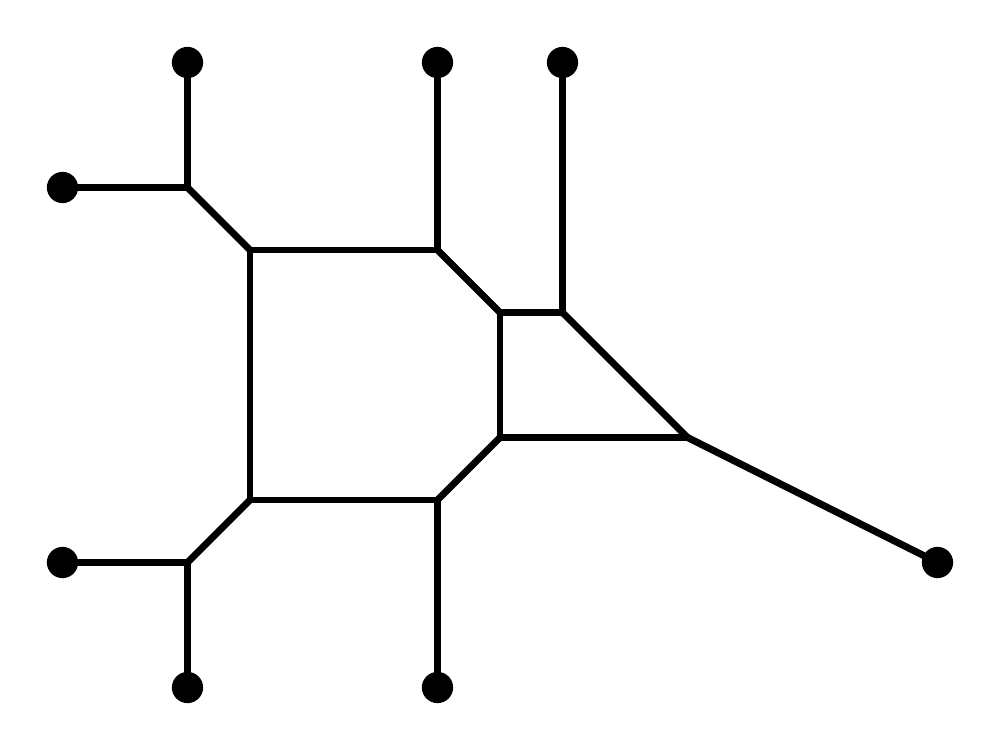}
\caption{$[SU(2)+2\bF]-SU(2)_0$}
\label{Fig:SU2-SU2-2F-0}
\end{minipage}
\end{figure}
\begin{figure}
\begin{minipage}{0.45\hsize}
\centering
\includegraphics[width=4cm]{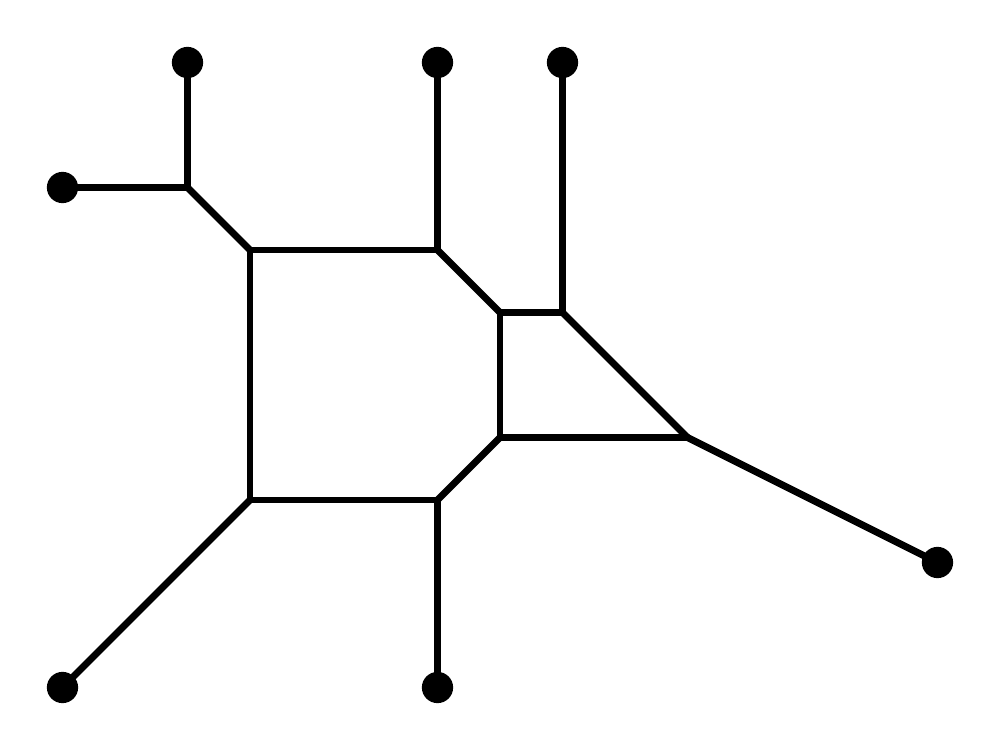}
\caption{$[SU(2)+1\bF] \times SU(2)_0$}
\label{Fig:SU2-SU2-1F-0}
\end{minipage}
%\end{figure}
%%
%\begin{figure}
\begin{minipage}{0.45\hsize}
\centering
\includegraphics[width=4cm]{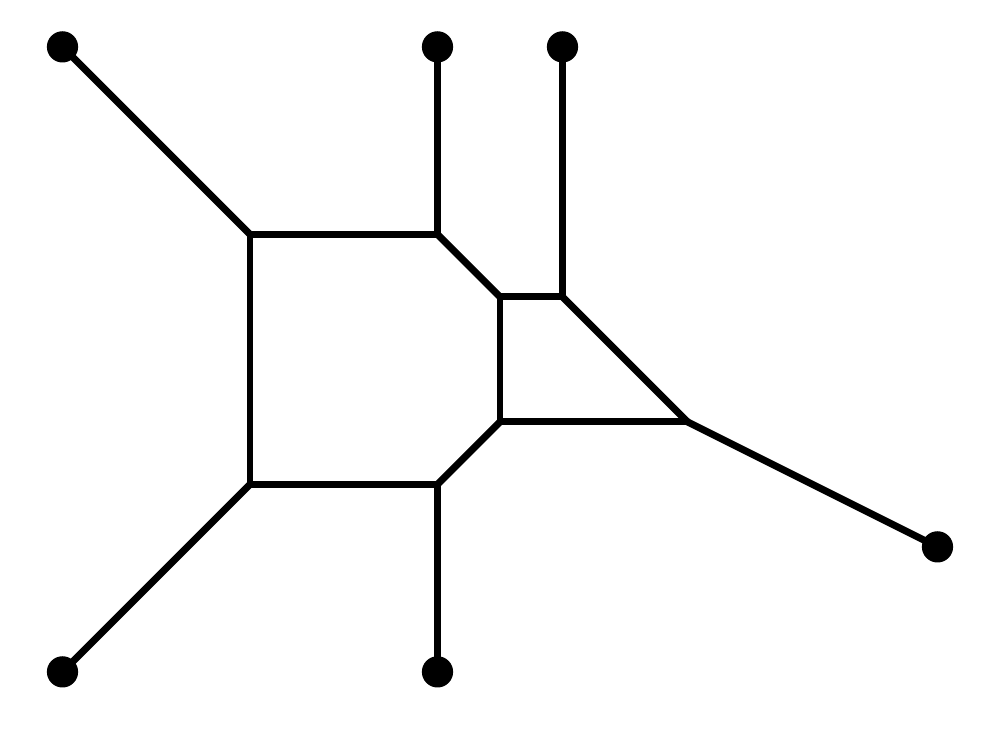}
\caption{$SU(2)_{\pi}\times SU(2)_{0}$}
\label{Fig:SU2-SU2-00}
\end{minipage}
\end{figure}
\begin{figure}%[H]
\begin{minipage}{0.45\hsize}
\centering
\includegraphics[width=4cm]{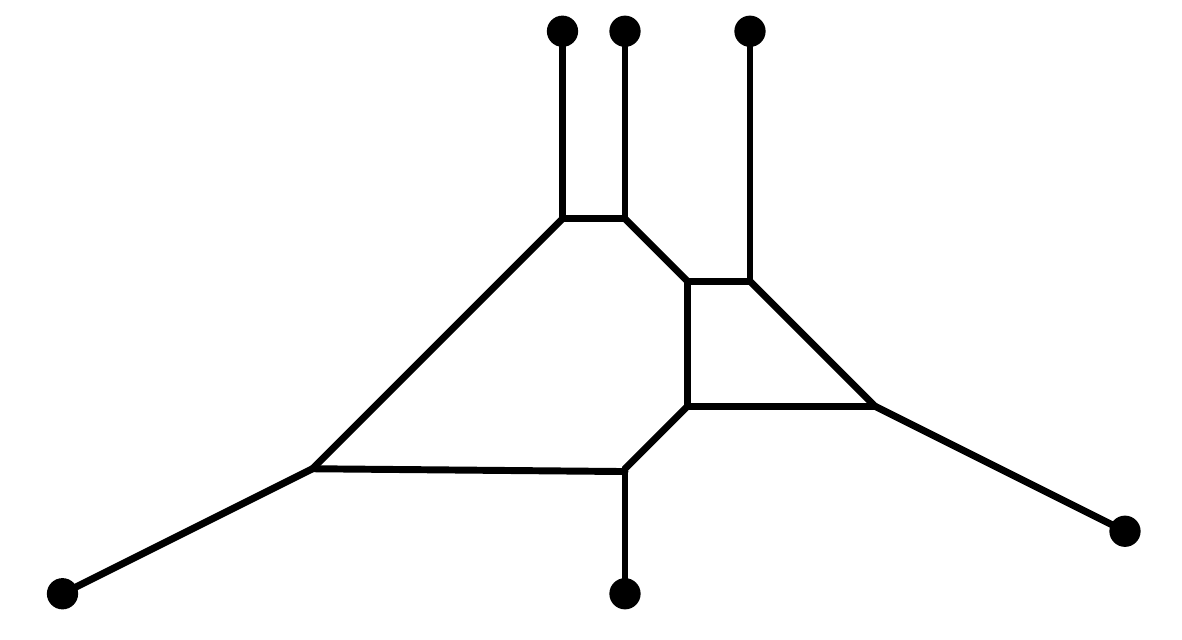}
\caption{$SU(2)_0\times SU(2)_0$}
\label{Fig:SU2-SU2-pi0}
\end{minipage}
%\end{figure}
%%
%\begin{figure}
\begin{minipage}{0.45\hsize}
\centering
\includegraphics[width=4cm]{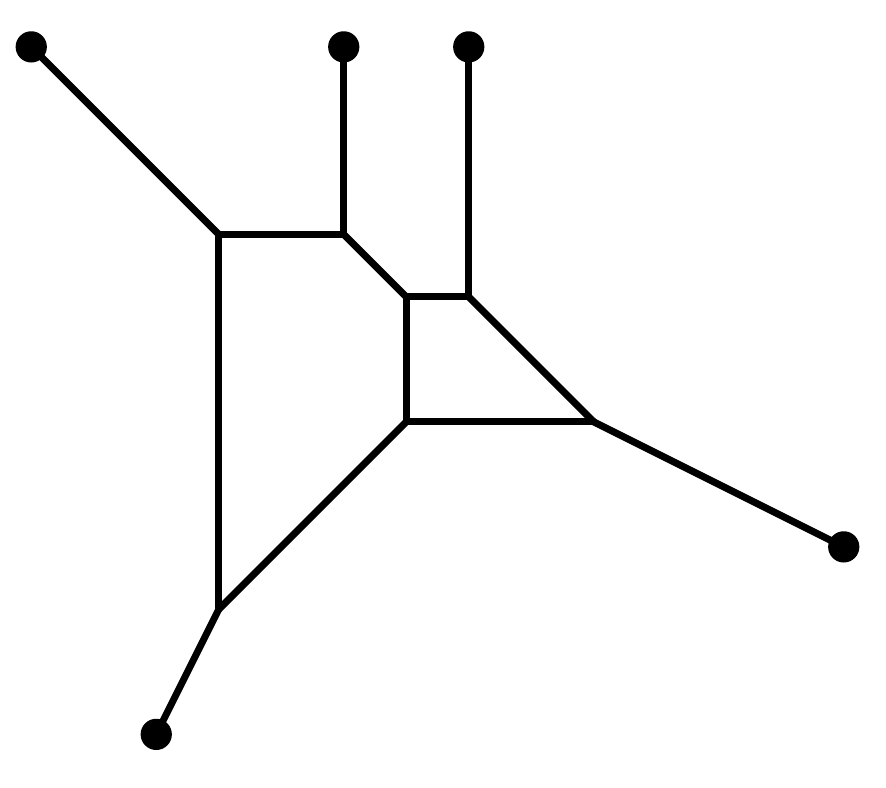}
\caption{$\mathbb{F}_1 \cup$ dP$_2$}
\label{Fig:F1-dP2}
\end{minipage}
\end{figure}
\begin{figure}%[H]
\begin{minipage}{0.45\hsize}
\centering
\includegraphics[width=4cm]{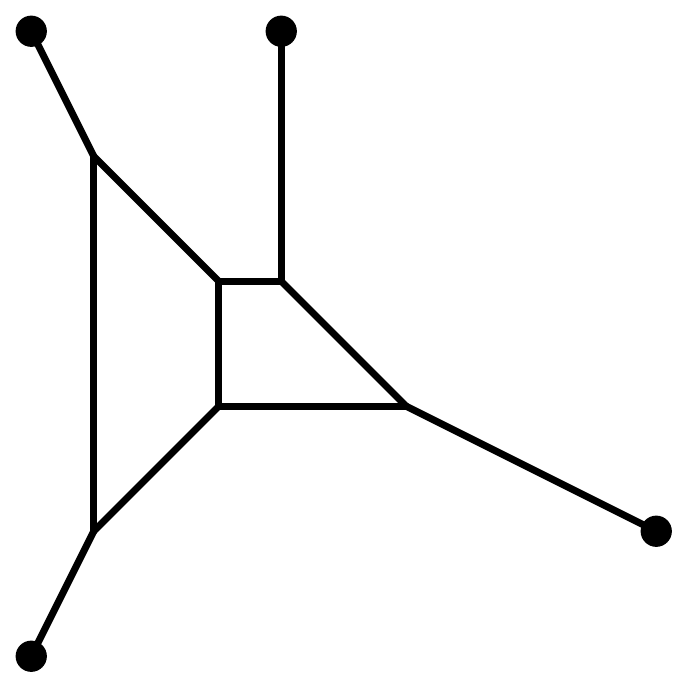}
\caption{$\mathbb{F}_2 \cup$ dP$_1$}
\label{Fig:F2-dP1}
\end{minipage}
%\end{figure}
%%
%\begin{figure}
\begin{minipage}{0.45\hsize}
\centering
\includegraphics[width=4cm]{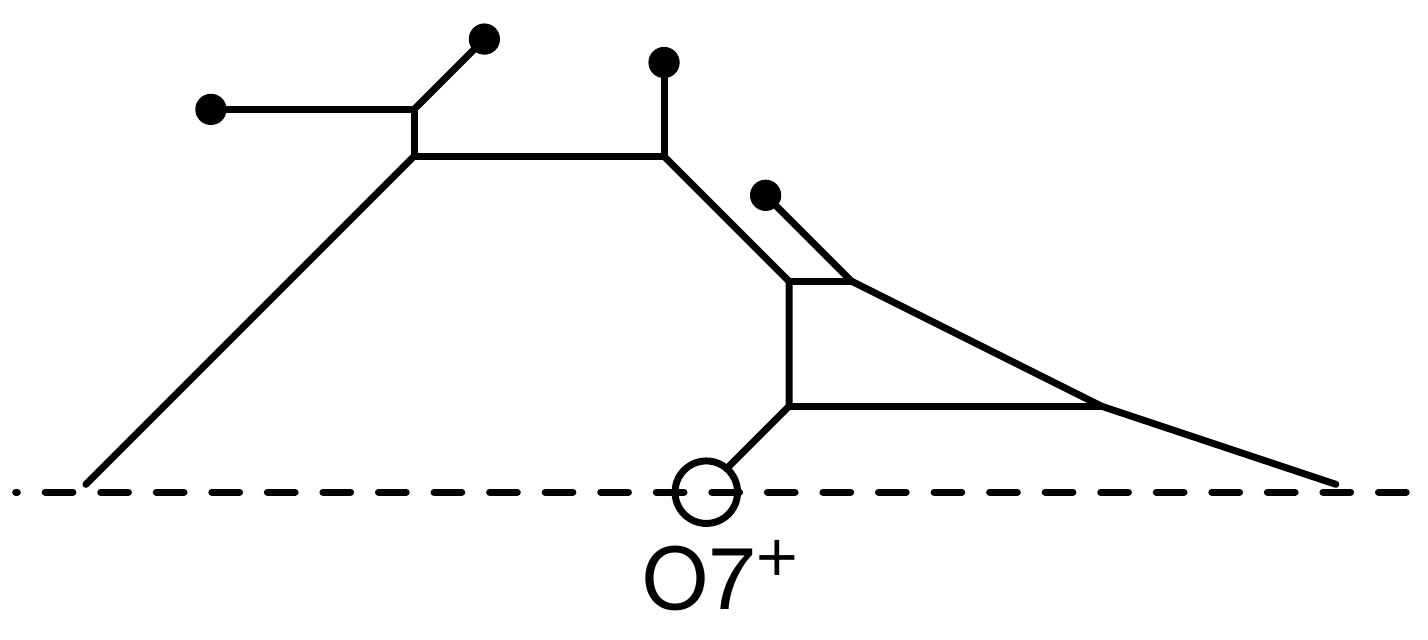}
\caption{$SU(3)_{0} + 1 \mathbf{Sym} + 1 \mathbf{F}$}
\label{Fig:SU3-1Sym-1F}
\end{minipage}
\end{figure}
\begin{figure}%[H]
\begin{minipage}{0.45\hsize}
\centering
\includegraphics[width=4cm]{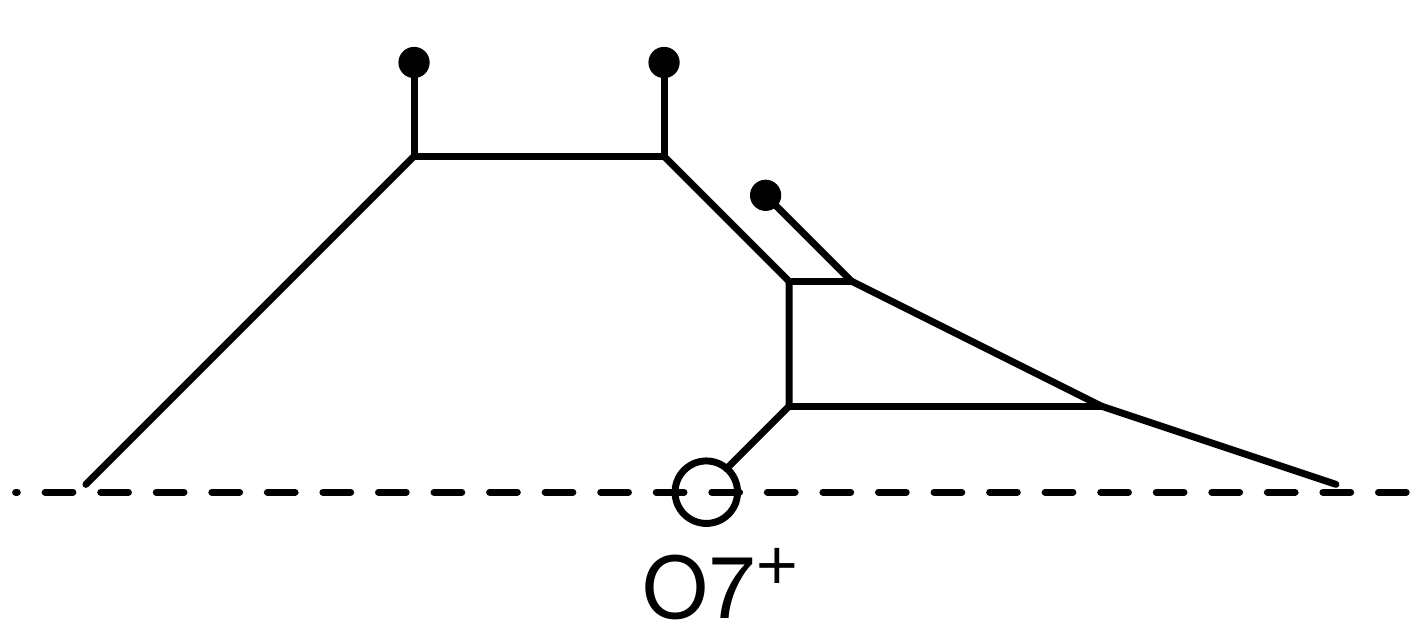}
\caption{$SU(3)_{\frac{1}{2}} + 1 \mathbf{Sym}$}
\label{Fig:SU3-1Sym}
\end{minipage}
%\end{figure}
%%
%\begin{figure}
\begin{minipage}{0.45\hsize}
\centering
\includegraphics[width=5.5cm]{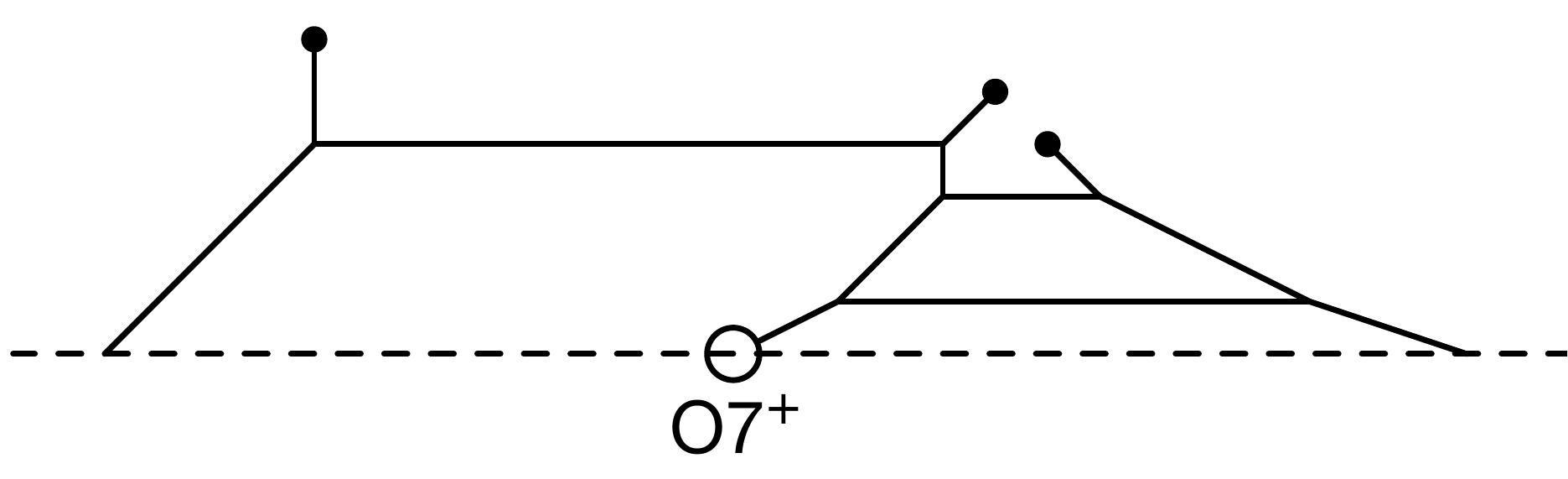}
\caption{$SU(3)_{\frac{3}{2}} + 1 \mathbf{Sym}$}
\label{Fig:SU3-1Sym-3/2}
\end{minipage}
\end{figure}
\begin{figure}
\centering
\includegraphics[width=4cm]{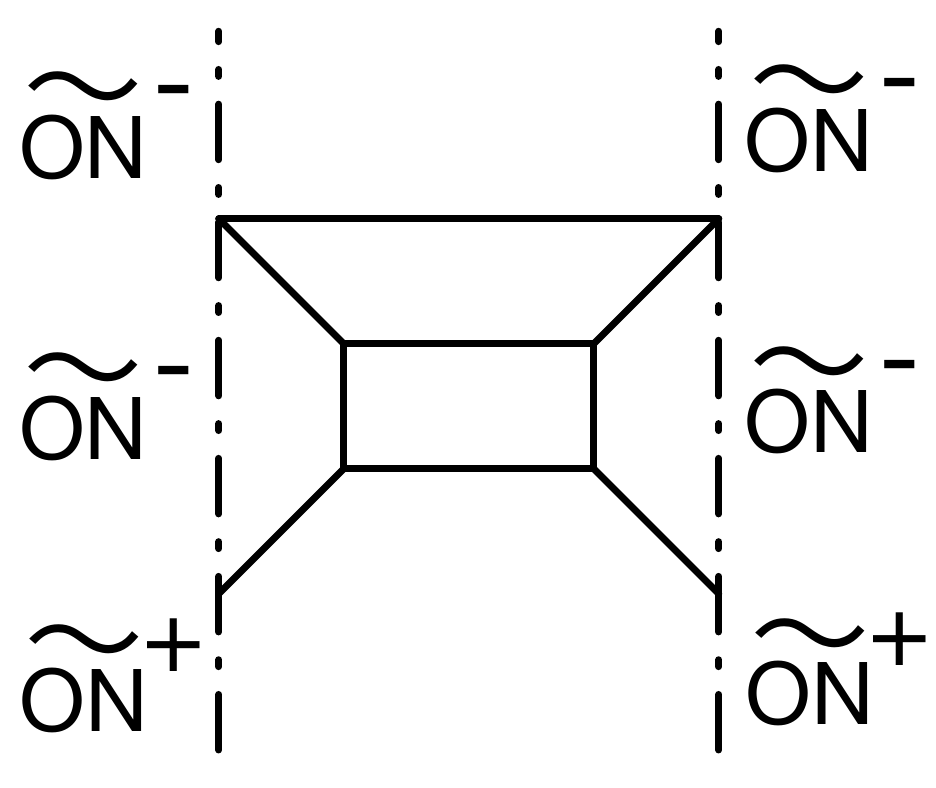}
\caption{Pure $SU(3)_9$}
\label{Fig:SU3-9}
\end{figure}